\RequirePackage[T1]{fontenc}
\RequirePackage[utf8]{inputenc} 
\RequirePackage[dvipsnames]{xcolor}
\RequirePackage[hyphens]{url}
\documentclass[12pt,a4paper,british]{article} 
\pdfoutput=1

\newif\ifshowkeys
\showkeysfalse

\ifshowkeys
   \usepackage[notref,notcite,color]{showkeys}  
\fi

\newif\ifshowmarginalia
\showmarginaliatrue 

\definecolor{refkey}{rgb}{0,0,1}
\definecolor{labelkey}{rgb}{1,0,0}
\usepackage{graphicx}
\usepackage[export]{adjustbox}
\usepackage{lineno,enumerate} 

\usepackage[top=80pt,bottom=85pt,left=70pt,right=60pt]{geometry} 
\usepackage[utf8]{inputenc} 
\usepackage[T1]{fontenc}

\usepackage{epsfig}
\usepackage{amssymb}
\usepackage{amsfonts}
\usepackage{amsbsy}
\usepackage[all]{xy}
\usepackage{amsmath}
\usepackage{amsthm}
\usepackage{float}
\usepackage{accents}
\usepackage{empheq}
\usepackage{bm}
\usepackage{tikz-cd}
\usetikzlibrary{matrix,intersections}
\pgfdeclarelayer{fg}
\pgfdeclarelayer{crossing over}
\pgfsetlayers{main,crossing over,fg}
\usepackage{eso-pic} 
\usepackage[export]{adjustbox}
\usepackage{soul}
\usepackage{mathrsfs}
\usepackage{bbm}
\usepackage{slashed}
\usepackage{stmaryrd}
\makeatletter
\newcommand{\xMapsto}[2][]{\ext@arrow 0599{\Mapstofill@}{#1}{#2}}
\def\Mapstofill@{\arrowfill@{\Mapstochar\Relbar}\Relbar\Rightarrow}
\makeatother

\def\be{ \begin{equation} }
\def\ee{ \end{equation}}

\newtheorem{theorem}{Theorem}
\numberwithin{theorem}{section}
\newtheorem{definition}{Definition}
\numberwithin{definition}{section}
\newtheorem{corollary}{Corollary}
\theoremstyle{definition}
\newtheorem{remark}{Remark}[section]

\AtBeginDocument{%
}

\makeatletter
\DeclareRobustCommand{\tightfootnote}[1]{\unskip\footnote{#1}}
\makeatother

\numberwithin{equation}{section}

\makeatletter
\newcommand*\bigbullet{\mathpalette\bigbullet@{1.8}}
\newcommand*\bigbullet@[2]{\mathbin{\vcenter{\hbox{\scalebox{#2}{$\m@th#1\bullet$}}}}}
\makeatother

\ifshowmarginalia
 \newcommand{\cg}[1]{\marginpar{\raggedright \tiny \rem #1 \rem}}

 \newcommand{\cvs}[1]{\marginpar{\raggedright \tiny \textcolor{blue}{\rem VS: #1 \rem}}}    

\else
  \newcommand{\cg}[1]{}
  \newcommand{\cvs}[1]{}
 \fi

\def\be{ \begin{equation} }
\def\ee{ \end{equation}}

\def\rem{$\clubsuit$}

\def\Aut{\mathsf{Aut}}
\def\Bun{\mathsf{Bun}}

\def\Det{{\mathsf{Det}}}
\def\det{{\mathsf{det}}}
\def\dim{\mathsf{dim}}
\def\End{\mathsf{End}}
\def\exp{{\rm exp}}
\def\Ext{\mathsf{Ext}}
\def\Tor{\mathsf{Tor}}
\def\Tors{\mathsf{Tors}}
\def\Free{\mathsf{Free}}

\def\Hom{\mathsf{Hom}}

\def\im{\mathsf{im}}

\def\ker{\mathsf{ker}}
\def\log{{\rm log}}

\def\tr{{\mathsf{Tr\,}}} 
\def\Tr{{\mathsf{Tr\,}}}

\def\vol{\mathsf{vol}}

\def\half{\frac{1}{2}}
\def\p{\partial}

\def\rem{$\clubsuit$}

\def\rem{$\clubsuit$}

\def\one{{\hbox{ 1\kern-.8mm l}}}

\def\CA{{\cal A}}
\def\CB{{\cal B}}
\def\CC {{\cal C}}
\def\CD {{\cal D}}
\def\CE {{\cal E}}
\def\CF {{\cal F}}
\def\CG {{\cal G}}
\def\CH {{\cal H}}
\def\CI {{\cal I}}

\def\CL {{\cal L}}

\def\CN {{\cal N}}
\def\CO {{\cal O}}
\def\CP {{\cal P}}
\def\CR {{\cal R}}
\def\CV {{\cal V}}
\def\CW {{\cal W}}
\def\CX {{\cal X}}
\def\CO {{\cal O}}
\def\CZ {{\cal Z}}
\def\CE {{\cal E}}
\def\CG {{\cal G}}
\def\CH {{\cal H}}
\def\CI {{{\cal I}}}
\def\CB {{\cal B}}
\def\CQ {{\cal Q}}
\def\CS {{\cal S}}
\def\CT {{\cal T}}
\def\CU {{\cal U}}
\def\CX {{\cal X}}

\def\CZ{{\cal Z}}

\def\IA{\mathbb{A}}

\def\IC{\mathbb{C}}

\def\IF{\mathbb{F}}

\def\IM{\mathbb{M}}

\def\IP{\mathbb{P}}
\def\IQ{\mathbb{Q}}
\def\IR{{\mathbb{R}}}
\def\IS{{\mathbb{S}}}
\def\IT{{\mathbb{T}}}

\def\IZ{{\mathbb{Z}}}

\def\rmk#1{\bigskip\noindent{\bf Remarks} }

\usepackage[hyphens]{url} 
\usepackage[most]{tcolorbox}

\newtcolorbox[auto counter,number within=section]{redbox}[2][]{
colframe=red!60!black,
enhanced,
breakable,
fonttitle=\bfseries,
title={\redboxicon Box \thetcbcounter: {#2}},
label=#1
}

\newtcolorbox[auto counter]{cbox}{
colframe=green!75!black,
enhanced,
breakable,
fonttitle=\bfseries,
title={\bexclaim Remark}
}

\newtcolorbox[auto counter,number within=section]{exbox}[2][]{
colframe=gray!95!black,
enhanced,
breakable,
fonttitle=\bfseries,
title={\exerciseicon Exercise \thetcbcounter: {#2}},
label=#1
}

\newtcolorbox[auto counter]{rbox}[1]{
colframe=red!75!black,
enhanced,
breakable,
fonttitle=\bfseries,
title={#1}
}

\newcommand{\eqa}[1]{\begin{alignat}{4}#1\end{alignat}} 
\newcommand{\beqa}[1]{\begin{empheq}[box=\fbox]{alignat=4}#1\end{empheq}}
\newcommand{\eqas}[1]{\begin{equation}\begin{alignedat}{3}#1\end{alignedat}\end{equation}}

\newcommand{\Met}{\mathsf{Met}}
\newcommand{\Riem}{\mathsf{Riem}}
\newcommand{\Diff}{\mathsf{Diff}}

\newcommand{\Sq}{\mathsf{Sq}}

\renewcommand\thefootnote{\textcolor{red}{\arabic{footnote}}} 
\newcommand*{\dt}[1]{%
  \accentset{\mbox{\large\bfseries .}}{#1}}

\DeclareUnicodeCharacter{0393}{\Gamma}
\DeclareUnicodeCharacter{0394}{\Delta}
\DeclareUnicodeCharacter{0398}{\Theta}
\DeclareUnicodeCharacter{039B}{\Lambda}
\DeclareUnicodeCharacter{039E}{\Xi}
\DeclareUnicodeCharacter{03A0}{\Pi}
\DeclareUnicodeCharacter{03A3}{\Sigma}
\DeclareUnicodeCharacter{03A5}{\Upsilon}
\DeclareUnicodeCharacter{03A6}{\Phi}
\DeclareUnicodeCharacter{03A8}{\Psi}
\DeclareUnicodeCharacter{03A9}{\Omega}
\DeclareUnicodeCharacter{03B1}{\alpha}
\DeclareUnicodeCharacter{03B2}{\beta}
\DeclareUnicodeCharacter{03B3}{\gamma}
\DeclareUnicodeCharacter{03B4}{\delta}
\DeclareUnicodeCharacter{03B5}{\epsilon}
\DeclareUnicodeCharacter{03B6}{\zeta}
\DeclareUnicodeCharacter{03B7}{\eta}
\DeclareUnicodeCharacter{03B8}{\theta}
\DeclareUnicodeCharacter{03D1}{\vartheta}
\DeclareUnicodeCharacter{03B9}{\iota}
\DeclareUnicodeCharacter{03BA}{\kappa}
\DeclareUnicodeCharacter{03BB}{\lambda}
\DeclareUnicodeCharacter{03BC}{\mu}
\DeclareUnicodeCharacter{03BD}{\nu}
\DeclareUnicodeCharacter{03BE}{\xi}
\DeclareUnicodeCharacter{03C0}{\pi}
\DeclareUnicodeCharacter{03C1}{\rho}
\DeclareUnicodeCharacter{03C3}{\sigma} 
\DeclareUnicodeCharacter{03C4}{\tau}
\DeclareUnicodeCharacter{03C5}{\upsilon}
\DeclareUnicodeCharacter{03C6}{\phi}
\DeclareUnicodeCharacter{03D5}{\varphi}
\DeclareUnicodeCharacter{03C7}{\chi}
\DeclareUnicodeCharacter{03C8}{\psi}
\DeclareUnicodeCharacter{03C9}{\omega}
\DeclareUnicodeCharacter{21D0}{\Leftarrow}
\DeclareUnicodeCharacter{0212B}{\AA}
\DeclareUnicodeCharacter{00B7}{\cdot}
\DeclareUnicodeCharacter{00B0}{^{\circ}}
\DeclareUnicodeCharacter{266A}{\eighthnote}
\DeclareUnicodeCharacter{266B}{\twonotes}
\DeclareUnicodeCharacter{2202}{\partial}

\newcommand{\nn}{\nonumber\\}

\newcommand{\n}{\bm{\nabla}}

\newcommand{\sfA}{\mathsf{A}}
\newcommand{\sfB}{\mathsf{B}}
\newcommand{\sfC}{\mathsf{C}}
\newcommand{\sfE}{\mathsf{E}}
\newcommand{\sfF}{\mathsf{F}}
\newcommand{\sfG}{\mathsf{G}}
\newcommand{\sfJ}{\mathsf{J}}

\newcommand{\wt}{\widetilde}
\newcommand{\wh}{\widehat }

\usepackage[normalem]{ulem}
\usepackage{cancel}
\usepackage[font={small,singlespacing}]{caption}
\usepackage{setspace}
\usepackage{stackrel}

\newcommand{\ov}[1]{\overline{#1}}

\usepackage{multirow}

\usepackage[most]{tcolorbox}
\usepackage{framed}
\definecolor{shadecolor}{gray}{0.9}

\usepackage{cancel}

\usepackage{listings}

\usepackage{fontawesome5}
\newcommand{\bexclaim}{\textcolor{blue}{\faExclamationCircle}\,\,}

\newcommand{\redboxicon}{\textcolor{Green}{\faBookReader}\,\,}
\newcommand{\exerciseicon}{\textcolor{blue}{\faDumbbell}\,\,}

\usepackage[bottom,hang,flushmargin]{footmisc} 

\usetikzlibrary{shapes.geometric,tqft}
\usetikzlibrary{decorations.markings,decorations.text,decorations.pathmorphing}
\usepackage{tqft}

 \usetikzlibrary{arrows.meta,bending,positioning} 

\tikzset{
  tqft/use nodes=false,
}

\tikzset{
    marrow/.style={decoration={markings,mark=at position 0.5 with {\arrow{#1}}}, postaction=decorate}
}

\usepackage{subcaption}

\usepackage{enumitem}

\def\imag{\mathsf{i}}

\DeclareFontFamily{U}{mathx}{}
\DeclareFontShape{U}{mathx}{m}{n}{<-> mathx10}{}
\DeclareSymbolFont{mathx}{U}{mathx}{m}{n}
\DeclareMathAccent{\widecheck}{0}{mathx}{"71}

\usepackage[useregional,showdow]{datetime2}

\definecolor{linkc}{rgb}{.8,.15,1}
\usepackage[pagebackref=true,colorlinks=true,linkcolor=linkc,citecolor=linkc,urlcolor=linkc,linktoc=all,pdfusetitle=true]{hyperref}
\renewcommand*{\backref}[1]{}
\renewcommand*{\backrefalt}[4]{{%
		\ifcase #1 
		\or [Cited on: pg.~#2.]%
		\else [Cited on: pgs. #2.]%
		\fi%
	}}

 \usepackage{fancyhdr} 
\renewcommand{\sectionmark}[1]{}

\newcommand{\SectionWithHeader}[3]{%
  \section{#1}\label{#3}
  \markboth{Section \thesection}{#2}
}

\usepackage{aliascnt}
\newcommand{\Q}{_{\IQ}}
\newcommand{\R}{_{\IR}}
\newcommand{\mynewtheorem}[2]{
  \newaliascnt{#1}{equation}
  \newtheorem{#1}[#1]{#2}
  \aliascntresetthe{#1}
  \expandafter\def\csname #1autorefname\endcsname{#2}
}
\theoremstyle{plain}
\mynewtheorem{lemma}{Lemma}
\newcommand\mybigwedge{\raisebox{1pt}{\scalebox{.9}{$\bigwedge$}}}
\DeclareRobustCommand{\mstrut}{^{\vphantom{1*\prime y\vee M}}}
\newcommand{\hneg}{\mkern-.5\thinmuskip}
\newcommand{\plu}[1]{\pi\mstrut _{\hneg#1}}
\newcommand{\tplu}[1]{\widetilde{\pi}\mstrut _{\hneg#1}}



\makeatletter
\newcommand\customfootnote[2]{%
  \begingroup
  \renewcommand\thefootnote{\text{#1}}%
  \footnotemark
  \footnotetext{#2}%
  \endgroup
}
\makeatother

\begin{document}

\begin{titlepage}  

    \vskip .5in 
	\noindent
 
       
    \vskip 1.5in

    \begin{center}
        \textcolor{blue}{\LARGE \bf{TASI LECTURES ON}
        \\ \vspace{-2mm}
        \bf{TOPOLOGICAL FIELD THEORIES}
           \\ \vspace{2.5mm}
           \bf{AND DIFFERENTIAL COHOMOLOGY}}

        \bigskip\medskip

        {Gregory W. Moore$^1$, Vivek Saxena$^{1}$}

        \vspace{2.2mm}
       \textcolor{purple}{{with an \hyperref[App:TorsionAndTorsion]{\textcolor{purple}{Appendix}} by Daniel S. Freed}}

        \bigskip\medskip

        {\small $^1$ New High Energy Theory Center and Department of Physics and Astronomy,\\Rutgers University, 126 Frelinghuysen Rd., Piscataway, NJ 08855-0849, USA 
        }

        \vskip 5mm
        {\small \tt \href{mailto:gwmoore@physics.rutgers.edu}{gwmoore@physics.rutgers.edu}, \href{mailto:vivek.hepth@gmail.com}{vivek.hepth@gmail.com}}
        
        \vskip 9mm

        \hrule height 0.1cm
        
        \vskip 9mm
        {\bf Abstract}
        \vskip .1in
    \end{center}

    \noindent 
    These are lecture notes expanding upon a set of lectures given by G.M. at the TASI 2023 School. Part I is an introduction to topological field theory, including extended topological field theory. Part II is an introduction to generalized Abelian gauge theories and their relation to differential cohomology. \\$\,$\\
    \DTMnow
    \vfill
    \eject
    
\end{titlepage}

\pagestyle{empty}
\tableofcontents

\pagestyle{fancy}



\SectionWithHeader{General Plan}{General Plan}{sec:GeneralPlan}

This is a write-up based on lectures given by one of us at the Theoretical Advanced Study Institute (TASI) 2023 summer school \cite{TASI2023}. The lectures are meant to be very elementary introductions to two distinct (but closely related) subjects, both of which play an important role in modern discussions of generalized notions of symmetry and topological effects in quantum field theory, supergravity, and string/M-theory. Accordingly, the notes are divided into two parts. The first part addresses topological field theory. This was the topic of the first two TASI lectures. The videos are at  \cite{Moore:TASIVideoLec1,Moore:TASIVideoLec2}.
The second part concerns generalized Abelian gauge theories and differential cohomology. This was the topic of the last two TASI lectures. The videos are at \cite{Moore:TASIVideoLec3,Moore:TASIVideoLec4}.

We comment here on some other pedagogical sources for this material. 
For the part on topological field theory, some older reviews include   \cite{Thompson:1991qt,Birmingham:1991ty}. At some points, we will closely follow the review   \cite{Moore:2006dw} and the closely related chapter in  \cite{Aspinwall:2009isa}. Three other useful reviews are   \cite{Bartlett:2005qy,Davydov:2011kb,Carqueville:2017fmn,Carqueville:2023jhb}.
 A recent set of lectures, the Simons Lectures On Categorical Symmetry \cite{Costa:2024wks}, has much useful material.

For part two on differential cohomology, a key reference is the original paper  
by J. Cheeger and J. Simons \cite{CheegerSimons:1985}, which still makes for good reading. An important foundational paper by M. Hopkins and I. Singer is \cite{Hopkins:2002rd}. Pedagogical material and reviews can be found in  \cite{Freed:2000ta,Freed:2006ya,Freed:2006yc,Moore:2011SCGP,Szabo:2012hc,Cordova:2019jnf,Debray:2023kvh}. There are two books on the subject:  \cite{Amabel:2021wbk} and \cite{BarBeckerBook}.    
Some other references with summaries or alternative viewpoints 
are \cite[App. A]{Freed:2021anp}, \cite[Sec. 4]{Freed:2024heu}, \cite{Bunke:2012rsi,Schreiber:2013pra,BunkeNikolausVolkl:2013} and \cite[Sec. 2]{Hsieh:2020jpj}. 

We supplement the text with several exercises for the reader. These exercises can be found in gray bounded boxes labeled by the rather suggestive weightlifting icon (\exerciseicon\!\!). Numerous important facts appear in red bounded boxes labeled by the book icon (\redboxicon\!\!) -- a green signal for some reflection.

\section*{Acknowledgments}
G.M. would like to thank the organizers I. Bah, T. DeGrand, O. DeWolfe, K. Intriligator, E. Neil, and S.-H. Shao, of the TASI 2023 School, for the opportunity to give the TASI lectures. He would also like to thank the participants for asking lots of good questions during the lectures. 
We thank T. Banks, D. Berwick-Evans, A. Cattaneo, 
I. Garc{\'i}a Etxebarria, D. Freed, D. Friedan, P. Gorantla, D. Harlow, M. Hopkins, T. Johnson-Freyd, C. LeBrun, J. Maldacena, G. Segal, N. Seiberg, S.-H. Shao, S. Stolz, C. Teleman, and E. Witten for useful remarks and discussions on various aspects of the content of these lectures. G.M. particularly thanks D. Freed for numerous suggestions while these notes were being written. We thank R. K. Singh for comments on 
the draft. 
The work of  G.M. and V.S.  was supported by the US Department of Energy under grant DE-SC0010008. The work of V.S. was supported in part by NSF grant PHY-2210533. G.M. thanks the Institute for Advanced Study for hospitality while much of this manuscript was written. In particular, G.M. was supported by the IBM Einstein Fellow Fund. V.S. gratefully acknowledges the support and hospitality of the C.N. Yang Institute for Theoretical Physics and the Simons Center for Geometry and Physics during the final stages of this manuscript.
$\,$\\$\,$\\
\centerline{\emph{No part of this paper was written by AI.}}


\SectionWithHeader{Preliminary: Quantum Mechanics}{Preliminary: Quantum Mechanics}{sec:QM}

Quantum mechanics is based on a pairing of states and observables known as the Born rule. 
 Physical states $\rho$ are represented by traceclass positive operators $\rho$ on a $\IC$-Hilbert space $\CH$ with trace $\mathsf{tr}(\rho) = \mathbbm{1}$. They are often referred to as density matrices. Physical observables $\CO$ are self-adjoint operators on $\CH$.
 The Born rule is a pairing of states and observables valued in probability distributions on $\IR$. To define it, we need to recall the spectral theorem for self-adjoint operators \cite{Halmos1963,ReedSimon:1972,Rudin1990-od}. 

The spectral theorem asserts that for a self-adjoint operator $\CO$, there is an associated projection-valued measure $P_{\CO}$.
If $\CO$ has a discrete spectrum of  eigenvalues $\{\lambda_i\}$, then for a Borel-measurable set   $E\subset \IR$, we have
\eqa{
  P_{\CO}(E) &= \sum_{\lambda_i \in E} P(\lambda_i) ~,
}
where $P(\lambda_i)$ is the projector onto the eigenspace for $\lambda_i$.
The Born Rule can then be stated as: 
\eqa{
  p_{\rho,\CO}(E) &= \mathsf{Tr}_{\CH}\big( P_{\CO}(E)\rho\big) ~. \label{eq:BornRule}
}

When discussing physical states in quantum mechanics, attention is often focused on the 
pure states. These are extremal points of the convex set of physical states and are given by rank-one projection operators. For this reason, states are often heuristically identified with vectors in Hilbert space. However, this identification should be used with caution, and can lead to confusion. 
For more about the relation between the projective structure of quantum mechanics and its generalization to field theory, see \cite{Freed:2023snr}. 

Pure states are rank-one projection operators. But such operators are in 1--1 correspondence with one-dimensional subspaces (often called ``lines'')  in Hilbert space. Thus, the set of pure states can be identified with 
  \eqa{
    & \IP\CH &&:= ( \CH - \{0\})/\IC^{\times} ~.
  }
Now, a \underline{non-zero} vector $\psi \in \CH$ determines a line via: 
  \eqa{
    L_{\psi} := \{ z \psi ~|~ z \in \IC  \} \subset \CH ~ . 
  }
This line depends neither on the normalization nor the phase of $\psi$, i.e., $L_{\psi} = L_{\xi \psi}$ for any nonzero complex number $\xi \in \IC^{\times}$. This is another way of understanding that the space of pure states is the projective Hilbert space. 
 Note that the projective Hilbert space is \underline{not} a vector space. 
Given any line $L \in \CH$, one can define the corresponding rank-one projection operator
onto that line as the unique projection operator. One can write an explicit formula for this projection operator by choosing \underline{any} nonzero vector $\psi \in L$ 
and writing in Dirac's bra-ket notation: 
  \eqa{
    P_{L} &:= \frac{|\psi\rangle\langle\psi|}{\langle\psi|\psi\rangle} ~.
  }
Note that a rescaling $\psi \mapsto \xi \psi$ for $\xi \in \IC^{\times}$ produces the same projector.  

In quantum mechanics textbooks, physical states are often represented by nonzero vectors $\psi\in \CH$. They are taken to be normalized $\langle \psi, \psi \rangle = 1$, which eliminates some of the scaling ambiguity $\psi \mapsto \xi \psi$, but retains the ambiguity of scaling by phases. The equivalence class of normalized vectors under rescaling by a phase is sometimes referred to -- rather confusingly -- as a ``ray in Hilbert space.'' We would discourage this terminology. 

In quantum mechanics,  time evolution is modeled by the action of a coherent family of 
unitary operators $U(t_1, t_2)$ where $U(t_1, t_2)U(t_2,t_3)=U(t_1,t_3)$. 
In physical systems with a time-translation symmetry, $U(t_1,t_2) = U(t_1-t_2)$ 
for a one-parameter family of operators with $U(t) = e^{-\imag  t H}$, 
where $H$ is a self-adjoint operator, the Hamiltonian. The action of time translation is: 
\be 
\rho \mapsto  U(t) \rho U(t)^{-1} ~,
\ee
and, if we represent $\rho$ by a nonzero vector $\psi$ in Hilbert space, 
we have the evolution 
\be\label{eq:TimeTranslation}
\psi \mapsto U(t) \psi ~.
\ee
In Euclidean signature, one ``Wick-rotates'' these operators to $U(t) = e^{-tH}$, where $t$ is real. Topological field theory should be thought of as taking place in Euclidean signature, even when there is no metric dependence and hence no choice of signature. 

Because the action of time translation is linear in \eqref{eq:TimeTranslation}, quantum states can be superposed. The principle that physical states can be linearly combined is a fundamental aspect of quantum information theory and leads to the distinctive quantum phenomenon of entanglement. On the other hand, quantum theory is projective, and projective space is nonlinear, so one might wonder how one describes linear superposition from the projective viewpoint.
\tightfootnote{We thank D. Freed for raising this interesting point and for discussions about it.}
Given two pure states corresponding to  linearly independent lines $\ell_1, \ell_2 \subset \CH$, we obtain a two-dimensional subspace $S\subset \CH$ spanned by the lines. A choice of linear superposition is then simply a choice of a pure state corresponding to a line $\ell$ in $S$. If we choose nonzero vectors $\psi_i\in \ell_i$ (for $i=1,2$), then 
the traditional textbook description of linear superposition says that such a superposition
``is''  a vector $z_1 \psi_1 + z_2 \psi_2$, where $z_i \in \IC$.  The same vector can also 
be represented by $\wt z_1 \psi_1 + \wt z_2 \psi_2$ provided that 
$z_i = \lambda \wt z_i$ for some $\lambda \in \IC^{\times}$. Thus we recover the set of linear combinations as points in $\IC \IP^1$. Put another way, two linearly independent pure states define points $\ell_1, \ell_2 \in \IP(\CH)$. Standard projective geometry tells us that 
two such points determine a line $\IC \IP^1 \subset \IP(\CH)$. This is the line of quantum superpositions. It is also interesting to describe the linear combination of states in a choice-free way in terms of rank-one projection operators. Suppose then that we are given two different rank-one projection operators $P_i$, $i=1,2$.  ($P_i$ are the projectors onto the lines $\ell_i$.) One then notes that 
the equations 
\be\label{eq:OperatorLine}
P_1 Q = Q  ~, \qquad \qquad  \text{ and }   \qquad \qquad  Q P_2 = Q ~,
\ee
define a \underline{one-dimensional subspace} of $\End(\CH)$. The choice of linear 
combination is -- essentially -- a choice of operator $Q$ in this one-dimensional subspace. 
In the case where $P_1,P_2$ are orthogonal projection operators, so $P_1 P_2 = P_2 P_1 = 0$,
we can form a projector of the form 
\be 
P = \alpha P_1 + Q + Q^\dagger + \delta P_2 ~,
\ee
where $0 \leq \alpha, \delta \leq 1$,  provided $\tr(Q Q^\dagger) \leq \frac{1}{4}$. 
Relative to an orthonormal (ON) basis for $\ell_1, \ell_2$, the resulting $\IC\IP^1$ family of projection operators can be written as the Bott projector:
\tightfootnote{So named because this family of projection operators plays a central role in discussions of Bott periodicity in algebraic $K$-theory.}
\be 
P(z) = \frac{1}{1+ \vert z \vert^2} \begin{pmatrix} 1 &  z^* \\ z & \vert z \vert^2 \end{pmatrix} ~,
\ee
and the choice of $Q$ corresponds to the choice of $z\in \IC \cup \{ \infty \}$. 
\tightfootnote{We record here the formula in the case where $P_1 P_2 \not=0$. Again, we choose 
an operator $Q$ in the line determined by equation \eqref{eq:OperatorLine}, and then 
take $P= \alpha_{\pm} P_1 + Q  + Q^\dagger + \alpha_{\mp} P_2$, where, 
$$\alpha_\pm = \half\left( (1- 2~\mathsf{Re}(t)) \pm \sqrt{(1- 2~\mathsf{Re}(t))^2 - 4x }\right) ~,$$
and 
$t = \Tr(Q)$ and $x = \Tr(Q Q^\dagger)$ and $x \leq \frac{1}{4}(1-2~\mathsf{Re}(t))^2$. The reader might be puzzled by the use of both roots $\alpha_\pm$ as the coefficient of $P_1$. Together, they parametrize the entire $\IC\IP^1$ of families corresponding to a ``linear superposition of states''. To recover the case of orthogonal projectors, set $t=0$ and $x = \frac{1}{1+\vert z\vert^2}$ or $x = \frac{\vert z\vert^2}{1+ \vert z \vert^2}$.}

Finally, an important aspect of quantum mechanics concerns how to ``combine'' two physical systems described by operator algebras associated with two Hilbert spaces $\CH_1$ and $\CH_2$. The ``combined'' system has Hilbert space $\CH_1\otimes \CH_2$, and not $\CH_1\oplus \CH_2$. This simple fact will lead to many interesting aspects of topological field theory. More generally, the Hilbert spaces of quantum mechanics should be super-Hilbert spaces, that is, they should be $\IZ_2$-graded. Then one uses the graded tensor product. The $\IZ_2$-gradings are always present when the physical system involves fermions, and conversely, a nontrivial $\IZ_2$-grading is a hallmark of the presence of fermions. In these lectures, for simplicity, we will generally restrict to the ungraded case.

\eject
\thispagestyle{empty}
\phantomsection
\vspace*{\fill}
\addcontentsline{toc}{section}{\hspace{6em}\textcolor{red}{PART I: TOPOLOGICAL FIELD THEORY}}
\begin{center}
\label{part1}
\textcolor{red}{
\LARGE\textbf{PART I: TOPOLOGICAL FIELD THEORY }
}
\end{center}
\vspace*{\fill}
\eject

\SectionWithHeader{Basic Picture In TFT: Heuristic Motivation}{Basic Picture In TFT: Heuristic Motivation}{sec:basicpicture-tft-heuristic}

What is physics? Physics means many things to many people, but 
one common and central goal one will often encounter is the desire to 
describe or predict the future, given a knowledge of the present. 
Of course, in cosmology, one might wish to predict the past, given a knowledge of the present, but in both cases, we are trying to describe some kind of time evolution. In quantum systems, such predictions are based on the computation of amplitudes. In topological field theory, 
one abstracts certain primitive aspects of such amplitudes in a system of axioms and then describes the resulting mathematical structure defined by just those axioms. 

\begin{figure}[H]
  \centering
  \includegraphics[width=3in]{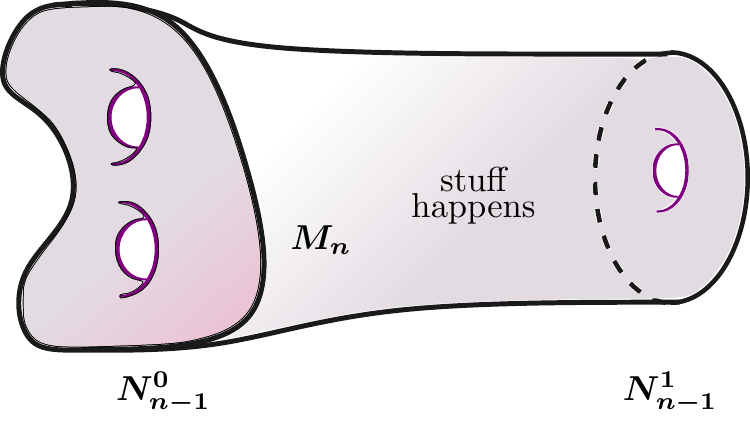}
  \caption{Picture of ``time'' evolution. Note that $M_n$ need not 
  be a cylinder, i.e., diffeomorphic to $N_{n-1} \times [0,1]$, and there might not be a globally well-defined vector field of time. }
  \label{fig:1}
\end{figure}

An important aspect of topological field theory, and indeed much of the literature of the past few decades, is an emphasis on defining theories in spacetimes of different topologies. (This is in contrast with most textbooks on quantum field theory (QFT), which generally assume spacetime to be Minkowski space.) Thus, in a theory defined on $n$-dimensional spacetimes, a spatial manifold would be of dimension $(n-1)$. If we are engaged in prediction, then we might have a past manifold $N_{n-1}^{0}$ and a future manifold $N_{n-1}^{1}$, and given a knowledge of a quantum state in the past, we wish to say something about the quantum state in the future (or vice versa). We should not assume that $N_{n-1} \cong \IR^{n-1}$. Indeed, the past and future spatial manifolds are allowed to have different topologies. There must be some $n$-dimensional manifold that 
``interpolates'' between the past and the future, somewhat as in \autoref{fig:1}.
%
In quantum theory, there is a Hilbert space associated with the initial and final spatial manifolds:
\tightfootnote{The phrase ``Hilbert space of states'' is an abuse of terminology: a state in quantum theory is a density matrix: it is a positive operator of trace one. Given a nonzero vector $|\psi\rangle$ in $\CH$, then $P = \frac{|\psi\rangle\langle\psi|}{\vert\!\vert\psi\vert\!\vert^2}$ is a projection operator, which is a more appropriate descriptor of the state.}
\begin{table}[H]
\centering
\begin{tabular}{ccccl}
$N_{n-1}^{0}$ & $\rightsquigarrow$ & $\CH_0$ & : & \text{Hilbert space from which we describe initial states.}\\
$N_{n-1}^{1}$ & $\rightsquigarrow$ & $\CH_1$ & : &\text{Hilbert space from which we describe final states.}
\end{tabular}
\end{table}
In other words, part of the description of a ``theory'' is to give a 
``function'' on $(n-1)$-dimensional manifolds valued in Hilbert spaces.
\tightfootnote{Our final definition of a topological field theory will be more elastic and allow $F_{\rm states}$ to take values which are not vector spaces.   See \autoref{sec:GeneralizeDomainCodomain} below.} 
We could call this ``function'' $F_{\rm states}$. Given an $(n-1)$-dimensional manifold $N_{n-1}$, there should be a corresponding Hilbert space $F_{\rm states}(\CH(N_{n-1}))$,
the ``Hilbert space of states on the spatial manifold $N_{n-1}$.'' So we have: 
\be 
\begin{split}
\CH_0 & = F_{\rm states}(N_{n-1}^{0}) ~,\\
\CH_1 & = F_{\rm states}(N_{n-1}^{1}) ~.\\
\end{split}
\ee

Next, the   interpolating history $M_n$ should define a linear map, 
\eqa{
& F_{\rm ampl}(M_{n}) : F_{\rm states}(N_{n-1}^{0} ) &&\to F_{\rm states}(N_{n-1}^{1})~. \label{eq:interpol}
}
The linear map depends on interpolating spacetime and the stuff that happened in between. In QFT, $F_{\rm ampl}$ is often computed from a path integral, and indeed, the axioms for  $F_{\rm ampl}$ below are intended to capture properties one expects such path integrals to satisfy. Most path integrals of interest are extremely difficult to compute. It can be helpful to extract or axiomatize key properties of path integrals. In topological quantum field theory (TQFT), we axiomatize to retain the key properties of \emph{locality}.

Topological field theory (TFT) is meant to capture this very basic idea in a way that expresses locality but eliminates almost all the complications of typical quantum systems. Accordingly in 
TFT, we take as axiomatic that the values of $F_{\rm states}$ 
and $F_{\rm ampl}$ only depend (up to a suitable notion of isomorphism) on the  \underline{diffeomorphism class} of $N^0_{n-1}$, $N^1_{n-1}$, and $M_n$, respectively. 

The axiomatic definition for TFT we will give below turns out to define an amazingly rich mathematical structure, one that continues to be explored and refined in current research. Moreover, it naturally suggests a formulation of field theory in general. This formulation, 
sometimes known as \emph{geometrical field theory} or \emph{functorial field theory} 
is discussed briefly in \autoref{sec:FunctorialApproachQFT}.   One of the main motivating examples is G. Segal's formulation of 2d conformal field theory (CFT) \cite{Segal:2002ei}. Giving a fully satisfactory definition of quantum field theory 
along such lines is an extremely ambitious goal for the future.
Somewhat surprisingly, in addition to playing an important role in mathematics, topological field theory has also played an important role in some mathematical discussions of condensed matter physics (such as anyons, topological materials, and phases of matter). It has even played a notable role in some discussions of quantum gravity.

The statement that  $F_{\rm states}$ 
and $F_{\rm ampl}$ only depend on diffeomorphism classes implies, 
among other things, that they do not depend on a choice of metric on these manifolds. 
 Since there is no choice of metric, there is no choice of a signature of the metric, so the terms ``spatial'' and ``time evolution'' above are being used rather loosely. We should think of the axioms of TFT as capturing the behavior of  Euclidean, i.e., Wick-rotated field theory. Since there is no metric dependence, we are throwing away the dependence on length scales and energy scales. This has the virtue of clarifying the essential physical ideas related to locality, as well as making things computable.

%
%
%
%


We now come to our first axiomatization of locality. In quantum mechanics, the statespace of a ``union'' of two separate systems 
is the tensor product of the constituent statespaces. For example, 
the statespace of one qubit is $\IC^2$, and that of $N$ qubits is 
$(\IC^2)^{\otimes N}$. In our context, that means we should say there is an isomorphism of statespaces: 
\tightfootnote{In fermionic theories, the spaces will be $\IZ_2$-graded, i.e., ``super-vector spaces'' (in the mathematical sense), and the tensor product will be a $\IZ_2$-graded product.}
\eqa{
& F_{\rm states}\left(N_{n-1} \coprod N_{n-1}'\right) &&\cong  F_{\rm states}(N_{n-1}) \otimes F_{\rm states}(N_{n-1}') ~. \label{eq:LOC1} \tag{$\textcolor{Blue}{\bm{\mathsf{LOC\,1}}}$}
}
\begin{remark}
$\,$
\begin{enumerate}\itemsep 0pt
%
%
\item \textbf{N.B.!}  It follows from \eqref{eq:LOC1} that $F_{\rm states}(\emptyset_{n-1}) \cong \IC$.
\tightfootnote{Take $N_{n-1}' = \emptyset_{n-1}$. Then \eqref{eq:LOC1} implies that $F_{\rm states}(N_{n-1}) = F_{\rm states}(N_{n-1}\coprod\emptyset_{n-1}) = F_{\rm states}(N_{n-1}) \otimes F_{\rm states}(\emptyset_{n-1})$, and thus $F_{\rm states}(\emptyset_{n-1}) \cong \IC$, as we are working over the category of complex vector spaces. Here $\cong$ stands for ``is isomorphic to,'' and implies an isomorphism of $\IC$-vector spaces.} 
\item The isomorphism class of the vector space $F_{\rm states}(N_{n-1})$  depends only on the diffeomorphism class of $N_{n-1}$. It is in this sense that $F_{\rm states}(N_{n-1})$ is a (smooth) ``topological invariant.'' It is interesting to compare with traditional topological 
invariant vector spaces associated to manifolds. For example, for homology we have  
\be 
H_{k}\left(M \coprod M'\right) \cong  H_{k}(M) \oplus H_{k}(M') ~,
\ee
and similarly for cohomology. Note that for the fundamental group, 
one is at a loss to define $\pi_1(M \coprod M')$ since the definition of $\pi_1$ requires the choice of a basepoint. 
 Classical invariants of algebraic topology depend very differently on underlying manifolds than what quantum mechanics would dictate. For this reason, the spaces $F_{\rm states}(N_{n-1})$ are  called ``Quantum Invariants.''
\end{enumerate}
\end{remark}

\begin{figure}[H]
\centering
\includegraphics[width=2.5in]{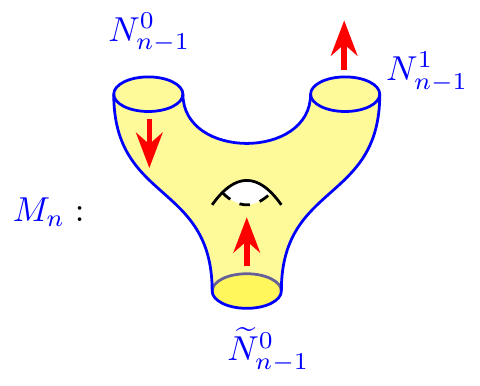}
\caption{An $n$-manifold $M_{n}$ with boundary. The different connected components of the boundary carry an arrow indicating whether they are ``ingoing'' (i.e., ``initial'' or ``past'') spatial components or ``outgoing'' (i.e., ``final'' or ``future'') spatial components. Note well that these arrows, sometimes referred to as a \emph{co-orientation}, do not imply that $M_n$ or its boundaries are oriented, or even orientable.  }\label{fig:2}
\end{figure}

Now consider an $n$-manifold $M_{n}$ with boundary, see \autoref{fig:2}.
We want to think of this as a spacetime connecting past and future states. The boundary has several connected components.  We need to choose which spatial slices are components of the ``past'' spatial manifold and which are components of the ``future'' spatial manifold. 
We indicate this with 
\textcolor{red}{red arrows} pointing in (past) or out (future) of the manifold.  

So, for the example of \autoref{fig:2}, our quantum amplitudes $F_{\rm ampl}$ will give a linear map:
\be
F_{\rm ampl}(M_n) : F_{\rm states}\big(N_{n-1}^{0} \coprod \widetilde{N}_{n-1}^{0}\big) \longrightarrow F_{\rm states}(N_{n-1}^{1}) ~. \label{eq:qinterpol}
\ee

The next aspect of locality that we wish to axiomatize is the idea that a sum over a complete set of states is tantamount to an insertion of the identity. This leads to the idea of ``gluing.''

To motivate this, consider the field theory of maps into a target space $\mathscr{X}$ (some topological space). 
\tightfootnote{For $N$ scalar fields, $\mathscr{X}$ will be $\IR^{N}$. In K. Intriligator's TASI 2023 lectures \cite{Intriligator:TASIVideoLec1,Intriligator:TASIVideoLec2,Intriligator:TASIVideoLec3,Intriligator:TASIVideoLec4}, the target space was $\IC\IP^{1}$.}
For example, one might consider a sigma model with target $\mathscr{X}$. 
%
We denote $\mathscr{X}^{N_{n-1}} = \mathsf{Map}(N_{n-1} \to \mathscr{X})$ as the space of maps from $N_{n-1}^{0,1}$ into the target. 
Let $\phi_{i}, \phi_{f} \in \mathscr{X}^{N_{n-1}}$. Then we expect 
$F_{\rm states}(N_{n-1})$ should be something like ``$L^{2}(\mathscr{X}^{N_{n-1}})$.''
So a quantum ``statevector'' would be a functional $\Psi$ assigning a complex number 
$\Psi[\phi]$ to every $\phi \in \mathscr{X}^{N_{n-1}}$ subject to some kind of normalizability constraint.

\begin{figure}[h]
  \centering
  \includegraphics[width=3.5in]{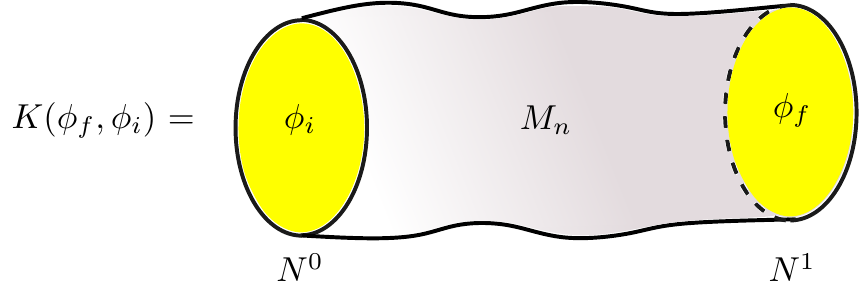}
  \caption{The propagator.}
  \label{fig:propagator}
  \end{figure}

In a QFT, the Feynman path integral over field configurations on $M_n$ defines a ``propagator'' or kernel map on initial and final field configurations, see \autoref{fig:propagator}.

Then the linear map $F_{\rm ampl} (M_n)$ would map the initial 
statevector $\Psi_i$ to a final one $\Psi_f= F_{\rm ampl} (M_n)(\Psi_i)$, 
according to a formula that looks like:  
\eqa{
& \left( F_{\rm ampl}(M_n)(\Psi_i) \right)[\phi_f] &&= \int_{\phi_i \in \mathcal{X}^{N^0}} K(\phi_f, \phi_i) \Psi_i[\phi_i] \CD\phi_i ~. \label{eq:FeynAmp}
}

\begin{redbox}[box:FeynAmp]{Interpretation Of The Feynman Amplitude}
Here $F(M_n)$ is, by \eqref{eq:interpol}, a linear transformation that acts on the initial statevector $\Psi_i[\phi_i]$ (as a functional of the initial field configuration $\phi_i$) to produce a vector in the final state Hilbert space. To form an amplitude (a $c$-number), one has to project this statevector onto the final field configuration $\phi_f$, which is achieved by multiplying with the kernel $K(\phi_f, \phi_i)$, and summing over all initial configurations $\phi_i$. This is the content of \eqref{eq:FeynAmp}. This is simply the QFT version of the transition amplitude in QM, but now one is summing over \underline{field configurations}. In general, the integral over the initial field configurations $\phi_i$ will itself be a path integral.  
\end{redbox}

\begin{figure}[h]
  \centering
  \includegraphics[width=3.5in]{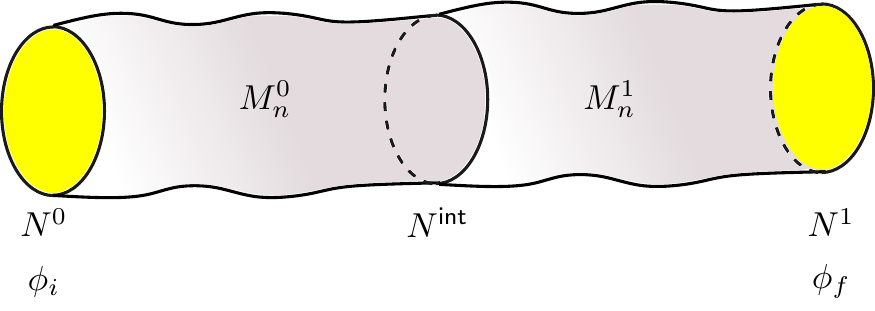}
  \caption{Gluing.}
  \label{fig:4}
  \end{figure}

But we expect that if we cut $M_n$ along some intermediate $(n-1)$-manifold, see \autoref{fig:4},

%
we ought to have
\eqa{
& K(\phi_f, \phi_i) &&= \int \underbrace{K(\phi_f, \phi_{\text{int}})}_{M_n^1} \underbrace{K(\phi_{\text{int}},\phi_{i})}_{M_n^0} \CD\phi_{\text{int}} ~.
}

So, without trying to make direct mathematical sense of the path integral, we simply postulate that the \underline{result} of such a path integral should satisfy  the gluing axiom:
\eqa{
   F_{\rm ampl}(M_n^1) \circ F_{\rm ampl}(M_n^0) = F_{\rm ampl}(M_n^1 \circ_{N^{\text{int}}} M_n^0) ~, \label{eq:LOC2} \tag{$\textcolor{Blue}{\bm{\mathsf{LOC\,2}}}$}
}
where $\circ_{N^{\text{int}}}$ denotes gluing along $N^{\text{int}}$, and on the LHS, $\circ$  denotes composition of linear maps.

We repeat: We might not yet be able to give rigorous mathematical definitions to the most interesting path integrals for quantum field theory, but we can certainly axiomatize certain properties we would definitely want these path integrals to satisfy. The above gluing axiom is an example of such a property.

We will next put these general ideas on a nice and precise mathematical foundation. Our first step in this direction is to  introduce the idea of ``bordism.''

\SectionWithHeader{Bordisms}{Bordisms}{sec:Bordisms}

For much more about bordism theory (with a view to applications in TFT), see \cite{FreedBordism,FreedCBMS}. We also recommend the classic textbook \cite{MilnorStasheff}.

In our heuristic discussion, we did not distinguish between compact and noncompact manifolds. Noncompact manifolds are very natural in physics, but do introduce complications of choosing boundary conditions at infinity. For this reason, the rigorous mathematical theory is generally restricted to compact manifolds. 
We will abbreviate ``compact manifold without boundary'' to ``closed manifold.''

\begin{figure}[H]
  \centering
  \includegraphics[width=2in]{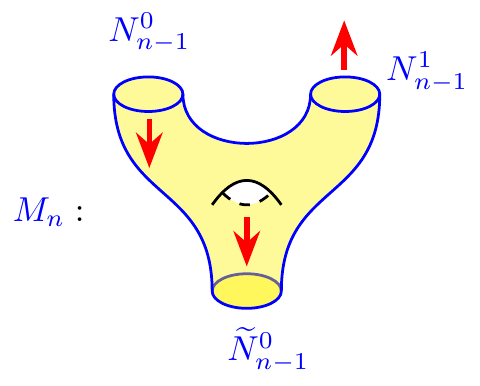}
    \includegraphics[width=2in]{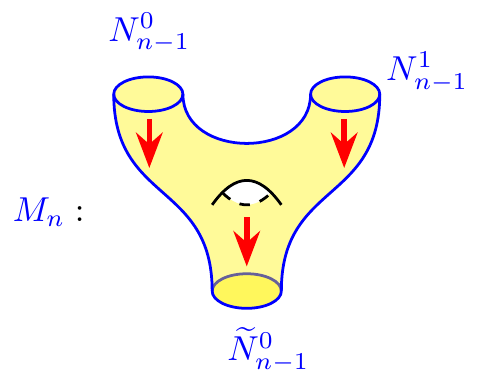}
  \caption{Same manifold, different bordism!!}\label{fig:5}
  \end{figure}

\begin{definition}[\textcolor{red}{Bordism}] Let $N_{n-1}^{0}$ and $N_{n-1}^{1}$ be closed $(n-1)$-manifolds. A \textbf{bordism from $N_{n-1}^{0}$ to $N_{n-1}^{1}$} is a triple $(M_n, \theta_{\text{in}}, \theta_{\text{out}})$ consisting of the following collection of data:
\begin{itemize}\itemsep 0pt
 \item[$(a)$] A compact $n$-manifold with boundary $M_n$.
 \item[$(b)$] A decomposition of boundary components into ``in'' and ``out'':
 \eqa{
   &\partial M_n &&= (\partial M_{n})^{{\rm in}} \coprod (\partial M_{n})^{{\rm out}} ~.
 }
 \item[$(c)$] Diffeomorphisms of collar neighborhoods:
 \tightfootnote{A collar is an open neighborhood which is the Cartesian product of a boundary component with a half-open interval. The collar neighborhood theorem (Brown \cite{Brown1962})
 states that the boundary of any manifold admits a collar.}
 \eqa{
    \theta_{\text{in}} \quad : \quad & N_{n-1}^{0} \times [0,\varepsilon) &&\xlongrightarrow{\text{embedding}} M_{n} ~, \\
                     &  N_{n-1}^{0} \times \{0\} &&\xmapsto{\text{diffeomorphism}} (\partial M_{n})^{{\rm in}} ~, \\
    \theta_{\text{out}}\quad : \quad & N_{n-1}^{1} \times (1-\varepsilon,1] &&\xlongrightarrow{\text{embedding}} M_{n} ~,\\
                       & N_{n-1}^{1} \times \{1\} &&\xmapsto{\text{diffeomorphism}} (\partial M_{n})^{{\rm out}} ~.   
 }
 \end{itemize}
 \end{definition}
 The ``in'' and ``out'' designations -- introduced in \autoref{sec:basicpicture-tft-heuristic} via arrows -- are important, as illustrated in \autoref{fig:5}.
 \begin{figure}[H]
  \centering 
  \includegraphics[width=2in]{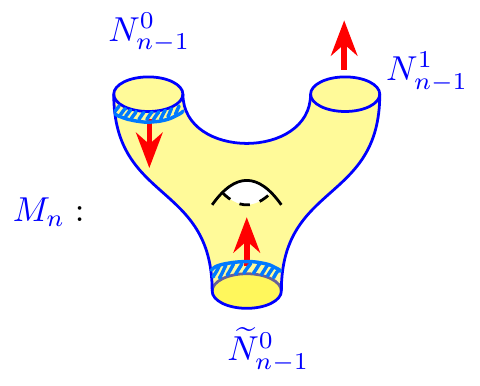}
  \caption{In this example, $N_{n-1}^{0}$ is a disjoint union of two (ingoing) circles.}\label{fig:6}
  \end{figure}

 $\theta_{\text{in}}$ is an embedding of a little collar neighborhood into $M_{n}$ (see \autoref{fig:6}):
  \eqa{
    \theta_{\text{in}} : N^{0} \times [0, \varepsilon) &\xrightarrow{\text{embedding}} M_{n} ~.
   }
   $\theta_{\text{in}}$ takes these circles times $[0, \varepsilon)$ (a little interval), and embeds them into $M_{n}$ as shown by the shaded parts in \autoref{fig:6}. $\theta_{\text{in}}$ has to satisfy the restriction that at $\{0\} \in [0, \varepsilon)$ it becomes a diffeomorphism with the ingoing components:
   \eqa{
      N^{0} \times \{ 0 \} &\xrightarrow[\text{diffeomorphism}]{\cong} (\partial M_{n})^{{\rm in}} ~.
   }
   $\theta_{\text{out}}$ is similar.

\bigskip
\begin{remark} The manifolds discussed here need not be connected. Moreover, we must stress that at this point we have said nothing about orientation! Our manifolds are unoriented. They might, or might not, be orientable. 
\end{remark}
\bigskip
\begin{remark}\label{rem:cobordism-vs-bordism} Some authors use the term ``cobordism'' for what we are calling ``bordism.'' The former term is more traditional but the 
``co'' prefix carries technical implications that are misleading, so we prefer the term ``bordism.'' 
\end{remark}
\bigskip
\bigskip

\noindent A \textbf{diffeomorphism of bordisms},
\be
\begin{tikzcd}[row sep=huge, column sep=huge, text height=1.5ex, text depth=0.25ex]
N_{n-1}^{0} \arrow[rr,bend left, "{(M_n,\,\theta_{\text{in}},\,\theta_{\text{out}})}"] \arrow[rr,bend right, "{(M_n',\,\theta_{\text{in}}',\,\theta_{\text{out}}')}",swap] & \Bigg\downarrow{\psi}  & N_{n-1}^{1} 
\end{tikzcd}
\ee
is a diffeomorphism $\psi: M_{n} \to M_{n}'$, so that the diagrams
\be
\begin{tikzcd}
 & M_n \arrow[dd, "\psi"]\\
 N_{n-1}^{0} \times [0, \varepsilon) \arrow[ru, "\theta_{\text{in}}"] \arrow[rd, "\theta_{\text{in}}'",swap] \\
 & M_{n}'
\end{tikzcd}
\quad \text{and} \quad
\begin{tikzcd}
 & M_n \arrow[dd, "\psi"]\\
 N_{n-1}^{1} \times (1-\varepsilon, 1] \arrow[ru, "\theta_{\text{out}}"] \arrow[rd, "\theta_{\text{out}}'",swap] \\
 & M_{n}'
\end{tikzcd}
\ee
\underline{commute}.

One of the reasons it is useful to incorporate the data of $\theta_{\text{in}}$, $\theta_{\text{out}}$ in the definition of a bordism is that it allows us to \underline{glue} two bordisms:
\eqas{
& (M_n, \theta_{\text{in}}, \theta_{\text{out}})\quad &:\quad N_{n-1}^0 \to N_{n-1}^1 ~,\\
& (M_n', \theta_{\text{in}}', \theta_{\text{out}}')\quad &:\quad N_{n-1}^1 \to N_{n-1}^2 ~,
}
into a single bordism $N_{n-1}^0 \to N_{n-1}^2$. If we just considered manifolds with boundary, we could not account for the possible ``twists'' in gluing the two bordisms together.
\\\\
\noindent \textbf{Example:} Consider the following $0$-manifolds: $N^0 = N^1 =$ disjoint union of two points.
\paragraph{Question:} How many bordisms 
\begin{figure}[H]
\centering
\begin{tikzpicture} 
\coordinate (A) at (0, 0); 
\coordinate (B) at (1, 0);
\coordinate (C) at (1, 1);
\coordinate (D) at (0, 1);

\node [fill=blue,circle,minimum size=2pt,inner sep=-2] at (A) {};
\node [fill=blue,circle,minimum size=2pt,inner sep=-2] at (B) {};
\node [fill=blue,circle,minimum size=2pt,inner sep=-2] at (C) {};
\node [fill=blue,circle,minimum size=2pt,inner sep=-2] at (D) {};

\draw[->,thick] (0.2,0.5) -- (0.8,0.5);

\end{tikzpicture}
\caption{Two disjoint unions of two points.}
\end{figure}
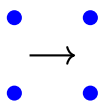
are there (up to diffeomorphism)? 

\paragraph{Answer:} We have three obvious bordisms, see \autoref{fig:three-obvious-bordisms}: 

\begin{figure}[H]
\centering
\begin{minipage}{0.2\textwidth}
\begin{tikzpicture}
\coordinate (A) at (0, 0);
\coordinate (B) at (1, 0);
\coordinate (C) at (1, 1);
\coordinate (D) at (0, 1);

\node [fill=blue,circle,minimum size=2pt,inner sep=-2] at (A) {};
\node [fill=blue,circle,minimum size=2pt,inner sep=-2] at (B) {};
\node [fill=blue,circle,minimum size=2pt,inner sep=-2] at (C) {};
\node [fill=blue,circle,minimum size=2pt,inner sep=-2] at (D) {};

\draw[blue, very thick, -] (A) -- (B);
\draw[blue, very thick, -] (D) -- (C);

\draw[->,red,thick] (0.1,0.25) -- (0.45,0.25);
\draw[->,red,thick] (1.1,0.25) -- (1.45,0.25);
\draw[->,red,thick] (0.1,1.25) -- (0.45,1.25);
\draw[->,red,thick] (1.1,1.25) -- (1.45,1.25);

\end{tikzpicture}
\end{minipage}%
\begin{minipage}{0.25\textwidth}
$\,$\\ 
\begin{tikzpicture}
\coordinate (A) at (0, 0);
\coordinate (B) at (1.5, 0);
\coordinate (C) at (1.5, 1);
\coordinate (D) at (0, 1);

\node [fill=black,circle,minimum size=2pt,inner sep=-2] at (A) {};
\node [fill=black,circle,minimum size=2pt,inner sep=-2] at (B) {};
\node [fill=black,circle,minimum size=2pt,inner sep=-2] at (C) {};
\node [fill=black,circle,minimum size=2pt,inner sep=-2] at (D) {};


\draw[->,red,thick] (0.1,-0.25) -- (0.45,-0.25);
\draw[->,red,thick] (1.7,0) -- (2.05,0);
\draw[->,red,thick] (0.1,1.25) -- (0.45,1.25);
\draw[->,red,thick] (1.7,1) -- (2.05,1);

\draw[ultra thick,blue] (0,0) arc (-90:90:0.5);
\draw[ultra thick,blue] (1.5,0) arc (270:90:0.5);

\end{tikzpicture}
\end{minipage}%
\begin{minipage}{0.2\textwidth}
\begin{tikzpicture}
\coordinate (A) at (0, 0);
\coordinate (B) at (1, 0);
\coordinate (C) at (1, 1);
\coordinate (D) at (0, 1);
\coordinate (E) at (0.45, 0.45);
\coordinate (F) at (0.55, 0.55);

\node [fill=blue,circle,minimum size=2pt,inner sep=-2] at (A) {};
\node [fill=blue,circle,minimum size=2pt,inner sep=-2] at (B) {};
\node [fill=blue,circle,minimum size=2pt,inner sep=-2] at (C) {};
\node [fill=blue,circle,minimum size=2pt,inner sep=-2] at (D) {};

\draw[blue, very thick, -] (A) -- (E);
\draw[blue, very thick, -] (F) -- (C);
\draw[blue, very thick, -] (B) -- (D);

\draw[->,red,thick]  (0,0.25) -- (0.20,0.45);
\draw[->,red,thick] (0.75,1.05) -- (0.95,1.25);
\draw[->,red,thick] (0.05,1.30) -- (0.25,1.10);
\draw[->,red,thick] (1.05,0.35) -- (1.25,0.15);

\end{tikzpicture}
\end{minipage}%
\caption{Three obvious bordisms from a disjoint union of two points to a disjoint union of two points.}
\label{fig:three-obvious-bordisms}
\end{figure}
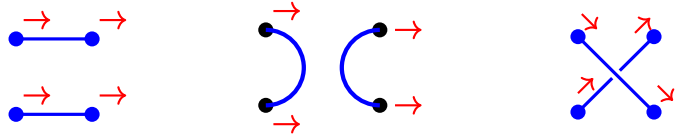

\noindent But since a circle is a bordism from $\emptyset$ to $\emptyset$, we can take a disjoint union with any number of circles, e.g., see \autoref{fig:infinitely-many-bordisms}:

\begin{figure}[H]
\centering
\begin{minipage}{0.15\textwidth}
\begin{tikzpicture}
\coordinate (A) at (0, 0);
\coordinate (B) at (1, 0);
\coordinate (C) at (1, 1);
\coordinate (D) at (0, 1);

\node [fill=blue,circle,minimum size=2pt,inner sep=-2] at (A) {};
\node [fill=blue,circle,minimum size=2pt,inner sep=-2] at (B) {};
\node [fill=blue,circle,minimum size=2pt,inner sep=-2] at (C) {};
\node [fill=blue,circle,minimum size=2pt,inner sep=-2] at (D) {};

\draw[blue, very thick, -] (A) -- (B);
\draw[blue, very thick, -] (D) -- (C);

\draw[->,red,thick] (0.1,-0.25) -- (0.45,-0.25);
\draw[->,red,thick] (1.1,-0.25) -- (1.45,-0.25);
\draw[->,red,thick] (0.1,1.25) -- (0.45,1.25);
\draw[->,red,thick] (1.1,1.25) -- (1.45,1.25);

\filldraw[color=blue!60,fill=white,very thick] (0.18,0.65) circle (0.1);
\filldraw[color=blue!60,fill=white,very thick] (0.65,0.65) circle (0.1);
\filldraw[color=blue!60,fill=white,very thick] (0.45,0.35) circle (0.1);
\filldraw[color=blue!60,fill=white,very thick] (0.85,0.35) circle (0.1);

\end{tikzpicture} 
\end{minipage}%
\begin{minipage}{0.2\textwidth}
\end{minipage}
\caption{Infinitely many bordisms upon including disjoint union with arbitrarily many circles.}
\label{fig:infinitely-many-bordisms}
\end{figure}
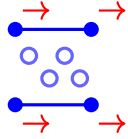
\noindent 
and hence there are infinitely many inequivalent bordisms. 
%

%
%

Now we can give a more precise version of \eqref{eq:LOC1} and \eqref{eq:LOC2}  above. 
A TFT is an association to any bordism $(\{M_n, \theta_{\text{in}}, \theta_{\text{out}}\})$ of a corresponding   linear map 
$F(\{M_n, \theta_{\text{in}}, \theta_{\text{out}}\})$ from $F(N_{n-1}^{0})$ to $F(N_{n-1}^{1})$. Here, and henceforth, we will usually drop the 
subscript ``states'' and ``amplitude'' on $F$ because it should be clear which one is meant by context. Moreover, the value of $F$  only depends on the ``diffeomorphism equivalence class'' of the bordism.
%
%

Returning to TFT, the rules are:
\eqa{
& F_{\rm states} : \left\{ \begin{array}{c} \text{compact $(n-1)$-fold $N$} \\ \partial N = \emptyset \end{array} \right\} &&\longrightarrow \left\{\begin{array}{c} \text{Vector spaces} \\ \text{(over $\IC$, in these lectures)} \end{array}\right\} ~,
}
\eqa{
& F_{\rm states}(N \coprod N') &&\cong F_{\rm states}(N) \otimes F_{\rm states}(N')  \tag{$\textcolor{Blue}{\bm{\mathsf{LOC\,1}'}}$} ~,\\
( &\implies F_{\rm states}(\emptyset_{n-1}) &&= \IC ~) \nonumber
}
\eqa{
& F_{\rm ampl}: \left( \begin{array}{c}\text{Bordism} \\ M_{n}: N^{0} \to N^{1} \end{array} \right) &&\longmapsto F_{\rm ampl}(M_{n}) \in \Hom( F_{\rm states}(N^{0}), F_{\rm states}(N^{1}) ) ~, \label{eq:Fbord}
}
such that $F$ maps a composition of bordisms to a composition of linear transformations:
\eqa{
& F_{\rm ampl}(M_{n}' \circ M_{n}) &&= F_{\rm ampl}(M_{n}') \circ F_{\rm ampl}(M_{n}) ~.  \tag{$\textcolor{Blue}{\bm{\mathsf{LOC\,2}'}}$} \label{eq:LOC2prime}
}
Recall that given two vector spaces $V$ and $W$, $\Hom(V, W)$ is the vector space of all linear transformations from $V$ to $W$. Note that $F_{\rm states}(N_{n-1})$ is some $\IC$-vector space only depending on the diffeomorphism type of $N_{n-1}$. Also, note that we have suppressed $\theta_{\text{in}}$ and $\theta_{\text{out}}$ in the gluing of bordisms encoded in 
\eqref{eq:LOC2prime}, but in fact, the details of the gluing will rely on these diffeomorphisms. 
\paragraph{Corollaries:}

\begin{enumerate}
\item[1.)] For \underline{any} one-dimensional complex vector space $L$, $\Hom(L, L) \cong \IC$ \underline{canonically}: every linear transformation $L \to L$ is of the form $v \mapsto T(v) = z_0 v$ where $v\in L$ and   $z_0 \in \IC$. So we can naturally associate to $T$ a complex number, and vice versa.\\
$\implies$ A closed $n$-manifold $M_n$ can be viewed as a bordism from the empty $(n-1)$-manifold to the empty $(n-1)$-manifold, i.e., $M_{n}: \emptyset_{n-1} \to \emptyset_{n-1}$, and therefore, by \eqref{eq:Fbord},
\eqa{
& F(M_n) &&\in \Hom(F(\emptyset_{n-1}), F(\emptyset_{n-1})) 
\underset{\text{\eqref{eq:LOC1}}}{\overset{\text{Locality}}{=\joinrel=\joinrel=\joinrel=\joinrel=}} \Hom(\IC, \IC) \cong \IC ~,
}
and the corresponding complex number $F(M_n) \in \IC$ is called ``the partition function.''
\item[2.)] Consider a bordism from nothing to something, see \autoref{fig:bordism-from-nothing-to-something}: 
\begin{figure}[h]
\centering
\begin{tikzpicture}
\node (limage) [inner sep = 0pt]{\includegraphics[width=1.8in]{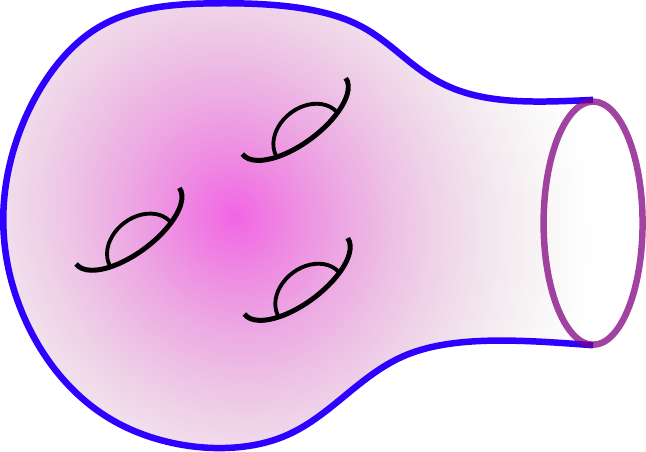}};
\node[] at (1.95,-1.3) {\scalebox{1}{$N$}};
\draw[->,red,thick] (1.9,0) -- (2.7,0);
\end{tikzpicture}
\caption{A bordism from nothing to something.}\label{fig:bordism-from-nothing-to-something}
\end{figure}
\eqa{
 & M_{n} : \emptyset_{n-1} &\to N_{n-1} ~.
}
Then $F(M_{n}) \in \Hom(\IC, F(N_{n-1}))$ is the map $1 \mapsto \text{Vector}$. This can be visualized as
\begin{align}
\hspace{-0.1in}F\Bigg(
\begin{tikzpicture}[scale=0.5,baseline=-0.5ex]
\node (limage) [inner sep = 0pt]{\includegraphics[width=1.1in]{images/bordism_from_nothing_to_something.pdf}};
\node[] at (2.2,-1.6) {\scalebox{1}{$N$}};
\draw[->,red,thick] (2.4,0) -- (3.3,0);
\end{tikzpicture}
\Bigg) 
&\in \Hom\big(F(\IC), F(N) \big) \cong \Hom\big( \IC, F(N) \big) \cong F(N) ~.
\end{align}
We call $F(M_n) \in F(N)$ the ``state created by the bordism.'' It formalizes the idea of the Hartle-Hawking state \cite{HartleHawking:1983} from quantum gravity. 
(In the latter context, one might, or might not, wish to sum over topologically distinct bordisms with the same $N$, with a suitable weighting of the topologies.)

\item[3.)] Consider any cylinder $M_{n} = N_{n-1} \times [0,1]$ with $\theta_{\text{in}} = \theta_{\text{out}}$ = \text{identity}. There are different bordisms that we can associate with this. Since we're working up to diffeomorphism type, we can take the diffeomorphism to be the gluing of two cylinders to make one cylinder. This yields
%
\def\cylinderone{
\begin{tikzpicture}[baseline={([yshift=-.5ex]current bounding box.center)},vertex/.style={anchor=base,
    circle,fill=black!25,minimum size=18pt,inner sep=2pt}]
\node[cylinder, cylinder uses custom fill,draw=black,text = purple,cylinder body fill=magenta!05,cylinder end fill=magenta!05,minimum width = 1.5cm,
    minimum height = 3.5cm,shape border rotate=180,ultra thick] (c) at (0,0) {$\vphantom{\substack{\textcolor{black}{\bullet}\\ [g_0]}}$};
    \draw[->,red,thick] (-1.9,0) -- (-1.6,0);
    \draw[->,red,thick] (1.0,0) -- (1.3,0);
   \node[] at (-1.6,1.1) {$0$};
      \node[] at (1.1,1.1) {$2$};
    \node[] at (-1.6,-1.1) {$N_{n-1}$};
    \node[] at (1.1,-1.1) {$N_{n-1}$};
    \node[] at (-0.1,-1.1) {$1$};
    \draw[dashed,black] (1.16,0.74) arc (90:270:0.38 and 0.74);
    \draw[dashed,black] (-0.1,0.74) arc (90:270:0.38 and 0.74);
    \draw[ultra thick,black,rotate=180] (0.2,0.74) arc (90:270:0.38 and 0.74);
\end{tikzpicture}
}
\def\cylindertwo{
\begin{tikzpicture}[baseline={([yshift=-.5ex]current bounding box.center)},vertex/.style={anchor=base,
    circle,fill=black!25,minimum size=18pt,inner sep=2pt}]
\node[cylinder, cylinder uses custom fill,draw=black,text = purple,cylinder body fill=magenta!05,cylinder end fill=magenta!05,minimum width = 1.5cm,
    minimum height = 3.5cm,shape border rotate=180,ultra thick] (c) at (0,0) {$\vphantom{\substack{\textcolor{black}{\bullet}\\ [g_0]}}$};
    \draw[->,red,thick] (-1.9,0) -- (-1.6,0);
    \draw[->,red,thick] (1.0,0) -- (1.3,0);
   \node[] at (-1.6,1.1) {$0$};
      \node[] at (1.1,1.1) {$1$};
    \node[] at (-1.6,-1.1) {$N_{n-1}$};
    \node[] at (1.1,-1.1) {$N_{n-1}$};
    \draw[dashed,black] (1.16,0.74) arc (90:270:0.38 and 0.74);
\end{tikzpicture}
}
\def\cylinderthree{
\begin{tikzpicture}[baseline={([yshift=-.5ex]current bounding box.center)},vertex/.style={anchor=base,
    circle,fill=black!25,minimum size=18pt,inner sep=2pt}]
\node[cylinder, cylinder uses custom fill,draw=black,text = purple,cylinder body fill=magenta!05,cylinder end fill=magenta!05,minimum width = 1.5cm,
    minimum height = 3.5cm,shape border rotate=180,ultra thick] (c) at (0,0) {$\vphantom{\substack{\textcolor{black}{\bullet}\\ [g_0]}}$};
    \draw[->,red,thick] (-1.9,0) -- (-1.6,0);
    \draw[->,red,thick] (1.0,0) -- (1.3,0);
    \draw[dashed,black] (1.16,0.74) arc (90:270:0.38 and 0.74);
\end{tikzpicture}
}
\eqa{
& F(M_n) \circ F(M_n) &&= F\Bigg(\cylinderone\Bigg) \\
& && 
\underset{\text{}}{\overset{\text{topological !!}}{=\joinrel=\joinrel=\joinrel=\joinrel=\joinrel=}}F\Bigg(\cylindertwo\Bigg) \\
&\implies F(M_n) &&\in \Hom\big( F(N_{n-1}), F(N_{n-1}) \big)  ~.
}
This is a projector: $P = F(\raisebox{2pt}{\scalebox{0.4}{\cylinderthree}})$ satisfies $P = P^2$. In particular, any amplitude $F(N_{n-1})$ can be written by gluing on a little cylinder to the bordism, and 
\eqa{
  F(N_{n-1}) &= P F(N_{n-1}) \oplus (1-P)F(N_{n-1}) ~.
}
All amplitudes vanish on the second term, i.e., on $\ker(F(M_n))$. Therefore, one can assume without loss of generality that $F\left(\raisebox{2pt}{\scalebox{0.4}{\cylinderthree}}\right)$ $=$ Identity.

\item[4.)]\label{item:dualizability} \emph{Dualizability}. Now consider another bordism on the cylinder, now with both ingoing circles.
\def\dualizabilityone{
\includegraphics[]{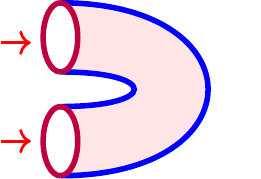}
}
\def\dualizabilityoneNOARROW{
\includegraphics[]{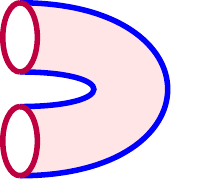}
}
\def\dualizabilitytwo{
\includegraphics[]{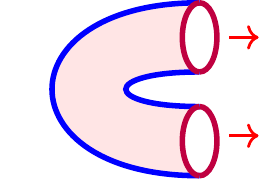}
}
\def\dualizabilitytwoNOARROW{
\includegraphics[]{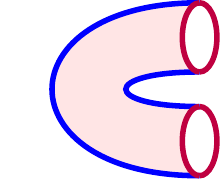}
}
\eqa{
 & F \Bigg(\vcenter{\hbox{\dualizabilityone}} \!\!\!\!\!\!\!\Bigg) : V \otimes V &\longrightarrow \IC ~.
}
We could make a diffeomorphism where the two ingoing circles are exchanged.
Therefore $\Diff$ action $\implies$ this is a \emph{symmetric} bilinear form $V \otimes V \xlongrightarrow{b} \IC$.

One could also consider the circles to be outgoing. This yields the following picture:
\eqa{
 & F \Bigg( \!\!\!\!\!\!\!\vcenter{\hbox{\dualizabilitytwo}}\,\,\Bigg) : \IC &\xlongrightarrow{\widetilde{b}} V\otimes V ~,
}
yielding another symmetric bilinear map $\widetilde{b}$, this time from $\IC$ to  $V \otimes V$.

Now consider the ``$S$-diagram,'' see \autoref{fig:sdiag}.
\begin{figure}[h]
  \centering
  \begin{tikzpicture}[scale=0.8,baseline=-0.5ex]
    \node (limage) [inner sep = 0pt]{\includegraphics[width=3in,valign=c]{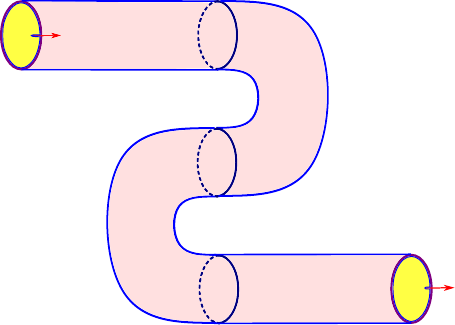}};
    \node at (0.6,2.7) {$\bm{N}$};
    \node at (0.6,0) {$\bm{N}$};
    \node at (0.6,-2.7) {$\bm{N}$};
    \node at (-5.2,2.7) {$\textcolor{violet}{\bm{N}}$};
    \node at (5.2,-2.7) {$\textcolor{violet}{\bm{N}}$};
  \end{tikzpicture}
  $\cong$ \quad 
  \begin{tikzpicture}[scale=0.8,baseline=-0.5ex]
    \node (limage) [inner sep = 0pt]{\includegraphics[width=2in,valign=c]{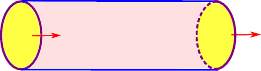}};
    \node at (-2.7,-1.35) {$\bm{N}$};
    \node at (2.3,-1.35) {$\bm{N}$};
  \end{tikzpicture}
  \caption{The $S$-diagram.}
  \label{fig:sdiag}
  \end{figure}
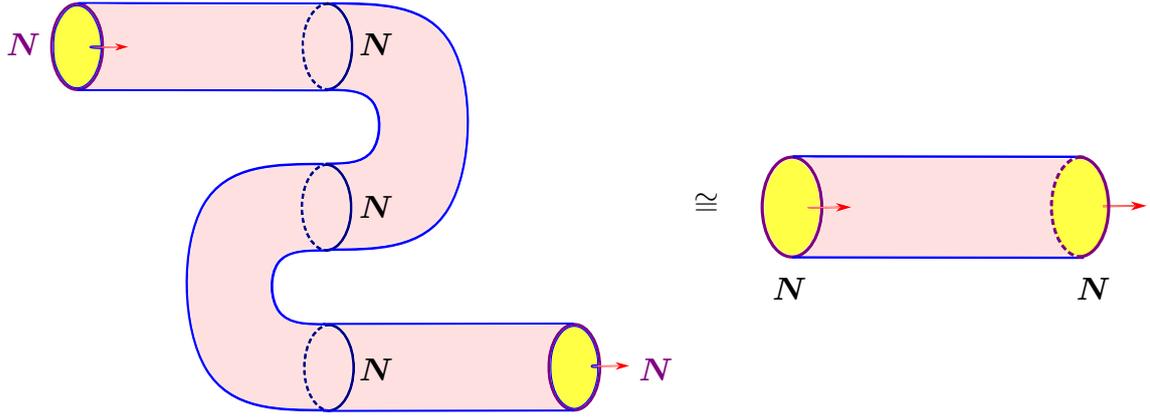
This gives a map (recall $V = F(N)$):
\eqa{
& V \xrightarrow{\mathsf{Id}\,\otimes\,\widetilde{b}} V \otimes (V \otimes V) &&\cong (V \otimes V) \otimes V  \xrightarrow{b\,\otimes\,\mathsf{Id}} V ~.
}
Because of the equivalence in \autoref{fig:sdiag}, the composition must be the identity. This implies that $b$ is nondegenerate and $V = F(N)$ is \underline{finite-dimensional}. This is part of something called \emph{dualizability}.
\tightfootnote{See, however, remarks in \autoref{sec:FunctorialApproachQFT}, where we relax dualizability.}

\begin{proof} Choose a basis $\{v_i\}$ for $V$ and define $b(v_i, v_j) = b_{ij}$. Then $\widetilde{b}(1) = \widetilde{b}^{ij}v_{i}\otimes v_j$. We learn that $\widetilde{b}^{ij}b_{jk} = \delta^{i}_{k}$ $\implies$ $b_{ij}$ is invertible. If $V$ is infinite-dimensional, the sums defining $b$ and $\wt{b}$ are infinite sums and might not correspond to well-defined vectors. For example, if $V$ is a Hilbert space, we could choose an ON basis $\{ v_i\}$. Then $b_{ij}$ must decay rapidly as $i,j\to \infty$, and consequently $\wt{b}$ will not converge. 
\end{proof}

\item[5.)] It follows that
\eqa{
 & F(N \times \IS^1) &&= \dim_{\IC}(F(N)) ~,\\
 & F\bigg(\!\!\!\!\!\!\!\vcenter{\hbox{\dualizabilitytwoNOARROW\!\!\!\!\!\!\!\!\!\!\dualizabilityoneNOARROW}}\!\!\!\!\!\!\!\bigg) &&= \widetilde{b}^{ij}b_{ji} = \sum_i \delta_{i}^{i} =\dim_{\IC}(F(N))  ~.
}

\item[6.)] We claim that $F(N_{n-1})$ is a representation of $\Diff(N_{n-1})$ which  factors through a representation of the mapping class group $\pi_0\big(\Diff(N_{n-1})\big)$. To see this, consider 
$F(\{M_n, \theta_{\text{in}}, \theta_{\text{out}}\})$ for the bordism 
$M_n = N_{n-1} \times [0,1]$ with $\theta_{\text{in}}=\mathsf{Id}$, but  
$\theta_{\text{out}}$ some general element of $\Diff(N_{n-1})$. For this 
bordism, $F(\{M_n, \theta_{\text{in}}, \theta_{\text{out}}\}):=\rho(\theta_{\text{out}})$ 
is a linear map on $F(N_{n-1})$, and moreover, composition of such bordisms for 
two choices of $\theta_{\text{out}}$ shows that these maps form a representation 
of $\Diff(N_{n-1})$. If there is a smooth path of diffeomorphisms connecting 
$\theta_{\text{out}}$ to the identity then we can use that path in the definition of a 
diffeomorphism of bordisms to show that for such topologically trivial 
diffeomorphisms, $\rho(\theta_{\text{out}})$ is the identity transformation on 
$F(N_{n-1})$. Therefore, the representation factors through to a representation of the 
quotient, i.e., the mapping class group (see \autoref{rem:bordism-remark}(3) below). 
\end{enumerate}

\begin{remark}\label{rem:bordism-remark}
$\,$
\begin{enumerate}
\item The finite-dimensionality result of item \ref{item:dualizability} above 
leads to the inconvenient fact that many ``topological field theories'' found in the physics literature are not strictly topological field theories as we have defined them. In particular, the original example of Witten \cite{Witten:1988ze}, his field-theoretical formulation of the Donaldson invariants does not fit the above axiom system. Another notable example is Chern-Simons theory for noncompact groups. See \autoref{subsec:nonexamples} on ``non-examples'' below for further discussion.  

\item  In many discussions of QFT, the overall normalization of the path integral gets no respect. This is not the case in TFT, where the overall normalization has a definite meaning, and for some manifolds, it is even quantized. This is clear from the result 
\be
F_{\rm ampl}(N \times \IS^1) = \dim_{\IC}(F_{\rm states}(N)) ~.
\ee
Note that the RHS is an integer. 
One of the many applications is to produce topological invariants. 
The partition function $F(M)$ is a topological invariant and often an 
enumerative invariant, counting geometrical objects of some kind. 
For examples:
(1) In supersymmetric quantum mechanics (SQM), the partition function on the circle (with periodic boundary conditions for fermions) is the 
Euler character of the target space. (2) In Donaldson-Witten theory, the partition function (without insertions of observables) is a signed sum over instanton solutions. (3) In topological string theory, the partition function counts the number of holomorphic maps into a target space.

\item \emph{Mapping class groups.}  For a general topological space $\mathfrak{X}$, $\pi_0(\mathfrak{X})$ does not admit any natural group structure. However, one can give a topology on the space of diffeomorphisms of a closed manifold so that $\mathsf{Diff}(N)$ is a topological group. Now, for any topological group $G$, the connected component of the identity, $G_0$, is a normal subgroup, a.k.a. an invariant subgroup. It is an easy, but good,  exercise to prove that. Therefore, one has an exact sequence 
\eqa{
& 1 &&\to G_0 \to G \to \pi_0(G) \to 1 ~, \label{eq:conn-comp-sec}
}
so $\pi_{0}(G) \simeq G/G_0$ is a group. For $G=\mathsf{Diff}(N)$, it is known as the \emph{mapping class group} of $N$.

A good example of a nontrivial mapping class group which has played a very notable role in many physical applications is the case where $N$ is a two-dimensional torus. We can exhibit it as  $N = \IT^2= (\IR \oplus \IR)/\IZ \oplus \IZ$, i.e., as the quotient of the plane by translations along the $x$ and $y$ directions. If $(\sigma^1, \sigma^2)$ denote the coordinates on $\IT^2$, identified as $\sigma^{i} \sim \sigma^{i} + n$ for $n \in \IZ$ and $i = 1, 2$. The linear transformation on $\IR \oplus \IR$,
\eqa{
& \begin{pmatrix} \sigma^1 \\ \sigma^2 \end{pmatrix} &&\mapsto \begin{pmatrix} a & b \\ c & d \end{pmatrix} \begin{pmatrix} \sigma^1 \\ \sigma^2 \end{pmatrix} ~, \qquad \begin{pmatrix} a & b \\ c & d \end{pmatrix} \in \mathsf{GL}(2,\IZ) ~,
}
defines a diffeomorphism, which is compatible with the $\IZ \oplus \IZ$ group action, and descends to a diffeomorphism on the torus.2 The projection to the quotient group $\pi_0(\Diff(\IT^2))$ is nontrivial because it acts nontrivially on $H_1(\IT^2; \IZ)$.  The reader should be warned that, although this example is pervasive, it is somewhat \uwave{atypical} and unrepresentative since we have presented a $\mathsf{GL}(2,\IZ)$ \underline{subgroup} of $\Diff(\IT^2)$ rather than as a \underline{quotient group}. In general, $\pi_0(G)$ is \underline{not} a subgroup of $G$ and the sequence \eqref{eq:conn-comp-sec} above does not split.

\item \emph{Bordism groups.} Using the data of $\theta_{\text{in}}$, $\theta_{\text{out}}$ we can \underline{glue} together bordisms:
\be
\begin{tikzcd}
N^{0} \arrow[r, "{(M,\,\theta,\,\theta)}"] \arrow[rr, bend right, "{(M,\,\theta,\,\theta)\circ(M,\,\theta,\,\theta)}", swap] & N^1 \arrow[r, "{(M,\,\theta,\,\theta)}"] & N^2 ~,
\end{tikzcd}
\ee
to prove that bordism is an equivalence relation on $(n-1)$-manifolds. 
Now, the disjoint union of manifolds can be used to define a 
disjoint union of bordism classes. Therefore, the bordism classes 
of $n$-dimensional manifolds -- denoted $\Omega_n$ -- form a monoid. 
The identity element of this monoid is the empty $n$-dimensional manifold $\emptyset_n$. Now consider the picture of \autoref{fig:identitybordism}.
\begin{figure}[H]
\centering
\includegraphics[width=1.8in]{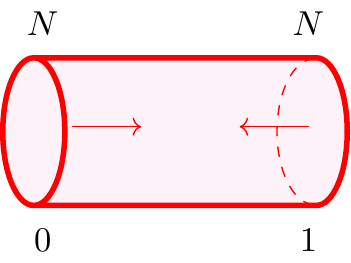}
\caption{The bordism from $N \amalg N$ to $\emptyset$. }\label{fig:identitybordism}
\end{figure}
This shows that  
\eqa{
 & [N] \amalg [N] &&\cong [\emptyset_{n}] ~.
}
Therefore, $\Omega_n$ is an Abelian group (which is completely 2-torsion). 
Note that $\Omega_{1} = \{0\}$ is the trivial group because every circle bounds a disk. But $[\IR\IP^2] \in \Omega_2$ is nontrivial: if $\exists$ $M_3$ such that $\partial M_3 = \IR\IP^2$, then gluing yields $M_3 \bigcup\limits_{\IR\IP^2} M_3$ $=$ closed $3$-fold with Euler characteristic $2\chi(M_3)-1$, but this must vanish. In fact, by the classification theorem for surfaces \cite{Zeeman1962,Gallier:2008,GallierXu:2012}, $\Omega_2 \cong \IZ/2\IZ$. By taking products of manifolds, we can turn $\oplus_n \Omega_n$ into a ring. 

It turns out that we can equip manifolds with topological and geometrical structures and still define bordism groups. One simple example is orientation. Another is (s)pin structure. So there are also oriented bordism groups, (s)pin bordism groups, etc. 
Bordism groups play an especially important role in the application of TFT to the mathematical theory of topological phases of matter.

\end{enumerate}

\end{remark}

\SectionWithHeader{Example 1: Topological Field Theory In One Dimension}{Example 1: Topological Field Theory In One Dimension}{sec:1dTFT}

What data do we need to define a one-dimensional topological field theory?  We first address the case when our bordisms are unoriented. 
Then we comment on how the picture changes when they are oriented.

\bigskip 
\noindent \underline{Unoriented One-Dimensional TFT} 

One-dimensional bordisms will connect zero-dimensional manifolds on their boundary. There exists a unique connected $0$-dimensional unoriented manifold, namely, the humble point. All other zero-dimensional manifolds are finite disjoint unions of points. 
(They are finite because we assume our manifolds to be compact.) 
By the rule \eqref{eq:LOC1}, 
we know all statespaces associated to zero-dimensional manifolds once we know the value of $F_{\rm states}$ on an unoriented point. That value 
must be some finite-dimensional vector space $V$, i.e., $F_{\rm states}({\rm pt}) = V$. We could work over any field, but for definiteness, we will take it to be a complex vector space. 
Thus, the statespaces are: 
\eqa{
 F\big(\underbrace{{\rm pt}\coprod {\rm pt} \cdots \coprod{\rm pt}}_{n \text{ times}}\big) & \cong  V^{\otimes n} ~.
}
Note that the diffeomorphism group is the symmetric group $\mathsf{S}_{n}$ and indeed, $V^{\otimes n}$ is a representation of $\mathsf{S}_{n}$ acting as a permutation of the factors. 

The vector space $V$ has a nondegenerate bilinear form because of the bordism: 
\def\onedexample{
\begin{tikzpicture}[baseline={([yshift=-.5ex]current bounding box.center)}]
\coordinate (A) at (0, 0);
\coordinate (D) at (0, 1);

\node [fill=violet,circle,minimum size=2pt,inner sep=-2] at (A) {};
\node [fill=violet,circle,minimum size=2pt,inner sep=-2] at (D) {};


\draw[->,red,thick] (-0.15,0.25) -- (0.2,0.25);
\draw[->,red,thick] (-0.15,0.75) -- (0.2,0.75);

\draw[ultra thick,blue] (0,0) arc (-90:90:0.5);

\end{tikzpicture}
}

\eqa{
& F\bigg(\,\,\onedexample\,\,\bigg) &: V \otimes V \xlongrightarrow{b} \IC ~.
}
Running the bordism the other way produces a state  $\wt{b}(1) \in V \otimes V$. Moreover, 
by the $S$-diagram of \autoref{fig:sdiag}, $b$ must be nondegenerate. Indeed, $\wt{b}$ can be computed from $b$. Recall that if we choose a basis, then $\widetilde{b}^{ij}b_{jk} = \delta^{i}_{k}$.  

\begin{figure}[H]
\centering
\includegraphics[width=3in]{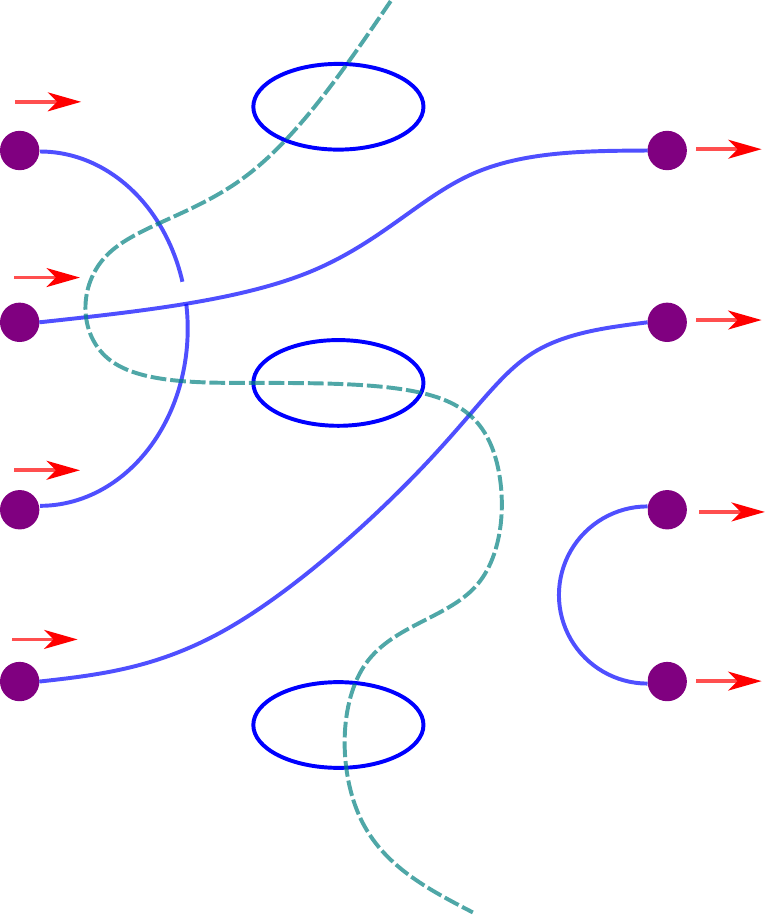}
\caption{A bordism $B$ leading to a linear map $F_{\rm ampl}(B): V^{\otimes 4} \to V^{\otimes 4}$. The dotted line shows one way of cutting up the bordism so that $F_{\rm ampl}(B)$ can be computed in terms of the defining data $(V,b)$. It is a nontrivial fact, proven using Morse-theoretic ideas described below, that no matter how we cut along intermediate channels, we will produce the \underline{same} linear operator $F_{\rm ampl}(B)$.}
\label{fig:V4toV4}
\end{figure}

That's all the data we need to specify!   
We now have the basic data to compute any amplitude we like, such as $V^{\otimes 4} \rightarrow V^{\otimes 4}$, see \autoref{fig:V4toV4}.

\bigskip 
\noindent \underline{Oriented One-Dimensional TFT} 

It is interesting to contrast the above description with the topological field theory we obtain using \underline{oriented} bordisms. In one-dimensional oriented bordism theory, the one-dimensional manifolds carry an orientation. There are therefore two kinds of points: A positively oriented point ${\rm pt}_+$ is a \underline{source} and a negatively oriented point ${\rm pt}_-$ is a \underline{sink}. The basic bordisms we can draw are: 
\def\onedexampleNOARROWBENDRIGHT{
\begin{tikzpicture}[baseline={([yshift=-.5ex]current bounding box.center)}]
\coordinate (A) at (0, 0);
\coordinate (D) at (0, 1);

\node [fill=violet,circle,minimum size=2pt,inner sep=-2] at (A) {};
\node [fill=violet,circle,minimum size=2pt,inner sep=-2] at (D) {};



\draw[ultra thick,blue] (0,0) arc (-90:90:0.5);
\node [] at (-0.5, 0) {{${\rm  {pt}}_-$}};
\node [] at (-0.5, 1) {{${\rm  {pt}}_+$}};

\end{tikzpicture}\label{eq:1d-Oriented-DualityData}
}
\def\onedexampleNOARROWBENDLEFT{
\begin{tikzpicture}[baseline={([yshift=-.5ex]current bounding box.center)}]
\coordinate (A) at (0, 0);
\coordinate (D) at (0, 1);

\node [fill=violet,circle,minimum size=2pt,inner sep=-2] at (A) {};
\node [fill=violet,circle,minimum size=2pt,inner sep=-2] at (D) {};



\draw[ultra thick,blue] (0,0) arc (-90:-270:0.5);
\node [] at (0.5, 0) {{${\rm  {pt}}_+$}};
\node [] at (0.5, 1) {{${\rm  {pt}}_-$}};

\end{tikzpicture}
}
\def\onedexamplegeneratingleft{
\begin{tikzpicture}[baseline={([yshift=-.5ex]current bounding box.center)}]
\coordinate (A) at (1, -1);
\coordinate (D) at (-1, 1);
\node [fill=violet,circle,minimum size=2pt,inner sep=-2] at (A) {};
\node [fill=violet,circle,minimum size=2pt,inner sep=-2] at (D) {};

\draw[ultra thick,blue] (0,0) arc (-90:90:0.5);
\draw[ultra thick,blue] (0,-1) arc (-90:-270:0.5);
\draw[ultra thick,blue] (-1,1) -- (0,1);
\draw[ultra thick,blue] (0,-1) -- (1,-1);

\node [] at (-1.6, 1) {{${\rm  {pt}}_+$}};
\node [] at (1.6, -1) {{${\rm  {pt}}_+$}};

\node [] at (2.5,0) { {$=$} };

\draw[ultra thick,blue] (4.5, 0) -- (6.5, 0);
\coordinate (E) at (4.5, 0);
\coordinate (F) at (6.5, 0);
\node [fill=violet,circle,minimum size=2pt,inner sep=-2] at (E) {};
\node [fill=violet,circle,minimum size=2pt,inner sep=-2] at (F) {};

\node [] at (4, 0) {{${\rm  {pt}}_+$}};
\node [] at (7, 0) {{${\rm  {pt}}_+$}};

\end{tikzpicture}
}
\def\onedexamplegeneratingright{
\begin{tikzpicture}[baseline={([yshift=-.5ex]current bounding box.center)}]
\coordinate (A) at (-1, -1);
\coordinate (D) at (1, 1);
\node [fill=violet,circle,minimum size=2pt,inner sep=-2] at (A) {};
\node [fill=violet,circle,minimum size=2pt,inner sep=-2] at (D) {};

\draw[ultra thick,blue] (0,0) arc (-90:-270:0.5);
\draw[ultra thick,blue] (0,-1) arc (-90:90:0.5);
\draw[ultra thick,blue] (1,1) -- (0,1);
\draw[ultra thick,blue] (0,-1) -- (-1,-1);

\node [] at (1.6, 1) {{${\rm  {pt}}_-$}};
\node [] at (-1.6, -1) {{${\rm {pt}}_-$}};

\node [] at (2.5,0) { {$=$} };

\draw[ultra thick,blue] (4.5, 0) -- (6.5, 0);
\coordinate (E) at (4.5, 0);
\coordinate (F) at (6.5, 0);
\node [fill=violet,circle,minimum size=2pt,inner sep=-2] at (E) {};
\node [fill=violet,circle,minimum size=2pt,inner sep=-2] at (F) {};

\node [] at (4, 0) {{${\rm  {pt}}_-$}};
\node [] at (7, 0) {{${\rm  {pt}}_-$}};
\end{tikzpicture}
}
\be
\begin{split}
 \text{Generating objects:}  \quad & { {\rm pt}_+} \text{ and } { {\rm pt}_-} \\
 \text{Generating 1-morphisms:} \quad &    \onedexampleNOARROWBENDRIGHT \quad \onedexampleNOARROWBENDLEFT 
 %
\end{split}
\ee
The generating relations are:
\be
\begin{split}
  &\onedexamplegeneratingleft ~,\\
  \text{ and } \\
  &\onedexamplegeneratingright ~.
\end{split}
\ee

We can now define 
$F_{\rm states}({\rm pt}_+) = V$. 
In contrast to the unoriented case, this is the only data we need. 
It follows, again from the $S$-diagram argument of  
\autoref{fig:sdiag}, that  $F_{\rm states}({\rm pt}_-) = V^\vee$ is the linear dual. All amplitudes can be computed just in terms of those data. 

\begin{exbox}{Bordisms to linear maps and amplitudes} 
\begin{itemize}
\item[(a))] Show that the linear map associated with the bordism $B$ of \autoref{fig:V4toV4} is given by 
\be 
F(w_1 \otimes w_2 \otimes w_3 \otimes w_4) = 
(\dim\,V)^3 b(w_1,w_3) w_2 \otimes w_4 \otimes \wt{b}(1) ~,
\ee
where $w_i\in V$, $i=4$ is a choice of four vectors. One then extends to arbitrary elements of $V^{\otimes 4}$ by linearity. 

\item[(b))] Now endow the bordism of \autoref{fig:V4toV4} with some orientations, and compute the resulting amplitude. 
\end{itemize}

\end{exbox}

\SectionWithHeader{Example 2: Oriented TFT In Dimension Two}{Example 2: Oriented TFT In Dimension Two}{sec:Example2-TFT-Dim2}

\subsection{Defining Data And Frobenius Algebras}

For $n=2$, to avoid complications with the classification of unoriented surfaces, we work with oriented bordisms. There exists a unique connected oriented $1$-manifold without boundary, namely, an oriented $\IS^1$. Suppose $F(\IS^1) = V$. There are two algebraic structures: 
\begin{enumerate}
\item Canonical bordism $\IS^1 \to \emptyset$:
\eqa{
&\text{the disk } &&= 
\vcenter{\hbox{\includegraphics[]{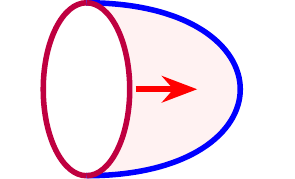}}}
\theta : V \to \IC ~.
}
\item Multiplication: Pair of pants:
\eqa{
\vcenter{\hbox{\includegraphics[]{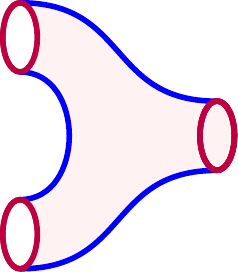}}}
\qquad
& m : V \otimes V &&\to V ~.
}
\end{enumerate}

\begin{figure}[h]	
\centering
\includegraphics[width=2.8in]{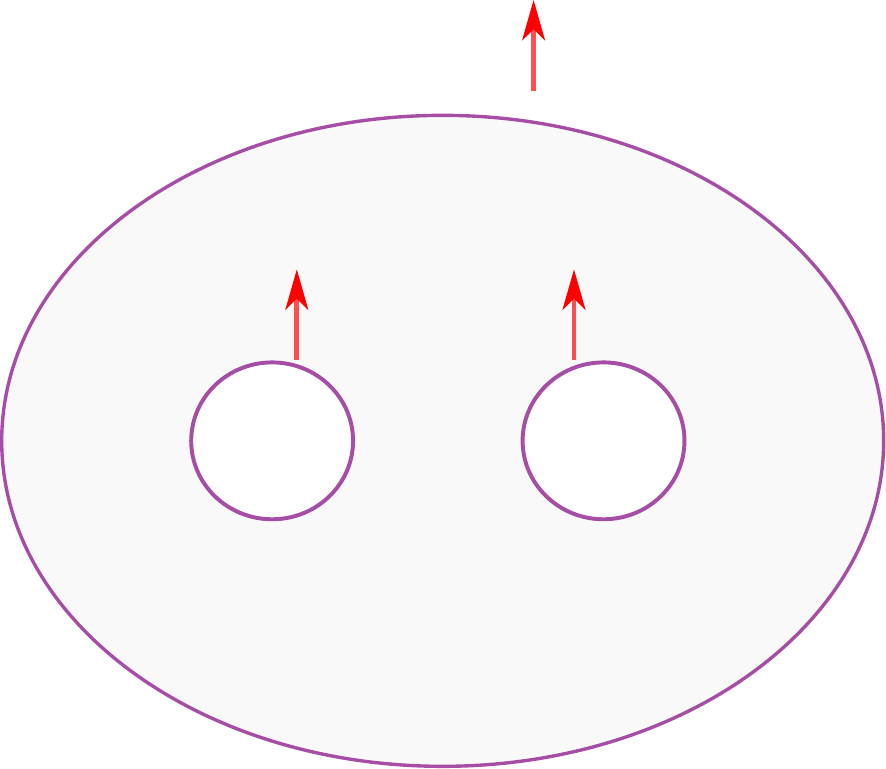}
\caption{Disks within disks.}
\label{fig:diskswdisks}
\end{figure}

A useful way to view the pair of pants is shown in \autoref{fig:diskswdisks}. We have disks within disks. An advantage of this viewpoint is that it readily generalizes to higher dimensions. 

\autoref{fig:diskswdisks} has some nice immediate corollaries: 

\paragraph{Corollary 1:} $m: V \otimes V \to V$ is a commutative and associative multiplication.

\paragraph{Corollary 2:} 
In the $n =2$ case, the bilinear form $b(v_1, v_2) = \theta(v_1 \cdot v_2)$ is nondegenerate. Proving this is a good exercise. 
\tightfootnote{Answer to the exercise: $\theta$ corresponds to capping off the maps at the waist. Then the surface is homeomorphic to a cylinder. Apply the $S$-diagram argument.  }

\paragraph{Corollary 3:} As we noted above, \autoref{fig:diskswdisks} generalizes to $n$ dimensions. From this, we deduce that for   \underline{any} $n$-dimensional TFT, $F(\IS^{n-1})$ is a commutative and associative algebra.

\begin{definition}[\textcolor{red}{Frobenius algebra}] An associative algebra with $\theta: V \to \IC$ such that $b(v_1, v_2) = \theta(v_1 \cdot v_2)$ is nondegenerate, is a \textbf{Frobenius algebra}.
\end{definition}

For an elementary exposition of Frobenius algebras, see for example, \cite{Kock2003}.

\subsection{Sewing Theorem And Morse Theory}

\begin{figure}[H]
\centering
\includegraphics[]{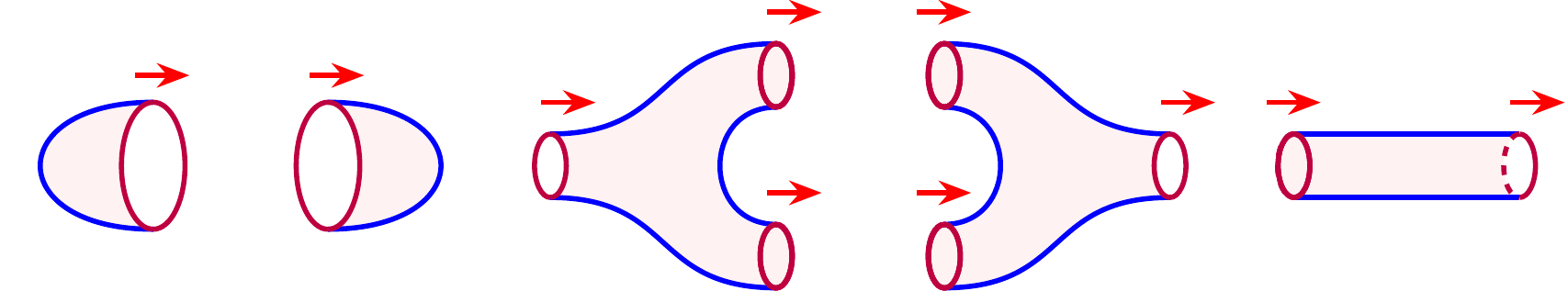}
\caption{Elementary pieces.}\label{fig:elementarypieces}
\end{figure}

Using the data $(V, m, \theta)$ one can compute \underline{any} amplitude
by cutting into the elementary pieces shown in \autoref{fig:elementarypieces}.


\subsection{The Sewing Theorem: Morse Theory Proof}

Now a given bordism $B$, it can be cut up into elementary bordisms in many ways. No matter how we cut it up, we should get the same answer for $F(B)$. This would seem to place constraints on the linear transformations we associate to the elementary bordisms.

\begin{theorem}[\textcolor{red}{Sewing Theorem}]\label{thm:SewingTheorem}
If $(V,m,\theta)$ is a commutative Frobenius algebra, then all amplitudes are well-defined and can be computed by cutting up the bordisms in any way. Put differently: To give an $n=2$ dimensional TFT is to give a commutative finite-dimensional Frobenius algebra. 
\end{theorem}

For a long time, the Sewing Theorem was a folk theorem attributed variously to D. Friedan, R. Dijkgraaf, G. Segal, and perhaps others. It appears, for example, in R. Dijkgraaf's Ph.D. thesis \cite{DijkgraafPhD}.
A careful proof can be found in the appendix of the expository paper  \cite{Moore:2006dw}. We now briefly review the proof from \cite{Moore:2006dw}.  The basic idea is to make use of Morse-Cerf theory. 

Recall that a \emph{Morse function} on a manifold $M_{n}$ is a smooth real-valued function on $M_{n}$ that has isolated nondegenerate critical points. A \emph{critical point} $p\in M_{n}$ is a point where $df(p)=0$. The associated \emph{critical value} is $f(p)$. A critical point is called \emph{nondegenerate} if the \emph{Hessian at $p$}, defined to be  
\be 
\left.\frac{\partial^2 f}{\partial x^i \partial x^j}\right\vert_{p} ~,
\ee
in local coordinates $x^i$ near $p$, is a nondegenerate symmetric matrix. The nondegeneracy condition does not depend on what coordinate system we use near $p$. Because $p$ is a critical point,  we don't need to use covariant derivatives to define the second derivative. 

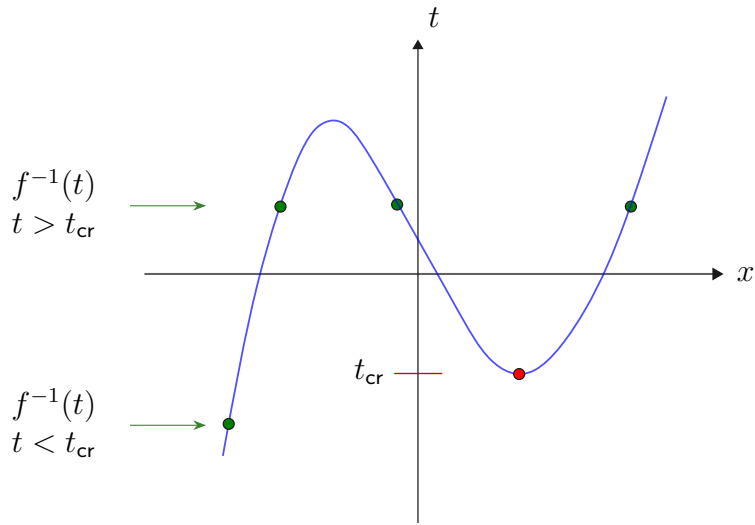
\begin{figure}[h]
  \centering
  \begin{tikzpicture}
  \node (graph) [inner sep = 0pt] {\includegraphics[width=3in]{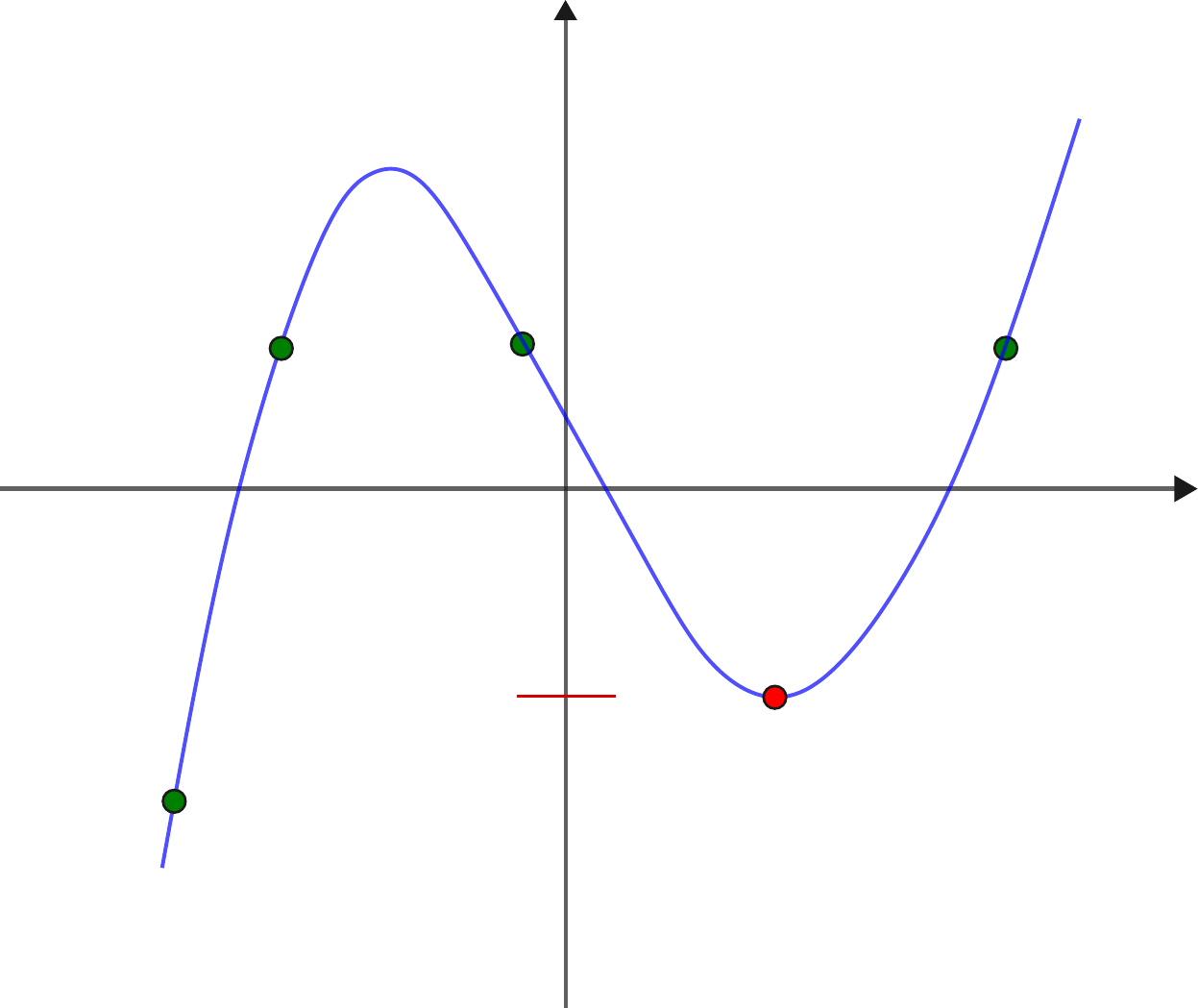}};
  \node at (4.1,0.1) {$x$};
  \node at (0,3.5) {$t$};
  \node at (-0.85,-1.2) {$t_{\mathsf{cr}}$};
  \draw (-4,-1.91) edge[-Stealth,OliveGreen] (-3,-1.91);
  \node at (-5,-1.91) {$\begin{array}{c} f^{-1}(t)\\ t < t_{\mathsf{cr}}\end{array}$};
  \draw (-4,1) edge[-Stealth,OliveGreen] (-3,1);
  \node at (-5,1) {$\begin{array}{c} f^{-1}(t)\\ t > t_{\mathsf{cr}}\end{array}$};
  \end{tikzpicture}
  \caption{A one-dimensional bordism between two points labeled by a 
  Morse function whose value is $t$. }
  \label{fig:ex1}
\end{figure}

\begin{figure}[h]
  \centering
  \begin{tikzpicture}
  \node (graph) [inner sep = 0pt] {\includegraphics[width=3in]{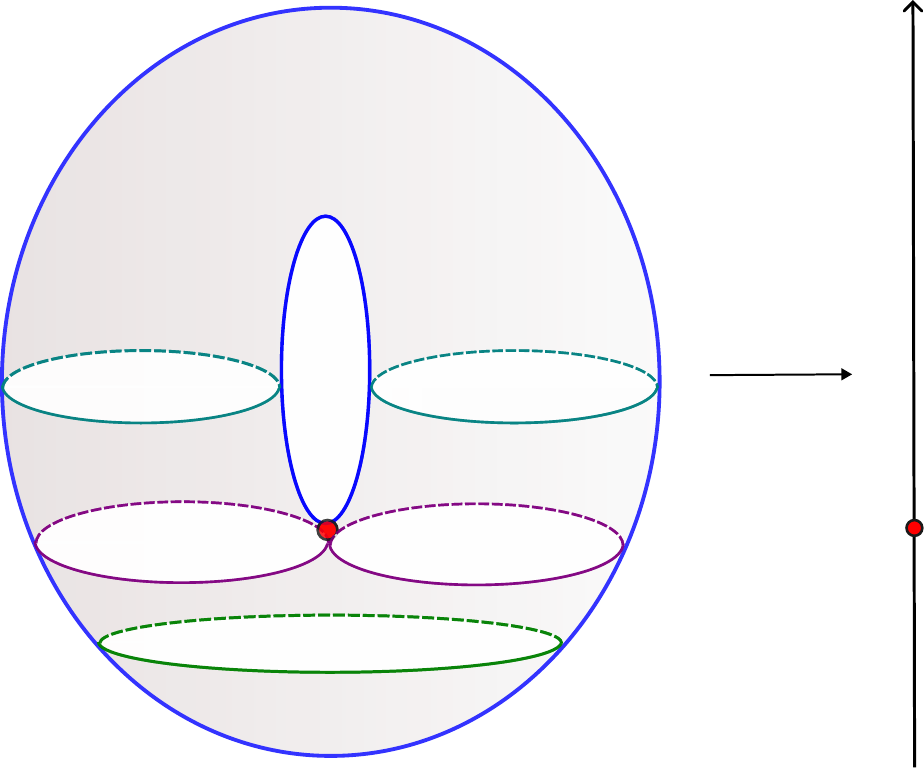}};
  \node at (4.1,3.1) {$\IR_{t}$};
  \node at (2.7,0.4) {$f$};
  \node at (5.4,-0.8) {\textcolor{red}{critical value}};
  \end{tikzpicture}
  \caption{A two-dimensional bordism from $\emptyset$ to $\emptyset$ labeled by a 
  Morse function whose value is $t$. }
  \label{fig:ex2}
\end{figure}
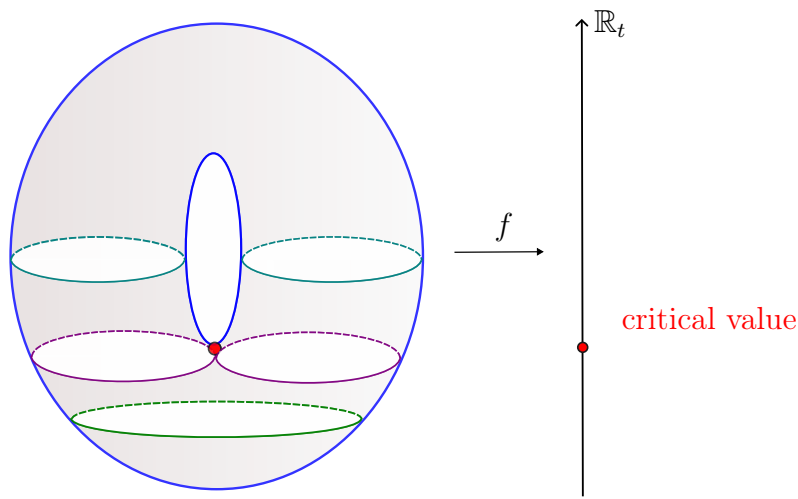

A Morse function gives  a decomposition of $M_{n}$ into level sets 
 $f^{-1}(t) = N_{t} \subset M_{n}$, for $t\in \IR$,  to be thought of as ``time slices,'' 
or spatial slices.   Now $f^{-1}(t)$ will be a nice smooth manifold \underline{unless} $t$ is a critical value. 
A Morse function on a bordism $M_{n}: N_{n-1}^{0} \to N_{n-1}^{1}$ is \emph{excellent} if it is constant on $N_{n-1}^0$, $N_{n-1}^1$,  and the critical points can be ordered so that the critical values are
\eqa{
& c_0 = f(N_{n-1}^0) &&< c_1 < \cdots < c_N < c_{f} = f(N_{n-1}^1) ~.
}
The spatial slices $f^{-1}(t)$ are all diffeomorphic for all $t$ in an interval $c_i < t < c_{i+1}$. However, the slices for values of 
$t$ in two successive intervals, for example $c_{i-1} < t_1 < c_i$ and 
$c_i < t_2 < c_{i+1}$ will differ by a (simple) change of topology. 
For example, in \autoref{fig:ex1} and \autoref{fig:ex2}, the topology change as $t$ increases through a critical value are readily seen.
$\,$\\
\noindent \textbf{Example: $n=2$.} See \autoref{fig:ex2}.

Now we can \underline{change} time-slicings by considering a path of smooth functions $f_s$ which are excellent Morse functions for generic $s$. A crucial theorem of Cerf 
\cite{ Cerf1970} states that in a natural topology
\tightfootnote{The topology is known as the \emph{Whitney topology}. See \cite[Ch. 20]{FreedBordism}.} 
on the set of infinitely differentiable real functions on $M_{n}$, i.e., on  $C^\infty(M_{n} \to \IR)$, the set of excellent Morse functions is open and dense, but \underline{disconnected}. Moreover, we can define a function $f: M_n \to \IR$ to be ``good'' if it is Morse everywhere except for one or two critical points, and, (in the case that $M$ is two-dimensional) one has:
\begin{itemize}\itemsep 0pt
\item One critical point with a function locally of the form $\pm y^2 + x^3$,
\item Two critical points with the same value.
\end{itemize}
Then, a remarkable result \cite{Cerf1968,Cerf1970}    states that the set of excellent and good functions is a \emph{connected set}. The good but not excellent functions form a real codimension-one subset. 

It follows from the results quoted above that a generic path of $\CC^\infty$ functions, $f_s$ with path parameter $s$,  in the space of excellent and good functions, will cross a finite set of critical values $s_1, \ldots, s_k$ where the functions are good but not excellent.

$\,$\\
\noindent \textbf{Example:} Consider the case $n=1$, and consider the path of functions  $f_s(x) = \frac{x^3}{3} - s x$, for $s\in (-1, 1)$. 
For $s\not=0$, the function is excellent, but for $s=0$, it is only good.
Note that as $s$ changes from positive to negative the nature of the function changes: 
\begin{figure}[H]
\centering
\includegraphics[width=2.8in]{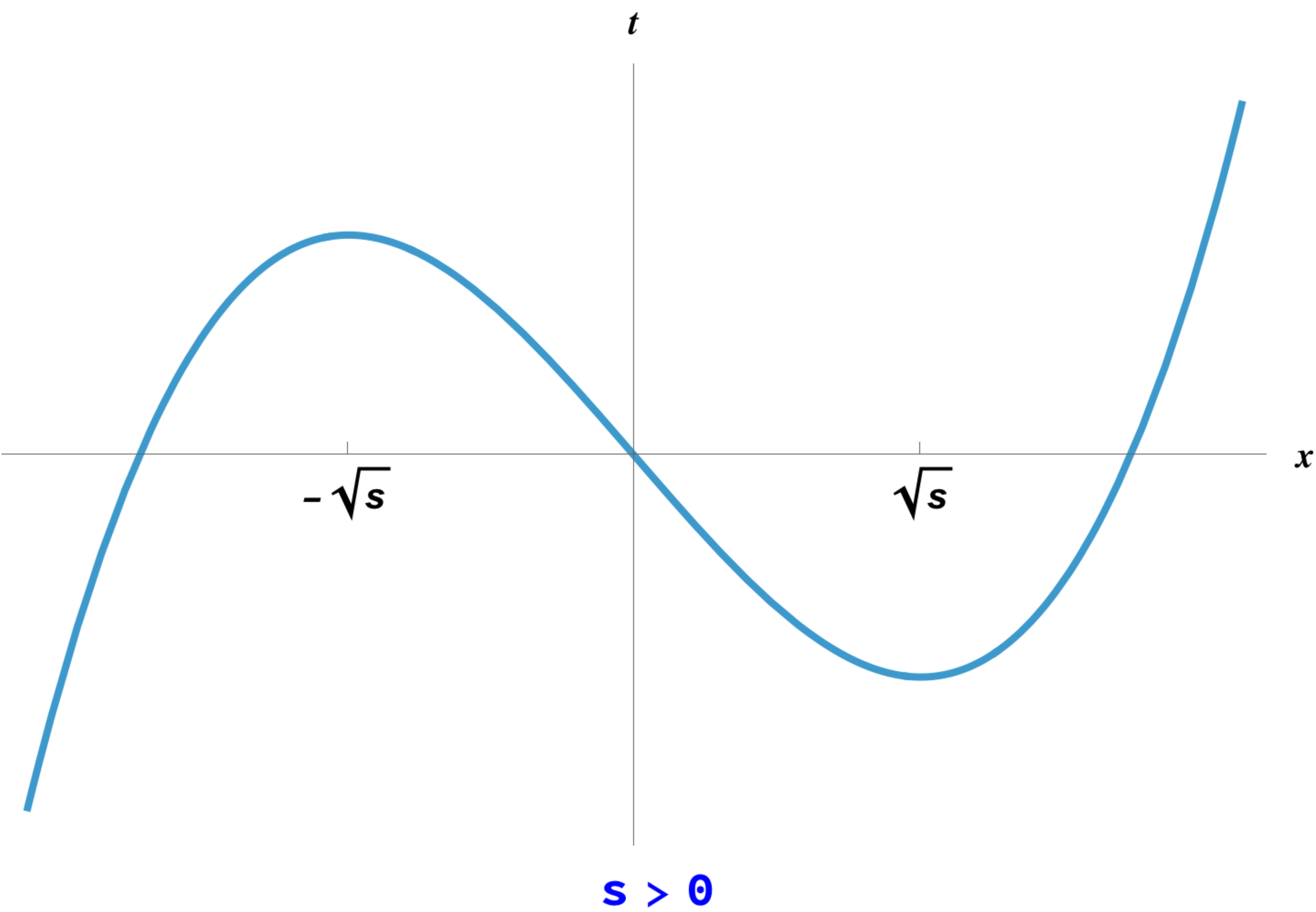}\qquad
\includegraphics[width=2.8in]{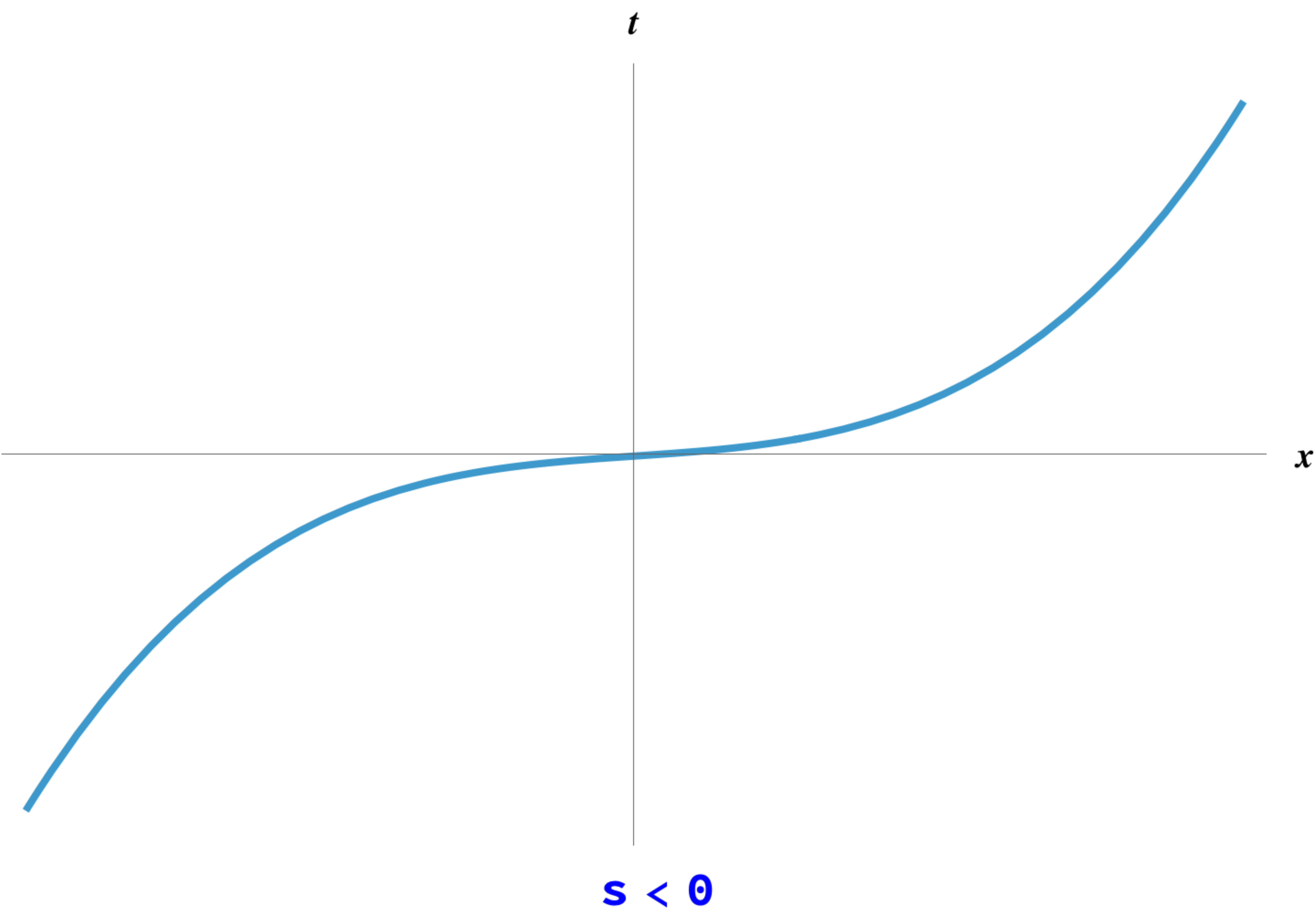}
\caption{The function $f_{s}(x)$ for $s >0$ (left), and $s < 0$ (right).}
\end{figure}
It now follows from Cerf-Morse theory that in the 2d field theory, any two changes of time slicings are related by a sequence of elementary changes in the decomposition of the bordism into the the elementary pieces of \autoref{fig:elementarypieces}. (Each of these elementary pieces corresponds to an excellent Morse function with a single critical point.) 
Invariance under those elementary changes is guaranteed by the algebraic axioms of a commutative Frobenius algebra. This is how the sewing theorem is proven. See  \cite{Moore:2006dw} for more complete details.

\subsection{Semisimplicity}\label{subsec:SemiSimplicity}

 Choose an ordered basis $\{v_i\}$ for an algebra $V$. Consider the operator $L_i$ defined by left-multiplication by $v_i$. It has matrix elements:
\eqa{
& L_i(v_j) &&= v_i v_j = N_{ij}{}^{k} v_{k} ~.
}
If the algebra $V$ is commutative, then   $[L_i, L_j] = 0$. 
If  the $L_i$ are all diagonalizable, we say the algebra $V$ is \emph{semisimple}. In this case, there is a basis of idempotents $\{\varepsilon_i\}$:
\eqa{
& \varepsilon_i \varepsilon_j &&= \delta_{ij}\varepsilon_i ~.
}
Therefore, for a semisimple commutative Frobenius algebra, the  
only invariants of the Frobenius algebra are the number of 
idempotents (i.e., the dimension), and the values of the trace 
applied to the idempotents are $\theta(\varepsilon_i) = \theta_i$.

\begin{remark}
$\,$
\begin{enumerate}

\item If we view this model as a baby model of string theory with a zero-dimensional target ``spacetime,'' then the target spacetime is a finite union of points, the Frobenius algebra $V$ is the space of functions on spacetime, and $\theta_i$ can be interpreted as defining the value of the dilaton/string coupling on the $i^{th}$ component ``universe.''

\item We can also use the $n=1, 2$ theories as topological models of quantum gravity. In this context, they are useful playgrounds for exploring the role of topology change and ``baby universes'' in Euclidean Quantum Gravity (EQG). An important paper on this is by D. Marolf and H. Maxfield \cite{Marolf:2020xie}, with clarifications, generalizations and further extensions in the paper by A. Banerjee and G. W. Moore \cite{Banerjee:2022pmw}, which in turn inspired a general framework for EQG laid out by D. Friedan \cite{Friedan:2023vxx}.
\end{enumerate}

\end{remark}

\begin{exbox}[exercise:frobenius-0]{Frobenius algebra} Suppose $(V,\theta)$ is a semisimple Frobenius algebra and let $F$ be the associated oriented two-dimensional topological field theory. 

\begin{itemize}\itemsep 0pt
\item[(a)] Show that the state produced by a handle is
\eqa{
& F\bigg(\includegraphics[width=1in,valign=c]{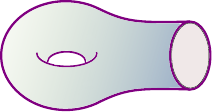}\bigg) &&= \sum_{i}\theta_{i}^{-1}\varepsilon_i ~.
}
\item[(b)] Suppose $\Sigma_g$ is a connected genus $g$ surface without boundary. Show that
\eqa{
& F(\Sigma_g) &&= \sum_{i}\theta_{i}^{1-g} ~. \label{eq:PF-FrobeniusAlgebra}
}
\item[(c)] Deduce that the sum over connected vacuum to vacuum amplitudes, i.e., the sum over connected bordisms  $\emptyset \to \emptyset$   is
\eqa{
& F_{\mathsf{connected}} &&= \sum_{i} \frac{\theta_i}{1-\theta_i^{-1}} = \sum_{i} \lambda_i ~.
}
Show that \underline{with a suitable weighting of topologies}, the full vacuum to vacuum amplitude, that is, the sum over all bordisms 
$\emptyset \to \emptyset$, is
\eqa{
& F_{\mathsf{vac} \to \mathsf{vac}} &&= \prod_{i} e^{\lambda_i} ~.
}
\end{itemize}
\end{exbox}

 \begin{exbox}{Unit element for algebra multiplication}Show that $F\bigg(\!\!\!\! 
 \vcenter{\hbox{\includegraphics[]{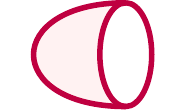}}}
\!\!\!\bigg)$ defines the unit element for the algebra multiplication in $V$ by illustrating a suitable change of Morse function.
\end{exbox}

\begin{exbox}{Associativity of algebra multiplication}Illustrate the change of Morse function that implies the multiplication on $V$ is associative.\end{exbox}

\begin{exbox}[exercise:frobenius]{Non-semisimple Frobenius algebra} 
  Consider a compact orientable manifold $X$ with all odd Betti numbers $b_i(X) = 0$. Consider the commutative algebra given by the cohomology group $H^{\bullet}(X;^{\bullet} \IC)$. 
\begin{itemize}
\item[(a)] Show that the trace defined by integral with the fundamental class defines a Frobenius algebra structure,  but that it is not semisimple. 

\item[(b)]Show that for   $X = \IC\IP^1$ we get  $\IC[x]/(x^2)$
with $\theta(x)=1$. Compute all the amplitudes in the topological field theory associated to this Frobenius algebra.

\item[(c)] Which bordisms $\emptyset \to \IS^1$ create ``states'' which are the zero-vector?   These give examples where there is not a quantum mechanical interpretation of the ``states'' created by some topology. 
\end{itemize}
\end{exbox}

\begin{remark}\label{rem:zerovector} The last exercise above brings up an important point. If we follow the traditional mathematical definition of an $n$-dimensional topological field theory, the vector in $F(N_{n-1})$ created by a bordism with $N_{n-1} = \partial M_n$  might well be the \underline{zero vector}: $F(M_n) =0$.  As we stressed in 
\autoref{sec:QM}, in traditional quantum mechanics a (pure) state is represented by a rank-one projection operator $P$. If $\psi \in \mathsf{im\,} P$ and $\psi$ is nonzero, we can make the projector onto the line through $\psi$:   $P = \frac{|\psi\rangle\langle \psi|}{ \vert\!\vert\psi\vert\!\vert^2 }$. But this cannot be done if $\psi=0$.  So the term ``statespace'' for $F(N_{n-1})$ is inaccurate. There are further examples in three-dimensional Chern-Simons theory on $M_3$, where $F(M_3)=0$ for interesting bordisms $\emptyset \to \IS^2$ \cite{Freed1991}. (The partition function vanishes for certain Lens spaces and one can cut out 
a 3-dimensional ball from such Lens spaces to produce an example.) 
There is very interesting literature applying 
Chern-Simons theory to quantum information theory \cite{Salton:2016qpp,Balasubramanian:2016sro}.
We expect that further research in this direction will likely need to address this phenomenon. 
\end{remark}
 
\SectionWithHeader{Open-Closed Oriented Two-Dimensional TFT And The Emergence Of Categories}{Open-Closed Oriented 2d TFT And Categories}{sec:open-closed-TFT-Categories}

If we think of TFT for oriented two-dimensional manifolds as a baby model of topological string theory with zero-dimensional target ``spacetime,''  it is natural to ask about the extension to open strings.
It turns out that this generalization is extremely interesting and introduces several conceptually new ingredients. 

At first, it might seem like it is no big deal. For example, the spatial manifolds we consider will simply consist of disjoint unions of both circles and intervals. But the intervals call for boundary conditions. 
Let us simply label the boundary conditions by symbols $a,b,c,\dots$ in 
some set $\CB_0$ of boundary conditions.  

Now our field theory must assign ``open string statespaces'' to oriented intervals labeled by a pair of boundary conditions: 
\eqa{
& F\Bigg(
  \begin{tikzpicture}[baseline=-0.5ex]
    \node (graph) [inner sep = 0pt] {\includegraphics[width=0.15cm,valign=c]{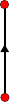}};
    \node at (-0.25,-0.70) {$a$};
    \node at (-0.25,0.70) {$b$};
    \end{tikzpicture}~
\Bigg) &&:= \mathcal{O}_{ab}  = \left(\begin{array}{c} \text{a vector space of open string states} \\ \text{with boundary conditions $a$ and $b$} \end{array}\right) ~.
}

\begin{figure}[h]
\centering
  \scalebox{0.88}{
\begin{tikzpicture}[baseline=-0.5ex]
  \node (graph) [inner sep = 0pt] {\includegraphics[width=1.5in,valign=c]{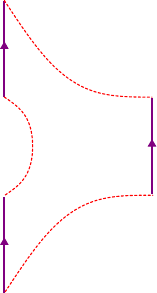}};
  \node at (-2.1,-3.5) {$a$};
  \node at (-2.1,-1.2) {$b$};
  \node at (-2.1,+1.2) {$b$};
  \node at (-2.1,3.5) {$c$};
  \node at (+2.1,1.2) {$c$};
  \node at (+2.1,-1.2) {$a$};
  \end{tikzpicture}
  }
\caption{Associativity.}\label{fig:tqft_associativity}
\end{figure}
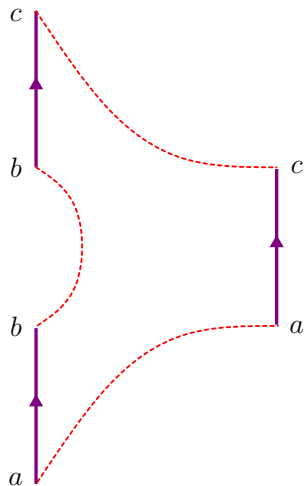

\noindent It follows from consideration of the bordism shown in  
\autoref{fig:tqft_associativity}, that 
\begin{enumerate}
\item $\CO_{aa}$ is an associative (but not necessarily commutative) \emph{algebra}.
\item $\CO_{ab}$ is a \emph{bimodule} for $\CO_{aa}\times \CO_{bb}$.
\item There is an \emph{associative} multiplication 
\eqa{
& \CO_{ab} \times \CO_{bc} &&\to \CO_{ac} ~.
}

The proof of associativity is given in \autoref{fig:tqft_proof_associativity}.
\begin{figure}[h]
  \centering
  \scalebox{0.88}{
  \begin{tikzpicture}[baseline=-0.5ex]
    \node (graph) [inner sep = 0pt] {\includegraphics[width=2in,valign=c]{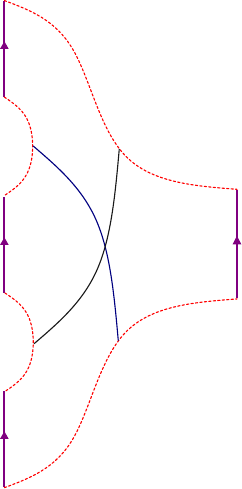}};
    \node at (-2.8,-5.1) {$a$};
    \node at (-2.8,-3.1) {$b$};
    \node at (-2.8,-1.2) {$b$};
    \node at (-2.8,+1.2) {$b$};
    \node at (-2.8,+3.2) {$c$};
    \node at (-2.8,+5.2) {$d$};
    \node at (+2.8,1.2) {$d$};
    \node at (+2.8,-1.2) {$a$};
    \end{tikzpicture}
    }
  \caption{Proof of associativity.}
  \label{fig:tqft_proof_associativity}
  \end{figure}
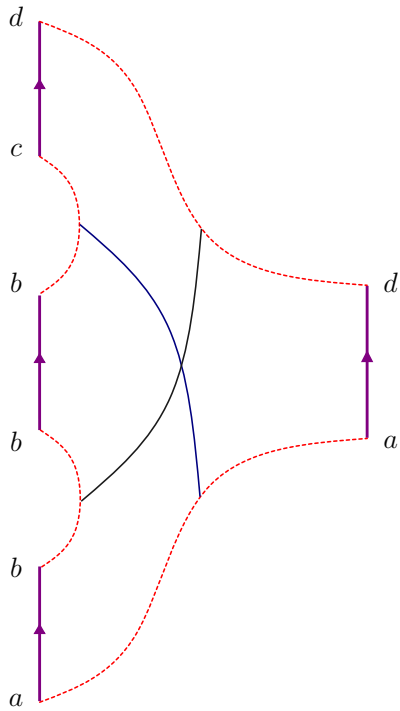
\end{enumerate}

\begin{figure}[h]
\centering
  \scalebox{0.88}{
\begin{tikzpicture}[baseline=-0.5ex]
  \node (graph) [inner sep = 0pt] {\includegraphics[width=0.5in,valign=c]{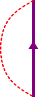}};
  \node at (0.7,+1.7) {$a$};
  \node at (0.7,-1.7) {$a$};
  \end{tikzpicture}
  }
\caption{Bordism creating $1_{aa} \in \CO_{aa}$.}\label{fig:bordism_1aa}
\end{figure}

Moreover for any boundary condition $a$, the half-disk with boundary condition $a$ creates a state $1_{aa}\in \CO_{aa}$, which is a unit 
in the algebra $\CO_{aa}$, see \autoref{fig:bordism_1aa}.

\begin{figure}[H]
\centering
\includegraphics[valign=c,width=0.8in]{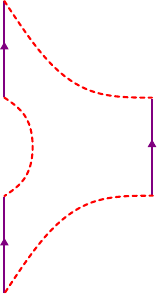}
\caption{Multiplication.}\label{fig:multchaps}
\end{figure}

The structure we have encountered is precisely that of a \emph{category}. The formal definition 
of a category is given in \autoref{sec:CategoryBackground}. Comparing with the 
definition given there, we see that for $2d$ open-closed TFT, there is a category $\CC$ of boundary conditions
whose set of objects $C_0$ is the set of boundary conditions, labeled $a,b,c,\dots$, and whose hom spaces
$\Hom(a,b)$ are the vector spaces of open string states $\CO_{ab}$. The composition in the category is the multiplication provided by the pair of chaps in \autoref{fig:multchaps}.

A natural question arises as to the analog of a Frobenius algebra structure 
and the sewing theorem (cf. \autoref{thm:SewingTheorem})
for the open-closed case. This was provided in \cite[Sec. 2.2]{Moore:2006dw}. The proof of the sewing theorem 
(in Appendix A.2 of that paper) again makes use of Morse theory, but now for manifolds with boundary. A general formulation of Morse theory for manifolds with boundary has been discussed in \cite[Sec 2.4]{Kronheimer2007}, and 
in \cite{Borodzik2016}.

\SectionWithHeader{Some Background On Categories}{Some Background On Categories}{sec:CategoryBackground}

Now that we have seen how naturally categories emerge from 
TFTs with boundary, it is useful to take a step back and review some basic background mathematical material about categories. Some pedagogical references are \cite{MacLane1978,Gelfand2003,Aspinwall:2009isa,Lurie2009-el,FreedBordism,Riehl:2017category,FreedCBMS}.
\tightfootnote{Ironically, in the 2023 lectures by G.M. at the TASI school, the definition of a category was first introduced 
in a lecture on square dancing \cite{MooreRabe:TASISquareDancing}.}

We begin with the formal definition: 
\begin{definition}[\textcolor{red}{Category}] A category $\CC$ is a collection of data $(C_0, C_1, p_0, p_1, m)$ where
\tightfootnote{If we restrict $C_0$ to be a set, it is called a \emph{small category}. That will always be the case in these notes. We will also sometimes write $\mathsf{Obj}(\CC)$ for $C_0$.}
\begin{itemize}
\item[$(a)$] $C_0, C_1$ are sets.\\
$C_0 : $ ``the set of objects'' $($also denoted as $C_0 = \mathsf{Obj}(\CC))$,\\
$C_1 : $ ``the set of morphisms.''
\item[$(b)$] $C_1 \substack{\xrightarrow{p_1} \\ \xrightarrow[p_0]{}} C_0$ the codomain and domain maps.
\end{itemize}
Denote $\{f \in C_1 ~|~ p_1(f) = y \text{ and } p_0(f) = x \} =: C(x,y) := \Hom_{C}(x,y)$.
\begin{itemize}
\item[$(c)$] Define $C_2 := C_1 \vphantom{xx}_{p_1}\!\!\times_{p_0} C_1 = \{(f,g) \in C_1 \times C_1  ~|~ p_0(f) = p_1(g)\}$, the set of composable pairs of morphisms $m: C_2 \to C_1$. Denote $m(f,g) := f \circ g$ satisfying conditions:
\begin{itemize}
\item[$(\alpha)$] $\forall$ $x \in C_0$, $\exists$ morphism $1_x \in C(x,x)$ such that,
\eqas{
 & \forall \, f \in \Hom(y, x), \qquad && 1_x \circ f = f ~, \\
 & \forall \, g \in \Hom(x, y), \qquad && g \circ 1_x = g ~.
}
\item[$(\beta)$] Consider the set of $3$ composable morphisms,
\eqa{
& C_3 &&= \{ (f, g, h) \in C_1 \times C_1 \times C_1 ~|~ p_0(f) = p_1(g) \text{ and } p_0(g) = p_1(h) \} ~.
}
The diagram
\begin{center}
\begin{tikzcd}
& C_2 \arrow[rd, "m"] & \\
C_3 \arrow[ru, "m \times \mathsf{ Id}"] \arrow[rd, "\mathsf{Id }\times m",swap] & & C_1 \\
& C_2 \arrow[ru, "m",swap] &
\end{tikzcd}
\end{center}
commutes, i.e.,
\eqa{
& (f \circ g)\circ h &&= f \circ (g \circ h) ~.
\label{eq:StrictAssociative}
}
\end{itemize}
\end{itemize}
\end{definition}

\begin{remark}In \autoref{sec:SimplicialSets}  and 
\autoref{subsec:Fields-FHT-def}, we will encounter the notion of the \textbf{opposite category} to a category $\CC$, denoted $\CC^{\rm op}$. The object and morphism sets are the same: 
\begin{align}\label{eq:OppCat}
C^{\rm op}_0 &= C_0 ~, \quad 
C^{\rm op}_1 = C_1 ~.
\end{align}
But the source and target maps are reversed: 
\be 
p_0^{\rm op} = p_1 ~, \qquad  p_1^{\rm op} = p_0  ~,
\ee
which has the effect that all arrows are reversed. If $(f,g)$ is a composable pair  of morphisms in 
$\CC$, then $(g,f)$ is a composable pair in $\CC^{\rm op}$ and 
\be 
m^{\rm op}(g,f) := m(f,g) ~.
\ee
\end{remark}

In general, it is often useful to think of a category as a directed graph, with
\eqas{
& \text{Objects} \quad &&:\quad \text{Vertices of the graph.}\\
& \text{Morphisms} \quad &&:\quad \text{Oriented edges of the graph.} \nonumber
}
For us, a very important category is the \textbf{bordism category} $\mathsf{Bord}_{\langle n-1, n\rangle}$ with
\eqas{
& \text{Objects} \quad &&:\quad \text{Smooth, closed, $(n-1)$-manifolds.}\\
& \text{Morphisms} \quad &&:\quad \text{Bordisms (up to diffeomorphism).}\\
& \text{Compositions\,\,} m \quad &&:\quad \text{Gluing of bordisms.} \nonumber
}
\begin{exbox}{Identity morphism in $\mathsf{Bord}_{\langle n-1, n\rangle}$}
  What is the identity morphism in $\mathsf{Bord}_{\langle n-1, n\rangle}$?\end{exbox}

Another important category for us is $\mathsf{VECT}$, the category of finite-dimensional complex vector spaces:
\eqas{
&\mathsf{VECT} \quad : \quad &&\text{Objects} &&: \text{finite-dimensional $\IC$-vector spaces.}\\
& &&\text{Morphisms} &&: \text{$\IC$-linear transformations between vector spaces.} \\
& &&\text{Compositions\,\,} m &&: \text{composition of linear maps.} \nonumber
}

\begin{remark} \emph{Categories and Algebras.} While a directed graph is one way to visualize a category, another helpful way to think about categories is in terms of algebras. 
Recall that an associative \emph{algebra} $A$ over  a field $\kappa$ is a $\kappa$-vector space $A$ equipped with a bilinear pairing $A \times A \to A$, denoted $(a, b) \mapsto ab$, which is associative: $(ab)c = a(bc)$ for all $a, b, c \in A$, and (if $A$ is \emph{unital}) consists of a unit element $1 \in A$ such that $1a = a 1 = a$ for all $a\in A$. An algebra can be viewed as a   category with one object: $*$, such that $\mathsf{End}(*) \cong A$, and the composition of morphisms corresponds to multiplication in the algebra. This leads to a 
natural generalization:     A  \emph{linear category} (also known as an \emph{algebroid}) is a category whose morphisms are vector spaces (or more generally, modules) and where the composition is bilinear. More precisely, given a field $\kappa$, for two objects $X, Y$ in a $\kappa$-linear category $\CC$, the set of morphisms $\Hom_{\CC}(X, Y)$ is a vector space over $\kappa$, and the composition of morphisms is a $\kappa$-bilinear map: $\circ: \Hom_{\CC}(Y, Z) \times \Hom_{\CC}(X,Y) \to \Hom_{\CC}(X,Z)$. Every object $X \in \mathsf{Obj}(\CC)$ has an identity morphism $\mathsf{id}_{X} \in \Hom_{\CC}(X, X)$. Two  familiar examples are:
\begin{enumerate} 
\item The category of vector spaces over $\kappa$ is a $\kappa$-linear category -- in our notation above, this is $\mathsf{VECT}_{\kappa}$ (above, we took $\kappa = \IC$). This is essentially just saying that for vector spaces $V_1, V_2$, the set of linear transformations 
$\Hom(V_1,V_2)$ is itself a vector space such that composition of linear transformations is 
bilinear. 
\item The category of modules over a $\kappa$-algebra $A$ is a $\kappa$-linear category.
\end{enumerate} 

Conversely, any linear category with a finite set of objects may be viewed as an algebra. The endomorphism algebra of each object provides a subalgebra, but the full algebra need not be a direct sum of these subalgebras.

\end{remark}


%
With one more idea from category theory, we can nicely formalize one key aspect of TFT:
\begin{definition}[\textcolor{red}{Functor}] Let $\CC, \CD$ be two categories. A \textbf{functor} $F: \CC \to \CD$ is a pair of maps:
\eqas{
& F_0 &&: C_0 \to D_0 ~,\\
& F_1 &&: C_1 \to D_1 ~,
}
such that $\forall$ $x, y \in C_0$,
\eqa{
& F_1 &: \Hom_{\CC}(x,y) &&\to \Hom_{\CD}(F_0(x), F_0(y)) ~,
}
and $\forall$ $f, g \in C_1$, 
\eqa{
& \text{\underline{either}}  \quad && F_1(f \circ g) &&= F_1(f) \circ F_1(g) \quad \text{(covariant functor)} ~,\\
& \text{\underline{or}}  \quad && F_1(f \circ g) &&= F_1(g) \circ F_1(f) \quad \text{(contravariant functor)} ~.
}
\end{definition}

With the concept of functor, we can summarize a lot of what we have said so far about topological field theory by simply saying that an $n$-dimensional  TFT \underline{is} a functor,
\eqa{
& F : \mathsf{Bord}_{\langle n-1, n\rangle} &&\to \mathsf{VECT} ~.
}
The equation
\eqa{
& F(f \circ g) &&= F(f) \circ F(g) ~,
}
captures \eqref{eq:LOC2}. But we have not yet captured everything we said above in the math definition. What about \eqref{eq:LOC1}? That is,
\eqa{
& F\big(N \coprod N'\big) && \cong  F(N) \otimes F(N') ~.
}
To incorporate \eqref{eq:LOC1} into our math definition of a field theory, we need four more definitions from category theory:
\begin{definition}[\textcolor{red}{Natural Transformation}] Given categories $C, D$ and two functors,
\begin{center}
\begin{tikzcd}
\CC \arrow[rr,bend left, "{F}"] \arrow[rr,bend right, "{G}",swap] & \big\Downarrow{\,\tau}  & \CD
\end{tikzcd}
\end{center}
a \emph{natural transformation} (a.k.a. ``morphism of functors'') denoted
\eqa{
& \tau : F \implies G ~,
}
is a collection of maps $\tau_x$, indexed by $x \in C_0 = \mathsf{Obj}(\CC)$, such that, for all $x, y \in C_0$ and all $f \in \Hom_{C}(x,y)$, there is a commutative diagram
\begin{center}
\begin{tikzcd}
F(x) \arrow[r, "F(f)"] \arrow[d, "\tau_x", swap] & F(y) \arrow[d, "\tau_y"]\\
G(x) \arrow[r, "G(f)"] & G(y)
\end{tikzcd}
\end{center}
\end{definition}

\noindent \textbf{Example:} The $k^{th}$ integral cohomology is a contravariant functor:
\eqa{
& H_{\IZ}^{k} : \mathsf{TOP} &&\to \mathsf{ABGROUP} ~,
}
where $\mathsf{TOP}$ is the category of topological spaces (as objects) and continuous maps between topological spaces (as morphisms), and $\mathsf{ABGROUP}$ is the category of Abelian groups (as objects) and group homomorphisms (as morphisms). On morphisms, for $f: X \to Y$ a continuous map between topological spaces $X$ and $Y$,
\eqa{
& H_{\IZ}^{k}(X \xrightarrow{f} Y) &&= f^{*}: H^{k}(Y; \IZ) \to H^{k}(X; \IZ) ~.
}
Then the cup product is a \textbf{natural} transformation between
\tightfootnote{The tensor product $A\otimes B$ of two Abelian groups $A, B \in \mathsf{Obj}(\mathsf{ABGROUP})$ is defined as the Abelian group equipped with a bilinear map $\otimes: A \times B \to A\otimes B$ satisfying the following universal property: for every Abelian group $C \in \mathsf{Obj}(\mathsf{ABGROUP})$ and every bilinear map $f: A \times B \to C$, there exists a unique group homomorphism $\widetilde{f}: A \otimes B \to C$ such that $f(a,b) = \widetilde{f}(a \otimes b)$, $\forall$ $a \in A$, $b\in B$. This property characterizes the tensor product up to unique isomorphism. Alternatively, consider the free Abelian group $\IZ[A\times B]$ generated by pairs $(a,b) \in A \times B$, and the subgroup $\mathcal{R} \subset \IZ[A\times B]$ generated by relations that encode bilinearity: (1) $(a + a', b) \sim (a,b) + (a',b)$, (2) $(a, b+b') \sim (a,b) + (a,b')$, and (3) $(ka, b) \sim k(a,b) \sim (a, kb)$ for all $k \in \IZ$, $a, a' \in A$, $b,b' \in B$. Then, the tensor product is the quotient group $A\otimes B:= \IZ[A\times B]/\mathcal{R}$, a typical element of which is the \underline{class} $a\otimes b$ of $(a,b) \in A \times B$. More formally, the tensor product of Abelian groups is well-defined because Abelian groups are $\IZ$-modules (see footnote \ref{foot:R-module}).}
$H_{\IZ}^{k_1} \otimes H_{\IZ}^{k_2}$ and $H_{\IZ}^{k_1 + k_2}$:
\eqa{
& H^{k_1}(X; \IZ) \otimes H^{k_2}(X; \IZ) &&\to H^{k_1 + k_2}(X; \IZ) ~,
}
where $\tau_X = $ cup product. Similarly, Steenrod squares are natural transformations.

\begin{exbox}{Natural transformations and $\mathsf{VECT}$} 
  For $V \in \mathsf{Obj}(\mathsf{VECT})$ define a functor $F_V: \mathsf{VECT} \to \mathsf{VECT}$ by 
\eqa{
& F_V(W) &&:= \Hom(V, W) \oplus V ~,\\
& F_{V}(W_1 \xrightarrow{T} W_2) &&: \Hom(V, W_1) \oplus V \xrightarrow{\circ T \oplus \mathsf{Id}} \Hom(V, W_2) \oplus V ~.
}
Show that the evaluation map,
\eqas{
  \tau_{W} : F_{V}(W) &\longrightarrow W \\
  A \oplus v &\longmapsto A(v) ~,
}
is a natural transformation of $F_V$ to the identity functor $\mathsf{Id}: \mathsf{VECT} \to \mathsf{VECT}$.
\end{exbox}

\begin{definition}[\textcolor{red}{Isomorphism of functors}]
  An isomorphism of functors $\tau: F_1 \to F_2$ is a natural transformation $\tau$, such that there is a natural transformation $\tau': F_2 \to F_1$ with commutative diagrams:
  \begin{equation}
   \hspace{-0.5cm} \begin{tikzcd}[row sep=large]
        & F_2(X) \arrow[rd, "\tau_X'"] \\
        F_1(X) \arrow[ru,"\tau_X"] \arrow[rr, "\mathsf{Id}_{F_1(X)}",swap] & & F_1(X)
    \end{tikzcd} 
    \quad \text{ and } \quad
    \begin{tikzcd}[row sep=large]
      & F_1(X) \arrow[rd, "\tau_X'"] \\
      F_2(X) \arrow[ru,"\tau_X"] \arrow[rr, "\mathsf{Id}_{F_2(X)}",swap] & & F_2(X)
  \end{tikzcd}
  \end{equation}
\end{definition}

\begin{definition}[\textcolor{red}{Equivalence of categories}] An equivalence of categories \textbf{$C$} and $D$ is a pair of functors
  \be
      F : \CC \to \CD ~, \quad \text{ and } \quad G : \CD \to \CC ~,
  \ee
  with isomorphisms of $F \circ G$ and $G \circ F$ to the identity functors $\mathsf{Id}: \CC \to \CC$ and $\mathsf{Id}: \CD \to \CD$ respectively.
\end{definition}

\noindent Many, many, important results in mathematics are statements of equivalence of categories. See, e.g.,  \cite[p. 556-7]{MooreGroupTheory:2023}, for a list of examples.

\begin{definition}[\textcolor{red}{Tensor category (a.k.a. monoidal category)}] A tensor category is a category with a functor
  \be
     \otimes: \CC \times \CC \to \CC ~,
  \ee
  \underline{and} an isomorphism of the functors:
  \be
  \begin{tikzcd}[column sep=large]
    & \CC \times \CC \arrow[rd, "\otimes"] \\
    \CC \times \CC \times \CC \arrow[ru, "\otimes_{12}\times\mathsf{Id}"] \arrow[rd, "\mathsf{Id}\times \otimes_{23}",swap] & \Bigg\Downarrow{\,\CA} & \CC \\
    & \CC \times \CC \arrow[ru, "\otimes",swap]
  \end{tikzcd}
  \ee
  $\CA$ is known as the \textbf{associator}. It is a morphism, 
  \be
     \CA_{x,y,z} : (x\otimes y) \otimes z \longrightarrow x \otimes (y \otimes z) ~,
  \ee
  and it must satisfy the \textbf{pentagon identity},
  \be
  \hspace{-0.5in}\label{eq:pentagon-identity}
  \begin{tikzcd}[column sep=tiny,row sep=large]
    & \big( (x_1 \otimes x_2)\otimes x_3 \big) \otimes x_4 \arrow[ld] \longrightarrow \big(x_1 \otimes x_2\big)\otimes \big(x_3 \otimes x_4\big) \arrow[rd]\\
    \big(x_1 \otimes (x_2 \otimes x_3)\big)\otimes x_4 \arrow[rd] & & x_1 \otimes \big(x_2 \otimes (x_3 \otimes x_4)\big) \\
    & x_1 \otimes\big( (x_2\otimes x_3)\otimes x_4 \big) \arrow[ru]
  \end{tikzcd}
  \ee
  Finally, there is an identity object $\mathbbm{1}_{C} \in \mathsf{Obj}(\CC)$, and natural transformations:
  \eqa{
    \iota_{L} : \mathbbm{1}_{\CC} \otimes ( \bullet ) &\longrightarrow \mathsf{Id} ~,\\
    \iota_{R} : ( \bullet ) \otimes \mathbbm{1}_{\CC}  &\longrightarrow \mathsf{Id} ~,
  }
  satisfying some natural compatibility conditions. See  \cite{Etingof2016-za} for a complete treatment of the subject. 
\end{definition}

A \emph{monoidal functor} (a.k.a. \emph{tensor functor})  between monoidal (tensor) categories $F: \CC \to \CD$ is a functor that preserves structure in the sense that there are isomorphisms
\eqa{
 F(X \otimes Y) &\xlongrightarrow{\theta_{X,Y}}  F(X) \otimes F(Y) ~,\\
 F(\mathbbm{1}_{\CC}) &\xlongrightarrow{\eta_{\CC}} \mathbbm{1}_{\CD} ~,
}
satisfying some natural conditions. Again, the reader should consult 
\cite{Etingof2016-za}  for details. The words ``monoidal'' and ``tensor'' are synonymous.

\begin{remark} \label{rem:braiding} A \textbf{braiding} is an isomorphism of $\otimes: \CC \times \CC \to \CC$ with $\otimes \circ \sigma: \CC \times \CC \to \CC$ where $\sigma: (X,Y) \to (Y,X)$ is the exchange functor. This amounts to the data of isomorphisms:
\be\label{eq:Braiding}
  \Omega_{X,Y} : X\otimes Y \longrightarrow Y \otimes X ~.
\ee
If $\Omega_{Y,X}\circ \Omega_{X,Y} = \mathsf{Id}_{X\otimes Y}$, the category is a \textbf{symmetric monoidal category}. 
\end{remark}

Now, $\mathsf{VECT}$ is a tensor category, using the tensor product of vector spaces. The associator is trivial, so this is a 
symmetric tensor category. Less trivially, the category of super-vector spaces is also a symmetric tensor category. Moreover   $\mathsf{Bord}_{\langle n-1, n\rangle}$ is a symmetric tensor category where the monoidal structure is defined using the disjoint union.  
\tightfootnote{Categories in which there is a braiding which is not necessarily symmetric, known as \emph{braided tensor categories}, play an important role in physical mathematics, 
particularly in the mathematics of 2+1 dimensional anyons and three-dimensional topological field theories. We will not go into this subject in any detail in these notes, confining ourselves to just the briefest remarks in \autoref{sec:BriefAnyons}. } 

\bigskip
\begin{exbox}{Monoidal unit} What is the monoidal unit $\mathbbm{1}_{C}$ in $\mathsf{VECT}$ and in $\mathsf{Bord}_{\langle n-1, n\rangle}$?\end{exbox}
\bigskip

We can now summarize concisely in a simple definition the notion of 
topological field theory we have been using as follows: 

\bigskip
\begin{definition}[\textcolor{red}{Topological Field Theory (TFT)}] An $n$-dimensional TFT is a symmetric tensor functor
  \be\label{eq:Definition-TopTwo-TFT}
    F : \mathsf{Bord}_{\langle n-1, n\rangle} \longrightarrow \mathsf{VECT} ~.
  \ee
\end{definition}

\begin{remark}
$\,$
\begin{enumerate}

\item We will eventually generalize this definition to a much more elaborate statement in equation \eqref{eq:FullyExtended} below. 

\item An important example of topological field theory (in any dimension) is provided by finite group gauge theory. After reviewing some background material on principal $G$-bundles in \autoref{sec:G-Bundles}, we begin describing these in \autoref{sec:FinGrpGT-Part1}. 

\item   In a topological field theory $F$ of any dimension $n>1$, the value $F(\IS^{n-1})$ on a sphere is always a commutative algebra. The basic multiplication is obtained simply by considering the bordism 
\be 
\IS^{n-1} \coprod \IS^{n-1} \rightarrow \IS^{n-1} ~,
\ee
given by cutting out two $n$-disks within a larger $n$-disk. 
This observation goes back, at least, to  \cite{Abrams:1996ty}.
These are always commutative algebras and in closely related contexts, they have more structure and are  known as $E_n$ algebras.  For a review of $E_n$ algebras, see \cite{Lurie2014en}.

\item There is a very useful analogy of this definition of a topological field theory 
with a group representation. Recall that a group representation of a group $G$ with 
carrier space $V$ -- a vector space -- is a homomorphism $\phi: G \to \End(V)$. The homomorphism 
property means that $\phi$ is compatible with group multiplication and composition of linear transformations. For topological field theory, we replace the group $G$ by a bordism category, the carrier space $V$ by a category $\mathsf{VECT}$, and the homomorphism 
by a functor. Once again the functor is -- by definition -- compatible with the composition in the bordism category and in the target category. 

\end{enumerate}

\end{remark}

\SectionWithHeader{Sums And Products Of Theories}{Sums And Products Of Theories}{sec:SumsAndProducts}

Now that we have a definition of a topological field theory 
we can describe a notion of sum and products of theories.

\bigskip 
\noindent 
\underline{Products Of Theories}. A familiar notion from Lagrangian field theory is that if one has two sets of fields $\Phi_1$ and $\Phi_2$ with actions $S[\Phi_1]$ and $S[\Phi_2]$, one can define a new theory with fields $\Phi_1$, $\Phi_2$ and exponentiated action: 
\be\label{eq:Stacking}
e^{ - S[\Phi_1]  - S[\Phi_2] } ~.
\ee
The amplitudes will, rather obviously, be products of the amplitudes from the constituent theories. (Of course, this will \underline{not} be true if we deform \eqref{eq:Stacking} by adding interaction terms between $\Phi_1$ and $\Phi_2$.) This procedure defines a ``multiplication'' of Lagrangian theories. It is also sometimes referred to as \emph{stacking} theories. The procedure naturally 
generalizes to a notion of a product of field theories. 
In the case of topological field theories, $F_1$ and $F_2$, we can define $F_1 \otimes F_2$ by taking the tensor products of the statespaces and amplitudes in the obvious way. 

An important consequence of the notion of multiplication 
is the notion of an \emph{invertible field theory}. We define 
the trivial field theory $F$ to have statespace $F(N) = \IC$ on all spatial manifolds and $F(M): \IC \to \IC$ is the identity map on all bordisms $M$. (In particular $F(N)$ is a trivial representation of the diffeomorphism group of $N$.)   We say $F$ is 
\emph{invertible} if there is another field theory 
$G$ such that $F \otimes G$ is the trivial field theory. 
Note that $G$ inverts $F$. An immediate consequence is that for an invertible field theory, $F(N)$ is a one-dimensional vector space on any spatial manifold $N$, and $F(M)$ is always invertible, hence can be expressed as multiplication of $\IC$ by a nonzero complex number, for any bordism $M$. 

A natural source of invertible theories is given by 
classical actions of Lagrangian field theories. These are not in general topological field theories, so our bordisms will be endowed with 
field configurations we denote as $\Phi$. We can take the  Hilbert space associated to spatial manifolds to be $\IC$. A bordism 
$M_n: N^0_{n-1} \to N^{1}_{n-1}$ will be endowed with fields $\Phi$ and the amplitude involves multiplication by the exponentiated action 
$e^{-S[\Phi]}$. 
Local actions will satisfy the crucial locality condition \eqref{eq:LOC2}. Nonlocal actions will not. Invertible \underline{topological} field theories are therefore local topological terms in field theory actions. It was exactly in this context that the notion of invertible field theory was introduced in 
\cite{Freed:2004yc}. A rather thorough classification of invertible theories has been described by Freed and Hopkins \cite{Freed:2016rqq}. 

Although classical actions furnish natural examples of invertible field theories, not all invertible theories are of this type. 
A simple example is the exponentiated eta invariant of the Dirac operator, which satisfies the locality properties of an invertible theory \cite{Dai:1994kq}. For further discussion, see \cite{Witten:1999eg, Witten:2015aba,Witten:2019bou}.

\bigskip 
\noindent 
\underline{Sums of theories}. A more subtle notion is the \underline{sum} of field theories $F_1$ and $F_2$. If 
$N_{n-1}$ is a \underline{connected} spatial manifold, we define: 
\be\label{eq:SumStateSpaces}
(F_1+F_2)(N_{n-1}):= F_1(N_{n-1}) \oplus F_2(N_{n-1}) ~.
\ee
Note that the formula for $F_1+F_2$ on disconnected manifolds will be more complicated. Moreover, if  $M_n: N^0_{n-1} \rightarrow N^1_{n-1}$ is a 
\underline{connected} bordism between connected spatial manifolds, then we likewise have 
\be 
(F_1+F_2)(M_n):= \begin{pmatrix} F_1(M_n) & 0 \\
0 &   F_2(M_n)  \\  \end{pmatrix} ~,
\ee
in the natural block diagonalization provided by 
\eqref{eq:SumStateSpaces}.

Here are some examples of sums of theories: 

\begin{enumerate} 

\item Given a nonzero complex number $\theta \in \IC^{\times}$, we can define an invertible 2d orientable topological field theory by the rule that 
\be\label{eq:EulerTheory}
\Sigma \longmapsto F_{\theta}(\Sigma) = \theta^{\half \chi(\Sigma)} ~,
\ee
on closed oriented Riemann surfaces $\Sigma$. We can call these 
\emph{Euler character theories}. It follows from the discussion of \autoref{subsec:SemiSimplicity} that the general semisimple 2d oriented TFT is a direct sum of Euler character theories. 

\item A closely related (but nontopological) example is that of 2d Yang-Mills theory described in \autoref{subsec:nonexamples}, where the sum is over the unitary dual of the gauge group. 
\tightfootnote{The \emph{unitary dual} of a group is the set of unitary irreducible representations of the group.}

\item A simple example of a sum of theories is provided by a scalar field sigma model, where the target space is a disjoint union of connected manifolds.

\end{enumerate}
 
Sums of theories naturally arise when describing long-distance limits, or IR limits of theories. Thus they appear naturally in discussions of spontaneous symmetry breaking. Another example is 
in long-distance limits of renormalization group (RG) flow as in \cite[Fig. 1]{Harvey:2001wm}.
The concept has also been useful in elucidating the long-distance behavior of 2d adjoint QCD \cite{Komargodski:2020mxz}.
Finally, we mention that the notion of sums of theories is closely related to the subject of ``decomposition'' of theories, especially in the examples of theories admitting a non-faithful action of a (generalized) symmetry group. For a sampling of the literature on this, see \cite{Pantev:2005zs,Hellerman:2006zs,Caldararu:2010ljp,Gu:2018fpm,Sharpe:2019ddn,Tanizaki:2019rbk,Eager:2020rra,Katz:2022lyl,Sharpe:2022ene,SharpeTalk:2023,Sharpe:2023lfk,Santilli:2024dyz}.

\SectionWithHeader{Some Brief Remarks On Anyons}{Some Brief Remarks On Anyons}{sec:BriefAnyons}

The notion of braiding introduced in equation \eqref{eq:Braiding}
turns out to be particularly well suited to the description of 
\emph{anyons} -- particles with exotic statistics that can exist in 2+1 dimensional systems. One common way of describing such particles 
uses 3d Chern-Simons theory as a description of the long-distance, low-energy physics of the anyonic excitations. General ideas of low energy effective theory predict that the low-energy physics of a gapped system that breaks parity symmetry should be described by some Chern-Simons theory (which has only one derivative in the action). Chern-Simons theory turns out to be   particularly well-suited to fractional quantum Hall systems.
\tightfootnote{For some reviews, see \cite{Bieri:2010za,Tong2016QuantumHall,Simon2023-jn,Cheng:2025ube}.}
In this description, anyons can be ``fused'' and the mathematical embodiment of that is to model anyons as objects in a braided  tensor category.   In general, for anyons, $\Omega_{y,x}\circ\Omega_{x,y}$ is not the identity.

The ``fusion of the anyons'' can be pictured as:
\be
\begin{tikzpicture}[scale=0.9,baseline=-0.5ex]
  \node (graph) [inner sep = 0pt] {\includegraphics[width=1in,valign=c]{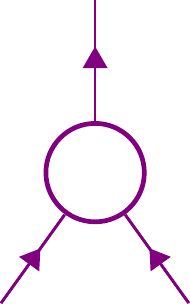}};
  \node at (0,-0.25) {$t$};
  \node at (-1.5,-2.5) {$a$};
  \node at (1.5,-2.5) {$b$};
  \node at (0,+2.5) {$c$};
  \node at (3.0,-0.25) {$t \in \Hom(a\otimes b, c) ~.$};
  \end{tikzpicture}
\ee
The associator is:
\be
 \begin{tikzpicture}[scale=1,baseline=-0.5ex]
   \node (limage) [inner sep = 0pt]{\includegraphics[width=1.8in,valign=c]{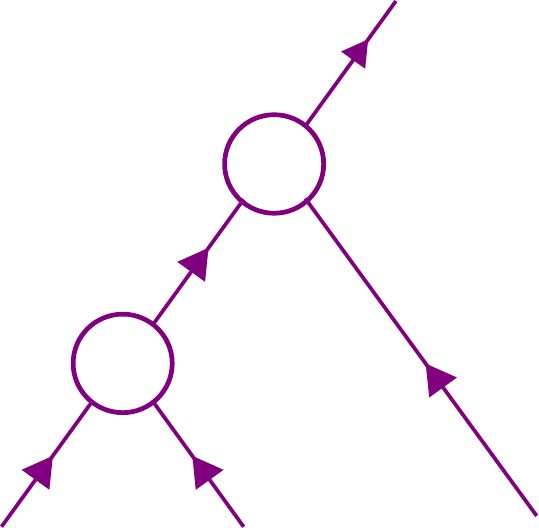}};
   \node at (-1.2,-0.85) {$t_1$};
   \node at (0.1,0.85) {$t_2$};
   \node at (-2.4,-2.5) {$a$};
   \node at (0,-2.5) {$b$};
   \node at (2.5,-2.5) {$c$};
   \node at (1.3,2.5) {$d$};
 \end{tikzpicture}
 = \sum_{t_3, t_4} F_{t_1 t_2}^{t_3 t_4}
 \begin{tikzpicture}[scale=1,baseline=-0.5ex]
  \node (rimage) [inner sep = 0pt]{\includegraphics[width=1.8in,valign=c]{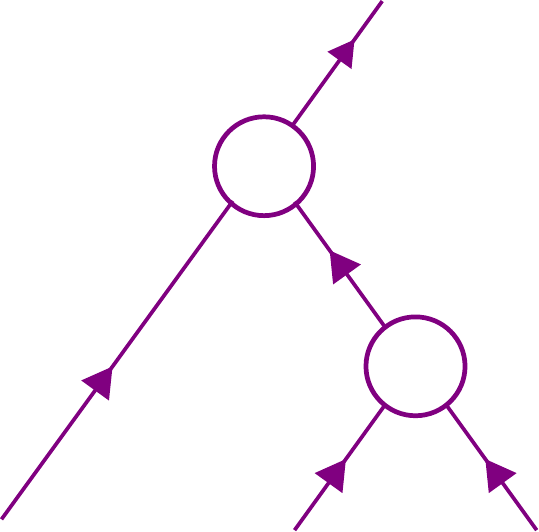}};
  \node at (1.28,-0.85) {$t_3$};
  \node at (0,0.85) {$t_4$};
  \node at (-2.4,-2.5) {$a$};
  \node at (0,-2.5) {$b$};
  \node at (2.5,-2.5) {$c$};
  \node at (1.22,2.5) {$d$};
\end{tikzpicture}
\ee

\bigskip
\begin{exbox}{Pentagon diagram} Write out the pentagon diagram using the above notation.\end{exbox}
\bigskip 

The physics of anyons is believed to be described by a special kind of braided tensor category, 
 known as a \emph{modular tensor category}. 
The connection of modular tensor categories to anyons in gapped materials is that at long distances, the dominant term in an effective action describing the system with anyons should be a Chern-Simons term, but 3d Chern-Simons theory is ``holographically dual'' to 2d rational conformal field theory, and the monodromy behavior of 2d rational conformal field theories is governed by a modular tensor category
\cite{Moore:1989vd,Moore:1988qv}. This was the argument originally used in \cite{Moore:1991ks}.

We comment briefly on a (rough) definition of a modular tensor category (MTC). 
An MTC is a semisimple braided tensor category 
equipped with   an extra piece of data 
known as a ``ribbon structure'' (related to the spin of the anyons) and a compatible notion of 
duals (related to the CPT conjugate of the anyon) allowing one to define a trace on the endomorphism of any object in the category. The ribbon structure and the ``$S$-matrix,'' 
\be 
S_{ij} := \Tr(\Omega_{X_i, X_j}\circ \Omega_{X_j, X_i} ) ~,
\ee
is invertible. 
\tightfootnote{In the original definition, $S_{ij}$ and a matrix $T$ derived from the ribbon structure 
were constrained to satisfy the relations of the modular group $\mathsf{SL}(2,\IZ)$. 
Later, it was realized \cite{Turaev:1992,Turaev:2016} that one only need require the invertibility of the ``$S$-matrix,'' and then the relations of the modular group would follow. This same observation was made by I. Frenkel in collaboration with G.M., and announced by G.M. at a conference in Princeton in November 1989, but was not published.}
For more discussion of modular tensor categories, see \cite{Walker1991_TQFTNotes,Bakalov2000-hq,Rowell:2005hv,Kitaev:2005hzj,Stirling:2008bq,Rowell:2016lrv,Rowell:2018wnv}.

Modular tensor categories  played a crucial role in combinatorial 
constructions of the invariants of 3-manifolds predicted by Chern-Simons-Witten theory 
\cite{Reshetikhin:1990,Reshetikhin:1991} and indeed the three-dimensional topological field 
theory associated with a modular tensor category is often referred to  as \emph{Reshetikhin-Turaev} theory, or \emph{Reshetikhin-Turaev-Witten} theory. 

There are numerous reviews and books on this topic. A very useful one 
is \cite{Kong:2022cpy}. See also the recent book \cite{Simon2023-jn}.
For some historical remarks about the role of (modular) tensor categories in anyon physics, see \cite{Moore:RCFT-To-MTC,Moore:PlecticsVideo}.

\SectionWithHeader{Some Background On Principal Bundles}{Some Background On Principal Bundles}{sec:G-Bundles}

For a group $G$, a \emph{$G$-torsor} or a \emph{principal homogeneous space} is a set $T$ with a free and transitive $G$-action on $T$. The points of $G$ can be put into one-to-one correspondence with those of $T$ by picking a point $t_0\in T$ and identifying the point $g\cdot t_0$ with $g\in G$, but to make this correspondence, one needs to 
\underline{choose} a point $t_0\in T$, so the one-to-one correspondence is not natural. 

\bigskip 
\noindent \textbf{Example 1:} 
A good example is the set of points $T=\theta + \IZ\subset \IR$, where $\theta$ is a non-integer real number. The group $G=\IZ$ acts by translations on the set $T$, but there is no natural basepoint in $T$. 

\bigskip
\noindent \textbf{Example 2:} Another good example is affine space $\IA^n$. 
This is a space with a transitive action of $\IR^n$ without fixed points. Thus, if $p\in \IA^n$ and $v\in \IR^n$, then there is a notion of translating $p$ by $v$, so $p \mapsto p+v$, and given any two points 
$p_1, p_2 \in \IA^n$, there is a unique vector $v\in \IR^n$ such that 
$p_2 = p_1 + v$, but there is no natural origin. In the case of $\IA^2$, think of the map of the world flattened to a plane. There is no natural choice of capital city which one could call ``the origin.''

For $G$ a topological group and $X$ a topological space, a \uline{principal $G$-bundle over $X$} is a continuous map of topological spaces $\pi: P \to X$ such that 
\begin{enumerate}
  \item $P$ admits a continuous and free right $G$-action such that $\pi(p\cdot g) = \pi(g)$ and the fibers $\pi^{-1}(x)$ are $G$-torsors.
  \item $\pi: P \to X$ is locally trivial: for all $x \in X$, there exists $\CU_{x} \subset X$ such that the diagram
  \be
   \begin{tikzcd}
    \pi^{-1}(\CU_{x}) \arrow[rr, "\phi_{\CU_{x}}"] \arrow[rd, "\pi",swap] && \CU_{x} \times G \arrow[ld]\\
    & \CU_{x}
   \end{tikzcd}
  \ee
  commutes, and $\phi_{\CU_x}$ is $G$-equivariant.
\end{enumerate}

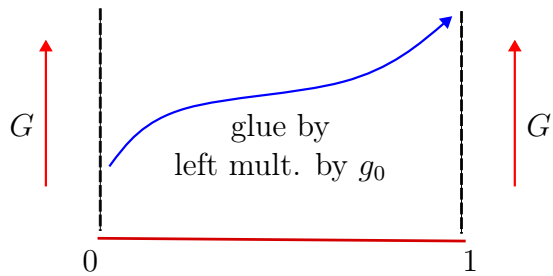
\begin{figure}[h]
\centering
\begin{tikzpicture}[scale=1,baseline=-0.5ex]
  \node (image) [inner sep = 0pt]{\includegraphics[width=2.5in,valign=c]{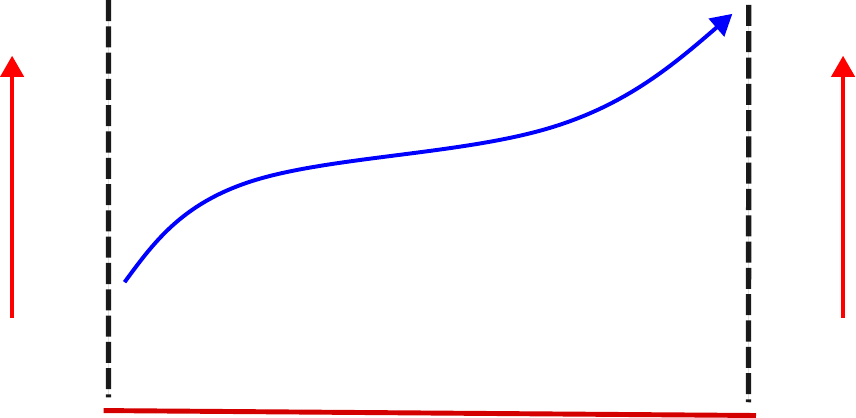}};
  \node at (0,-0.3) {$\begin{array}{c}\text{glue by}\\\text{left mult. by $g_0$}\end{array}$};
  \node at (-2.5,-1.8) {$0$};
  \node at (2.5,-1.8) {$1$};
  \node at (-3.4,0) {$G$};
  \node at (3.4,0) {$G$};
\end{tikzpicture}
\caption{Constructing a bundle over $\IS^1$.}\label{fig:Pg0-Gluing}
\end{figure}

A key example for us is the following: choose $g_0 \in G$ and let $\IZ$ act on $\IR \times G$ by $n: (x, g) \mapsto (x + n, g_{0}^{n}g)$.
Taking equivalence classes of $\IZ$-orbits on $\IR\times G$ defines  a space we can call $P_{g_0}$. Similarly, taking equivalence classes of the $\IZ$-action on $\IR$ gives the quotient space $\IS^1$, and we have a continuous map:
\eqas{
  P_{g_0} := \big(\IR \times G\big)/\IZ &\xlongrightarrow{\pi} \IR/\IZ = \IS^1\\
  [(x, g)] &\longmapsto [x]  ~.
}
An intuitive way of thinking of $P_{g_0}$ is that we are considering $[0,1]\times G$ and gluing the ``boundary'' $\{0\}\times G$ to 
$\{ 1\} \times  G$  by left multiplication by $g_0$, as shown in \autoref{fig:Pg0-Gluing}. 
 
Denote this principal $G$-bundle over $\IS^1$ by $\pi: P_{g_0}\to \IS^1$, or just $P_{g_0}$ for short. 

\bigskip
\begin{definition}[\textcolor{red}{Bundle Map}] A bundle map, or morphism of principal $G$-bundles over $X$ is a fiber-preserving $G$-equivariant map:
  \be
  \begin{tikzcd}
     P_1 \arrow[rr, "\psi"] \arrow[rd, "\pi_1", swap] && P_2 \arrow[ld, "\pi_2"]\\
    & X
  \end{tikzcd}
  \ee
\end{definition}

\begin{exbox}{Principal $G$-bundles over $\IS^1$} Show that the bundle map
\be
\begin{tikzcd}
   \IR \times G \arrow[rr, "\psi_h"] \arrow[rd] &&
   \IR \times G \arrow[ld]\\
  & \IR
\end{tikzcd}
\ee
given by $\psi_{h}: (x, g) \mapsto (x, hg)$ induces an isomorphism of bundles over the circle $\psi_{h}: P_{g_0} \cong P_{h g_{0} h^{-1}}$.
Therefore, isomorphism classes of principal $G$-bundles over $\IS^1$ are labeled by \emph{conjugacy classes} of elements of $G$. The automorphism group of $P_{g}$ is $Z(g) = \{h \in G ~|~ h g h^{-1} = g\}$. Moreover, show that the automorphism group of $P_{g_0}$ 
(that is the group of bundle isomorphisms of $P_{g_0}$ with itself) 
is just: 
\be 
 \mathsf{Aut}(P_{g_0}) \cong Z(g_0) ~,
\ee
where $Z(g_0)$ is the centralizer of $g_0$ within $G$. 
\end{exbox}
\bigskip

\begin{remark}
$\,$
\begin{enumerate} 

\item\label{item:G-eqvt-map-is-invertible} Note that a $G$-equivariant map $\psi: X \to Y$ between two $G$ torsors must be invertible: Equivariance implies that 
$\psi(xg) = \psi(x)g$, so if we choose a basepoint $x_0\in X$ so that 
every element of $X$ is expressible as $x_0g$ for a unique group element $g$, then we can use $\psi(x_0)$ as a basepoint for $Y$. Then 
$\psi^{-1}(\psi(x_0)g) = x_0 g$ defines the inverse map. When upgraded to families, we learn that   every bundle map of principal $G$-bundles has an inverse bundle map, so it defines an isomorphism of bundles. 

\item In general, let $G$ be a finite group. Isomorphism classes of principal $G$-bundles over a topological space $X$ are in one-to-one correspondence with elements of
\be
  \Hom\big( \pi_{1}(X, x_0) , G \big)/G ~,
\ee
where the quotient identifies $\phi \sim \phi'$ if there exists $g \in G$ such that $\phi(\gamma) = g \phi'(\gamma)g^{-1}$ for all $\gamma \in \pi_1(X, x_0)$. Note that setting $X = \IS^1$, we recover the claim that isomorphism classes are in one-to-one correspondence with conjugacy classes of $G$.

\end{enumerate}

\end{remark}

\SectionWithHeader{Groupoids}{Groupoids}{sec:Groupoids}

A special class of categories known as \emph{groupoids} is extremely useful when working with gauge theories, and will be used quite frequently in the material below. We begin with the formal definition:

\begin{definition}[\textcolor{red}{Groupoid}] A groupoid is a category $\CC = (C_0, C_1, p_0, p_1, m)$ in which every morphism (i.e., every element of $C_1$) is invertible (an isomorphism).
\end{definition}

\noindent Here are some  key examples:

\begin{enumerate}

\item A group $G$ can be viewed as an example of a category.  
The category has one object, call it $*$. The morphisms $* \rightarrow *$ are in one-to-one correspondence with the group elements $g\in G$, that is, $C_1 = G$. The  composition of morphisms $m: C_1 \times C_1 \to C_1$ is group multiplication. 
The identity and associativity axioms are part of the definition of a category. The fact that group elements have inverses implies that this category is in fact a groupoid. 

\item The previous example can be generalized as follows: Consider a set $X$ with a (left) $G$-action. We can define the category denoted as $X/\!/G$ whose objects are elements $x\in X$, so $C_0(X/\!/G) = X$, and whose morphisms are pairs $(x,g) \in X \times G$. The morphism $(x,g)$ is 
considered as a morphism $(x,g): x \mapsto g \cdot x$. It is a good exercise to draw a picture of this category for a finite group acting on a finite set and to write the general group law for composition of morphisms. The previous example is the case of a totally ineffective action of a group $G$ on a single point: ${\rm pt}/\!/G$. This category turns out to be important in the theory of principal $G$-bundles. The origin of the name ``groupoid'' comes from this generalization of the notion of group. Categories 
presented in the form $X/\!/G$ are sometimes called \emph{action groupoids} or 
\emph{transformation groupoids}. 

\item Given $G$ and a topological space $X$, we can form the category 
$\Bun_G(X)$ of principal $G$-bundles over $X$. The objects are principal $G$-bundles and the morphisms are bundle maps of principal $G$-bundles. It follows from Remark \ref{item:G-eqvt-map-is-invertible}  above that this category is a groupoid. It is sometimes called the \emph{stack of principal $G$-bundles over $X$}.
\tightfootnote{\label{foot:stack}A stack is an equivalence class of groupoids. Equivalently, a groupoid is a presentation of a stack. For example, a smooth manifold is a (smooth) stack, and an open covering by coordinate charts furnishes a presentation as a groupoid. See \cite[App. A]{Freed:2007wja} and \cite{Hopkins:1999coctalos} for details.}

\item Let $X$ be a topological space. We can define a groupoid known as the \emph{fundamental groupoid} of $X$. The objects are the points $x\in X$, and the morphisms are continuous paths $\gamma:[0,1]\to X$ considered up to homotopy with fixed endpoints. The source and target maps are $p_0(\gamma)=\gamma(0)$ and $p_1(\gamma) = \gamma(1)$. It is necessary to consider the paths up to homotopy (with fixed endpoints) in order for the strict associativity condition to hold.

\end{enumerate}

\begin{exbox}{Groupoids and automorphism groups} 

\begin{enumerate}
\item Show that in a groupoid $\CG$, the homsets $\Hom(x,x)$ are 
\underline{groups} for every object $x\in C_0(\CG)$. This is called 
the \emph{automorphism group of the object $x$}. 

\item  Show that if $\CG = X/\!/G$, the category determined by the group action of $G$ on a set $X$, then the automorphism group of an object in $X/\!/G$ can be identified with the stabilizer group of a point of $X$. 
\end{enumerate}

\end{exbox} 

\SectionWithHeader{Finite Group Gauge Theory}{Finite Group Gauge Theory}{sec:FinGrpGT-Part1}
 
Let $G$ be a finite group. 
In general, the functor defining finite group gauge theory for gauge group $G$ is defined by ``quantizing'' in 
some way the groupoid of  principal $G$-bundles over $X$, denoted $\mathsf{Bun}_G(X)$,
and introduced in \autoref{sec:Groupoids}. Here $X$ is a smooth manifold, of a dimension that depends on the physical quantity under consideration. The fact that $\mathsf{Bun}_G(X)$ is a groupoid is very important 
because some principal $G$-bundles have nontrivial automorphisms, and we should ``quotient'' by that group of automorphisms in a gauge theory.  

We can illustrate the general outlook of the previous paragraph by first writing the formula for the partition function. This will be a sum over the set of isomorphism classes of principal $G$-bundles. 
When $G$ is a finite group and $X$ is a compact space, the set of isomorphism classes in $\mathsf{Bun}_G(X)$ has finite cardinality. We denote the set of isomorphism classes 
by $[\mathsf{Bun}_G(M_n)] $  (sometimes we will also denote it 
by $\pi_0( \mathsf{Bun}_G(M_n)) $ below). For each isomorphism class, choose a representative 
principal $G$-bundle $P$, and let $\mathsf{Aut}(P)$ denote the automorphism group of $P$. Isomorphic 
bundles $P$ will have isomorphic groups $\mathsf{Aut}(P)$, so the order $\vert \mathsf{Aut}(P)\vert$
only depends on the isomorphism class of $P$. With these notations in place, the partition function 
of the theory on a compact $n$-dimensional manifold without boundary, $M_n$  is: 
\be\label{eq:FinGroupPF}
F(M_{n}) := \sum_{[\mathsf{Bun}_G(M_n)] }\frac{1}{|\mathsf{Aut}(P)|} ~,
\ee

Similarly, we can also write down the statespaces on closed $(n-1)$-dimensional manifolds $N_{n-1}$ 
as a linear space obtained by ``quantizing''  $\mathsf{Bun}_G(N_{n-1})$. The explicit formula is:
\tightfootnote{\label{foot:fun-map-compactopen}We will use the notation $\mathsf{Fun}(X\to Y)$ for the \underline{vector space} of 
continuous functions from $X \to Y$, when $Y$ is a topological vector space. Usually, we are taking $Y=\IC$, and sometimes, 
we will simply write $\mathsf{Fun}(X)$ for the vector space of continuous complex-valued functions on $X$. We use the notation $\mathsf{Map}(X\to Y)$ for the topological space of continuous maps between topological spaces $X$ and $Y$ with compact-open topology. The compact-open topology 
on the space of continuous maps $\mathsf{Map}(X \to Y)$ between two topological spaces $X,Y$ can be defined as the topology derived from a basis of open sets $\CO(K,U)$ labeled by a compact set 
$K\subset X$ and an open set $U\subset Y$. The formula for these open sets is   
$\CO(K,U)= \{ f: X \to Y ~|~ f(K) \subset U\}$. It is the topology such that 
a sequence of functions   converges uniformly on compact sets. It is also the topology so that 
$\mathsf{Map}(X\times Y \to Z)$ is homeomorphic to $\mathsf{Map}(X \to \mathsf{Map}(Y \to Z))$ for all topological spaces $X,Y,Z$.}
\be\label{eq:FinGroupStateSpace}
 F(N_{n-1})  := \mathsf{Fun}\left([\mathsf{Bun}_G(N_{n-1})] \to \IC \right) ~.
\ee
The discussion of how to compute amplitudes in this theory -- which is quite necessary to define the 
functor and check the gluing properties -- is best deferred to after we introduce $BG$ below. 

\begin{remark} A useful observation is that for a finite group $G$, there is a unique connection on 
$P$ and the isomorphism class of the bundle is completely determined by the holonomy. (See \autoref{sec:Connections-on-G-Bundles} below for discussions of connections and holonomy.) So, for $X$ a connected space, we can choose any basepoint $x_0 \in X$, and there is a 1--1 correspondence, 
\be 
[\mathsf{Bun}_G(X)] \cong  \Hom(\pi_1(X,x_0) , G)/G ~,
\ee
where the quotient by $G$ is conjugation action on the homomorphism. That is, if 
$\varphi:\pi_1(X,x_0) \rightarrow G $ is a group homomorphism and $g\in G$, then 
$(g\cdot \varphi)([\gamma]):= g \varphi([\gamma]) g^{-1}$, for all homotopy classes $[\gamma]$ of 
closed loops $\gamma$ based at $x_0$. 
\end{remark}

\bigskip 
\noindent \textbf{Example 1:}  Let us consider $G$-gauge theory for $n = 1$.
According to the above, the partition function on the circle is a sum over isomorphism classes of $G$-bundles over $\IS^1$. We have just seen that the isomorphism classes are in 1--1 correspondence with conjugacy classes. The Boltzmann weight $B(P_{g_0})$ of a bundle $P_{g_0}$ should therefore be a class function of $g_0$, and can be written as  
\eqa{
  B(P_g) &= \frac{\chi_\rho(g)}{|Z(g)|} ~,
}
where $\chi_\rho$ is the character for some element $\rho$ in the representation ring of $G$, and $Z(g)$ is the centralizer of $g$. 

The orthogonality relations for characters imply that the sum in the partition function projects to the identity isotypical component of $\rho$. Therefore, without loss of generality, we can take $\rho$ to be some multiple of the trivial representation. Taking that multiple to be one, for simplicity we have  $\chi_{\rho}(g) = 1$. This recovers 
the definition in \eqref{eq:FinGroupPF}. 
Note that we can rewrite the sum as: 
\eqa{
  F^{1d}_{\rm ampl}(\IS^1) &= \sum_{\mathsf{C.C.}}\frac{1}{|Z(g)|}\\
  &= \sum_{P_g}\frac{1}{|G|} = 1 ~.
}
In the second line, we are summing over all bundles $P_{g_0}$ and dividing by the ``volume of the gauge group.''  Since the result is 
simply $1$, we expect that  $F^{1d}_{\rm statespace}({\rm pt}) = \IC$.
This too, is in accord with \eqref{eq:FinGroupStateSpace}, since there is a unique isomorphism class of principal $G$-bundles over a point. Therefore, the statespace is the vector space of functions on a point, and is hence a copy of $\IC$.

\bigskip 
\noindent \textbf{Example 2:} For now,  we just note that for the two-dimensional case since $\mathsf{Bun}_G(\IS^1)$ is in 1--1 correspondence with the conjugacy classes on $G$, we can identify $F^{2d}_{\rm statespace}(\IS^1)$ with the 
space of class functions on $G$.  

Using the formulation of \autoref{sec:FinHomTheory} below, we will see that the composition of \autoref{fig:composition-m-psi1-psi2}, 
\begin{figure}[H]
\centering
\includegraphics[]{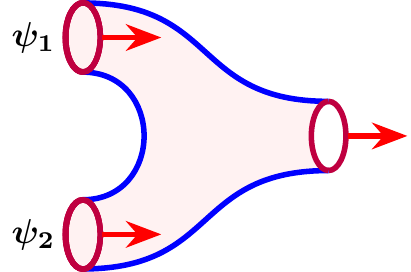}
\caption{Composition $m(\psi_1, \psi_2)$.}\label{fig:composition-m-psi1-psi2}
\end{figure}
works out to be  (by equation \eqref{eq:SigXcomp} below) 
the convolution product,
\be\label{eq:ConvolutionProduct}
m(\psi_1, \psi_2)(g)  = \frac{1}{\vert G\vert}  \sum_{g_1 g_2 = g}\psi_1(g_1) \psi_2(g_2) ~.
\ee

A natural basis of the vector space of class functions is given by the characters $\chi_{\mu}$ in the irreps $\mu \in \mathsf{Irrep}(G)$. 
The orthogonality relations for characters of irreps of $G$, denoted  $\chi^\mu$
(where $\mu$ labels the distinct irreps), is
\be
 \frac{1}{\vert G\vert}   \sum_{h\in G } \chi^\mu(h) \chi^\nu(h^{-1}g) = 
 \delta_{\mu\nu}  \frac{\chi^\nu(g)}{\chi^\nu(1)} ~.
\ee
It follows that a basis of idempotents of the Frobenius 
algebra is $\varepsilon_{\mu} = \chi_{\mu}(1)\chi_{\mu}$.

The Frobenius trace must be proportional to the evaluation 
of the character function at $g=1$: 
\eqa{
  \theta(\psi) &= \lambda\cdot\psi(1) ~.
}
Any $\lambda \in \IC^{\times}$ defines a Frobenius structure. 
(The definition \eqref{eq:SigXcomp} below gives 
$\lambda = 1/\vert G\vert$.)  Applying equation \eqref{eq:PF-FrobeniusAlgebra} from Exercise \ref{exercise:frobenius-0}, 
we obtain the partition function of the finite group gauge theory 
on a genus $g$ Riemann surface: 
\eqa{
  F(\Sigma_g) &= \lambda^{2-2g}\sum_{\substack{\text{irreps $\mu$}\\\text{of $G$}}}\big( \mathsf{dim\,}V_{\mu}\big)^{2-2g} ~,
}
it can be shown that this is, 
\eqa{
  \big( \lambda|G|\big)^{2-2g} \frac{\#\Hom\big( \pi_1(\Sigma_g, x_0), G \big)}{|G|} ~,
}
so is indeed a sum of the Boltzmann weights $1/|G|$ over isomorphism classes of $G$-bundles, up to an overall factor $\big(\lambda|G|\big)^{2-2g}$. We can attribute this factor to multiplication by an invertible TFT, namely, the Euler theory introduced in \eqref{eq:EulerTheory} above. Again, with the 
definition   \eqref{eq:SigXcomp}, this overall factor is simply $1$.

\SectionWithHeader{Classifying Spaces}{Classifying Spaces}{sec:ClassifyingSpace}

We will now present another perspective on the finite group gauge theory discussed above. 

For any topological group $G$, there is a topological space (only defined up to homotopy equivalence) denoted $BG$ with the property that there is a natural one-to-one correspondence:
\eqa{
&\left\{ \begin{array}{c} \text{homotopy classes} \\ \text{ of maps } \\ f : M \to BG \end{array}\right\} &&\longleftrightarrow  \left\{ \begin{array}{c} \text{isomorphism classes} \\ \text{ of principal $G$-bundles } \\ P \to M \end{array} \right\} ~.
}
One way to construct $BG$ is to find a $G$-space (i.e., a space with a $G$ action on it) which is
a ``CW complex,'' 
\tightfootnote{\label{foot:CWComplex}At several points in these notes when making technical comments, we refer to the notion of a CW complex \cite[Sec. 5]{Whitehead1949}. Informally, a CW complex is a generalization of both a simplicial complex and of a manifold. The definition states -- roughly -- that it is a space that can be obtained from a finite set of points by successively attaching open balls, or cells,  (of various dimensions) by continuous maps. The ``C'' stands for ``closure finite,'' and the ``W'' stands for ``weak topology.'' } 
such that it

\begin{itemize}\itemsep 0pt
\item[(a)] is weakly contractible,
\tightfootnote{\label{foot:weaklycontractible}A space $X$ is \emph{weakly contractible} if all its homotopy groups are trivial, i.e., $\pi_{i}(X) = \{1\}$ for all $i$. This is equivalent to the statement that the canonical map $X \to {\rm pt}$ is a weak homotopy equivalence. The space $X$ is \emph{contractible} if the map $X \to {\rm pt}$ is a homotopy equivalence. This technical distinction is important in situations where the Whitehead theorem (introduced in footnote \ref{foot:whitehead}) may not apply.
 By the Whitehead theorem, a weakly contractible CW complex is contractible. Two examples of spaces that are weakly contractible but not contractible are the so-called long line, and the double comb space. Both these spaces do not have the homotopy type of a CW complex. For discussions of such examples, see for example, \cite{Steen1978}.} 
 and,
\item[(b)] admits a free $G$-action.
\end{itemize}

For any topological group $G$, such a space exists, although it is not unique. It is, however, unique up to homotopy equivalence, 
and any such space is often denoted $EG$. Given $EG$, since the $G$ action (say, on the right) is free, we can define $BG:=EG/G$, so that there is a principal $G$-bundle, known as the \emph{universal bundle} $\pi: EG \to BG$, where $\pi$ is the quotient map. 
\tightfootnote{Note that, if a space $X$  satisfies properties $(a)$ and $(b)$, then so does any finite product of $X$ with itself. This is in part the reason that classifying spaces are only defined up to homotopy equivalence.}
The correspondence above in one direction is then simply the pullback map $f \mapsto f^*(EG)$, which defines a principal $G$-bundle over the domain of $f$. Homotopic maps $f$ define isomorphic bundles over $M$.

\bigskip 
\noindent \textbf{Example 1:} Take $G = \IZ$. Then we can take $EG = \IR$ with $\IZ$ acting by translations. This clearly satisfies properties $(a)$ and $(b)$, so $BG$ is any topological space homotopy equivalent to $\IS^1$. Similarly, 
if $G=\IZ^n$, then we can model $BG$ with an $n$-dimensional torus. 
This example is misleadingly simple, as we see by contemplating the next case. 
\bigskip

\bigskip 
\noindent \textbf{Example 2:}  Consider $G = \IZ_2$. The simplest model for $EG$ is the unit sphere in an infinite-dimensional real Hilbert space, and $B\IZ_2 \cong \IR\IP^{\infty}$. A heuristic argument for that is the following: There is a natural free action of $\IZ_2$ on the $k$-dimensional sphere. Of course, $\IS^k$ is not contractible. However note that the homotopy groups $\pi_j(S^k,*) =0$ for $j<k$. (We say that 
$\IS^k$ is $(k-1)$-connected.)  ``So,'' the ``limit''  $\lim\limits_{k\to \infty}\IS^k$ should, in fact, have vanishing homotopy groups and be connected. In fact, this turns out to be rigorously true if we interpret the limit  $\lim\limits_{k\to \infty}\IS^k$ as the unit sphere in an infinite-dimensional countable Hilbert space.

Example 2 above shows that, given a topological group $G$, it is not completely trivial to construct a space $EG$ or $BG$, so it is good to have a systematic construction. Such a construction exists and is due to Milnor \cite{Milnor1,Milnor2}. 
The construction  begins by identifying a group $G$ as a category with one object, all of whose morphisms are invertible: 
\eqa{
& C_{0} &&= \left\{ \textcolor{Orange}{\bigbullet} \right\} ~, \qquad C_{1} = \Hom(\textcolor{Orange}{\bigbullet}, \textcolor{Orange}{\bigbullet}) = G ~.
}
The composition of arrows in $C_1$ is defined by group multiplication:
\be
\begin{tikzpicture}[scale=1,baseline=-0.5ex]
  \node (image) [inner sep = 0pt]{\includegraphics[width=1in,valign=c]{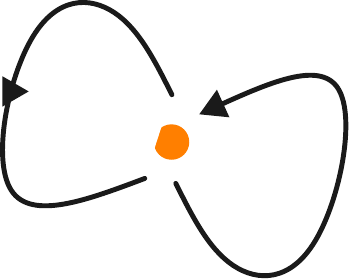}};
  \node at (-0.5,1.25) {$g_2$};
  \node at (1.55,0.25) {$g_1$};
\end{tikzpicture} \qquad g_1 \cdot g_2 = g_1 g_2 ~.
\ee
Now, assume $G$ is a finite group. We construct a CW complex:
\eqa{
 \text{0-skeleton} &= * ~,\\
 \text{1-skeleton} &= G ~.
}
Next, the aim is to attach disks of higher dimension so that all the higher homotopy groups vanish, and $\pi_1(BG,*) \cong G$, but $\pi_j(BG) = 0$ for $j > 1$.

We view $C_2$ as defining triangles which we wish to fill in:
\be
\begin{tikzpicture}[scale=1,baseline={1cm-0.5*height("$=$")},>=Stealth,decoration={
  markings,
  mark=at position 0.5 with {\arrow{>}}}]
  \node (A) at (0,0) [draw,circle,fill=orange] {};
  \node (B) at (1,1.732) [draw,circle,fill=orange] {};
  \node (C) at (2,0) [draw,circle,fill=orange] {};
  \draw[thick,purple,postaction={decorate}] (B)--(A);
  \draw[thick,purple,postaction={decorate}] (A)--(C);
  \draw[thick,purple,postaction={decorate}] (C)--(B);
  \draw[fill=gray!30,color=gray!20!blue!10] (A.north east) -- (C.north west) -- (B.south) -- (A.north east);
  \node at (1,-0.4) {$g_1$};
  \node at (2,0.9) {$g_2$};
  \node at (-0.45,0.9) {$(g_1 g_2)^{-1}$};
\end{tikzpicture}
\qquad \qquad 
\left(
  \begin{array}{cc}
    \text{arrow} & \begin{tikzpicture}[scale=1,baseline=-0.5ex,>=Stealth,decoration={
      markings,
      mark=at position 0.5 with {\arrow{>}}}]
      \node (A1) at (0,0) [draw,circle,fill=orange] {};
      \node (A2) at (2,0) [draw,circle,fill=orange] {};
      \draw[thick,purple,postaction={decorate}] (A1) -- (A2);
      \node at (1,0.27) {$g$};
    \end{tikzpicture} \\
    \multicolumn{2}{c}{\text{with reversed orientation}}\\
    \text{is} & 
    \begin{tikzpicture}[scale=1,baseline=-0.5ex,>=Stealth,decoration={
      markings,
      mark=at position 0.5 with {\arrow{>}}}]
      \node (A1) at (0,0) [draw,circle,fill=orange] {};
      \node (A2) at (2,0) [draw,circle,fill=orange] {};
      \draw[thick,purple,postaction={decorate}] (A2) -- (A1);
      \node at (1,-0.34) {$g^{-1}$};
    \end{tikzpicture}
  \end{array}
\right)
\ee
It is important that the three orange dots above are meant to represent the \underline{same} object in our category. They have been written as distinct points on the page for visual purposes. If we do not fill in these triangles then $\pi_1$ would be more complicated than the group $G$. 

\begin{figure}[h]
\centering
\includegraphics[width=3.5in]{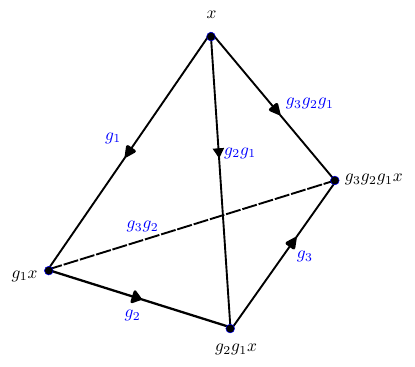}
\caption{\label{fig:oriented-3-simplex}An oriented 3-simplex determined by $g_1, g_2, g_3$. Edges are labeled in \textcolor{blue}{blue} color.}
\end{figure}

Now we have built an oriented simplicial complex with a single $0$-vertex, $1$-simplices corresponding to group elements, and $2$-simplices corresponding to arbitrary pairs of group elements. This complex will have a nontrivial second homotopy group $\pi_2$, and will not be homotopic to $BG$. We can start with this complex and add 3-simplices to produce a space with $\pi_2=0$. To do that, we take triples of group elements and consider the $3$-simplex they define, see \autoref{fig:oriented-3-simplex}.
Filling that in will define a new complex with $\pi_2=0$, but will have nontrivial $\pi_3$. So we keep going. Some readers will wish to have a more systematic description of the construction we have just described.   We tackle that in \autoref{sec:SimplicialSets}.

\bigskip 
\noindent
\begin{remark}
Given the existence of $BG$, we can view finite $G$ gauge theory as  
a homotopy-theoretic version of a nonlinear $\sigma$ model with 
target space $BG$.  The definitions \eqref{eq:FinGroupPF} and \eqref{eq:FinGroupStateSpace} above  can now be  written as: 
\begin{align}
F(M_{n}) &= \sum_{\pi_{0}\left( \mathsf{Map}(M_{n} \to BG) \right)}\frac{1}{|\mathsf{Aut\,}\phi|} ~,\label{eq:BG-Fmn}\\
F(N_{n-1}) &= \mathsf{Fun}\left(\pi_{0}\left( \mathsf{Map}(N_{n-1} \to BG) \right)\right) ~.
\end{align}
where in the sum in \eqref{eq:BG-Fmn}, $\phi$ is any representative ``sigma model map $\phi:M_n \to BG$'' taken from each component. 

This is part of a larger story of the theory of homotopy sigma models that we will sketch in \autoref{sec:FinHomTheory} below. 
\end{remark}

\SectionWithHeader{Simplicial Sets And The Construction of $BG$}{Simplicial Sets And The Construction of $BG$}{sec:SimplicialSets}

In this section, we give a more systematic description of the construction of $BG$ outlined in the previous section. As a bonus, we will learn about \emph{simplicial sets}.  
Simplicial sets play an important role in algebraic topology and are needed to incorporate the higher-form fields and differential cohomology in the general definition of background fields.    Our treatment will be very brief. For more details, see the Wikipedia article \cite{WikiSimplicialSet}, the nLab article \cite{nlab:simplicial_set}, and \cite{Gelfand2003,Friedman:2008,RiehlSimplicialSets,Porter2021}. For a very 
thorough presentation, see \cite{Lurie:Kerodon}.

We begin by reviewing the standard $n$-simplex, identified as a subset of $\IR^{n+1}$: 
\be\label{eq:Standard-n-siplex}
  \Delta_n = \{ (t_0, \ldots, t_n) ~|~ t_{i} \geq 0 \text{ and } \sum_{i=0}^{n}t_i = 1 \} \subset \IR^{n+1} ~.
\ee
There is an ordered set of $n+1$ vertices $v_i$ given by putting $t_i=1$ and all other coordinates to zero.
For each $j=0,\dots, n$, there is the $j^{th}$ \emph{face map}: 
\be 
\delta_{n,j}: \Delta_{n-1} \to \Delta_n ~.
\ee
Note that in $\Delta_n$, the set of points with $t_{j}=0$ is a copy of an $(n-1)$-simplex. It is 
the simplex ``opposite'' to the $j^{th}$ vertex, and we call it the $j^{th}$ face. 
The map $\delta_{n,j}$ maps $\Delta_{n-1}$ to the $j^{th}$ face. In equations, we have:  
\be\label{eq:FaceMaps}
\delta_{n,j}:  (t_0, \dots, t_{n-1})  \mapsto 
\begin{cases}  
(0,t_0, \dots, t_{n-1}) ~, & j =0 ~, \\
(t_0, \dots, t_{j-1},0,t_{j},\dots, t_{n-1}) ~, & 0 <  j < n ~, \\ 
(t_0, \dots, t_{n-1},0) ~,  & j= n ~.
\end{cases}
\ee
  On the other hand, for $0\leq j \leq n$  we also have the degeneracy maps:
\be 
\sigma_{n,j}: \Delta_{n+1} \to \Delta_{n} ~,
\ee
given by ``composing'' the $j^{th}$ and $(j+1)^{th}$ coordinates: 
\be\label{eq:DegeneracyMaps}
\sigma_{n,j}: (t_0, \dots, t_{n+1} ) \mapsto 
\begin{cases} 
(t_0, \dots, t_j + t_{j+1}, t_{j+2}, \dots, t_{n+1}) ~, & 0 \leq j < n ~,\\ 
(t_0, \dots, t_{n-1}, t_n + t_{n+1} ) ~, & j = n ~.
\end{cases}
\ee

It is now straightforward, if tedious, to check the following relations on 
the maps $\delta_{n,j}$ and $\sigma_{n,j}$: 

\be\label{eq:SimplexMapRels-1}
\begin{split}
\delta_{n+1,j} \circ \delta_{n,i}  &  = \delta_{n+1,i}\circ \delta_{n,j-1} ~, \qquad  i < j ~,\\ 
\sigma_{n,j} \circ \sigma_{n+1,i} & = \sigma_{n,i} \circ \sigma_{n+1,j+1} ~, \qquad i \leq j ~,\\
\sigma_{n,j} \circ \delta_{n+1,i} & = 
\begin{cases}
\delta_{n,i}\circ \sigma_{n-1,j-1} ~, & i< j ~,\\ 
\mathsf{Id} ~,  & i = j, j+1 ~,\\ 
\delta_{n,i-1} \circ \sigma_{n-1,j} ~, & i > j+1 ~.
\end{cases}  
\end{split}
\ee

Next, in order to motivate the general definition of a simplicial set, let us recall how one begins the definition of singular homology of a topological space   $M$. The set of singular $n$-simplices $\CS(M)_n$ is the set of all continuous maps $\phi:\Delta_{n} \to M$. In singular homology theory, one defines $C_n(M)$ to be the free Abelian group 
generated by $\CS(M)_n$, and the space of singular cochains is $C^n(M):= \Hom(C_n(M),\IZ)$. These will play some role in \hyperref[part2]{Part II}.
Note that precomposition of $\phi$ with   $\delta_{n,j}$ defines maps: 
\be\label{eq:SimplexMaps-X(M)}
d_{n,j}: \CS(M)_n \to \CS(M)_{n-1} ~ ,  \qquad 0 \leq j \leq n ~.
\ee
That is, for $\phi \in\CS(M)_n $, we have  $d_{n,i}(\phi) := \phi \circ \delta_{n,i}$.  
\tightfootnote{Note that if we extend $d_{n,i}$ to $C_n(M)$ by linearity, then the boundary map 
$\partial: C_n(M) \to C_{n-1}(M)$ is just $\partial = \sum_j (-1)^j d_{n,j}$. Then, $\partial^2=0$, and the homology of the resulting complex is the $n^{th}$ singular homology of $M$. }
Similarly, precomposition with $\sigma_{n,j}$ defines maps, 
\be 
s_{n,i}: \CS(M)_n \to \CS(M)_{n+1}   ~ ,  \qquad 0 \leq j \leq n ~.
\ee
The relations \eqref{eq:SimplexMapRels-1} above on $\delta_{n,j}$ and $\sigma_{n,j}$ imply the relations,
\be\label{eq:SimplexMapRels-2}
\begin{split}
d_i d_j   &  = d_{j-1} d_i  ~,  \qquad  i < j ~,\\ 
s_i s_j  & = s_{j+1} s_i  ~, \qquad i \leq j  ~,\\
d_i s_j  & = 
\begin{cases}
s_{j-1} d_i ~,  & i< j ~,\\ 
\mathsf{Id} ~, & i = j, j+1 ~,\\ 
s_j d_{i-1} ~, & i > j+1 ~,\\
\end{cases}  
\end{split}
\ee
where we have intentionally streamlined the notation and dropped the subscript $n$.

The mathematical structure provided by the sets of simplices in a topological space turns out to be a useful construction in other contexts and so we have the definition:

\bigskip 
\begin{definition}[\textcolor{red}{Simplicial set}] A \emph{simplicial set} $\CX$ consists of the following data: 
\begin{enumerate}

\item A collection of sets $\CX_n$, labeled by nonnegative integers $n$. The sets $\CX_n$ are known as the ``set of $n$-simplices in the simplicial set.'' 

\item A collection of maps, known as {face maps}: 
\be 
d_{n,j}: \CX_n \to \CX_{n-1} ~, \qquad 0 \leq j \leq n ~.
\ee

\item A collection of maps, known as {degeneracy  maps}: 
\be 
s_{n,j}: \CX_n \to \CX_{n+1} ~, \qquad 0 \leq j \leq n ~,
\ee
%


\end{enumerate}
such that the maps satisfy the relations of equation \eqref{eq:SimplexMapRels-2}.  
\end{definition}

\begin{remark}
$\,$
\begin{enumerate}

\item A simplicial set is not a set with special properties. It is the collection of data above satisfying the degeneracy relations. 
Nevertheless, an ordinary set can be considered as a special case of a simplicial set. Suppose $S$ is a set. We can give it the \emph{discrete topology} where every subset is open. Then $\CX(S)_n$ 
is naturally identified with $S$ itself, since any continuous map 
from an $n$-simplex to $S$ must map to a single element of $S$. 
Note that all the degeneracy and face maps are simply the identity 
map on $S$. 

\item There is a much more elegant categorical definition of a simplicial set. To motivate it, let us introduce the ordered set $[n]$ of integers from $0$ to $n$:
\be 
[n] := \{ 0,1,\dots, n\} ~.
\ee
Suppose we have an ordered set $S\subset [n]$ of $k$ integers between $0$ and $n$, where $k\leq n$. 
Then there is a sub-simplex $\Delta_S \subset \Delta_{[n]}$, where all coordinates with $j\notin S$ are equal to zero. Thus, the $j^{th}$ vertex is $\Delta_{\{ j\}}$, and $\Delta_{[n]}= \Delta_n$.  The image of $\delta_{n,j}(\Delta_{[n-1]})$ in $\Delta_{[n]}$ is the sub-simplex $\Delta_{S_{n,j}}$, where  $S_{n,j}$ is the image of the ``coface map'' $\delta^{n,j}: [n-1] \to [n]$ that preserves order and ``misses'' $j$. More is true: If $T \subset [n-1]$, then 
\be\label{eq:delta-nj-prop}
\delta_{n,j}( \Delta_T) = \Delta_{\delta^{n,j}(T)} \subset \Delta_{[n]} ~.
\ee
Similarly, we can define the ``co-degeneracy map'':
\be 
\sigma^{n,j}: [n+1] \to [n] ~,
\ee
that preserves order, and maps   $j$ and $j+1$ in $[n+1]$ to $j$ in $[n]$, for $0\leq j \leq n$. 
Analogous to \eqref{eq:delta-nj-prop}, we have 
\be\label{eq:delta-nj-prop-2}
\sigma_{n,j}( \Delta_T) = \Delta_{\sigma^{n,j}(T)} \subset \Delta_{[n]} ~.
\ee
These maps satisfy the relations \eqref{eq:SimplexMapRels-1}. Now it turns out that \underline{any}
order-preserving map $[n] \to [m]$ can be written as a composition of the coface and 
codegeneracy maps $\delta^{\ell,j}$ and $\sigma^{\ell,j}$ 
for various $\ell$'s and $j$'s. Therefore, we can define a category $\Delta$ whose objects are the sets $[n]$
and whose morphisms are order-preserving maps. We can now define a 
simplicial set to be a functor $\CX: \Delta^{\mathsf{op}} \to \mathsf{SET} $.  (Recall the definition of the 
opposite category in equation \eqref{eq:OppCat} et. seq.) 
Thus, $\CX([n]) = \CX_n$, and 
\be 
\CX(\delta^{n,j})=d_{n,j}: \CX_{n} \to \CX_{n-1} ~,
\ee
and so on. It is now possible to define a category of 
simplicial sets $\mathsf{sSET}$, where the objects are 
simplicial sets and the morphisms are natural transformations 
$\tau: \CX \Rightarrow \CX'$. Concretely, a morphism is a set of 
maps $\tau_n: \CX_n \to \CX_n'$ commuting with the face and degeneracy maps. The category $\mathsf{sSET}$ will be useful in \autoref{sec:BackgroundFields}.

\item Given a topological space $M$, our motivating example of the 
spaces of $n$-simplices $\CS(M)_n$ form a simplicial set $\CS(M)$
known as the \emph{singular set of $M$}.

\end{enumerate}
\end{remark}

\bigskip

\bigskip 
\noindent \textbf{Example 1:} Let $G$ be a group. We form the simplicial set $\CX(G)$ where 
$\CX(G)_n := G^n$, is the Cartesian product of $n$ copies of $G$, and the set is 
often written as: 
\be
\bullet \begin{array}{c}\xlongleftarrow{}  \\[-2.5mm] \xlongleftarrow{} \end{array} G \begin{array}{c} \xlongleftarrow{d^0} \\[-2.5mm] \xlongleftarrow{d^1} \\[-2.5mm] \xlongleftarrow{d^2} \end{array} G\times G \begin{array}{c} \xlongleftarrow{} \\[-3mm] \xlongleftarrow{} \\[-3mm] \xlongleftarrow{} \\[-3mm] \xlongleftarrow{} \end{array} G\times G \times G \begin{array}{c} \xlongleftarrow{} \\[-3mm] \xlongleftarrow{} \\[-3mm] \xlongleftarrow{} \\[-3mm] \xlongleftarrow{} \\[-3mm] \xlongleftarrow{} \end{array} \cdots 
\ee

The face maps are 
\be\label{eq:SimplicialG-facemaps}
\begin{split} 
d_0(g_1, \dots, g_n) & = (g_2, \dots, g_n) ~, \\
d_1(g_1, \dots, g_n) & = (g_1g_2,g_3, \dots, g_n) ~,\\
d_j(g_1, \dots, g_n) & = (g_1, \dots, g_{j-1}, g_j g_{j+1}, g_{j+2}, \dots, g_n ) ~,\qquad 1 < j < n-1 ~,\\
d_{n-1}(g_1, \dots, g_n) & = (g_1,g_2,  \dots,g_{n-2}, g_{n-1} g_n) ~, \\
d_n(g_1, \dots, g_n) & = (g_1, \dots, g_{n-1}) ~,
\end{split}
\ee
and the degeneracy maps are: 
\be 
s_j(g_1, \dots, g_n) = (g_1, \dots, g_j, 1_G, g_{j+1}, \dots, g_n) ~.
\ee

\bigskip 
\noindent \textbf{Example 2:}  Given a category $\CC$, we can form the \emph{nerve of the category}, 
$N(\CC)$ as the simplicial set whose set of $n$-simplices is the set of strings of $n$ composable morphisms. More precisely,  $N(\CC)_0$ is the set of objects of $\CC$, $N(\CC)_1$ is the set of morphisms of $\CC$, and for $n>1$, we let $N(\CC)_n$ to be the set of collections of $n$ morphisms which can be concatenated to make a morphism. The degeneracy maps $s_j : N(\CC)_n \to N(\CC)_{n-1}$ insert the identity morphism in the $j^{th}$ slot, while the face maps $d_j: N(\CC)_n \to N(\CC)_{n-1}$ are defined by:
\be 
d_j (f_1, \dots, f_n) = 
\begin{cases}  
(f_2, \dots, f_n) ~, & j=0 ~,  \\
(f_1, \dots, f_j \circ f_{j+1}, \dots , f_n) ~, & 0 < j < n ~,\\ 
(f_1, \dots, f_{n-1}) ~, & j = n ~.
\end{cases}
\ee
Notice that if we regard a group $G$ as a category with one object, 
that is, as the groupoid ${\rm pt}/\!/G$, then $N({\rm pt}/\!/G)$ is just the simplicial set $\CX(G)$ constructed in the previous example.

Now, to any simplicial set $\CX$, we can associate the \emph{geometric realization}. This is defined as a quotient of the disjoint union of $\CX_n \times \Delta_n$ over all $n\geq 0$ modulo relations for the face and degeneracy maps: 
\be\label{eq:GeometricRealization}
\vert \CX \vert =   
\left( \coprod_{n=0}^{\infty} \CX_n \times \Delta_n \right)\bigg/ \sim ~,
\ee
where the relations are given by:
\be 
(x, \delta_{n+1,j}(t) ) \sim (d_{n+1,j}(x), t) ~,
\ee
for all $x\in \CX_{n+1}$, $t\in \Delta_n$ and $0 \leq j \leq n+1$, together with 
\be 
(x, \sigma_{n+1,j}(t) ) \sim (s_{n+1,j}(x), t) ~,
\ee
for all $x\in \CX_{n+1}$, $t\in \Delta_{n+1}$ and $0 \leq j \leq n+1$. We make this a topological space by taking the discrete topology on $\CX_n$, the standard topology on $\Delta_n$, and then using the quotient topology. 

The construction of the geometric realization raises an obvious 
question: Given a topological space $Y$, consider the simplicial set 
given by the singular set $\CS(Y)$. The singular set has a geometric realization $\vert \CS(Y)\vert$ producing another topological space. 
How are these topological spaces related? 
Note that there is a canonical map:
\be 
\vert \CS(Y)\vert \to Y ~,
\ee
given by evaluation: $(\phi, t) \mapsto \phi(t)$, where $\phi$ is 
a map $\phi: \Delta_n \to Y$ and $t\in \Delta_n$. (The map respects the equivalence relation in \eqref{eq:GeometricRealization}.) 
The references above show that the map $\vert \CS(Y)\vert \to Y$ 
induces an isomorphism of homotopy groups, and when $Y$ is a CW complex (see footnote \ref{foot:CWComplex}), the map is a homotopy equivalence. 
\emph{Thus, in homotopy theory, we may pass freely between simplicial sets to topological spaces.}

Finally, given a group $G$, we form the simplicial set $\CX(G)$ and take its geometric realization to form a concrete model for $BG$: 
\be 
BG = \vert \CX(G) \vert ~ . 
\ee
Note that if $G$ is a topological group, $\CX(G)$ is not the 
same simplicial set as $\CS(G)$.

\SectionWithHeader{Group Cohomology}{Group Cohomology}{sec:GroupCohomology}

The cohomology of $BG$ defines the group cohomology. Concretely,
\eqa{
  C^{n}(G; A) &= \{\phi ~:~ G^{n} \to A \} ~,
}
where $A$ is Abelian. The differential
\eqa{
  \delta: C^{n}(G; A) \longrightarrow C^{n+1}(G; A) ~,
}
is given by
\eqa{
  (\delta\phi)(g_1, \ldots, g_{n+1}) &= \phi(g_2, \ldots, g_{n+1}) - \phi(g_{1}g_{2}, g_{3}, \ldots, g_{n+1}) + \phi(g_{1}, g_{2}g_{3}, \ldots, g_{n+1}) \nn
  &\quad \pm \cdots + (-1)^{n+1}\phi(g_{1}, \ldots, g_{n}) ~.
}
\begin{exbox}{Differential in group cohomology}
\begin{itemize}\itemsep 0pt
  \item[(a)] Check $\delta^2 = 0$.
  \item[(b)] Write out $\delta \phi = 0$ for the first few cases.
\end{itemize}
\end{exbox}

\begin{definition}[\textcolor{red}{Group Cohomology}]$H^{n}(G; A) := \mathsf{ker\,}\delta/\mathsf{im\,}\delta$.
\end{definition}

\SectionWithHeader{Digression: Projective Representations In Quantum Mechanics}{Digression: Projective Representations In Quantum Mechanics}{sec:DigressionProjectiveQM}

One of the most important group cohomologies in physics is $H^{2}(G; \mathsf{U(1)})$, which classifies isomorphism classes of central extensions. It plays an important role in the discussion of symmetries in quantum systems. 

Recall the Born rule summarized in \autoref{sec:QM}, in particular equation 
\eqref{eq:BornRule}. These are the physical probability distributions. They must be preserved by a ``symmetry.'' 
An \emph{automorphism} of a quantum system is a bijective map of states and observables preserving the Born rule. One can show that it is determined by a bijective correspondence on \emph{pure states} preserving the overlap function on pure states: 
\eqa{
  \CO(P_1, P_2) &:= \mathsf{Tr}(P_1  P_2)  ~.
}
It is called the overlap function because if we represent the pure states by 
vectors $\psi_1$ and $\psi_2$, so that, 
\eqa{
  P_{i} &= \frac{|\psi_i\rangle\langle \psi_i|}{\vert\!\vert\psi_i\vert\!\vert^2} ~,
}
then 
\be 
\CO(P_1, P_2)  = \frac{|\langle \psi_1 | \psi_2 \rangle|^2}{\vert\!\vert\psi_1\vert\!\vert^2 \vert\!\vert\psi_2\vert\!\vert^2} ~ . 
\ee

As discussed in \autoref{sec:QM}, the pure states form a \emph{projective} Hilbert space and the overlap function is related to the Fubini-Study metric on projective space:
\eqa{
  \theta(P_1, P_2) &= \cos^2 \frac{d_{\mathsf{FS}(P_1, P_2)}}{2} ~.
}
So the automorphism group of a quantum system is the group of isometries of complex projective space. Wigner's Theorem relates this to (anti-)linear operators on $\CH$.

Let $\mathsf{Aut}(\CH)$ be the group of unitary and antiunitary operators on $\CH$. We define,
\eqas{
  \pi: \mathsf{Aut}(\CH) &\longrightarrow \mathsf{Aut}(\text{QM}) \\
  g &\longmapsto \pi(g): P \mapsto g P g^{-1} ~.
}
Wigner's theorem asserts that $\pi$ is surjective and the kernel is the group $\mathsf{U(1)}$ acting as scalars on $\CH$.
\be
  \begin{tikzcd}[row sep=large]
    1 \arrow[r] & \mathsf{U(1)} \arrow[r] & \mathsf{Aut}(\CH) \arrow[r, "\pi"] & \mathsf{Aut}(\text{QM}) \arrow[r] & 1\\
    & & & G \arrow[u, "\varphi",swap]
  \end{tikzcd}
\ee

Dynamical evolution is described in terms of a 1-parameter group 
of automorphisms $\alpha_t$ on states and observables. For example, given a Hamiltonian, $\alpha_t(\rho) = e^{-\imag  t H} \rho e^{\imag  t H}$. 
Only a subgroup of $\mathsf{Aut}(\text{QM})$ will commute with the flows on the states and observables. Thus, we have a homomorphism 
$\varphi: G \to \mathsf{Aut}(\text{QM})$. By Wigner's theorem, $\forall$ $g \in G$, we can choose a $U(g) \in \mathsf{Aut}(\CH)$, such that
\be
\pi(U(g)) = \varphi(g) ~.
\ee
Because $\pi$ has a kernel, the choice is not unique. Two choices 
$U(g)$ and $\wt U(g)$ are related by a phase: $\wt U(g) = b(g) U(g)$, where $b(g) \in \mathsf{U(1)}$. 
Now, since $\varphi(g_1) \varphi(g_2) = \varphi(g_1 g_2)$, and 
since $\pi$ is a group homomorphism, we deduce: 
\eqa{
  \pi\big( U(g_1) U(g_2) \big) &= \pi\big( U(g_1 g_2)\big) ~.
  \label{eq:Ug1Ug2}
}
Equation \eqref{eq:Ug1Ug2}  does \underline{not} imply that 
$U(g_1) U(g_2)$ is the same as $U(g_1 g_2)$, again because $\pi$ has 
a kernel. All we can conclude is that: 
\eqa{
  U(g_1)U(g_2) &= c(g_1, g_2) U(g_1 g_2) ~,
}
for some function $c: G\times G \to \mathsf{U(1)}$.
Note that if we make a different choice $U(g) \to \wt U(g)$, then the function $c$ changes accordingly. 

Now assume (for simplicity) that the $U(g)$ are $\IC$-linear for $g \in G$. Then, $c \in Z^{2}(G; \mathsf{U(1)}):= \mathsf{ker}(\delta)\cap C^{2}(G;\mathsf{U(1)})$.

\begin{exbox}{Cocycle in group cohomology} Show that $c \in Z^{2}(G; \mathsf{U(1)})$.
\end{exbox}

The pullback group, which is, as a set: 
\eqa{
  \widetilde{G} &= \mathsf{U(1)} \times G ~,
}
has the group law:  
\eqa{
  (z_1, g_1)\cdot(z_2, g_2) := (z_1 z_2 c(g_1,g_2), g_1g_2) ~.
}
The group $\widetilde{G}$ is the group which is  linearly represented on $\CH$: 
\eqa{
  T\big( (z, g) \big) &= z\, U(g) ~.
}
A good example is a spin-$1/2$ qubit where the $\mathsf{SO(3)}$ isometry of $\IC\IP^1$ is represented on the Hilbert space $\IC^2$ by the central extension
\be
\begin{tikzcd}
  1 \arrow[r] & \IZ_2 \arrow[r] & \mathsf{SU(2)} \arrow[r] & \mathsf{SO(3)} \arrow[r] & 1 ~.
\end{tikzcd}
\ee

\begin{exbox}{$\IC$-antilinear symmetries} Generalize the above discussion to include 
symmetries which are $\IC$-antilinear. 
\end{exbox}

\SectionWithHeader{Dijkgraaf-Witten Theory}{Dijkgraaf-Witten Theory}{sec:DW-Theory}

Dijkgraaf and Witten \cite{Dijkgraaf:1989pz} gave a ``lattice gauge theory model'' of topological finite $G$ gauge theory, and an important generalization thereof based on group cohomology.

Let us just describe it for $2$ dimensions, and we just explain how to compute the partition function. The construction generalizes to ``fully extended'' (see below) $n$-dimensional theories.

Let $M_2$ be a compact oriented surface. \underline{Choose} a triangulation $\Delta$ on $M_2$. Require that the gauge field be flat, so that the plaquette Boltzmann weight is only determined by two group elements, as in \eqref{eq:plaquetteBoltzmannWeight}.

\be\label{eq:plaquetteBoltzmannWeight}
\begin{tikzpicture}[scale=1.5,baseline={1cm+1.5*height("$=$")},>=Stealth,decoration={
  markings,
  mark=at position 0.5 with {\arrow{>}}}]
  \node (A) at (0,0)  {};
  \node (B) at (1,1.732)  {};
  \node (C) at (2,0)  {};
  \draw[thick,postaction={decorate}] (B.mid)--(A.mid);
  \draw[thick,postaction={decorate}] (A.mid)--(C.mid);
  \draw[thick,postaction={decorate}] (C.mid)--(B.mid);
  \node at (1,-0.35) {$g_1$};
  \node at (2,0.9) {$g_2$};
  \node at (-0.4,0.9) {$(g_1 g_2)^{-1}$};
\end{tikzpicture}
\quad \text{has weight } W(g_1, g_2) \in \IC^{\times} ~.
\ee
Some choices must be made for this to be unambiguous.
\eqa{
  F(M_2) &= \sum_{\mathsf{Map}(\mathsf{Edges} \to G)}\prod_{\Delta}W(g_1, g_2) ~.
}
Now demand invariance under triangulation:
\be
\begin{tikzpicture}[scale=1.5,baseline={1.5cm-0.5*height("$=$")},>=Stealth,decoration={
  markings,
  mark=at position 0.5 with {\arrow{>}}}]
  \node (A) at (0,0)  {};
  \node (B) at (2,0)  {};
  \node (C) at (2,2)  {};
  \node (D) at (0,2) {};
  \draw[thick,postaction={decorate}] (A.mid)--(B.mid);
  \draw[thick,postaction={decorate}] (B.mid)--(C.mid);
  \draw[thick,postaction={decorate}] (C.mid)--(D.mid);
  \draw[thick,postaction={decorate}] (A.mid)--(D.mid);
  \draw[thick,postaction={decorate}] (A.mid)--(C.mid);
  \node at (1,-0.35) {$g_1$};
  \node at (2.4,1) {$g_2$};
  \node at (-0.75,1) {$g_1 g_2 g_3$};
  \node at (1,2.35) {$g_3$};
  \node at (1.2,0.7) {$g_1 g_2$};
\end{tikzpicture} 
= 
\begin{tikzpicture}[scale=1.5,baseline={1.5cm-0.5*height("$=$")},>=Stealth,decoration={
  markings,
  mark=at position 0.5 with {\arrow{>}}}]
  \node (A) at (0,0)  {};
  \node (B) at (2,0)  {};
  \node (C) at (2,2)  {};
  \node (D) at (0,2) {};
  \draw[thick,postaction={decorate}] (A.mid)--(B.mid);
  \draw[thick,postaction={decorate}] (B.mid)--(C.mid);
  \draw[thick,postaction={decorate}] (C.mid)--(D.mid);
  \draw[thick,postaction={decorate}] (A.mid)--(D.mid);
  \draw[thick,postaction={decorate}] (B.mid)--(D.mid);
  \node at (1,-0.35) {$g_1$};
  \node at (2.4,1) {$g_2$};
  \node at (-0.75,1) {$g_1 g_2 g_3$};
  \node at (1,2.35) {$g_3$};
  \node at (1.4,1.25) {$g_2 g_3$};
\end{tikzpicture} ~,
\ee
which implies that $W(g_1, g_2)$ is a group cocycle. Then  one checks, using the cocycle identity, that we have invariance under refinement of the triangulation: 
\be
\begin{tikzpicture}[scale=1.5,baseline={1.5cm-0.5*height("$=$")},>=Stealth,decoration={
  markings,
  mark=at position 0.6 with {\arrow{>}}}]
  \node (A) at (0,0)  {};
  \node (B) at (1,1.732)  {};
  \node (C) at (2,0)  {};
  \draw[thick,postaction={decorate}] (B.mid)--(A.mid);
  \draw[thick,postaction={decorate}] (A.mid)--(C.mid);
  \draw[thick,postaction={decorate}] (C.mid)--(B.mid);
  \node at (1,-0.35) {$g_1$};
  \node at (1.9,0.9) {$g_2$};
  \node at (-0.1,0.9) {$g_1 g_2$};
\end{tikzpicture}
= 
\begin{tikzpicture}[scale=1.5,baseline={1.5cm-0.5*height("$=$")},>=Stealth,decoration={
  markings,
  mark=at position 0.6 with {\arrow{>}}},dot/.style={circle,fill,minimum size=#1,inner sep=0pt,outer sep=0pt},dot/.default=6pt]
  \node (A) at (0,0)  {};
  \node (B) at (1,1.732)  {};
  \node (C) at (2,0)  {};
  \node (D) at (1,0.866) [dot] {};
  \draw[thick,postaction={decorate}] (B.mid)--(A.mid);
  \draw[thick,postaction={decorate}] (A.mid)--(C.mid);
  \draw[thick,postaction={decorate}] (C.mid)--(B.mid);
  \draw[thick,postaction={decorate}] (B.mid) -- (D.mid);
  \draw[thick,postaction={decorate}] (D.mid) -- (A.mid) node [midway,below,sloped] {$g_1 g_2 g_3$};
  \draw[thick,postaction={decorate}] (C.mid) -- (D.mid) node [midway,below,sloped] {$g_2 g_3$};
  \node at (1,-0.35) {$g_1$};
  \node at (1.9,0.9) {$g_2$};
  \node at (-0.1,0.9) {$g_1 g_2$};
  \node at (1.2,1.2) {$g_3$};
\end{tikzpicture}
\ee

It is a mathematical theorem \cite{PACHNER1991129} that all triangulations can be obtained by these two moves. So $F(M_2)$ is independent of triangulation. Moreover, $F(M_2)$ depends only on the group cohomology class determined by $W(g_1, g_2)$.

In $n$-dimensions, we use a simplicial decomposition and the Boltzmann weights are an $n$-cocycle over $G$ valued in $\IC^{\times}$. In particular, in $3$ dimensions, the theory is determined by an element of $H^{3}(G; \mathsf{U(1)})$. Using techniques 
discussed in \autoref{subsec:SeveralMathPrelim} below, one can prove 
that for a finite group one has 
\be 
H^{3}(G; \mathsf{U(1)}) \cong H^{4}(G; \IZ) ~ . 
\ee

Now, 3d Chern-Simons-Witten theory can be formulated for any compact gauge group $G$. Quite generally, the Chern-Simons action is determined by a ``level'' which is an element of $k\in H^4(BG;\IZ)$. It follows that $n=3$ Dijkgraaf-Witten theory is simply Chern-Simons-Witten theory for a finite gauge group. This was, in fact, the original motivation of 
Dijkgraaf and Witten \cite{Dijkgraaf:1989pz}.

\SectionWithHeader{Higher Categories}{Higher Categories}{sec:HigherCategories}

Now we want to start describing \emph{extended} TFT. The idea is to take locality to its logical limit.
Some physics students will be resistant to the need to extend to higher categories, so we give \underline{five} motivations for doing so. 

\bigskip
\bigskip

The \underline{first way} proceeds by considering the gluing formula relating partition functions to pairings of vectors in statespaces. The basic idea is summarized by the picture: 
\be
F\bigg(\includegraphics[valign=c,width=2in]{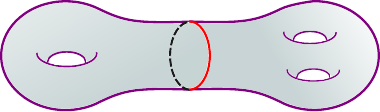}\bigg) = \left\langle F\bigg(\includegraphics[valign=c,width=1.25in]{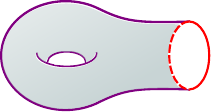}\bigg), F\bigg(\includegraphics[valign=c,width=1.25in]{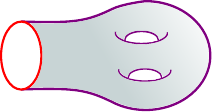}\bigg) \right\rangle ~.
\ee
In this picture, the closed manifold on the left-hand side is $n$-dimensional in an $n$-dimensional quantum field theory. We could ask if the same picture also applies when 
the closed manifold on the left-hand-side is  $(n-1)$-dimensional. So, suppose the compact manifold $N_{n-1}$ without boundary can likewise be glued together from pieces:  
\eqa{
   N_{n-1} = N_{n-1}^{0} \bigcup\limits_{\Sigma_{n-2}}N^1_{n-1}  ~ . 
   \label{eq:CutSpatialSlice}
}
Here $N^i_{n-1}$ are compact manifolds with boundary $\Sigma_{n-2}$, and 
we are gluing them along $\Sigma_{n-2}$ to produce the compact manifold without boundary $N_{n-1}$. Now, we  know that $F(N_{n-1}) = F_{\rm statespace}(N_{n-1})$ is the space of states on the closed spatial manifold $N_{n-1}$. It is natural to expect that there should be mathematical objects $F(N^i_{n-1})$ and $F(\Sigma_{n-2})$, which 
can be ``glued'' to produce the vector space $F(N_{n-1})$.

\bigskip
\bigskip

\begin{figure}[H]
  \centering
  \begin{tikzpicture}[>=Stealth,baseline=-0.5ex]
    \node (graph) [inner sep = 0pt] {\includegraphics[width=1.5in,valign=c]{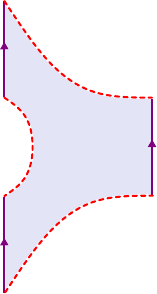}};
    \node at (-2.1,-3.5) {$\textcolor{OliveGreen}{\bm{a}}$};
    \node at (-2.1,-1.2) {$\textcolor{OliveGreen}{\bm{b}}$};
    \node (pnode) at (-4.5,0) {\text{1-morphism}};
    \draw[->] (pnode.mid east) to [out=45,in=180] (-2.2,+2.5);
    \draw[->] (pnode.mid east) to [out=-45,in=180] (-2.2,-2.5);
    \node (rnode) at (4,0) {\text{1-morphism}};
    \draw[->] (rnode.mid west) to [out=0,in=0] (2.1,0);
    \node (nlow) at (2.4,-2.5) {\begin{tabular}{l}\text{2-morphisms between} \\ \text{1-morphisms}\end{tabular}};
    \draw[->] (nlow.mid west) to [out=180,in=270] (0,0);
    \node at (-2.1,+1.2) {$\textcolor{OliveGreen}{\bm{b}}$};
    \node at (-2.1,3.5) {$\textcolor{OliveGreen}{\bm{c}}$};
    \node at (+2.1,1.2) {$\textcolor{OliveGreen}{\bm{c}}$};
    \node at (+2.1,-1.2) {$\textcolor{OliveGreen}{\bm{a}}$};
    \end{tikzpicture}
  \caption{Higher categories from 2d open/closed string theory.}\label{fig:tqfthighercat}
  \end{figure}
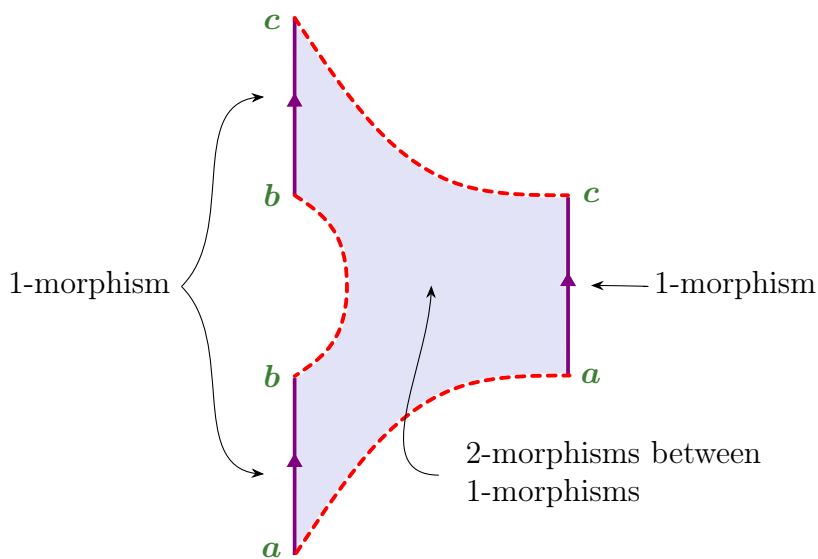

A \underline{second way} of motivating higher categories comes from considering 2d open/closed string theory: We have already seen that 
intervals are morphisms -- which we now call ``1-morphisms'' -- in a 
bordism category, and the field theory $F$ associates to these ``1-morphisms,'' open string state spaes $\CO_{ab} = \Hom_{\mathfrak{B}}(a,b)$, which are ``1-morphisms'' in a category $\mathfrak{B}$ of boundary conditions. It is natural to ask if we can view bordisms such as that shown in \autoref{fig:tqfthighercat}, as ``2-morphisms'' between the intervals (``1-morphisms'') in an extension of the bordism category, and whether those are represented by the field theory as some kind of ``2-morphisms'' in a target category. 

\bigskip
\bigskip

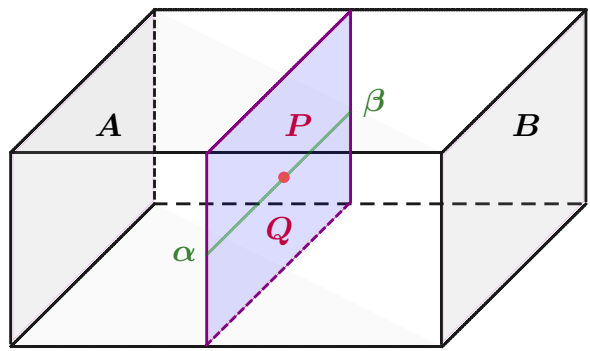
\begin{figure}[H]
    \centering
    \begin{tikzpicture}[>=Stealth,baseline=-0.5ex]
      \node (graph) [inner sep = 0pt] {\includegraphics[width=3in,valign=c]{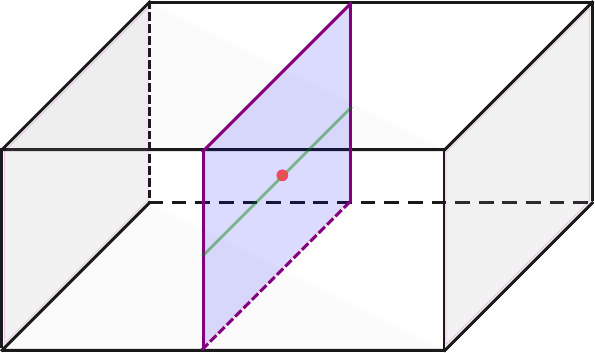}};
      \node at (-1.5,-1) {$\textcolor{OliveGreen}{\bm{\alpha}}$};
      \node at (1,+1) {$\textcolor{OliveGreen}{\bm{\beta}}$};
      \node at (0,0.7) {$\textcolor{purple}{\bm{P}}$};
      \node at (-0.25,-0.7) {$\textcolor{purple}{\bm{Q}}$};
      \node at (+3,0.7) {$\bm{B}$};
      \node at (-2.5,0.7) {$\bm{A}$};
      \end{tikzpicture}
    \caption{Defects within defects.}\label{fig:tqfthighercatbox}
    \end{figure}

  A \underline{third way} of motivating these ideas comes from thinking of defects within defects, see \autoref{fig:tqfthighercatbox}. See the ICM 2010 lecture by Kapustin \cite{Kapustin:2010ta} for a discussion of this. Briefly, in the $n$-dimensional spacetime of \autoref{fig:tqfthighercatbox}, there are theories in the regions to the left and right of the codimension $1$ (i.e., $(n-1)$-dimensional) domain wall in the middle. We can consider the 
  boundary conditions $A$ and $B$ as domain walls between a theory and the empty or trivial theory. The domain walls are labeled by $P,Q,...$
  and can be thought of as morphisms between a category of theories. But there can be codimension $2$ (i.e., $(n-2)$-dimensional) domain walls within domain walls separating $P$ and $Q$ as shown in the picture. These ``2-morphisms'' will carry labels $\alpha, \beta,\dots$. Then there can be codimension $3$ domain walls between $\alpha$ and $\beta$. And so on. 

  For a concrete realization of the above idea in the context of supersymmetric 2d Landau-Ginzburg theory, see \cite[Sec. 3.3.3]{Gaiotto:2015zna}.

\bigskip
\bigskip

A \underline{fourth way} comes from thinking about the proper categorical interpretation of the fundamental group: Let $X$ be a topological space. We form a \underline{category} whose objects are the points of $X$. The morphisms $x_1 \to x_2$ are paths in $X$, $\gamma: x_1 \to x_2$ considered up to homotopy with fixed endpoints: $\Hom(x_1, x_2) = \CP(x_1,x_2)/\mathsf{homotopy}$. Then $\pi_1(X, x_0) = \Hom(x_0, x_0) := \mathsf{Aut}(x_0)$. This (important) category is called the \emph{fundamental groupoid} $\pi_{\leq 1}(X)$. But we could make a more elaborate object if we decline to consider paths only up to homotopy. We could consider $\pi_{\leq 2}(X)$ where $1$-morphisms between objects ($x_1 \to x_2$) $\CP(x_1, x_2)$ and ``$2$-morphisms'' are homotopies of paths. And so on, up to $\pi_{\infty}(X)$.

\bigskip
\bigskip

\begin{figure}[h]
  \qquad
  \begin{tikzpicture}[>=Stealth,baseline=-0.5ex]
  \node (graph) [inner sep=0pt]{\includegraphics[width=3.5in]{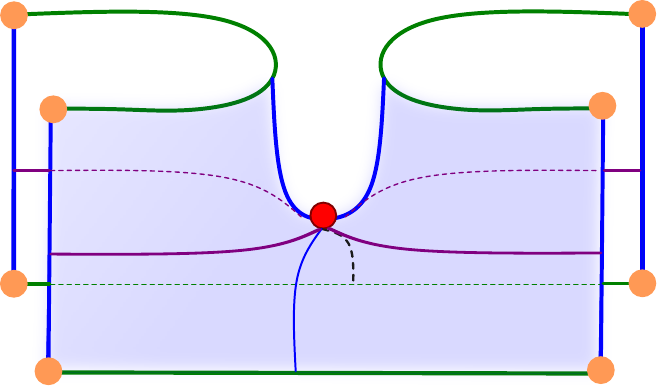}};
  \draw[->,ultra thick,red] (4.5,2.8) -- (5.2,2.8); 
  \draw[->,ultra thick,red] (4.0,1.4) -- (4.7,1.4); 
  \draw[->,ultra thick,red] (4.5,2.8) -- (4.5,3.5); 
  \draw[->,ultra thick,red] (4.0,1.4) -- (4.0,2.1); 
  \draw[->,ultra thick,red] (4.5,-1) -- (5.2,-1); 
  \draw[->,ultra thick,red] (4.0,-2.2) -- (4.7,-2.2); 
  \draw[->,ultra thick,red] (4.5,-1) -- (4.5,-0.3); 
  \draw[->,ultra thick,red] (4.0,-2.2) -- (4.0,-1.5); 
  \draw[->,ultra thick,red] (-4.0,-1) -- (-3.3,-1); 
  \draw[->,ultra thick,red] (-3.5,-2.2) -- (-2.8,-2.2); 
  \draw[->,ultra thick,red] (-4.0,-1) -- (-4.0,-0.3); 
  \draw[->,ultra thick,red] (-3.5,-2.2) -- (-3.5,-1.5); 
  \draw[->,ultra thick,red] (-4.0,2.8) -- (-3.3,2.8); 
  \draw[->,ultra thick,red] (-3.5,1.4) -- (-2.8,1.4); 
  \draw[->,ultra thick,red] (-4.0,2.8) -- (-4.0,3.5); 
  \draw[->,ultra thick,red] (-3.5,1.4) -- (-3.5,2.1); 
  \node at (0,-3) {$\xrightarrow{\hspace*{3cm}}$};
  \node at (0,-3.5) {$t_1 \in [0,1]$};
  \node at (-5.8,0) {$t_2 \in [0,1]$};
  \node at (-4.8,0) [rotate=90]{$\xrightarrow{\hspace*{3cm}}$};
  \end{tikzpicture}
  \caption{Topology induced by a saddle.}
  \label{fig:manifoldwithcorners-v2}
\end{figure}
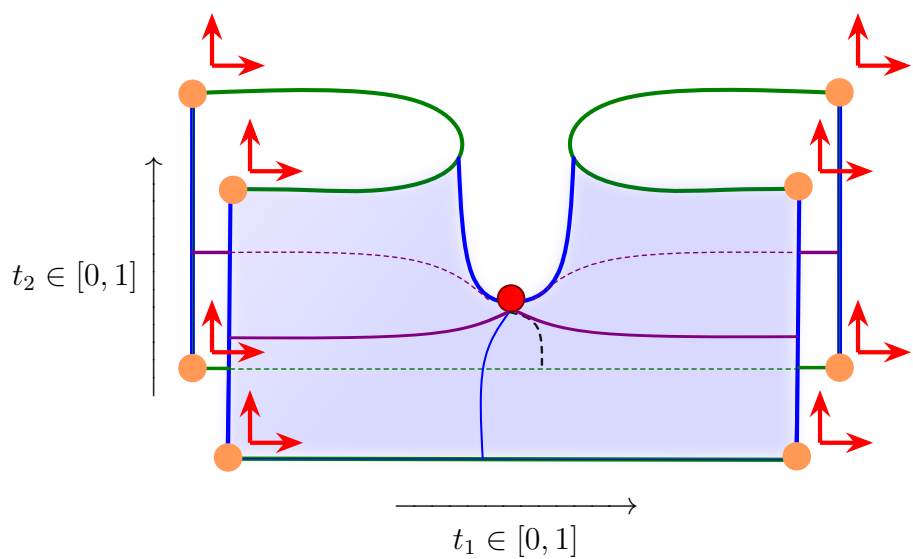

A \underline{fifth way} originates from Morse theory. Let us revisit topology change induced by a saddle, such as the one shown in \autoref{fig:ex2} with a single critical point of Morse index $1$. Focusing on the neighborhood of this critical point, we see that it resembles \autoref{fig:manifoldwithcorners-v2}.

We can start to see the structure of a higher category here: 
The objects (now called ``0-morphisms'')  are the orange points. At time $t_2 = 0$, we have a bordism from two points to two points colored in green. It is the evolution along $t_1$. Recall that the bordism is a $1$-morphism in the category $\mathsf{Bord}_{\langle 0,1\rangle}$.
At time $t_2 = 1$, we have another bordism between the same pair of objects we had at $t_2 = 0$, but the green bordism at $t_2=0$ is a different $1$-morphism in $\mathsf{Bord}_{\langle 0,1\rangle}$ from that at $t_2=1$. The saddle itself is a $2$-morphism between these two $1$-morphisms. The saddle in this figure is an example of a 2-morphism in a $2$-category one could denote by $\mathsf{Bord}_{\leq 2}$. Note that this example shows that 
in extending the bordism category to a higher category, we must make use of manifolds with corners. For a foundational discussion of the theory of manifolds with corners, see \cite{JoyceCorners:2009,JoyceCorners:2015}.  
 
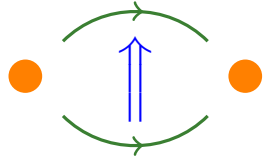
\begin{figure}[h]
\centering
  \adjustbox{scale=1.5,valign=c}{%
  \tikzcdset{arrow style=tikz, diagrams={>= ... }}
  \begin{tikzcd}[>=Classical TikZ Rightarrow,column sep=tiny,decoration={
    markings,
    mark=at position 0.55 with {\arrow{>}}}]
  \textcolor{orange}{\bigbullet} \arrow[rr,bend left=45, OliveGreen,thick,dash,postaction={decorate}] \arrow[rr,bend right=45,OliveGreen,thick,dash,postaction={decorate}] & \textcolor{blue}{\Big\Uparrow}  & \textcolor{orange}{\bigbullet} 
  \end{tikzcd}
  }
  \caption{2-morphisms.}\label{fig:2-morphism-Picture}
\end{figure}

In general, if we think of categories in terms of directed graphs, when we add 2-morphisms, we introduce a new kind of arrow, see \autoref{fig:2-morphism-Picture}.

\begin{figure}[h]
\centering
  \adjustbox{scale=1.5,valign=c}{%
  \tikzcdset{arrow style=tikz, diagrams={>= ... }}
  \begin{tikzcd}[>=Classical TikZ Rightarrow,column sep=tiny,decoration={
    markings,
    mark=at position 0.55 with {\arrow{>}}}]
  \textcolor{orange}{\bigbullet} \arrow[rr,bend left=45, OliveGreen,thick,dash,postaction={decorate}] \arrow[rr,bend right=45,OliveGreen,thick,dash,postaction={decorate}] & \textcolor{blue}{\Big\Uparrow}  & \textcolor{orange}{\bigbullet} \arrow[rr,bend left=45, OliveGreen,thick,dash,postaction={decorate}] \arrow[rr,bend right=45,OliveGreen,thick,dash,postaction={decorate}] & \textcolor{blue}{\Big\Uparrow} & \textcolor{orange}{\bigbullet}
  \end{tikzcd}
  }
  \quad 
  \adjustbox{scale=1.5,valign=c}{%
  \tikzcdset{arrow style=tikz, diagrams={>= ... }}
  \begin{tikzcd}[>=Classical TikZ Rightarrow,column sep=tiny,decoration={
    markings,
    mark=at position 0.55 with {\arrow{>}}}]
  \textcolor{orange}{\bigbullet} \arrow[rr,bend left=45, OliveGreen,thick,dash,postaction={decorate}] \arrow[rr,bend right=45,OliveGreen,thick,dash,postaction={decorate}]  \arrow[rr,OliveGreen,thick,dash,postaction={decorate},"\textcolor{blue}{\bm{\Uparrow}}","\textcolor{blue}{\bm{\Uparrow}}"'] & \textcolor{white}{\Big\Uparrow}  & \textcolor{orange}{\bigbullet} 
  \end{tikzcd}
  }
  \caption{Two ways of composing 2-morphisms.}\label{fig:TwoComps-2-morphisms}
\end{figure}
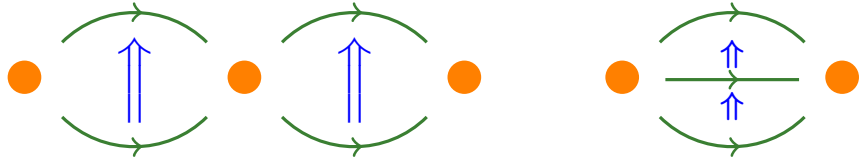

These can be composed in several different ways, for example, two ways of composing 2-morphisms can be illustrated as shown in \autoref{fig:TwoComps-2-morphisms}.

Given the above heuristic ideas, one can give technically precise 
definitions of higher categories. The simplest one uses an inductive procedure: 
 
\begin{definition}[\textcolor{red}{Strict $n$-category}] A \textbf{strict $n$-category} $\CC$ is a collection of objects $C_0(\CC)$  (the ``$0$-morphisms'') and 
for each pair of objects $x,y\in C_0(\CC)$, an $(n-1)$-category $\Hom(x,y)$, such that, for all $x,y,z\in C_0(\CC)$, there exist  
(strict) 
$(n-1)$-bifunctors: 
\be 
c_{x,y,z}: \Hom(x,y) \times \Hom(y,z) \to \Hom(x,z) ~,
\ee
which are (strictly) associative.
\tightfootnote{A ``bi-functor'' means that   $c_{x,y,z}$ is an $(n-1)$-functor if we hold all the $k$-morphisms in $\Hom(x,y)$ fixed 
and consider it as an $(n-1)$-functor $\Hom(y,z) \to \Hom(x,z)$, 
 or if we hold all the $k$-morphisms in  $\Hom(y,z)$ fixed and 
 consider an $(n-1)$-functor $\Hom(x,y) \to \Hom(x,z)$. } 
\end{definition}

If we understand a ``$0$-category'' to be a set, then a $1$-category as defined in \autoref{sec:CategoryBackground} satisfies \eqref{eq:StrictAssociative}, and is 
hence a strict $1$-category. We can then proceed to define 
$n$-categories inductively using the above definition once we agree 
that a   \emph{strict functor} between $n$-categories $\CC$ and $\CD$ is a map of objects $\phi_0: \CC_0 \rightarrow \CD_0$, such that for all $x,y\in \CC_0$, 
\be 
\phi_1: \Hom(x,y) \rightarrow \Hom(\phi_0(x), \phi_0(y)) ~,
\ee
which is a (strict) $(n-1)$-functor. 

\begin{remark}
$\,$
\begin{enumerate}

\item The inductive definition thus provides a collection of 
sets $C_0(\CC), \dots, C_{n}(\CC)$, where $C_k(\CC)$ is the 
set of $k$-morphisms in $\CC$, for $0 \leq k \leq n$. There are
source and target maps: 
\be 
p_0,p_1:  C_{k}(\CC) \to C_{k-1}(\CC) ~,
\ee
and many kinds of compositions. If we choose 
\be\label{eq:HighCatExplSet1}
\begin{split}
x_0,y_0 & \in C_0(\CC) ~,\\ 
x_1,y_1 & \in C_0(\Hom(x_0,y_0)) ~,\\
x_2,y_2 & \in C_0(\Hom(x_1,y_1) ) ~,\\ 
\vdots & \vdots \\ 
x_{n-1},y_{n-1} & \in C_0(\Hom(x_{n-2}, y_{n-2})) ~,
\end{split}
\ee
then $\Hom(x_k,y_k)$ is an $(n-k-1)$-category and the $\ell$-morphisms in this category are such that 
\be\label{eq:HighCatExplSet2}
\begin{split}
C_\ell(\Hom(x_k,y_k)) \subset C_{\ell+1}(\Hom(x_{k-1},y_{k-1})) 
 & \subset C_{\ell+2}(\Hom(x_{k-2},y_{k-2}))    \subset \cdots \\
\cdots \subset C_{\ell+k}(\Hom(x_0,y_0))&  \subset C_{\ell+k+1}(\CC) ~.
\end{split}
\ee

\item The adjective ``strict'' used above refers to the associativity of the compositions. It means that all the many associativity relations between $k$-morphisms hold as equalities. This notion turns out to have limited utility.  
One should relax the strict associativity conditions to allow for 
suitable equivalences between different orders of compositions (much like one allows for an associator in a tensor category). When this is done, much new data and many new conditions are introduced.
The subject of higher categories can get quite technical. 
For a sampling of the literature,   see the works of Barwick et al. \cite{Barwick2016}, Bergner \cite{Bergner2006},
Lurie's kerodon.net \cite{Lurie:Kerodon} and \cite{Lurie2009-el},  Schommer-Pries \cite{Schommer-Pries:2013}, Rezk \cite{Rezk2000}, Riehl et al. \cite{Riehl:2017category,RiehlCategoricalHomotopyTheory,RiehlInfinityCategoryTheory} and references in the surveys by Freed \cite{Freed:CobordismHypothesis,FreedCBMS},   and Teleman \cite{Teleman:2016}.

\item It makes sense to take $n=\infty $  in the definition of an $n$-category. There is then  a useful refinement of the notion of a higher category to a $(p, q)$ category:
\eqas{
  &\text{All $k$-morphisms with $k > p$ are the identity.}\\
  &\text{All $k$-morphisms with $k > q$ are invertible.}\nonumber
}
In this notation, a $(0,0)$ category is a set. A $(1,1)$ category is a category in the normal sense. One commonly encounters the term ``$(\infty, n)$-category''. So it is an $\infty$-category where all $k$-morphisms for $k > n$ are invertible. A good example is $\pi_{\infty}(X)$, which  is an $(\infty,0)$-category. An important notion when discussing the cobordism hypothesis below is the idea of 
truncating to invertible morphisms. If $\CC$ is an $(\infty,n)$-category and $k< n$ then $\CG_k(\CC)$ is the $(\infty,k)$ category 
where we simply delete the noninvertible $\ell$-morphisms with $\ell>k$.

\end{enumerate}

\end{remark}

\bigskip
\noindent \textbf{Example 1:} A good example of a $2$-category is the category $\CC= \mathsf{ALG}(\mathsf{VECT})$. This is the prime example of what is sometimes called a 
\emph{Morita $2$-category}. The notation is meant to indicate that we are considering the category of ``algebra objects'' in the category $\mathsf{VECT}$. In this case, an ``algebra object'' is 
just a finite-dimensional complex algebra, so the   set $C_0(\mathsf{ALG}(\mathsf{VECT}))$ of ``0-morphisms'' is the set of finite-dimensional complex algebras. Given two such algebras $A_1, A_2$, we have a $1$-category $\Hom(A_1,A_2)$.
The objects of $\Hom(A_1,A_2)$ are $A_1-A_2$ bimodules. 
\tightfootnote{An $A_1-A_2$ bimodule is a vector space $M$ which is a left $A_1$-module and a right $A_2$-module where the actions of the two algebras commute. In other words it is a left $A_1 \times A_2^{\mathsf{op}}$-module.   }
The objects of $\Hom(A_1,A_2)$ are the $1$-morphisms of 
$\mathsf{ALG}(\mathsf{VECT})$. The composition of these 
$1$-morphisms is the composition of bimodules:  
\be 
\Hom(A_1, A_2) \times \Hom(A_2,A_3) \to \Hom(A_1,A_3) ~,
\ee
defined by  $\mathsf{M}_1 \circ \mathsf{M}_2:=  \mathsf{M}_1 \otimes_{A_2} \mathsf{M}_2$.
The $1$-morphisms of the $1$-category $\Hom(A_1,A_2)$
are bimodule maps. Indeed, given two $A_1-A_2$ bimodules, 
$\mathsf{M}_1,\mathsf{M}_2$ 
\be 
\Hom_{\Hom(A_1,A_2)}(\mathsf{M}_1,\mathsf{M}_2) ~,
\ee
is simply the linear space of $A_1-A_2$ bimodule maps 
$T: \mathsf{M}_1 \to \mathsf{M}_2$. This means that $T$ is a linear map 
which commutes with the 
left and right $A_1$, $A_2$ actions, respectively. 
These bimodule maps are $2$-morphisms in the $2$-category 
$\mathsf{ALG}(\mathsf{VECT})$. 
It is a good exercise to work out the two different compositions of 2-morphisms shown in \autoref{fig:TwoComps-2-morphisms}. 
\bigskip 

\bigskip
\noindent \textbf{Example 2:} A second useful example of a $2$-category is $\mathsf{CAT}$. The set of $0$-morphisms is the set of (small) $\IC$-linear categories. Given two such categories $C_1$ and $C_2$, the homset between $C_1$ and $C_2$ is the set of functors from $C_1$ to $C_2$. So functors are 1-morphisms. Given two functors $F_{1,2}:C_1 \to C_2$,  we have a \underline{set}  
$\Hom(F_1,F_2)$. It is the   set of natural transformations 
$\tau: F_1 \Rightarrow F_2$. So the 2-morphisms in 
$\mathsf{CAT}$ are natural transformations. 
\bigskip

\bigskip
\noindent \textbf{Example 3:} \emph{The extended bordism category}.
By generalizing the discussion of the saddle shown in \autoref{fig:manifoldwithcorners-v2} above, we can define an $n$-category $\mathsf{Bord}_{\leq n}$ taking the $k$-morphisms to be bordisms (which are manifolds with corners)  with $k$-times. This can even be generalized to an $(\infty,n)$-category by taking $(n+1)$-morphisms to be diffeomorphisms of bordisms preserving all the initial and final $k$-bordisms. Note that these are invertible, whereas the $k$-morphisms with $k\leq n$ are in general not invertible. 
The $(n+2)$-morphisms are isotopies of diffeomorphisms, etc. 

\paragraph{Monoidal Structure:} The notion of a monoidal (tensor) category can be extended to $n$-categories. 
For the $2$-category $\mathsf{ALG}(\mathsf{VECT})$, the monoidal structure is the usual $\otimes$ product of algebras, bimodules, and linear maps. For   $\mathsf{Bord}_{\leq n}$, the monoidal structure is disjoint union. In general, if done honestly, the definition of the monoidal structure beyond the first two levels requires hard work. 

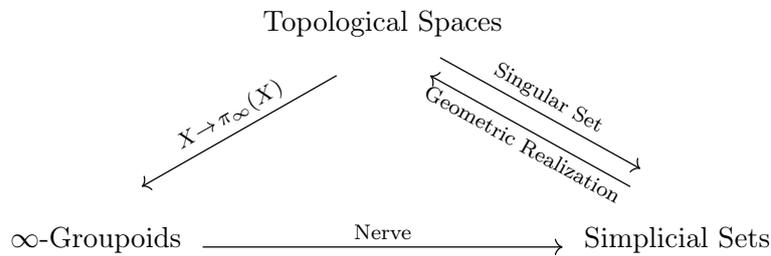
\begin{figure}[h]
\be\nonumber 
  \begin{tikzcd}[row sep=huge,column sep=small]
       & \ar[phantom, "\begin{tabular}{c}\text{\small Topological Spaces}\end{tabular}"above=0pt] \text{\textcolor{white}{Topological Spaces}} \arrow[ld, "X \to \, \pi_{\infty}(X)", sloped] \arrow[rd, "\text{Singular Set}",sloped, shift left=1.5ex] \\
   \text{\textcolor{white}{$\infty$-Groupoids}} \ar[phantom, "\begin{tabular}{c}\text{\small $\infty$-Groupoids} \end{tabular}"below=0pt] \arrow[rr, "\text{Nerve}", shift right=2.5ex] & & \text{\textcolor{white}{Simplicial Sets}} \ar[phantom, "\begin{tabular}{c}\text{\small Simplicial Sets}\end{tabular}"below=0pt] \arrow[lu, "\text{Geometric Realization}",sloped,swap]
  \end{tikzcd}
\ee
\caption{Triangle of concepts.}\label{fig:GroupoidTriangle}
\end{figure}

\paragraph{$\infty$-Groupoids:} An \emph{$\infty$-groupoid} is a
higher category in which all the $k$-morphisms are invertible. Thus it is an $(\infty, 0)$-category in the above notation. A good example is the extension of the fundamental groupoid $\pi_{\leq 1}(X)$ of a topological space $X$ to the $\infty$-category  $\pi_{\infty}(X)$   mentioned above. In some sense, this is the only kind of example. The reason is that from an $\infty$-groupoid, we may produce 
a simplicial set. We take the nerve of the category $\pi_{\leq 1}(X)$ but also add simplices associated with the higher morphisms. 
For example, the 2-morphism of \autoref{fig:2-morphism-Picture} defines 
a $2$-simplex in a natural way. We have already seen that 
in homotopy theory we can freely pass between simplicial sets and 
topological spaces, so we have a triangle of concepts -- see \autoref{fig:GroupoidTriangle} -- equivalent in the eyes of a homotopy theorist.

Thus for example, given a simplicial set $\CX$, we can go around the 
above triangle: 
\be 
\CX \to \vert \CX \vert \to \pi_{\infty}(\CX) \to N(\pi_{\infty}(\vert \CX\vert)) ~,
\ee
and $\CX$ should be equivalent to $N(\pi_{\infty}(\vert \CX\vert))$
in the category of simplicial sets. 
For (much) more information, see \cite{Lurie:Kerodon}.

\SectionWithHeader{Background Fields}{Background Fields}{sec:BackgroundFields}

In physics, we often study quantities like partition functions as functions of other physical parameters. In quantum theory, we might 
also study operator algebras as functions of parameters. These parameters often obey locality properties, and when they do they are 
known as \emph{background fields}, or more simply as \emph{fields}.

\subsection{Some Motivating Examples}\label{subsec:Fields-Motivation}

In order to explain the mathematical definition of what a field is 
it is useful to begin with some examples.

\bigskip
\noindent\textbf{Example 1:}\label{example:scalarfields} \emph{Scalar field theories}. 
Consider a compact Riemannian manifold $(X,g_X)$, known as the 
\emph{target space}. A scalar field theory is a theory 
whose dynamical fields are 
maps $\xi: M_n \to X$, where the domain $M_n$ is a Riemannian 
manifold with metric $g_{M_n}$, to be thought of as the spacetime 
of the field theory. 

The action is a function of the dynamical field $\xi$ and also of the background fields $(g_{M_n}, g_X)$. In order to write it, we note that the  differential 
$d\xi: T_p M_n \to T_{\xi(p)}X$ is an element of $\Hom(T_p M_n, T_{\xi(p)}X)$, and the metrics on the ``worldvolume'' $M_n$ and the ``target'' $X$ define a metric on $\Hom(T_p M_n, T_{\xi(p)}X)$. The action is then,  
\be\label{eq:ScalarFieldAction}
S[\phi;g_{\mu\nu},g_X] = \int_{M_n}  \vert\!\vert d\xi \vert\!\vert^2 \vol(g_{M_n}) ~.
\ee
We have chosen length scales both on the target and the domain 
manifold and set them to $1$ so that all quantities here are dimensionless. The action we have written is sometimes called 
the \emph{nonlinear sigma model action} for historical reasons. 
\tightfootnote{It goes back to the work of M. Gell-Mann and M. Levy \cite{GellMann1960} 
and F. G\"ursey \cite{Gursey1960} on early field theory models of pion dynamics. }

A simple case is where the target space $X$ is a circle of radius $R$, 
so $g_X = R^2(d\phi)^2$ on $X = \IS^1$, where $\phi \sim \phi + 2\pi$ is a periodic target space coordinate. It is customary to choose a local logarithm $\xi = e^{\imag  \phi}$, 
and then the action,  written out with indices, is:
\be 
S[\phi;g_{\mu\nu},R]  = R^2 \int_{M_n} \sqrt{g(x)} g^{\mu\nu}(x) \p_\mu \phi(x) \p_\nu \phi(x) d^n x ~.
\ee

 At least formally, the   partition function is a function of the background fields $g_{M_n}$ and $g_X$: 
\eqa{
  Z[g_{M_n},g_X] = \int d\phi\, e^{-S[\phi; g_{\mu\nu},g_X]} ~.
}
In general, there will be difficulties in giving a sensible definition 
to this partition function -- even by the standards of physicists. 
But there are two notable exceptions: 

First, if $n=2$ so $M_n$ is a 
two-dimensional manifold, then using ideas of the renormalization 
group it can be nicely defined   \cite{Friedan:1980jf,Friedan:1980jm}.

Second, if the target space is flat (so $X$ is a torus), then 
we are working with free field theory, and rigorous definitions are available. Returning to the simple case where $X=\IS^1$, we can express the function explicitly as 
\be 
Z[g_{\mu\nu},R] = \frac{R}{\sqrt{{\det}'\Delta}} \sum_{\omega \in \CH^1_{\IZ}} e^{- R^2 \vert\!\vert\omega \vert\!\vert^2} ~,
\ee
where the sum is over the lattice $\CH^1_{\IZ}$ or harmonic 1-forms on $(M_n, g_{M_n})$ with integral periods, and ${\det}'\Delta$ is the 
determinant of the scalar Laplacian acting on the space of functions orthogonal to the constant functions. Note that in the $R\to\infty$ limit, we expect to obtain the real-valued scalar field. The theta function goes to $1$ exponentially fast in $R$, but there is an overall factor of $R$. In this noncompact limit of the target space, 
the finite quantity of interest is the partition function per unit volume of the target space.

In QFT textbooks, another set of background fields is also typically introduced. Let $\CO$ be the algebra of local operators, choose a 
linear basis $\CO_n(x)$ of local operators, and let 
$J\in \mathsf{Map}(M_n \to \CO^\vee)$. Then we can extend the action by 
\be\label{eq:ScalarFieldActionSourced}
S[\phi;g_{\mu\nu},g_X;J] = \int_{M_n}  \vert\!\vert d\xi \vert\!\vert^2 \vol(g_{M_n})  + \int_{M_n} \sum_n J^n(x)\CO_n(x) \vol(g_{M_n}) ~,
\ee
where $J^n(x)$ are the components relative to a dual basis. 
This is known as \emph{adding source terms}. These too, are background fields, although in the quantum theory, it might be necessary to study the partition function only as a formal series in $J^n$.

\bigskip
\noindent\textbf{Example 2:}\label{example:orientation} \emph{The Quantum Rotor}. 
We consider a one-dimensional scalar field theory, where the 
target space is a circle with round metric, but there is an additional 
parameter $B$, which is a real number. The action is 
\be\label{eq:PartCircle}
\begin{split}  S  
& =\int_{M_1}  \half I \dt{\phi}^2 dt + { B\over 2\pi } \dt{\phi}
dt ~.\\
\end{split}
\ee
Here, $I$ plays the role of the radius squared and the second term does not affect the classical theory but plays a very important role in the quantum theory. There are many physical interpretations of this action, including that of a charged massive particle constrained to move on a circle that links a solenoid carrying flux $B$, and the dimensional reduction of Yang-Mills theory with a $\theta$-term. 
The reader should note that the two terms in the action behave differently under time-reversal $t \mapsto - t$. Thus, the action 
depends not only on $I$ and $B$ but on another quantity,  the 
\underline{orientation} $\mathfrak{o}$ of the worldvolume $M_1$. The quantum 
partition function is thus a (very interesting) function of 
$(I,B,\mathfrak{o})$. As we will argue below, $\mathfrak{o}$ should also be considered as a background field. As in the previous case, it can also be coupled to other background fields. For expository discussions of this very interesting and crucial example, we refer the reader to \cite[App. D]{Gaiotto:2017yup}, \cite[Sec. 2.1]{Moore:2019TASI}, and \cite[Sec. 15.4]{MooreGroupTheory:2023}.

\bigskip
\noindent\textbf{Example 3:}\label{example:spin} \emph{Spin Theories}. Theories that involve fermions typically require a choice of spin or spin$^{\mathsf{c}}$ structure on the spacetime. A simple example is supersymmetric quantum mechanics, where the dynamical fields are maps of a $(1|1)$ supermanifold
\tightfootnote{For mathematical expositions of supermanifolds and supergeometry, see, for example, \cite{Leites1980,DeWitt1992,Freed:1999mn,DeligneFreedSupersymmetry:1999,Deligne:1999ur,Manin1997,Teichner2011,Witten:2012bg}. For physical expositions, see \cite{Gates:1983nr,Wess1992-ab,Terning2005}.}
to a Riemannian target space $(X,g_X)$.  The partition function on the circle depends on the choice of spin structure. If it is the nonbounding (a.k.a. Ramond, a.k.a. periodic) spin structure, then the partition function is the \emph{Witten index}. It is a topological invariant of the target space $X$. In fact, it is just the Euler character of $X$ \cite{Witten:1982df,Witten:1982de,Witten:1982im}. 
By contrast, the partition function with the other spin structure, the bounding (a.k.a. Neveu-Schwarz, a.k.a. anti-periodic) spin structure, then the partition function is a complicated function of $(X,g_X)$. Thus, supersymmetric quantum mechanics is an example of a \emph{spin theory}, namely, a theory that 
depends on a choice of spin structure on the spacetime of definition. 
Once again, spin structures will be considered to be examples of 
background fields. 

Another, important, example of a spin theory is the \emph{Arf theory}. This is a two-dimensional TFT defined on spin manifolds. The partition function for a closed, oriented, spin manifold $(M_2, \mathfrak{s})$ 
is the \emph{Arf invariant}, which is valued in $\{ \pm 1 \}$. 
The Arf invariant can be defined in many ways. One way \cite{Atiyah1971},  identifies it as the mod-two index of the Dirac operator. (In terms of holomorphic geometry, the spin structure defines a square root of the canonical bundle $K$, and the invariant
can be defined as $\dim\,H^0_{{\rm hol}}(M_2; K^{1/2})$, since the Riemannian metric on $M_2$ defines a complex structure.)   Another way identifies it as the Arf invariant of the quadratic function 
$q: H_1(M_2;\IZ_2) \to \IZ_2$, where $q$ on a circle is $0$ or $1$, 
according to whether the circle inherits bounding or nonbounding spin structure from $\mathfrak{s}$, respectively. The Hilbert space on 
the circle with bounding spin structure is the super-Hilbert space 
$\IC^{1\vert 0}$, and on the circle with non-bounding spin structure is the super-Hilbert space $\IC^{0\vert 1}$.
For much more information about this theory, see \cite{Gunningham2016,Debray2018,Karch:2019lnn}.
\tightfootnote{The Arf theory was first used in quantum field theory in the theory of the 2d boson/fermion correspondence \cite{Alvarez-Gaume:1986nqf,Alvarez-Gaume:1987wwg}. 
It was found that this term needed to be included in the theory of the 2d scalar field to make it a spin theory with a partition function equal to that of a fermion with a fixed spin structure. }

\subsection{Freed-Hopkins-Teleman Definition Of  A Field }\label{subsec:Fields-FHT-def}

In all the above examples we have encountered various fields. 
These include traditional fields like the scalar field $\xi: M_n \to X$ and the Riemannian metric $g_{M_n}$ in \hyperref[example:scalarfields]{Example 1}, but, as has been argued by D. Freed, M. Hopkins and C. Teleman, in 
\cite{Freed:2012bs,Freed:2013gc} that, for example, 
 the orientation $\mathfrak{o}$ of \hyperref[example:orientation]{Example 2}, and the spin structure $\mathfrak{s}$ of \hyperref[example:spin]{Example 3}  
should equally be considered to be ``fields.'' Moreover, the parameters like $I$ and $B$ in \hyperref[example:orientation]{Example 2} could also be rendered spacetime dependent -- an old idea known as the \emph{Janus construction} -- and these two should be considered to be ``fields.'' 
Usually, they are constrained to be constant as functions of spacetime. Quite generally, given any symmetry, one should specify 
background fields associated with that symmetry. For example, 
if a theory has a global symmetry $G$, then it is natural to 
make the theory equivariant by defining the theory coupled to a principal $G$-bundle with connection over spacetime.

Some general locality properties we wish our fields to satisfy are: 
\begin{enumerate} 

\item   They should pull back under smooth maps. In order to incorporate nondegenerate metrics and orientations in the definition, 
we restrict this condition to the condition that they pull back (or push forward) under local diffeomorphisms. 

\item They should satisfy a sheaf property: if they are defined on open sets $\CU$ and $\CV$, and agree on $\CU \cap \CV$, then there is a unique extension to $\CU \cup \CV$.

\end{enumerate}

In general, the TFT functor gives the answer to the path integral: the dynamical fields (if there are any) have been integrated out and are not present in the definition. However, the background fields are present and we study how the partition functions, statespaces, amplitudes, etc. depend on these background fields. It is useful to 
begin with a classical  field theory (which, as we have noted in \autoref{sec:SumsAndProducts} above, defines an invertible field theory) with fields $\CF$ then declare another set of fields $\CF^{{\rm bck}}$ to be the background fields. We must have some kind of 
map 
\be 
\pi: \CF \to \CF^{{\rm bck}} ~.
\ee
The quantization procedure (e.g., by path integration) then integrates over the fiber of this map to produce a functor that 
depends on $\CF^{{\rm bck}}$.

We can give a precise mathematical definition of ``fields'' following 
Appendix A of \cite{Freed:2012bs} and section 3 of \cite{Freed:2013gc} by formalizing them as a ``sheaf on $\mathsf{Man}_{n}$.'' This is the 
same thing as a functor: 
\eqa{
   \mathcal{F} : \mathsf{Man}_{n}^{\mathsf{op}} \longrightarrow \mathsf{sSET}  ~,\label{eq:FunctorDefBckFlds}
}
where $\mathsf{Man}_{n}$ is the category whose objects are smooth 
$n$-dimensional manifolds and morphisms are local diffeomorphisms 
between $n$-manifolds. 
The superscript $\mathsf{op}$ means the opposite category (see equation \eqref{eq:OppCat} et. seq.)  
and $\mathsf{sSET}$ is the category of simplicial sets and maps of simplicial sets as described in \autoref{sec:SimplicialSets}.
(We ignore paradoxes of set theory here.)

In \hyperref[example:scalarfields]{Example 1} above, the fields $\xi, g_{M_n}$ and $g_X$ are all 
set-valued. For example, the field $\xi$ is defined by a sheaf 
$\CF_{\text{sigma model field}}$ whose value on an $n$-manifold is 
\tightfootnote{A more complete discussion would specify precisely what kind of maps. We typically use $C^\infty$ maps, but this is not sufficient for quantum theory.}
\be 
\CF_{\text{sigma model field}}(M_n):= \mathsf{Map}(M_n \to X) ~.
\ee
Here $\mathsf{Map}(M_n \to X)$ is a set, but as we have seen, a set can be turned into a simplicial set in a natural way, so these examples fit the 
FHT definition \eqref{eq:FunctorDefBckFlds}. 

The use of the codomain of simplicial sets (rather than sets)  in the definition 
\eqref{eq:FunctorDefBckFlds} becomes  important for fields 
with gauge symmetries if one wants to preserve the notation of locality. A simple example will illustrate the main point.
\tightfootnote{We thank D. Freed for pointing out this illuminating example.}
Let us consider principal $\IZ_2$-bundles in one-dimension (equivalently, double-covers). View $\IS^1$ as the union of two 
open sets $D_\pm$ corresponding to the ``northern and southern hemispheres.'' Then $D_\pm$ are intervals and $D_+ \cap D_-$ is a 
disjoint union of two intervals around the equator. Suppose we 
tried to restrict the field so that we only consider 
\underline{isomorphism classes} of principal bundles. Then we would 
have exactly one point associated with each 1-manifold
$D_+$, $D_-$, and $D_+ \cap D_-$. 
This is not enough data to capture the two inequivalent double covers of the circle. If on the other hand, the target space of the field is the groupoid of principal $\IZ_2$-bundles, then associated to $D_\pm$ is a category with one object but two invertible morphisms between the objects: The identity map of the double cover and the deck transformation. Now on $D_\pm$, we can compare these invertible morphisms with a 2-morphism and recover the distinct double covers of the circle. 

Moving on to gauge theories, for a topological group  $G$,  we can define a field $\CF$ that assigns to an $n$-manifold $M_n$ the groupoid $P_G(M_n)$ of principal $G$-bundles over $M_n$. In order to fit the FHT definition 
of a field, we should take $\CF$ to be the nerve of that category so we take: 
\be\label{eq:Field-Gbundle}
\CF(M_n) = N(P_G(M_n)) ~.
\ee
This will suffice as the definition of a field in $G$-gauge 
theory for a finite group $G$, since principal $G$-bundles come with a unique connection. In general, for $G$-gauge theory, we wish to include the connection as one of the fields. Therefore, we consider $P_G^{\n}(M_n)$, the groupoid of 
principal $G$-bundles over $M_n$ equipped with connection. 
So, in general the field of a gauge theory in the FHT definition is:
\be\label{eq:Field-Gbundle-With-Connection}
\CF(M_n) =  N(P_G^{\n}(M_n)) ~.
\ee

As we will see in \hyperref[part2]{Part II} when studying higher-form fields whose isomorphism classes on $M_n$ are the differential cohomology groups 
$\widecheck{H}^\ell(M_n)$, these groups are the group of isomorphism classes of a (higher) groupoid. (See \autoref{subsec:HopkinsSingerCocycles}, especially 
\autoref{subsubsec:DiffCoh-HigherGroupoids}, for further discussion.)  
Once again, the field should be viewed in terms of a sheaf valued in 
simplicial sets, for the same reason as for gauge theories -- locality. However,  in this case, it is even more important to do so because of the phenomena of ``ghosts for ghosts'' in theories of higher-form fields. One can take the nerve of that (higher) groupoid and that is the value of $\CF(M_n)$.

\subsection{Decorating Bordisms With Fields}\label{subsec:Fields-bordisms}

The definition of background fields allows us to put  $\CF$-structures on our bordisms to define an enhanced category $\mathsf{Bord}_{\langle n-1, n\rangle}^{\CF}$. In practice, this means that all the manifolds encountered in the definition of bordisms are equipped with orientation, (s)pin structure, Riemannian metric, etc. in a compatible way. To define the structures on the boundary components of the bordism, we must evaluate $\CF$ on germs of the boundary components. Thus, for example, 
if $\CF$ is a spin structure, we only work with spin structures in $n$-dimensions. Then, we define a TFT with such background fields as:
\eqa{
  \CF : \mathsf{Bord}_{\langle n-1, n\rangle}^{\CF} \longrightarrow \mathsf{VECT} ~.
}

This works well for discrete structures like orientations, and principal $G$-bundles with finite $G$. 
This viewpoint was quite useful in a recent paper clarifying some issues about global symmetries in Class $\CS$ theories and topological twisting of $d=4$, $\CN=2$ theories \cite{Moore:2024vsd}.

\bigskip 

\begin{remark}
$\,$
\begin{enumerate}

\item In example 3 of \autoref{sec:HigherCategories}, we encountered the $(\infty,n)$-category $\mathsf{Bord}_{\leq n}$. If we endow the bordisms with the structure of a field $\CF$, we  define $(\infty, n)$ categories $\mathsf{Bord}_{\leq n}^{\CF}$. These will be important in the definition of fully local TFT in \autoref{sec:ExtendedTFT} below.

\item The background fields of a \underline{topological} field theory 
should be -- in some sense -- invariant under homotopies. Examples include orientations and spin structures. One technical way to express this is to use the concept of a \emph{tangential structure}. 
See \cite{FreedCBMS}, Definition 1.40 and section 24 of \cite{Freed:QTGV} for expositions. 
 In particular in \cite{Freed:QTGV},
a notion of locally constant $\CF$ is developed and proven to be equivalent to tangential structure. On the other hand, if the background fields include, for example,  scalar fields, gauge fields, Riemannian metrics, conformal structures,.... the theory will no longer be topological. 
We will return to this in \autoref{sec:FunctorialApproachQFT} below.

\end{enumerate}

\end{remark}

\SectionWithHeader{Generalizing Domain And Codomain}{Generalizing Domain And Codomain}{sec:GeneralizeDomainCodomain}

Let us return to our preliminary definition 
\eqref{eq:Definition-TopTwo-TFT} of a topological field theory.
In this point of view, a field theory \underline{is} a functor. 
The codomain of the functor is $\mathsf{VECT}$. That is natural because the traditional viewpoint in physics is that partition functions should be complex numbers and spaces of states should be constructed from compact operators on Hilbert spaces. However, 
the mathematical definition of a field theory admits natural generalizations where one changes either the domain or the codomain category.   

One way in which the domain category can be generalized is by including fields as in \autoref{sec:BackgroundFields}. Another way arises in the work of Freed and Hopkins \cite{Freed:2016rqq} on invertible field theories. These theories factor through a quotient of 
the bordism category where all morphisms become invertible. That allows them to be viewed as morphisms
of ``spectra'' (see \autoref{sec:generalizedcohomology} below), which in turn opens up all the tools of stable homotopy theory. 

Instead of taking a quotient of the domain bordism category, one 
can try to extend the domain to include non-smooth manifolds. 
The use of $C^{\infty}$ manifolds in the definition of the bordism category is intended to guarantee that the gluing axioms work smoothly.   Of course, with extended topological field theory, one works with manifolds with corners, but one could ask if these manifolds could admit singularities of some kind or even be CW complexes. One early contribution in this direction is the lecture series by Frank Quinn in \cite{Quinn:1991kq}. Another recent contribution to this question is \cite{Khovanov:2023}. 
Moreover, section 4.3 of Lurie's famous paper \cite{Lurie:2009keu}
extends the cobordism hypothesis to include manifolds with certain allowed singularities. Finally, we note that there is a small but interesting literature where one uses something very like a  ``one-dimensional topological field theory,'' but with graphs rather than one-manifolds to describe categories related to the representation theory of  exceptional groups. For examples, see \cite{F4Diagrammatics} and references therein.
Finally, we remark that it might be fruitful to interpret the ``$XZ$-calculus'' of quantum information theory \cite{Coecke:2008lcg,Duncan:2009ocf,vandeWetering:2020giq,Gorantla:2024ocs} in terms of a 1d topological field theory with a domain consisting of colored graphs instead of manifolds.  

Another kind of generalization of the notion of field theory is obtained by changing the codomain category, which until now has been $\mathsf{VECT}$. Of course, one could consider vector spaces over different ``fields'' (in the sense of abstract algebra). Moreover, one should generalize from the category of vector spaces to super-vector spaces, especially when working with theories with fermions. A more drastic example of a change of codomain category arises in the study of Higgs branches of Class $\CS$ supersymmetric 4d QFTs \cite{Moore:2011ee}. The codomain category has as objects complex algebraic groups. The $\Hom$ space between two such groups   $G_1$ and $G_2$ is the set of holomorphic symplectic manifolds with a holomorphic $\IC^{\times}$ action scaling the symplectic form and a holomorphic Hamiltonian $G_1 \times G_2$ action. 
This is very far from the linear categories we have thus far met.

\SectionWithHeader{Fully Local TFT: The Cobordism Hypothesis }{Fully Local TFT: The Cobordism Hypothesis }{sec:ExtendedTFT}

We have thus far stressed that the most primitive physical notions 
of locality are captured in the definition of an $n$-dimensional TFT   by taking it to be a functor $F: \mathsf{Bord}_{\langle n-1,n\rangle}\to \mathsf{VECT}$.  As we have seen, we can generalize the bordism category by including background fields to 
$\mathsf{Bord}_{\langle n-1,n\rangle}^{\CF}$. We can also generalize the codomain to a be a general symmetric monoidal $n$-category $\CC$, so we can consider a more general viewpoint of a field theory as a monoidal functor, 
\be 
F: \mathsf{Bord}_{\langle n-1,n\rangle}^{\CF} \rightarrow \CC  ~ . 
\ee
But we have also seen that we can generalize $\mathsf{Bord}_{\langle n-1,n\rangle}^{\CF}$ to an $(\infty,n)$-category 
$\mathsf{Bord}_{\leq n}^{\CF}$ and hence we are led to a final generalization of a fully extended field theory as a monoidal $(\infty,n)$-functor with a codomain $\CC$ which is a symmetric 
monoidal $(\infty,n)$-category: 

\begin{definition}[\textcolor{red}{Fully Local Field Theory}] Let $\CC$ be a symmetric monoidal $(\infty,n)$-category
 and  $\CF$ be a sheaf on $\mathsf{Man}^{\mathsf{op}}$. Then a fully local field theory  with background fields $\CF$ is a monoidal functor between $(\infty,n)$-categories:
\be\label{eq:FullyExtended}
  F : \mathsf{Bord}_{\leq n}^{\CF} \to \CC ~.
\ee
\end{definition}

\begin{remark}
$\,$
\begin{enumerate}
    \item The terms ``extended field theory'' and 
    ``fully extended field theory'' are  also used for a ``fully local field theory.'' 

    \item The theory will be topological if $\CF$ is ``locally constant'' in the sense described in \cite[Sec. 24.4]{Freed:QTGV}. The definition is only fully rigorous in the topological case. 
    
\end{enumerate}
\end{remark}

Monoidal $n$-categories have a distinguished $0$-morphism $1_{\CC}$, the unit under $\otimes$, and one defines the looped category:
\be
  \Omega\CC := \Hom(1_{\CC}, 1_{\CC}) ~,
\ee
which is a monoidal $(n-1)$-category, and therefore has a monoidal unit $1_{\Omega\CC}$, so we can define,
\be
  \Omega^2\CC := \Hom(1_{\Omega\CC}, 1_{\Omega\CC}) ~,
\ee
which is a monoidal $(n-2)$-category. We can iterate the  
looping process to produce an $(n-k)$-category $\Omega^k\CC$. At each stage, we get a monoidal category with a unit 
$1_{\Omega^{k-1}\CC}$ for that category. 

Let $\emptyset_{k-1}$ be the empty $(k-1)$-manifold.  Then 
$F(\emptyset_{k-1}) = 1_{\Omega^{k-1}\CC}$. If   $M_{k}$ is a compact $k$-manifold without boundary, then it defines a $k$-morphism  
$\emptyset_{k-1} \to \emptyset_{k-1}$. Therefore, $F(M_k)$ must be a $k$-morphism in $\CC$. Indeed, consulting \eqref{eq:HighCatExplSet1}
and \eqref{eq:HighCatExplSet2}, we see that it should be a 
$0$-morphism, i.e., an object, in  the $(n-k)$-category: 
\be 
 \Hom( 1_{\Omega^{k-1}\CC}, 1_{\Omega^{k-1}\CC}) = 
\Omega^{k}\CC ~ . 
\ee
The right hand side is an $(n-k)$-category. That is,  
\be
  F(M_{k}) \in C_0(\Omega^{k}\CC) = \mathsf{Obj}(\Omega^{k}\CC) ~.
\ee
For a sanity check, let us suppose that $\Omega^n\CC = \IC$. 
(Remember that a monoidal ``$0$-category'' is just a set with a multiplication, hence a monoid.) Then for a compact $n$-manifold without boundary,  $F(M_n)$ -- the partition function -- is indeed a complex number. Moreover, if we take 
$\Omega^{n-1}\CC = \mathsf{VECT}$, then for a compact $(n-1)$-manifold without boundary $N_{n-1}$, we learn that $F(N_{n-1})$ is an object in 
$\mathsf{VECT}$, that is, a vector space, in harmony with the previous discussions.

\begin{remark}
As we have just seen, it is very natural to assume that 
$\CC$ satisfies: 
\be\label{eq:TopTwoLevel}
 \Omega^{n}\CC = \IC ~, \qquad  \text{ and } \qquad   \Omega^{n-1}\CC = \mathsf{VECT}   ~.
\ee
However, when proceeding to the next categorical level, there can be many different choices of $\CC$  that satisfy \eqref{eq:TopTwoLevel}. 
For example at the next level, two natural choices are: 
\begin{enumerate} \itemsep 0pt

\item $\Omega^{n-2}\CC  = \mathsf{ALG}(\mathsf{VECT})$. 

\item $\Omega^{n-2}\CC = \mathsf{CAT}$.  

\end{enumerate} 
\end{remark}

For example, when discussing finite gauge theory in $n=2$ dimensions,
in the first case, $F({\rm pt})$ will be an object in the Morita $2$-category of algebras, bimodules, and bimodule maps. In other words, it will be an algebra.   In the second case, $F({\rm pt})$ will be an object in the $2$-category of categories. In other words, it will be a category.  
In the case of two-dimensional finite group gauge theory based on a finite group $G$, one can show that (see \autoref{sec:FinHomTheory} below for more details): 
\eqas{
    F({\rm pt}) &= \IC[G] ~,\\
    F({\rm pt}) &= \mathsf{Rep}(G)  ~, \label{eq:TwoChoicesCodomain}
}
where $\IC[G]$ is the group \underline{algebra}, and 
$\mathsf{Rep}(G)$ is the \underline{category} of representations of $G$.

The main idea behind \eqref{eq:FullyExtended} is that one has taken 
locality to its logical conclusion by chopping up spacetime into smaller spaces of increasingly high codimension. The most primitive level of locality says that we can ``insert a complete set of states along a spatial slice,'' if we cut an $n$-dimensional spacetime into ``past'' and ``future'' parts by cutting along an $(n-1)$-dimensional spatial slice. Now, we can consider the space of states on an $(n-1)$-dimensional spatial slice and divide that into two pieces by cutting along an $(n-2)$-dimensional manifold, as in the discussion around \eqref{eq:CutSpatialSlice} above. And so on. This is the reason for the terminology  \emph{fully local} used above. 

One of the main results in the study of extended TFTs is the 
proof of the ``cobordism hypothesis.''  The ``cobordism hypothesis'' is an idea going back to J. Baez and J. Dolan \cite{Baez:1995xq}. It states, very roughly, that a fully extended TFT is ``completely determined by its value on a point''. Recall that $F({\rm pt}) \in \mathsf{Obj}(\CC)$ is a $0$-morphism in an $n$-category.
A good example is the case of one dimensional oriented TFT where 
all amplitudes are determined by a single piece of data: $F({\rm pt}_+) = V$. 
A precise version of the cobordism hypothesis was stated and proved by J. Lurie \cite{Lurie:2009keu} generalizing some important unpublished work of M. Hopkins and J. Lurie. 
Here is a very brief sketch of that result. We follow here the 
expositions \cite{Freed:CobordismHypothesis,Teleman:2016,FreedCBMS}.  

An important part of the cobordism hypothesis is to study 
\underline{all} the fully extended theories at once, so we should study the 
\emph{category of theories} with fixed dimension $n$, background fields $\CF$, and codomain $\CC$. 
%
%
Thus, we   consider the category of theories, which we denote as 
\be\label{eq:CategoryOfTheories}
\Hom^{\otimes}(\mathsf{Bord}^{\CF}_{\leq n}, \CC) ~.
\ee
It turns out that this category is itself an $(\infty,n)$-category. 
The $0$-morphisms are the different theories. A $1$-morphism 
between two theories $F_0$ and $F_1$, written
$\eta: F_0 \Rightarrow F_1$, assigns to every $k$-dimensional 
bordism $W_k$, a $(k+1)$-morphism in $\CC$ in the hom-space: 
\be 
\eta(W_k) \in \Hom_{k+1}(F_0(W_k), F_1(W_k) ) ~,
\ee
where $\Hom_{k+1}$ are the $(k+1)$-morphisms in $\CC$. 
In the simplest case of $n=1$, a field theory is a functor between 
$1$-categories (thought of as $(\infty,1)$-categories) and a $1$-morphism is just a natural transformation between those functors. 
The $k$-morphisms in \eqref{eq:CategoryOfTheories} for $k>1$ are defined similarly. See the above references for details. 

The next step is a remarkable claim: All the morphisms in the category of theories \eqref{eq:CategoryOfTheories} are \underline{invertible}! 
Thus \eqref{eq:CategoryOfTheories} is actually an $(\infty,0)$-category, i.e., an $\infty$-groupoid. As we have seen in \autoref{fig:GroupoidTriangle} above, 
an $\infty$-groupoid $\CG$ determines a space, via the geometric realization of its nerve: $\CG \to \vert N(\CG)\vert$, and this space in turn determines the $\infty$-groupoid, up to a suitable notion of equivalence. In our case the space, 
\be\label{eq:ModuliSpaceTheories}
\vert N(\Hom^{\otimes}(\mathsf{Bord}^{\CF}_{\leq n}, \CC) ) \vert ~,
\ee
is the \underline{moduli space of theories}  of dimension $n$ with 
background fields $\CF$ and codomain $\CC$. 

In order to get some sense of why \eqref{eq:CategoryOfTheories} is a 
groupoid, we need to introduce the notion of dualizability. 
Suppose first that $D$ is a symmetric monoidal category and $x\in \mathsf{Obj}(D)$. Then $x$ is said to be \emph{dualizable}, if there exists another object $x^\vee$ in $D$, and morphisms  
\be 
e: x^\vee \otimes x \rightarrow  1_{D} ~,
\ee
\be 
c: 1_D \rightarrow x \otimes x^\vee ~,
\ee
such that the composition of morphisms, 
\be 
x \rightarrow 1_D \otimes x \xlongrightarrow{ c\otimes \mathsf{Id}_x}  (x \otimes x^\vee) \otimes x \cong x \otimes (x^\vee \otimes x) 
\xlongrightarrow{ \mathsf{Id}_x \otimes e} x \otimes 1 \rightarrow x ~,
\ee
just produces $\mathsf{Id}_x: x \to x$. Similarly, we require the 
composition of morphisms: 
\be 
x^\vee \rightarrow   x^\vee \otimes 1_D  \xlongrightarrow{ \mathsf{Id}_{x^\vee} \otimes c} x^\vee \otimes (x \otimes x^\vee)  \cong (x^\vee \otimes x)  \otimes x^\vee 
\xlongrightarrow{  e\otimes \mathsf{Id}_{x^\vee}} 1_D\otimes x^\vee \rightarrow x^\vee ~,
\ee
just produces $\mathsf{Id}_{x^\vee}: x^\vee \to x^\vee$.
\tightfootnote{Careful readers will recognize that there is a distinction between left- and right- dualizable. We will ignore that subtlety here.} 
The morphisms $c$ and $e$ are called the \emph{coevaluation} and \emph{evaluation}, respectively. 
The data $(x^\vee,e, c)$ are called \emph{duality data for $x$}.
These data are unique up to canonical isomorphism. 
We have $(x\otimes y)^\vee \cong x^\vee \otimes y^\vee$. Moreover, if $f: x \to y$ is a morphism between dualizable objects there is a dual morphism (unique up to canonical isomorphism) $f^\vee: y^\vee \to x^\vee$. 

\bigskip 
\noindent \textbf{Example 1:} 
As a good example, in $\mathsf{VECT}$, only the finite-dimensional spaces have duality data and the dual to a finite-dimensional vector space is  the standard dual space  $ V^\vee= \Hom (V,\IC)$. The evaluation morphism is the canonical linear transformation $e: V^\vee \otimes V \to \IC$. 
To express the coevaluation, we choose a basis $\{v_i\}$ for $V$, which 
determines a dual basis $\{ v_i^\vee \}$ for $V^\vee$, and then, the linear transformation $c: \IC \to V \otimes V^\vee$ is determined by 
its value on $1\in \IC$: 
\be 
c(1) = \sum_i v_i \otimes v_i^\vee  ~ . 
\ee
Note that this will only give a well-defined element of $V\otimes V^\vee$  if $V$ is \underline{finite-dimensional}. 

\bigskip 
\noindent \textbf{Example 2:} For another example of dualizability, consider the category of oriented bordisms $\mathsf{Bord}_{\langle 0,1\rangle}^{{\rm or}}$. A positively oriented point, $x={\rm pt}_+$ is dualizable. Its dual object $x^\vee$ is the oppositely oriented point $x^\vee= {\rm pt}_-$. The duality data $e,c$ are provided by the standard bordisms given by the pictures \eqref{eq:1d-Oriented-DualityData} above.  

At this point, we can indicate why $\Hom^{\otimes}(\mathsf{Bord}_{\langle 0,1\rangle}^{{\rm or}}, \CC)$ is a groupoid. Here $\CC$ is an ordinary tensor $1$-category. If $F$ is a theory, i.e., an object in $\Hom(\mathsf{Bord}_{\leq 1}^{{\rm or}}, \CC)$, then $F({\rm pt}_+)$ is a dualizable object in $\CC$. Moreover, if $\eta: F_0 \Rightarrow F_1$ is a natural transformation of the tensor functors, then there is a morphism 
$\eta({\rm pt}_+): F_0({\rm pt}_+) \to F_1({\rm pt}_+)$. It is shown in \cite[Lemma A.44]{FreedCBMS} that $\eta({\rm pt}_-)^\vee$ is inverse to $\eta({\rm pt}_+)$. 
From this, it follows that $\eta: F_0 \Rightarrow F_1$ is invertible. 
 
Dualizability has a generalization to symmetric monoidal $(\infty,n)$-categories known as \emph{full dualizability}. The basic idea 
is that we demand that the dualizability data itself is dualizable in the appropriate category. More technically, recall that for an 
$(\infty,n)$-category $\CC$, we let $\CG_k(\CC)$ be the $(\infty,k)$-category where we delete the noninvertible $\ell$-morphisms with 
$\ell>k$. Then, if $\CC$ is a symmetric monoidal $(\infty,n)$-category 
an object $x$ (``$0$-morphisms'') of $\CC$ is said to be \emph{fully dualizable} if:

\begin{enumerate}\itemsep 0pt

\item $x$ is dualizable in $\CG_1(\CC)$.  

\item The evaluation and coevaluation morphisms $e,c$ of $x$ are 
dualizable in $\CG_2(\CC)$.

\item And so forth up to, but not including, the $n$-morphisms. 

\end{enumerate}

A key observation is that the image of a TFT only contains fully-dualizable objects. This follows from the same style of argument we used in the $S$-diagram of \autoref{fig:sdiag} above, to show that statespaces in a TQFT are finite-dimensional. Indeed, ``fully dualizable'' is just a generalization of this finite-dimensional property. Thus, the field theory functor factorizes through a functor to the subcategory $\CC^{ {\rm fd} }$ of fully dualizable objects and morphisms. 

The next step is to recognize that there is a very special kind of 
background field $\CF$ known as an \emph{$n$-framing}. In general, if 
$\pi: V \to X$ is a rank $r$ vector bundle, then a \emph{framing} is an 
isomorphism of $V$ with $X \times \IR^r$. It is the same as a trivialization of $\pi: V \to X$.  If $k<n$, an $n$-framing of a smooth $k$-manifold $W_k$ is an framing of $TW_k \oplus \IR^{n-k}$, that is, 
we have a trivialization, that is an isomorphism: 
\eqa{
    TW \oplus \big(W \times \IR^{n-k}\big) & \rightarrow W \times \IR^{n} ~.
    \label{eq:n-framing}
  }
Thus, an $n$-framing of a point is a vector space $\IR^n$ with the point at the origin. One should think of this vector space as being diffeomorphic to a neighborhood of a point in an $n$-manifold. The neighborhood can be arbitrarily small so it really represents a germ of a point within an  $n$-dimensional manifold. In general, all the $k$-dimensional bordisms $W_k$ in the extended bordism category  $\mathsf{Bord}_{\leq n}^{ {\rm fr} }$ of framed manifolds should be thought of as having such germs of $n$-dimensional neighborhoods. In $\mathsf{Bord}_{\leq n}^{ {\rm fr} }$, the framings are only considered up to homotopy. Consider again the case 
where $W_0$ is a point. The isomorphism 
of $T\IR^n \oplus T({\rm pt}) \to T\IR^n$ has a determinant of either sign, and the sign is the only homotopy invariant of the framing. 
Thus, there are two points ${\rm pt}_\pm$ in $\mathsf{Bord}_{\leq n}^{ {\rm fr} }$.

We can now state part 1 of the cobordism hypothesis: There is an equivalence $\tau$ of $\infty$-groupoids between  the 
groupoid of $n$-dimensional field theories defined on bordisms with $n$-framing and codomain $\CC$, and the $\infty$-groupoid $\CG_0(\CC^{ {\rm fd} })$:
\tightfootnote{ $\CG_0(\CC^{{\rm fd}})$ is sometimes called the \emph{core} of $\CC$.} 
\be 
\tau: \Hom^{\otimes}(\mathsf{Bord}^{ {\rm fr} }_{\leq n}, \CC) \to 
\CG_0(\CC^{{\rm fd}}) ~.
\ee
$\tau$ is determined on $0$-morphisms simply by: 
\be\label{eq:CobordHyp-Map1}
\tau_0: F \mapsto  F({\rm pt}_+) ~.
\ee
This suffices to determined how $\tau_k$ acts on the $k$-morphisms with $k>0$. (We saw an example above for $n=1$.) 
As we have stressed, $\infty$-groupoids and spaces are ``the same'' in the eyes of a homotopy theorist. At the level of spaces, 
the map \eqref{eq:CobordHyp-Map1} (and its higher $k$ analogs) determines a continuous map of spaces: 
\be\label{eq:CobordHyp-Spaces}
\vert N(\Hom^{\otimes}(\mathsf{Bord}^{ {\rm fr} }_{\leq n}, \CC)) \vert 
\to \vert \CG_0(\CC^{{\rm fd}} ) \vert ~,
\ee
and the cobordism hypothesis asserts this to be a homotopy equivalence. 
As noted in \eqref{eq:ModuliSpaceTheories}, the LHS is the moduli space of theories. 

Let us test \eqref{eq:CobordHyp-Spaces} in the case of 1-dimensional theories. All 1-manifolds admit $1$-framings. If we choose $\CC = \mathsf{VECT}$,  then $\CC^{{\rm fd}}$ will pick out the subcategory of finite-dimensional vector spaces. Next, the groupoid $\CG_0(\CC^{{\rm fd}})$ 
is obtained by dropping the noninvertible morphisms. Thus $\CG_0(\CC^{{\rm fd}})$ is the category whose objects are finite-dimensional vector spaces and morphism spaces $\Hom_{\CG_0(\CC^{{\rm fd}})}(V_1,V_2)$ are zero, if 
$V_1$ and $V_2$ have different dimensions, while 
$\Hom_{\CG_0(\CC^{{\rm fd}})}(V_1,V_2)$ is the $\mathsf{GL}(r,\IC)$-torsor of 
linear isomorphisms $V_1 \to V_2$ if  $\dim\, V_1 = \dim\, V_2 = r$. 
The category $\CG_0(\CC^{ {\rm fd} })$ is thus equivalent to the category whose objects are nonnegative integers $r$, with morphism spaces, 
\be 
\Hom(r_1, r_2) = \begin{cases} 0 ~, & r_1 \not= r_2  \\ 
\mathsf{GL}(r,\IC) ~, &  r_1 = r_2 = r ~.
\end{cases}
\ee
The space associated to this groupoid is just: 
\be\label{eq:ModuliSpace1d}
\coprod_{r\geq 0}  \mathsf{BGL}(r,\IC)  ~ . 
\ee
Recall that this is supposed to be the moduli space of theories. Being a moduli space means that any family of theories parametrized by a space $S$ is equivalent to a continuous map $S \to \coprod_{r\geq 0}  \mathsf{BGL}(r,\IC)$.  Such a map is the same thing as a vector bundle over $S$. Recall that a single theory is determined by $F({\rm pt}_+)\in \mathsf{Obj}(\mathsf{VECT})$, so a family of theories is just a vector bundle over $S$. So indeed,  
\eqref{eq:ModuliSpace1d} is the right answer for the moduli space of one-dimensional theories! 

\bigskip 
\noindent 
\begin{remark} An analogy (\cite[Thm. 4.40]{FreedCBMS}) is quite helpful. The oriented bordism group in $0$ dimensions is 
$\Omega_0^{\mathsf{SO}} \cong \IZ$, generated by ${\rm pt}_+$ (or ${\rm pt}_-$). 
Consider the $0$-dimensional oriented bordism invariants valued in 
an Abelian monoid $\mathsf{M}$, written additively. Such bordism invariants should be viewed as elements $F$ in the set of monoid homomorphisms $ \Hom(\Omega_0^{\mathsf{SO}}, \mathsf{M})$. Now 
$F({\rm pt}_+) + F({\rm pt}_-) = F(\emptyset) = 0 $, so $F({\rm pt}_+)$ is an invertible element of $\mathsf{M}$. It follows that 
$F$ factors through a monoid homomorphism to the set   $\CG(\mathsf{M}) \subset \mathsf{M}$ of invertible elements of $\mathsf{M}$. 
This is the analog of the statement that a field theory factors through $\CC^{ {\rm fd} }$.  Now note that $\Hom(\Omega_0^{\mathsf{SO}}, \mathsf{M})$ is itself a commutative monoid: $(F+G)(Y) := F(Y) + G(Y)$, but since $F$ is invertible it is an 
Abelian group. This is the analog of the claim that 
$\Hom^{\otimes}(\mathsf{Bord}^{ {\rm fr} }_{\leq n}, \CC) $ is an 
$\infty$-groupoid. Likewise the invertible elements $\CG(\mathsf{M})\subset \mathsf{M}$ form an Abelian group. The analog of the cobordism hypothesis is that 
$F \to F({\rm pt}_+)$ defines an isomorphism of Abelian groups, and indeed this isomorphism is equivalent to the statement that 
$\Omega_0^{\mathsf{SO}}$ is the free Abelian group generated by an oriented point.  One statement of part 1 of the cobordism hypothesis that one will encounter is that $\mathsf{Bord}_{\leq n}^{{\rm fr}}$ is the ``free symmetric monoidal $n$-category generated by a fully dualizable object.'' 
\end{remark}

Of course, we are interested in manifolds which are not framed and might not even admit a framing. The cobordism hypothesis can be extended to theories defined on such manifolds as follows. 
Suppose we are given a homomorphism $\varphi: G \to \mathsf{O}(n)$
and let $\mathsf{Bord}^{\varphi}_{\leq n}$ be the extended bordism category where the bordisms have a reduction,  along $\varphi$,  of the $\mathsf{O}(n)$ structure on their tangent bundle to a $G$-structure. In particular, if $G=\mathsf{O}(n)$ and  $\varphi$ is the identity homomorphism, then $\mathsf{Bord}^{\varphi}_{\leq n}$ is the extended bordism category of unoriented $n$-manifolds.  If $\varphi: \mathsf{SO}(n) \to \mathsf{O}(n)$ is the standard embedding, then 
$\mathsf{Bord}^{\varphi}_{\leq n}$ is the oriented bordism category.  
If $\varphi: \mathsf{Spin}(n) \to \mathsf{O}(n)$ is the standard double cover of the $\mathsf{SO}(n)$ subgroup of  $\mathsf{O}(n)$, then $\mathsf{Bord}^{\varphi}_{\leq n}$ is the extended spin bordism category. In this notation, the framed bordism 
category corresponds to the case where $G = \{ 1\}$ is the 
trivial group.

An $n$-dimensional theory $F$ defined on manifolds with tangential structure determined by $\varphi: G \to \mathsf{O}(n)$ can be presented as 
a factorization of a theory $F^{{\rm fr}}$  defined on framed manifolds as in the following diagram:  
\be \label{fig:FactorFromFramed}
  \begin{tikzcd}
            & \mathsf{Bord}_{\leq n}^{\varphi} \arrow[rd,"F"] \\
      \mathsf{Bord}_{\leq n}^{{\rm fr}} \arrow[ru] \arrow[rr, "F^{{\rm fr}}"] & & \CC
  \end{tikzcd}
\ee

To state part 2 of the cobordism hypothesis, note that 
the group $\mathsf{O}(n)$ acts on $n$-framings: The action by 
$g\in \mathsf{O}(n)$ is  post-composition of the trivialization 
\eqref{eq:n-framing}  
with $g:  W_k \times \IR^n \to W_k \times \IR^n$ for $g\in \mathsf{O}(n)$. 
Part 2 of the cobordism hypothesis says that factorizations
\eqref{fig:FactorFromFramed}   correspond to 
``$\mathsf{O}(n)$ fixed point structures on  $F^{{\rm fr}}({\rm pt}_+)$.''  
In the case of $n=1$ with $\CC=\mathsf{VECT}$, the $\mathsf{O(1)} \cong \IZ_2$ action 
takes $V \to V^\vee$. An ``$\mathsf{O(1)}$ fixed point structure  on  $V$'' is equivalent to a  nondegenerate bilinear form on $V$. In this way, we recover the description of one-dimensional unoriented theories of \autoref{sec:1dTFT}. In general, there is a lot to unpack in making sense of 
``$\mathsf{O}(n)$ fixed point structures on  $F^{{\rm fr}}({\rm pt}_+)$.'' 
One will find both data and conditions. For a detailed discussion 
for ``low'' values of $n$, see \cite{Schommer-Pries:2013}.

\begin{remark}
$\,$
\begin{enumerate}
    \item A great deal of the content of the cobordism hypothesis, as formulated above, is buried in the elusive notions of ``fully dualizable object'' and ``$\mathsf{O}(n)$ fixed point structure.'' It might be a fruitful exercise to explain the meaning of these terms more fully in ``physical'' terminology. 

\item Equation \eqref{eq:ModuliSpaceTheories} 
realizes, at least in the setting of topological field theory, an old dream of many physicists, namely,  that there is a well-defined    ``space of $n$-dimensional quantum field theories''. This notion is essential to the Wilsonian viewpoint on QFT and renormalization group flow \cite{Wilson:1973jj}.  
In the case of $n=2$, it exhibits 
extremely interesting properties as described by Friedan, Zamolodchikov et. al. It is therefore interesting to note that in the case of 
$\CN=(0,1)$ supersymmetric field theories, S. Stolz and P. Teichner \cite{StolzTeichner1,StolzTeichner2} (following work of Segal \cite{Segal88,Segal2007})
have proposed another definition of a ``space of theories''. They proved the 1d version of their conjecture for SQM with time-reversal symmetry, identifying the space of theories with an $\Omega$ spectrum which turns out to be $\mathsf{KO}$. (See \autoref{subsec:Spectra}  
below for a brief discussion of the meaning of ``spectrum'' in topology.) %
\tightfootnote{In particular, the SQM is equipped with an antilinear operator $\mathsf{T}$ that squares to the identity: $\mathsf{T}^2 = \mathsf{1}$ on the Hilbert space. This corresponds to a $\mathsf{Pin}_{-}$ structure \cite{FreedCBMS}. In view of the Freed-Hopkins classification of invertible phases \cite{Freed:2016rqq}, the gravitational anomaly in such theories is characterized by an element in the Anderson dual of the three-dimensional $\mathsf{Pin}_{-}$ bordism homology group, i.e., $n \in \big(I_{\IZ}\Omega^{\mathsf{Pin}_{-}}\big)^{3} \cong \IZ/8\IZ$, reflecting the mod$-8$ periodicity of the real $\mathsf{KO}$ spectrum.}
In the case of two dimensional $\CN=(0,1)$ SQFTs, the  $\Omega$ spectrum 
is conjecturally the spectrum of Topological Modular Forms ($\mathsf{TMF}$) \cite{Hopkins95,Hopkins2002,Goerss:2009} (see also \autoref{subsec:elliptic-cohomology}). 
The theories studied by Stolz and Teichner are not topological. 
Nevertheless, It would be quite interesting to relate these two notions of a ``space of theories.'' 
    
\end{enumerate}
  
\end{remark}   

\SectionWithHeader{Homotopy Sigma Models (Of Finite Type)}{Homotopy Sigma Models (Of Finite Type)}{sec:FinHomTheory}

Homotopy Sigma Models 
\tightfootnote{These theories have other names such as ``finite homotopy theories'' and ``finite total homotopy TQFTs.'' See, for example, \cite{Freed:2009qp,MartinsPorter:2023}.} 
are a very useful class of topological field theories that generalize finite group gauge theory and Dijkgraaf-Witten theory.
These theories prove to be excellent laboratories for exploring general ideas in topological field theory. They also have applications to a wide variety of other theories by providing ``higher'' symmetry structures of those theories, as we will briefly indicate in \autoref{sec:Quiche}. Closely related to this, these spaces also serve to define background fields for theories with ``higher group symmetry.'' Thus, a thorough study of these theories is well worth the effort. At least in the case of four-dimensional TQFT, they essentially exhaust the possible 
fully extended TQFTs according to \cite{Douglas:2018qfz,Johnson-Freyd:2020usu,Decoppet:2024htz}.

The study of these TFTs goes back 
to \cite{Kontsevich:1988br,Quinn:1991kq} and the study of 
``finite path integral'' quantization of these theories goes back to \cite{Freed:1994ad}. 
  For the case where the codomain $\CC$ is a ``Morita $n$-category'' (constructed from algebra objects), a fairly complete sketch of the fully local theory can be found in Freed-Hopkins-Lurie-Teleman \cite{Freed:2009qp}. A full description of the quantization procedure is work to appear by   C. Scheimbauer and T. Walde \cite{scheimbauer-walde-wip}. Here, we are following Appendix A of \cite{Freed:2022qnc}.
 Some useful remarks can also be found in    \cite{Turaev2010}
and \cite{Costa:2024wks}.

 Before briefly describing these theories, we review some relevant mathematical background in \autoref{subsec:Pi-Finite-Space} and \autoref{subsec:CorrespCourse} below. 

\subsection{$\pi$-finite Spaces}\label{subsec:Pi-Finite-Space}

When we discussed finite group gauge theory we introduced the space $BG$. It has a basepoint $*$, and the property that
\be
 \pi_{q}(BG,*) \cong \left\{
   \begin{array}{ll}
    \{1\} ~, & q > 1 ~,\\
    G ~, & q = 1 ~.
   \end{array}
 \right.
\ee
There is a generalization available when $G = A$ is an Abelian group.

For every Abelian group $A$ and integer $n > 1$, we can define an ``Eilenberg-MacLane space'' $K(A,n)$  as a path-connected space, such that 
\be
  \pi_{q}\big( K(A,n) \big) \cong \left\{
   \begin{array}{ll}
    \{1\} ~, & q \neq n \geq 0  ~,\\
    A ~, & q = n ~.
   \end{array}
  \right.
\ee
One can show that such spaces exist and are unique up to homotopy equivalence 
\cite{Spanier1981}. In fact, they can be constructed explicitly. For $A=\IZ$, one attaches successively higher-dimensional cells to $\IS^n$ to kill the higher homotopy groups.  

\bigskip 
\noindent \textbf{Example 1:}   $K(\IZ, 1) = \IS^1$. Again, this is deceptively simple and atypical, as we see in the next example.  
\bigskip 

\noindent \textbf{Example 2:}  Note that we cannot take $K(\IZ, 2) \stackrel{?}{\cong}  \IS^2$.
After all, $\pi_{3}(\IS^2) \cong \IZ$!
This and the higher homotopy groups, 
\be
\pi_{j\geq 4}(\IS^2) = \IZ_2, \IZ_2, \IZ_2, \IZ_2, \ldots 
\ee
show that $\IS^2$ will not do. But it is a good start. To construct $K(\IZ,2)$, one can inductively attach higher and higher cells to kill the higher homotopy groups. Another model for $K(\IZ,2)$ is the following:  Consider the set of pure states in an $n$-dimensional Hilbert space:
\be
   \IC\IP^{n} = \IS^{2n+1}/\mathsf{U(1)} ~.
\ee
The long exact sequence of homotopy groups implies:
\eqa{
  \pi_2(\IC\IP^{n}) &\cong \IZ ~,\\
  \pi_j(\IC\IP^{n}) &= 0 \quad j=3, \ldots, 2n+1 ~.
}
Once again, as in our heuristic description of $B\IZ_2$, one would like to take a limit as $n\to \infty$.  Indeed, it turns out that we can identify $K(\IZ, 2)$ with the space of pure states in $\infty$-dimensional complex Hilbert space.

In general, $K(A, n)$ for a general Abelian group $A$ and integer $n$ will   have some kind of infinite-dimensional model. While the $K(A,n)$ are not groups, they are ``almost'' groups. They are known as ``homotopy groups'' or ``groups up to homotopy'' or simply ``$H$-groups.'' In much of the literature of the past few years, they have been used in attempts to generalize the notion of symmetry. 

To define   an \emph{$H$-group} 
\tightfootnote{We follow here the very nice exposition of \cite[p. 33]{Spanier1981}.} 
let us first recall the definition of a group. We can define a group 
formally as a set $G$ with a map  $\mu: G\times G \to G$  such that we have a 
commutative diagram: 
%
%
\be
 \begin{tikzcd}
     G \times G \times G \ar[r,"\mu \times \mathsf{Id}"] \ar[d,"\mathsf{Id} \times \mu",swap]
&   G\times G\ar[d, "\mu"] \\
G\times G \ar[r, "\mu"] &  G 
 \end{tikzcd}
\ee
This defines an associative multiplication map $\mu$, we have 
$\mu\circ (\mathsf{Id} \times \mu) = \mu \circ (\mu \times \mathsf{Id})$. Next, $G$ has a basepoint $e\in G$, so that we can define a map $c: G\to G$ that takes every element to this basepoint. 
This map has the property that  the composite maps given by 
\be 
 \begin{tikzcd}
      G \ar[r,"{(\mathsf{Id}, c)}"] & G \times G \ar[r, "\mu"] & G ~,
   \end{tikzcd}
\ee
and
\be 
\begin{tikzcd}
     G \ar[r, "{(c, \mathsf{Id})}"] & G \times G \ar[r, "\mu"] & G ~,
\end{tikzcd}
\ee
are equal to the identity map $\mathsf{Id}: G \to G$. Finally, to encode inverses, we assume there is a 
map $\varphi: G \to G$ such that the composite  maps 
\be 
\begin{tikzcd} 
G \ar[r,"{(\mathsf{Id}, \varphi)}"] & G \times G \ar[r, "\mu"] & G ~,
\end{tikzcd}
\ee
and
\be 
\begin{tikzcd} 
G \ar[r, "{(\varphi, \mathsf{Id})}"] & G \times G \ar[r, "\mu"] & G ~,
\end{tikzcd}
\ee
are equal to the map $c: G\to G$.

Now we generalize. An \emph{$H$-group} (we also use the synonymous term \emph{$H$-space})  is a based topological space $P$ such that there exist continuous maps $c:P \to P $ (defined by taking all points to the basepoint), $\mu: P \times P \to P$, and 
$\varphi: P \to P$, so that the diagrams above defining a group are all true up to homotopy. That is, we replace ``are equal to'' by ``are homotopic to'' in the above definition of a group.

A crucial property of an $H$-group is that if $X$ is a pointed space, then the set of basepoint preserving homotopy classes of maps $[X, P]$ is a group. Conversely, if $P$ is a space with basepoint 
such that the sets $[X,P]$ admit a group structure for all based spaces $X$,  then $P$ is an \emph{$H$-space}. 
A key example of an $H$-space is the based loop space $\Omega X$ of a pointed space $X$. 
One can generalize the notion of a principal fiber bundle with group $G$ to a fibration by $H$-spaces. 
The Eilenberg-Maclane spaces are $H$-spaces, so fibrations by Eilenberg-Maclane spaces
are generalizations of principal bundles. One can show that  
\tightfootnote{\label{foot:adjointsuspension}Proof: $\Omega$ is adjoint to suspension. If $X$ and $Y$ are pointed spaces
then $[\Sigma X, Y]\cong [X, \Omega Y]$, where $\Sigma$ is the suspension. Apply 
this to $X = \IS^k$, so $\Sigma X \cong \IS^{k+1}$ and note that  
$\pi_q(X)/\pi_1(X) \cong [\IS^q, X]$. }
\be 
\Omega K(A,n) \cong K(A, n-1) ~.
\ee

Now, for a pointed topological space $Z$, we define $PZ$ to be the space of 
continuous paths $[0,1]\to Z$, and then we have the path space fibration $\pi:PZ \to Z$, given by evaluation of the path at $t=1$. Note that $PZ$ is contractible and that the 
fibers are homotopy equivalent to $\Omega Z$. Apply this to $Z=K(A,n+1)$. 
Then the fibers of   $\pi: PK(A,n+1) \to K(A,n+1)$ are Eilenberg-Maclane spaces $K(A,n)$
and hence, the path fibration is the analog of the universal bundle for the $H$-space $K(A,n)$. Therefore, the classifying space is: 
\be\label{eq:KAn-classifying}
BK(A,n) \cong K(A,n+1) ~.
\ee
It follows from \eqref{eq:KAn-classifying} that  fibrations over $X$ with fiber $K(A,n)$ are classified by homotopy classes of maps, 
\eqa{
  X \longrightarrow K(A, n+1) ~.  \label{eq:ClassifyK(A,n)Bun}
}
Since $K(A,1) = BA$ is the classifying space of principal $A$-bundles, equation \eqref{eq:KAn-classifying} shows that we can write 
$K(A,n) = B^n A$. This notation emphasizes the role of the Eilenberg-MacLane spaces as classifying spaces and we will sometimes use it. 

\begin{definition}[\textcolor{red}{$\pi$-finite space}] A $\pi$-finite space $\CX$ is a topological space with a finite set of connected components, each of which has a finite set of nonzero homotopy groups $\pi_{j}(\CX )$, each of which is a finite group.
\end{definition}

When $\CX$ is connected, it is homotopy equivalent to a  ``Postnikov tower,'' which is  an iterated fibration of Eilenberg-MacLane spaces (see Bott-Tu, p. 250-251 \cite{Bott1982}):
\be
\begin{tikzcd}
 & \vdots \\
   K(\pi_3, q_3) \arrow[r] & \CX^{(3)} \arrow[d] \\
   K(\pi_2, q_2) \arrow[r] & \CX^{(2)} \arrow[r] \arrow[d] &  BK(\pi_3, q_3) 
    \cong K(\pi_3, q_3+1) \\ 
                           & \CX^{(1)} = K(\pi_1, q_1) \arrow[r] & BK(\pi_2, q_2) \cong K(\pi_2, q_2+1)   \\
                           & \vdots
\end{tikzcd}
\ee
The horizontal arrows are classifying maps of the $K(\pi_k,q_k)$ 
bundle over $\CX^{(k-1)}$. So from \eqref{eq:ClassifyK(A,n)Bun}, we 
see that the fibration $\CX^{(k)}\to \CX^{(k-1)} $ is classified by the homotopy classes of 
maps $\CX^{(k-1)} \to K(\pi_{k+1}, q_{k+1} + 1)$. A (defining) property of Eilenberg-MacLane spaces implies that the set of homotopy classes is just the cohomology group $H^{q_{k+1}+1}(\CX^{(k-1)}; \pi_{k+1})$.
\tightfootnote{\label{foot:homotopy-class-abelian}In general the homotopy classes of maps $S\to K(A,n)$ form the Abelian group $H^n(S,A)$. This is part of the theory of spectra and generalized cohomology theories mentioned briefly in 
\autoref{sec:generalizedcohomology}. }
The classes are known as \emph{Postnikov invariants}. For a $\pi$-finite space, the tower will terminate at some finite stage. 
The full tower $\CX$ has $j^{th}$ homotopy group $\pi_j$. If all the Postnikov invariants vanish then $\CX$ is just $\prod_{j} K(\pi_j, q_j)$. Otherwise, it is a twisted product of Eilenberg-MacLane spaces, and in general is not homotopy equivalent to a direct product of Eilenberg-MacLane spaces. 
\tightfootnote{\label{foot:whitehead} Two spaces can have the same homotopy groups and fail to be homotopy equivalent. The standard example is $\IS^2 \times \IR \IP^3$ and $\IR \IP^2 \times \IS^3$. Both are free $\IZ_2$ quotients of $\IS^2\times \IS^3$, so the homotopy groups can be computed by the standard long exact sequence 
  and are the same. Other simple and interesting examples of this phenomenon are discussed in \cite{MO53399}.  A foundational theorem in algebraic topology, the Whitehead theorem, 
  states that if a map $f: X \to Y$ between CW complexes induces an isomorphism 
  of  homotopy groups  then $X$ is homotopy equivalent to $Y$. }

As the simplest nontrivial example, a  $2$-stage Postnikov decomposition of the form
\be\label{eq:2Group}
\begin{tikzcd}
  K(A,2) \arrow[r] & \CX \arrow[d]\\
  & K(G,1) \arrow[r] & K(A,3) ~,
\end{tikzcd}
\ee
where $A$ is an Abelian group and $G$ is a (possibly) non-Abelian group.  The data that classifies $\CX$ is therefore $G,A$ and an element of $H^3(G; A)$. 
\tightfootnote{In general, given a homomorphism $G\to \Aut(A)$ one can define group cohomology with twisted $A$-coefficients. Thus, the Postnikov invariant also includes the data of this twisting.  } 
These spaces -- sometimes referred to as ``two-groups'' -- are
closely related to (non)-extensions of $G$ by $A$ and are important in the  subject of ``crossed modules''  \cite{Brown1982}.
A 2-stage Postnikov tower plays the role of a classifying space for 
crossed modules.  In theories with ``two-group symmetry,'' the background fields will be homotopy classes of maps to \eqref{eq:2Group}.

More generally, just as $H$-spaces, and the $K(A,n)$ in particular, generalize the notion of groups, 
so too   $\pi$-finite spaces generalize groups. Indeed they are sometimes referred to as 
  ``higher groups.'' These spaces will be ``target spaces'' 
  for the finite homotopy sigma models. Those theories are in turn 
  quite useful in the   ``quiche picture'' of finite 
  symmetry in QFT described very briefly in 
  \autoref{sec:Quiche} and in \cite{Freed:2022qnc}.  
  In this sense these ``higher groups''   serve to generalize standard symmetry groups in quantum field theory.

\subsection{A Correspondence Course}\label{subsec:CorrespCourse}
 
In general, a \emph{correspondence} between two sets $R_1$ and $R_2$ is a space $S$ and a pair of maps:
\be
\begin{tikzcd}
& S \arrow[ld,"f_1",swap] \arrow[rd, "f_2"]\\
R_1 & & R_2
\end{tikzcd}
\ee
It generalizes the notion of a function from $R_1$ to $R_2$. In the case of a function $F$ from $R_1$ to $R_2$,   $S$ would be the graph of $F$ and $f_1$, and $f_2$ would be the projections to the domain and codomain respectively. To check things like the gluing axioms for the amplitudes we shall write below, we would like to be able to compose correspondences. That is,  we want to go
from two concatenated correspondences,
\be
\begin{tikzcd}
& S_{12} \arrow[ld,"f_1",swap] \arrow[rd, "f_{12}"] & & S_{23} \arrow[ld, "f_{23}",swap] \arrow[rd, "f_{3}"]\\
R_1 & & R_2 & & R_3
\end{tikzcd} 
\ee
to construct a new correspondence based on a set $S_{13}$ and 
maps $g_1, g_2$, such that
\be\label{eq:ConcatenatedCorresp}
\begin{tikzcd}
  & & S_{13} \arrow[ld, "g_1",swap] \arrow[rd, "g_2"]\\
  & S_{12} \arrow[ld, "f_1", swap] \arrow[rd, "f_{12}"] &  & S_{23} \arrow[ld, "f_{23}", swap] \arrow[rd, "f_3"]\\
  R_1 & & R_2 & & R_3
\end{tikzcd}
\ee
Therefore, given 
\be
\begin{tikzcd}
& S_{12}   \arrow[rd, "f_{12}"] & & S_{23} \arrow[ld, "f_{23}",swap]  \\
  & & R_2 & &  
\end{tikzcd} 
\ee
one wants to generate a correspondence between $S_{12}$ and $S_{23}$ 
\be
\begin{tikzcd}
& S_{13} \arrow[ld,"g_1",swap] \arrow[rd, "g_2"]\\
S_{12}  & & S_{23}
\end{tikzcd}
\ee
so that we can form the diagram  \eqref{eq:ConcatenatedCorresp}.

In the world of topological spaces and continuous maps, a natural way to complete concatenated correspondences nicely to a new correspondence is via the \emph{homotopy fiber product}. 
Given continuous maps $f_1: S_1 \to R$ and $f_2: S_2 \to R$ between topological spaces, one defines the 
homotopy fiber product $S_{1}\times_{h} S_2$ to be   the set of triples,
\be 
(s_1, s_2, \gamma)\in S_1 \times S_2 \times \mathsf{Map}([0,1] \to R) ~,
\ee
where  $\gamma$ is a path  in $R$ (i.e., a homotopy)  from $f_1(s_1)$ to $f_2(s_2)$.  
\tightfootnote{The space of continuous maps $\mathsf{Map}([0,1] \to R)$ is given the compact-open topology. See footnote \ref{foot:fun-map-compactopen}. }
The subscript $h$ stands for ``homotopy'' in the homotopy fiber product. 
The homotopy fiber product fits into the following diagram: 
\be
\begin{tikzcd}
  & S_{1}\times_{h} S_2 \arrow[ld, "p_1", swap] \arrow[rd, "p_2"]\\
  S_1 \arrow[rd,"f_1",swap] & & S_2 \arrow[ld, "f_2"] \\
  & R
\end{tikzcd}
\ee
with $p_1, p_2$ the obvious projection maps. This can then be used to concatenate correspondences in a useful way.

The homotopy fiber product should be contrasted with the 
 ordinary fiber product, where one demands equality $f_1(s_1) = f_2(s_2)$, i.e., 
\eqa{
  S_{1}\times_{f_1 f_2}S_{2} &= \{ (s_1, s_2) ~|~ f_1(s_1) = f_2(s_2) \} ~.
}
The extra data of the homotopy $\gamma$ in the homotopy fiber product gives a much bigger -- but much more useful --  space that behaves better than the ordinary fiber product in some standard constructions in algebraic topology. Since there can be different paths in 
$R$ from $ f_1(s_1) $  to $f_2(s_2)$, the homotopy fiber product can have some interesting topology. 
For example, if $S_1 = S_2 = \{ r_0 \}$, where $r_0$ is an element of $R$,  the homotopy fiber product is just the loop space of $R$ based at $r_0$. 

The homotopy fiber product is closely related to a standard manipulation in algebraic topology that associates a fibration to \underline{any} continuous map $f: S\to R$. See, e.g.,  \cite[p. 249]{Bott1982}. One defines the space, 
\be 
E_f := \{ (s, \gamma) \} \subset S \times \mathsf{Map}([0,1] \to R) ~,
\ee
such that $\gamma(0) = f(s)$. It is not difficult to show that if $S \subset R$, then $E_f$ is homotopic to $S$. Then, the fibration associated to the map $f:S\to R$ is the fibration: $\pi: E_f \to R$ defined by $\pi(s,\gamma) = \gamma(1)$. 
The fiber of this map above a fixed point $r \in R$ -- known as the \emph{homotopy fiber of $f$} -- is the set of pairs $(s,\gamma)$ so that $\gamma$ is a path from $f(s)$ to $r$. It therefore only changes by homotopy equivalence as $r$ is varied within a path-connected component of $R$. Therefore the homotopy fiber is only defined up to homotopy equivalence. Since $\pi: E_f \to R$ is a fibration, there is a long exact sequence of homotopy groups. This is an example of how the homotopy fiber behaves well in homotopy theory.  Finally, note that if we take $S_1=S$ and 
$S_2 = \{ r\} \in R$ in the definition of the homotopy fiber product, we reproduce the homotopy fiber 
above $r$ of a map $f:S \to R$. 

\bigskip 
\noindent \textbf{Example:} Consider $S=[0,\pi]$, $R=\mathsf{U(1)}$ and the map $f:[0,\pi] \to \mathsf{U(1)}$ given by $f(x)=e^{\imag  x}$. 
The image of the map is the upper hemisphere of the unit circle in the complex plane. The ordinary fiber of $f$ above 
a point in the unit circle is a single point $s\in S$,  determined by the principal value of the logarithm, for 
the fiber above a point $r$ in the upper hemisphere,  but is the  empty set above points $r$ in the lower hemisphere. This discontinuous behavior will wreak havoc on constructions in algebraic topology relying on continuity. By contrast, the homotopy fiber above any point
$r\in \mathsf{U(1)}$ is the set of pairs $(s,\gamma)$ where $s\in [0,\pi]$ and 
$\gamma$ is a continuous path in  $\mathsf{U(1)}$ from $e^{\imag s}$ to $r$. This set of pairs is always homotopic to the space of paths between two points on the circle so the homotopy fiber above $r$ varies continuously with $r$.  Notice that in this example, the homotopy fiber has infinitely many connected components labeled by winding number and is thus topologically interesting.

\begin{exbox}

For topological spaces $X$ and $Y$, let $Y^X$ be the space of maps from $X$ to $Y$ equipped with the compact-open topology. Let $T$ be the the two-dimensional torus, and slice that torus into two pieces $T_L $ and $T_R$
both diffeomorphic to cylinders. The left and right parts are separated by $S$ which is a disjoint union of circles. 
Consider the concatenation of correspondences: 
\be\label{eq:BundleOnTorusCorrespondence}
\begin{tikzcd}
  & & \CX^T \arrow[ld, "g_1",swap] \arrow[rd, "g_2"]\\
  & \CX^{T_L} \arrow[ld, "f_1", swap] \arrow[rd, "f_{12}"] &  & \CX^{T_R} \arrow[ld, "f_{23}", swap] \arrow[rd, "f_3"]\\
  \CX^{\emptyset} & & \CX^{S} & & \CX^{\emptyset}
\end{tikzcd}
\ee
\begin{itemize}
\item[(a)] Work through the natural maps in this correspondence and verify that the diagram commutes. 
 
\item[(b)] Let $\CX = BG$. Explain how the correspondence 
\eqref{eq:BundleOnTorusCorrespondence} describes the construction of a principal $G$-bundle over the torus from gluing two principal $G$-bundles over two cylinders. 
\end{itemize}
\end{exbox}

\subsection{Statespaces And Amplitudes For The Finite Homotopy Sigma Models}

The data needed to define a finite homotopy sigma model consists of: 

\begin{enumerate} \itemsep 0pt

\item A nonnegative integer $n$, the spacetime dimension. 

\item A $\pi$-finite space $\CX$. (See \autoref{subsec:Pi-Finite-Space}.)

\item A cocycle $\lambda \in Z^n(\CX;\IC^{\times})$. 
\tightfootnote{The isomorphism class of the theory only depends on the cohomology class of $\lambda$.}

\item A symmetric monoidal $n$-catgeory $\CC$. 
We will assume that $\Omega^n\CC = \IC$ and $\Omega^{n-1}\CC = \mathsf{VECT}$.

\end{enumerate} 

We denote the resulting fully local theory by $\sigma^{(n)}_{\CX,\lambda,\CC}$. An important point about these theories is that their ``quantization'' always involves some version of a finite path integral over some ``dynamical fields.'' The functorial description of field theories we have described thus far involves a description of ``the answer'' to the path integral, and its expected properties, without concerning ourselves with how path integrals are to be done. But in the homotopy sigma models, we do have a notion of what the ``dynamical fields'' are. The  homotopy models should be thought of as ``sigma models'' where the fields are only distinguished up to homotopy -- hence the name. Thus there will by a description by 
``dynamical fields.''  (In \cite{Freed:2022qnc} the dynamical fields are called  ``semiclassical data.'')  A useful notation for what follows is the following: For any manifold  $M_{n}$,  let $\CX^{M_{n}}$ be the space of all continuous maps $M_{n} \to \CX$. The ``finite path integral'' -- at every categorical level -- will be considered to be some kind of ``quantization'' of 
mapping spaces of the form $\CX^{M_{n}}$. The exact meaning of this ``quantization'' is somewhat involved and can be found in \cite{Freed:2009qp} and the upcoming work of Scheimbauer and Walde \cite{scheimbauer-walde-wip}. (For a telegraphic description, see equation (A.8) of 
\cite{Freed:2022qnc}.)

Let us begin by describing the spaces of states on closed spatial manifolds $N_{n-1}$. In order to motivate the definition consider the standard quantum field theory of an $n$-dimensional sigma model 
with target space $X$. Typically, we take $X$ to be a finite-dimensional Riemannian manifold. If $N_{n-1}$ is a closed   spatial manifold, the sigma model fields at fixed time are elements of the mapping space $X^{N_{n-1}}$. If $N_{n-1}$ and $X$ are endowed with Riemannian metrics, the mapping space inherits a Riemannian structure: 
A tangent vector $\alpha$ to a sigma model field $\phi \in X^{N_{n-1}}$ is a section of $\phi^*(TX)$, and $\vert\!\vert\alpha\vert\!\vert^2 = \int_{N_{n-1}} \vert \alpha\vert^2 ~\vol$. The   Hilbert space of the theory on $N_{n-1}$ is morally something like $L^{2}(X^{N_{n-1}})$. States would be derived from normalizable wavefunctionals $\Psi[\phi(x)]$ for $\phi \in \CX^{N_{n-1}}$. In the TFT setting, with target space $\CX$,  the fields are only considered up to homotopy and so our Hilbert space should consist of 
wavefunctions defined on the set of \underline{homotopy classes} of maps $N_{n-1} \to \CX$. If $\CX$ is a $\pi$-finite space and $N_{n-1}$ is closed, the set of homotopy classes of maps will be a finite set and the Hilbert space will be finite dimensional. 
\tightfootnote{Note that in a homotopy sigma model, $\CX$ need not be finite dimensional and it is typically not finite dimensional. }
The set of homotopy classes of ``sigma model field configurations'' 
is just $\pi_{0}(\CX^{N_{n-1}})$, and so we have the definition of the statespace:
\be\label{eq:FinHomThy-StateSpace}
 \sigma_{\CX,\CC}^{(n)}(N_{n-1}) := \mathsf{Fun}\big( \pi_{0}(\CX^{N_{n-1}}) \big) ~.
\ee
where the notation on the right-hand side indicates the vector space of complex-valued functions.

\begin{remark}
In equation \eqref{eq:FinHomThy-StateSpace}, we have taken $\lambda$ to be a trivial cocycle. 
When it is nontrivial, the data of $\lambda$ defines a line bundle 
$L_\lambda \to \CX^{N_{n-1}}$, and instead of a space of functions, one considers a space of sections up to homotopy. For simplicity, we will generally take $\lambda$ to be trivial, and just denote the theory by 
$\sigma_{\CX,\CC}^{(n)}$. 
\end{remark}

\noindent \textbf{Examples:}
\begin{itemize}
  
\item[(1)] Let $G$ be a finite group (not necessarily Abelian) and take  $\CX = K(G,1) = BG$. The finite homotopy sigma model $\sigma_{\CX,\CC}^{(n)}$ is just the finite group gauge theory studied in \autoref{sec:FinGrpGT-Part1}. In this case, we have: 
  \be
    \pi_{0}\big( \CX^{N_{n-1}}\big) = \left\{
      \begin{array}{c}
        \text{isomorphism classes of }\\
        \text{principal $G$-bundles} \\
        \text{over spatial manifold $N_{n-1}$}
      \end{array}
      \right\} ~.
  \ee
and equation \eqref{eq:FinHomThy-StateSpace} reduces to 
\eqref{eq:FinGroupStateSpace}. If we consider a nontrivial cocycle $\lambda$, then the finite gauge theory is generalized to Dijkgraaf-Witten theory.  
  \item[(2)] Let $A$ be a finite Abelian group, and take $\CX = K(A,p)$. Then, as noted in footnote \ref{foot:homotopy-class-abelian},  $\pi_{0}(\CX^{N_{n-1}}) \cong H^{p}(N_{n-1};A)$. In this case, one can think of 
  $\CX$ as classifying a ``higher'' notion of principal bundle 
  sometimes called a ``$p$-principal bundle.'' See \cite{Brylinski1996,Bunke:2012rsi}. So the ``space of states of $\sigma_{\CX,\CC}^{(n)}$'' on the spatial manifold $N_{n-1}$ is the vector space of functions from the finite Abelian group $H^{p}(N_{n-1};A)$ to the complex numbers.

\end{itemize}

We leave it as a challenge to the reader to work out similarly explicit formulations of the space of states when the target space is a higher group.

\begin{figure}[h]
  \centering
  \begin{tikzpicture}[>=Stealth,baseline=-0.5ex]
  \node (graph) [inner sep=0pt]{\includegraphics[width=2.5in]{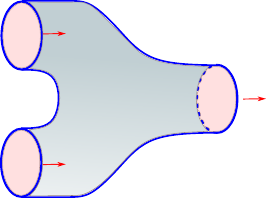}};
  \node at (0,0) {$M_{n}$};
  \node at (-3.5,0) {$N_{n-1}^{0}$};
  \node at (3.8,0) {$N_{n-1}^{1}$};
  \end{tikzpicture}
  \caption{A bordism.}
  \label{fig:ampbord}
\end{figure}
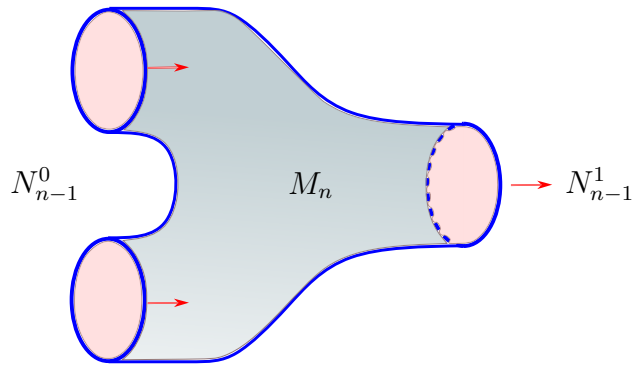

Now, in order to define amplitudes associated with a bordism (see \autoref{fig:ampbord}),
we form a correspondence:
\be
\begin{tikzcd}
  & \CX^{M_{n}} \arrow[ld, "p_0",swap] \arrow[rd, "p_1"]\\
  \CX^{N_{n-1}^{0}} & & \CX^{N_{n-1}^{1}}
\end{tikzcd}
\ee
where $p_0$ and $p_1$ are given by restricting the field $\phi \in \CX^{M_{n}}$ to the in- and out- boundaries. The idea is that the linear map,
\be
  \sigma_{\CX,\CC}^{(n)}(M_{n}) : \sigma_{\CX,\CC}^{(n)}(N_{n-1}^{0}) \to \sigma_{\CX,\CC}^{(n)}(N_{n-1}^{1}) ~,
\ee
is given by the pullback and  pushforward:
\be\label{eq:Sig-XL-Ampl}
  \sigma_{\CX,\CC}^{(n)}(M_{n}) := (p_{1,*})\circ p_{0}^{*} ~,
\ee
The map  $p_{0}^{*}$ is straightforward pullback: 
\be 
p_{0}^{*}: \mathsf{Fun}(\pi_0(\CX^{N^0_{n-1}})) \to \mathsf{Fun}(\pi_0(\CX^{M_n})) ~,
\ee
by the standard equation $p_0^*(F)(\phi)  = F(p_0(\phi))$. Here $\phi:M_n \to \CX$ is 
a ``sigma model map'' on the bulk space $M_n$, and $F$ is a locally constant functional on the sigma model maps 
on the boundary. Note that evaluation of the pullback function $\pi_0^*(F)$ only 
involves the values of $F$  on sigma model maps on the boundary 
that extend to the bulk. 

The map, 
\be 
p_{1,*} :   \mathsf{Fun}(\pi_0(\CX^{M_n})) \to \mathsf{Fun}(\pi_0(\CX^{N^1_{n-1}})) ~,
\ee
is more involved. It uses   the ``homotopy cardinality.'' If 
$\Psi$ is a locally constant function on $\pi_0(\CX^{M_n})$, then the pushforward function is defined by its values: 
\eqa{
  p_{1,*}(\Psi)(\phi_\partial)\quad & := \sum_{[\phi] \in \pi_{0}(p_{1}^{-1}(\phi_{\partial}))}\left(\prod_{i=1}^{\infty}\left| \pi_{i}\big( p_{1}^{-1}(\phi_{\partial}), \phi \big) \right|^{(-1)^{i}} \right)\Psi(\phi) ~.
\label{eq:SigXcomp} }
Here $\phi_{\partial}: N^1_{n-1} \to \CX$ is a sigma model field on the outgoing boundary.
We are defining a locally constant function $p_{1,*}(\Psi)$ on these sigma model fields. 
To evaluate it, we   consider the fiber $p_1^{-1}(\phi_{\partial})$, that is, the sigma model 
fields $\phi: M_n \to \CX$ that extend $\phi_{\partial}$ to a field on the bulk. For each connected component 
of the set $p_1^{-1}(\phi_{\partial})$, we choose a representative $\phi$ in that component 
and evaluate the locally constant function $\Psi$ on that $\phi$, and weight the result by 
the homotopy cardinality -- that is, by the alternating product of orders of homotopy groups for that component.

Using properties of the homotopy fiber product described in \autoref{subsec:CorrespCourse}, 
one can check the crucial gluing properties.
(See comments between (A.12) and (A.13) of \cite{Freed:2022qnc} and references therein.
See also \cite{MartinsPorter:2023}.) 
It is essential that one uses the homotopy cardinality for this to work.

\begin{remark} Taking $N^0 = N^1 = \emptyset$ gives the partition function on a compact $n$-manifold without boundary as a corollary:
\eqa{
  \sigma_{\CX,\CC}^{(n)}(M_{n}) &= \sum_{[\phi] \in \pi_{0}(\CX^{M_{n}})}\left( \prod_{i=1}^{\infty}\left| \pi_{i}\big(\CX^{M_{n}},\phi\big)\right|^{(-1)^{i}}\right) ~.
\label{eq:sig-XL-PartFun}
}
If $\CX$ is $\pi$-finite, the factors in the infinite product eventually all become equal to $1$. The alternating product of these factors -- the homotopy cardinality -- is a manifestation of the ``ghosts for ghosts'' phenomenon discussed later in the case of generalized Abelian gauge theories in \autoref{sec:PartitionFunctionsGenMax}. 
\end{remark}

\noindent \textbf{Examples:} 

\begin{enumerate} 

\item For finite $G$ gauge theory (with $G$ not necessarily Abelian) with $\CX = K(G,1)$, the partition 
function \eqref{eq:FinGroupPF} is a special case of \eqref{eq:sig-XL-PartFun}. The sum over $\pi_{0}(\CX^{M_{n}})$ is a sum over the 
principal $G$-bundles over $M_n$. The weighting factor accounting for automorphisms is only nonzero for $\pi_1$ and $\vert \pi_1(K(G,1)^{M_n})\vert = \vert G\vert$ when $M_n$ is connected. These are just the ``global gauge transformations,'' so we are correctly dividing by the ``order of the gauge group.''

\item For the ``higher'' forms of bundles with $\CX = K(A,p)$, $p>1$, 
the partition function is also sum over isomorphism classes of ``higher'' bundles weighted by automorphisms. The bundles are classified by, 
\be 
\pi_0(K(A,p)^{M_n}) \cong H^p(M_n; A) ~. 
\ee
Moreover, because the Eilenberg-MacLane spaces form a spectrum, we have 
\be 
\begin{split}
    \pi_{\ell}(K(A,p)^{M_n}) & = [S^\ell M_n, K(A,p)] \\ 
    & = [M_n, \Omega^{\ell} K(A,p)] = [M_n, K(A,p-\ell)] \\
    & = H^{p-\ell}(M_n; A) ~.
\end{split}
\ee
So,
\be\label{eq:KAp-PF} 
\begin{split}
\sigma_{\CX,\CC}^{(n)}(M_n) & = \sum_{H^p(M_n;A)} 
\frac{ 1 }{ \vert H^{p-1}(M_n;A)\vert }
\cdot \frac{ \vert H^{p-2}(M_n;A)\vert}{\vert H^{p-3}(M_n;A)\vert }
\cdots   \\
&  = \prod_{j=0}^p \vert H^{p-j}(M_n;A)\vert^{(-1)^j} ~.
\end{split}
\ee
%

\end{enumerate}

\bigskip 
\begin{exbox}{Convolution product from Finite Homotopy $\sigma$ Models} Derive equation \eqref{eq:ConvolutionProduct}
as a special case of \eqref{eq:Sig-XL-Ampl} and \eqref{eq:SigXcomp}. 
\end{exbox}
\bigskip

Let us go to the next categorical level, that is, we consider the value of the field theory on  closed manifolds of dimension $M_{n-2}$.
In this case, we need to know more about the codomain category $\CC$. Among the many choices for $\Omega^{n-2}\CC$, two natural choices are often adopted in the literature:  

\begin{enumerate}\itemsep 0pt

\item $\Omega^{n-2} \CC    = \mathsf{ALG}(\mathsf{VECT})$. 

\item $\Omega^{n-2} \CC  = \mathsf{CAT}$.  

\end{enumerate}

The answers are expressed in terms of the finite groupoid 
$\pi_{\leq 1}(\CX^{M_{n-2}})$ whose objects are the points 
of $\CX^{M_{n-2}}$ (i.e., maps $M_{n-2} \to \CX$) and whose 
morphisms are paths between points. 

When $\Omega^{n-2} \CC  = \mathsf{CAT}$  is the $2$-category of linear categories then $\sigma^{(n)}_{\CX,\CC}(M_{n-2})$ is the category of ``locally constant'' vector bundles over $\CX^{M_{n-2}}$.
Recall that we only consider 
the maps in $\CX^{M_{n-2}}$ and paths of maps up to homotopy. This implies the local constancy. Thus, we get the  category of locally constant vector bundles over the  groupoid $\pi_{\leq 1}(\CX^{M_{n-2}})$. More concretely, locally constant bundles are   vector bundles  over $\pi_0(\CX^{M_{n-2}})$ with an action of $\pi_1(\CX^{M_{n-2}},\phi_0)$, where, at the cost of some naturalness, we have chosen a basepoint $\phi_0$   in each component.

When $\Omega^{n-2} \CC    = \mathsf{ALG}(\mathsf{VECT})$, 
we consider the groupoid path algebra on 
$\pi_{\leq 1}(\CX^{M_{n-2}})$. Paths generate the algebra and 
the multiplication is nonzero only when paths compose, in which case the product is the composed path. So we write: 
\be 
\sigma^{(n)}_{\CX,\CC}(M_{n-2}) = \mathsf{Alg}( \pi_{\leq 1}(\CX^{M_{n-2}})) ~ ,
\ee
(where $\mathsf{Alg}$ denotes a specific algebra in the category $\mathsf{ALG}$). 
This algebra can be expressed (noncanonically) as a direct sum of 
group algebras: 
\be 
\mathsf{Alg}( \pi_{\leq 1}(\CX^{M_{n-2}})) = \bigoplus_{[\phi_0] \in \pi_0(\CX^{M_{n-2}}) }  \IC[\pi_1(\CX^{M_{n-2}}, \phi_0)]  ~.
\ee
Again, this presentation is slightly noncanonical because we chose a basepoint $\phi_0$ in each connected component of $\CX^{M_{n-2}}$.

It is worth unpacking the above answers a bit more in the case 
where $\CX = BG$. Note that   $BG^{M_{n-2} }= \Bun_G(M_{n-2})$ is the 
moduli \underline{stack} of principal $G$-bundles over $M_{n-2}$.

\begin{enumerate} 

\item 
If   we take $\Omega^{n-2} \CC  = \mathsf{CAT}$, then 
 the category $\sigma^{(n)}_{BG,\CC}(M_{n-2})$ is the category of vector bundles over $\pi_0\big(\mathsf{Bun}_G(M_{n-2})\big)$ together with an action of 
$\Aut(P_x)$ on the fibers over $P_x$, where $x$ runs over 
the finite set $\pi_0\big(\mathsf{Bun}_G(M_{n-2})\big)$. Specializing further to $\sigma^{(2)}_{BG}$, we have $\Bun_G({\rm pt}) = {\rm pt}/\!/G$, and $\pi_1 \cong G$, so the vector bundle is just a $G$-representation, and we get $\mathsf{Rep}(G)$, thus recovering \eqref{eq:TwoChoicesCodomain} above. 
On the other hand, for $\sigma^{(3)}_{BG}$, if $M_{n-2}=\IS^1$, then 
\be
\sigma^{(3)}_{BG}(\IS^1) = 
\mathsf{Vect}(\pi_{\leq 1}(BG^{\IS^1})) = \mathsf{Vect}(G/\!/G) := \mathsf{Vect}_{G}(G) ~,
\ee
where the $G$ action on $G$ is by conjugation. Generating objects in this category consist of a group element $g$ and a representation of the centralizer of $g$. If we include a cocycle $\lambda$, then vector bundles are replaced by twisted vector bundles.

\item 
The algebra is therefore, 
\be 
\sigma^{(n)}_{BG,\CC}(M_{n-2}) = \bigoplus_{[P] \in \pi_0(\mathsf{Bun}_{G}(M_{n-2})) } \IC\big[ \pi_1\big(\mathsf{Bun}_G(M_{n-2}\big), P\big] ~,
\ee
specializing this further to $n=2$ dimensions we have $\Bun_G({\rm pt}) = {\rm pt}/\!/G$ and 
\be 
\pi_1\big(\mathsf{Bun}_G(M_{n-2}), P\big) \cong G ~,
\ee
so we recover 
$\sigma^{(2)}_{BG}({\rm pt}) = \IC[G]$ quoted in \eqref{eq:TwoChoicesCodomain} above. If we include a cocycle $\lambda$ then  the group algebra products are twisted.

\end{enumerate} 

Let us see how the statespace of $M_{n-1}$ ``factorizes'' when 
divided into two pieces  by cutting along $M_{n-2}$. 
In this discussion we take $\Omega^{n-2} \CC  = \mathsf{CAT}$. 
It is   useful first to rederive \eqref{eq:FinHomThy-StateSpace} from the correspondence: 
\be
\begin{tikzcd}
  & \CX^{M_{n-1}} \arrow[ld, "p_0",swap] \arrow[rd, "p_1"]\\
  \CX^{\emptyset_{n-2}} & & \CX^{\emptyset_{n-2}}
\end{tikzcd}
\ee
The evaluation of the field theory on the bordism $M_{n-1}: \emptyset_{n-2} \to \emptyset_{n-2}$ is an object  in $\Hom(1_{\Omega^{n-2}\CC}, 1_{\Omega^{n-2}\CC})= \mathsf{VECT}$, 
that is, it is a vector space. The formula for the vector space is 
$p_{1,*}\circ p_0^*$ applied to the generating vector bundle over 
$\CX^{\emptyset_{n-2}}$. That is a vector bundle over a point, so we 
wish to evaluate $p_{1,*}\circ p_0^*(\IC)$.  The pull-back $p_0^*(\IC)$ gives the rank-one vector bundle which is the trivial rank-one bundle over each component of $\CX^{M_{n-1}}$. The pushforward gives 
a vector space. It is the sum of $\IC$ over the components $\pi_0( \CX^{M_{n-1}})$.  A function on  $\pi_0( \CX^{M_{n-1}})$ gives a 
complex number for each connected component of $\CX^{M_{n-1}}$ 
and hence the pushforward bundle is naturally identified with the 
statespace  \eqref{eq:FinHomThy-StateSpace}.

Now suppose that we can write $M_{n-1}$ by gluing together 
two bordisms $M_{n-1}^0: \emptyset_{n-2} \to M_{n-2}$ and 
$M_{n-1}^1: M_{n-2} \to \emptyset_{n-2}$. The vector space we get is 
obtained by composing the push-pull morphisms in: 
\be
\begin{tikzcd}
& \CX^{M^1_{n-1}} \arrow[ld,"p",swap] \arrow[rd, "r_L"] & & \CX^{M^2_{n-1}} \arrow[ld, "r_R",swap] \arrow[rd, "p"]\\
\CX^{\emptyset_{n-2}} & & \CX^{M_{n-2}} & & \CX^{\emptyset_{n-2}}
\end{tikzcd} 
\ee
That is, the vector space is: 
\be 
(p_*\circ r_R^*) \circ (r_{L,*}\circ p^*)(\IC) ~.
\ee
This makes sense: $p^*(\IC)$ is the trivial bundle over
 $\CX^{M^1_{n-1}} $   and $r_{L,*}$  will push it forward 
 to those maps in $\CX^{M^1_{n-2}}$ which can be restrictions from maps on $M^1_{n-1}$. Then $r_R$ pulls back those maps 
 to those which are restrictions of maps $M^2_{n-1}$. Altogether 
 we get a trivial bundle over the components of maps from 
 $M_{n-1}$.  

If instead $\Omega^{n-2} \CC    = \mathsf{ALG}(\mathsf{VECT})$, then  $F(M_{n-2})$ is an algebra, $F(M^1_{n-1})$ is a left-module, and $F(M^2_{n-1})$ is a right-module for the algebra and the vector space factorizes as: 
\be 
F(M_{n-1}) =  F(M^1_{n-1})\otimes_{F(M_{n-2})} F(M^2_{n-1}) ~.
\ee
The modules over the path algebra of the groupoid $\pi_{\leq 1}(\CX^{M_{n-2}})$ are bundles over that groupoid, so the discussion is similar to the previous case. 

Alternatively, one can choose not to think, and just rely on the formal theory of correspondences. 

We will not continue to higher codimension systematically. 
Such a systematic procedure  is given in the literature for proceeding to higher categorical levels. Here we simply mention one very important special case: Consider finite group gauge theory in $3$ dimensions. In this case one must choose a $3$-category $\CC$ for the codomain. A natural choice is the $3$-category of tensor categories. Tensor categories are algebra objects in the $2$-category of categories. The canonical procedure produces 
\be 
\sigma^{(3)}_{BG,\CC}({\rm pt}) = \mathsf{Vect}[G] ~,
\ee
where $\mathsf{Vect}[G]$, the category of vector bundles over $G$, is 
turned into a tensor category as follows: For $G$ finite, a vector bundle over $G$ amounts to a choice of a finite-dimensional vector space $V_g$ for each $g\in G$. Then the tensor product of two such vector spaces is defined by the convolution product: 
\be\label{eq:ConvolutionProduct-2}
(V_1 \otimes V_2)_g \cong \bigoplus\limits_{\substack{(g_1,g_2) \\[2pt] g \,=\, g_1 g_2}}
V_{1,g_1} \otimes V_{2,g_2} ~,
\ee
where the direct sum on the right hand side is over pairs $(g_1,g_2)$
such that $g= g_1 g_2$. If we include a twisting by $\lambda\in Z^3(BG;\IC^{\times})$, then it is shown in \cite[Sec. 4.1]{Freed:2009qp} that $\lambda$ is equivalent to a choice of complex line bundle over $L \to G\times G$, and \eqref{eq:ConvolutionProduct-2} is replaced by a twisted convolution product: 
\be\label{eq:TwistConvolutionProduct}
(V_1 \otimes V_2)_g \cong \bigoplus\limits_{\substack{(g_1,g_2) \\[2pt] g \,=\, g_1 g_2}} L_{g_1, g_2} \otimes 
V_{1,g_1} \otimes V_{2,g_2} ~.
\ee

\begin{remark}
\textbf{Electromagnetic Duality}. There is a finite group analog of the electromagnetic duality discussed in more detail in \hyperref[part2]{Part II} of these lectures. The theory $\sigma^{(n)}_{\CX,\CC}$ for $\CX = B^pA$ is equivalent to the theory 
with $\CX = B^{n-p-1}A^\vee$, where $A^\vee$ is the Pontryagin dual group $A^\vee = \Hom(A,\mathsf{U(1)})$, at least, up to a product with an invertible Euler theory. The need for the Euler theory can already be seen by considering the partition function \eqref{eq:KAp-PF}, since it shows that for $M_n$ closed 
\tightfootnote{To prove \eqref{eq:ComparePF}, note first that the perfect pairing, 
\be 
H^{n-j}(M_n;A^\vee) \times H^j(M_n;A) \to \mathsf{U(1)} ~,
\ee
implies that $\vert H^{n-j}(M_n;A^\vee) \vert = \vert H^j(M_n;A)\vert$. 
Therefore, we can write: 
\be
\sigma_{K(A^\vee,n-p-1),\CC}^{(n)}(M_n) = 
\vert H^{p+1}(M_n;A)\vert \vert H^{p+2}(M_n;A)\vert^{-1}\cdots 
\ee
Now, we claim that: 
\be\label{eq:AltSumId}
\prod_{j=0}^n  \vert H^j(M_n;A) \vert^{(-1)^j} = \vert A \vert^{\chi(M_n)} ~.
\ee
This is easily proved by 
%
using a finite chain complex to model the cohomology of $M_n$
(e.g., by use of a finite CW decomposition)  and invoking the 
Euler-Poincar\'e principle to see that 
$|C^j(M_n) \otimes A|  = \vert A \vert^{{\rm rank}\, C^j(M_n)} $.} 
\be \label{eq:ComparePF}
\frac{\sigma_{K(A,p),\CC}^{(n)}(M_n) }
{\sigma_{K(A^\vee,n-p-1),\CC}^{(n)}(M_n)} =
\left( \vert A\vert^{\chi(M_n)}\right)^{(-1)^p} ~.
\ee
The paper \cite{Liu:2021hhy} shows that, in fact, for the fully extended 
TFTs, we have: 
\be\label{eq:FinHomType-EM-Duality} 
\sigma_{K(A,p),\CC}^{(n)} 
\cong \CE_{\theta} \otimes \sigma_{K(A^\vee,n-p-1),\CC}^{(n)} ~,
\ee
where $\CE_{\theta}$ is the Euler theory introduced in 
\eqref{eq:EulerTheory} with $\theta = \vert A \vert^{(-1)^p}$. 
This electric-magnetic duality is closely related to 
that of the $\mathsf{BF}$ theories studied in \autoref{sec:QuantBF-Theory} below.  
Moreover, the duality has  an elegant extension to arbitrary finite homotopy theories. See \cite{Liu:2021hhy} for the details.

\end{remark}

\SectionWithHeader{Defects}{Defects}{sec:Defects}

\subsection{General Remarks On Defects}\label{subsec:DefectGenRemark}

An important further generalization of the notion of field theory we have thus far explained is the inclusion of \emph{defects}. Intuitively, these are modifications of a theory obtained by inserting (possibly nonlocal) observables or by enforcing (dynamical) fields to have certain singularities. If $\CD$ is the defect, then the modified 
theory $F_{\CD}$ will have new partition functions, statespaces, etc., on closed manifolds. 

A simple example of a defect in standard Lagrangian field theory is obtained by inserting a local field at a point in spacetime. Another famous class of examples in gauge theory are the \emph{Wilson defects}. These are defined by choosing a representation $R$ of the gauge group and a closed curve $\gamma$ in spacetime. Denote the defect by $\CD = W_{R,\gamma}$. In terms of the path integral of Yang-Mills theory, the modification by $\CD$ changes the measure in the Yang-Mills path integral by multiplying by the character in the representation $R$ of the holonomy around $\gamma$. The path integral of Yang-Mills theory associated to a principal bundle $P\to M_{n}$ can be written as: 
\be 
F^{\mathsf{YM}}(P\to M_{n}; g_{\mu\nu}) =  \int_{\CA(P)/\CG(P)} e^{- \int_{M_{n}} \Tr F\wedge \star F } ~,
\ee
where $\CA(P)/\CG(P)$ is the (stack) of isomorphism classes of connections
\tightfootnote{\label{foot:connection-space-notation}Here and below, we will use the notation $\CA(P)$ for the affine space of gauge fields (connections) on a principal $G$-bundle $P \to M_{n}$. See \autoref{sec:Connections-on-G-Bundles} below for a discussion of the meaning of connections. The group of automorphisms of $P$ covering the identity on $M_{n}$, i.e., $\mathsf{Aut}(P)$, is known in physics as the group of gauge transformations, denoted $\CG(P)$. The quotient (stack) $\CA(P)/\CG(P)$ is the ``space'' of gauge-inequivalent Yang-Mills field configurations.}
on $P$,  $F$ is the fieldstrength of a connection, $\mathsf{Tr}$ is a positive definite Ad-invariant bilinear form on the Lie algebra (whose normalization defines the Yang-Mills coupling constant), 
and $\star$ is the Hodge star. The modification of the path integral 
for the defect $\CD = W_{R,\gamma}$ is simply:
\be 
F^{\mathsf{YM}}_{\CD}(P\to M_{n}; g_{\mu\nu}) =  \int_{\CA(P)/\CG(P)} e^{- \int_{M_{n}} \mathsf{Tr\,} F\wedge \star F } \mathsf{Tr}_{R} \mathsf{Hol}(\n, \gamma) ~.
\ee

\begin{figure}[H]
\centering
  $\CZ$ : \includegraphics[valign=c,width=3in]{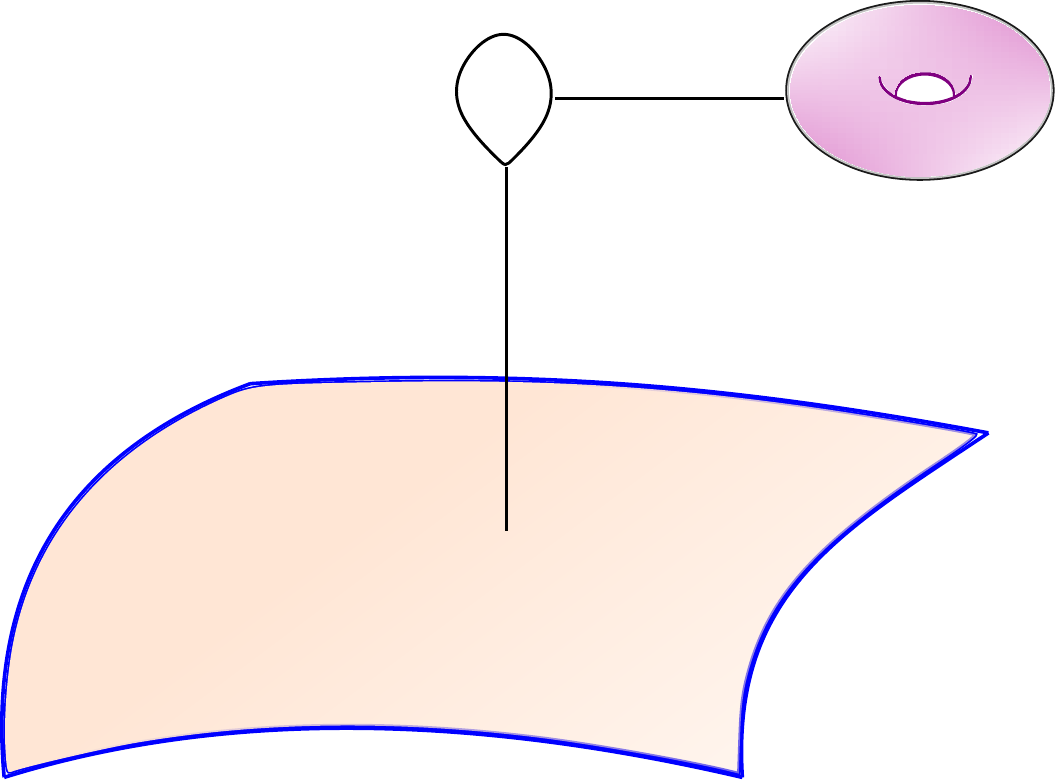}
  \caption{A stratified space.}\label{fig:StratifiedSpace}
\end{figure}

In general, a defect has a support on a subset  $\CZ$ in a spacetime. 
For local fields, the defect would be supported on the spacetime points where the local fields are included in the path integral. For a Wilson defect, the support would be along $\gamma$, etc. 
The support $\CZ$ need not be smooth -- it could be a stratified space, such as that illustrated in \autoref{fig:StratifiedSpace}.
When working with defects whose support is a stratified space, one first describes defects whose support $\CZ$ is a manifold, then for the stratified space, one pieces them together working \underline{up} in codimension. The various strata of the 
defects carry tangential structures, and the issue of which tangential structures are allowed is complex. For example, one could have oriented defects in unoriented theories, or the other way around. A general discussion can be found in \cite[Sec. 2.5]{Freed:2022qnc}. For some recent progress, see \cite{Muller:2025ext}.
 
In the case of topological field theories, we study \emph{topological defects}. The new amplitudes with $\CD$ only depend on the \underline{isotopy class} of the support $\CZ$ 
of the defect. In fully extended field theory, there are two ways of labeling defects using \emph{local labels} and \emph{global labels}. These are terms introduced in 
\cite{Freed:2022qnc}.

To describe the global label of a defect $\CD$ with connected support $\CZ$, we consider a neighborhood $\CU_{\CZ}$ of $\CZ$ inside spacetime which, moreover, does not intersect any other defects. The boundary of $\CU_{\CZ}$ is a closed $(n-1)$-dimensional manifold, and the bordism $\emptyset \to \partial \CU_{\CZ}$ creates a 
vector $\psi_{\CD} \in F(\CU_{\CZ})$. In a topological theory, $\psi_{\CD}$ will not depend on the precise choice of $\CU_{\CZ}$. 
This is a very natural way to describe defects: Surround them by a small neighborhood, and describe them by the kind of state created on the boundary of that neighborhood. 
\tightfootnote{ 
In a nontopological theory, one would wish to choose some kind of 
inverse limit on a system of neighborhoods shrinking to the support $\CZ$, 
along the lines of equation \eqref{eq:ObsDirectLimit} below.}
The vector $\psi_{\CD}$ is the \emph{global label} of the defect $\CD$. One can use the global label to define the partition function 
$F_{\CD}(M_n)$ on a closed $n$-dimensional manifold in the presence of the defect $\CD$ as follows: The complement $W_n = M_n-\CU_{\CZ}$ can be used to construct a bordism $W_n: \partial\CU_{\CZ} \rightarrow \emptyset$. 
Therefore, $F(W_n)$ is a linear functional on the vector space $F(\partial \CU_{\CZ} )$. Then we say, 
\be\label{eq:PF-With-Defect}
F_{\CD}(M_n) :=  \langle F(W_n), \psi_{\CD} \rangle ~.
\ee
For the finite homotopy sigma models $\sigma^{(n)}_{\CX,\CC}$, one can 
evaluate $F_{\CD}$ fairly explicitly using \eqref{eq:Sig-XL-Ampl} with 
$N^0_{n-1}= \partial \CU_{\CZ}$ and $N^1_{n-1}= \emptyset$. 
One can extend \eqref{eq:PF-With-Defect} to define $F_{\CD}$ on manifolds with corners of lower dimension.

\begin{figure}[H]
  \centering
  \begin{tikzpicture}[>=Stealth,baseline=-0.5ex]
  \node (graph) [inner sep=0pt]{\includegraphics[width=1.8in]{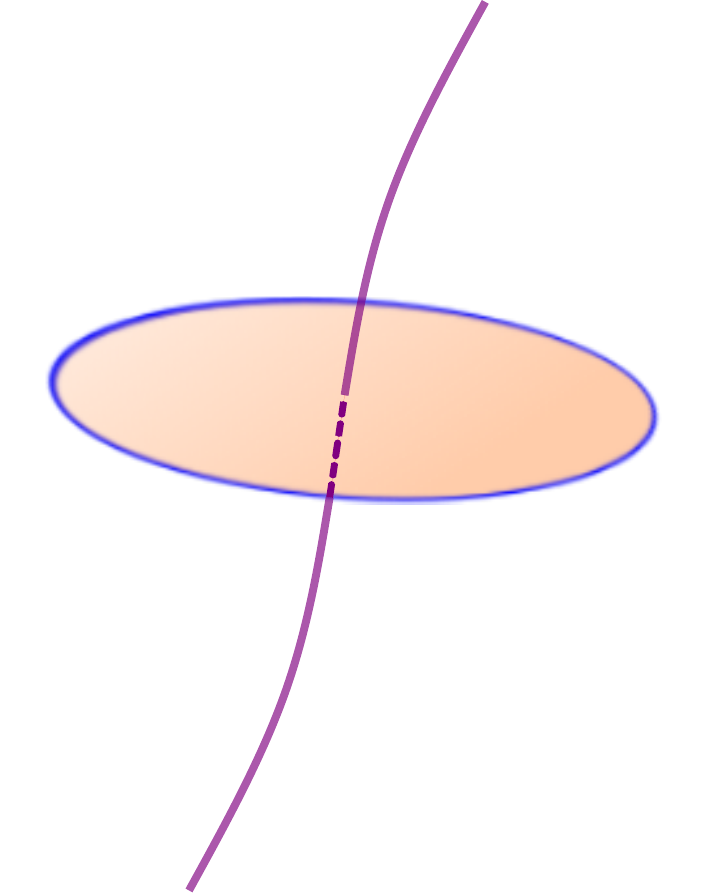}};
  \node at (1.3,0.25) {$\bm{D^{\ell}}$};
  \node (nlow) at (3.8,-0.5) {$\bm{\IS^{\ell-1}}$};
  \node at (-0.7,-3) {$\textcolor{purple}{\CZ}$};
    \draw[->,very thick,blue] (nlow.north west) to [out=120,in=0] (2.1,0);
  \end{tikzpicture}
  \caption{A linking sphere.}
  \label{fig:linkingsphere}
\end{figure}
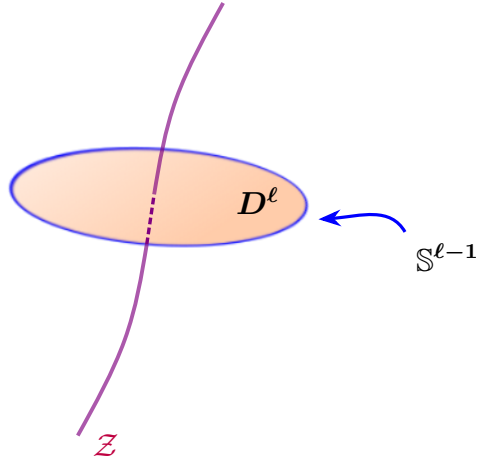

In the spirit of the cobordism hypothesis, one would wish to have a more local description of defects. Let us now assume that $\CZ$ is 
smooth and of codimension $\ell$, as in \autoref{fig:linkingsphere}. Then there is a disk bundle $\nu$ of 
$\CZ \subset M_{n}$ with fibers diffeomorphic to the closed disk $D^{\ell}$. The boundary of a fiber above a point $p\in \CZ$ is the linking sphere at $p$. It is diffeomorphic to   $ S^{\ell-1}$ as in \autoref{fig:linkingsphere}. The definition of the local label at $p$ is motivated by (the TFT version of) Kaluza-Klein reduction. The field theory $\mathsf{KK}(F)$ defined by $\mathsf{KK}(F)(M_{n}) = F(M_{n}\times \IS^{\ell-1})$ is an $n-(\ell-1)$-dimensional field theory. Taking the 
radial direction to be in the interval $r\in [0,\epsilon]$, the effect of the defect is to define a \underline{boundary theory} $\CB(\CD)$ in 
the $n-(\ell-1)$-dimensional theory $\mathsf{KK}(F)$. Therefore, 
\be
\mathsf{KK}(F)([0,\epsilon] \times \CZ) ~,
\ee
in the presence of the boundary $\CB(\CD)$ at $r=0$, is an object in: 
\be\label{eq:DefectCreatesBoundary}
\Hom\left( \mathsf{KK}(F)(\emptyset) , \mathsf{KK}(F)(\CZ) \right) ~.
\ee
Recalling the definition of $\mathsf{KK}(F)$, we are motivated to \underline{define} the \emph{local label} of $\CD$ to be the morphism, 
\be 
\psi_{\CD,p} \in \Hom(1_{\Omega^{\ell-1}\CC}, F(\partial \nu_p) ) ~,
\ee
determined by \eqref{eq:DefectCreatesBoundary}. This is the definition found in \cite{Freed:2022qnc}.

As in the cobordism hypothesis, one would like to deduce the 
global label, and indeed the entire functor $F_{\CD}$ from the local labels $\psi_{\CD,p}$. This is, in general, a highly nontrivial task, especially when one takes into account background fields. This has yet to be done in all its glory. But a very detailed sketch of how it should be done can be found in section 2.5 of \cite{Freed:2022qnc}. The story is summarized in figure 8 of that paper. 

When defects are ``in parallel,'' they satisfy an analog of the operator product expansion. Thus, there is a whole calculus of 
defects, and in some cases, these can result in very interesting algebraic structures. Some examples include: 

\begin{enumerate}

\item  Multiplication of Wilson/'t Hooft line defects in GL-twisted $d=4$, $\CN=4$ super-Yang-Mills (reproducing results in Geometric Langlands (GL) theory) \cite{Kapustin:2006pk,Witten:2009at,Witten:2015dta}. 

\item Multiplication of defects in Class $\CS$ theories \cite{Gaiotto:2010be,Neitzke:2020jik,Gaiotto:2023ezy}.  

\item Multiplication of interfaces between $d=2$ QFTs with $\CN=(2,2)$ supersymmetry \cite{Gaiotto:2015zna,Gaiotto:2015aoa,Khan:2020hir,Khan:2024yiy}.  

\item In very interesting work, M. Dedushenko and N. Nekrasov have described how algebras derived from multiplication of supersymmetric domain walls include interesting quantum algebras such as Yangians, and generalizations thereof. See \cite{Dedushenko:2021mds,Dedushenko:2023qjq}.   

\end{enumerate}
 
In the next section, we specialize to the finite homotopy theories and briefly describe defects in this context.

\subsection{Defects And Domain Walls In The Theories $\sigma_{\CX,\CC}^{(n)}$ }\label{subsec:DefectsFiniteHomotopyTheory}

As we have mentioned, the finite homotopy theories introduce 
notions of dynamical fields. As a consequence, the reference  \cite{Freed:2022qnc} defines 
three new kinds of labels for defects in these theories: 

\begin{enumerate} \itemsep 0pt

\item Classical labels.

\item Semiclassical local labels. 

\item Semiclassical global labels. 

\end{enumerate}   

The classical label of a defect with smooth support $\CZ$ of codimension $\ell$  is an element of $\pi_0(\CX^{\IS^{\ell-1}})$. 
Although often used to describe defects in the literature, these 
labels often leave out many important aspects of $\CD$. We will see 
an example below. 

We leave the description of the local and global semiclassical labels to Appendix A of \cite{Freed:2022qnc}. One of the main virtues of the formalism is that it defines an unambiguous product on parallel defects. The central idea generalizes the use of the pair of pants to define a product in 2d TFT. Given defects $\CD_1$ and $\CD_2$ of 
codimension $\ell$ with ``parallel'' support, one considers disk bundles around each of these defects and embeds the $\ell$-dimensional disk into a larger disk of dimension $\ell$. From this, one constructs a correspondence that defines a product. Some interesting examples can be found in Section 4 and Appendix A of \cite{Freed:2022qnc}. 

An especially important class of defects 
are the \emph{interfaces} between theories. As with all defects, these can be given local and global semiclassical descriptions. An interface between a trivial theory and a theory $\sigma^{(n)}_{\CX,\CC}$ is a \emph{boundary theory}. These play an especially important role in the literature on TFT. 

We conclude by examining the classical and global labels of some defects in the finite homotopy theories defined by $\CX = K(A,p)$, where $A$ must be an Abelian group if $p>1$. Recall that $p=1$ is the case of finite gauge theory, where the group need not be Abelian. We assume $p>0$. 

The classical labels are determined by
\be 
\pi_0(\CX^{\IS^{\ell-1}}) = H^p(\IS^{\ell-1}; A)
= 
\begin{cases}  A ~, & \text{for } p = \ell-1 ~,\\ 
0 ~,& \text{otherwise}. \\ 
\end{cases}
\ee
Note that for $p=1$, we only find interesting classical labels for codimension $\ell=2$. The label is a group element: This is the classic way of describing an 't Hooft defect, namely, by specifying the holonomy created by the defect on a small linking circle. 
The Wilson defects are not visible with classical labels. 

Let us now move on to global labels of defects supported on a smooth 
closed codimension $\ell$ submanifold of spacetime $\CZ \subset M_n$.
We assume the normal bundle can be trivialized, so that we can identify the boundary of a small neighborhood $\CU_{\CZ}$ with 
\be
\CU_{\CZ} \cong \CZ \times \IS^{\ell-1} ~.
\ee
Referring back to \eqref{eq:FinHomThy-StateSpace}, we see that the global labels are vectors $\psi_{\CD}$ in the vector space of functions, 
\be 
\psi_{\CD}: H^p(\CZ \times \IS^{\ell-1}; A) \rightarrow \IC ~.
\ee
The K\"unneth theorem states that this group sits in an extension, 
\be\label{eq:KunnethLables}
\begin{split}
0 \rightarrow &  H^p(\CZ; A) \otimes_{\IZ} A \oplus H^{p-(\ell-1)}(\CZ; A)\otimes_{\IZ} A \rightarrow  \\ 
& \rightarrow H^p(\CZ \times \IS^{\ell-1}; A) \rightarrow  \\
& \rightarrow \Tor(H^{p+1}(\CZ;A),A) \oplus \Tor(H^{p+2-\ell}(\CZ;A),A) \rightarrow 0 ~.
\end{split}
\ee

As we will discuss in \autoref{sec:QuantBF-Theory}  below, if we take $A=\IZ/N\IZ$, then the finite homotopy sigma model $\sigma^{(n)}_{B^pA}$ is closely related to the $n$-dimensional ``$\mathsf{BF}$ theory'' with a $p$-form and an $(n-p-1)$-form gauge field. The latter theory has Wilson and 't Hooft defects, as explained in \autoref{sec:QuantBF-Theory}. 
Suppose that $\CZ$  is an orientable cycle of dimension $p$, then it is natural to expect that functions concentrated on the component 
\be 
H^p(\CZ;\IZ/N\IZ)\otimes_{\IZ} \IZ/N\IZ  \cong \IZ/N\IZ ~,
\ee
correspond to states created by the Wilson defect measuring the holonomy of the $p$-form 
gauge field along $\CZ$. Similarly, if $\CZ$ is a cycle of dimension 
$n-p-1$, then the codimension is $\ell = p+1$, and it is natural to expect that functions 
supported on the component 
\be 
H^{p-(\ell-1)}(\CZ; A) \otimes A \cong H^0(\CZ; A)\otimes A \cong A ~,
\ee
correspond to states created by 't Hooft defects measuring the holonomy of an $(n-p-1)$-form gauge field along $\CZ$. The can, in principle, be checked by evaluating correlation functions of such defects using equation \eqref{eq:PF-With-Defect} and comparing with the 
$\mathsf{BF}$ correlator \eqref{eq:LinkingNumberCorrelator} below. We will not attempt this rather interesting exercise here.

Equation  \eqref{eq:KunnethLables} suggests the existence of defects other than the standard 't Hooft and Wilson defects. They do not appear to have been investigated. 
\tightfootnote{The Kontsevich-Segal inverse limit procedure discussed in equation 
\eqref{eq:ObsDirectLimit} suggests that every global label should come from some defect.}

\SectionWithHeader{Symmetry Action Of A  TQFT On A QFT: The Quiche Picture}{Symmetry Action Of A  TQFT On A QFT: The Quiche Picture}{sec:Quiche}

One of the central themes of $20^{th}$ century physics has been that   \underline{symmetry} is a fundamental concept in formulating the laws of nature, and that symmetry as a tool is one of the most effective ways we have to understand physical systems. 
The past ten years have witnessed a strong movement in theoretical physics to expand notions of symmetry in ways which -- one hopes -- will prove to be equally fundamental and powerful. These expanded notions include ``generalized symmetries,'' ``categorical symmetries,'' and ``noninvertible symmetries.''  The TASI 2023 lectures of S.-H. Shao \cite{Shao:TASIVideoLec1,Shao:TASIVideoLec2,Shao:TASIVideoLec3,Shao:TASIVideoLec4} and T. Dumitrescu \cite{Dumitrescu:TASIVideoLec1,Dumitrescu:TASIVideoLec2,Dumitrescu:TASIVideoLec3,Dumitrescu:TASIVideoLec4} addressed these ideas. Some recent reviews include \cite{Cordova:2022ruw,Freed:FiniteSymmetry2022,Freed:2022iao,Schafer-Nameki:2023jdn,Shao:2023gho,Brennan:2023mmt,Bhardwaj:2023kri,Costa:2024wks,Shao2025:PT}.
 
Generalized/Categorical/Noninvertible symmetries have thus far proven to be useful in constraining dynamics of some nontrivial physical systems, but have not achieved the kind of fundamental status analogous to that of symmetry in the traditional sense. The field is evolving vigorously and future developments -- especially nontrivial and surprising applications --  might alter this judgment.

The paper \cite{Freed:2022qnc} provides a unified view of the generalized, categorical, and noninvertible symmetries, at least for symmetries ``of finite type.'' In practice, ``finite type'' means that the symmetry is implemented using the finite homotopy sigma models studied in \autoref{sec:FinHomTheory}. One of the main ideas of 
\cite{Freed:2022qnc} goes back to D-brane physics and holography from the 1990's in which a physical system is divided into   ``bulk'' and ``boundary'' systems,  and a gauge symmetry in the ``bulk'' is related to a global symmetry on the ``boundary.'' For a relatively recent discussion in the context of holography, see \cite{Harlow:2018jwu,Harlow:2018tng}.
A second key idea in \cite{Freed:2022qnc} is that, as we have seen in examples above, for a fully local topological field theory, $F({\rm pt})$ plays the role of some kind of algebra (perhaps at a higher categorical level). An outcome of the theory of defects sketched in \autoref{sec:Defects} above is that a boundary theory then plays the role of a module. 

\begin{figure}[h]
\centering
\includegraphics[width=3in,height=2.5in]{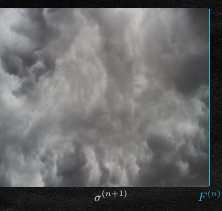} 
\caption{\label{fig:quiche1}Left-half plane with half-plane decorated with TFT $\sigma^{(n+1)}$   and quantum field theory $F^{(n)}$ on the right boundary. }
\end{figure}

In the history of group theory \cite{Wussing1984-wp}, symmetries were first studied in the context of examples. It was only slowly, and with some difficulty, that the abstract notion of a group emerged. It is useful to separate groups from their actions on modules because universal relations can be proven using the structure of groups, which then applies to all modules. Combining this observation with the second idea mentioned above, one is led to try to identify an $(n+1)$-dimensional TFT $\sigma^{(n+1)}$ as a generalization of a group and an $n$-dimensional QFT $F^{(n)}$ -- not necessarily topological! -- as an analog of a module. The preliminary picture (which will need to be modified) is  shown in \autoref{fig:quiche1}.

\begin{figure}[h]
\centering
\includegraphics[width=4in]{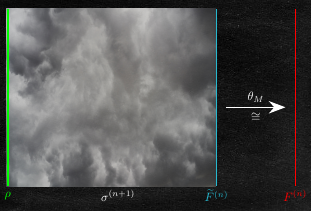} 
\caption{\label{fig:QuicheAction} A cylindrical manifold diffeomorphic to $[0,1] \times M$, where $M$ is of dimension $\leq n$. An $(n+1)$-dimensional \underline{topological} field theory $\sigma^{(n+1)}$ is supported on $[0,1] \times M$. There is a \underline{topological}  boundary theory $\rho$ on the left boundary and a not necessarily topological boundary theory $\widetilde{F}^{(n)}$ on the right. Because $\sigma^{(n+1)}$ is topological, the entire configuration is isomorphic to the value of some $n$-dimensional field theory 
$F^{(n)}$ evaluated on $M$. This picture is known as the ``sandwich'' picture of a finite symmetry action on an $n$-dimensional field theory. }
\end{figure}

This picture is not entirely satisfactory: The system is really $(n+1)$-dimensional, and we are attempting to describe symmetries of $n$-dimensional quantum field theories. To make the system $n$-dimensional, we should ``cap off'' the left half-plane in \autoref{fig:quiche1}  with a boundary theory, and in order to avoid introducing new local degrees of freedom, we introduce a 
\underline{topological} boundary theory on the left. Thus a more accurate picture is that shown in \autoref{fig:QuicheAction}. It is sometimes referred to as ``the sandwich.''

In \autoref{fig:QuicheAction}, note that the boundary theory on the right 
$\widetilde{F}^{(n)}$ is a boundary theory for $\sigma^{(n+1)}$. Because the ``bulk'' theory 
is topological, the interval can be shrunk to an arbitrarily small size, and so the 
system shown is an $n$-dimensional field theory. In the ``action'' on a field theory 
$F^{(n)}$, there should be an isomorphism $\theta_M$ to $F^{(n)}$. This isomorphism is more data in the definition of the action. The reader may well ask, ``the action of what?'' 

The proposal of  \cite{Freed:2022qnc} is that the action of a group is 
replaced in field theory by the action of a \emph{quiche}: 
\bigskip
\begin{definition}[\textcolor{red}{Quiche}]
An $n$-dimensional \emph{quiche} is a pair $(\rho,\sigma)$ where $\sigma$ is an $(n+1)$-dimensional topological field theory and $\rho$ is a topological boundary theory.
\tightfootnote{ 
We will write $\rho$ on the left and the ``physical'' theory on the right. Many authors do the reverse.} 
\end{definition}

\begin{figure}[h]
\centering
\includegraphics[width=3in,height=2.5in]{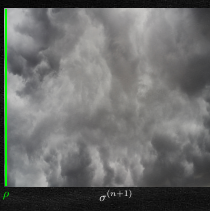} 
\caption{\label{fig:QuichePicture} A quiche. The   TFT $\sigma^{(n+1)}$ 
is evaluated on $\IR_+ \times M$ with a topological boundary theory $\rho$ on the left.}
\end{figure}

A quiche may then be pictured as in \autoref{fig:QuichePicture}.
The central proposal of  \cite{Freed:2022qnc} is that in field theory, it is the quiche $(\rho, \sigma)$ which should be regarded as the generalization of the notion of a group. The boundary theory $\wt{F}$ together with the isomorphism $\theta_M$ provide the notion of a module action. In this way, one separates the algebraic structure from its action on different modules. Thus the second crucial definition from   \cite{Freed:2022qnc} is:

\begin{definition}[\textcolor{red}{Action of quiche on field theory}] Let $F$ be an $n$-dimensional field theory (not necessarily topological) and let $(\rho,\sigma)$ be an $n$-dimensional quiche. We say that \textbf{ $(\rho,\sigma)$ acts on $F$} if there is a boundary theory $\wt{F}$ for $\sigma$ and a coherent set of 
isomorphisms 
for all manifolds $M$ of dimension $\leq n$,
\be
\theta_M: \sigma_{\rho,\wt{F}}([0,1] \times  M   )  \rightarrow F(M) ~ . 
\ee
Here  $\sigma_{\rho,\wt{F}}([0,1] \times  M   )$ means that the theory $\sigma$ is evaluated on $[0,1] \times  M$ with boundary 
theories $\rho$ on $\{0\} \times M$ and $\wt{F}$ on $\{ 1\}\times M$. 
%
%
\end{definition}

 This viewpoint becomes especially powerful when one starts to include topological defects in the quiche, and thus we replace the preliminary picture of \autoref{fig:quiche1} with \autoref{fig:QuicheAction}.

\begin{remark} The picture  of \autoref{fig:QuicheAction} is commonly called the ``sandwich'' picture of symmetry action, and the TFT $\sigma$ is commonly called the ``Symmetry TFT'' or ``SymTFT.'' While the use is very common, it is sub-optimal. First, the term ``SymTFT'' refers to $\sigma$, and does not account for the important data $\rho$. Second, the terminology does not clearly distinguish between the abstract algebraic symmetry and the module it acts upon. The term ``quiche'' is meant to evoke an open sandwich, but one which resides in the quantum world.  
\end{remark}
\bigskip
\begin{remark} The relation of the sandwich construction to symmetries of quantum field theories has many precedents. 
An incomplete list includes 
\cite{Witten:1998wy,Frohlich:2004ef,Belov:2004ht,Moore:2006dw,Gaiotto:2014kfa,Kong:2017hcw,Freed:2018cec,Gaiotto:2020iye,Apruzzi:2021nmk}. 
\end{remark}

Two examples where the above picture has been worked out quite thoroughly are:

\begin{itemize}
  \item A reworking of the standard notion \cite{Wigner2013-xq}
  of $G$-symmetry in quantum mechanics, for $G$ a finite group.  
  In this case, $\sigma$ is the finite homotopy sigma model  
  $\sigma^{(2)}_{BG,\CC}$, where $\Omega^2\CC$ is the Morita $2$-category of algebras, bimodules, and bimodule maps. 
  \item  A description of the different four-dimensional Yang-Mills theories with gauge Lie algebra $\mathfrak{g}$ (and adjoint matter). In this case, we have $\sigma = \sigma^{(5)}_{B^2 A} $,  where $A \subset Z(\widetilde{G})$ and $\widetilde{G}$ is the compact, connected, and simply connected gauge group with Lie algebra $\mathfrak{g}$. The data $\rho$ contains crucial information usually summarized via topological couplings in the gauge theory. 
 For further details, see \cite{Gaiotto:2010be,Aharony:2013hda,Kapustin:2014gua,Gaiotto:2014kfa,Freed:2022qnc,Freed:2022iao}. 

\end{itemize}

An important open problem for the future is the generalization of the quiche construction from finite symmetries to continuous symmetries. 
Some notable attempts in this direction include \cite{Brennan:2024fgj,Gagliano:2024off,DelZotto:2024ngj,Bonetti:2024cjk,Apruzzi:2025hvs,Bonetti:2025dvm}.

\SectionWithHeader{A Survey Of Some Famous Examples Of TFTs}{A Survey Of Some Famous Examples Of TFTs}{sec:survey-famous-examples-TFTs}

\subsection{Some Brief Historical Remarks} 

The subject of topological quantum field theory began with 
Edward Witten's reformulation of the Donaldson invariants of smooth four-dimensional manifolds in the framework of quantum field theory  \cite{Witten:1988ze}. Witten, following some promptings from Michael Atiyah, showed that, with a suitable choice of background fields, certain correlation functions in $d=4$, $\CN=2$ supersymmetric gauge theory with gauge group $\mathsf{SU(2)}$ reproduce the Donaldson invariants. See \cite{Labastida:2005zz,Moore:2012SCGP,Moore:2017SCGP} for expository reviews, and \cite{Moore:2024vsd} for a description of topological twisting in the spirit of these lectures. 

After Witten's insight \cite{Witten:1988hf} that the Jones polynomial of knots in $\IS^3$ can also be formulated using Wilson defects in three-dimensional Chern-Simons theory, Atiyah \cite{Atiyah:1989vu} gave a famous definition of topological field theory equivalent to our definition \eqref{eq:Definition-TopTwo-TFT} of a topological field theory.  Atiyah's formulation was strongly influenced by Graeme Segal's formulation of conformal field theory \cite{Segal:1987sk,Segal:1988zk,Segal1988,Segal:2002ei} as a functor on a bordism category on two-manifolds equipped with conformal structure. Segal's definition of conformal field theory may be regarded as the beginning of the functorial formulation of field theory.  
\tightfootnote{In a previous paper by the present authors, it was pointed out to us by Y. Tachikawa that there were important precedents for the functorial formulation of field theory going back some time. In particular, a perusal of the original papers of J. Schwinger \cite{Schwinger1948} and S. Tomonaga \cite{Tomonaga:1946zz}
is most illuminating. Schwinger and Tomonaga formulated relativistic field theory in terms of families of unitary operators $U(\sigma, \sigma')$ associated to spatial slices $\sigma,\sigma'$ and satisfying gluing axioms which we recognize as being essentially the rules of gluing that define a functor. An important advance relative to the older works is that Schwinger and Tomonaga were working in a very restricted (but important) example of spacetime, namely, Minkowski space. Thus, all the bordisms they considered were in fact invertible. It is very important that, in general, bordisms are not invertible. Another important difference is that the Atiyah-Segal axioms are aiming to capture the properties of Euclidean signature field theory. The entire question of Wick rotation is a nontrivial one. Atiyah and Segal were most directly influenced by the Feynman path integral approach to field theory.} 

If one attempts to fit the original example of Donaldson-Witten theory (topologically twisted super-Yang-Mills) into the Atiyah-Segal framework, one runs into trouble. The statespaces should be instanton Floer homology, but the Floer homology groups present technical difficulties and must come in several flavors \cite{Donaldson2002}.
\tightfootnote{This is also true infrared description using Seiberg-Witten theory, although in that case, much more can be said \cite{Kronheimer2007,Lin:MonopoleFloerHomology}.   } 
Ironically, the Donaldson-Witten theory is \underline{not} a topological field theory, and certainly not a fully extended topological field theory in the mathematical sense discussed above. Similar remarks apply to other famous topological field theories, such as the Vafa-Witten \cite{Vafa:1994tf} and  Rozansky-Witten \cite{Rozansky:1996bq} theories. The opposite appears to be the case for the 4d GL-twisted $\CN=4$ super-Yang-Mills theory, that is, for Kapustin-Witten theory  \cite{Kapustin:2006pk}. Here, the theory can be defined in low dimensions, but not in dimensions $3$ and $4$. 
These theories are all examples of \emph{partially defined topological field theories}. They satisfy some of the axioms, but 
are not defined on all elements of the bordism category. 

Another important class of topological field theories in the development of the subject is the topological sigma models introduced in \cite{Witten:1988xj,Witten:1991zz}. 
The subject is vast and central to topological string theory and (homological) mirror symmetry, among many other things. It is beyond the scope of these lectures. For some expository material, see \cite{Greene:1997ty,Hori:2003,Aspinwall:2009isa}.

A common theme of the above examples is that they are 
\emph{cohomological topological field theories}. They are based on $\CQ$-closed sectors of nontrivial nontopological field theories, where $\CQ$ is an odd operator with $\CQ^2=0$. For some expository discussion 
of cohomological topological field theories, see 
\cite{vanBaal:1989aw,Witten:1990bs,Birmingham:1991ty,Roca:1992ry,Cordes:1994sd,Cordes:1994fc,Moore1995,Szabo2000,Labastida:2005zz,Pestun:2016jze,Pestun:2016qko,Jiang:2024nnk}.

\subsection{Chern-Simons Theories}\label{subsec:Chern-Simons-Theory}

One of the most famous and important examples of a topological field theory is pure 3d Chern-Simons theory. This is a huge topic and itself could be the focus of an entire book. It is a topic that continues to influence a great deal of current research. We only include a few remarks here for completeness. 
\tightfootnote{For more on 3d Chern-Simons theory, see the TASI 2019 lectures \cite{Moore:2019TASI} and the many references therein.}

Three-dimensional Chern-Simons theories were
made famous by the renowned and profound paper of 
E. Witten \cite{Witten:1988hf} on the quantum field theory interpretation of the Jones polynomial, and related topological invariants of knots, links, and three-manifolds.

\subsubsection{Abelian Chern-Simons Theory} 

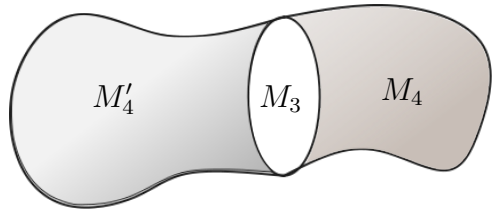
\begin{figure}[h]
  \centering
  \begin{tikzpicture}[scale=1,baseline=-0.5ex]
    \node (image) [inner sep = 0pt]{\includegraphics[width=2.5in,valign=c]{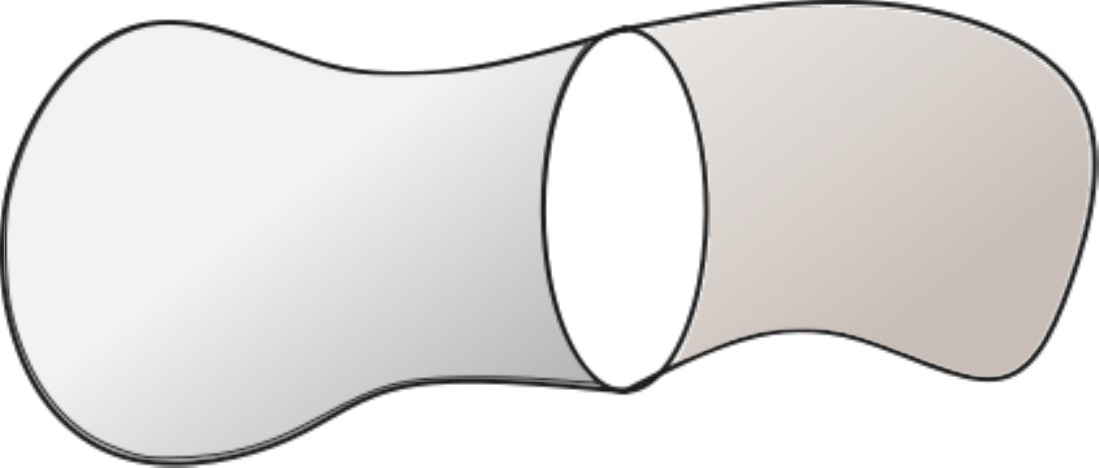}};
    \node at (0.4,0.1) {$M_3$};
    \node at (-1.8,0.1) {$M_4'$};
    \node at (2,0.2) {$M_4$};
  \end{tikzpicture}
  \caption{Non-uniqueness of the extension of $M_3$ to a four-manifold.}
  \label{fig:nounext}
  \end{figure}

Perhaps the simplest example of a Chern-Simons theory is that with gauge group $\mathsf{U(1)}$, so we begin by describing this theory. 
The dynamical field is a gauge theory with a $\mathsf{U(1)}$ gauge field, i.e., a connection on a principal $\mathsf{U(1)}$-bundle $P \to M_3$, with $M_3$ a 3-dimensional oriented manifold.
\tightfootnote{See \autoref{sec:G-Bundles} and \autoref{sec:Connections-on-G-Bundles} for some mathematical background on principal $G$-bundles and connections (and \autoref{sec:DegreeTwo-Diff-Characters} for the $G = \mathsf{U(1)}$ case). We will take $A$ (and hence, $F$) to be antihermitian, which is at variance with some physics conventions -- this point is explained in much more detail in footnote \ref{foot:ConventionClash}.}
(So one of the background fields is an orientation of $M_3$.)  Locally, the gauge field is described by a real $1$-form $A$ with globally defined fieldstrength $F \in \Omega^{2}(M_3)$.
\tightfootnote{\label{foot:space-of-forms}Notation: For any manifold $M_{n}$, $\Omega^p(M_{n})$ is the vector space of all globally defined smooth $p$-forms. We write 
$\Omega^p_{\IZ}(M_{n})$ for those with integral periods. If the forms are valued in a vector bundle  $E \rightarrow M_{n}$, we write $\Omega^p(M_{n};E)$ .} 
Locally, $F = dA$. The exponentiated action in the path integral is
\eqa{
  \exp\left(\imag  \frac{k}{4\pi}\int_{M_3}A\wedge dA \right) ~,
  \label{eq:U1-CS-Act}
}
where we have normalized $A$ so that $F$ has periods in $2\pi \IZ$. The action does not look gauge invariant, but under ``small'' gauge transformations $A \mapsto A + d\epsilon$, 
\eqa{
  A\wedge dA &\longmapsto A\wedge dA + d(\epsilon dA) ~,
}
so if $M_3$ is compact and without boundary, $\int_{M_3}A\wedge dA$ is gauge-invariant. But under ``large'' gauge transformations $A \mapsto A+\omega$, with $\omega \in \Omega_{\IZ}^{1}$, it is   not   well-defined. 
\tightfootnote{We will say more about small and large gauge transformations in \autoref{sec:DegreeTwo-Diff-Characters}.}
One way to attempt to deal with that problem is to exhibit $M_3$ as a boundary of a four-manifold $\partial M_4 = M_3$.
One can prove that it is possible to extend $P \to M_3$ and its connection to a principal $\mathsf{U(1)}$-bundle $P \to M_4$.
\tightfootnote{\label{foot:obstruction-Omega3-SO-BU1}The obstruction to such an extension would be a nontrivial element of the bordism group $\Omega_{3}^{\mathsf{SO}}(B\mathsf{U(1)})$. (Bordism groups were introduced towards the end of \autoref{sec:Bordisms}.) However, as we discuss in footnote \ref{foot:bordism-3-SO}, $\Omega_{3}^{\mathsf{SO}}(B\mathsf{U(1)}) \cong H_{3}(B\mathsf{U(1)};\IZ)$. Now $B\mathsf{U(1)} \cong K(\IZ,2) \cong \mathsf{CP}^{\infty}$, the homology groups of which are trivial in odd degree and isomorphic to $\IZ$ in even degree, since $\mathsf{CP}^{\infty}$ has zero cells in every odd dimension and one cell in every even dimension. In particular, it follows that $\Omega_{3}^{\mathsf{SO}}(B\mathsf{U(1)}) \cong \{0\}$. }
 Once the bundle extends topologically, there is no problem extending the connection (as one proves using a partition of unity).   Then the identity $F \wedge F = d(A\wedge dA)$ and Stokes' Theorem motivates a 
provisional definition: 
\eqa{
  \int_{M_3}A\wedge dA &\stackrel{?}{=} \int_{M_4} F\wedge F ~.
  \label{eq:ProvisionalDefCS-Invt}
}
There is a fundamental problem with this definition: The extension 
from $M_3$ to $M_4$ is not unique. There can be different extensions as illustrated in  \autoref{fig:nounext}.

In general, 
  \eqa{
    \int_{M_4} F\wedge F &\neq \int_{M_4'}F \wedge F ~.
  }
  What saves the day is that on the closed manifold $M_4 \cup M_4'$, 
\be\label{eq:M4-ambiguity}
    \int_{M_4 \cup M_4'} F \wedge F \in (2\pi)^2 \IZ ~,
\ee
  and hence
  \eqa{
   \left(  \frac{1}{2\pi}\int_{M_4}F\wedge F\right) \mathsf{~mod~} 2 \pi \IZ  ~,
  }
  is well-defined. Therefore, if $k$ is an even integer, the action 
  appearing in equation \eqref{eq:U1-CS-Act} \underline{is not well-defined as a real number, but only as an element of $\IR/2\pi \IZ$}. Even though the action is not a well-defined number, the exponentiated action,
  \eqa{
  \exp\left(\imag  \frac{k}{4\pi}\int_{M_3}A\wedge dA \right) ~,
  }
  is a well-defined phase. So, at least formally, the exponentiated action makes sense as a weighting factor in the path integral.
  
Note that the path integral ``measure'' $\exp\left(\imag  \frac{k}{4\pi}\int_{M_3}A\wedge dA \right)$ is metric-independent. So we expect the path integral,
  \eqa{
    F(M_3) &= \int_{\CA/\CG}[dA] \exp\left(\imag  \frac{k}{4\pi}\int_{M_3}A\wedge dA \right) ~, \label{eq:U1-CS-PathIntegral}
  }
to be a topological invariant, and to define a topological field theory.

 \begin{remark}
$\,$
\begin{enumerate}

\item In the above discussion, we have been forced to take $k$ to be an even integer. However, in some cases (e.g., applications to the quantum Hall effect), one wishes to make sense of such expressions for odd integers $k$. One can do this if one endows $M_3$  with a spin structure. The price one pays is that  the value of the Chern-Simons action depends on the spin structure.  
When $M_3$ has a spin structure, we can find a bordant spin manifold $M_4$, and again the 
principal $\mathsf{U(1)}$-bundle can be extended to $M_4$. The key observation is that 
equation \eqref{eq:M4-ambiguity} can be refined to the statement that on the  closed manifold $M_4 \cup M_4'$,  we have 
\be 
\int_{M_4 \cup M_4'} \frac{1}{2!} \left( \frac{F}{2\pi}\right)^2  \in \IZ ~.
\ee
This follows from the Atiyah-Singer index theorem. The extra factor of $2$ renders 
the path integral weight in \eqref{eq:U1-CS-PathIntegral} well-defined for $k$ an odd integer, but the weight will depend on the spin structure on $M_3$. If we change the spin structure by $\epsilon \in H^1(M_3; \IZ/2\IZ)$, then the exponentiated action changes by 
a sign: 
\be 
\exp\left[ \imag \pi k \int_{M_3} \epsilon \cup \frac{F}{2\pi} \right] ~.
\ee
(For an explanation, see \cite[Sec. 2.1.1]{Belov:2005ze}.) 
Thus, the background fields in the theory with $k$ an odd integer include both an orientation and a spin structure.  

\item The path integral \eqref{eq:U1-CS-PathIntegral} is formally metric-independent, but in fact is not truly metric-independent. 
One way to see this is that one must use gauge fixing to deal with the gauge symmetry. The determinants of the resulting Faddeev-Popov ghosts turn out to have metric dependence. Another, rather elegant way to derive the metric dependence, is to view the Chern-Simons theory as the strong coupling (equivalently, long-distance) limit of Maxwell-Chern-Simons theory with Euclidean action 
\be 
\int_{M_3} \left(\frac{1}{2e^2} F\wedge \star F + \frac{\imag k}{4\pi} A \wedge dA\right) ~.
\ee
One finds an overall dependence on the metric 
 \cite{Witten:1988hf}  
  \eqa{
    F(M_3) &= e^{2\pi \imag  \frac{c}{24}\omega_{\mathsf{CS}}(g)} \times \underbrace{\widetilde{F}(M_3)}_{\substack{\text{metric}\\\text{independent}}} ~,
  }
  with $c = {\rm sign}(k)$, and $\omega_{CS}(g)$ is the 
  ``gravitational Chern-Simons term,'' with 
\be 
d\omega_{\mathsf{CS}}(g) = \tr R\wedge R ~.
\ee

One would like to consider the gravitational Chern-Simons term to be a ``local counterterm,'' but   now one has difficulty defining precisely the gravitational Chern-Simons term (again because the Chern-Simons term is not well-defined as a real number). 

\item The metric dependence of the previous remark is known as the 
``framing anomaly.'' There are various approaches to dealing with it. 
The upshot is that the topological Chern-Simons theory depends on more background fields.
These extra background fields can be taken to be a ``2-framing'' in the sense of 
Atiyah \cite{Atiyah:1989vu,Atiyah1990} or a  \emph{$p_1$-structure} \cite[App. A]{Blanchet1995}.
\tightfootnote{In \cite{Segal:2002ei}, this is referred to as a \emph{rigged structure}. See also \cite[Sec. 2.3]{BunkeNaumann}.}
We briefly explain here some details of what these terms mean.
\tightfootnote{V.S. thanks D. Freed for a discussion about these issues.}
Fix a positive integer $n \in \IZ_{\geq 0}$, and let $M_{n}$ be an oriented $n$-manifold. A $p_1$-structure on $M_{n}$ is a lift of the classifying map $M_{n} \to B\mathsf{GL}^{+}(n,\IR)$ of the oriented tangent bundle through the map $X_{n} \to B\mathsf{GL}^{+}(n,\IR)$, where $X_{n}$ is the homotopy fiber  
\tightfootnote{See \autoref{subsec:CorrespCourse} for a definition of the homotopy fiber.}
 of the map $B\mathsf{GL}^{+}(n) \xrightarrow{p_1} K(\IZ,4)$ that realizes the first Pontryagin class, such that the diagram 
\begin{equation}
\begin{tikzcd}[row sep=large]
& X_{n} \arrow[d, "\pi"] \\
M_{n} \arrow[ur, dashed, "\widetilde{\tau}_{M_n}", blue] \arrow[r, "\tau_{M_n}"] & B\mathsf{GL}^{+}(n,\IR) \arrow[r, "p_1"] & K(\IZ, 4)
\end{tikzcd} \label{eq:p1-lift}
\end{equation}
commutes. A point of $X_{n}$ is a pair $b: {\rm pt} \to B\mathsf{GL}^{+}(n,\IR)$ with $H: [0, 1] \to K(\IZ,4)$ such that $H(0) = p_1 \circ b$ and $H(1) = {\rm pt}$, i.e., $H$ is a homotopy from $p_1 \circ b$ to the basepoint in $K(\IZ, 4)$. 
This lift exists if and only if the composition $p_1 \circ \tau_{M_{n}} : M_{n} \to K(\IZ, 4)$ is nullhomotopic. The set of homotopy classes of lifts \eqref{eq:p1-lift} is a torsor over $H^{3}(M_{n}; \IZ)$.

For a closed, oriented $3$-manifold $M_{3}$, a $p_1$-structure always exists. In this case, for connected $M_{3}$, $H^{3}(M_{3}; \IZ) \cong \IZ$, so the set of $p_1$-structures (up to homotopy) forms a torsor over the integers. Projectively, 3d Chern-Simons theory is a theory of oriented manifolds, i.e., it does not depend on the $p_1$-structure. More precisely, on a closed, connected $3$-manifold $M_3$, shifting the $p_1$-structure by an integer $k \in \IZ$ multiplies the Chern-Simons partition function by $\exp\left(2\pi\imag\frac{c}{24} k\right)$ where $c$ is the central charge of the associated modular tensor category. For a discussion of why $p_1$ structures may be considered more natural in defining, for instance, a fractional Chern-Simons theory,\tightfootnote{By ``fractional'' Chern-Simons theory, we mean a situation where the level is allowed to be a rational or real number. Recall that the standard exponentiated Chern-Simons action functional is well-defined for integer level.} see \cite[Sec. 36.4]{Freed:QTGV}.

\item The basic example of 3d $\mathsf{U(1)}$ Chern-Simons theory can be generalized to include multiple $\mathsf{U(1)}$ gauge fields $A^I$, $I = 1, \ldots, r$, with action
    \eqa{
       S &= \frac{1}{4\pi}\int k_{IJ}A^I \wedge dA^J ~.
    }
    These theories prove to be very useful long-distance descriptions of the physics of various fractional quantum Hall states.  The matrix $k_{IJ}$ must be a symmetric integral matrix and it determines an integral lattice. The lattice must be even. (With ``spin topological fields theories'' -- where we include a spin structure as one of the background fields -- the lattice can also be an odd lattice.) The quantum amplitudes can be expressed in terms of invariants of this lattice \cite{Belov:2005ze,Stirling:2008bq,Freed:2009qp,Kapustin:2010hk}.

\end{enumerate} 

\end{remark}

 \subsubsection{Non-Abelian Chern-Simons Theory} 
  
A much richer theory emerges when we generalize to 3d Chern-Simons theories with non-Abelian gauge group.    Now take a connection on a non-Abelian principal $G$-bundle $P \to M_3$ for $G$ a compact simple group. Locally (i.e., on a contractible patch $\CU_{\alpha} \subset M_{3}$), the connection is $d + A$, with $A \in \Omega^{1}(\CU_\alpha, \mathfrak{g})$, where $\mathfrak{g} = \mathsf{Lie}(G)$. We can construct the Chern-Simons form:
\eqa{
      d\mathsf{Tr}\left(A dA + \frac{2}{3}A^3\right) &= \mathsf{Tr}(F \wedge F) ~ . 
}
In the case where $(M_3, A)$ can be extended to a $G$-bundle with 
connection on a bounding four-manifold $(M_4, A')$, we can define 
the Chern-Simons action as: 
    \eqa{
      \frac{k}{4\pi}\int_{M_3}\mathsf{Tr}\left(A dA + \frac{2}{3}A^3\right) & := \frac{k}{4\pi}\int_{M_4}\mathsf{Tr}(F\wedge F) ~.
    }
As in the Abelian case, there can be different extensions leading to different values for the right-hand-side. 
For a suitable notion of trace (e.g., $\mathsf{Tr} = \mathsf{Tr}_{N}$ for $G = \mathsf{\mathsf{SU}(N)}$), $k$ must be an integer, and then (for $M_3$ compact and without boundary) the 
\underline{exponentiated} action,
  \eqa{
    \exp\left(\imag \frac{k}{4\pi}\int_{M_3}\mathsf{Tr}\left(A dA + \frac{2}{3}A^3\right)\right) &= \exp\left(\imag \frac{k}{4\pi}\int_{M_4}\mathsf{Tr}(F\wedge F) \right) ~,
  }
is a well-defined function on the space of 
gauge equivalence classes of connections. The logarithm is not well-defined.

  \begin{exbox}{Non-Abelian Chern-Simons 3-form} Compute the change of $\mathsf{Tr}(A dA + \frac{2}{3}A^3)$ under a gauge transformation $d + A \mapsto g^{-1}(d+A)g$ to see why the $3$-form $\mathsf{Tr}(A dA + \frac{2}{3}A^3)$ is not globally well-defined on $M_3$.
  \end{exbox}

\begin{remark}
$\,$

\begin{enumerate}

\item The above definition is not entirely satisfactory because there can be situations in which the principal $G$-bundle and $M_3$ cannot be extended to a principal $G$-bundle over an oriented bounding four-manifold $M_4$. 
\tightfootnote{\label{foot:bordism-3-SO}Technically, the bordism group $\Omega^{\mathsf{SO}}_3(BG)$ can be nonzero for some compact groups. See, for example, \cite{Johnson-Freyd:2017ble} for 
a discussion of interesting properties of $H^3(BG;\IR/\IZ)$ where $G$ is the Monster group. The exact sequence \eqref{eq:UCT-Cohomology} below and the fact that $H_2(BG;\IZ)=0$ for $G$ a compact group shows that $H_3(BG;\IZ)$ is nonzero. Then the Atiyah-Hirzebruch Spectral Sequence shows that 
$\Omega_3^{\mathsf{SO}}(BG) \cong H_3(BG;\IZ)$ for a compact group $G$.  }
A better definition, intrinsic to the 3-manifold $M_3$ is provided 
by differential cohomology. See \autoref{sec:WZW-CS-Term} below.

\item The action also makes sense when $M_3$ has a nonempty boundary, but in this case, the exponentiated action must be considered as a section of a line bundle over the space of isomorphism classes of gauge fields, rather than a well-defined function on that space. 
Again, the viewpoint of differential cohomology turns out to be very useful. See \autoref{sec:DegreeTwo-Diff-Characters} below for details.

\item  One can define a nice Chern-Simons-Witten TFT for \underline{any} compact group, and indeed, the theories for finite groups $G$ are special cases of the finite homotopy theories studied above. (Note that the discussion above using local connections  does not immediately apply.)   In general, a 3d Chern-Simons theory is determined simply by a choice of a compact Lie group $G$,  a ``level'' $k \in H^{4}(BG; \IZ)$, and an orientation.

\item It is generally expected that 3d Chern-Simons theory for a compact group can be formulated as a fully local extended TQFT. One associates to a closed one-manifold $\IS^1$,
the relevant modular tensor category which can be viewed as a $2$-category.  The $3$-category associated to a point is more subtle. Some comments for the case of an Abelian 
group appear in \cite{Freed:2009qp}. Remarks for the non-Abelian case appear in \cite{Henriques:2015xxa} and \cite{Teleman2022Simons}.

\item  One can also extend 3d Chern-Simons theory to noncompact groups. Here the flavor is quite different and they are typically \underline{not} TFTs for the simple reason that statespaces are infinite-dimensional. But they are very similar to topological field theories.   For more about this very active research topic, see \cite{Witten:2010cx,EllegaardAndersen:2011vps,Dimofte:2016pua,Andersen:2018pnw,Bah:2022wot}.
More references can be found in \cite[Sec. 8.2]{Bah:2022wot}. 
One of the many reasons Chern-Simons theory for noncompact groups is of great interest is 
its relation to 3d quantum gravity. See, for example, \cite{Achucarro:1986uwr,Witten:1988hc,Witten:1989sx,Carlip:1994tt}. For some of the most recent developments on this idea, we refer to \cite{Collier:2023fwi,Collier:2024mgv}.

\item Chern-Simons theories can also be ``upgraded'' to higher-dimensional theories by using extra background fields: for example, for a suitably normalized closed   form $\omega$, we can contemplate a higher-dimensional theory defined by an action such as,
  \eqa{
    \int \omega \mathsf{Tr}\left(A dA + \frac{2}{3}A^3\right) ~.
  }
  One needs to be careful here to get a well-defined propagator and we will not go into details. This generalization goes back at least as far as the famous paper of A. Schwarz \cite{Schwarz:1979ae} (see p. 13). When $\omega$ is a K\"ahler form on a 4d K\"ahler manifold, one obtains a 5-dimensional theory which was  studied in \cite{Losev:1995cr,Losev:1995zf,Losev:1996up,Losev:1997hx} and further developed in N. Nekrasov's Ph.D. thesis \cite{NekrasovPhD}.  More recently, theories of this nature 
  have played a major role in works by K. Costello and collaborators, and have proved especially useful in providing new insights into integrable models. For examples, see \cite{Costello:2013zra,Costello:2013sla,Costello:2017dso,Costello:2018gyb}. Curiously, a similar expression makes an appearance in an effective action for 4d topological insulators with nontrivial first Chern class of the band structure bundle \cite{Moore:2017byz}.

\item Another way to generalize Chern-Simons theory to higher dimensions is to replace the closed $2$-form $F$ of Maxwell theory by an $\ell$-form $F \in \Omega^{\ell}(M_{n})$. Now $dF = 0$ implies $F = dA$, locally,  where $A$ is a locally defined $(\ell-1)$-form. In $n$ dimensions, we can introduce another locally defined higher form ``gauge potential'' $B$ of degree $n-\ell$ and consider the renowned ``$\mathsf{BF}$-action'':
\be\label{eq:BF-action-FirstPass}
  \exp\left(2\pi \imag N \int_{M_{n}}B\wedge F \right) ~.
\ee
If the periods of $F$ and a corresponding field strength associated to $B$ have 
quantized periods, then $N$ must be quantized.  This action can be used to define 
different theories depending on how one treats the zeromodes. 
One natural way to make sense of such $\mathsf{BF}$ theories   
makes use of differential cohomology, studied in \hyperref[part2]{Part II}. 
See in particular, \autoref{subsec:DiffCohoPairing}  and \autoref{sec:QuantBF-Theory}.

\item The $\mathsf{BF}$ theories also admit a non-Abelian generalization. 
These arise quite naturally in studying 2d non-Abelian Yang-Mills theory \cite{Witten:1991we}. An interesting relation of non-Abelian $\mathsf{BF}$ theory to 
Koszul duality is 
indicated in \cite[App. A]{Khan:2024yiy}.
There is a relation of 2d non-Abelian $\mathsf{BF}$ theory to 2d quantum gravity that is 
analogous to the relation of 3d non-Abelian Chern-Simons theory to 3d quantum gravity. 
It appears this was first noticed in a remarkably prescient paper of T. Fukuyama and 
K. Kamimura \cite{Fukuyama:1985gg}. 
The connection was further explored in  \cite{Chamseddine:1989yz,Chamseddine:1989wn}. More recently, the subject has 
risen to importance in studies of 2d gravity and holography. See  \cite[Sec. 3.3]{Saad:2019lba} and  \cite{Stanford:2019vob} for more extensive discussion.

\end{enumerate}

\end{remark}

\subsection{Two-Dimensional Yang-Mills Theory}\label{subsec:nonexamples}

A very interesting example of a completely solved field theory which is not topological is two-dimensional Yang-Mills theory. The exact solution began with the work of A. Migdal \cite{Migdal:1975zg}.
In modern terms, one can define the theory 
as follows. One chooses a compact Lie group $G$ and defines two fields: A connection on a principal $G$-bundle over a two-manifold $P\to \Sigma$, together with a zero-form $\phi\in \Gamma[\mathsf{ad~}P]$. The theory has a background field $\mu$, which is an area form on the two-dimensional spacetime $\Sigma$. With a suitably normalized bilinear form on the Lie algebra, the action is,
  \eqa{
    S &= \int \mathsf{tr}(\phi F) + \mu\,\mathsf{tr}(\phi^2) ~.
  }
This will be defined on a bordism category where each connected component $\Sigma_c$ of the two-dimensional morphisms $\Sigma$ is equipped with a total area $A_c$ for that component, where $A_c$ is a positive real number. The areas are additive under gluing. 

The Hilbert space of this theory on the circle turns out to be the 
space of $L^2$ class functions on $G$, that is 
\be 
F(\IS^1) = L^2(G)^G ~,
\ee
where the $G$ action is by conjugation. One should think of this Hilbert space as the space of functions of the one gauge-invariant 
quantity we can assign to the circle, namely, the holonomy. Since the holonomy is only defined up to conjugation, we must use class functions of the holonomy. A natural basis for $F(\IS^1)$  is 
the set of characters $\chi^{\mu}$ in the irreducible representations of $G$, labeled by $\mu$. 

For a gauge group of positive dimension, there will be infinitely many irreducible representations, so $\CH(\IS^1)$ is \underline{not} a topological field theory in the sense we have discussed. 

Nevertheless, 2d Yang-Mills theory for a positive dimensional gauge group   behaves very much like a topological field theory. 
For example, the amplitude associated to a cylinder with one ingoing and one outgoing circle,  in the case where
$G$ is finite, is,
\be 
F( [0,1]\times \IS^1) = \sum_{\mu} \chi^{\mu,\vee} \otimes \chi^{\mu} ~,
\ee
and for a gauge group with positive dimension, is, 
\be 
F( [0,1]\times \IS^1; A) = \sum_{\mu} \chi^{\mu,\vee} \otimes \chi^{\mu}
e^{- A c_2(\mu)} ~,
\ee
where $c_2(\mu)$ is the quadratic Casimir, and $A$ is a dimensionless measure of the area. There is a multiplication defined by the pair of pants with total area $A$.
(The picture is the same as in \autoref{fig:composition-m-psi1-psi2}.) 

In equation \eqref{eq:ConvolutionProduct} above, we found that for $G$ finite, the pants multiplication of characters $\chi_{\mu}$ (where $\mu$ indexes the distinct irreducible representations of $G$) is:
\be 
m_P(\chi^\mu, \chi^\nu) = \delta_{\mu\nu} \frac{\chi^\mu}{\chi^{\mu}(1)} ~.
\ee
The generalization to the case where $G$ has finite dimension is:
\be 
m_{P,A} (\chi^\mu, \chi^\nu) = \delta_{\mu\nu} \frac{\chi^\mu}{\chi^{\mu}(1)}e^{- A c_2(\mu)} ~.
\ee
One can show that the  partition function can be expressed as: 
  \eqa{
    F(\Sigma_g, A) = \sum_{R: \mathsf{irrep}}(\mathsf{dim\,}R)^{2-2g} e^{-A C_2(R)} ~.
  }
Note that this expression does not have  a good $A \to 0$ behavior for $g = 0$. So the limit $\mu\to 0$ only gives a partially defined topological field theory.  See \cite{Witten:1992xu,Cordes:1994sd,Cordes:1994fc} for more details 
(or \cite{Moore1995} for a short summary). 


Some time ago, D. Gross and W. Taylor made a very interesting attempt to express two-dimensional 
Yang-Mills theory as a string theory \cite{Gross:1993hu}. 
Details of the $1/N$ expansion of $\mathsf{\mathsf{SU}(N)}$ theory are given in 
\cite{Cordes:1994sd,Cordes:1994fc}.
For a very careful description of the theories in the spirit of these notes, taking careful account of the infinite-dimensional topological vector space theory needed in a rigorous description of these theories, see  \cite{Runkel:2018uls,Wedeen:2023llo}.

\SectionWithHeader{Geometric Approach To Field Theory}{Geometric Approach To Field Theory}{sec:FunctorialApproachQFT}

We now return to some of the motivating remarks from \autoref{sec:basicpicture-tft-heuristic} 
concerning geometric field theory. 
The question we address here is: Can the basic formalism of TFT be generalized to nontopological field theories? Put differently, we ask: Can a general QFT be viewed as a ``representation'' of some geometric bordism category?  

The reason the above questions are interesting is that, while quantum field theory is the most successful scientific framework for describing the laws of nature ever conceived, nevertheless, a completely satisfactory description of what quantum field theory ``is'' remains a subject of much debate and little research. See, for example, \cite[Sec. 2.1]{Bah:2022wot}, and the Snowmass whitepaper  \cite{Dedushenko:2022zwd}, and references therein for some overview remarks.   Among the various approaches to a mathematical formulation of QFT described there, the one closest to Part I of these lectures seeks to define a QFT as some kind of functor out of some kind of bordism category. One might call such a description of QFT  ``geometric field theory,'' in contrast to ``topological field theory,'' because one now includes fields such as metrics in the definition of the bordism category. (Technically, the sheaf on $\mathsf{Man}_n$ of background fields should not be flat.) It has also been called ``functorial field theory'' because, ultimately, a ``field theory'' is defined to be a certain kind of functor. Another name which has been suggested is ``dynamical field theory,'' because typical theories have local dynamics of locally defined degrees of freedom. One of the most significant papers exploring this approach to general quantum field theory is the 
paper of M. Kontsevich and G. Segal \cite{Kontsevich:2021dmb}. Another important approach, also with a geometric (and sheaf-theoretic) flavor, is that of factorization algebras \cite{Costello2016,Costello2021}. Some overlapping ideas can also be found in 
\cite{Losev:2019bel,Mnev:2025skb,Gritskov:2025mee}.

In order to appreciate some of the issues involved, it is worth starting with the one-dimensional case, namely, quantum mechanics, where we include a Riemannian metric 
in the fields defining the bordism category. Note that  the only invariant of a 
Riemannian metric on a one-dimensional manifold is the length of each connected component. 

At first sight, what one should assign to a point would appear to be a complex Hilbert space $\CH$. Or should it be the dual Hilbert space $\CH^\vee$? These are isomorphic, by the 
$\IC$-antilinear Riesz isomorphism, 
\be 
R: \CH \to \CH^\vee ~,
\ee
where if $\psi\in \CH$, then $R_{\psi}$ is the linear functional, 
\be 
R_\psi(\xi):= \langle \psi , \xi \rangle ~.
\ee
However, when the Hilbert space is graded by a Hamiltonian, the isomorphism will not be natural in families (such as manifolds of control parameters, i.e., background fields, deforming the theory). This is one (of the many) motivations to replace the spatial point by a ``germ'' of a point in a 1-dimensional manifold. For the moment, we identify a germ with an infinitesimal open neighborhood of the point in some 1-manifold. Because the neighborhood is infinitesimal, it can be any 1-manifold. The 1-manifold is oriented (forward time evolution). The point needs to carry a co-orientation $\pm \frac{\p}{\p t}$.  Then we can say, 
\be 
F\bigg({\rm pt}, \frac{\p}{\p t}\bigg) = \CH ~, \qquad\qquad F\bigg({\rm pt}, - \frac{\p}{\p t}\bigg) = \CH^\vee ~.
\ee

 If the bordism is diffeomorphic to an interval of length $L$, then composition of bordisms implies that for $t_2 - t_1 = L>0$, if the bordism $[t_1, t_2]: t_1 \rightarrow t_2 $ has a metric of length $L$ then 
 it follows from gluing that, 
\be\label{eq:EucTimeEvolution}
F([t_1, t_2]) = e^{- L H} ~,
\ee
where $H$ is an operator on $\CH$. For Wick-rotated versions of unitary theories $H$ will be self-adjoint (not necessarily bounded), and identified with the Hamiltonian.
\tightfootnote{For some interesting examples where $H$ is not self-adjoint see 
\cite{Losev:2019bel,Losev:2022tzr,Losev:2023bhj,Beck:2024xtd}.
}

(Here we have limited attention to systems with time-translation invariance, corresponding to a time-independent Hamiltonian.) In topological quantum mechanics, the Hamiltonian is zero. But for general quantum mechanics, it is nonzero, and hence there are dynamics. 

An important remark is that, as we have repeatedly stressed, the $S$-diagram argument based on \autoref{fig:sdiag} continues to hold when we include metrics. However, the cylinder must have positive length $L>0$ and therefore, the compositions of the cup and cap morphisms will 
\underline{not} produce an identity operator. Thus, the ``dualizability'' condition that the statespace is finite-dimensional will no longer hold. Of course, many quantum mechanical systems of interest have infinite-dimensional statespaces. 

When $\CH$ is infinite-dimensional and $H$ is an unbounded operator, there are further important subtleties because the domain of $H$ cannot be the entire Hilbert space. 
Rather, there will be a dense subspace $\check E\subset \CH$ on which expectation values 
of $H$ and its powers will make sense. Technically, $\check E$ will be a 
``nuclear Fr\'echet space.'' 
\tightfootnote{See \cite[App. 2]{Costello2011} and \cite[App. A]{Wedeen:2023llo} for quick summaries of the theory of nuclear topological vector spaces and nuclear maps.}
When combined with the idea that a field theory should be a map out of a bordism category where the    objects are germs of points and the morphisms are germs of bordisms,  what is naturally associated to a point is a ``rigged Hilbert space,'' 
\be\label{eq:GelfanTriple}
\widecheck{E} \subset \CH \subset \widehat{E} ~,
\ee
where  $\widehat{E}$ is the dual vector space to $\widecheck{E}$. 
This is an important part of the treatment of  geometric field theory given by 
Kontsevich and Segal \cite{Kontsevich:2021dmb}. The triple \eqref{eq:GelfanTriple} is 
known as a \emph{Gelfand triple} and a Hilbert space equipped with a Gelfand triple is known as a ``rigged Hilbert space.'' 
\tightfootnote{The notion of ``rigged Hilbert space'' was introduced by Gelfand \cite{Gelfand2016-eg} in order to give precise mathematical meaning to the bra- and ket- eigenstates of unbounded operators in quantum mechanics. For an overview see, e.g., \cite{Roberts1966:Rigged,Roberts1966:BraKet,Madrid2005}.
The name ``rigged Hilbert space'' is an unfortunate translation from the Russian. The term has, 
regrettably, become standard.  The name ``equipped Hilbert space'' would be more accurate. See \cite{MO43313} for additional references.
}

In general dimensions, Kontsevich and Segal argue that rather than incorporating general Riemannian metrics as background fields on the bordisms (as in \autoref{sec:BackgroundFields}), it is important to consider a class of complex metrics which have the nice property that the real part of the   action for all nonzero real generalized Abelian gauge fields (see equation \eqref{eq:actionh} below) is positive. For any manifold $M$, this defines an infinite-dimensional complex manifold $\Met_{\IC}(M)$, and Kontsevich and Segal propose replacing $\Riem(M)$ in $\CF(M)=\Riem(M)$ by $\Met_{\IC}(M)$ when defining the space of background metrics. Moreover, they   require that the field theory ``functor'' should depend holomorphically on the complex metric. 

As an example, returning  to the one-dimensional case   $\Riem([t_1, t_2])\cong \Diff^+([t_1,t_2])$ and the metric is $f^*((dt)^2)$. The Kontsevich-Segal domain $\Met_{\IC}([t_1,t_2])$ can be identified with the infinite-dimensional manifold of smooth embeddings $f:[t_1,t_2] \to \IC$ such that $f(t_1)=0$ and $\mathsf{Re}(f'(t))>0$ for all $t\in [t_1, t_2]$. The ``holomorphy'' of the field theory functor is, in this case, related to the fact that the time-evolution operator of quantum mechanics,  
$U(t) = e^{- \imag t H/\hbar}$, admits an analytic continuation to the lower-half of the complex plane as a function of $t$ (for $H$ which is bounded below). The complex metric on the interval is $f^*((dz)^2)$. 

In general, a complex domain has a notion of boundary known as a \emph{Shilov boundary} (it enforces a maximum modulus principle). A very remarkable property of the 
Kontsevich-Segal domain is that while the Shilov boundary of $\Met_{\IC}(M)$ contains real metrics, it only contains those real metrics of \underline{Lorentzian} signature! 

When one tries to define the bordism ``category'' $\mathsf{Bord}_{\langle n-1, n\rangle}^{\CF}$, with the sheaf $\CF$ on $\mathsf{Man}_n$   given by $\CF(M) = \Met_{\IC}(M)$, one encounters a subtlety with composition of morphisms. To do this smoothly, one needs 
to know \underline{all} the normal derivatives of the spatial metric along the $(n-1)$-dimensional spatial slice $N_{n-1}$ along which the gluing takes places. 
\tightfootnote{ Indeed, on the field theory side, it is a property already seen in
the quantization of free fermion field theories that the quantization procedure depends on the first $[n/2]$ normal derivatives of the metric  \cite{Segal:DiracOperators}. }
The upshot is that the objects in the ``category'' 
$\mathsf{Bord}_{\langle n-1, n\rangle}^{\CF}$ should be \emph{germs} of 
compact $(n-1)$-dimensional (``spatial'') manifolds $N_{n-1}$. The germ $[N_{n-1}]$ of 
$N_{n-1}$ is an equivalence class of open $n$-dimensional manifolds $M_n$, 
containing $N_{n-1}$ in their interior, with the equivalence relation that $M_n \sim M_n'$ if there is an $M_n''$ with a diffeomorphism (equal to the identity on $N_{n-1})$ 
to open submanifolds of $M_n$ and $M_n'$. As we have remarked, the 
germ should be equipped with a co-orientation indicating the direction of forward ``time'' evolution. The morphisms in the ``category'' should likewise be defined in terms of 
germs. We have been using scare-quotes on the word ``category'' thus far because of the very 
crucial fact that objects do \underline{not} have an identity morphism. This is closely related to the fact that the S-duality diagram of \autoref{fig:sdiag} does not imply that the statespaces are fully dualizable. A careful formulation of the category with germs of manifolds can be found in section 2 of the paper of R. Wedeen \cite{Wedeen:2023llo}.
\tightfootnote{A criticism of this approach to defining a bordism category with metrics can be found at the beginning of   \cite{Berwick-Evans:2015ria}.
 }

A crucial step in the geometric approach to field theory is the formulation of local operators and their correlation functions. The Kontsevich-Segal definition of local operators at a point $x\in M_n$ is derived from standard intuition in field theory and is particularly famous in the formulation of vertex operators in 2d conformal field theory. 
\tightfootnote{A similar discussion in the context of a functorial formulation of field theory was 
described by A. Losev in \cite{Losev:2019bel} and further developed in \cite{Gritskov:2025mee}. See also 
\cite{Mnev:2025skb}. }
It is worth taking a moment to recall this intuition 
before giving the formal definition. Consider an amplitude associated with a bordism 
$M_n: N^0_{n-1}  \to N^1_{n-1}$, but where a local operator $\CD_x$ has been inserted at a point $x$ in the interior of $M_n$. We can denote this amplitude in the presence of the defect as,
\be 
F(M_n; \CD_x): F(N^0_{n-1}) \to F(N^1_{n-1}) ~.
\ee
Now, we can surround $x$ by an $n$-dimensional ball $D \subset \mathsf{Int}(M_n)$. 
The field theory associates an amplitude,
\be 
F(M_n-D):  F(N^0_{n-1}) \otimes F(\p D) \to F(N^1_{n-1}) ~,
\ee
to the bordism  $M_n - D :  N^0_{n-1} \amalg \p D \to N^1_{n-1}$. 
On the other hand, the effect of the local operator is to create a state $\psi_D \in F(\p D)$, 
and we can say that, 
\be
F(M_n; \CD_x)(\xi) = F(M_n-D)(\xi \otimes \psi_D) ~,
\ee 
for all $\xi \in F(N^0_{n-1})$. 
Now, the state $\psi_D$ does not intrinsically characterize the defect $\CD_x$  because it contains irrelevant information about the details of what goes on in the interior of $D$ away from $x$. However, if $D'$ is a smaller ball around $x$ in the interior of $D$ then the amplitude associated to the  bordism $D-D':  \p D' \to  \p D $
takes $\psi_{D'} \to \psi_D$. More formally, we have a directed system of (germs of) small balls $x\in D \subset M_n$ with 
$D < D'$ if $x\in \mathsf{Int}(D') \subset D $. (Note the ordering, which some readers might find confusing.)   The co-orientation on $\p D$ is directed outward. Then
the (germ of the) bordism $D- \mathsf{Int}(D')$ from $\p D' \to \p D$ is associated by the field theory to a  family of   (nuclear) linear maps $F_{D',D}: \CH(\p D') \to \CH(\p D)$
which are coherent in the sense that if $D < D' < D''$, then $F_{D'',D} = F_{D'',D'} \circ F_{D',D}$. Therefore, we can take the inverse limit: 
\be\label{eq:ObsDirectLimit}
\CO_x = \varprojlim_D \CH(\p D) ~,
\ee
to obtain a vector space of observables. This is the Kontsevich-Segal definition of local obervables in the field theory $F$. 

Let us spell  out the meaning of the Kontsevich-Segal definition in the case of 
quantum mechanics. The ball surrounding a point $t$ has $F(D) \cong \CH \otimes \CH^\vee$, which can be identified with the space $\mathsf{HS}(\CH)$ of Hilbert-Schmidt operators on $\CH$. For $t_1 < t_2$, the map from $F( [-t_1, t_1) ) \to F([-t_2,t_2])$ considered as a map of Hilbert-Schmidt operators is the map, 
\be
\CO \mapsto \CO = e^{-\Delta t H } \CO e^{-\Delta t H} ~,
\ee
where $\Delta t = t_2-t_1$, and $H$ is the Hamiltonian. 
An element of the direct limit \eqref{eq:ObsDirectLimit} in this case is equivalent to a map $\Phi: [0,1] \to \mathsf{HS}(\CH)$ such that, 
\be\label{eq:InverseLimitCriterion}
\Phi(t_2) =  e^{-\Delta t H}  \Phi(t_1) e^{-\Delta t H} ~ .
\ee
Thus, the evolution of $\Phi(t)$ to larger times is essentially evolution by the 
heat kernel $e^{-t H}$, with $t>0$. In the (typical) case where the spectrum of 
$H$ is bounded below and unbounded above the heat kernel operator $e^{-t H}$ is 
a smoothing operator. To gain some intuition for what this means let $H$ be the Laplacian acting on functions on a smooth Riemannian manifold. Under evolution by the heat kernel, the eigenfunctions of the Laplacian with eigenvalue $\lambda$ scale to zero exponentially fast like $e^{-\lambda t}$. Distributions at time $t=0$ will evolve to smooth functions at any positive time. By the same token, inverse heat flow will take a typical smooth function and make it increasingly oscillating. Applying this intuition to \eqref{eq:InverseLimitCriterion}, 
we see that if we fix a typical Hilbert-Schmidt operator $\Phi(t)$ at some time $t_2$ 
then at earlier times $t_1$ is will become less-well behaved. The net effect is that the 
Kontsevich-Segal definition of the space of observables at time $t=0$ is a nuclear Frechet space of unbounded operators on the Hilbert space. This is consistent with standard ideas in quantum mechanics. The paper \cite{Kontsevich:2021dmb} goes on to make contact with the Wightman axioms of QFT. While the definition is in accord with physical intuition and makes 
good mathematical sense, fundamental questions remain to be resolved. 
For example, in Segal's mathematical definition of two-dimensional conformal field theory  \cite{Segal:1987sk,Segal:1988zk,Segal:2002ei}, it is an important open question whether one can prove that the tangent space to moduli spaces of CFTs can be identified with exactly marginal operators. 

Another approach to defining field theories based on some notion of a bordism category equipped with Riemannian metrics has been explored in  \cite{Grady:2020sxl}. Moreover, the 
approach has been extended to a definition of field theory, together with an analog of the 
cobordism hypothesis in  \cite{Grady:2021kii}. The issue of the lack of dualizability in 
general quantum mechanics and quantum field theories is addressed by considering freely generated symmetric monoidal categories -- in essence attaching duals to statespaces even when they do not exist in the domain of functional analysis. Whether this more formal approach will be fruitful in defining general interacting quantum field theories remains to be seen. 

\bigskip
\bigskip
\begin{remark}
$\,$
\begin{enumerate}

\item  Geometric field theory offers a natural framework for describing the renormalization group. If $(M, g_{\mu\nu})$ is a bordism equipped with metric, then we can scale the metric 
by $g_{\mu\nu} \to \lambda^2 g_{\mu\nu}$ where $\lambda\in \IR$. Then we define a famiy of field theories: 
\be 
\CF_{\lambda}(M, g_{\mu\nu} ) := \CF(M,\lambda^2 g_{\mu\nu}) ~ . 
\ee
The long-distance ``limit'' is, formally, $\lim\limits_{\lambda \to \infty} \CF_\lambda$. Even in the simplest case of free field theories, it is clear that taking such a limit is not entirely trivial. There is a standard piece of folklore that if $\CF$ is a ``gapped'' theory, then the long-distance limit is a topological field theory. 
In fact, strictly speaking, there are counterexamples to this claim \cite{FMT:Unpublished}. So a more precise formulation of the folklore is needed. 

\item An interesting challenge to the standard Wilsonian paradigm of the renormalization group and effective field theories is presented by the phenomenon of statistical models with fracton excitations \cite{Nandkishore:2018sel,Pretko:2020cko,Gromov:2022cxa,You:2024zyk}.  
It has been proposed that theories with fracton excitations can be defined by defining theories equipped with foliations \cite{Shirley:2017suz}. 
Defining the bordism category 
for manifolds equipped with foliations is an interesting open problem. 

\item The general framework of geometric field theory proved to be quite useful in clarifying how one defines a topological twisting of a general $d=4$, $\CN=2$ field theory 
\cite{Moore:2024vsd}. 

\item It is worth noting that the standard model of fundamental interactions based on 
gauge group $(\mathsf{SU(3)} \times \mathsf{SU(2)} \times \mathsf{U(1)})/\Gamma$, where $\Gamma \subset \IZ_6$, which describes the strong and electroweak interactions -- so far as we know -- cannot fit into the framework of geometric field theory because the theory is not UV complete.

\end{enumerate}

 \end{remark}

We conclude that the geometrical approach to QFT is promising but much more remains to be done, even for free field theories. In \hyperref[part2]{Part II} of these lectures, we turn to a description of an interesting but tractable class of field theories. Even though these theories are free field theories, they present a large number of interesting and instructive features. These are the Generalized Abelian Gauge Theories. They are all generalizations of Maxwell's formulation of electricity and magnetism and these generalizations appear very naturally in supergravity and string theory, especially when one considers branes in these theories. 
Giving a complete formulation of these theories along the lines sketched above is a 
very interesting problem for future research. It is nontrivial, but should be tractable.

\eject
\thispagestyle{empty}
\phantomsection
\vspace*{\fill}
\addcontentsline{toc}{section}{\hspace{6em}\textcolor{red}{PART II: DIFFERENTIAL COHOMOLOGY}}
\begin{center}
\label{part2}
\textcolor{red}{
\LARGE\textbf{PART II: DIFFERENTIAL COHOMOLOGY }
}
\end{center}
\vspace*{\fill}
\eject

\SectionWithHeader{Preliminary Math Background: Hodge Theory}{Preliminary Math Background: Hodge Theory}{sec:HodgeTheory}

Before we turn to the study of higher form gauge fields and generalized Abelian gauge theory, it is useful to review the Hodge star and some basic statements of Hodge theory. 
  
Let $(M_n, g_{\mu\nu})$ be an $n$-dimensional manifold with a  
metric $g_{\mu\nu} $ defining a section in  $\Gamma\big(\mathsf{Sym}_{+}^2 T^* M_{n}\big)$, where the subscript $+$ stands for positive definite. Assume $M_{n}$ is oriented, and choose a nowhere zero volume form $\mathsf{vol}(g)$. In local coordinates around a point $p \in M_{n}$,
\eqa{
  \mathsf{vol}(g) &= \sqrt{|\det~g_{\mu\nu}|}\underbrace{dx^{1} \wedge \cdots \wedge dx^{n}}_{\substack{\text{determined by orientation}\\\text{of $T_{p}^*M_{n}$}}} ~.
}
Now we define a linear operator called the ``Hodge star,'' as a map,
\tightfootnote{\label{foot:Hodge-Unoriented}Although we have assumed that $M_{n}$ is oriented, the Hodge star can, in fact, be defined on an unoriented manifold by twisting with the orientation bundle -- a real rank-one vector bundle built as follows. At any point $p \in M_{n}$, the tangent space $T_{p}M_{n}$ has two possible orientations: these choices are the connected components of the \underline{set} $\Lambda^{n}T_{p}^{*}M_{n} \setminus \{0\}$. The set of these orientation choices over each point of $M_{n}$ forms the orientation double cover $\wt{M}_{n}$ of $M_{n}$, which is a 2-sheeted cover of $M_{n}$, or equivalently, a  principal $\IZ_2$ bundle (see \autoref{sec:G-Bundles}) over $M_{n}$. The orientation bundle $\mathsf{Or}(M_{n})$ then is the real line bundle associated to this $\IZ_2$ bundle by letting $\IZ_2$ act on the fiber $\IR$ by multiplication by $\pm 1$. The Hodge star can then be defined as $\star: \Omega^{\ell}(M_{n}) \to \Omega^{n-\ell}(M_{n}; \mathsf{Or}(M_{n}))$, where $\Omega^{k}(M_{n}; \mathsf{Or}(M_{n}))$ (see also footnote \ref{foot:space-of-forms}) is the space of $k$-forms on $M_{n}$ twisted by the orientation bundle, or equivalently, sections of $\Lambda^{k}T^{*}M_{n}\otimes \mathsf{Or}(M_{n})$.}
\eqa{
 \star : \Omega^{\ell}(M_{n}) \longrightarrow \Omega^{n-\ell}(M_{n}) ~.
}
To do this, we first introduce the local inner product on $\Lambda^{\ell}T_{p}^{*}(M_{n})$. For
\eqa{
\alpha &= \frac{1}{\ell!}\alpha_{\mu_1\cdots\mu_\ell}dx^{\mu_1}\wedge \cdots \wedge dx^{\mu_{\ell}} \in \Lambda^{\ell}T_{p}^{*}(M_{n}) ~,
}
we define
\eqa{
 (\alpha, \beta)_{p} &:= \frac{1}{\ell!} g^{\mu_1\nu_1}\cdots g^{\mu_{\ell}\nu_{\ell}}(p) \alpha_{\mu_1\cdots\mu_{\ell}}(p)\beta_{\nu_1\cdots\nu_{\ell}}(p) ~.
}
Then the formula
\eqa{
 \left.\alpha \wedge \star \beta\right|_{p} &:= (\alpha, \beta)_{p} \mathsf{vol}(g) ~,
}
defines $\star\beta$ since it holds for all $\alpha$.

\begin{exbox}{Hodge Star $\star$}
Here are eight exercises on $\star$. Solve them, and the Hodge star will never strike fear into your heart:\\

\textbf{Exercise 1.} Show that $\star^2: \Omega^{\ell} \rightarrow \Omega^{\ell}$ acts as multiplication by the sign
\eqa{
 \left.\star^2\right|_{\Omega^{\ell}} &= (-1)^{\ell(n-\ell)}\cdot\text{sign}(\det~g_{\mu\nu}) ~, \label{eq:HodgeStar-squared}
}
and this works for any signature.
\tightfootnote{\textbf{Solution to Exercise 1:}
Choose an ON basis for $T_{p}^{*}M_{n}$, namely, $e^1, \ldots, e^n$, so that the orientation volume form is $e^1 \wedge \cdots \wedge e^n$ and $(e^\alpha, e^\beta) = \eta^{\alpha}\delta^{\alpha\beta}$, where $\eta^\alpha \in \{\pm 1\}$. For a multi-index $I= (\alpha_1 < \alpha_2 < \cdots < \alpha_\ell)$, let $I_c = (\beta_1 < \beta_2 < \cdots < \beta_{n-l})$. Define a sign $s(I, I_c)$ by
\begin{align}
    e^I \wedge e^{I_c} = s(I, I_c) e^1 \wedge \cdots \wedge e^n ~.\nonumber
\end{align}
Note that $e^I \wedge e^{I_c} = (-1)^{\ell(n-\ell)} e^{I_c} \wedge e^I$. So, 
\begin{align}
    s(I, I_c)s(I_c, I) = (-1)^{\ell(n-\ell)} ~.\nonumber
\end{align}
Now 
\begin{align}
    \star e^{I} &= \eta^{\alpha_1}\wedge \cdots \wedge \eta^{\alpha_\ell} s(I, I_c) e^{I_c} ~,\nonumber\\
    \star e^{I_c} &= \eta^{\beta_1}\wedge \cdots \wedge \eta^{\beta_\ell} s(I_c, I) e^{I} ~.\nonumber
\end{align}
So,
\begin{align}
    \star^2 e^{I} &= \eta^{I} \wedge \cdots \wedge \eta^{n} (-1)^{\ell(n-\ell)}e^{I} = \text{sign}(\det~g_{\mu\nu})(-1)^{\ell(n-\ell)}e^{I} ~.\nonumber
\end{align}
}
\\$\,$\\
\textbf{Exercise 2.} Show that in local coordinates,
\eqa{
\star\big( dx^{\mu_1} \wedge \cdots \wedge dx^{\mu_k} \big) &= \frac{ |\det\,g_{\mu\nu} |^{1/2}}{(n-k)!}g^{\mu_1\mu_1'}\cdots g^{\mu_k\mu_k'}\varepsilon_{\mu_1'\cdots \mu_k' \nu_1\cdots \nu_{n-k}} dx^{\nu_1}\wedge \cdots \wedge dx^{\nu_{n-k}} ~,
}
with $\varepsilon_{12\cdots n} = + 1$. Thus $\varepsilon_{\mu_1'\cdots \mu_k' \nu_1\cdots \nu_{n-k}}$ takes values in $\{ 0, \pm 1\}$.
\\$\,$\\
\textbf{Exercise 3.} For a product metric $g_{1} \oplus g_{2}$ on $M_1 \times M_2$ and product orientation $\mathsf{vol}(g_1) \wedge \mathsf{vol}(g_2)$, with $\omega_1 \in \Omega^{k_1}(M_1)$ and $\omega_2 \in \Omega^{k_2}(M_2)$, show that
\eqa{
  \star_{g_1 \oplus g_2}(\omega_1 \wedge \omega_2) &= (-1)^{k_2(n_1-k_1)}(\star_{g_1}\omega_1) \wedge (\star_{g_2}\omega_2) ~,
}
where $n_1 = \mathsf{dim\,}M_1$.
\\$\,$\\
\textbf{Exercise 4.} Show that under a conformal transformation,
\eqa{
  g_{\mu\nu}' &= \lambda^2 g_{\mu\nu} \implies \left.\star_{g_{\mu\nu}'}\right|_{\Omega^{k}} = \lambda^{n-2k}\star_{g_{\mu\nu} }~.
  \label{eq:HodgeStarScaling}
}
A corollary of this result is that the Hodge star is conformally invariant when acting on the middle cohomology of an even-dimensional manifold.
\\$\,$\\
\textbf{Exercise 5.} On $\mathbb{M}^{1,1}$ with line element $g_{\mu\nu}dx^{\mu}dx^{\nu} = -dx^0 \otimes dx^{0} + dx^1 \otimes dx^{1}$ and volume form $\mathsf{vol}(g) = dx^1 \wedge dx^0$, show that $\star dx^{\pm} = \pm dx^{\pm}$, where $x^{\pm} := x^{0} \pm x^{1}$.
\\$\,$\\
\textbf{Exercise 6.} On $\mathbb{M}^{1,3}$ with metric $g_{\mu\nu} = \mathsf{diag}(-1,+1,+1,+1)$ and volume form $\mathsf{vol}(g) = dx^{0}\wedge dx^{1}\wedge dx^{2} \wedge dx^{3}$, show that
\eqa{
   \star\big( dx^{0} \wedge dx^{i} \big) &= -\frac{1}{2}\varepsilon^{ijk}dx^{j} \wedge dx^{k} ~,\\
   \star\big( dx^{i} \wedge dx^{j} \big) &= \varepsilon_{ijk}dx^{0} \wedge dx^{k} ~.
}
\textbf{Exercise 7.} On Euclidean $\mathbb{R}^{D}$ with volume form
\eqa{
   dx^{1} \wedge \cdots \wedge dx^{D} &= r^{D-1} dr \wedge \Omega_{D-1} ~,
}
show that
\be 
   \star d\left(\frac{1}{r^{D-2}}\right) = -(D-2)\Omega_{D-1} ~.
\ee
Introduce the unit volume form on $\IS^{D-1}$,
\eqa{
\omega_{D-1} &= \frac{\Omega_{D-1}}{\nu_D} ~, \label{eq:NrmSphrVol}
}
where 
\eqa{
 \nu_D &= \frac{2 \pi^{D/2} }{\Gamma(D/2)} ~,
}
Then 
\be
   d\omega_{D-1} =  \delta^{(D)}(0) dx^{1} \wedge \cdots \wedge dx^{D} ~.
\ee
In general, we will write a $\delta$-function supported form 
Poincar\'e dual to $W \subset M$ as $\eta(W\hookrightarrow M)$. 
So we can write this equation as
\eqa{
  d\left( \frac{1}{\nu_D} \star \frac{dr}{r^{D-1}}\right) &= \eta\left( \{0\} \hookrightarrow \mathbb{R}^{D} \right) ~.
}   
\textbf{Exercise 8.} In $D = 4$, show that the variation of the Hodge $\star$ (acting on a 2-form) with respect to an arbitrary infinitesimal variation $\delta g_{\mu\nu}$ of the metric is given by
\tightfootnote{For a solution, see \cite[App. E]{Cushing:2023rha}.}
\begin{align}
    (\delta_{g}\star)_{\mu\nu\rho\sigma} &= |\det~g_{\mu\nu}|^{1/2} \varepsilon_{\mu\nu\rho\sigma}\big( \delta_{g} g^{\alpha\rho'}\big)g_{\rho\rho'} - \frac{1}{4}|\det~g_{\mu\nu}|^{1/2} \varepsilon_{\mu\nu\rho\sigma}g_{\rho'\sigma'}\delta_{g}g^{\rho'\sigma'} \\
    &= \left(-|\det~g_{\mu\nu}|^{1/2} \varepsilon_{\mu\nu\alpha\sigma}g^{\alpha\alpha'}\delta_{\rho}{}^{\beta'} + \frac{1}{4}|\det~g_{\mu\nu}|^{1/2} \varepsilon_{\mu\nu\rho\sigma}g^{\alpha'\beta'}\right)\delta g_{\alpha'\beta'} ~.
\end{align}
 In the second equality, we used $\delta g^{\mu\nu} = -g^{\mu\mu'}g^{\nu\nu'}\delta g_{\mu'\nu'}$. 
\end{exbox}

 Let $M_n$ be an $n$-dimensional smooth manifold. The de Rham complex is defined to be: 
\eqa{
 \Omega^{\bullet}(M_{n}) &= \bigoplus_{k} \Omega^{k}(M_{n}) ~,
}
with a degree one operator $d$ that squares to zero.
If $M_n$ is equipped with a Euclidean signature Riemannian metric and an orientation, then we can define a   nondegenerate and nonnegative inner product on $\Omega^{\bullet}(M_{n})$ by,
\eqa{
 (\alpha, \beta) &= \int_{M_{n}}\alpha \wedge \star \beta ~.
}
Now, one can compute the adjoint $d^\dagger$ of the exterior derivative with respect to this inner product. One finds that the adjoint of $d:\Omega^k \to \Omega^{k+1}$ is 
$d^\dagger: \Omega^{k+1} \to \Omega^k$, given by 
\tightfootnote{If one performs the same computation, formally, in Lorentzian signature, one finds $d^\dagger = (-1)^{n(k+1)}\star d\star $. Also, note that the definition of $d^\dagger$ in \eqref{eq:d-dagger} holds even if $M_{n}$ is unorientable, because $\star$ appears twice in it and $d^\dagger$ maps even forms to odd forms and vice versa. See also footnote \ref{foot:Hodge-Unoriented}.
}
\eqa{ \label{eq:d-dagger}
  d^\dagger &= (-1)^{n(k+1)+1} \star d\star  ~ . 
}
Note that since Hodge star squares to a constant, we have $(d^\dagger)^2=0$. 
 
Now assume in addition that $M_n$ is compact. Then there is an  \underline{orthogonal decomposition}, known as the  Hodge decomposition of $\Omega^{\bullet}$, where: 
\tightfootnote{See \cite[Ch. 6]{Warner1983} a proof.} 
\eqa{ \label{eq:HodgeDecomposition}
   \Omega^{k}(M_{n}) &= \CH^{k}(M_{n}) \oplus \mathsf{im}(d: \Omega^{k-1} \to \Omega^{k}) \oplus \mathsf{im}(d^\dagger: \Omega^{k+1} \to \Omega^{k}\big) ~,
}
where $\CH^{k}(M_{n})$ is the real vector space of \emph{harmonic forms}:
\eqa{
\CH^{k}(M_{n}) : = \{ \alpha ~|~ d\alpha = 0 \text{ and } d^\dagger \alpha = 0 \} ~.
}

\begin{remark}
The Hodge decomposition leads to a nice connection to K. Intriligator's TASI 2023 lectures \cite{Intriligator:TASIVideoLec1,Intriligator:TASIVideoLec2,Intriligator:TASIVideoLec3,Intriligator:TASIVideoLec4}. Consider the quantum mechanics of a particle moving on a Riemannian manifold $(M_n, g_{\mu\nu})$. The wavefunctions are in $\Omega^0(M_n)\otimes \IC$ 
    and should be $L^2$-normalizable with respect to the Riemannian measure.
    In the language of \hyperref[part1]{Part I} of these lectures, we are considering an  $n=1$ field theory based on maps,
    \eqa{
     \phi: \mathbb{R}_{t} \to (M_{n}, g_{\mu\nu}) ~.
    }
In the supersymmetric case, the target space is replaced by a supermanifold built from the cotangent bundle of $M_n$, and the Hilbert space of the supersymmetric quantum mechanics is 
$\big(\Omega^{\bullet}(M_{n}) \otimes \mathbb{C}\big)_{L^2}$, where the subscript $L^2$ means normalizability with respect to the inner product introduced above.  The supersymmetry operators are $d$ and $d^\dagger$. 
The Hamiltonian is the Laplacian on forms: $H= dd^\dagger + d^\dagger d$. The vector space of supersymmetric ground states -- also known as BPS states --   is the (complexified) 
vector space of normalizable harmonic forms.  The connection to \hyperref[part1]{Part I} of these lectures (and to K. Intriligator's lectures \cite{Intriligator:TASIVideoLec1,Intriligator:TASIVideoLec2,Intriligator:TASIVideoLec3,Intriligator:TASIVideoLec4}) is that in this one-dimensional field theory, $F({\rm pt})=\big(\Omega^{\bullet}(M_{n}) \otimes \mathbb{C}\big)_{L^2}$. In \hyperref[part2]{Part II} of these lectures, we are studying dynamical fields on a spacetime $M_{n}$. Same mathematics. Very different physics.

\end{remark}

\begin{exbox}[exercise:hodge-laplacians]{\emph{Spectrum Of The Hodge Laplacians}}
The material in this exercise will be useful when we discuss partition functions 
in \autoref{sec:PartitionFunctionsGenMax} below.  
Suppose $M_n$ is smooth, compact, and oriented. For $0 \leq k \leq n-1$, define: 
\be 
V_k: = \mathsf{im}(d^\dagger: \Omega^{k+1} \to \Omega^{k}\big) ~,
\ee
and for $1\leq k \leq n$, define 
\be 
\wt{V}_k := \mathsf{im}(d: \Omega^{k-1} \to \Omega^{k}) ~.
\ee
\begin{enumerate}
\item[(1)] Show that 
\be\label{eq:HodgeSusyIsoms}
\begin{split}
d: V_s \to \wt{V}_{s+1} &  \qquad 0 \leq s \leq n-1 ~,\\
d^\dagger: \wt{V}_{s+1} \to V_s &  \qquad 0 \leq s \leq n-1 ~,\\
\end{split}
\ee
are isomorphisms. Note that the proof of these assertions is the same as the proof that the Witten index in SQM with target space $(M_n, g)$ is invariant under deformations of the metric. 

It is not difficult to show that spectrum $\big\{ \mu_i^{(k)} \big\}_{i=1,2,\dots } $   of $d^\dagger d$ on $V_k$ is positive, discrete, and unbounded.  

\item[(2)] Show that $\big\{ \mu_i^{(k)} \big\}_{i=1,2,\dots } $ is identical to the spectrum of 
$dd^\dagger$ on $\wt{V}_{k+1}$. 

\item[(3)] The Hodge Laplacian on $\Omega^k$ is  $\bm{\Delta}_k := d d^\dagger + d^\dagger d$. 
Show that the spectrum of $\bm{\Delta}_k$ is: 
\begin{itemize}
\item[(a)] $\{ \mu^{(0)}_i \}\cup \{ 0 \} $ on $\Omega^0$. 

\item[(b)] The union of $\{ \mu^{(k)}_i \} \cup \{ \mu^{(k-1)}_i \}$  (and union   zero if there are harmonic forms) on $\Omega^k$ for 
$1\leq k \leq n-1 $. 

\item[(c)]  $\big\{ \mu^{(n-1)}_i \big\}$ on $\Omega^n$.
\end{itemize}
 \end{enumerate}

\end{exbox}

\SectionWithHeader{Classical Generalized Maxwell Theory}{Classical Generalized Maxwell Theory}{sec:Classical-Generalized-Maxwell-Theory}
%
%
We begin with Maxwell theory on $\mathbb{M}^{1,3}$ ($(3+1)$-dimensional Minkowski spacetime). Start with the classical fieldstrength $F \in \Omega^{2}(\mathbb{M}^{1,3})$. This is a globally defined $2$-form. The spacetime splitting leads to  a splitting given by a field decomposition
\eqa{
  F & = \frac{1}{2}\varepsilon_{ijk}B_{i}dx^{j}\wedge dx^{k} + E_{i}dx^{0}\wedge dx^{i} ~.
}
We define the electric and magnetic fields separately:
\be 
\begin{split} 
F_m & := \frac{1}{2}\varepsilon_{ijk}B_{i}dx^{j}\wedge dx^{k} ~, \\ 
F_e & := E_{i}dx^{0}\wedge dx^{i} ~,
\end{split} 
\ee
so that $F= F_m + F_e$. The components $E_i$ and $B_i$ determine 
vectors and (pseudo-) vectors $\bm{E}$ and $ \bm{B}$ respectively, valued in $\IR^3$, which are functions of space and time so that,  so that 
we can write  half of the  vacuum Maxwell equations on   $\IM^{1,3}$ as:
\eqa{
  dF &= 0 &\iff \left\{\begin{array}{l} \n\cdot \bm{B} = 0 ~, \\ \frac{\partial\bm{B}}{\partial x^{0}} + \n\times\bm{E} = 0 ~.  \end{array} \right.
}

In order to write the other half of the Maxwell equations in terms of $F$, we must use the Hodge $\star$ operator reviewed in 
\autoref{sec:HodgeTheory}. Then,  the other half of Maxwell's equations in vacuum are:
\eqa{
  d\star F &= 0 &\iff \left\{\begin{array}{l} \n\cdot \bm{E} = 0 ~, \\ \frac{\partial\bm{E}}{\partial x^{0}} - \n\times\bm{E} = 0 ~. \end{array}\right.
}

There is a natural generalization of Maxwell theory, which 
we will call ``\textbf{Classical Generalized Maxwell theory}'', which has the following equations of motion: 
Consider a manifold (of any dimension $n$)   with nondegenerate metric (of any signature) $(M_n, g_{\mu\nu})$. The physically observable field of the theory is an $\ell$-form,  $F \in \Omega^{\ell}(M_n)$, 
and the on-shell field satisfies: 
\eqa{
dF &= 0 ~,\\
d\star F&= 0 ~. \label{eq:GenMaxwell-EOM}
}
Note that if $\ell=0$, then $dF=0$ implies that $F$ is a constant, whereas 
if $\ell=n$, then $d\star F=0$ implies that $F$ is a constant. So, the most interesting theories arise for $0<\ell< n$. 

\bigskip 
\noindent
\textbf{Electric-Magnetic Duality}: 
Some authors refer to $dF=0$ as the ``Bianchi identity'' and 
$d\star F=0$ as the ``equation of motion.'' But the properties of 
Hodge $\star$ imply that there is a symmetry of the equations. 
If we set $\widetilde{F} = \star F \in \Omega^{n-\ell}(M_{n})$, then $\widetilde{F}$ satisfies the same form of the equations with $\ell \leftrightarrow (n-\ell)$, where ``equation of motion'' and ``Bianchi identity'' have been exchanged. 

\begin{exbox}{Dual Fieldstrength in Maxwell Theory}Returning to $\ell=2$ on Minkowski space $\mathbb{M}^{1,3}$, work out $\widetilde{F}$ in terms of $\bm{E}$ and $\bm{B}$.
\end{exbox}

\begin{exbox}{$\bm{E}$ and $\bm{B}$ fields in Generalized Maxwell Theory}
  When there is a global space-time splitting, define the analogs of electric and magnetic fields for classical generalized Maxwell theory, and write the 
transformation $F\to \wt{F}$ in terms of these fields. 
\end{exbox}

\paragraph{Solutions of the equations of motion.} 
In Minkowski space $\mathbb{M}^{1,n-1}$, a natural ansatz is given by 
\eqa{
   F &= f e^{\imag k \cdot x} ~, 
\label{eq:PlaneWave}}
with $f\in \Lambda^\ell (\IR^n)^\vee$ a constant tensor. 
The equations are linear, so the superposition principle applies, and we can make real combinations of such solutions. Assume $k\not=0$. Then plugging  \eqref{eq:PlaneWave} into $dF=0$
shows that  $k\wedge f =0$. Since we could take $k$ to be 
the first vector in a basis, it follows that $f=k\wedge a$ for some rank $(\ell-1)$ constant tensor $a$. 
\tightfootnote{This provides a proof of the Stokes lemma.}
Then, the equation $d\star F=0$ implies $k\wedge \star(k\wedge a)=0$. Wedging this with $a$ shows that $(k,k)(a,a)=0$. 
Assuming (as we may) that $(a, a)\not=0$ it follows that $k^2=0$. 
 The condition $k^2=0$ implies that in Lorentzian signature, waves propagate at the speed of light.  The space of solutions is an infinite-dimensional linear space. What that space exactly is depends on the boundary conditions at infinity.

It follows from the Hodge decomposition of \autoref{sec:HodgeTheory} 
that the space of solutions in Euclidean space is
the space of harmonic $\ell$-forms: $\CH^{\ell}(M_n)$. Note that this space is isomorphic 
to $\CH^{n-\ell}(M_n)$ by Poincar\'e duality. 
This already shows that there is a close relation between 
Poincar\'e duality and electromagnetic duality.


\begin{remark} \emph{Self-dual theories.}

\begin{enumerate} 

\item 
There is an important generalization of Classical Generalized Maxwell Theory called the theory of the self-dual (SD) or anti-self-dual (ASD) field. 
Suppose $n = 2\ell$. Then
\eqa{
        \star: \Omega^{\ell} &\rightarrow \Omega^{\ell} ~, \quad \star^2 = (-1)^{\ell}\text{sign}(\det\,g_{\mu\nu}) ~.
        \label{eq:StarSquare}
    }
When $\star^2 =  1$,  we can impose (anti-)self-duality equations on real fieldstrengths:
    \eqa{
         F &= \star F \quad (\text{SD field}) \quad \text{ or } \quad F = -\star F \quad (\text{ASD field}) ~.
    }
    It follows from \eqref{eq:StarSquare} that $\star^2 = 1$ for 
    \eqa{
       \begin{array}{lll}
           \text{Euclidean signature}   & :  & \quad n = 0 \text{ mod } 4 ~, \\
            \text{Lorentzian signature} & :  & \quad n = 2 \text{ mod } 4 ~. \\
       \end{array}
    }
   If we also impose $dF = 0$, then the other Maxwell equation comes for free. This is the classical theory of the (anti-)self-dual field.
 
\item As an example of a self-dual theory, in  two-dimensional Minkowski space $\mathbb{M}^{1,1}$ with line element $ds^2 = -(dx^0)^2 + (dx^1)^2$ and volume element $dx^1 \wedge dx^0$, we can apply the above with $n = 2$ and $\ell = 1$:
\eqa{
  F &= \pm \star F &&\implies F = \underbrace{\phi(x^0, x^1)}_{\substack{\text{$\mathbb{R}$-valued}\\\text{function}}} dx^{\pm} ~,\\
  dF &= 0  &&\implies \partial_{\mp}\phi = 0 ~.
}
So the theory of the $(1+1)$d self-dual field is the theory of the chiral self-dual scalar field. Higher-dimensional SD (ASD) fields are generalizations of the theory of the chiral self-dual scalar field.

\end{enumerate} 

\end{remark}

So far we have introduced generalized Abelian gauge theories via their equations of motion. It is natural to ask about an action principle. We address this next. 

\SectionWithHeader{Action Principle For Generalized Maxwell Fields}{Action Principle For Generalized Maxwell Fields}{sec:ActionPrinciple}

Let us return to the non-self-dual field. To write an action principle, we must break manifest electromagnetic duality. We want to use Stokes' lemma, so if we prefer $F \in \Omega^{\ell}(M_{n})$ to $\widetilde{F} \in \Omega^{n-\ell}(M_{n})$, then,
\eqa{
dF &= 0 ~, \label{eq:max1}
}
implies that \underline{locally}, we can write,
\eqa{
F &= dA ~, \quad A \in \Omega^{\ell-1}(\mathcal{U}) ~,
}
in some contractible neighborhood (patch) $\mathcal{U} \subset M_{n}$. The obstruction to writing $F = dA$ globally is the degree $\ell$ de Rham cohomology: 
\eqa{
H^{\ell}_{\mathsf{dR}}(M) &= \mathsf{ker}\big(d: \Omega^{\ell} \to \Omega^{\ell+1}\big)/\mathsf{im}\big(d: \Omega^{\ell-1} \to \Omega^{\ell}\big) ~.
}
For a fixed cohomology class $h \in H^{\ell}_{\mathsf{dR}}(M_n)$, let us choose a representative $F_{0}$ with $h = [F_0]$. Then  every closed form $F$ with $[F]=h$ can be
written as 
\eqa{
F &= F_{0} + dA ~,  \label{eq:fclass}
}
where $A \in \Omega^{\ell-1}(M_{n}) $ is a  globally well-defined $(\ell-1)$-form.
We can now write an action principle for the field $A$. The action depends on a 
choice of  de Rham cohomology class $h$:
\eqa{
S_{h}[A] &= \int \pi \lambda F \wedge \star F ~. \label{eq:actionh}
}
where $\pi \lambda$ is a coupling constant, with the factor of $\pi$ introduced for later convenience in \autoref{sec:PartitionFunctionsGenMax}. 
Stationarity, $\delta S_{h}[A] = 0$, yields the other Maxwell equation: 
\eqa{
d\star F &= 0 ~. \label{eq:max2}
}

Now a theorem due to Hodge states that for every de Rham cohomology class, there is a unique harmonic representative in that class and consequently, we have an isomorphism 
of real vector spaces: 
\eqa{
 \underbrace{H^{\ell}_{\mathsf{dR}}(M_{n})}_{\substack{\text{$\ell^{th}$ de Rham}\\\text{cohomology}}} &\cong \underbrace{\CH^{\ell}(M_{n})}_{\substack{\text{space of} \\ \text{harmonic forms}\\\text{of degree $\ell$}}} ~.
}

One proves this by combining the Hodge decomposition with the existence of a Green's function (a.k.a. Green function \cite{Wright2006}) 
for the Laplacian $\{ d, d^\dagger \}$. Again, see 
\cite[Ch. 6]{Warner1983} for a proof. 

Now, if $F_h$ is the harmonic representative of $h$ then we have the Hodge 
decomposition 
\be 
F = F_0 + d A = F_h + d \alpha ~,
\ee
and Hodge theory guarantees that for any $F_0$,  there exists an $A$ that sets $\alpha=0$.  On the other hand, substitution into the action gives,
\be 
S_{h}[A] = \pi \lambda \int F_h \wedge \star F_h  + \pi \lambda (d\alpha, d\alpha) ~.
\ee
Now $(d\alpha, d\alpha) \geq 0$ and is equal to zero only when $d\alpha=0$ (almost everywhere). Therefore, there exists a unique minimum of the action \eqref{eq:actionh} on a compact Euclidean $M_{n}$, and the minimum is the unique harmonic form in the specified cohomology class $h$.

One benefit of the action \eqref{eq:actionh} is that it gives a coupling to gravity (i.e., a coupling to the background field given by a Riemannian metric). This coupling defines an energy-momentum tensor. In general, the energy-momentum tensor $T_{F}$ is defined by the variation of the action with respect to the background metric: in local coordinates,
\be 
\delta S = \int_{M_n } \vol(g) \half \delta g^{\mu\nu}T_{F,\mu\nu} ~.
\ee
In the present case, one can show that,  
\eqa{ \label{eq:TF-home}
T_F &\in \Gamma\big( \mathsf{Sym}^2 T^*M_{n} \big) ~,
}
is defined by saying that for all $v \in T_{p}M_{n}$, 
\eqa{
    (\pi\lambda)^{-1}T_{F}(v\otimes v)\vert_{p}  = \big(\iota_{v}F, \iota_{v}F\big)_{p} - \frac{1}{2}(v,v)_{p}(F,F)_{p} ~.
\label{eq:GAGT-Tmunu}
}

\begin{exbox}{$T_{F}$ in Generalized Maxwell Theory}
    Derive \eqref{eq:GAGT-Tmunu}.
\end{exbox}

\begin{remark}
$\,$
\begin{enumerate}
    

    \item In the case $\ell = 1$, we can write $F = d\phi$ locally, but the scalar field $\phi$ might not be globally defined. If we take $\phi \in \mathbb{R}/2\pi\mathbb{Z}$, then the factor $\lambda$ has the meaning of the squared radius of the target space circle of $\phi$. Recall that, in general, the form of the nonlinear $\sigma$-model action,
    \eqa{
        \int g_{ij}(X) h^{\mu\nu}\partial_{\mu}X^{i}\partial_{\nu}X^{j}\,\mathsf{vol}(h) ~,
    }
    shows that the \underline{kinetic term} defines a $(\text{length})^2$ on the target space.

    \item More generally, without constraints on the periods of $F$, $\lambda$ is meaningless and can be absorbed via a redefinition of $F$. But if periods of $F$ are quantized, $\lambda$ has physical meaning. 
    For example for $\ell=1$, it is proportional to the squared radius of the target circle. For $\ell=2$, it is proportional to the inverse square of the electric coupling, and so forth.

\item \textbf{Electric-Magnetic Duality:}
Even though the action principle \eqref{eq:actionh} has broken manifest electromagnetic duality, physical quantities such as the energy-momentum tensor are electromagnetically dual.  In ordinary Maxwell theory, the familiar equations for energy and momentum density are  $\bm{E}^2 + \bm{B}^2$ and $\bm{E} \times \bm{B}$ respectively, and are thus clearly electric-magnetic dual. Equation \eqref{eq:EM-Dual-Tmunu} below generalizes that statement to all generalized Maxwell fields. 
 
\item The above remark raises the obvious question of whether 
there is an action which is manifestly electromagnetically dual. This is a subtle and difficult problem. If we write 
an action like 
\be 
\int_{M_{n}} \big(  F \wedge \star F + \wt{F} \wedge \star \wt{F}  \big) ~,
\ee
then we have doubled the number of degrees of freedom, and thereby changed the theory. If we try to introduce Lagrange 
multipliers to enforce the duality relation, we introduce again new degrees of freedom. The issue is closely related 
to writing a Lorentz invariant action for the self-dual field, something we comment on in \autoref{subsec:SelfDualAction} below.

\end{enumerate}

\end{remark}

\begin{exbox}{Electric-Magnetic duality invariance of $T_{F}$}Show that equation 
\eqref{eq:GAGT-Tmunu} is invariant under electric-magnetic duality: 
\be\label{eq:EM-Dual-Tmunu}
 T_{F} = T_{\widetilde{F}} ~ .  
\ee
\end{exbox}

\subsection{An Action Principle For The Self-Dual Field? }\label{subsec:SelfDualAction}

 In the self-dual case, $F = \pm \star F$ and $n = 2 \text{ mod } 4$, so we have $\ell = 1 \text{ mod } 2$ \underline{odd}, hence $\int F \wedge \star F = \int F \wedge F = 0$. Certainly there is no obvious local and Lorentz invariant action for a self-dual field. Many people infer 
 that there is no action principle for the self-dual field. This is not true. In fact, there exist (many) actions for the self-dual field, but in each case, more data needs to be provided. 

 There is a very large literature on attempts to write 
 an action for the self-dual field. For some recent papers 
 addressing this issue, together with extensive references to the many previous works on the subject, see \cite{Sen:2019qit,Lambert:2023qgs}.

One of the many approaches to formulating a Lagrangian for the self-dual field uses the ``holographic'' duality between Chern-Simons theory on a manifold with boundary 
and the chiral edge modes on the boundary  \cite{Belov:2006jd}. There is a close analogy to a theta 
function: The function exists and is independent of duality frame, but to write an explicit formula one must choose a 
suitable Lagrangian decomposition of a cohomology space. 
We sketch the idea of \cite{Belov:2006jd} in the simplest case where there is no nontrivial topology, that is 
we consider the theory  on $\IR^{1,4s+1}$. Similar considerations work in Euclidean signature and on manifolds with nontrivial topology. The argument proceeds along the following lines:

\begin{enumerate}

\item One begins by observing that 
$V:=\Omega^\ell(M_n)$ with $\ell=2s+1$ has a symplectic structure:
\be
\omega(\phi_1,\phi_2) := \int_{M_n} \phi_1 \wedge \phi_2 ~.
\ee

\item First, as in the ordinary non-self-dual theory, to formulate an action, we restrict
attention to fields $R \in V_{{\rm cl}}:=\Omega^\ell_d(M)$. We will vary within this space.
Note that it is a Lagrangian subspace of $V$. We call the fieldstrength $R$ so that
it will not be confused with the classical self-dual field we see in the semiclassical
physics of this theory. We will do the path integral over closed fields $R$ modulo gauge transformations.

\item Next, we \emph{choose} another Lagrangian subspace $V_m\subset V$, assumed to be
maximal Lagrangian and transversal to $V_{{\rm cl}}$ (i.e., $V_{{\rm cl}}\cap V_m= \{0\}$)  and moreover,
\be
\label{eq:lag-spl-i}
V = V_m \oplus \star V_m ~,
\ee
is a decomposition into maximal Lagrangian subspaces.
It is important to demand that $V_m$ and $\star V_m$ are transverse. (This condition can be slightly relaxed.)
This is the extra piece of data one must introduce to obtain 
an action principle. 

\item  Now, given \eqref{eq:lag-spl-i}, there is a unique decomposition of any $R\in V_{{\rm cl}}$ as:
\be
R = R_m + R_e ~,
\ee
with $R_m\in V_m$ and $R_e\in V_e:=\star V_m$.

\item The Lorentzian signature action for the $\epsilon$-self-dual field is then:
\be
\label{eq:sd-action}
S = \pi \int \left( R_e \star R_e + \epsilon R_e R_m \right) ~.
\ee
There are two nice features of this action:

\item First, the action is stationary iff the $\epsilon$-self-dual field $\CF := R_e - \epsilon \star R_e$
is closed:
\be
\label{eq:stat-eq-sd}
d \CF = 0 ~.
\ee
\emph{Thus, the set of stationary points of the action is the set of solutions of the
self-dual equations of motion for $\CF$. }

\item The proof that the variation of the action gives \eqref{eq:stat-eq-sd} goes as follows:
\begin{align}
\delta S & = \pi \int  2 \delta R_e \star R_e + \epsilon \delta R_e R_m + \epsilon R_e \delta R_m \nn
& = \pi \int  2 \delta R_e \star R_e + 2\epsilon \delta R_e R_m   \nn
& = 2\pi \int \delta R (\star R_e + \epsilon R_m) \nn
& = 2\pi \int d(\delta c) (\star R_e + \epsilon R_m) ~,
\end{align}
where in the second line, we used the fact that both $R$ and $\delta R$ are in $V_{{\rm cl}}$, which is
Lagrangian. In the third line, we notice that $\star R_e + \epsilon R_m\in V_m$, and hence, we can
replace $\delta R_e$ by $\delta R$. In the fourth line, we use the fact that variations of $R$
in $V_{{\rm cl}}$ are exact. Now integration by parts gives $d( \star R_e + \epsilon R_m)=0$. Finally,
$dR = dR_e + dR_m=0$, so $d( \star R_e + \epsilon R_m)=0$ is equivalent to $d( \star R_e - \epsilon R_e)=0$, 
which is equivalent to $d(R_e - \epsilon \star R_e)=0$.

\end{enumerate}

A nice   feature of the above action is that if we consider the action as a functional of both the
metric and the field $R$ then, varying the metric holding $R$ fixed, the action varies into
\be
\delta S = \frac{\pi}{2} \int \vol(g) \delta g^{\mu\nu} T(\CF)_{\mu\nu} ~,
\ee
where $T(\CF)_{\mu\nu}$ is the standard energy-momentum tensor for the $\epsilon$-self-dual
field $\CF := R_e + \epsilon \star R_e$.

  A common choice of $V_m$ is given by choosing some timelike direction $\xi$, 
and letting $V_m$ be the forms anniliated by $\iota(\xi)$.  It is commonly
said that there is no Lorentz invariant action for the self-dual field, but this
is not really true, as the following example shows. Let us consider the self-dual
scalar on $\IR^{1,1}$ so $\star dx^\pm = \pm dx^\pm$. We can choose $V_m$ to be of the
form,
\begin{equation}
V_m = \{ R = f(x) dx^+ + (G\cdot f)(x) dx^- \} ~,
\end{equation}
where $f(x)$ is any suitably normalizable function and  $(G\cdot f)(x) := \int G(x,y) f(y) d^2 y$.
Then $V_m$ will be Lagrangian if $G(x,y) = G(y,x)$ is symmetric, and it will be Lorentz
invariant if $G(\phi_\lambda(x),y) = \lambda^{-2} G(x, \phi_\lambda^{-1}(y)) $,
where $\phi_\lambda(x^+,x^-) = (\lambda x^+, \lambda^{-1} x^-)$. An example of
such a kernel function would be $G(x,y) = (x^+ - y^+)^{-2}$. This leads to the
action for a self-dual scalar field,
\begin{equation}\label{eq:LorentzInvNonlocal}
S = \pi \int \p_+\phi \, \p_- \phi\, d^2 x - \pi \int d^2 x\, d^2 y \,\p_-\phi(x)\, G^{-1}(x,y) \, \p_- \phi(y) ~.
\end{equation}

While equation \eqref{eq:LorentzInvNonlocal} is Lorentz invariant, it is nonlocal.   It is possible that with a suitable definition of ``local,'' there is
no local and Lorentz invariant Lagrangian for the self-dual field. A first step to proving such a no-go theorem would be showing that there is no ``local'' and Lorentz invariant choice for $V_m$.

\begin{remark}
$\,$
\begin{enumerate} 

\item  Standard folk wisdom states that there is no description of a \underline{non-Abelian} analog in terms of elementary fields and a local action. (There are local fields in the non-Abelian analog, i.e., the energy-momentum tensor.) There is, however, a hypothetical quantum theory of a non-Abelian self-dual field. Current folklore asserts that it is impossible to write fundamental fields that describe this theory and there is no generally accepted description in terms of an action principle.
 About twenty years ago, people said similar things about the non-Abelian M2-brane theory, but then along came the ABJM description \cite{Aharony:2008ug}. One moral from \cite{Aharony:2008ug} is that folklore can be overturned. It would be good to state and prove a rigorous no-go theorem about classical non-Abelian self-dual actions. The hypotheses of no-go theorems merely define the limits of our imagination. 

\item In unpublished work, I. Nidaiev \cite{Nidaiev:PathIntegralUnpublished,Nidaiev:2019mzt}
began with the formulation of \cite{Belov:2006jd} for the self-dual field and refined it into a formulation very close to that of the work of A. Sen \cite{Sen:2019qit}. For 
further developments, see \cite{Hull:2023dgp,Hull:2025bqo,Hull:2025rxy}.

\end{enumerate}

\end{remark}

\SectionWithHeader{Many Maxwell Fields}{Many Maxwell Fields}{sec:ManyMaxwell}

\subsection{Kaluza-Klein Reduction Of The Generalized Maxwell Field }

One way to motivate the consideration of theories with several Maxwell fields is to consider the reduction of the theory of a single Maxwell field 
$F\in \Omega^\ell(M_n \times K)$, where $K$ is a compact manifold and 
we assume there is a direct sum metric on $M_n \times K$.

We will show that if we consider a family of background fields where the 
metric of $K$ is scaled to zero with a real parameter $t^2$, and we take the limit $t\to 0$, then 
the correlation functions of the generalized Maxwell theory coincides -- up 
to exponential accuracy in $t$ -- with 
those of a generalized Maxwell theory in $n$ dimensions with fields,
\be\label{eq:KK-Redux1}
F \in \bigoplus_{p+q = \ell }   \Omega^p(M_n)\otimes \CH^q(K) ~,
\ee
where $\CH^q(K)$ is the real vector space of harmonic $q$-forms on $K$. 

In order to justify the claim above, we introduce an ON eigenbasis of the Laplacian on $\Omega^q(K)$, denoted $\phi_s^{q}\in \Omega^q(K)$, where $s$ labels the eigenmodes. It is convenient to change the index $s$ to reflect the Hodge decomposition on $K$, so we separate the eigenmodes into:
\be 
\phi_h^{q} ~, \quad h=1, \dots, b_q(K) ~, \qquad\qquad  \phi_e^{q}~, \quad e=1\dots, \infty ~,  \qquad\qquad 
\phi_c^{q}~, \quad  c=1\dots, \infty ~,
\ee
corresponding to the harmonic, exact, and co-exact eigenmodes, respectively. 
Then the general section $F\in \Omega^\ell(M_n \times K)$ can be written as: 
\be\label{eq:Fourier}
\begin{split}
F &  = \sum_{p=0}^\ell \sum_{s} F_s^p   \phi_s^{\ell-p}   \\
& =  \sum_{p=0}^\ell \sum_{h=1}^{b_{q}(K)} F_h^p \phi_h^{q} + \sum_{p=0}^{\ell-1}\sum_{e=1}^\infty  F_e^p \phi_e^{q}
 + \sum_{p=1}^{\ell} \sum_{c=1}^\infty F_c^p \phi_c^{q} ~.
\end{split}
\ee
Here $F_s^p$ are pulled back from $M_n$, so we can regard them as 
$F_s^p \in \Omega^{p}(M_n)$. We stress that $F_h^p, F_e^p, F_c^p$ are 
general sections and the index $h,e,c$ indicates the nature of the form 
(harmonic, exact, or coexact) it multiplies in the Fourier decomposition \eqref{eq:Fourier}. 

Now, imposing that $F$ is exact shows that  $dF_h^p=0$ and 
\be
F_c^p = - \frac{1}{\sqrt{\mu_e^{(\ell-p+1)}}} d F_e^{p-1} ~,
\ee
where we are using the isomorphisms \eqref{eq:HodgeSusyIsoms} to 
establish a 1--1 correspondence between the indices $c$ and $e$.

 %
%

Substitution into the action \eqref{eq:actionh} gives an action, 
\be 
\begin{split}
S &  = \pi \lambda \int_{M_n} \sum_{p=0}^{\ell} \int F_h^p \star F_h^p  + \pi \lambda  \sum_{p=0}^{\ell-1} \sum_{e=1}^\infty 
\int_{M_n}  \left( \frac{1}{\mu_e^q} dF_e^p \star dF_e^p + F_e^p \star F_e^p \right) ~,
\end{split}
\ee
subject to the constraint that $dF_h^p=0$. We stress again that 
the fields $F_e^p$ are general sections of $\Omega^p(M_n)$. The path 
integral for $F_e^p$ computes a determinant of a Laplacian of 
\be 
d^\dagger d + \mu_e^q ~,
\ee
on $\Omega^p_{L^2}(M_n)$ and will be $1$, up to exponentially small 
terms in the scale $t$ of the metric on $K$. Similar arguments apply other observables. This justifies the Kaluza-Klein reduction to a Maxwell field 
valued in \eqref{eq:KK-Redux1} and satisfying the equations of motion 
$dF=0$ and $d\star F=0$.

\subsection{Coupling Constants} 

The previous section has taught us that it is natural to broaden the 
definition of a generalized Maxwell field to a field, 
\be\label{eq:SingleV}
F \in  \Omega^{\bullet}(M_n; V)  ~,
\ee
where $V$ is a graded vector space, and one can impose a condition on the total grading.  One could also generalize further to the case where $V$ is a flat bundle. For simplicity, in the following discussion, we will simply take $V$ to have degree zero and,
\be 
F \in  \Omega^\ell(M_n; V)  ~,
\ee
where the fixed vector space $V$ has grading zero. 

In order to write an energy momentum tensor, we should assume that $V$ is endowed with a positive definite metric. This also allows us to write a Lagrangian. By choosing a basis $e_I$ for $V$ we can express the theory as a theory of several 
generalized Maxwell fields $F_I$, where $F= F^I e_I$. The Lagrangian is: 
\be 
S = \pi  \int_{M_n} \lambda_{IJ} F^I \star F^J ~.
\ee

\begin{remark}
$\,$
\begin{enumerate} 

\item 
The $\lambda_{IJ}$ is a positive definite real symmetric metric on $V$. If we make no assumptions about the periods of the $F^I$, it can be diagonalized and scaled to the unit matrix. On the other hand, if, as in string theory, one assumes some quantization conditions on the periods, then $\lambda_{IJ}$ becomes meaningful. For example, in the Kaluza-Klein example above, it is more natural 
to replace the ON basis $\phi_h^q$ (which depends on a metric) to a metric-independent basis of harmonic forms with integral periods. 

\item When there are several Maxwell fields an additional coupling constant 
can be added so that the action reads: 
\be 
S = \pi  \int_{M_n} \left(  \lambda_{IJ} F^I \star F^J + \theta_{IJ} F^I F^J \right) ~,
\ee
where $\theta_{IJ}$ is an anti-symmetric form on $V$ when $\ell$ is odd and a
symmetric form when $\ell$ is even. These terms do not affect the equations of motion but have profound effects on the quantum system. 

\item It is natural to generalize $\lambda_{IJ}$ and $\theta_{IJ}$ to be functions on $M_n$. When the generalized Maxwell field is embedded in a larger theory, these fields often have their own kinetic terms. The $\theta_{IJ}$ are known as axion fields. 

\end{enumerate}
\end{remark}

\subsection{Manifestly Electromagnetically Dual Equations Of Motion}

On application of the generalization of Maxwell theory to fields of the 
form \eqref{eq:SingleV} is that it gives a nice framework for writing 
equations of motion that are manifestly electric-magnetic duality-invariant. 
For simplicity, we consider the case of $n=0 \mathsf{~mod~} 4 $ dimensions with Lorentzian signature with $\ell = n/2$. (This includes the original case of Maxwell theory.)  We are following the discussion in sections 2.5.3 -- 2.5.4 of \cite{Moore:2010PiTP}. 

We will introduce a field, 
\be 
\IF \in \Omega^{\ell}(M_n) \otimes V ~,
\ee
where $V$ is a real vector space. In order to be able to write an energy-momentum tensor, $V$ will need to be endowed with a positive definite symmetric metric.

Our field will be closed, $d \IF =0$ and will satisfy a self-duality condition of the form, 
\be 
\IF =( \star \otimes J)  F ~,
\ee
where 
\be 
J : V \to V ~,
\ee
is a linear transformation. Under the conditions specified above 
with $\ell = n/2$ and even integer and Lorentzian signature consistency 
requires $J^2=-1$. In other words, the real vector space $V$ admits a 
complex structure. It follows that $V$ must have even dimension:
\be 
\dim_{\IR} V = 2r ~,
\ee
for some $r \in \IZ_{\geq 0}$. We claim that the equations: 
\be\label{eq:DualityInvariantEqs}
\begin{split} 
d\IF & = 0 ~,\\ 
\IF & = (\star \otimes J ) \IF ~,
\end{split}
\ee
describe a manifestly electric-magnetically dual formulation of a collection of $s$ independent generalized Maxwell fields of degree $\ell$.

For compatibility of the energy-momentum tensor and the 
self-duality condition, the metric should be compatible with the metric: 
\be 
(J(v), J(v') ) = (v,v')  \qquad \forall \,\, v,v'\in V ~,
\ee
where $(\cdot, \cdot )$ denotes the inner product on $V$. If follows 
that $V$ has a symplectic structure $\omega$ defined by 
\tightfootnote{In general when a real vector space has a compatible  metric, complex structure and symplectic structure, any two structures will imply the third. }
\be 
\omega(v, v') :=  (v, J(v')) ~.
\ee
To lighten the notation, we write $\langle v, v' \rangle := \omega(v, v')$.  
We can therefore introduce a symplectic basis (aka a Darboux basis)  
denoted $\{ \alpha_I, \beta^I\}$ where $I=1,\dots, r$. Our
convention is that
\begin{equation}\label{eq:DarbouxBasis}
\begin{split}
\langle \alpha_I, \alpha_J \rangle & = 0 ~,\\
\langle \beta^I  , \beta^J \rangle & = 0 ~, \\
\langle \alpha_I, \beta^J \rangle & = \delta_{I}^{~J} ~.
\end{split}
\end{equation}
The  linear span of $\alpha_I$ is a maximal Lagrangian
subspace $L_1\subset V $, while that for $\beta^I$ is another $L_2 \subset V$, and we
have a Lagrangian decomposition:
\begin{equation}
V \cong  L_1 \oplus L_2 ~.
\end{equation}
Upon complexification, the $\IC$-linear extension of $J$ has 
eigenvalues $\pm \imag$. We let    
$V^{0,1}$ denote the $-\imag$ eigenspace and  $V^{1,0}$ the   $+\imag$ 
eigenspace, so: 
\be 
V\otimes_{\IR}\IC \cong V^{0,1}
\oplus V^{1,0} ~.
\ee

Given a Darboux basis, we can define a basis for $V^{0,1}$ of the form, 
\begin{equation}\label{eq:fI-def}
f_I := \alpha_I + \tau_{IJ} \beta^J ~, \qquad I = 1, \dots, r ~.
\end{equation}
while $V^{1,0}$ is then spanned by,
\begin{equation}\label{eq:fIbar-def}
\ov{f}_I := \alpha_I + \ov{\tau}_{IJ} \beta^J ~, \qquad I = 1, \dots, r ~.
\end{equation}
Compatibility if $J$ with the symplectic structure now implies
$\langle f_I, f_J \rangle =0$ and hence $\tau_{IJ} = \tau_{JI}$.
It is useful to split $\tau_{IJ}$ into its real and imaginary parts:
\begin{equation}
\tau_{IJ} = X_{IJ} + \imag\,Y_{IJ} ~.
\end{equation}
Positive definiteness of $(\cdot, \cdot) $
implies $Y_{IJ}$ is positive definite. It will be convenient to
denote the matrix elements of the inverse by $Y^{IJ}$, so,
\begin{equation} \label{eq:Inv-Of-Y}
Y^{IJ}Y_{JK} = \delta^{I}_{~K} ~.
\end{equation}

\begin{exbox}{Darboux basis and symplectic structure}
\begin{enumerate}
\item Find the change of basis between $f_I, \ov{f}_I$ and $\alpha_I, \beta^I$ in terms of the matrices $X,Y$. 

\item Compute the action of $J$ in the basis $\alpha_I, \beta_I$ in terms of the matrices $X,Y$. 
\end{enumerate}
\end{exbox}

Now, given a Darboux basis we can expand: 
\be 
\IF   = \alpha_I F^I - \beta^I G_I ~.
\ee
Then, the self-dual equation allows one to solve for $G_I$ in terms of $F^I, \star F^I$, and the matrices $X,Y$: 
\be 
G_J = - Y_{JK} \star F^K - X_{JK} F^K ~.
\ee
Then, the equations of motion $dF^I =0 $ and $dG_I =0$ can be derived from a 
Lagrangian: 
\be 
\int_{M_n} Y_{JK} F^J \star F^K + X_{JK} F^J F^K ~.
\ee

\begin{remark}
$\,$
\begin{enumerate}

\item Of course, one could change the sign of the duality condition 
in \eqref{eq:DualityInvariantEqs} to an anti-self-duality condition. 
The change in the Lagrangian is a change in the axion coupling. 

\item The exchange of duality basis $\alpha_I, \beta^J \mapsto 
\beta^J, - \alpha_I $ corresponds to the electromagnetic duality we have 
thus far described. In fact, we see that in this theory, there is a natural duality group $\mathsf{Sp}(V)$, which is the group of symplectic linear transformations of $V$ and 
the equations \eqref{eq:DualityInvariantEqs} are manifestly $\mathsf{Sp}(V)$-invariant
because they do not refer to a specific Lagrangian decomposition. 
If one makes a \underline{choice} of Lagrangian decomposition of $V$, then it is possible to write an action principle for the equations. This is compatible with the considerations of \autoref{subsec:SelfDualAction}. 

\item If one imposes a quantization condition on the periods of $F^I$, then 
in general, only a discrete subgroup of $\mathsf{Sp}(V)$ will be compatible with that 
quantization condition. This typically happens in string theory, where the theory of branes implies quantization conditions. 

\item If we specialize to a single self-dual $3$-form $H\in \Omega^3(M_4 \times \Sigma)$, where $\Sigma$ is a closed Riemann surface, then Kaluza-Klein reduction 
gives a family of Maxwell fields $F^I$, $I= 1, \dots, b_1(\Sigma)$. 
The space of harmonic one forms $\CH^1(\Sigma)$ carries a natural geometric 
symplectic structure: The intersection form. The matrix $\tau_{IJ}$ relative to a Darboux basis has the geometric interpretation of a period matrix for the 
Riemann surface $\Sigma$. If we require that the periods of $F^I$ are integral, then the duality group becomes $\mathsf{Sp}(2b_1, \IZ)$, and thus we have 
\underline{geometrized} the duality transformations. This geometric interpretation of duality symmetry appears naturally in string theory, in F-theory, and has been used in precisely this context in \cite{Witten:1995gf,Verlinde:1995mz}.
    
\end{enumerate}
 
\end{remark}

\SectionWithHeader{Electric And Magnetic Currents}{Electric And Magnetic Currents}{sec:Currents}

Recall that in vacuum, we had
\eqa{
   dF &=0 ~, \quad \emph{ and }  \quad d\star F = 0 ~.
}
Now, let us introduce magnetic and electric currents as external sources in these equations. The currents are:
\eqa{
   \begin{array}{lll}
       J_{m} &\in \Omega^{\ell+1}(M_{n}) & \text{ magnetic} ~,\\
       J_{e} &\in \Omega^{n-\ell+1}(M_{n}) & \text{ electric} ~.
   \end{array}
}
The response of fields to background currents is given by Maxwell's equations with sources
\eqa{
   d\star F &= J_{e} ~,\\
   dF &= J_{m} ~. \label{eq:TrivializeCurrents}
}

\begin{remark}
$\,$
\begin{enumerate}
    \item A nonzero magnetic current obstructs the existence of a gauge potential: $J_{m} \neq 0$ implies $F \neq dA$, even locally, at least not where the support of $J_{m}$ is. So, if one does not excise the region of support of $J_m$, it is not possible to regard Maxwell theory as a theory  of a connection on a $\mathsf{U(1)}$ line bundle. One will have to start talking about \emph{gerbes} and \emph{gerbe connections}, best described in the context of differential cohomology. See \autoref{subsubsec:GerbeConnections} below.  

    \item Current conservation automatically follows from these equations, as the currents are closed:
    \eqa{
         dJ_{m} &= 0 ~, \quad dJ_{e} = 0 ~.
    }
    \item If $F$ is smooth, then equations \eqref{eq:TrivializeCurrents} say that  $J_{m}$ and $J_{e}$ are cohomologicaly trivial, i.e., $[J_{m}] = 0$ and $[J_{e}] = 0$. This may seem confusing: don't the cohomology classes have something to do with the charge? The answer is that they do, but the charge is measured in a \underline{relative} cohomology group, see Box \ref{box:relcoh}.
\end{enumerate}

\end{remark}

\begin{redbox}[box:relcoh]{Relative Cohomology}
\begin{center}
\begin{tikzpicture}[>=Stealth,baseline=-0.5ex]
\node (graph) [inner sep=0pt]{\includegraphics[width=2in]{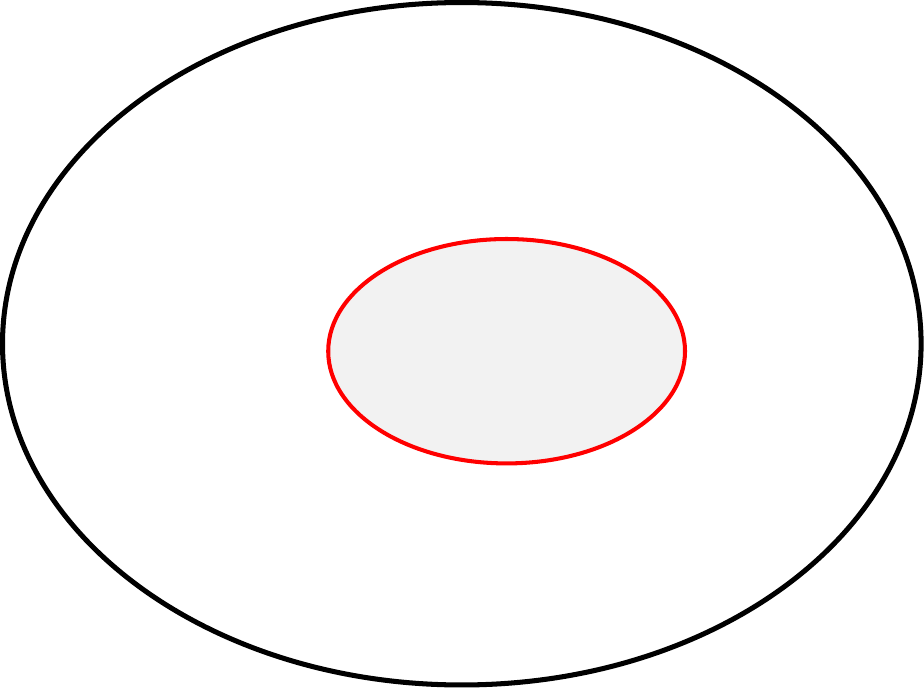}};
\node at (0.5,0) {$\bm{S}$};
\node at (-1.5,1) {$\bm{X}$};
\end{tikzpicture}
\captionof{figure}{A topological space $X$ and a subspace $S \subset X$.}
\label{fig:relativecohomology}
\end{center}
(See also Bott-Tu \cite{Bott1982}.) Fix a positive integer $k \in \IZ_{\geq 1}$. Given an inclusion $\iota: S \hookrightarrow X$ (see \autoref{fig:relativecohomology}), the \emph{relative cochain complex} is
\eqa{
    C^{k}_{\mathsf{dR}}(X, S) &= \Omega^{k}(X) \oplus \Omega^{k-1}(S) ~, \label{eq:relcomplex}
}
with differential
\eqa{\label{eq:relcoh-differential}
   d(\omega, \theta) &:= (d\omega, \iota^*\omega -d\theta) ~.
}
 Define the \emph{$k^{th}$ relative cohomology group of the pair $(X,S)$}:
 \eqa{
      H^{k}_{\mathsf{dR}}(X, S) &:= \mathsf{ker\,}d/\mathsf{im\,}d ~.
 }
 (This is not to be confused with cohomology with a nontrivial coefficient group.) 
 Now, the short exact sequence
\begin{equation}
     \begin{tikzcd}[column sep=small,row sep=0]
         0 \arrow[r] & \Omega^{k-1}(S) \arrow[r] & C^{k}(X,S) \arrow[r] & \Omega^{k}(X) \arrow[r] & 0 \\
         & \theta \arrow[r,mapsto] & (0, \theta) \\
         & & (\omega, \theta) \arrow[r,mapsto] & \omega 
     \end{tikzcd}
 \end{equation}
 induces a long exact sequence in cohomology
\begin{equation}
    \begin{tikzcd}
        \cdots \arrow[r] & H^{k-1}(S) \arrow[r] & H^{k}(X,S) \arrow[r] & H^{k}(X) \arrow[r, "{\text{\tiny{restriction}}}"] & H^{k}(S) \arrow[r] & \cdots 
    \end{tikzcd}
\end{equation}
\end{redbox}

\bigskip
\begin{exbox}{Relative Cohomology}
For the differential \eqref{eq:relcoh-differential}:
 \begin{itemize}\itemsep 0pt
     \item[(a)] Check that $d^2 = 0$.
     \item[(b)] Show that ``closed'' means $d\omega = 0$ and $\left.\omega\right|_{S}$ is trivialized.
 \end{itemize}
\end{exbox}
\bigskip

Let us now apply the above ideas about relative cohomology to the question of how the charges are related to cohomology classes of currents. 
The following discussion comes from \cite{Moore:1999gb}.
\tightfootnote{In this section, we are working with de Rham cohomology, i.e., with coefficients in $\IR$. However, these results hold more generally for integral cohomology (i.e., coefficients in $\IZ$), and there is an analogous version in $K$-Theory, as discussed in \cite{Moore:1999gb}.}
Note that $(J_e, \star F)$ is exact, but $[(J_e, 0)]$ could be a nontrivial class in $H^{n-\ell+1}_{\mathsf{dR}}(M_{n}, M_{n}^{-J_{e}})$, where $M_{n}^{-J_{e}} := M_{n} - \mathsf{Supp}(J_{e})$.
There is a long exact sequence in cohomology,
 \begin{equation}\label{eq:les-cohomology}
    \begin{tikzcd}[column sep=small,row sep=0]
        \cdots \arrow[r] & H_{\mathsf{dR}}^{n-\ell}(M_{n}) \arrow[r, "\iota^*"] & H_{\mathsf{dR}}^{n-\ell}(M_{n}^{-J_{e}}) \arrow[r, "\delta"] & H_{\mathsf{dR}}^{n-\ell+1}(M_{n}, M_{n}^{-J_e}) \\
        \arrow[r, "\psi"] & H_{\mathsf{dR}}^{n-\ell+1}(M_{n}) \arrow[r] & \cdots
    \end{tikzcd}
\end{equation}
Therefore,
    \eqa{
       \mathsf{ker\,}\psi &\cong \mathsf{im\,}\delta \cong H_{\mathsf{dR}}^{n-\ell} (M_{n}^{-J_e})/\iota^* H_{\mathsf{dR}}^{n-\ell}(M_{n}) ~.
    }
    The electric charge group is by definition the kernel of $\psi$, and by the long exact sequence,
    \eqa{
     Q_{e} &\cong H_{\mathsf{dR}}^{n-\ell}(M_{n}^{-J_{e}})/\iota^{*}H_{\mathsf{dR}}^{n-\ell}(M_{n}) ~.
    }
    Here $H^{n-\ell}_{\mathsf{dR}}(M_{n})$ is the classical flux group: these are the fluxes not sourced by charge.
  This is the mathematical expression of the idea that the charge is measured by the ``flux at $\infty$''. If $M_{n} = \mathbb{R}_{t}\times N_{n-1}$ and the support of $J_{e}$ is compact for all time, we can identify $Q_{e}$ as the kernel of the map,
\eqa{
    H_{\mathsf{cpct},\mathsf{dR}}^{n-\ell+1}(N_{n-1}) \to H_{\mathsf{dR}}^{n-\ell+1}(N_{n-1}) ~,
}
where the subscript `$\mathsf{cpct}$' denotes compactly supported cohomology.
\begin{redbox}[box:relcoh-compactcoh]{Relative cohomology and compactly supported cohomology}
The relative cohomology group $H_{\mathsf{dR}}^{n-\ell+1}(M_{n}, M_{n}^{-J_{e}})$ is isomorphic to the compactly supported cohomology group $H_{\mathsf{cpct},\mathsf{dR}}^{n-\ell+1}(M_n)$.
\end{redbox}

\SectionWithHeader{Branes}{Branes}{sec:Branes}
As before, $n \in \IZ_{\geq 1}$ denotes the spacetime dimension, and fix $p \in \IZ_{\geq 0}$ such that $p \leq n-1$. $p$-branes are \underline{extended objects} generalizing point particles with worldvolumes:
\eqa{
  \mathcal{W} &= \mathcal{S}_{p} \times \mathbb{R}_{t} \subset N_{n-1} \times \mathbb{R}_{t} = M_{n} ~.
}
In a generalized Maxwell theory, electrical branes can be viewed as objects that produce an electric current:
\eqa{
J_{e} &= q_{e}\!\!\!\!\!\!\underbrace{\eta(\mathcal{W}_{e} \hookrightarrow M_{n})}_{\substack{ \text{$\delta$-function representative}\\ \text{$n-(p_{e}+1)$-form} \\ \text{Poincar\'{e} dual to $\mathcal{W}_{e}$} }}     ~, \label{eq:elec-brane-Je}
}
Then $d\star F = J_{e}$ implies electrically charged branes have 
$p_{e} = \ell - 2$. 

\begin{figure}[h]
\centering
\begin{tikzpicture}[>=Stealth,baseline=-0.5ex]
\node (graph) [inner sep=0pt]{\includegraphics[width=1.5in]{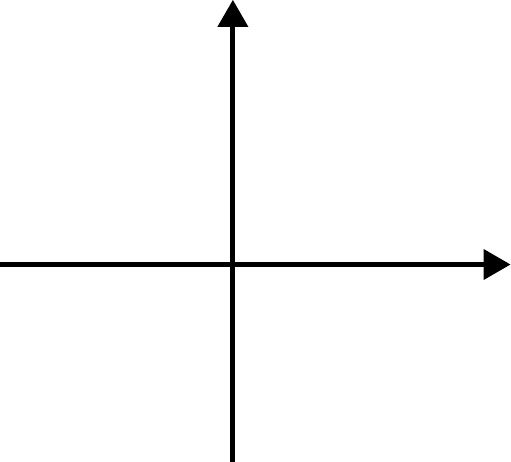}};
\node at (3.0,-0.3) {$H_{e}^{\perp} \cong \IR^{D_e}$};
\node at (4.7,-1.0) {$D_{e} = (n-1)-p_{e} = n-\ell+1$};
\node at (0.8,1.5) {$H_{e} \cong \IR^{p_{e}}$};
\end{tikzpicture}
\caption{An flat electric $p$-brane as a hyperplane.}
\label{fig:electric-p-brane}
\end{figure}
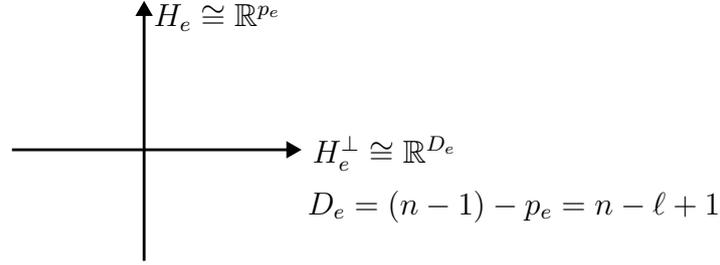

A crucial solution exists when $\mathcal{S}_{p_e} = H_{e}$ is a hyperplane in spatial $\mathbb{R}^{n-1}$, see \autoref{fig:electric-p-brane}.

The field due to this configuration is given by,
\eqa{
  F &= \frac{q_e}{V_e} \frac{dt \wedge dr}{r^{D_e-1}} \wedge \mathsf{vol}(H_e) ~.
}
From \eqref{eq:les-cohomology},
\begin{align}
\begin{split}
    H^{0}(\mathcal{W}_e) &\xlongrightarrow{\iota^*} H^{n-\ell+1}(M_{n}, M_{n} - \mathcal{W}_{e}) \\
    \mathbb{R} \ni q_{e} &\longmapsto [(J_e, 0)] ~.
\end{split}
\end{align}
Similarly, \underline{magnetic branes} would have a worldvolume:
\eqa{
   \mathcal{W}_{m} &= \mathcal{S}_{m} \times \IR_{t} \subset \mathcal{N}_{n-1} \times \IR_{t} = M_{n} ~,
}
and produce a magnetic current:
\eqa{
  J_{m} &= q_{m}\!\!\!\!\!\underbrace{\eta(\mathcal{W}_{m} \hookrightarrow M_{n})}_{\substack{ \text{$\delta$-function representative}\\ \text{$(\ell+1)$-form} \\ \text{Poincar\'{e} dual to $\mathcal{W}_{m}$} }}     ~,
}
Then $dF = J_{m}$ implies $p_{m} = n-\ell-2$.

In $\mathbb{M}^{1,n-1}$,
\eqa{
  F &= 2\pi q_{m}\omega_{\perp} ~,
}
where $\omega_{\perp}$ is the unit volume form for $\IS^{\ell} \subset H_{m}^{\perp} \cong \mathbb{R}^{\ell+1}$.

\SectionWithHeader{Quantum Mechanics Of A Test Brane In An External Classical Generalized Maxwell Field}{QM Of Test Brane In External Classical Generalized Maxwell Field}{sec:ActionTestBranes}

Let us recall a key insight from the theory of branes \cite{Polchinski:1995mt,Callan:1991ky,Callan:1991dj,Callan:1991at,Johnson2002}. Branes are not just solitonic solutions in supergravity, but are dynamical objects which must be included in the description of the space of states in string theory. Unlike \underline{defects}, branes can wiggle and move. These are among their internal degrees of freedom. So -- similar to the AdS/CFT correspondence \cite{Maldacena:1997re,Witten:1998qj,Aharony:1999ti} -- we can change our point of view and consider the worldvolume theory of a test brane moving in a \underline{prescribed} field configuration $F$.

Let's start with a charged particle moving in the field of an  $\ell=2$ Maxwell theory in a Lorentzian signature spacetime $M_{n}$. The gauge field is a nondynamical background field. In this context, one sometimes refers to the particle as a ``test'' particle, because the backreaction on the electromagnetic field on $M_n$ is ignored.  For an uncharged 
but massive particle with worldline $\CW_1 \subset M_n$,  the action is given 
by,
\eqa{
   S =  \int_{\mathcal{W}_{1}}T\,ds  ~, \quad ds = \sqrt{ -\left(\frac{dx}{d\tau}\right)^2} d\tau ~.  \label{eq:ParticleAction} 
}
Here $\tau$ is the time coordinate along a worldline, and $T$ is the mass of the test particle. Choosing boundary conditions so that the boundary variation vanishes, 
the equation of motion is the geodesic equation,
\tightfootnote{Note that 
%
%
\eqref{eq:geodesic} is equivalent to 
\begin{align*}
    \frac{d}{d\tau}\left( \frac{T g_{\mu\nu}}{\sqrt{-\left(\frac{dx}{d\tau}\right)^2}} \frac{dx^\nu}{d\tau}  \right) &= 0 ~,
\end{align*}
which, after using the chain rule and standard expressions for the Christoffel symbols, reduces to the conventional form of the geodesic equation
\begin{align*}
    \frac{d^2 x^\mu}{d\tau^2} + \Gamma^{\mu}{}_{\alpha\beta}\frac{dx^\alpha}{d\tau}\frac{dx^\beta}{d\tau} = 0 ~,
\end{align*}
for constant $T$.
}
\eqa{ \label{eq:geodesic}
  \frac{d}{d\tau}\left(T \frac{dx_\mu}{ds} \right) &= 0 ~.
}

The action \eqref{eq:ParticleAction} has a natural generalization to arbitrary branes known as the \emph{Dirac-Born-Infeld} (DBI) action.
\tightfootnote{The DBI action was originally introduced as a formulation of nonlinear electrodynamics intended to cure the divergence of electron self-energy. See \cite{nlab:dirac-born-infeld_action} for more information.}
It is simply the induced volume 
of the worldvolume times $T$, where $T$ -- known as the \emph{tension} -- has units of energy divided by spatial worldvolume. So in units where $\hbar = c =1$, it has units of energy to the $(p+1)^{th}$ power, so the action is dimensionless, as it must be.

Now let us return to the particle in $\ell=2$ Maxwell theory, and let us suppose that 
the test particle also has electric  charge. If the   particle has charge $q_{e}$, the equations 
of motion must be modified to account for the Lorentz force: 
\eqa{
   \frac{d}{d\tau}\left( T \frac{dx_\mu}{ds} \right) &= q_{e}F^{\mathsf{phys}}_{\mu\nu}\frac{dx^\nu}{d\tau} ~. \label{eq:testeom}
}
When   $F^{\mathsf{phys}} = dA^{\mathsf{phys}}$, where $A^{\mathsf{phys}} \in \Omega^{1}(M_n)$ is a globally defined $1$-form, then these equations can again be derived from an action where the DBI action is modified to: 
\eqa{
   S =  \int_{\mathcal{W}_{1}}T\,ds   + \int_{\CW_1} q_e A^{\mathsf{phys}} ~.  \label{eq:ParticleAction-With-A} 
}

Let us now consider the \underline{quantum theory} of a charged particle moving in 
a classical background electromagnetic field. The path integral over the possible trajectories of the particle  will be weighted by:  
\eqa{
 \exp\left( \frac{\imag }{\hbar}\int_{\mathcal{W}_{1}}T\,ds\right ) \cdot \underbrace{\chi(\mathcal{W}_1)}_{\substack{ \text{some $\mathsf{U(1)}$ factor} \\ \text{that depends on} \\ \text{the worldline} }} ~.
 \label{eq:GenChPrtAct}
}
This $\mathsf{U(1)}$ factor is a map $\chi: \CW_1 \to \mathsf{U(1)}$. If $F^{\mathsf{phys}} = dA^{\mathsf{phys}}$, where $A^{\mathsf{phys}} \in \Omega^{1}(M_n)$ is a globally defined $1$-form, then,
\eqa{
\chi(\mathcal{W}_1) &= \exp \left( \frac{\imag }{\hbar}\int_{\mathcal{W}_1}q_{e}A^{\mathsf{phys}} \right) ~,
}
produces the equation of motion \eqref{eq:testeom}. Now, we note two key properties which will be defining properties of the map,
\eqa{
 \chi: \left\{ \text{worldlines} \right\} &\longrightarrow \mathsf{U(1)} ~.
}
\begin{itemize}
    \item[\textbf{(A)}] Suppose we have two particles. Their worldlines are $\mathcal{W}_1$ and $\mathcal{W}_1'$ so the combined worldline is $\mathcal{W}_1 \cup \mathcal{W}_1'$. Their actions should add if they are not interacting:
    \eqa{
        \chi(\mathcal{W}_1 \cup \mathcal{W}_1') &= \chi(\mathcal{W}_1)\chi(\mathcal{W}_1') ~.
    }
    So if we restrict to $1$-cycles, it makes sense to consider $\chi$ as a group homomorphism from the infinite-dimensional Abelian group of $1$-cycles on $M$ to $\mathsf{U(1)}$:
    \eqa{
        \chi : Z_1(\mathcal{M}_n) \longrightarrow \mathsf{U(1)} ~,
        \label{eq:EM-Coupling1}
    }
    i.e., $\chi \in \Hom(Z_1(\mathcal{M}_{n}), \mathsf{U(1)})$ is a homomorphism of Abelian groups.
    \item[\textbf{(B)}] If we have a closed worldline and it is of the form $\mathcal{W}_1 = \partial \CB_2$ for some disc $\CB_2$, then
    \eqa{
        \chi(\mathcal{W}_1) &= \exp\left( \frac{\imag }{\hbar}\int_{\mathcal{B}_2}q_{e} F^{\mathsf{phys}} \right) ~.\label{eq:EM-Coupling2}
    }
    This gives an extension of $\chi$ to cases where $F^{\mathsf{phys}}$ defines a nontrivial cohomology class and there is no globally defined $1$-form $A^{\mathsf{phys}}$ with $F^{\mathsf{phys}}=dA^{\mathsf{phys}}$. At least it gives such an extension when $\CW_1$ is the boundary of some 2-chain $\CB_2$. We will see momentarily that equation \eqref{eq:EM-Coupling2} implies that the periods of $F^{\mathsf{phys}}$ are quantized. 
    
\end{itemize}

Equations \eqref{eq:EM-Coupling1} and \eqref{eq:EM-Coupling2} are in fact the two defining properties of a Cheeger-Simons character of degree two: 

\begin{definition}[\textcolor{red}{Degree 2 Cheeger-Simons Character}]\label{def:cheeger-simons-deg-2} A Cheeger-Simons character of degree $2$ is a group homomorphism
\eqa{
    \chi &\in \Hom\left( Z_{1}(M_{n}), \mathsf{U(1)} \right) ~,
}
with the special property that there exists $F \in \Omega^{2}(M_{n})$ (specific to $\chi$) such that whenever $\mathcal{W}_1 \in Z_1(M_{n})$ is the boundary of a $2$-chain, i.e., $\mathcal{W}_1 = \partial \mathcal{B}_2$, we have 
\eqa{
   \chi(\mathcal{W}_1) &= \exp\left( \imag\int_{\mathcal{B}_2} F \right) ~.
}
The set of such characters has a natural Abelian group structure:
\eqa{
 (\chi_1 \cdot \chi_2)(\mathcal{W}) &:= \chi_1(\mathcal{W})\chi_2(\mathcal{W}) ~,
}
and forms an infinite-dimensional Abelian group denoted $\widecheck{H}^{2}(M_{n})$, known as the Cheeger-Simons or Differential Cohomology group of degree $\ell = 2$.
\end{definition}

Note that, by definition, a Cheeger-Simons character is defined on all 1-cycles, not just those which are boundaries. 
As we will soon 
see, there is \underline{more} gauge-invariant information in the Cheeger-Simons character than is encoded in the fieldstrength $F$ of the character. Therefore we can extend the action principle for a charged particle in an electromagnetic field if we say that the electromagnetic field is actually specified not by $F$ but by a Cheeger-Simons character
$\chi$. Then we can use the action principle \eqref{eq:GenChPrtAct}.

\begin{remark}
$\,$    %
\begin{enumerate}
    \item We have absorbed the $q_{e}/\hbar$ into $F$. So ``$F^{\mathsf{phys}}$'' in our physical motivation is not exactly the same as ``$F$'' in the definition of the Cheeger-Simons group.

    \item Properties (A) and (B) (i.e., equations \eqref{eq:EM-Coupling1} and 
    \eqref{eq:EM-Coupling2}) above imply the quantization of the periods of $F$: Suppose spacetime has a closed $1$-cycle that bounds a $2$-chain $\mathcal{B}_2$. Suppose it were to bound another $2$-chain $\mathcal{B}_2'$ so that $\Sigma_2 = \mathcal{B}_2 \cup \overline{\mathcal{B}}_2'$ is a closed $2$-cycle, see \autoref{fig:W-bounded-2-chains}. 
    \begin{figure}[h]
      \centering
      \scalebox{0.7}{
        \begin{tikzpicture}[>=Stealth,baseline=-0.5ex]
        \node (graph) [inner sep=0pt]{\includegraphics[width=3in]{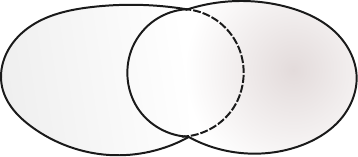}};
         \node at (0.25,-2) {$\bm{\CW}_1$};
         \node at (-2,0) {$\bm{\CB}_2$};
         \node at (2,0) {$\bm{\ov{\CB}}_2'$};
        \end{tikzpicture}
        }
       \caption{Worldline $\CW_1$ bounded by two 2-chains $\CB_2$ and $\ov{\CB}_2'$.}\label{fig:W-bounded-2-chains}
    \end{figure}

    Then,
 \eqa{
    \chi(\mathcal{W}_1) &= \exp\left( \imag  \int_{\mathcal{B}_2} F \right) = \exp\left( \imag  \int_{\mathcal{B}_2'} F \right) ~.
 }
This implies that the fieldstrength satisfies:
\eqa{
    \exp\left( \imag\int_{\Sigma_2}F \right) &= 1 ~,
}
otherwise, the two expressions would lead to different answers. Therefore,
\eqa{
   F &\in \Omega_{\mathbb{Z}'}^{2}(M_{n})  \hookrightarrow \Omega_{d}^{2}(M_{n}) := \mathsf{ker}(d) ~,
}
where $\Omega_{\mathbb{Z}'}^{2}$ denotes the space of 2-forms with periods valued in $\mathbb{Z}' := 2\pi \mathbb{Z}$.
\end{enumerate}
\end{remark}

\begin{remark}
$\,$
\begin{enumerate}
    \item Any $2$-form which has all periods of $2$-cycles in $2\pi\mathbb{Z}$ must be \underline{closed}. Here is a proof:  
        Suppose $\omega \in \Omega^{2}_{\IZ'}(M_{n})$ is a 2-form with all periods in $\IZ' := 2\pi\IZ$. Let $\Sigma_2$ be a 2-cycle. Any other cycle homologous to it is 
        of the form $\Sigma_2 \mapsto \Sigma_2 + \partial \CB_3$. If we choose $\partial \CB_3$ to be a small deformation then, since $\IZ'$ is discrete, the period cannot 
        change. Therefore, using Stokes' theorem,
        \be
            \int_{ \CB_3}d\omega = 0 ~.
        \ee 
    Now, we can choose $\CB_3$ to be a degenerate chain 
    whose image in $M_n$ is $2$-dimensional everywhere 
    except for a small $3$-ball around any point $x\in M_n$ (presuming $M_n$ is connected). Since we can make the ball arbitrarily small, it must be that $\left.d\omega\right|_{x} =0$.
    \item The argument above is closely related to Dirac's quantization argument for the product of electric and magnetic charge:
    \eqa{
        \frac{g^2}{\hbar c}\frac{e^2}{\hbar c} &= \left( \frac{S}{2} \right)^2 ~, \quad S \in \mathbb{Z} ~.
    }
    Reinstating $\hbar$ and $q_{e}$, the worldline action on $\IR^3 \setminus \{0 \}$ has exponentiated action
    \eqa{
       e^{\frac{\imag }{\hbar}\int_{\gamma}q_{e}A} &= e^{\frac{\imag }{\hbar}\int_{D_+}q_{e}F} = e^{-\frac{\imag }{\hbar}\int_{D_-}q_{e}F} ~,
    }
    but $D_{+}\cup D_{-}$ is a closed $2$-cycle enclosing the magnetic source $F = q_{m}\omega_{2}$, where $\omega_2$ is the volume form \eqref{eq:NrmSphrVol}. So,
    \eqa{
      e^{\frac{\imag }{\hbar}\int_{D+\cup D_-} q_{e}q_{m}\omega_{2}} &=1 \implies \frac{q_e q_m}{\hbar} \in 2\pi\mathbb{Z} ~.
    }
    The $q_e$ and $q_m$ are proportional to charges $e, g$ that appear in Coulomb's law by factors depending on $2$ and $\pi$.
\end{enumerate}
\end{remark}

\SectionWithHeader{Connections On Principal $G$-bundles}{Connections On Principal $G$-bundles}{sec:Connections-on-G-Bundles}

    Recall from \autoref{sec:G-Bundles} that 
    a principal $G$-bundle is a continuous family of $G$-torsors over $M_{n}$. It is a space $P$ with a continuous map $\pi: P \to M_{n}$ 
    invariant under a free right-action of $G$. 
      For a point $p \in M_{n}$, the fiber over the point, $\pi^{-1}(p)$ is a $G$-torsor. Near any such point, there exists a neighborhood $U$ such that $\pi^{-1}(U) \cong U \times G$, where $\cong$ denotes a homeomorphism.

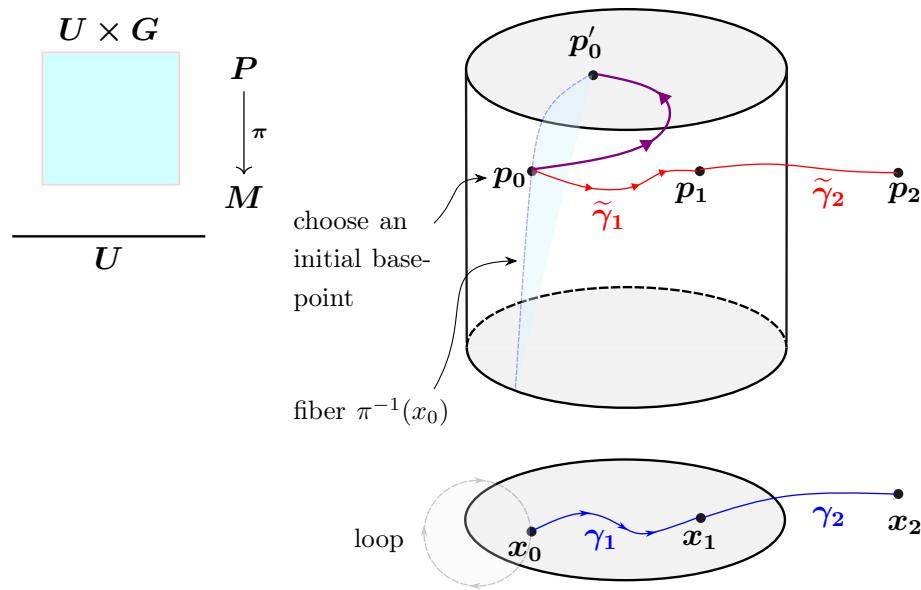
\begin{figure}[H]
  \centering
  \begin{subfigure}[T]{0.15\textwidth}
  \centering
  \begin{tikzpicture}
  \node (graph) [inner sep=0pt]{\includegraphics[width=1in]{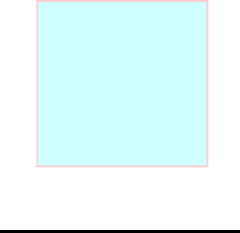}};
  \node at (0,1.5) {$\bm{U \times G}$};
  \node at (0,-1.5) {$\bm{U}$};
  \end{tikzpicture}
  \end{subfigure}
  \centering
  \begin{subfigure}[T]{0.05\textwidth}
  \vspace{+0.5cm}
  \begin{tikzcd}[row sep=large]
      \bm{P} \arrow[d,"\bm{\pi}"] \\ \bm{M}
  \end{tikzcd}
  \end{subfigure}
  \begin{subfigure}[T]{0.4\textwidth}
  \begin{tikzpicture}[>=Stealth,baseline=-0.5ex]
  \node (graph) [inner sep=0pt]{\includegraphics[width=2.5in]{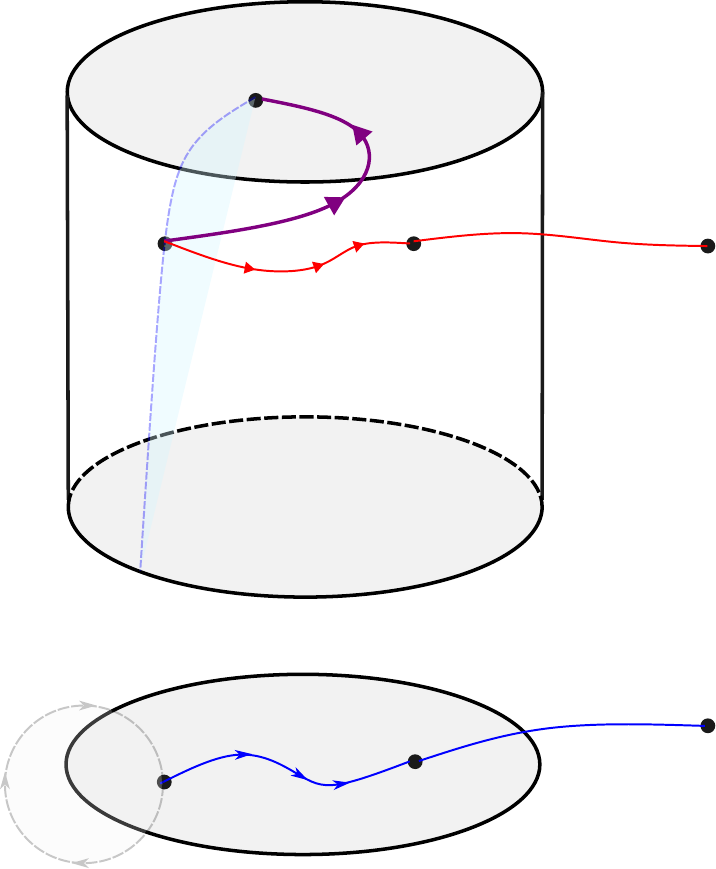}};
  \node at (-1.8,-3.35) {$\bm{x_0}$};
  \node at (-2,1.6) {$\bm{p_0}$};
  \node at (-1,3.4) {$\bm{p_0}'$};
  \node at (3.2,-3) {$\bm{x_2}$};
  \node at (3.2,1.4) {$\bm{p_2}$};
  \node at (0.5,-3.2) {$\bm{x_1}$};
  \node at (0.4,1.4) {$\bm{p_1}$};
  \node at (-0.8,-3.2) {$\textcolor{blue}{\bm{\gamma_1}}$};
  \node at (-0.7,1.1) {$\textcolor{red}{\bm{\wt{\gamma}_1}}$};
  \node at (2.2,-2.9) {$\textcolor{blue}{\bm{\gamma_2}}$};
  \node at (2.2,1.4) {$\textcolor{red}{\bm{\wt{\gamma}_2}}$};
  \draw[->] (-3, 1) to [out=45,in=180,looseness=1.5] (-2.3,1.6);
  \node at (-3.8,0.5) [text width=2.1cm]{\footnotesize{choose an initial basepoint}};
  \draw[->] (-3, -1.2) to [out=60,in=180,looseness=1.5] (-1.9,0.5);
  \node at (-3.8,-1.5) [text width=2.1cm]{\footnotesize{fiber $\pi^{-1}(x_0)$}};
  \node at (-3,-3.2) [text width=2.1cm]{\footnotesize{loop}};
  \end{tikzpicture}
  \end{subfigure}
  \caption{Coherent path lifting.}
  \label{fig:cohpathlifting}
\end{figure}

     A \emph{connection} is a rule for coherently lifting paths from paths in 
     $M$ to paths in $P$, see \autoref{fig:cohpathlifting}.
    Consider a path $\gamma: [0,1] \to M_{n}$, such that $\gamma(0) = x_0 \in M_{n}$. Pick an initial point $p_0 \in \pi^{-1}(x_0)$ in the fiber over $x_0$. The lifted path $\wt{\gamma}$ depends on the initial path $\gamma$ and the choice of initial point $p_0$.
    Since the path is lifted, we have 
    \be 
    \pi(\wt{\gamma}(t)) = \gamma(t) ~.
    \ee
      The connection needs to be coherent with respect to gluing: the concatenation of paths $\gamma_1$ and $\gamma_2$ is the composition $\gamma_1 \circ \gamma_2$. For reasons that will be explained soon, the connection is denoted $\n$. 
          
      Now, suppose we are given a connection $\n$, and we consider a  loop from $x_0$ to itself where $x_0$ is a point in $M_{n}$.  In doing so, we go from an initial point $p_0$ in the fiber $\pi^{-1}(x_0)$ above $x_0$ to another point $p_0'$ in the fiber $\pi^{-1}(x_0)$. Since these are two points in the same fiber of a $G$-bundle, they will be related by a $G$-action: 
    \be  
    p_0' = p_0 \cdot g ~.
    \ee
    If we change the initial point of the lifted path to $\wt{p}_0= p_0\cdot h$ then the group element $g$ is changed by conjugation with $h$. However, the conjugacy class of $g$ is a well-defined, function of the connection $\n$ and the path $\gamma$. 
    Thus a connection defines a function: 
    \be 
     \mathsf{Hol}_{\n} : Z_{1}(M_{n}) \longrightarrow \mathsf{ConjClass}(G) ~,
     \ee
known as the \emph{holonomy of the connection}. The value of a particular cycle $\mathsf{Hol}_{\n}(\gamma)$ is called the 
\emph{holonomy around $\gamma$}.  (See the thick \textcolor{violet}{violet} curve connecting $p_0$ and $p_0'$ in \autoref{fig:cohpathlifting}).
We also write $\mathsf{Hol}(\n, \gamma)$
when $\n$ is not fixed. 

Note that if the group $G$ is Abelian, the set of conjugacy classes is the same as the group and we can view the holonomy as valued in $G$. 

Two connections are said to be \emph{gauge equivalent} if one is 
obtained from the other by a bundle automorphism. The holonomy function is a gauge-invariant function of the connection. But it does not determine the connection up to gauge equivalence. 
\tightfootnote{For a counterexample, consider flat $G$-connections on a torus. The isomorphism classes of such connections are pairs of group elements $(g_1, g_2)$ up to equivalence by simultaneous conjugation $(g_1, g_2) \sim (h g_1 h^{-1}, h g_2 h^{-1})$. $g_1, g_2$ can be thought of as the holonomy around $a,b$ cycles on the torus. One cannot lift a pair of 
conjugacy classes to such an equivalence class of $(g_1, g_2)$.}
However, if we choose a basepoint $x_0 \in M_n$ and trivialize the fiber above $P_0$, then 
the holonomy becomes a function from loops based at $x_0$ to $G$. This function 
\underline{does} determine the equivalence class of the bundle and connection.  
\tightfootnote{See \cite{Kobayashi1954,Anandan1986,Barrett:1991aj,Caetano:1993zf,Lewandowski:1993qy,Meneses:2019wyy} for discussions. Some of these papers restrict to the case where the structure group $G$ is compact. Some do not specify conditions on $G$ for validity of the theorem.  The references \cite{Barrett:1991aj,Caetano:1993zf} only require that $G$ be a Lie group.} 
  
    \begin{theorem}\label{thm:holonomy-theorem} Let $G$ be a Lie group. 
        Let $P \to M_{n}$ be a principal $G$-bundle over a path-connected space $M_{n}$ with connection $\n$ trivialized at a point $x_0\in M_n$, and consider the 
        holonomy map from loops based at $x_0$:
        \eqa{
           \mathsf{Hol}_{\n, x_0} : \Omega_{x_0}(M_n)  \longrightarrow G ~.
        }
    If $\mathsf{Hol}_{\n,x_0} = \mathsf{Hol}_{\n',x_0}$ for two connections $\n$ and $\n'$, then $P \cong P'$, and $\n$ and $\n'$ are gauge-equivalent.
    \end{theorem}

The basic idea of the proof is the following: We will use the path lifting provided by the 
connections to define a bundle isomorphism $\varphi: P \to P'$. By assumption, $P$ and $P'$ have a trivialization over $x_0$. So we take the map $\varphi$ from  $P_{x_0} \cong \{x_0 \} \times G$ to $P'_{x_0} \cong \{x_0 \} \times G$ to be the identity map. To define the map on the fibers above a point $x\in M_n$, we 
choose a path $\gamma$ from $x_0$ to $x$. Choosing a point $(x_0,g_0)\in P_{x_0}$, the connection 
$\n$ provides a unique lifted path $\wt{\gamma}_{\n,g_0}(t): [0,1] \to P$ such that 
$\wt{\gamma}_{\n,g_0}(0) = (x_0,g_0)$. Now, 
$\varphi$ takes $p =\wt{\gamma}_{\n,g_0}(1) \in P_{x}$ to $p' =\wt{\gamma}_{\n',g_0}(1) \in P'_{x} $.
This will be smooth and provide the desired bundle isomorphism so that the connections are isomorphic. 
The main thing to check is that $\varphi$ is well-defined. This is so because 
$\mathsf{Hol}_{\n,x_0} = \mathsf{Hol}_{\n',x_0}$. Note that the key property that is used is continuity and ordinary differential equations. Thus the argument applies to the case that the structure group $G$ is any finite-dimensional Lie group. 

\begin{remark} One can try to weaken the hypotheses of \autoref{thm:holonomy-theorem}, so that we only require that,
\be
\chi\circ \mathsf{Hol}_{\n,x_0} = \chi\circ \mathsf{Hol}_{\n',x_0} ~,
\ee
for all characters $\chi$ on $G$. In this case, the modified theorem with weakened hypotheses is still true for compact groups $G$, but for noncompact groups, there are counterexamples \cite{Sengupta1994}.
One could also consider generalizing the theorem to infinite-dimensional Lie groups, but this would require further considerations of functional analysis beyond the scope of this discussion. For some purposes, it is often useful to generalize the notion of 
holonomy of a connection (with irregular singularities) to include Stokes data. However in this case, the fact that extra conditions are needed to solve the Riemann-Hilbert problem suggests that a potential generalization of \autoref{thm:holonomy-theorem} along such lines would be false. 
\end{remark}

There are other equivalent descriptions of connections which will be useful to us. We will be brief here. See textbooks on the differential geometry of fiber bundles for more details, e.g., \cite{Kobayashi1996-kb,Kobayashi1996-fy,Taubes2011}. 

\begin{enumerate}

\item \textbf{Horizontal Subspaces.} The map $\pi: P\to M_{n}$ defines for each $p\in M_{n}$ a surjective linear map, 
\be 
d\pi: T_{p}P \to T_x M_{n} ~,
\ee
where $x=\pi(p)$. The kernel of the map $V_p := \ker\, d\pi \subset T_p P$ is 
the \emph{vertical tangent space}. We define a \emph{horizontal tangent space} to be a choice of a complementary subspace 
$H_p\subset T_p P$, so that, 
\be 
T_p P \cong V_p \oplus H_p ~.
\ee
A connection, then, is a smooth choice of horizontal subspaces. The path lifting is defined by the condition that, 
\be\label{eq:HorizontalLifting} 
\widetilde{\gamma}_* \left( \frac{d}{dt} \right) \subset H_{\widetilde{\gamma}(t) } ~.
\ee
For a connection on a principal bundle, the horizontal spaces should be equivariant under the right $G$ action on $P$. 

\item \textbf{Global 1-form.} A connection can also be defined by a 
globally well-defined 1-form $\Theta$ on $P$, valued in the tangent space of the fiber,  such that for vertical 
vector fields $v \in V_p$, we have $\Theta(v) = v$. For principal $G$-bundles, the tangent space of the fiber can be globally identified with the Lie algebra $\mathfrak{g}$ of $G$, and we can view 
$\Theta \in \Omega^1(P;\mathfrak{g})$. Then for $X\in \mathfrak{g}$, the right $G$-action defines a vertical vector field $\xi(X)$ on $P$, 
and we require that $\Theta(\xi(X)) = X$. We also require that $\Theta$ be $G$-equivariant: $R_g^*(\Theta) = \mathsf{Ad}(g)(\Theta)$. 

\item \textbf{Covariant Derivative.} If $E$ is a vector bundle associated to $P$ via a representation of $G$, then the path-lifting condition 
\eqref{eq:HorizontalLifting} can be expressed in terms of a 
first order differential operator, 
\be 
\n : \Gamma(E) \to \Omega^1(M_{n}; E) ~,
\ee
that satisfies the Leibniz rule: 
\be 
\n(f s) = s \otimes df + f \n s ~,
\ee
where $s$ is a section of $E$ and $f$ is a function on $M_{n}$. 
The local differential equation defining a lifting of a path in $M_{n}$ to a path in $E$ is:
\be 
\n_{\dt{\gamma}(t)} s(t) = 0 ~.
\ee

\item \textbf{Local Gauge Fields.} In local patches, the covariant 
derivative can be expressed relative to a coordinate basis $dx^\mu$ for the cotangent bundle as
\tightfootnote{We take our gauge fields to be ``real,'' as is standard in the physics literature. Many factors of $\imag$ could be removed by taking them to be antihermitian, but we will not do this.}
\be 
\partial_{\mu}  + \imag A_{\mu} ~, \label{eq:cot}
\ee
where $A_\mu(x) \in \mathfrak{g}$ for the principal bundle and is in the endomorphisms of the representation space for an associated bundle.  The $A_{\mu}$ are different in different patches and are related by gauge transformations on patch overlaps. If the bundle 
has been locally trivialized in patches 
\begin{align}\begin{split}
\varphi_\alpha: \pi^{-1}(\CU_\alpha) \cong \CU_\alpha \times G  ~,\\
\varphi_\beta: \pi^{-1}(\CU_\beta) \cong \CU_\beta \times G ~,
\end{split}
\end{align}
and the covariant derivatives are $d+\imag A^{(\alpha)}$ and 
$d+\imag A^{(\beta)}$ in the two patches, then on the patch overlap 
\be 
d+ \imag A^{(\beta)} = g_{\alpha\beta}^{-1} \left( d+ \imag A^{(\alpha)}\right) 
g_{\alpha\beta} ~,
\ee
where $g_{\alpha\beta}:\CU_{\alpha\beta} \to G$ is the transition 
function. So  the connections on the patch overlap are gauge equivalent.

\end{enumerate}

\noindent In terms of the gauge fields, the holonomy can be expressed as a 
path-ordered exponential: 
\tightfootnote{For a review of path ordering, see \cite[App. H]{Moore:2019TASI}.}
\be 
\mathsf{Hol}_{\n}(\gamma) = P\,\exp \left( \imag \oint_{\gamma} A \right) ~.
\ee
Restricting now to $G = \mathsf{U(1)}$, the holonomy function 
of a principal $\mathsf{U(1)}$-bundle with connection 
defines a character: 
\be 
\mathsf{Hol}_{\n} ~~:~~ Z_{1}(M_{n})  \longrightarrow \mathsf{U(1)} ~,
\ee 
which can be expressed as: 
\be
\mathsf{Hol}_{\n}(\gamma) = \exp\left( \imag \oint_{\gamma} A \right) ~, \label{eq:hol}
\ee
(where we have reverted to expressing the gauge field as a locally-defined  real $1$-form). 
Stokes' theorem implies that this character on $Z_1(M_n)$ satisfies the definition of a differential character. Thus, the holonomy function of a principal $\mathsf{U(1)}$-bundle with connection defines a degree two differential character. 
The converse is also true: see Box \ref{box:nontrivial-cheeger-simons-fact}.

   \begin{redbox}[box:nontrivial-cheeger-simons-fact]{Nontrivial Fact}
    Every Cheeger-Simons character is the holonomy function of some connection on some principal $\mathsf{U(1)}$-bundle on $M_{n}$, that is,
    \eqa{
       \chi(\mathcal{W}) &= \mathsf{Hol}\big(\n, \mathcal{W} \big) ~,
    }
    for some connection $\n$ on some principal $\mathsf{U(1)}$-bundle $P \xrightarrow{\mathsf{U(1)}} M_{n}$. Informally, we can write $\n = d + \imag A$, and $\mathsf{Hol}(\n, \mathcal{W}) = \exp\left( \imag\int_{\mathcal{W}}A\right)$, but $A$ is not globally well-defined. Conversely, if we have a principal $\mathsf{U(1)}$-bundle with connection, then the holonomy can be viewed as a map $\chi: Z_{1}(M_{n}) \to \mathsf{U(1)}$ as elaborated upon below.
\end{redbox}

\SectionWithHeader{Degree Two Differential Characters And Connections On Principal $\mathsf{U(1)}$ Bundles}{Degree Two Characters, Connections on $\mathsf{U(1)}$ Bundles}{sec:DegreeTwo-Diff-Characters}   

    Using the interpretation of degree two characters in terms of the holonomy function of some connection on a line bundle, we can describe the full Abelian group $\widecheck{H}^{2}(M_{n})$. This group contains the gauge-invariant information of all possible principal $\mathsf{U(1)}$-bundles with connection. We now unpack that information: 
    
    For each $x \in H^{2}(M_{n}; \IZ)$, there exists a principal $\mathsf{U(1)}$-bundle $P_{x} \to M_{n}$ with $c_1(P_{x}) = x$.
    \tightfootnote{\label{foot:chern-class-principal-bundle}The Chern class is defined for a complex vector bundle, which, after being endowed with a Hermitian metric on its fibers, may equally well be viewed as a $\mathsf{U(1)}$ bundle. What we have in mind here is the Chern class of the complex line bundle associated to the principal $\mathsf{U(1)}$-bundle $P_{x} \to M_{n}$ by the fundamental representation. See \autoref{subsec:CechModel} for a description of the first Chern class in terms of transition functions.} 
    Let $(\CA/\CG)_{x}$ be the set of gauge equivalence classes of connections on $P_{x}$. (See footnote \ref{foot:connection-space-notation}.)
    %
    %
    We have,
    \eqa{
        \widecheck{H}^{2}(M_{n}) &= \coprod_{x \in H^{2}(M_{n}; \IZ)}(\CA/\CG)_{x} ~.
    }
    Specifying a point in $\CA/\CG$ is precisely the same as specifying a holonomy function.

    Now quite generally, for \underline{any} principal $G$-bundle $P \xrightarrow{G} M_{n}$ for \underline{any} Lie group  $G$, the space of connections on $P$, 
    \eqa{
       \CA(P) := \mathsf{Conn}(P \xrightarrow{G} M_{n}) ~,
    }
    is an affine space modeled on $\Omega^{1}(M_{n}; \mathsf{ad\,}P)$.
    \tightfootnote{Here $\Omega^{k}(M_{n}; E)$ is the space of smooth $k$-forms on $M_{n}$ valued in a vector bundle $E \to M_{n}$.} 
     This means that we can choose (any) basepoint connection  $\n_{0}$ on $P$, and then  every other connection on $P$ is of the form
    \eqa{
       \n &= \n_{0} + \alpha ~,
    }
    for some suitable $ \alpha \in \Omega^{1}(M_{n}; \mathsf{ad\,}P)$. 

    Applying the previous paragraph to the case  $G = \mathsf{U(1)}$, the adjoint bundle is trivial, i.e., $\mathsf{ad\,}P = M_{n} \times \mathbb{R}$,
    and so, $\Omega^{1}(M_{n}; \mathsf{ad\,}P) = \Omega^{1}(M_{n})$, and $\n = \n_{0} + \imag  \alpha$, where $\alpha$ is a globally defined real $1$-form on $M_{n}$.
    \tightfootnote{\label{foot:ConventionClash}There is an unfortunate clash between standard math and physics conventions at this point. Strictly speaking, the fibers of $\mathsf{ad}(P)$ should be the Lie algebra of $G$, and for $G=\mathsf{U(1)}$, the real Lie algebra should be identified with $\imag  \IR$ so that the exponentiated Lie algebra elements are phases. }
    Gauge transformations are given by
    \eqa{
        \alpha &\mapsto \alpha + \omega ~, \quad \omega \in \Omega_{\IZ'}^{1}(M_{n}) ~.
    }
    Note that the periods have to be in $\IZ' := 2\pi \IZ$. The reason is that the shift $\n \mapsto \n + \imag  \omega$ shifts the holonomy 
    function by a factor  $\exp[ \imag  \oint_{\gamma} \omega]$. But for a gauge transformation, the holonomy function is gauge-invariant. 
    Therefore, $\omega$ must have periods in $\IZ'$.  
    If $\omega$ has nontrivial periods, they are called ``large gauge transformations,'' whereas if $\omega = d\epsilon$ for a globally well-defined $\epsilon \in \Omega^{0}(M_{n})$, they are called ``small gauge transformations.'' 

The previous paragraph implies that if we fix a Chern class $x = c_1(P)$, then  
\eqa{
       (\CA/\CG)_{x} &\cong \Omega^{1}(M_{n})/\Omega^{1}_{\IZ'}(M_{n}) ~.
    }

    \noindent Now let us recall the Hodge decomposition \eqref{eq:HodgeDecomposition}: $\Omega^{1}  \cong \CH^{1} \oplus \mathsf{im\,}d \oplus \mathsf{im\,}d^\dagger$. 
    %
    %
    Therefore,
    \eqa{
       \Omega_{\IZ'}^{1} &\cong \CH^{1}_{\IZ'} \oplus \mathsf{im\,}d ~,\\
       \Omega^{1}/\Omega_{\IZ'}^{1} &\cong \CH^{1}/\CH^{1}_{\IZ'} \oplus \mathsf{im\,}d^\dagger ~, \label{eq:DecTopTrivChar-2}
    }
    and hence, 
    \eqa{
       (\CA/\CG)_{x} &\cong \CH^{1}/\CH^{1}_{\IZ'} \oplus \mathsf{im\,}d^\dagger ~.
    }

The conclusion from the above discussion is that, as an Abelian group, 
$\widecheck{H}^{2}(M_{n})$ can be written as the direct product:
\begin{align}\label{eq:hcheck2-decomp}
      \widecheck{H}^{2}(M_{n}) &= \mathbb{T} \times V \times \Gamma ~,\\
      \mathbb{T} & = \CH^{1}/\CH^{1}_{\IZ'} \cong \mathsf{U(1)}^{b_1}= \text{a connected torus} ~,\\
      V & = \mathsf{im}(d^\dagger: \Omega^{2} \to \Omega^{1})
      = \text{an $\infty$-dimensional vector space} ~, \label{eq:DP-H2check}\\
      \Gamma & = H^{2}(M_{n}; \IZ)= \text{ a discrete Abelian group} ~.
\end{align}
The Abelian group $\Gamma$ specifies the topological sector. It is finitely generated if $M_{n}$ is 
compact. The connected torus $\mathbb{T}$ is the space of Wilson lines connected to the identity (these are Wilson lines for topologically trivial flat gauge fields), as opposed to ``discrete Wilson lines'' which 
cannot be smoothly deformed to $1$ in fieldspace. Finally, $V$ is the infinite-dimensional vector space of the oscillator modes of the Maxwell field, or more precisely, the gauge-invariant information in the oscillator modes. 

The decomposition \eqref{eq:hcheck2-decomp} of $\widecheck{H}^{2}(M_{n})$ is  noncanonical for two reasons: First, to identify the gauge equivalence class of connections on $P_x$ with a quotient of $\Omega^1(M_n)$, we needed to choose a basepoint connection $\n_{0,x}$. Second, to identify the quotient  of 
$\Omega^1(M_n)$ with a connected torus times a vector space, we need to introduce a gauge fixing condition. If we are willing to use the extra information of a Riemannian metric, then we can apply the Hodge decomposition, as we did above. Note that if $\alpha \in \mathsf{im\,}d^\dagger$, then $d^\dagger\alpha = 0$, which is equivalent to $\partial^{\mu}\alpha_{\mu} = 0$. This is the well-known Landau gauge fixing. So the use of the Hodge decomposition to write \eqref{eq:DP-H2check} is equivalent to choosing the Landau gauge. 

\SectionWithHeader{The Group Of Cheeger-Simons Characters}{The Group Of Cheeger-Simons Characters}{sec:CheegerSimonsGroup}

In \autoref{sec:ActionTestBranes}, we used quantum Maxwell theory to motivate the definition of Cheeger-Simons differential characters of degree two. Now we wish to extend that idea to characters of all nonnegative degrees. From the physical point of view, we generalize 
from the Maxwell field with $\ell=2$ to the generalized Abelian gauge field with any value of $\ell$.

One physics application of this mathematics to be stressed below, is to the Hamiltonian quantization of the generalized Maxwell field. This involves a construction of the Hilbert space that is manifestly $S$-duality invariant as in \autoref{sec:HilbertSpaceGAGT}
 and \autoref{sec:manifestEMduality}. It also illustrates the noncommutativity of electric and magnetic fluxes, which in turn provides a Hamiltonian interpretation of generalized symmetries and their 't Hooft anomalies as described in \autoref{sec:noncommutativity}.

    Our physical motivation in the case  $\ell = 2$  
    started by considering the Lorentz force of a test particle moving on a worldline $\CW_1$ of a charged particle moving in a background electromagnetic field configuration.  There is a natural generalization of the coupling $\exp\left(\imag \oint q_{e}A\right)$ to that for electric branes for a generalized Abelian gauge field.

As we have seen, for topologically trivial Maxwell fields, 
namely, fields where there is a globally well-defined one-form 
gauge potential $A\in \Omega^1(M_n)$ with $F=dA$, we have: 
\be 
 \chi(\CW_1)  =  \chi_A(\CW_1):= \exp\left( \imag  \int_{\CW_1} A \right) ~.  
\ee

    Now, consider an electrically charged $p$-brane for a generalized Maxwell field that is topologically trivial. Topological triviality 
    again means that we can write $F=dA$, where $A \in \Omega^{\ell-1}(M_{n})$ is globally well-defined. Then the electric coupling is just: 
\be \label{eq:chi-top-trivial}
 \chi(\CW_{p+1}) = \chi_A(\CW_{p+1}) := \exp\left( \imag\int_{\CW_{p+1}}A \right) ~.
\ee
Note that we recover the value of $p= \ell-2$ for an electrically charged $p$-brane because the above formulae only make sense for $p+1=\ell-1$. 
 
    But what do we do if $0 \neq [F] \in H^{\ell}_{\mathsf{dR}}(M_{n})$, i.e., if $F$ is cohomologically nontrivial? Now there is no globally defined $A$. If $\CW_{\ell-1}$ is the boundary of an $\ell$-chain (i.e., if the worldline is homologically trivial), then we can propose a definition:
    \eqa{
      \chi(\CW_{\ell-1}) &:= \exp\left( \imag  \int_{\CB_{\ell}} F \right) ~. \label{eq:chi-general}
    }
    Now in the case when $F$ is topologically trivial, \eqref{eq:chi-general} reduces to \eqref{eq:chi-top-trivial} by Stokes' Theorem. But in general, \eqref{eq:chi-general} is an extension of the definition of this term in the action to new field configurations where $F$ is not exact. Once again, since a given cycle $\CW_{\ell-1}$ will in general be bounded by several different $\ell$-chains which can be glued together to make a nontrivial $\ell$-cycle, we find that this proposal only works if $F$ has quantized periods, that is, if $F \in \Omega_{\IZ'}^{\ell}(M_{n}) \subset \Omega_{d}^{\ell}(M_{n})$.  (Recall that $\IZ' := 2\pi \IZ$.)
    %
    %

    But now, what does one do if $F$ is cohomologically nontrivial 
    \underline{and} $[\CW_{\ell-1}] $  is homologically nontrivial, 
    i.e., if $\CW_{\ell-1}$ cannot be written as the boundary of an $\ell$-chain $\CB_{\ell}$? Here, extra physical information is needed. One way to proceed is to assume that there is extra gauge-invariant information in the generalized Abelian gauge field background through which the $p$-brane moves.  One \underline{declares} the electromagnetic coupling to be a Cheeger-Simons character: 

\begin{definition}[\textcolor{red}{Cheeger-Simons Character}] \label{def:cheeger-simons-deg-ell} A Cheeger-Simons character of degree $\ell$ is a group homomorphism,
    \eqa{
       \chi &\in \Hom\left(Z_{\ell-1}(M_{n}), \mathsf{U(1)}\right) ~,
    }
    such that there exists $F \in \Omega^{\ell}(M_{n})$, known as the fieldstrength of the character $\chi$, such that if $\CW_{\ell-1} = \partial \CB_{\ell}$, then 
\be
\chi(\CW_{\ell-1}) = \exp\left(\imag  \int_{\CB_{\ell}}F \right) ~ . 
\ee
\end{definition}
    \noindent An argument similar to the one  around \autoref{fig:W-bounded-2-chains} implies that $F \in \Omega_{\IZ'}^{\ell}(M_{n}) \hookrightarrow \Omega_{d}^{\ell}(M_{n})$.

    \begin{definition}[\textcolor{red}{Differential Cohomology Group}]\label{def:differential-cohomology-group} Two Cheeger-Simons characters $\chi_1, \chi_2$ can be multiplied: 
    \be\label{eq:MultCheegerSimons}
    \left( \chi_1 \cdot \chi_2\right)(\CW_{\ell-1}) := \chi_1(\CW_{\ell-1}) \chi_2(\CW_{\ell-1})  \quad \forall \quad  \CW_{\ell-1} \in Z_{\ell-1}(M_{n}) ~. 
    \ee
    The Abelian group of all Cheeger-Simons characters of degree $\ell$ is denoted by $\widecheck{H}^{\ell}(M_{n})$. It is also known as the differential cohomology group.
    \end{definition}

 So we make a physical proposal that the proper way to describe the gauge-invariant information in generalized Maxwell theory is
 \tightfootnote{This physical proposal turns out to be correct in spirit but technically wrong for Ramond-Ramond (RR) fields in type II string theory, where the appropriate setup is in fact, Differential $K$-Theory. It is, however, correct modulo a subtlety, for the $\sfC$ field of M-theory, where the relevant object is $\widecheck{H}^{4}(M_{11})$. The subtlety has to do with a shift in the $\sfC$ field quantization, due to background topology.}
 %
    \eqa{
       \widecheck{H}^{\ell}(M_{n}) &= \left\{ \begin{array}{c} \text{gauge equivalence classes}\\ \text{of generalized Maxwell fields} \\ \text{with $\ell$-form fieldstrength on $M_{n}$} \end{array}     \right\} ~.
    }
Then, the action for the test brane in such a generalized Abelian gauge theory background whose gauge equivalence class is described by a differential character $\chi$, is: 
\be 
e^{- T \int_{\CW_{\ell-1}} \vol(\iota^*(g)) } \chi(\CW_{\ell-1}) ~,
\ee
where $T$ is the tension of the brane and $\iota: \CW \to M_n$ is the embedding of the worldvolume. 

We now start elucidating the gauge-invariant information contained in a character $\chi$ that is independent of its associated fieldstrength $F$.
We will explain this much more thoroughly in \autoref{sec:DancingHexagon} below.

If we just consider characters of the form $\chi_{A}$ for globally well-defined $A$, then we find 
\be 
\chi_{A+\omega} = \chi_{A} ~,
\ee
for $\omega \in \Omega^{\ell-1}_{\IZ'}(M_n)$. The shift $A\mapsto A+\omega$ can therefore be viewed as a gauge transformation. Indeed, when $\omega$ is exact, $\omega = d \Lambda$ for $\Lambda\in \Omega^{\ell-2}(M_n)$, these are standard gauge transformations in the theory of higher-form fields. The space of topologically trivial characters is the quotient space $\Omega^{\ell-1}/\Omega^{\ell-1}_{\IZ'}$. Thanks to the Hodge decomposition, we can write this as a connected torus times $\mathsf{im}(d^\dagger)$, just as in Maxwell theory. The torus 
\be\label{eq:HarmonicTorus}
\CH^{\ell-1}(M_n)/ \CH^{\ell-1}_{\IZ'}(M_n) ~,
\ee
describes the topologically trivial flat Wilson surfaces. These have $F=0$ but nonzero ``holonomy,''  generalizing the familiar situation in the Aharonov-Bohm effect for charged particles in an electromagnetic fields. 
Differential characters (i.e., gauge fields) described by such Wilson surfaces have nontrivial gauge-invariant information not encoded in the fieldstrength (which is zero). 

There is also a more subtle kind of gauge-invariant information that cannot be deduced from the fieldstrength. As we will see in \autoref{subsec:TopTrivialChars} below, to a character $\chi$, one can associate an integral cohomology class $c(\chi) \in H^{\ell}(M_n; \IZ)$ known as its \emph{characteristic class}. The general definition of $\chi \mapsto c(\chi)$ is best left to the ``chain complex'' descriptions, and we are not defining it precisely just now. When this class is nonzero, it is impossible to write $\chi$ in the form $\chi_{A}$ for some globally well-defined form $A\in \Omega^{\ell-1}(M_n)$. We can consider the embedding of $H^{\ell}(M_n; \IZ)$ into $H^{\ell}_{\IZ}(M_n; \IR)$, and the image of $c(\chi)$ under this embedding actually loses information. 
The image $c(\chi)_{\IR}$ can be captured by the fieldstrength through 
the de Rham cohomology class $[F/2\pi]$, but the kernel cannot be captured by the fieldstrength. The gauge-invariant information in $c(\chi)$
not captured by $[F/2\pi]$ is the more subtle gauge-invariant information we mentioned above. 
%
%




We now illustrate the two previous paragraphs with two examples of the group of differential characters: 
 
\bigskip
\noindent \textbf{Examples:}
   \begin{enumerate}
    \item $\bm{\ell = 1}$. A degree one Cheeger-Simons cohomology class is nothing other than a $\mathsf{U(1)}$-valued continuous function on the manifold. Note that $\CW_{\ell-1}$ is just a cycle of the form 
    $\sum_i n_i p_i$, where $p_i$ are points and given a $\mathsf{U(1)}$-valued function $f$ on $M_n$, we set 
\be 
\chi_f\bigg( \sum_i n_i p_i\bigg) :=  \prod_i f(p_i)^{n_i} ~,
\ee
which is well-defined because $n_i$ are integers. To find the fieldstrength, note that 
\be 
\chi_f(p_1 - p_2 ) = f(p_1) f(p_2)^{-1} = \exp\bigg[ \int_{\gamma} d \,\log(f) \bigg]  = 
\exp\bigg[ \imag  \int_{\gamma}  (-\imag  f^{-1} df) \bigg] ~, 
\ee
where $\gamma$ is a 1-chain such that $\p \gamma = p_1 - p_2$. 
We conclude that 
\be 
F(\chi_f)  = -\imag f^{-1}df \in \Omega^{1}_{\IZ'} ~ . 
\ee
To define the characteristic class, we take the canonical integral class 
$c\in H^1(\mathsf{U(1)}; \IZ)$ that integrates to one.  (So in de Rham theory, $c=d\theta/2\pi$, if we parametrize $\mathsf{U(1)}$ by $e^{\imag  \theta}$.) We set 
\be 
c(\chi):= f^*(c) ~.
\ee
Put differently, the evaluation of $c(\chi)\in H^1(M_n; \IZ)$ around any 
$1$-cycle is just the winding number. Now, one can write $f = e^{\imag \phi}$, where the scalar field  $\phi$ is not necessarily globally well-defined. Rather, it can  have nonzero periods
$ \oint_{\gamma}d\phi  \in \IZ'$. This is almost universally the way physicists write the theory of a periodic scalar field. Note that such a representation of $f$ makes it clear that the de Rham cohomology class of $c(\chi)$ corresponds to that of $[F(\chi)/2\pi]$. In this case, 
$H^1(M_n; \IZ)$ never has any nontrivial torsion subgroup, so there is no 
further information in a degree $1$ character. In conclusion, we have 
\be\label{eq:DifferentialH-one}
\widecheck{H}^{1}(M_{n}) \cong \mathsf{Map}(M_{n} \to \IR/\IZ) \cong \mathsf{Map}(M_{n} \to \mathsf{U(1)}) ~. 
\ee

\item $\bm{\ell = 2}$. This is the case described in our motivating example of Maxwell theory. Just to recapitulate: 
       To give a differential character of degree two is equivalent to giving an isomorphism class of a principal $\mathsf{U(1)}$-bundle with connection:
    \eqa{
      \chi &\longleftrightarrow [(P, \n)] ~.
    }
    Modeling the differential character as the holonomy function of a connection on a principal $\mathsf{U(1)}$ line bundle,
    \eqa{
       \chi(\CW_1) &= \mathsf{Hol}_{\n}(\CW_1)  \quad \forall \quad \CW_1 \in Z_{1}(M_{n}) ~,
    }
    for some $\n$ on $P \xrightarrow{\mathsf{U(1)}} M_{n}$. 
    We define the characteristic class of the character to be the 
    first Chern class of the $\mathsf{U(1)}$-bundle $P$ (recall footnote \ref{foot:chern-class-principal-bundle}): 
    \eqa{
        c(\chi) &:= c_1(P) \in H^{2}(M_{n}; \IZ) ~.
    }
    This brings up an important new point. The de Rham cohomology class associated with $c_1(P)$ can be captured by $[F/2\pi]$, but the latter expression does \underline{not} capture all the gauge-invariant information in $c_1(P)$. The reason for this will be described in \autoref{sec:DancingHexagon}. The reason is that there can be a nontrivial torsion subgroup of $H^{2}(M_{n}; \IZ)$. A simple example where such things appear is in the Lens spaces $\IS^{2k+1}/\IZ_n$. See \autoref{subsec:SeveralMathPrelim} for further details.

    \item $\bm{\ell > 2}$. This is best left to models of differential cohomology
    discussed in \autoref{sec:ModelDiffCoh}.

   \end{enumerate}

\begin{remark}
$\,$
\begin{enumerate} 

\item In the definition of differential characters one can replace $\mathsf{U(1)}$ by a higher rank Abelian group $\mathsf{U(1)}^d$ in the above (and subsequent) definitions. There is no universally accepted extension to the case of non-Abelian groups. 

\item \emph{Is differential cohomology the cohomology of some 
complex? } The answer is ``yes,'' and this   follows from the work of Hopkins and Singer \cite{Hopkins:2002rd}. This will be explained in 
\autoref{subsec:HopkinsSingerCocycles}.

\item \emph{Remark on notation for group structure on Abelian groups}. 
Whenever one works with Abelian groups, one must make a choice of whether to use multiplicative or additive notation. For example, when working with the cyclic group $\IZ/n\IZ$, it is natural to write the group product additively and think of the group operation as addition modulo $n$. On the other hand, working with the isomorphic group $\mu_n$ of $n^{th}$ roots of unity, it would be perverse to use an additive multiplication. Of course, the isomorphism between $\IZ/n\IZ$ and $\mu_n$ involves an exponential map. When working with a differential character, one simultaneously works with both additive and multiplicative notations. To begin, it is defined as the group of homomorphisms from $Z_{\ell-1}(M_n)$ to $\mathsf{U(1)}$. The codomain is naturally written in multiplicative notation. The domain is 
more naturally written with additive notation. Recall that to define the Abelian group 
$Z_{\ell-1}(M_n)$, we begin with the group of chains,    $C_{\ell-1}(M_{n})$, defined as the free Abelian group generated by continuous maps from $(\ell-1)$-simplices to the manifold $M_{n}$, i.e., $\phi : \Delta^{\ell-1} \to M_{n}$. Then one defines a boundary operation $\partial$ via a sequence of group homomorphisms,
    \be
    \begin{tikzcd}
        \cdots & \arrow[l] C_{\ell-2} & \arrow[l, "\partial",swap] C_{\ell-1} & \arrow[l] \cdots 
    \end{tikzcd}
    \ee
    such that $\partial^2 = 0$. Therefore, $Z_{\ell-1}(M_{n}) := \mathsf{ker\,}(\partial:  C_{\ell-1} \to C_{\ell})$ is a subgroup of the Abelian group $C_{\ell-1}$ and is naturally written in additive notation.  However, because the group structure on $\mathsf{U(1)}$ is multiplicative, in equation \eqref{eq:MultCheegerSimons} above, 
    we wrote (and will continue to write) the group structure on 
    characters multiplicatively. We will also introduce a notation 
    for (not necessarily topologically trivial) characters: 
    $\chi = \chi_{\check{A} }$ where $\check{A}$ is a set of data described in 
\autoref{subsec:CechModel}. Expressed this way, the natural group structure on $\check{A}$ is additive and again, the relation to $\chi$ involves an exponential map. 
An important pitfall to avoid is that there is a much more subtle (and very important)  multiplicative structure on characters which should not be confused with the Abelian group structure on characters. We will put a nontrivial   \underline{graded ring} structure on $\oplus_{\ell}\widecheck{H}^{\ell}(M_{n})$ in equation \eqref{eq:GradedRingProduct} below. 
    The Abelian group structure multiplies two characters $\chi_1, \chi_2$ of the \underline{same} degree $\ell$, to produce a character $\chi_1\cdot \chi_2$ of degree $\ell$. But the ring structure 
    multiplies two characters $\chi_1, \chi_2$ of arbitrary degrees 
    $\ell_1, \ell_2$, to produce a new character  $\chi_1 \odot \chi_2$ 
    of degree $\ell_1 + \ell_2$.  

\end{enumerate}

\end{remark}

    \begin{exbox}{$\IR$-valued characters}
      Show that if we repeat the definition of a Cheeger-Simons character but for $\Hom(Z_{\ell-1}(M_{n}), \IR)$ then the fieldstrength must have zero periods.\end{exbox}

\SectionWithHeader{The Dancing Hexagon}{The Dancing Hexagon}{sec:DancingHexagon}
  
  We now analyze the structure of $\widecheck{H}^{\ell}(M_{n})$ as an Abelian group through a number of interlocked exact sequences. Our goal in this section is to unpack and  understand the following diagram:
  \begin{equation}\label{eq:DancingHexagon}
  \adjustbox{scale=1,center}{%
   \begin{tikzcd}[row sep=large,column sep=large]
       \textcolor{OliveGreen}{0} \arrow[rd,OliveGreen] &  &  &  & \textcolor{Violet}{0} \\
        & \textcolor{OliveGreen}{\Omega^{\ell-1}(M_{n})/\Omega^{\ell-1}_{\IZ'}(M_{n})} \ar[phantom, "\boxed{\substack{\text{TOP.}\\\text{TRIVIAL}}}"above=13pt] \arrow[rd,OliveGreen] \arrow[rr, dashed, "d",blue] & & \Omega^{\ell}_{\IZ'}(M_{n}) \ar[phantom, "\boxed{\substack{\text{FIELD}\\\text{STRENGTH}}}"above=13pt] \arrow[ru,Violet] \arrow[rd, "F\mapsto\left\lbrack\frac{F}{2\pi}\right\rbrack",sloped,OrangeRed] \\
       \textcolor{OrangeRed}{H^{\ell-1}(M_{n}; \IR)} \arrow[ru,OrangeRed] \arrow[rd,OrangeRed] & & \textcolor{red}{\widecheck{H}^{\ell}(M_{n})} \arrow[ru, "\chi\mapsto F(\chi)",sloped,swap,Violet] \arrow[rd,"\chi \mapsto c(\chi)", sloped, OliveGreen] & & \textcolor{OrangeRed}{H^{\ell}_{\mathsf{dR}}(M_{n})} \\
       & \textcolor{Purple}{H^{\ell-1}(M_{n}; \IR/\IZ')} \ar[phantom, "\boxed{\substack{\text{FLAT}\\\text{FIELDS}}}"below=8pt] \arrow[ru,Purple] \arrow[rr, dashed, "\beta: \text{Bockstein}",swap,blue]& & \textcolor{OliveGreen}{H^{\ell}(M_{n}; \IZ)}  \ar[phantom, "\boxed{\substack{\text{TOP.}\\\text{CLASS}}}"below=8pt] \arrow[ru, "\otimes\IR", sloped,OrangeRed] \arrow[rd,OliveGreen] & \\
       \textcolor{Purple}{0} \arrow[ru,Purple] &  &  &  & \textcolor{OliveGreen}{0}
   \end{tikzcd}
   }
  \end{equation}
  A dance based on this diagram was choreographed by K. Borkin and A. Selisson \cite{stonybrook-differential-cohomology-premiere}, so we will refer to it as the ``dancing hexagon.'' The hexagon was introduced by 
  J. Simons and D. Sullivan in \cite{Simons2007}, 
 where it is shown that it forms the basis for an axiomatic definition of differential cohomology. 
 \tightfootnote{For a video of Jim Simons presenting the hexagon, see \cite{Simons:HexagonVideo}.}

\subsection{Several Mathematical Preliminaries}\label{subsec:SeveralMathPrelim}

  Before we get going, we need a few math preliminaries. (See texts on group theory and algebraic topology for proofs.)

\subsubsection{Abelian Groups} 
  \begin{enumerate}
      \item Abelian groups have a canonically defined subgroup called the \emph{torsion subgroup} 
      \eqa{
          \mathsf{Tors}(A) &= \{ a \in A ~|~ \exists~n \in \IZ \text{ such that } na = 0 \} ~,
      }
where we use additive notation. As an example, the torsion subgroup of $\mathsf{U(1)}$ is the subgroup 
of roots of unity. It is isomorphic to $\IQ/\IZ$. 

      \item If $A$ is a \underline{finitely generated} Abelian group, then $\mathsf{Tors}(A)$ is a finite Abelian group and,
      \begin{equation}\label{eq:ComponentsFinGenAbelGrp}
          \begin{tikzcd}
              0 \arrow[r] & \mathsf{Tors}(A) \arrow[r, "i"] & A \arrow[r, "\pi"] & \ov{A} \arrow[r] & 0 ~,
          \end{tikzcd}
      \end{equation}
      where $\ov{A} \cong \IZ^{b}$, where $b$ is the ``rank of $A$'' and $A \otimes_{\IZ} \IR \cong \IR^{b}$. The group of symmetries of a crystal is finitely generated. That group is a subgroup of 
      the affine Euclidean group, which is not finitely generated. 

      \item If $A$ is a \underline{compact} topological Abelian group
      \tightfootnote{Technically, we require the Pontryagin dual to be finitely generated.} 
      then, as we have noted in 
      equation \eqref{eq:conn-comp-sec}, we have a short exact sequence (SES):  
  \begin{equation}\label{eq:SES-compact-topological-Abelian-group}
  \begin{tikzcd}
      1 \arrow[r] & A_{0} \ar[phantom, "\substack{\textcolor{red}{\downarrow}\\[+1.0ex]\text{connected Abelian}\\\text{group $\cong \mathsf{U(1)}^{b}$}}"below=8pt] \arrow[r] & A \arrow[r] & \pi_{0}(A) \ar[phantom, "\substack{\textcolor{red}{\downarrow}\\[+1.0ex]\text{finite}\\\text{Abelian}\\ \text{group $\cong \oplus \IZ/n \IZ$} }"below=8pt] \arrow[r] & 1 ~,
   \end{tikzcd}
  \end{equation}
      where $A_0$ is the connected component of the identity. (We write $1$ and not $0$ at the ends of the SES because it is more natural in this context to think about Abelian groups multiplicatively.) This is a compact (normal) subgroup of $A$. 
      We have seen that $\pi_0(A) \cong A/A_0$ must be a group, and because $A$ is compact, it must be a finite group. So, as a topological space, $A$ is a disjoint union of $|\pi_{0}(A)|$ copies of an $r$-dimensional torus. As a simple example consider the homomorphism $\mathsf{U(1)} \times \mathsf{U(1)} \to \mathsf{U(1)}$ given by 
      $(\zeta_1,\zeta_2) \mapsto  (\zeta_1 \zeta_2)^n$ where $n>1$ is an integer. Let $A\subset \mathsf{U(1)}\times \mathsf{U(1)}$ be the kernel of this homomorphism.  The group of 
      components of $A$ is isomorphic to $\IZ/n\IZ$. The connected component of the identity is the 
      subtorus with $\zeta_1 \zeta_2=1$.  

\end{enumerate}

\subsubsection{Some Homological Algebra}

We begin by defining the important notion of the \emph{splitting of an short exact sequence}.
Suppose $0 \to \mathsf{M}_1 \xrightarrow{f} \mathsf{M}_2 \xrightarrow{g} \mathsf{M}_3 \to 0$ is a short exact sequence (SES) of modules over a commutative ring $R$.
\tightfootnote{\label{foot:R-module}In other words, we are working in the category $R-\mathsf{MOD}$ of left modules over the commutative ring $R$. Objects in this category are left $R$-modules and morphisms are $R$-linear maps. Of course, since $R$ is commutative, every right $R$-module can be regarded as a left $R$-module and vice versa. An $R$-module $\mathsf{M}$ is said to be \emph{free} if it has a basis $\{e_k\}$, i.e., if every element $m \in \mathsf{M}$ can be written uniquely as a finite linear combination: $m = \sum_{k} r_k e_k$, with $r_k \in R$ and all but finitely many $r_k = 0$. When $R = \IZ$, the category is isomorphic to the category $\mathsf{ABGROUP}$ of Abelian groups and group homomorphisms (which are $\IZ$-linear maps). An Abelian group is naturally a $\IZ$-module. However, note that any finite Abelian group with nonzero order is not a free $\IZ$-module. For example, for $n \in \IZ_{>0}$, the quotient group $\IZ/n\IZ$ is a $\IZ$-module but not a free $\IZ$-module: any element $x \in \IZ/n\IZ $ satisfies $n\cdot x = 0$, while also $0 = 0 \cdot x$, contradicting the uniqueness of representation in terms of a basis.} 
A \emph{left-splitting} (or \emph{splitting on the left}) of the SES is a homomorphism $r : \mathsf{M}_2 \to \mathsf{M}_1$ (called a \emph{retraction}) such that $r \circ f = \mathsf{id}_{\mathsf{M}_1}$. A \emph{right-splitting} (or \emph{splitting on the right}) of the SES is a homomorphism $s : \mathsf{M}_3 \to \mathsf{M}_2$ (called a \emph{section}) such that $g \circ s = \mathsf{id}_{\mathsf{M}_3}$. A standard theorem in homological algebra -- known as the \emph{splitting lemma} -- states that the existence of a left-splitting is equivalent to the existence of a right-splitting and either is equivalent to the existence of an isomorphism $\mathsf{M}_2 \to \mathsf{M}_1 \oplus \mathsf{M}_3$. Given a retraction $r$, we have $\mathsf{M}_2 \cong \im\,f \oplus \ker\,r$, whereas given a section, we have $\mathsf{M}_2 \cong \im\,s \oplus \ker\,g$. All subsequent statements in this section can be read assuming $R = \IZ$, i.e., we are working in the category of Abelian groups.

Recall that a \emph{chain complex} of Abelian groups is a series of Abelian groups labeled by the integers 
with a \emph{differential}, that is, a series of homomorphisms $\partial$
\be 
\cdots \leftarrow C_{k} \xleftarrow{\partial} C_{k+1} \xleftarrow{\partial}\cdots ~,
\ee
such that   $\partial^2 = 0$. One can define the \emph{homology with integer coefficients} 
$H_\bullet(C_{\bullet};\IZ)$ from the quotients of kernels modulo images of $\partial$:
\be 
H_j ( C_\bullet; \IZ):= \ker\left(\partial: C_j \to C_{j-1}\right)/\im\left(\partial: C_{j+1} \to C_j\right)  ~.
\ee
Then the homology with coefficients in an Abelian group $A$, whose graded components are 
denoted $H_j(C_{\bullet}; A)$, is the homology of the 
chain complex 
\be 
\cdots \leftarrow C_{k}\otimes A  \xleftarrow{\partial\otimes 1} C_{k+1}\otimes A \xleftarrow{\partial\otimes 1}\cdots 
\ee
We define the corresponding \emph{cochain complex} of Abelian groups 
\be  
\cdots \rightarrow C^{k} \xrightarrow{d} C^{k+1} \xrightarrow{d} \cdots
\ee
by taking $C^{k} = \Hom(C_{k}, \IZ)$ and the cohomology with coefficients in $A$ is defined by 
tensoring with $A$. 
The relation of  $H_{\bullet}(C_{\bullet}; A)$  to $H_{\bullet}(C_{\bullet})$
and of 
$H^{\bullet}(C^{\bullet}; A)$  to $H^{\bullet}(C^{\bullet})$ is 
not completely trivial and are governed by the Universal Coefficient Theorems, which say: 
\be
0 \longrightarrow H_{k}(C_{\bullet}; \IZ) \otimes_{\IZ} A \longrightarrow H_{k}(C_{\bullet}; A) \longrightarrow \Tor\big(H_{k-1}(C_{\bullet};\IZ), A\big) \longrightarrow 0 ~, \label{eq:UCT-Homology}
\ee
and 
\be
     0 \longrightarrow \Ext\big(H_{k-1}(C_{\bullet}; \IZ), A\big) \longrightarrow H^{k}(C^{\bullet}; A) \longrightarrow \Hom\big(H_{k}(C_{\bullet}; \IZ), A\big) \longrightarrow 0 ~. \label{eq:UCT-Cohomology}
\ee
   %
      For two finitely generated Abelian groups $A$, $B$, 
      one can define associated Abelian groups $\mathsf{Ext}(A,B)$ and $\mathsf{Tor}(A,B)$. See \cite{Bott1982} or any textbook on homological algebra (e.g., \cite{Weibel1994,Gelfand2003}) for the definitions. We summarize some of their useful properties in Box \ref{box:ExtAndTor}. 
       
\begin{redbox}[box:ExtAndTor]{Useful Properties Of $\Tor$ And $\Ext$ }
  $\Tor$ satisfies the following useful properties:
\begin{itemize}\itemsep -2pt
\item $\Tor(\text{Free}, G) = 0$ for any free group as one of the arguments.
\item $\Tor(\IZ_n, G) = \ker(G \xlongrightarrow{\times n} G)$.
\item $\Tor(A \oplus B, G) = \Tor(A, G) \oplus \Tor(B, G)$.
\item $\Tor(A, B) = \Tor(B, A)$.
\item $\Tor(\IZ_{m}, \IZ_{n}) \cong \IZ_{d}$, where $d = \mathsf{gcd}(m,n)$.
\item $\Tor$ detects the common torsion of its arguments.
\end{itemize}
 $\Ext$ satisfies the following useful properties:
\begin{itemize}\itemsep -2pt
\item $\Ext(\text{Free}, G) = 0$.
\item $\Ext(\IZ_n, G) = G/nG$.
\item $\Ext(A \oplus B, G) = \Ext(A, G) \oplus \Ext(B, G)$.
\item $\Ext(A, G) = 0$ if $G$ is a divisible group.
\tightfootnote{A divisible group is an Abelian group $(G, +)$ with the property that for every $g \in G$ and $n \in \IZ_{>0}$, there exists $x \in G$ such that $nx = g$. Examples of divisible groups are: the group $\IR$ of real numbers under addition, the group $\IQ$ of rational numbers under addition, $\IC^{\times}$ -- the multiplicative group of complex numbers, and $\mathsf{U(1)} \cong \IR/\IZ$. Any quotient of a divisible group is also divisible, so in particular, $\IQ/\IZ$ is a divisible group.}
\item $\Ext(\IZ_{m}, \IZ_{n}) \cong \IZ_{d}$, where $d = \mathsf{gcd}(m,n)$.
\item $\Ext$ also detects torsion.
\end{itemize}
Note that, unlike $\Tor$, the order of arguments in $\Ext$ is important.
\end{redbox}

It is worth noting that the sequences \eqref{eq:UCT-Homology} and \eqref{eq:UCT-Cohomology} split, but not naturally. Consequently, there are \underline{noncanonical} isomorphisms:
\begin{align}
	H_{k}(C_{\bullet}; A) &\cong H_{k}(C_{\bullet};\IZ) \otimes_{\IZ} A \oplus \mathsf{Tor}\big(H_{k-1}(C_{\bullet};\IZ), A\big) ~, \label{eq:UCT-Homology-Noncanonical}
\end{align}
and
\begin{align}
	H^{k}(C^{\bullet};A) &\cong \mathsf{Hom}\big(H_{k}(C_{\bullet};\IZ), A\big) \oplus \Ext\big(H_{k-1}(C_{\bullet};\IZ),A\big) ~. \label{eq:UCT-Cohomology-Noncanonical}
\end{align}

\begin{remark} 
$\,$

\begin{enumerate}
\item The splitting of \eqref{eq:UCT-Homology} can be explained as follows. Let $Z_{k} := \ker\left(\partial: C_k \to C_{k-1}\right)$ and $B_{k} := \im\left(\partial: C_{k+1} \to C_k\right)$. We have a short exact sequence:
\be
  \begin{tikzcd}
      0 \arrow[r] & Z_{k} \arrow[r] & C_{k} \arrow[r] & B_{k-1} \arrow[r] & 0 ~.
  \end{tikzcd}
\ee
Since $C_{k}$ is a free Abelian group (see footnote \ref{foot:R-module}), it has a basis $\{e_i\}_{i \in I}$ labeled by an index set $I$, so that $C_{k} \cong \bigoplus_{i \in I} \IZ \cdot e_i$. Now, without loss of generality, we can choose a basis for  $Z_{k}$, and extend this basis to a basis for  $C_{k}$. (This is always possible for a free Abelian group.) This allows us to write $C_{k} \cong Z_{k} \oplus B_{k-1}'$ where $B_{k-1}'$ is the subgroup generated by the remaining basis elements. It follows by construction that $B_{k-1}' \cong B_{k-1}$, and that every element $c \in C_{k}$ can be written uniquely as $c = z + b'$ with $z \in Z_{k}$ and $b' \in B_{k-1}'$. We now define $r: C_{k} \to Z_{k}$ by $z + b' \mapsto r(z+b') = z$.  The map $r$ is a group homomorphism, and furthermore, for any $z \in Z_{k}$, $r(z) = z$, therefore $\left. r\right|_{Z_k} = \mathsf{id}$. This function $r$ serves as a left-splitting or retraction (recall the splitting lemma above). Note that the choice of $r : C_{k} \to Z_{k}$ is dictated by the choice of complement $B_{k-1}'$ to $Z_{k}$ in $C_{k}$, which in turn depends on the choice of basis. Thus, the  map $r$  is highly noncanonical. 
%
%
%

Let $r: C_{k} \to Z_{k}$ be a splitting of $Z_{k} \hookrightarrow C_{k}$. Then we have a composition of maps:
\begin{align}
    Z_{k}(C_{\bullet} \otimes A) \subset C_{k} \otimes A \xrightarrow{r \otimes 1} Z_{k} \otimes A \to H_{k}(C_{\bullet};\IZ) \otimes A ~,
\end{align}
that takes $B_{k}(C_{\bullet}\otimes A)$ to zero, and induces a map,
\begin{align}
    \rho: H_{k}(C_{\bullet}\otimes A) \to H_{k}(C_{\bullet};\IZ) \otimes A ~,
\end{align}
which satisfies $\rho \circ \alpha = \mathsf{id}$, where $\alpha$ is the map,
\begin{align}
    \alpha: H_{k}(C_{\bullet};\IZ) \otimes_{\IZ} A \to H_{k}(C_{\bullet}\otimes A) \cong H_{k}(C_{\bullet}; A) ~,
\end{align}
in the UCT \eqref{eq:UCT-Homology}. The map $\rho$ yields a splitting of the UCT \eqref{eq:UCT-Homology}. A proof of the splitting of \eqref{eq:UCT-Cohomology} proceeds along similar lines. 

\item Here is an example to illustrate the noncanonical nature of the isomorphism \eqref{eq:UCT-Homology-Noncanonical} (equivalently, the non-natural or non-functorial splitting of \eqref{eq:UCT-Homology}), using the chain complex associated with a manifold (see \autoref{sec:SimplicialSets}). Consider the real projective plane $\IR\IP^2 \cong \IS^2/\IZ_2$, where the quotient by the $\IZ_2$ action identifies antipodal points. The integral homology groups of $\IR\IP^2$ and $\IS^2$ are:
    \begin{align}
        H_{k}(\IR\IP^{2};\IZ) &= \left\{
            \begin{array}{ll}
                 \IZ ~, & \text{for } k = 0 ~,\\
                 \IZ_{2} ~, & \text{for } k = 1 ~, \\
                 0 ~, & \text{otherwise} ~.
            \end{array}
        \right. 
        \quad
         H_{k}(\IS^{2};\IZ) = \left\{
            \begin{array}{ll}
                 \IZ ~, & \text{for } k = 0, 2 ~,\\
                 0 ~, & \text{otherwise} ~.
            \end{array}
        \right.
    \end{align}
    The UCT \eqref{eq:UCT-Homology} for $\IR\IP^2$ at degree $k=2$ with $A = \IZ_2$ reads 
    \begin{align}
        0 \to H_{2}(\IR\IP^2; \IZ) \otimes_{\IZ} \IZ_2 \xrightarrow{\phi} H_{2}(\IR\IP^2; \IZ_2) \xrightarrow{\psi} \mathsf{Tor}\big( H_{1}(\IR\IP^2; \IZ), \IZ_2 \big) \to 0 ~.
    \end{align}
    Upon substituting the homology groups, and using the fact that $\mathsf{Tor}(\IZ_2, \IZ_2) = \IZ_2$ (see Box \ref{box:ExtAndTor}), this yields
    \begin{align}
        0 \to 0 \to H_{2}(\IR\IP^2; \IZ_2) \xrightarrow{\psi} \IZ_2 \to 0 \implies H_{2}(\IR\IP^2; \IZ_2) \cong \IZ_2 ~.
    \end{align}
    Carrying out the same exercise for $\IS^2$, we find
    \begin{align}
        0 \to \IZ \otimes \IZ_2 \xrightarrow{\phi} H_{2}(\IS^2; \IZ_2) \xrightarrow{\psi} \mathsf{Tor}\big( 0, \IZ_2 \big) \to 0 ~,
    \end{align}
    which, upon substituting the homology groups, yields
    \begin{align}
        0 \to \IZ_2 \to H_{2}(\IS^2; \IZ_2) \to 0 \implies H_{2}(\IS^2; \IZ_2) \cong \IZ_2 ~.
    \end{align}
    Now there is a map -- called the pinch map -- denoted by 
    \begin{align}
        p: \IR\IP^{2} \to \IS^{2} ~, \label{eq:pinch-map}
    \end{align}
    which collapses the 1-skeleton $\IR\IP^{1} \cong \IS^{1} \subset \IR\IP^{2}$ to a point.
    \tightfootnote{This terminology was introduced in \autoref{sec:ClassifyingSpace}.} 
    Importantly, the pinch map is not nullhomotopic. But it induces a map on the homology,
    \begin{align}
        p_{*} : H_{2}(\IR\IP^2; \IZ_2) \to H_{2}(\IS^2; \IZ_2) ~, \label{eq:induced-pinch}
    \end{align}
    which turns out to be zero. Here is why: First, note that the $\IZ_2$ in $H_{2}(\IR\IP^2; \IZ_2)$ originated from the torsion in $H_{1}(\IR\IP^2; \IZ)$, whereas the $\IZ_2$ in $H_{2}(\IS^2; \IZ_2)$ originated from tensoring $\IZ$ with $\IZ_2$. So, if the isomorphism \eqref{eq:UCT-Homology-Noncanonical} (for $A = \IZ_2$) were to split canonically, the induced map \eqref{eq:induced-pinch} would have to send the torsion summand in the homology of $\IR\IP^2$ to the tensor summand in the homology of $\IS^2$, but there is no natural map that will do so. The pinch map \eqref{eq:pinch-map} induces a map of UCT short exact sequences:
    \begin{equation}
        \begin{tikzcd}
            0 \arrow[r] & \overbrace{H_{2}(\IR\IP^2; \IZ) \otimes_{\IZ} \IZ_{2}}^{=0} \arrow[d, "0"] \arrow[r] & H_{2}(\IR\IP^2; \IZ_2) \arrow[r, "\cong"] \arrow[d, "\cong"] & \overbrace{\mathsf{Tor}\big( H_{1}(\IR\IP^{2};\IZ), \IZ_2 \big)}^{\IZ_2} \arrow[r] \arrow[d,"0"] & 0 \\
            0 \arrow[r] & H_{2}(\IS^2;\IZ)\otimes_{\IZ} \IZ_2 \arrow[r, "\cong"] & H_{2}(\IS^{2}; \IZ_{2}) \arrow[r] & \underbrace{\mathsf{Tor}\big( H_{1}(\IS^2;\IZ), \IZ_2 \big)}_{0} \arrow[r] & 0 
        \end{tikzcd}
    \end{equation}
    from which it is evident that the splitting of the UCT short exact sequence is not functorial. Recall that $\IR\IP^2$ is built out of $\IR\IP^1 \cup e^{2}$, where $e^{2}$ denotes the 2-cell, and the pinch map \eqref{eq:pinch-map} collapses $\IR\IP^1$ mapping it to the base point of $\IS^2$ and retaining only the 2-cell, which is mapped homeomorphically to $\IS^2$. However, there is no integral 2-cycle in $\IR\IP^2$ in the first place, since $H_{2}(\IR\IP^2; \IZ) = 0$.
\end{enumerate}
\end{remark}

\begin{exbox}{Pinch map and UCT cohomology}
    Use the pinch map \eqref{eq:pinch-map} to exhibit the noncanonical nature of the UCT cohomology isomorphism \eqref{eq:UCT-Cohomology-Noncanonical}.
\end{exbox}

A cochain complex $B^\bullet$  of Abelian groups is sometimes called a \emph{differential graded Abelian group}. This means $B$ has an integer grading:
      \eqa{
        B &= \bigoplus_{k\in \IZ} B^{k} ~,
      }
      and there is a differential operator that raises degree by $1$ and squares to zero: $d^2 = 0$. In more physical terms, $B$ has a fermion number operator $F$ which has a spectrum that is integers and has a $\CQ$ symmetry, i.e., a $\CQ$ operator with charge $1$ satisfying $\CQ^2 = 0$ and $[\CQ, F] = \CQ$ (this means $\CQ$ has charge $1$ under the fermion number operator). This is entirely equivalent to the math description of a cochain complex.

When we have a differential graded subgroup $A^\bullet \subset B^\bullet$, there is an important 
construction known as the \emph{Bockstein map} that will be used below. To say that  $A^\bullet \subset B^\bullet$ is a differential subgroup means that  $d(A)\subset A$,
%
%
where $d(A)$ uses the restriction of $d$ to $A$. From this, we deduce that the quotient $B/A$ is also a differential Abelian group.  $B/A$ is the set of equivalence classes written as $(b+A) \in B/A$.
We can define a differential on $B/A$ by setting   $d(b+A) := db + A$. To see that this definition makes sense, one needs to check that 
the formula does not depend on representative $b$ of the equivalence class $b+A$, and this is where 
we use $d(A) \subset A$.  It follows immediately that $d^2=0$ on $B/A$. 
We have a sequence of differential-graded Abelian groups: 
      \begin{equation}
        \begin{tikzcd}
            0 \arrow[r] & A \arrow[r,"i"] & B \arrow[r] & B/A \arrow[r] & 0 ~.
        \end{tikzcd}
    \end{equation}

Now suppose that   $d(b+A) = 0 + A$, that is, suppose that $b+A$ is in the kernel of $d$ acting on $B/A$. 
In terms of a representative $b\in B$ of $b+A$, this means that   $db \in A$. Let us denote $s  = db$. 
It is an element of $A$ and need not be zero. In general, for an element $b+A$ in the kernel of $d$ on $B/A$, it might well happen that there is no representative $b$ that makes $s=0$. Moreover, we also have $ds=0$ because $d(db) = d^2 b =0$. Therefore, $s$ determines a cohomology class of the cohomology of $A$.  Note that if $b'\in B$ is another representative of $b+A$, then $b' = b +t$ where $t\in A$.  Therefore, $s' = s + dt$, so 
$s$ and $s'$ differ by an image of $d$ acting on $A$. Therefore, the cohomology classes of 
$[s]$ and $[s']$ in the cohomology of $A$ are the same, and therefore, $[s]$ only depends on $b+A \in B/A$, and not on the choice of representative. 
If $b+A$ has degree $k$ in $B/A$, then $b \in B$ has degree $k$, so $s$ has degree $(k+1)$. So, we have constructed a degree $1$ map,
    \eqas{
      \beta: H^{k}(B/A) &\longrightarrow H^{k+1}(A) \\
      [b+A] &\longmapsto [s] = [db] ~.
    }
    Note well that $s$ is in the image of $d$ acting on $B$, but not necessarily in the image of $d$ acting on $A$. So it might well define a nonzero cohomology class in   $H^{\bullet}(A)$!   The map $\beta$ is called the \emph{Bockstein map}.
      

\subsubsection{Some Algebraic Topology}\label{subsubsec:some-alg-top}

Recall that in \autoref{sec:SimplicialSets} we defined simplicial sets 
$\CX(M)$ associated with any manifold
\tightfootnote{One can generalize manifold to a much wider class of topological spaces.}
$M$. The groups of singular cochains $C^k(M)$ are the free Abelian groups generated by $\CX(M)_j$ and the face maps \eqref{eq:SimplexMaps-X(M)} can be extended by linearity 
to define a differential $\partial$. From this complex, we can define the homology and cohomology groups $H_k(M;\IZ)$ and $H^k(M;\IZ)$. 
It is shown in books on algebraic topology that for a compact manifold, these Abelian groups are finitely generated. 

Here is a simple example of some integral cohomology groups and corresponding groups with coefficients: 
 \begin{redbox}[box:lensspace]{Example: Lens Space}
         The Lens space $L_{m}$ is defined as the quotient $L_{m} := \IS^{3}/\IZ_{m}$ where $\IS^{3} = \{(z_1, z_2) \in \IC^2 ~|~ |z_1|^2 + |z_2|^2 = 1\}$ and the quotient by the $\IZ_{m}$ action identifies $(z_1, z_2) \sim (\omega z_1, \omega z_2)$ where $\omega \in \mu_{m}$ is an element of the multiplicative cyclic group of order $m$.
          \\$\,$\\
         We have the free action of a cyclic group on a simply connected space. So $\pi_{1}(L_{m}; x_0) \cong \IZ_{m}$. By the Hurewicz theorem 
         $\pi_{1}(L_{m};x_0) \cong H_{1}(L_{m}; \IZ)$. Note that earlier (just above equation \eqref{eq:DifferentialH-one} )  we asserted that there is no torsion in $H^{1}$, not $H_{1}$!  By Poincar\'e duality
          $H^{2}(L_{m}; \IZ) \cong H_{1}(L_{m}; \IZ) \cong \IZ_{m}$. 
          (One could also use  the Universal Coefficient Theorem with $\mathsf{Ext}(\IZ_{m}, \IZ) \cong \IZ_{m}$.)  This is an example where the second cohomology is pure torsion. 
         \\$\,$\\
         The group of isomorphism classes of $\mathsf{U(1)}$-bundles on any manifold is classified by $H^2(M;\IZ)$.
         Thus it is the finite group $\IZ_{m}$ in this case.  So the characteristic class of degree two differential characters on a Lens space is entirely torsion. 
         \\$\,$\\
         There are many ways of producing cochain complexes that compute the cohomology 
         of a manifold. One is singular cohomology as indicated above, but others are often more convenient. For example, one can use a cellular decomposition to produce much smaller cochain complexes with the same cohomology.  For example, in the case of a Lens space, one can use the symmetry 
         of the situation to produce an equivariant cell decomposition of $\IS^3$ well adapted to the quotient. For example, specializing to $m=2$, the $\IZ_{2}$-equivariant cell decomposition of $\IS^n$ is given by using the upper and lower hemispheres: 
        \begin{align}
          e_{j}^{\pm} &= \left\{ (x_1, \ldots, x_j, x_{j+1}, 0, \ldots, 0) \in \IS^{n} ~|~ \mathsf{sign}(x_{j+1}) = \pm \right\} ~.
        \end{align}
         The orientation on $e_{j}^{\pm}$ from $\pm dx^{1}\wedge \cdots \wedge dx^{j}$.
        \eqa{
         &\left.
           \begin{array}{l}
                \partial e_{j}^{+} = e_{j-1}^{+} + e_{j-1}^{-} \\
                \partial e_{j}^{-} = -\big( e_{j-1}^{+} + e_{j-1}^{-} \big) 
           \end{array}
         \right\} i \leq j \leq n ~,\\
         &\,\,\,\partial e_{0}^{\pm} = 0 ~.
        }
    We introduce dual cochains, satisfying
        \begin{align}
             c_{j}^{\alpha}( e_{k}^{\beta} ) &=\delta^{\alpha\beta}\delta_{jk} ~.
        \end{align}
         This yields the cochain model of $\IR\IP^{n} = \IS^{n}/\IZ_2$:
         \begin{align}
            \IZ(c_0^+ - c_0^-) \xrightarrow{0}  \IZ(c_1^+ + c_1^-) \xrightarrow{2} \IZ(c_2^+ - c_2^-) \xrightarrow{0}  \IZ(c_3^+ + c_3^-) \xrightarrow{2} \cdots
            \label{eq:EquiCochainRPn}
         \end{align}
        Therefore,
       \begin{align}
           H^{j}(\IR\IP^{n}; \IZ) &= \left\{
           \begin{array}{ll}
               \IZ ~, & j = 0 ~,\\
               0 ~, & j \text{ odd }, j < n ~,\\
               \IZ_2 ~, & j \text{ even }, 0 < j < n ~,\\
               \IZ ~, &j = n, \text{ odd } ,\\
               \IZ_2 ~, &j = n, \text{ even } ~.
           \end{array}
           \right.
       \end{align}
        To get the cohomology with $\IZ_2$ coefficients we tensor   the  complex \eqref{eq:EquiCochainRPn} with $\IZ_2$.  This simply makes all differentials $0$: 
      \begin{equation}
        \begin{tikzcd}[row sep=0]
          0 \arrow[r] & \IZ_2 \arrow[r, "0"] & \IZ_2 \arrow[r, "2"] & \IZ_2  \arrow[r,"0"] & \IZ_2 \arrow[r,"2"] & \cdots \\
          \text{cohomology:}    & \IZ_2 & \IZ_2 & \IZ_2 & \IZ_2 & \cdots
        \end{tikzcd}
      \end{equation}
       and we easily conclude that: 
       \begin{align}
          H^{\bullet}(\IR\IP^{n}; \IZ_2) &= \IZ_{2}[x]/(x^{n+1}) ~.
       \end{align}
 \end{redbox}

Next, the group $H^{k}(M_{n}; \IR/\IZ)$ will play an important role in our applications of differential cohomology to higher-form fields so it behooves us to understand the structure of this group. By the universal coefficient theorem,  we know that  $H^{k}(M_{n}; \IR/\IZ) \cong \Hom(H_{k}(M_{n}; \IZ), \IR/\IZ)$, and this is a compact Abelian group. Accordingly, it has a group of connected components. We now determine that group.

      Associated to the short exact sequence of coefficients,
      \begin{equation}
        \begin{tikzcd}
          0 \arrow[r] & \IZ \arrow[r] & \IR \arrow[r] & \IR/\IZ \arrow[r] & 0 ~,
        \end{tikzcd}
      \end{equation}
      is a long exact sequence in cohomology, which we now derive. 
Referring to the definition of the Bockstein map above, we take   $A = C^{\bullet}(M_{n})$ and $B = C^{\bullet}(M_{n}) \otimes \IR$. Now, $B/A = C^{\bullet}(M_{n}) \otimes \IR/\IZ$. We have
      \begin{equation}
         \hspace{-0.5cm} \begin{tikzcd}
              \vdots & \vdots & \vdots & \vdots & \vdots\\
              0 \arrow[r] & C^{k+1}(M_{n}; \IZ) \arrow[r] \arrow[u,"d",swap] & C^{k+1}(M_{n}; \IR) \arrow[r, "\pi"] \arrow[u,"d",swap] & C^{k+1}(M_{n}; \IR/\IZ) \arrow[r] \arrow[u,"d",swap] & \cdots \\
              0 \arrow[r] & C^{k}(M_{n}; \IZ) \arrow[u,"d",swap] \arrow[r] & C^{k}(M_{n}; \IR) \arrow[u,"d",swap] \arrow[r,"\pi"] & C^{k}(M_{n}; \IR/\IZ) \arrow[u,"d",swap] \arrow[r] & \cdots \\
               0 \arrow[r] & C^{k-1}(M_{n}; \IZ) \arrow[u,"d",swap] \arrow[r] & C^{k-1}(M_{n}; \IR) \arrow[u,"d",swap] \arrow[r,"\pi"] & C^{k-1}(M_{n}; \IR/\IZ) \arrow[u,"d",swap] \arrow[r] & \cdots \\
               \vdots & \vdots \arrow[u,"d",swap] & \vdots \arrow[u,"d",swap] & \vdots \arrow[u,"d",swap] & \vdots
          \end{tikzcd}
      \end{equation}
        For $\ov{a} \in Z^{k}(M_{n}; \IR/\IZ)$, lift to $a \in C^{k}(M_{n}; \IR)$, $\pi(da) = 0$, so $da$ has a lift to $b \in C^{k+1}(M_{n}; \IZ)$, but $db=0$, so $b \in Z^{k+1}(M_{n}; \IR/\IZ)$. With a little thought, one finds that we have a long exact sequence:
      \begin{equation}\label{eq:LES} 
         \hspace{-0.5cm} \begin{tikzcd}
  \cdots \rar & H^{k}(M_{n}; \IZ) \rar{\iota} & H^{k}(M_{n}; \IR) \rar{\psi}
             \ar[draw=none]{d}[name=X, anchor=center]{}
    & H^{k}(M_{n}; \IR/\IZ) \ar[rounded corners,
            to path={ -- ([xshift=2ex]\tikztostart.east)
                      |- (X.center) \tikztonodes
                      -| ([xshift=-2ex]\tikztotarget.west)
                      -- (\tikztotarget)}]{dll}[at end,swap]{\beta} \\      
 &   H^{k+1}(M_{n}; \IZ) \rar{\iota} & H^{k+1}(M_{n}; \IR) \rar & H^{k+1}(M_{n}; \IR/\IZ) \rar & \cdots 
\end{tikzcd}
      \end{equation}
      Here $\beta$ is the Bockstein map. Since the sequence is exact,
      \eqa{
        \mathsf{im}(\beta) \cong \ker(\iota)  & \cong \mathsf{Tors\,} H^{k+1}(M_{n}; \IZ) ~.
      }
      But 
      \eqa{
         \mathsf{im}(\psi) &= H^{k}(M_{n}; \IR)/\iota\big( H^{k}(M_{n}; \IZ) \big) \nonumber\\
         &= \text{Connected component of } H^{k}(M_{n}; \IR/\IZ) ~.
      }
      Therefore,
      \beqa{ \label{eq:connected-components-and-torsion-identity}
         \pi_{0}\left( H^{k}(M_{n}; \IR/\IZ) \right) &\cong \mathsf{Tors\,} H^{k+1}(M_{n}; \IZ) ~.
      }

Of course, we also have, by \eqref{eq:ComponentsFinGenAbelGrp}, 
\begin{equation}
    \begin{tikzcd}
    0 \arrow[r] & \mathsf{Tors\,}H^{k}(M_{n}; \IZ) \arrow[r] & H^{k}(M_{n}; \IZ) \arrow[r] & \overline{H}^{k}(M_{n}; \IZ) \cong \IZ^{b_{k}} \arrow[r] & 0 ~.
    \end{tikzcd}
\end{equation}
Note that $H^{k}_{\mathsf{dR}}(M_{n}) \cong H^{k}(M_{n}; \IZ) \otimes \IR \cong H^{k}(M_{n}; \IR) \cong \IR^{b_{k}}$.

\begin{exbox}{K\"unneth theorem}
Using the K\"unneth theorem (look it up), show that 
        \begin{equation}
            \begin{tikzcd}
                0 \arrow[r] & \IR/\IZ \oplus \IR/\IZ \arrow[r] & H^{3}\big( L_{m} \times L_{m}; \IR/\IZ \big) \arrow[r] & \IZ_{m} \arrow[r] & 0 ~.
            \end{tikzcd}
        \end{equation}
\end{exbox}

\begin{redbox}[box:useful-sec-torsion]{A Useful Short Exact Sequence}
   Another useful short-exact sequence that we will encounter below is
   \be
     0 \longrightarrow H^{k}(M_{n}; \IZ)\otimes \IR/\IZ \longrightarrow H^{k}(M_{n}; \IR/\IZ) \longrightarrow \mathsf{Tors\,}H^{k+2}(M_{n}; \IZ) \longrightarrow 0 ~,
   \ee
   where $H^{k}(M_{n}; \IZ)\otimes \IR/\IZ$ is the connected component of the identity of $H^{k}(M_{n}; \IR/\IZ)$, and $\mathsf{Tors\,}H^{k+2}(M_{n}; \IZ)$ is the group of connected components of $H^{k}(M_{n}; \IR/\IZ)$ as shown above.
\end{redbox}

\subsection{Flat Characters}\label{subsec:FlatCharacters}

We now return to the dancing hexagon. Let us begin with the group homomorphism
$\chi \mapsto F(\chi)$. The   very definition of a differential character assigns to a character $\chi$ a form $F \in \Omega_{\IZ'}^{\ell}(M_{n})$, such that,
\be
\chi(\p \CB) = \exp\left[ \imag  \int_{\CB} F \right] ~.
\ee
We will call the group homomorphism,
  \be
     \chi \mapsto F ~,
  \ee
  the ``fieldstrength map'': 
  \eqa{
          \widecheck{H}^{\ell}(M_{n}) &\xlongrightarrow[\text{strength}]{\text{field}} \Omega^{\ell}_{\IZ'}(M_{n}) \longrightarrow 0 ~.
   }
   We will see (e.g., from the \v{C}ech model in \autoref{subsec:CechModel}) that this map is surjective, hence the second arrow above.

   The homomorphism $\widecheck{H}^{\ell}(M_{n}) \to \Omega^{\ell}_{\IZ'}(M_{n})$   has a kernel. The kernel is called the set of \emph{flat differential characters}, i.e., characters with vanishing fieldstrength. Having a zero fieldstrength means that if two cycles $\Sigma_{\ell-1}$ and $\Sigma'_{\ell-1}$ are homologous, i.e., $\Sigma_{\ell-1} - \Sigma'_{\ell-1} = \p \CB$, then 
   \be
   \chi(\Sigma_{\ell-1})/\chi(\Sigma'_{\ell-1}) = \exp\left[\imag  \int_{\CB} F \right] = 1  ~,
   \ee
   so $\chi(\Sigma_{\ell-1})= \chi(\Sigma'_{\ell-1})$. It follows that flat characters define homomorphisms $\chi: Z_{\ell-1}(M_{n}) \to \mathsf{U(1)}$ that only depend on the homology class of $\Sigma_{\ell-1}$. One can show that
   \tightfootnote{Applying the UCT \eqref{eq:UCT-Cohomology} to a manifold $M_{n}$ at degree $j = \ell-1$, we get the noncanonical isomorphism \eqref{eq:UCT-Cohomology-Noncanonical}: $H^{\ell-1}(M_{n}; \mathsf{U(1)}) \cong \Ext(H_{\ell-2}(M;\IZ), \mathsf{U(1)}) \oplus \Hom(H_{\ell-1}(M;\IZ), \mathsf{U(1)})$. Since $\mathsf{U(1)}$ is a divisible group (see Box \ref{box:ExtAndTor}), $\Ext(-,\mathsf{U(1)})$ vanishes, and this reduces to \eqref{eq:hom-homology-cohomology-u1}.}
  \eqa{
     \Hom\big( H_{\ell-1}(M_{n}); \mathsf{U(1)} \big) &\cong H^{\ell-1}(M_{n}; \mathsf{U(1)}) ~, \label{eq:hom-homology-cohomology-u1} 
  }
so we identify the group of flat characters with $H^{\ell-1}(M_{n}; \mathsf{U(1)})$, 
or $H^{\ell-1}(M_{n}; \IR/\IZ')$, if we are using additive notation.  In general, 
if $\phi \in H^{\ell-1}(M_{n}; \IR/\IZ')$, we denote the corresponding flat differential 
character by $\widecheck \phi$.

We will need to understand the structure of the group $H^{\ell-1}(M_{n}; \IR/\IZ')$.
  This is a compact topological Abelian group.  
Recall that if $A$ is a compact topological Abelian group, then we have a short exact sequence \eqref{eq:SES-compact-topological-Abelian-group}, 
%
  %
  where $A_0$ is the connected component of the identity, and the sequence in general does not split.  In the present case,  $A = H^{\ell-1}(M_{n}; \IR/\IZ')$, and $A_{0} = H^{\ell-1}(M_{n}; \IZ) \otimes \IR/\IZ'$. 
  The latter group $A_{0} = H^{\ell-1}(M_{n}; \IZ) \otimes \IR/\IZ'$ should be identified 
  with the group of ``Wilson lines'' or ``Wilson surfaces'' that can be continuously 
  deformed to the identity character. In \autoref{subsec:TopTrivialChars},
  we will define topologically trivial characters and will see that $H^{\ell-1}(M_{n}; \IZ) \otimes \IR/\IZ'$  is the group of topologically trivial flat characters.

To understand the group of components of $H^{\ell-1}(M_{n}; \mathsf{U(1)})$, we use the exact sequence \eqref{eq:LES} with the Bockstein map,
  \eqa{\label{eq:degree-l-Bockstein}
     \beta: H^{\ell-1}(M_{n}; \IR/\IZ') &\longrightarrow \mathsf{Tors}\big( H^{\ell}(M_{n}; \IZ) \big) 
     \subset H^{\ell}(M_{n}; \IZ) ~,
  }
  to identify,
  \eqa{
    \pi_{0}(H^{\ell-1}(M_{n}; \IR/\IZ')) &\cong \mathsf{Tors}(H^{\ell}(M_{n}; \IZ)) ~.
  }
  It follows that there can be a nontrivial group of connected components of the group of flat fields, whenever there is torsion in the cohomology. We stress again that the group $H^{\ell-1}(M_{n}; \IR/\IZ')$ is in general \underline{not} a direct product of $H^{\ell-1}(M_{n}; \IR/\IZ')_{0}$ and $\pi_{0}(H^{\ell-1}(M_{n}; \IR/\IZ'))$. The topological 
  class of $\check \phi$ for $\phi \in H^{\ell-1}(M_{n}; \mathsf{U(1)})$ is: 
\be\label{eq:TopClassFlatCharacter}
c(\check \phi) = - \beta(\phi) ~ . 
\ee
This follows immediately from the cochain model of differential characters 
\eqref{eq:HS-Cocycle} and the definition of the Bockstein map. (The minus sign 
leads to the slightly awkward property that the arrow in the bottom triangle 
of \eqref{eq:DancingHexagon} is the map $\phi \mapsto - \check{\phi}$.)

As an example, consider a Lens space $M = L_{m} = \IS^{3}/\IZ_{m}$ discussed in Box \ref{box:lensspace}, and take $\ell=2$. In this case,  $H^{1}(M; \IR/\IZ') \cong H^{2}(M; \IZ) \cong \IZ/m\IZ$, and $\pi_{1}(L_{m}, x_{0}) = \IZ/m\IZ$. Let   $\gamma$ be a generator of $\pi_1(L_m,x_0)$. Every closed loop in $L_{m}$ is isotopic to $\gamma^{r}$ for some $r \in \IZ_{m}$. For each $k \in \IZ/m\IZ$, we define a character by saying that if a connected 1-cycle $\Sigma_1$ is isotopic to $\gamma^r$, then 
\eqa{
    \chi_{k}(\Sigma_1) &:= \exp\left( \frac{2\pi \imag  r k}{m} \right) ~, \quad k \in \IZ_{m} ~. \label{eq:lensflatcharacter}
  }
We then extend to all $1$-cycles by demanding that $\chi$ be a homomorphism. Clearly, 
the value only depends on the homology class of the $1$-cycle and therefore, the character is a flat character. The group of flat characters is isomorphic to $\IZ/m\IZ$.

\subsection{Topologically Trivial Characters}\label{subsec:TopTrivialChars}

As we have mentioned, there is another group homomorphism, 
\be 
c: \widecheck{H}^\ell(M_n) \rightarrow H^\ell(M_n;\IZ) ~,
\ee
known as the ``characteristic class,'' or ``topological class'' of the character. It is best defined using explicit models for differential characters in \autoref{sec:ModelDiffCoh}. 

The kernel of $\chi \mapsto c(\chi)$ is the subgroup of 
\emph{topologically trivial characters}. These are the characters for which there exists a   globally well-defined form $A \in \Omega^{\ell-1}(M_{n})$, such that
  \eqa{
    \chi(\CW_{\ell-1}) &= \exp\left(\imag  \int_{\CW_{\ell-1}} A \right) ~. \label{eq:TOpTrivChar}
  }
  We denote such characters as $\chi_A$, where, we stress,  $A$ is \underline{globally well-defined} on $M_n$. 

It is important to recall that a differential character only encodes   gauge-invariant information. 
Accordingly, for topologically trivial characters, there can be different forms $A, A'\in \Omega^{\ell-1}(M_n)$ so that $\chi_A = \chi_{A'}$. Plugging into \eqref{eq:TOpTrivChar}, we see that 
$\omega:= A'-A$ has the property that the periods of $\omega$ are in $\IZ'$. Conversely, 
shifting $A\mapsto A' + \omega$, where $\omega$ has periods in $\IZ'$, does not change the character. 
So we can think of these as \emph{gauge transformations}, and here we make contact with the standard literature on higher form gauge fields. We see that the kernel of the topological class homomorphism is:
\be 
\ker(c) = \Omega^{\ell-1}(M_n)/\Omega^{\ell-1}_{\IZ'}(M_n) ~.
\ee

As we have noted before, if $\omega \in\Omega^{\ell-1}_{\IZ'}(M_n) $, then it must be closed. 
Physicists tend to distinguish between two kinds of gauge transformations: The ``large gauge transformations'' are those which shift $A$ by an $\omega$ that has nonzero periods. Within 
the group of gauge transformations $\Omega^{\ell-1}_{\IZ'}(M_n)$, there is an important subgroup 
of ``small gauge transformations'' for which the periods of $\omega$ all vanish. Therefore, $\omega$ is not 
only closed but exact, and we can write the gauge transformation as: 
  \eqa{
     A &\mapsto A + d\Lambda ~, \quad \text{ for } \quad \Lambda \in \Omega^{\ell-2}(M_{n}) ~.
  }
We have now explained the short exact sequences on the south-west to north-east and north-west to south-east diagonals in  the dancing hexagon \eqref{eq:DancingHexagon} above. Next, we discuss the compatibilities of the homomorphisms.

\subsection{Compatibilities Of The Maps }

The blue dashed arrow at the top of the dancing hexagon \eqref{eq:DancingHexagon} expresses the 
evident fact that if we define a topologically trivial character $\chi_A$ then, by Stokes' theorem, $F=dA$.
Thus, the triangle in the northern part of the diagram commutes. Similarly, the topological class of a 
flat character $\widecheck \phi$ based on $\phi \in H^{\ell-1}(M_n; \IR/\IZ')$ is expressed in terms of the Bockstein map $c(\widecheck \phi) = -  \beta(\phi)$, so the diagram in the southern part also commutes. (Recall the arrow pointing to the northeast has the awkward sign $\phi \mapsto -\widecheck \phi$.)

Now, let us consider the square in the east. There is a map:
\be 
H^\ell(M_n; \IZ) \to H^\ell(M_n;\IZ)\otimes \IR \cong H^{\ell}(M_n; \IR) \cong H^\ell_{\mathsf{dR}}(M_n) ~,
\ee
where the last isomorphism follows from de Rham's theorem. The kernel of 
$H^\ell(M_n; \IZ) \to H^\ell(M_n;\IZ)\otimes \IR$ is the torsion subgroup of $H^\ell(M_n; \IZ) $.
The commutativity of the square says that 
the topological class of the character, reduced modulo torsion, is captured by the de Rham class 
of $F/2\pi$:
\be\label{eq:ClassesAgree}
\left[\frac{F}{2\pi}\right]_{\mathsf{dR}} = c(\chi)_{\IR} ~.
\ee

Finally, we consider the square in the west. A class $\alpha \in H^{\ell-1}(M_n; \IR)$ can, again by 
de Rham's theorem, be represented by a closed $\ell-1$ form $\alpha = [A]$. Once we choose $A$, we can define a topologically trivial character $\chi_A$, and $\chi_A$ is a flat character because, as we just said, $A$ is closed. Note that two different representatives $A$ and $A'$ of $\alpha$ differ by an exact form, and therefore define the same character $\chi_A = \chi_{A'}$, so the map, 
\be
H^{\ell-1}(M_n; \IR)\to \Omega^{\ell-1}(M_n)/\Omega^{\ell-1}_{\IZ'}(M_n) ~, \label{eq:map38p46}
\ee
is well-defined.
%
%
 In fact, if we shift $\alpha \mapsto \alpha + \omega$, where 
$\omega\in H^{\ell-1}(M_n;\IZ')$, 
we also produce the same character, so the map 
descends to a map, 
\be
H^{\ell-1}(M_n;\IR)/H^{\ell-1}(M_n;\IZ') \to \Omega^{\ell-1}(M_n)/\Omega^{\ell-1}_{\IZ'}(M_n) ~. \label{eq:38p49}
\ee
On the other hand, there is also a natural map, 
\be 
H^{\ell-1}(M_n;\IR)/H^{\ell-1}(M_n;\IZ') \to H^{\ell-1}(M_n; \IR/\IZ') ~,
\ee
which maps into the topologically trivial flat characters. 
It is easy to check that the square on the western part of the dancing hexagon \eqref{eq:DancingHexagon} also commutes. 

The discussion of equations \eqref{eq:DecTopTrivChar-2} and \eqref{eq:DecTopTrivChar-2} above 
generalizes immediately to higher degree forms, and so we have the general, noncanonical, decomposition of the group of differential characters into a product of a  connected torus, a finitely generated Abelian group, and an infinite-dimensional vector space: 

 In general, we have a noncanonical decomposition of the infinite-dimensional Abelian group $\widecheck{H}^{\ell}(M_{n})$ as:
\be\label{eq:hcheckl-decomp}
      \widecheck{H}^{\ell}(M_{n}) = \mathbb{T}_{\ell-1} \times V_{\ell} \times \Gamma_{\ell} ~,
\ee
where the factors are: 
\eqa{
      \mathbb{T}_{\ell-1} &= \text{Connected torus of Wilson lines} = H^{\ell-1}(M_{n}; \IZ)\otimes \IR/\IZ' \cong \mathsf{U(1)}^{b_{\ell-1}} ~,\\
      V_{\ell} &= \text{$\infty$-dimensional vector space} = \mathsf{im}(d^\dagger: \Omega^{\ell+1} \to \Omega^{\ell}) ~,\\
      \Gamma_{\ell} &= \text{Finitely generated Abelian group} = H^{\ell}(M_{n}; \IZ) \leftarrow \text{ topological classes}~.
   }
   This is a natural generalization of \eqref{eq:hcheck2-decomp}.

\bigskip
 \noindent \textbf{Example:} $\ell = 1$, $M = \IS^{1}$. We have,
  \eqa{
    \widecheck{H}^{1}(\IS^1) &= \mathsf{Map}\big( \IS^1 \to \mathsf{U(1)} \big) = \text{loop group } \mathsf{LU(1)} ~.
  }
  Let us identify the domain $\IS^1 \cong \IR/\IZ$ with coordinate $\sigma \sim \sigma + n$, $n \in \IZ$. An element of $\widecheck{H}^{1}(\IS^1)$ can be expanded as:
  \eqa{
    f(\sigma) &= \exp\left[ \imag  \phi_{0} + 2\pi \imag  w \sigma + \sum_{n\neq 0}\frac{1}{n}\phi_{n}e^{2\pi \imag  n \sigma} \right] ~.  \label{eq:ell=1-modes}
  }
  Here $\phi_{0} \in \IR/\IZ'$ represents the degree of freedom in the flat field, and $w \in \IZ$, which corresponds to the winding number, denotes the characteristic class. Since $f$ is a pure phase valued in 
  $\mathsf{U(1)}$, we must have $\phi_n^* = - \phi_{-n}$ for $n\not=0$.

\begin{remark} 
In the 2d conformal field theory of a single periodic scalar -- sometimes called the Gaussian 2d CFT --  the above decomposition corresponds to the mode decomposition of a free field. 
\end{remark}

\SectionWithHeader{Models Of Differential Cohomology}{Models Of Differential Cohomology}{sec:ModelDiffCoh}

As we have stressed, $\widecheck{H}^{\ell}(M_{n})$ is the set of \underline{gauge equivalence classes} of field configurations of an ``$(\ell-1)$-form gauge potential.'' In gauge theory, working all the time with $\CA/\CG$ is difficult. One wants to work with gauge potentials, i.e., elements of $\CA$. This is an affine linear space. We can write Lagrangians -- it is great for locality. There are a lot of good reasons to find a better local model for gauge equivalence classes. The same goes for differential cohomology. (As we have seen, for $\ell=2$,  $\CA/\CG$ is indeed the same thing as $\widecheck{H}^{2}(M_{n})$.)

We will discuss two kinds of models:
  \begin{enumerate}\itemsep 0pt
    \item \v{C}ech model.
    \item Hopkins-Singer model.
  \end{enumerate}

\subsection{\v{C}ech Model}\label{subsec:CechModel}

This formulation goes back to P. Deligne, and indeed in the context of complex geometry, Cheeger-Simons cohomology is also known as Deligne cohomology. It was first introduced into physics by O. Alvarez 
\cite{Alvarez:1984es} and independently, by K. Gawedzki 
\cite{Gawedzki:1987ak}. 

The basic idea is that on a contractible space, all field configurations must be topologically trivial, and there is no room for the extra gauge-invariant information encoded in the group of flat fields or characteristic classes. 
The extra gauge-invariant information arises from patching local data, but one must include multiple intersections of patches (to a degree determined by the degree of the differential character).  To implement this idea, we choose a good cover $\{\CU_{\alpha}\}$ of $M_{n}$, that is, a cover such that all $U_{\alpha_1\ldots\alpha_k} = \CU_{\alpha_1}\cap \cdots \cap \CU_{\alpha_k}$ are contractible. 
One can show (see, for example, \cite{Bott1982}) that such covers always exist on manifolds.

We describe the first 3 cases, for $\ell = 1, 2, 3$ quite explicitly.

\paragraph{\underline{$\ell=1$: The periodic scalar.}}

We begin with the globally defined fieldstrength $F \in \Omega^1(M_n)$, 
which is, moreover, closed:  $dF=0$. Restricting $F$ to each patch   $\CU_\alpha$, we have locally defined 1-forms $F_{\alpha} \in \Omega^{1}(\CU_{\alpha})$ and, because $dF_{\alpha}=0$, and because $\CU_{\alpha}$ is contractible $F_{\alpha} = d\phi_{\alpha} = - \imag  f_{\alpha}^{-1} df_{\alpha}$, where $f_{\alpha}: \CU_{\alpha} \to \mathsf{U(1)}$ has a well-defined logarithm $\log\,f_{\alpha} = \imag \phi_{\alpha}$. Now,
on the patch overlaps $\CU_{\alpha\beta}$, we have $F_{\alpha}-F_{\beta} = 0$ because $F$ is globally well-defined, but this implies that 
\eqa{
  f_{\alpha}^{-1}df_{\alpha} - f_{\beta}^{-1}df_{\beta} = 0 \implies  f_{\alpha}/f_{\beta}  = \text{const. on \,$\CU_{\alpha\beta}$.}
}

With this line of reasoning, we now come to an important choice. 
We can denote the constants in $\CU_{\alpha\beta}$ as: 
\be 
\phi_{\alpha} - \phi_{\beta} = 2\pi \imag  r_{\alpha\beta} ~,
\ee
with $r_{\alpha\beta} \in \IR$. From their definition,
\eqa{
\text{On $\CU_{\alpha\beta\gamma}$} : r_{\alpha\beta} + r_{\beta\gamma} + r_{\gamma\alpha} = 0 ~. \label{eq:ostar}
}
A collection of real numbers $r_{\alpha\beta}$ on $\CU_{\alpha\beta}$ satisfying \eqref{eq:ostar} is known as a ``\v{C}ech 2-cocycle''. It is shown in textbooks (e.g., see Bott-Tu \cite{Bott1982}) that such a cocycle determines a cohomology class in $H^{1}(M_{n}; \IR)$.

Now, we caution the reader that what we have described at this point is \underline{not yet} a Cheeger-Simons differential character. To get differential characters, we make the additional assumption that 
\be\label{eq:Quantize-ralphabeta}
r_{\alpha\beta} = n_{\alpha\beta} \in \IZ ~.
\ee
Note that this implies that on $\CU_{\alpha\beta}$, 
$f_\alpha/f_\beta =1$, that is $f_\alpha = f_\beta$, but this implies that there is a single globally defined function $f: M_n \to \mathsf{U(1)}$. We have already noted that $\widecheck{H}^1(M_n)$ can be identified with the Abelian group of continuous maps $M_n \to \mathsf{U(1)}$, so with the extra quantization condition \eqref{eq:Quantize-ralphabeta}, we determine an element of differential cohomology of degree $\ell=1$.  
So in the Cech description, an element of $\widecheck{H}^1(M_n)$ is a collection 
$(F_\alpha, n_{\alpha\beta})$ where $n_{\alpha\beta}$ define an integral Cech 1-cocycle, which in turn determines an integral cohomology class 
$H^1(M_n; \IZ)$.  This class is the topological class of the differential character.

\medskip

\paragraph{\underline{$\ell=2$: $\mathsf{U(1)}$-bundle with   connection.}}

We now apply the same line of reasoning for the case $\ell=2$. We 
begin with a globally defined closed $2$-form $F \in \Omega^2(M_n;\IR)$. 
Then on each patch, we have the pullback under the inclusion map:   $F_{\alpha} \in \Omega^{2}(\CU_{\alpha})$. Since the curvature is globally well-defined:
\eqa{
 \text{On $\CU_{\alpha\beta}$} : \quad F_{\alpha}-F_{\beta} = 0 ~.
}
On the other hand, $dF_{\alpha} = 0$ implies $F_{\alpha} = dA_{\alpha}$,  
where $A_{\alpha}$ is a globally well-defined 1-form on $\CU_{\alpha}$.
(Here we are using the properties of the good cover.) We should think of 
$A_\alpha$ as defining a connection on a line bundle $L_\alpha = \CU_{\alpha} \times \IC$. Now, 
\eqa{
   \text{On $\CU_{\alpha\beta}$} : \quad dA_{\alpha} - dA_{\beta} = 0 \implies A_{\alpha}-A_{\beta} = d\epsilon_{\alpha\beta} = -\imag  g_{\alpha\beta}^{-1}dg_{\alpha\beta} ~,\label{eq:ostar2}
}
where $g_{\alpha\beta}: \CU_{\alpha\beta} \to \mathsf{U(1)}$ again has a 
well-defined logarithm $\epsilon_{\alpha\beta}$. (Note that $g_{\beta\alpha} = g_{\alpha\beta}^{-1}$.) 
\eqa{
 \text{On $\CU_{\alpha\beta\gamma}$} : \quad \eqref{eq:ostar2} \implies g_{\alpha\beta}g_{\beta\gamma}g_{\gamma\alpha} = \text{ constant.}
}
At this stage, we introduce an important quantization condition and we \underline{impose} the condition that this constant is $1$:
\eqa{
   g_{\alpha\beta}g_{\beta\gamma}g_{\gamma\alpha} &= 1 \quad\text{ on \,$\CU_{\alpha\beta\gamma}$} ~.
}
Thus the $\{g_{\alpha\beta}\}$ are transition functions for a principal $\mathsf{U(1)}$-bundle over $M_{n}$. Note that on $\CU_{\alpha\beta\gamma}$,
\eqa{
  \log g_{\alpha\beta} - \log g_{\alpha\gamma} + \log g_{\beta\gamma} &= n_{\alpha\beta\gamma} \\
  \implies n_{\alpha\beta\gamma} - n_{\alpha\beta\gamma\delta} + n_{\alpha\gamma\delta} - n_{\beta\gamma\delta} &= 0 \,\,\text{ on \,$\CU_{\alpha\beta\gamma\delta}$} ~.
}
This implies that $\{n_{\alpha\beta\gamma}\}_{\CU_{\alpha\beta\gamma}}$ is a \v{C}ech 2-cocycle, which defines an element of \v{C}ech cohomology: 
$\{n_{\alpha\beta\gamma}\} \in H^{2}_{\text{\v{C}ech}}(M_{n}, \{\CU_{\alpha}\})$. 
As shown in Bott and Tu \cite{Bott1982}, the \v{C}ech cohomology is isomorphic to   $H^{2}(M_{n}; \IZ)$, and the resulting cohomology class is just the first Chern class of the principal $\mathsf{U(1)}$-bundle (see footnote \ref{foot:chern-class-principal-bundle}). We have recovered the description of 
$\widecheck{H}^2(M_n)$ as the space of isomorphism classes of principal $\mathsf{U(1)}$-bundles with connection over $M_n$.

The nice thing about this approach is that it is straightforward to extend it to $\ell > 2$.

\paragraph{\underline{$\ell=3$: ``Gerbe connection''.}}
We begin with a globally well-defined closed 3-form $H$. 
On $\CU_{\alpha}$, we have a local fieldstrength
\eqa{
  H_{\alpha} \in \Omega^{3}(\CU_\alpha) \quad \text{ with } \quad dH_{\alpha} = 0 ~.
}
Therefore, there exists $B_{\alpha} \in \Omega^{2}(\CU_{\alpha})$ such that $H_{\alpha} = dB_{\alpha}$.
\eqa{
  &\text{On $\CU_{\alpha\beta}$,} \quad B_{\alpha} - B_{\beta} = d\Lambda_{\alpha\beta} \quad  \text{ where } \Lambda_{\alpha\beta} \in \Omega^{1}(\CU_{\alpha\beta}) \\
  &\implies \text{On $\CU_{\alpha\beta\gamma}$,} \quad  d\big(\Lambda_{\alpha\beta} + \Lambda_{\beta\gamma} + \Lambda_{\gamma\alpha}\big) = 0 \\
  &\implies \text{On $\CU_{\alpha\beta\gamma}$,} \quad \Lambda_{\alpha\beta} + \Lambda_{\beta\gamma} + \Lambda_{\gamma\alpha} = -\imag  f_{\alpha\beta\gamma}^{-1}df_{\alpha\beta\gamma} \\
  &\implies \text{On $\CU_{\alpha\beta\gamma\delta}$,} \quad f_{\alpha\beta\gamma}f_{\alpha\gamma\delta}^{-1}f_{\alpha\gamma\delta}f_{\beta\gamma\delta}^{-1} = \text{constant} ~.
}
Now \underline{impose} quantization: On $\CU_{\alpha\beta\gamma\delta}$,
\eqa{  &f_{\alpha\beta\gamma}f_{\alpha\gamma\delta}^{-1}f_{\alpha\gamma\delta}f_{\beta\gamma\delta}^{-1} = 1 ~ .
}
Taking a logarithm, we find that 
\be 
 n_{\alpha\beta\gamma\delta} 
 := \frac{1}{2\pi \imag } \left( 
 \log f_{\alpha\beta\gamma} - \log f_{\alpha \beta \delta} + 
 \log f_{\alpha \gamma\delta} - \log f_{\beta\gamma\delta} 
\right) ~,
\ee
is a Cech cocycle, and it  defines an integral class in $H^{3}(M_{n}; \IZ)$. This is the topological class of the differential character.

\bigskip 
\paragraph{\underline{General Values of $\ell$.}}

Clearly, one can carry this discussion out for any $\ell$. One just has to use more indices. We begin with a globally well-defined closed form $F\in \Omega^{\ell}(M_n;\IR)$ from which we have pullbacks under the inclusion maps   $F_{\alpha} \in \Omega^{\ell}(\CU_{\alpha})$. Then, as before
\eqa{
    F_{\alpha} := \left.F\right|_{\CU_{\alpha}} = dA^{(\ell-1)}_{\alpha} \quad \text {where } A^{(\ell-1)}_{\alpha} \in \Omega^{\ell-1}(\CU_{\alpha}) ~.
  }
  On $\CU_{\alpha\beta}$, 
  \eqa{
    F_{\alpha}-F_{\beta} = 0 &\iff d(A^{(\ell-1)}_{\alpha}-A^{(\ell-1)}_{\beta}) = 0 \implies A^{(\ell-1)}_{\alpha}-A^{(\ell-1)}_{\beta} = dA^{(\ell-2)}_{\alpha\beta} ~,
  }
  because on the contractible subspace $\CU_{\alpha\beta}$, any closed form is exact. Therefore, in the \v{C}ech model, we have a collection of forms:  
  \eqa{
   \{F_{\alpha},  A_{\alpha}^{(\ell-1)}, A_{\alpha\beta}^{(\ell-2)}, A_{\alpha\beta\gamma}^{(\ell-3)}, \cdots, A_{\alpha_1\ldots\alpha_{\ell}}^{(0)},n_{\alpha_1 , \ldots, \alpha_{\ell+1}} \} ~, \label{eq:Check-ell-cocycle}
  }
where $A_{\alpha_1\ldots\alpha_{\ell}}^{(0)}$ are $\mathsf{U(1)}$-valued and 
$n_{\alpha_1,\dots , \alpha_{\ell+1}}$ are integral-valued, and 
define a cohomology class in $H^\ell(M_n;\IZ)$. That cohomology class is the topological class of the character.

\bigskip 
\noindent 
\textbf{\underline{Gauge Redundancy}}: It should be clear that the above description has a lot of gauge redundancy. We could make ``gauge transformations'' by a series of forms 
$(\epsilon^{(\ell-1)}_\alpha,  \epsilon^{(\ell-2)}_{\alpha\beta},\dots ) $, so that 
\be 
\begin{split} 
\wt A^{(\ell-1)}_\alpha & = A^{(\ell-1)}_\alpha + d\epsilon^{(\ell-1)}_\alpha ~,\\ 
\wt A^{(\ell-2)}_{\alpha\beta} & = A^{(\ell-2)}_{\alpha\beta} - \big( \epsilon^{(\ell-1)}_\alpha - \epsilon^{(\ell-1)}_\beta \big) + d\epsilon^{(\ell-2)}_{\alpha\beta}  ~,\\ 
\vdots \qquad & \qquad \vdots \\ 
\end{split}
\ee

\begin{remark}
$\,$
\begin{enumerate} 

\item We will sometimes denote the collection of data \eqref{eq:Check-ell-cocycle} by $\check{A}$, and its gauge equivalence class by $[\check{A}]$.

\item \textbf{Pros And Cons Of The \v{C}ech Formulation.}
   The \v{C}ech model is very good for physicists who like to think in terms of differential forms, and honest-to-God computations with towers of fields. The model is very explicit. It also makes clear that there is extra gauge-invariant data beyond the 
globally well-defined fieldstrength $F\in \Omega^{\ell}(M;\IR)$. Most notably, it gives a clear description of the topological class $c(\chi)\in H^\ell(M_n;\IZ)$ of the character. A disadvantage is that it is much harder to describe the holonomy of the character. (There is a formula for that, see \eqref{eq:CechHolonomyFormula} below.)  A related disadvantage is that it is hard to see the global structure of the differential cohomology group as the product of a finitely-generated Abelian group, a connected torus, and an infinite-dimensional vector space. A final disadvantage is that it does not readily generalize to arbitrary cohomology theories. These disadvantages are nicely addressed by a complementary model of differential characters, due to Hopkins and Singer \cite{Hopkins:2002rd}, to which we turn next.

\end{enumerate} 
\end{remark}

\begin{exbox}{\v{C}ech Model} 
\begin{itemize}
\item[(a)] Show how the periods of $F$ can be expressed in terms of the integers $n_{\alpha_1,\dots , \alpha_{\ell+1}}$ 
in \eqref{eq:Check-ell-cocycle}.

\item[(b)] A graded ring structure on differential characters will be defined in \autoref{subsec:GradedRingStructure} below. Show how to define it using the \v{C}ech model for 
differential characters described above. 
\end{itemize}

This exercise was inspired by an unpublished note by D. Harlow. 

\end{exbox}

\subsubsection{Gerbes And Gerbe Connections }\label{subsubsec:GerbeConnections}

As we have noted, the case of $\ell=1$ and $\ell=2$ in the 
above constructions have very commonly encountered physical 
manifestations: The periodic scalar field and an Abelian gauge field, 
respectively. It turns out that the case $\ell=3$ also has diverse 
applications and manifestations in physics. 

Equivalence classes of degree three characters are equivalent 
to ``gerbe connections.'' Mathematically, a ``gerbe'' can be defined 
in many different ways. One way, 
described for example in \cite{Hitchin:1999fh}, simply generalizes 
the description of $\mathsf{U(1)}$ lines bundles via $\mathsf{U(1)}$-valued transition functions $g_{\alpha\beta}$ by considering the data 
$\{f_{\alpha\beta\gamma} \} $ satisfying the cocycle condition as a gerbe. (One then imposes suitable equivalence relations.) 
Just the way complex line bundles over $M_{n}$ are classified by $H^2(M_{n}; \IZ)$, $\mathsf{U(1)}$ gerbes are classified by an element of 
by $H^3(M_{n}; \IZ)$.
Another approach to defining a gerbe is to consider a family of complex line bundles $\CL_{\alpha\beta}$ on patch overlaps together with 
isomorphisms on the triple overlaps 
\be 
\theta_{\alpha\beta\gamma}: \CL_{\alpha\beta}\otimes 
\CL_{\beta\gamma}\otimes \CL_{\gamma \alpha}\rightarrow 
\CU_{\alpha\beta\gamma} \times \IC ~.
\ee
On quadruple overlaps, the isomorphisms $\theta_{\alpha\beta\gamma}$ satisfy a cocycle condition. 
\tightfootnote{There are many other mathematical definitions of ``gerbes.'' For other accounts, see \cite{Brylinski1993, Murray:2007ps}.}

Now, turning to a \emph{gerbe connection}, the local data for 
$\check A$ for $\ell=2$ defines a principal $\mathsf{U(1)}$-bundle with 
connection. The local data for   $\check A$ for $\ell=3$ 
with specified $\{ f_{\alpha\beta\gamma} \}$ (defining a gerbe) 
defines a gerbe connection.  

One manifestation of a gerbe connection in physics is via the ``Neveu-Schwarz $B$-field'' of string theory, which is mathematically modeled as differential character of degree $\ell=3$. In the WZW model of 2d conformal field theory of 
maps from spacetime to a compact Lie group $G$, there is a family (parametrized by a ``level'') of canonical $B$-fields. The associated differential cohomology class is described in more detail in \autoref{subsec:WZW-TERMS} below. 
The precise nature of the differential cohomology group that models the $B$-field is different in different string models. See \cite{Distler:2010an} for a summary.

Another way in which gerbes and gerbe connections appear in physics 
is in the context of reduction of structure group. In general, 
given a principal $G$-bundle $P\to M_{n}$ and a homomorphism $\varphi: H\to G$, we have a \emph{reduction of structure group from $G$ to $H$ along $\varphi$ } if there is a principal $H$-bundle $Q\to M_{n}$, and 
an isomorphism of principal $G$-bundles, 
\be 
Q \times _{\varphi} G  \cong P ~.
\ee
This terminology applies even when the homomorphism 
$\varphi: H\to G$ is a covering map, such as the covering of 
$\mathsf{Spin}(n)$ over $\mathsf{SO}(n)$. When $\varphi$ is a covering map, 
a reduction of structure group is also commonly referred to as a 
``lifting'' of structure group.

Let $G$ be a topological group and $Z$ a subgroup of its center. 
Gerbes naturally arise when we consider obstructions to lifting the structure group of a  $G/Z$ bundle over $M_{n}$ to a $G$-bundle over $M_{n}$, i.e., to the reduction of structure group along the 
natural quotient homomorphism $\varphi: G \to G/Z$. 
The transition functions of a principal $G/Z$ bundle satisfy 
\be
g_{\alpha\beta}g_{\beta\gamma}g_{\gamma\alpha} = 1_{G/Z} ~,
\ee
on triple overlaps $\CU_{\alpha\beta\gamma}$. Note especially 
that this is an equation in the quotient group $G/Z$. On a good cover, 
one can choose lifts    $\widetilde{g}_{\alpha\beta}: \CU_{\alpha\beta} \to G$ such that $\pi(\widetilde{g}_{\alpha\beta}) = g_{\alpha\beta}$, then we conclude that,
\eqa{
\widetilde{g}_{\alpha\beta}\widetilde{g}_{\beta\gamma}\widetilde{g}_{\gamma\alpha} &= \zeta_{\alpha\beta\gamma} \quad \text{ on }\quad \CU_{\alpha\beta\gamma} ~,
}
where $\zeta_{\alpha\beta}: \CU_{\alpha\beta\gamma} \to Z$. The Cech cocycle $\{\zeta_{\alpha\beta\gamma}\}$ determines a class in $H^{2}(M_{n}; \underline{Z})$ where $\underline{Z}$ is a sheaf. 
If   $Z \subset \mathsf{U(1)}$, this is equivalent to a class in 
$H^3(M_{n}; \IZ)$.  

If $G$ is a compact Lie group with a finite center then 
the class in $H^3(M_{n}; \IZ)$. Then we can think of the obstruction to lifting a $G/Z$ bundle to a $G$-bundle as a flat degree three 
differential cohomology class.   Its characteristic class is:
\be
\beta(\zeta) \in H^3(M_{n};\IZ) ~,
\ee
where $\beta$ is the Bockstein map. Here are some significant 
special cases of the above construction:  

\begin{enumerate}

\item Suppose that $M_{n}$ is an oriented Riemannian $n$-manifold. Then its oriented frame bundle is a principal $\mathsf{SO}(n)$-bundle. But $\mathsf{SO}(n) \cong \mathsf{Spin}(n)/\IZ_2$ for a suitable $\IZ_2$ subgroup of the center. A reduction  of the 
bundle of oriented frames to a principal $\mathsf{Spin}(n)$ bundle is a spin structure. Evidently, from the above discussion, that reduction is obstructed by a degree two cohomology class with $\IZ_2$ coefficients. That obstruction is the second Stiefel-Whitney class 
of the tangent bundle $w_2(TM)$. It is a $\IZ_2$ class that depends 
continuously on the metric and hence is metric-independent. 

\item In applications to Yang-Mills theory for a compact Lie group of the 
form $\widetilde{G}/Z$, where $\widetilde{G}$ is a compact semisimple and simply connected gauge group the obstruction to lifting a principal 
$\widetilde{G}/Z$ to a principal $\widetilde{G}$-bundle is known as an 
't Hooft flux in honor of a very deep paper that studied the 
case of $\mathsf{\mathsf{PSU}}(N)$ bundles on the torus \cite{tHooft:1977nqb}.
In more modern language, background gerbe connections associated with this flux are called ``background 1-form gauge fields.''

\end{enumerate}

A very closely related construction appears when we consider bundles of simple algebras. These are sometimes referred to as 
\emph{Azumaya bundles} or \emph{Dixmier-Douady bundles}. 
Suppose $A$ is a simple algebra and $\pi: \mathscr{A} \to M_{n}$
is a bundle whose fibers are isomorphic as vector spaces to $A$
and whose transition functions are inner: $(x,a_\alpha) \in \CU_{\alpha} \times A$ is glued to $(x,a_{\beta}) \in \CU_{\beta} \times A$, via 
\be 
a_{\beta} = g_{\alpha\beta} a_{\beta} g_{\alpha\beta}^{-1} ~,
\ee
where $g_{\alpha\beta}$ is a continuous map from the overlap $\CU_{\alpha\beta}$ to the group $G(A)$ of invertible elements of $A$. 
Note that if $\rho_{\alpha\beta}: \CU_{\alpha\beta} \to G(A) \cap Z(A)$ is in the center, then the transition functions are unchanged. 
Therefore, the structure group of the bundle of algebras is: 
\be
PG(A):=G(A)/(G(A)\cap Z(A)) ~. 
\ee
Therefore, on triple overlaps, our bundle of algebras will satisfy: 
\be 
[g_{\alpha\beta}][g_{\beta\gamma}][g_{\gamma\alpha}] = 1_{PG(A)} ~.
\ee
We find ourselves precisely in the setting described previously: 
If we make a specific choice of maps $g_{\alpha\beta}$, then, 
\be\label{eq:DD-definition}
g_{\alpha\beta}g_{\beta\gamma}g_{\gamma\alpha} = \zeta_{\alpha\beta\gamma} ~.
\ee
For a simple algebra, $Z(A) \cong \IC$, and we obtain a class in 
$H^3(M_n; \IZ)$. In this setting, the class in  $H^{3}(M_{n}; \IZ)$ is called the \emph{Dixmier-Douady class}. For a nice description in terms of algebraic geometry, see \cite{SchochetDummies}.

\bigskip 
\noindent \textbf{Example 1:}
If we have a bundle $\mathscr{A} \to M_{n}$ of algebras with fiber $A=\mathsf{Mat}_{N}(\IC)$, their transition functions will be in $\mathsf{PGL}(N,\IC)$, and the $g_{\alpha\beta}$ above will be 
valued in $\mathsf{GL}(N,\IC)$. We can choose representatives of $g_{\alpha\beta}$ with unit determinant. Then, taking   the determinant of equation \eqref{eq:DD-definition}, we see that 
for finite-dimensional algebras, the Dixmier-Douady class must be 
torsion. However, if one considers infinite-dimensional $C^*$ algebras, it is entirely possible to achieve non-torsion classes 
\cite{Rosenberg1989,Brylinski1993}.

A natural way to obtain a bundle of algebras is to start with a 
vector bundle $E \to M_{n}$, and consider the bundle of endomorphisms 
$\End(E) \to M_{n}$. It is natural to ask if the converse is true. 
Given a bundle of algebras  $\mathscr{A} \to M_{n}$, is there a bundle 
of $A$-modules $E\to M_{n}$, so that $\End(E) \cong \mathscr{A}$? It turns 
out that the Dixmier-Douady class is precisely the obstruction 
to realizing such a bundle of $A$-modules. 
%
%

\bigskip
\noindent \textbf{Example 2:} An elegant example is provided by considering a
Riemannian manifold $(M, g_{\mu\nu})$. There is then a canonical 
bundle of Clifford algebras, where the fibers are the algebras 
generated by the Clifford relations on the tangent bundle: 
\be 
\{ e, e' \} = 2 g(e,e') ~.
\ee
For complex Clifford algebras, the Dixmier-Douady class is the Stiefel-Whitney class $W_3(TM)$ and if it is trivial, we can write the bundle of algebras as $\End(\CS)$ where $\CS$ is the spinor bundle of a spin$^{\mathsf{c}}$ structure \cite{Connes1984,Plymen1986}.

An important physical application of the Dixmier-Douady class 
is in the Hamiltonian formulation of anomalies 
\cite{FADDEEV198481,Faddeev1984,MR791642,Carey1996,Carey:1997xm,Mickelsson:1999td,Segal:FaddeevAnomalyUnpublished}. The standard setting for the discussion of the Hamiltonian formulation of anomalies is the case of gauge fields coupled to fermions. For a fixed gauge field, the quantization of the Clifford algebra canonical quantization rules for the fermions produces a \underline{projective} Hilbert space. 
One way to see that it is projective is that one must make a \underline{choice} of Clifford vacuum. 
For a nonanomalous theory, we wish to have a bundle of Hilbert spaces over the space of gauge fields 
on the spatial manifold. For a more general formulation of Hamiltonian anomalies see the 
discussion in Sections 4.1 and 5.4 of \cite{Freed:2003qx}.

Closely related to the above consideration is the topic 
of twisted $K$-theory which plays an important role in the 
study of D-branes in the presence of $B$-fields \cite{Witten:1998cd,Kapustin:1999di,Witten:2000cn,Bouwknegt:2000qt,Harvey:2000te,Moore:2006dw,Brodzki:2007hg}.

The mathematics of twisted $K$-theory goes back to \cite{Donovan1970,Karoubi:2007mg,Karoubi:2008eug}.
It was also studied in \cite{Mathai:2000iw,Bouwknegt:2001vu,Atiyah:2004jv,Atiyah:2005gu}.  A remarkable 
theorem in the subject is the Freed-Hopkins-Teleman theorem 
identifying twisted $G$-equivariant $K$-theory of $G$ (with the action by conjugation) with the Verlinde algebra \cite{Freed:2002gr,Freed:2003qx,Freed:2005qu,Freed:2007wja}, which can be interpreted physically in terms of the D-brane 
boundary conditions of the topological field theory obtained by gauging the $G$-WZW theory by $G$ \cite{Moore:2003vf}.
Twisted equivariant $K$-theory is also the proper setting for discussing RR fields and charges in type II string theory orientifolds \cite{Witten:1998cd,Gukov:1999yn,Distler:2009ri}.

Gerbes and gerbe connections have also played some role in 
condensed matter physics in recent years. Twisted equivariant $K$-theory is needed when discussing topological insulators in the 
framework of band structure \cite{Freed:2012uu,Thiang:2014fxa,Kellendonk:2015pda,Kruthoff:2016ver,Gomi:2017ymm,Bradlyn:2017pss,Stehouwer:2018xfs}.

The higher Berry 
curvatures studied in  
\cite{Kapustin:2020eby,Kapustin:2020mkl,Hsin:2020cgg,Artymowicz:2023erv} bear a strong resemblance to the curvatures of a gerbe connection. A recent paper making a more direct connection to the gerbe connection is \cite{Geiko:2024cra}.

\subsection{Hopkins-Singer Models}\label{subsec:HopkinsSingerCocycles}

The Hopkins-Singer model of differential cohomology is motivated by a ``homotopy pushout'' construction from homotopy theory. It has some advantages over the \v{C}ech description:

\begin{enumerate}

    \item It can be applied to any generalized cohomology theory. This is especially important in string theory which makes the use of ``differential $K$-theory.'' (See below.)

    \item It is easier to view the fields as forming a groupoid. Then the automorphism group of an object with isomorphism class in $\widecheck{H}^{\ell}(M_{n})$ is $H^{\ell-2}(M; \mathsf{U(1)})$. These automorphism groups play important roles in physics in formulating Gauss laws, defining charges, etc.

    \item The holonomy of the character is built into the description.

\end{enumerate}

 There are other models described in \cite{Hopkins:2002rd} as well as in the  book by Amabel, Debray, and Haine \cite{Amabel:2021wbk}.

  One way to motivate the Hopkins-Singer model is to consider a formula for the holonomy of a differential character in $\widecheck{H}^{\ell}$ on an $(\ell-1)$-cycle $\Sigma$ which is $n$-torsion. Thus $n\Sigma = \partial \CB$ for some integral $n$-chain $\CB$. Therefore,
\be
      \left( \chi(\Sigma) \right)^{n} = \exp\left( \imag\int_{\CB}F \right) ~.
\ee
  It would be \underline{wrong} to conclude that
  \eqa{
     \chi(\Sigma) &\stackrel{?}{=} \exp\left( \frac{\imag }{n}\int_{\CB}F \right) ~.
  }
  For one thing, $F \in \Omega_{2\pi\IZ}(M_{n})$ could have periods $\neq 0 \mathsf{\,\,mod\,\,} 2\pi \IZ$. Then the above formula becomes ill-defined. However,
  one can show that there exists an integral cochain $a \in C^{\ell}(M_{n}; \IZ)$ such that,
  \eqa{
    \chi(\Sigma) &= \exp\left[ \frac{\imag }{n}\left(\int_{\CB}F - 2\pi \langle a, \CB \rangle \right) \right] ~, \label{eq:ostarchar}
  }
  \underline{is} well-defined. Here $\delta$ is the coboundary operator and $\langle \cdot, \cdot\rangle$ denotes the natural pairing between cochains and chains on $M_{n}$. One can show that the class $a$ has $\delta a = 0$
  and hence, $[a] \in H^{\ell}(M_{n}; \IZ)$. This is the characteristic class (a.k.a. topological class)  of the character $\chi \in \widecheck{H}^{\ell}(M_{n})$. We can write \eqref{eq:ostarchar} heuristically by writing ``$\Sigma = \frac{1}{n}\partial \CB$,'' 
  taking the log of \eqref{eq:ostarchar}, and dividing by $2\pi \imag $ to get: 
\be
\biggl\langle  \frac{\log\,\chi}{2\pi \imag  } , \p\CB  \biggr\rangle
=    \biggl\langle \frac{F}{2\pi} - a, \CB \biggr\rangle ~,
\ee
which we rewrite (still heuristically) as: 
\be 
   \delta \left( \frac{\mathsf{log\,}\chi }{2\pi \imag } \right)   
     = \frac{F}{2\pi } - a_{\IR} ~.
\ee
  This is a motivating equation for the definition of Hopkins-Singer cocycles.

\subsubsection{The Cochain Model} 

  \begin{definition}[\textcolor{red}{Hopkins-Singer cochain and cocycle}] A Hopkins-Singer cochain is a triple,
  \be
     \chi = (a, h, \omega) \in C^{\ell}(M_{n}; \IZ) \times C^{\ell-1}(M_{n}; \IR) \times \Omega^{\ell}(M_{n}) ~. \label{eq:HS-Cochain}
  \ee
  Hopkins and Singer define a \emph{cocycle} to be a triple such that, 
  \be  \label{eq:HS-Cocycle}
  \begin{split}
      \delta a &= 0 ~,\\
      \delta h &= \omega - a_{\IR} ~,\\
      d\omega &= 0 ~,
  \end{split}
  \ee
  where $a_{\IR}$ is the image of $a$ under $C^{\ell}(M_{n}; \IZ) \to C^{\ell}(M_{n}; \IR)$, and $\omega$ is likewise embedded as $\Omega^{\ell}(M_{n}) \to C^{\ell}(M_{n}; \IR)$. The set of all Hopkins-Singer cocycles is denoted $\widecheck Z^{\ell}(M_n)$. 
  \end{definition}

Above, we used the language of a cochain complex.  
On the space of Hopkins-Singer cochains, one can define a differential $d_{\mathsf{HS}}$, which acts on a Hopkins-Singer cochain $(a, h, \omega)$ as:
\be \label{eq:HS-Differential}
   d_{\mathsf{HS}}(a, h, \omega) := (\delta a, \omega - a_{\IR} - \delta h, d\omega) ~.
\ee
The definition \eqref{eq:HS-Cocycle} of a Hopkins-Singer cocycle is equivalent to $d_{\mathsf{HS}}(a, h, \omega) = 0$. It is straightforward to verify that $d_{\mathsf{HS}}^2 = 0$.
However, if we just compute the cohomology of the differential $d_{\mathsf{HS}}$, we will not obtain differential cohomology. 
In order to obtain differential cohomology, we need to generalize 
to a double complex as follows.

We define a collection of cochain complexes $\widecheck{C}(\ell)^{\bullet}(M_{n})$ as:
%
%
%
  \be
        \widecheck{C}(\ell)^{k}(M_{n}) = \left\{
           \begin{array}{ll}
               C^{k}(M_{n};\IZ) \times C^{k-1}(M_{n}; \IR) \times \Omega^{k}(M_{n}) ~, & \text{for } k \geq \ell ~,\\
               C^{k}(M_{n};\IZ) \times C^{k-1}(M_{n}; \IR) ~, & \text{for } k < \ell ~,
           \end{array}
        \right.
  \ee
with the differential acting as:
  \be
   \begin{split}
      d_{\mathsf{HS}}(a, h, \omega) &= (\delta a, \omega - a_{\IR} - \delta h, d\omega) ~,\\
      d_{\mathsf{HS}}(a, h) &= \left\{
          \begin{array}{ll}
               (\delta a, -a_{\IR}-\delta h, 0) ~, & \text{ or } (a, h) \in \widecheck{C}(\ell)^{\ell-1} ~,\\
               (\delta a, -a_{\IR}-\delta h) ~, & \text{otherwise.}
          \end{array}
      \right.
   \end{split}
  \ee
The $k^{th}$ cohomology group of the cochain complex $\widecheck{C}(\ell)^{\bullet}(M_{n})$ is denoted by
\be
   \widecheck{H}(\ell)^{k}(M_{n}) := H^{k}\big( \widecheck{C}(\ell)^{k}(M_{n}), d_{\mathsf{HS}}\big) ~.
\ee
For $k>\ell$, the cohomology class is an integral class and a de Rham class with the constraint that they are the same in real cohomology, so we just get a copy of $H^k(M; \IZ)$. It is thus evident that \cite[Sec 3.2]{Hopkins:2002rd}: 
\be \label{eq:HS-Cohomology-Cochain-Complex}
   \widecheck{H}(\ell)^{k}(M_{n}) = \left\{
      \begin{array}{ll}
          H^{k}(M_{n};\IZ) ~, & \text{for } k > \ell ~,\\
          H^{k-1}(M_{n};\IR/\IZ') ~, & \text{for } k < \ell ~.
      \end{array}
   \right.
\ee
The Cheeger-Simons group of diferential characters of degree $(\ell-1)$ is identifed as
\be
   \widecheck{H}^{\ell}(M_{n}) \cong \widecheck{H}(\ell)^{\ell}(M_{n}) ~.
\ee

\noindent 
\begin{remark}
 We have chosen a different grading of the differential cohomology group from that used by Cheeger-Simons and Hopkins-Singer. For many purposes, it is more convenient to grade it by the degree of the fieldstrength and characteristic class.
\end{remark}

\begin{remark}
 Hopkins-Singer cocycles were also described by P. Deligne and D. Freed in \cite[Sec. 6.3]{DeligneFreed:1999QFT1}. Indeed, in the holomorphic context, a very similar construction is known as \emph{Deligne cohomology}. 

\end{remark}

\subsubsection{The Groupoid Model}

Another approach to modeling differential cohomology 
(also described in \cite[Sec. 2.3]{Hopkins:2002rd}) is to  define a groupoid whose objects are Hopkins-Singer 
cocycles and the isomorphism classes of these objects correspond to differential characters. 

We can motivate the groupoid approach of Hopkins and Singer by interpreting the standard cohomology group  $H^\ell(M_n;\IZ)$ as the set of isomorphism classes of a groupoid $\CH^n(M)$ as follows: The objects of 
$\CH^\ell(M_n)$ are smooth cocycles $c\in Z^\ell(M_n; \IZ)$. The morphism set  between two objects 
$\Hom(c_2,c_1)$ is identified with the set of elements $[b] \in C^{\ell-1}(M_n; \IZ)/\delta C^{\ell-2}(M_n; \IZ)$, such that,
\be\label{eq:HS-groupoid1}
c_2 = c_1 + \delta b ~.
\ee
Put differently: $\Hom(c_2,c_1)$ consists of $b\in C^{\ell-1}(M_n; \IZ)$, with an equivalence relation on $b$ defined by $b' \sim b + \delta v$, $v\in C^{\ell-2}(M_n; \IZ)$. This is a groupoid because 
we can bring $\delta b$ to the LHS of equation 
\eqref{eq:HS-groupoid1}.
The isomorphism classes of objects in $\CH^\ell(M_n)$
are in 1--1 correspondence with cohomology classes in $H^\ell(M_n; \IZ)$. What do we gain from this viewpoint? Now, we can understand that a cohomology class has an automorphism group. The automorphism group of $c$
is the equivalence class of $b\in C^{\ell-1}(M_n; \IZ)$, such that $\delta b = 0$. 
Thus, the automorphism group 
of any object in $\CH^\ell(M_n)$ is isomorphic to the cohomology group $H^{\ell-1}(M_n; \IZ)$.

The viewpoint of the previous section generalizes nicely to the differential setting. One can define a groupoid,
$\widecheck{\CH}^\ell (M)$ whose objects are Hopkins-Singer cocycles $(a,h,\omega)\in \widecheck Z^\ell(M)$  (and thus 
satisfy \eqref{eq:HS-Cocycle}). The morphism space    $\Hom(x,x')$ for $x=(a,h,\omega)$ and $x'=(a',h',\omega')$ consists of equivalence classes of pairs,  
\be 
(b, g) \in C^{\ell-1}(M_{n}; \IZ) \times C^{\ell-2}(M_{n}; \IR) ~,
\ee
such that, 
\be 
\begin{split} 
a'  & = a - \delta b ~,\\
h'  & = h + \delta g +  b ~.
\end{split}
\ee
(Note that since we are working with differential cocycles it follows that $\omega' = \omega$.) 
The equivalence relation on pairs $(b,g)$ is:
\be 
(b,g) \sim (b - \delta e, g + \delta v + e ) ~,
\ee
for all $(e,v) \in C^{\ell-2}(M; \IZ) \times C^{\ell-3}(M; \IR)$. 
See \cite{Hopkins:2002rd}, Definition 2.5.

As we have said, the isomorphism class of 
$x\in \widecheck{\CH}^\ell(M_n)$ can be identified with a 
differential character. The relation to the direct product description of $\widecheck{H}^\ell(M_n)$ is rather direct: $\omega$ is the fieldstrength $F/2\pi$, $h$ defines the holonomy, and $[a]$ is the topological class. Once again, the advantage of this viewpoint is that objects have a nontrivial automorphism group:

 \begin{exbox}[exercise:automorphism-hopkins-singer]{} Show that the automorphism group of an object $x\in \widecheck{\CH}^\ell(M_{n})$ can be identified with $H^{\ell-2}(M_{n}; \IR/\IZ')$.\end{exbox}

\begin{remark}
$\,$
\begin{enumerate}

\item When describing the $\sfC$-field of M-theory, it is very natural to work with a Hopkins-Singer model since, as we will see in 
\autoref{sec:MTheory}, the isomorphism classes of $\sfC$-fields form a 
\underline{torsor} for $\widecheck{H}^4(M_{11})$, and the $\sfC$-field itself must be defined using a differential cochain. Another model for the 
differential cocycles, used in \cite{Diaconescu:2003bm}, uses $\sfE_8$ bundles with connection. 

\item The automorphism group of a differential character plays an important role in some physical applications. For example, it is useful to give a quantum definition of electric charge. 
See \eqref{eq:QuantumGaussLaw} below. This viewpoint is very useful in giving a precise definition of $\sfC$-field electric charge in M-theory \cite{Diaconescu:2003bm}. 

\item There is a natural functor $F: \widecheck{\CH}^\ell(M_n) \to \CH^\ell(M_n)$. On objects, we simply have 
$F(a,h,\omega) = a$. The mapping of automorphism groups $F: \mathsf{Aut}(a,h,\omega) \to  \mathsf{Aut}(a)$
is the Bockstein map $H^{\ell-2}(M_n; \IR/\IZ') \to H^{\ell-1}(M_n; \IZ)$.

\end{enumerate}
\end{remark}

\subsubsection{Higher Groupoids}\label{subsubsec:DiffCoh-HigherGroupoids}

It is possible and useful to promote the groupoid $\widecheck{\CH}^\ell(M_{n})$ to a higher groupoid. We explain that here, following the nice discussion  in \cite[App. A.3]{Freed:2021anp}. 

Given a chain complex  $(A^\bullet, d)$, one can define an $\ell$-groupoid whose objects, or 0-morphisms, are the closed $\ell$-forms, 
$A^\ell_{\rm closed}:= \ker(d: A^\ell \to A^{\ell+1})$. The 1-morphisms are defined to be:
\be 
\Hom_1(\omega_0^\ell , \omega_1^\ell):= \{ \epsilon^{\ell-1}\in A^{\ell-1}~\vert~ \omega_1^\ell = \omega_0^\ell  + d \eta^{\ell-1} \} ~.
\ee
Note that the morphisms are all invertible. The 2-morphisms are:
\be 
\Hom_2(\eta_0^\ell , \eta_1^\ell):= \{ \zeta^{\ell-2}\in A^{\ell-2}~\vert~ \eta_1^\ell = \eta_0^\ell  + d \zeta^{\ell-2} \} ~,
\ee
and so on. Note that $\Hom_k$ between two $(k-1)$-morphisms is an $A^{k-3}_{\rm closed}$-torsor. 

We apply the above construction using the Hopkins-Singer model 
to enhance $\check \CH^\ell(M)$ to an $\ell$-groupoid. The objects 
are $\widecheck{Z}^\ell(M)$ as before. Now define:
\be 
\widecheck{C}^q(M):= \widecheck{C}(q)^q(M) ~,
\ee
and note that there is a natural inclusion of $\widecheck{C}^q(M) \hookrightarrow \widecheck{C}(q-1)^{q}(M)$. Now, the $1$-morphisms between 
$\check A_0$ and $\check A_1$ are: 
\be 
\Hom_1(\check A_0, \check A_1) := 
\{ \check \epsilon^{\ell-1}\in \widecheck{C}^{\ell-1}(M)| 
\check A_1 = \check A_0 + d_{\mathsf{HS}} \check \epsilon^{\ell-1} \} ~.
\ee
Note that the defining condition here takes place in 
$\widecheck{C}(\ell-1)^{\ell}(M)$. We can now proceed to define the 
2-morphisms as: 
\be 
\Hom_2(\check \epsilon^{\ell-1}_0, \check \epsilon^{\ell-1}_1) := 
\{ \check \eta^{\ell-2}\in \widecheck{C}^{\ell-2}(M)| 
\check \epsilon^{\ell-1}_1 = \check \epsilon^{\ell-1}_0 + d_{\mathsf{HS}} \check \eta^{\ell-2} \} ~,
\ee
and so on. From this higher groupoid, one can form a simplicial set. 
The generalized gauge field as a field in the sense of Freed-Hopkins-Teleman is a sheaf valued in this simplicial set. 

\SectionWithHeader{Graded Ring Structure, Integration, And Perfect Pairing Of Differential Cohomology}{Graded Ring Structure, Integration, And Perfect Pairing}{sec:ImptPropertiesDiffCoh}

No matter what model we use, the gauge-invariant information in differential characters satisfies some important properties. We summarize them briefly here:

\subsection{Graded Ring Structure}\label{subsec:GradedRingStructure}

 There is a graded associative product:
    \eqas{
      \widecheck{H}^{\ell_1} \times \widecheck{H}^{\ell_2} &\longrightarrow \widecheck{H}^{\ell_1+\ell_2} \\
      (\chi_1, \chi_2) &\longmapsto \chi_1 \odot \chi_2 := (-1)^{\ell_1\ell_2}\chi_2 \odot \chi_1 ~. \label{eq:GradedRingProduct}
    }
    This satisfies two nice properties:
    \eqa{
      F(\chi_1\odot\chi_2) &= F(\chi_1)\wedge F(\chi_2) ~, \quad \text{wedge product} ~, \\
      c(\chi_1\odot\chi_2) &= c(\chi_1) \cup c(\chi_2) ~, \quad\,\,\,\,\text{cup product} ~.
      \label{eq:HS-CochainPairing}
    }
    The formula for the holonomy is more complicated. One way to express it is to give the product in terms of Hopkins-Singer cocycles:
    \eqa{
        (a_1, h_1, \omega_1) \cup (a_2, h_2, \omega_2) := (a_1 \cup a_2, \pm a_1 \cup h_2 + h_1 \cup \omega_2 + H(\omega_1, \omega_2), \omega_1 \wedge \omega_2) ~,
    }
    where $H$ is a homotopy between the cup product $\cup$ and the wedge product $\wedge$ on $\Omega^{\ell}$, considered as defining smooth $\IR$-valued $\ell$-cochains.

\subsection{Integration} \label{subsec:integration-diff-cohomology}

    If $\Sigma_{\ell}$ is an $\ell$-cycle, then the holonomy on $\Sigma_{\ell}$ can be considered as an integration:
    \eqa{
      \int_{\Sigma_\ell}^{\widecheck{H}} &: H^{\ell+1}(\Sigma_{\ell}) \longrightarrow 
      \widecheck{H}^{1}({\rm pt})    ~. \label{eq:SingleDC-Int}
    }
If we think of the group $\widecheck{H}^{1}({\rm pt}) $ additively, we would write it as $\IR/\IZ'$, and if we think of it multiplicatively, we would write it as $\mathsf{U(1)}$. 


  In the \v{C}ech model, there is an explicit formula for \eqref{eq:SingleDC-Int}: Choose a triangulation of $\Sigma_{\ell}$ such that $\ell$-simplices sit in a definite $\CU_{\alpha}$. We write these as:  $\Sigma^{(i)}_{\ell,\alpha} \subset \CU_{\alpha}$. Then, arrange that the faces are in patch overlaps, so we write:  $\Sigma_{\ell,\alpha\beta}^{(i)} \subset \CU_{\alpha\beta}$, etc. Then, denoting the data of the \v{C}ech model by,
  \eqa{
      \wh{A} &= (A_{\alpha}, A_{\alpha\beta}, A_{\alpha\beta\gamma}, \cdots ) ~,
  }
  and the corresponding character by,
  \eqa{
      \chi &= [\check{A}] ~,
  }
  we have,
  \eqa{
      \exp\left(  \imag\int_{\Sigma_{\ell}}^{\widecheck{H}} [\check{A}] \right) := \prod_{\alpha,j}\exp\left(  \imag\int_{\Sigma_{\ell,\alpha}^{j}}A_{\alpha} \right) \cdot \prod_{\alpha < \beta,j}\exp\left(  \imag  \int_{\Sigma^{j}_{\ell,\alpha\beta}}A_{\alpha\beta}\right)\cdots
      \label{eq:CechHolonomyFormula}
  }

The integration generalizes to families. If we have a family $\CX$ of closed $n$-manifolds $M_{n}$ fibered over a manifold $\CS$ of control parameters:
  \begin{equation}
      \begin{tikzcd}
          M_{n} \arrow[r] & \CX\arrow[d]\\
          & \CS 
      \end{tikzcd}
  \end{equation}
then the notion of integration along fibers defines a homomorphism (for any $d$): 
  \eqa{
     \int^{\widecheck{H}}_{\CX/\CS} &: \widecheck{H}^{d}(\CX) \longrightarrow \widecheck{H}^{d-n}(\CS) ~. \label{eq:integ}
  }
 For generalizations of \eqref{eq:CechHolonomyFormula} to families, see B\"{a}r-Becker \cite{BarBeckerarXiv}.
  
\subsection{Pairing}\label{subsec:DiffCohoPairing}

Putting together the graded ring structure and the integration, we derive 
an important pairing on   differential cohomology: multiply and integrate:
    \eqas{
       \widecheck{H}^{\ell}(M_{n}) \times \widecheck{H}^{n-\ell+1}(M_{n}) &\longrightarrow \widecheck{H}^{1}({\rm pt}) = \IR/\IZ' \\
       (\chi_{1}, \chi_{2}) &\longmapsto \langle \chi_{1}, \chi_{2} \rangle := \int^{\widecheck{H}}\chi_{1} \odot \chi_{2} ~.  \label{eq:DC-Pairing}
    }
It follows from \eqref{eq:GradedRingProduct}, that, 
\be 
\langle \chi_1 , \chi_2 \rangle = (-1)^{\ell(n+1-\ell)} \langle \chi_2 , \chi_1 \rangle ~.
\ee

\noindent There are two important and useful special cases of the pairing \eqref{eq:DC-Pairing}:
    
    \begin{enumerate}
        \item If $\chi_{1}$ is topologically trivial, i.e., $\chi_{1} = \chi_{A_1}$ for $A_{1} \in \Omega^{\ell-1}(M_{n})$, then,
    \eqa{
       \langle \chi_{1}, \chi_{2} \rangle &= \int_{M_{n}}A_{1} \wedge F(\chi_{2}) ~,
    }
    which is an ordinary integral of differential forms. If $\chi_{2}$ is also topologically trivial, i.e., $\chi_{2} = \chi_{A_2}$ for $A_{2} \in \Omega^{n-\ell}(M_{n})$, then,
    \eqa{
       \langle \chi_{1}, \chi_{2} \rangle &= \int A_{1} \wedge dA_{2} ~, \label{eq:BF-Pairing}
    }
    
    \item If $\chi_{1}=\widecheck \phi_1$ is a flat field associated to $\phi_{1} \in H^{\ell-1}(M_{n}; \IR/\IZ')$, then, 
    \eqa{
        \exp\left( \imag \langle \chi_1, \chi_2 \rangle \right) &= \exp\left( \imag\int_{M_{n}}\phi_{1} \cup c(\chi_{2}) \right) ~, \label{eq:FlatPairing}
    }
    where we use the cup product on $H^{\ell-1}(M_{n}; \IR/\IZ')$ and $H^{n-\ell+1}(M_{n}; \IZ)$ to get an element of $H^{n}(M_{n}; \IR/\IZ')$. So in the pairing, if $\chi_{1}$ is a flat field, it only sees the topological class of $\chi_{2}$ and nothing else. This will be a useful observation for the definition of flux sectors in 
    \eqref{eq:ShiftFlatField} below.  

A good example of the pairing is the simple case of $\ell=1$ in two dimensions. 
We can use the mode decomposition  of equation \eqref{eq:ell=1-modes} to separate out the winding mode. So we write $f = \exp(2\pi \imag  \phi)$, and then, 
\be 
\langle \chi_{f^1} , \chi_{f^2} \rangle = - w^1 \phi^2(0) + 
\int_0^1 \phi^1 \frac{d}{d\sigma} \phi^2  d\sigma ~.
\ee
For a proof, see \cite{Freed:2006yc}.
\end{enumerate}

\begin{remark}

 Note that the pairing \eqref{eq:BF-Pairing} is none other than the $\mathsf{BF}$ or Chern-Simons action. Therefore the pairing can be viewed as one way of generalizing the actions, such as \eqref{eq:BF-action-FirstPass} of $\mathsf{BF}$ theory to topologically nontrivial situations. In particular, the 3d Chern-Simons action for Abelian gauge groups can be expressed as a pairing. A natural way to define an action for $\mathsf{BF}$ theory is via the pairing on differential cohomology. This point was first made in \cite[App. A]{Maldacena:2001ss}. Note that without a quantization condition on the periods, the level $k$ of the $\mathsf{BF}$ theory can always be scaled to one. 
    
\end{remark}

\subsection{Poincar\'e-Pontryagin Duality}

The pairings we have defined above satisfy an important property related to both Poincar\'e and Pontryagin duality. 

In general, a pairing of two Abelian groups $A$ and $B$ to $\mathsf{U(1)}$,
     \be
        \langle \cdot, \cdot\rangle : A \times B \to \IR/\IZ ~,
     \ee
     is \emph{perfect}, if,
     \begin{itemize}\itemsep 0pt
        \item $\langle a, x \rangle = 0$ $\forall$ $x \in B$ $\implies$ $a = 0$, and,
        \item $\langle y, b \rangle = 0$ $\forall$ $y \in A$ $\implies$ $b = 0$,
     \end{itemize}
      where we are working additively. A pairing $\langle a, \cdot\rangle$ is a homomorphism $B \to \IR/\IZ$. Being perfect means $a \mapsto \langle a, \cdot\rangle$ is an isomorphism of $\Hom(B, \IR/\IZ) \cong A$.

In general, given any Abelian group $A$, the \emph{Pontryagin dual of $A$ } is by definition the Abelian group: 
\be\label{eq:PontryaginDual}
\wt{A} = \Hom(A, \IR/\IZ) ~ . 
\ee
We will also regard the Pontryagin dual as $\Hom(A, \mathsf{U(1)})$, when working multiplicatively.  
For locally compact groups, the Pontryagin duality theorem says that the dual of the dual is isomorphic to the original group: $\widetilde{\!\widetilde{A}} \cong A$. For more on all this, see \autoref{sec:HeisenbergGroups}.

  Using the diagram of the dancing hexagon \eqref{eq:DancingHexagon} and standard Poincar\'e duality for compact oriented manifolds, one can show that the pairing \eqref{eq:DC-Pairing},
    \eqa{
      \widecheck{H}^{\ell}(M_{n}) \times \widecheck{H}^{n-\ell+1}(M_{n}) &\rightarrow \IR/\IZ' ~,
    }
    is a perfect pairing. That is,
    \eqa{
       \Hom\big( \widecheck{H}^{\ell}(M_{n}); \IR/\IZ' \big) &\cong \widecheck{H}^{n-\ell+1}(M_{n}) ~, \label{eq:ppduality}
    }
    and therefore, the Poincar\'e dual cohomology group is also the Pontryagin dual group. 
We call this property   Pontryagin-Poincar\'{e} duality of differential cohomology. The following diagram provides a proof: 
    \begin{equation}
    \hspace{-0.1in}
    \adjustbox{scale=0.85,center}{
      \begin{tikzcd}[column sep=large]
          0 \arrow[r] & H^{\ell-1}(M_{n}; \IR/\IZ') \arrow[r] \arrow[rrd,"\text{perfect pairing}",Red,sloped,leftrightarrow] & \widecheck{H}^{\ell}(M_{n}) \arrow[r] & \Omega_{\IZ}^{\ell}(M_{n}) \arrow[r] & 0 \\
          0 \arrow[r] & \Omega^{n-\ell}(M_{n})/\Omega^{n-\ell}_{\IZ'}(M_{n}) \arrow[r] \arrow[rru,"\text{perfect pairing}",Purple,sloped,swap,leftrightarrow] & \widecheck{H}^{n-\ell+1}(M_{n}) \arrow[r] & H^{n-\ell+1}(M_{n}; \IZ) \arrow[r] & 0
      \end{tikzcd}
      }
    \end{equation}
     
    \begin{remark}
    $\,$
    \begin{enumerate}

    \item Poincar\'e-Pontryagin duality is essential for the manifestly $S$-duality invariant formulation of the Hilbert space of generalized Maxwell theory. 

\item If $\Sigma$ is an $(n-\ell)$-cycle, then the evaluation map gives a homomorphism 
$\widecheck{H}^{n+1-\ell}(M_n) \to U(1)$ and hence by Pontryagin-Poincar\'e duality, 
defines a differential character $\check \delta(\Sigma) \in \widecheck{H}^{\ell}(M_n)$, via 
\be\label{eq:delta-check-sigma}
\langle \check \delta(\Sigma), \chi \rangle = \chi(\Sigma) ~.
\ee
The fieldstrength of this map is not a smooth differential form. It is a $\delta$-function 
supported representative of the Poincar\'e dual of $\Sigma$, also known as a ``current'' in the subject of geometric measure theory \cite{deRham1984}.

    \end{enumerate}

\end{remark}
     
\SectionWithHeader{The Stone-von Neumann Representation Of Heisenberg Groups}{The Stone-von Neumann Representation Of Heisenberg Groups}{sec:HeisenbergGroups}

Before returning to the physics of the generalized Abelian gauge field, we need to review one more mathematical topic. 

Let $A$ be an Abelian group with a   Haar measure. We can then define 
  the Hilbert space:
   \be
       \CH = L^{2}(A) ~.
   \ee
We will also assume in the following discussion that the measure is translationally invariant. That is, if $a\in A$ and $S$ is a subset of $A$, then the measure of $S$ is the same as the measure of the set, where we shift all elements of $S$ by $a$. Simple examples 
of such Hilbert spaces are:

\begin{enumerate} 

\item   If $A = \IR$, $L^{2}(\IR)$ is the space of square-integrable functions over the reals, familiar to us from quantum mechanics. 

\item If $A = \IZ$, $L^{2}(\IZ)$ is the set of wavefunctions on a lattice, e.g., a spin chain.

\item Finally, if $A = \mathsf{U(1)}$, then $L^{2}(\mathsf{U(1)})$ enters the quantum mechanics of a particle on a circle.

\item If $A$ is a cyclic group $\IZ_n$, then $L^2(A)$ is the Hilbert space for discrete quantum mechanics on the circle. This is the canonical example where there are clock and shift operators, as we will see below. 

\item We can take finite products of all the above groups. 

\end{enumerate} 
 
   Note that $L^{2}(A)$ is a representation of $A$ by translation operators. So for each $a_0 \in A$, there is a translation operator $T_{a_0}$, such that,
   \eqa{
      \big( T_{a_0}\psi\big)(a) &:= \psi(a + a_0) ~,\\
      T_{a_0}\circ T_{a_0'} &= T_{a_0 + a_0'} ~, \label{eq:TranslationOps-L2A}
   }

   Also, $L^{2}(A)$ is a representation of the Pontryagin dual $\wt{A}$ 
   defined in \eqref{eq:PontryaginDual}. Thinking of the Pontryagin dual 
   multiplicatively,  for a character $\chi \in \wt{A}$, there is a multiplication operator $M_{\chi}$, such that,
   \eqa{
     \big( M_{\chi}\psi \big)(a) &:= \chi(a)\psi(a) ~,\\
     M_{\chi_1}\circ M_{\chi_2} &= M_{\chi_1\cdot \chi_2} ~.\label{eq:MultiplicationOps-L2A}
   }
   However, famously, $L^{2}(A)$ is \underline{not} a representation of the direct product group $A \times \wt{A}$, because the group law on the direct product says the two factors commute. On the other hand, a simple computation shows that: 
   \be
      T_{a_0}M_{\chi} = \chi(a_0) M_{\chi} T_{a_0} ~. \label{eq:TeeEmmProd}
   \ee
   Rather, $L^2(A)$ is a representation of the Heisenberg extension, which, \underline{as a set}, is $\mathsf{U(1)} \times A \times \wt{A}$, but has a group law,
   \be
     \big( z_1, (a_1, \chi_1) \big) \cdot  \big( z_2, (a_2, \chi_2) \big) := \big( z_1 z_2\chi_1(a_2), (a_1+a_2, \chi_1\chi_2) \big) ~. \label{eq:prodrule}
   \ee
   (Here we are writing $A$ additively, and $\wt{A}$ multiplicatively.)
   Therefore the Heisenberg group sits in a short exact sequence,
   \be\label{eq:Heis-SES}
     \begin{tikzcd}
         1 \arrow[r] & \mathsf{U(1)} \arrow[r] & \mathsf{Heis}(A \times \wt{A}) \arrow[r] & A \times \wt{A} \arrow[r] & 1 ~.
     \end{tikzcd}
   \ee
   This is a \emph{central extension}, which means that the image of $\mathsf{U(1)}$ in $\mathsf{U(1)} \to \mathsf{Heis}(A \times \wt{A})$ is central. This is an example of an anomaly. One can say that there is an anomaly in $A \times \wt{A}$ and what is actually represented on the Hilbert space is the Heisenberg group.

   Because of \eqref{eq:TeeEmmProd}, 
   \eqa{
      \rho(z, (a, \chi)) &= z T_{a} M_{\chi} ~,
   }
   is an honest representation of the Heisenberg group. These operators are composed via the group law \eqref{eq:prodrule}.

   Now the key theorem on representations of $\mathsf{Heis}(A \times \wt{A})$ is:

   \begin{theorem}[Stone-von Neumann-Mackey] Up to isomorphism there is a \underline{unique}, unitary irrep of $\mathsf{Heis}(A \times \wt{A})$, such that the central $\mathsf{U(1)}$ acts by scalars:
   \be
      \rho(z) = z \cdot \mathbbm{1}_{\CH}  \quad \forall \, z \in \mathsf{U(1)} ~.
   \ee
   \end{theorem}
   \begin{proof} The proof can be found in many places. One of them (with further references) is \cite[Sec. 15.5.5]{MooreGroupTheory:2023}.
   \end{proof}

   One model for the SvN representation is $L^2(A)$ with $A$ and $\wt{A}$ acting as translation + multiplication operators. If $A$ is locally compact, then Pontryagin duality says $\wt{\wt{A}} \cong A$, so
   \be
      \mathsf{Heis}(A \times \wt{A}) = \mathsf{Heis}(\wt{A} \times A) ~.
   \ee
   Therefore another equivalent representation is $L^{2}(\wt{A})$, where $\wt{A}$ acts by translation and $A$ acts by multiplication. The isomorphism $L^{2}(A) = L^{2}(\wt{A})$ is Fourier transformation.

\begin{enumerate}

\item We stress that although $A$ and $\wt A$ are Abelian groups, the Heisenberg group 
$\mathsf{Heis}(A\times \wt A)$ is non-Abelian.  It fits in the short exact sequence 
\eqref{eq:Heis-SES} as a central extension.  But the feature of the Heisenberg extension that distinguishes it from all the other central extensions of $A \times \wt{A}$ is that this $\mathsf{U(1)}$ \underline{is} the entire center: $Z\big(\mathsf{Heis}(A \times \wt{A})\big) \cong \mathsf{U(1)}$, so it is as non-Abelian as it can possibly be, given the fact that it is a central extension by $\mathsf{U(1)}$.

\item \textbf{Generalizations Of The Pontryagin dual for non-Abelian groups}. 
What happens if we try to generalize the definition of the Pongtryagin dual group 
to the case where   $G$ is non-Abelian. First, we show that $\Hom(G, \mathsf{U(1)})$ factors through the Abelianization of $G$. Suppose $\chi \in \Hom(G, \mathsf{U(1)})$. Then $\chi: G \to \mathsf{U(1)}$ is a group homomorphism. Then, 
   \begin{align}
    \chi(g_1 g_2) &= \chi(g_1)\chi(g_2) \nn
                  &= \chi(g_2) \chi(g_1) \nn
                  &= \chi(g_2 g_1) ~,
   \end{align}
   which, of course, means that $\chi([g_1, g_2]) = 1$, where $[g_1, g_2]$ denotes the group commutator. Therefore, $\chi$ factors through the Abelianization:
   \be
   \begin{tikzcd}
      G \arrow[rr] \arrow[rd] &  & \mathsf{U(1)} \\
      & \begin{array}{c}G/[G,G] \\ =: \mathsf{Ab}(G)\end{array} \arrow[ru, "\widetilde{\chi}", dashed, swap]
   \end{tikzcd}
   \ee
   This takes us back to the theory of Abelian groups. So the definition of Pontryagin dual of $G$ as $\Hom(G, \mathsf{U(1)})$ does not give anything new. In fact, for some groups, the Abelianization is trivial, so in general, it does not give anything interesting. 
   
   There is, however, another definition that does better and goes further. 
   First, recall that for Abelian $A$, the Pontryagin dual $\wt{A} $ can be identified with the group of isomorphism classes of irreducible unitary representations of $A$.  So if we have an Abelian group, one can diagonalize the operators simultaneously. So one might as well make them one-dimensional, and the irreducible ones are in $\mathsf{U(1)}$. So one can equivalently think of $\wt{A}$ as the set of irreducible unitary representations of $A$.  For \underline{non-Abelian} $G$, we can identify $\wt{G}$ as the \underline{set} of unitary irreps of $G$:
   \be
        \wt{G} = \{ \text{set of unitary irreps of $G$} \} ~.
   \ee
   Now we do not have a group anymore. But we do have a category. In fact, we have a \underline{monoidal category}, which is called $\mathsf{Rep}(G)$. The objects are unitary irreps, the morphisms are $G$-equivariant maps, and the tensor product is the usual tensor product of representations.  What replaces Pontryagin duality is something called Tannaka-Krein duality \cite{Joyal1991}. Given a fusion category with a fiber functor (i.e., a monoidal functor to the category of vector spaces), one can actually reconstruct a group. The construction reproduces the group $G$ from the fusion category $\mathsf{Rep}(G)$.   
    
\end{enumerate}

\SectionWithHeader{The Hilbert Space Of A Generalized Abelian Gauge Theory}{The Hilbert Space Of A Generalized Abelian Gauge Theory}{sec:HilbertSpaceGAGT}
  %
  %
  The set of gauge equivalence classes of Generalized Abelian Gauge Theory (GAGT) on spacetime $M_{n}$ is $\widecheck{H}^{\ell}(M_{n})$. We now assume a spacetime splitting,
  \be
     M_{n} = N_{n-1} \times \IR ~,
  \ee
  with $\IR$ parametrized by the time coordinate, and an action \eqref{eq:actionh}:
  \be
    \pi \int_{M_{n}}\lambda F \wedge \star F ~.
  \ee
  As we have noted before, since $F$ has quantized periods, the coupling $\lambda$ is physically relevant. For example, the periodic scalar $\lambda$ is related to the radius $R$ of the target circle by $\lambda = R^2$.

  The classical momentum is the $(n-\ell)$-form,
  \be
     \pi = 2\pi\lambda \left.(\star_{M_n}F)\right|_{N_{n-1}} ~.
  \ee
  The classical phase space is the cotangent space of field configurations on the spatial slice:
  \be
     T^* \widecheck{H}^{\ell}(N_{n-1}) = \widecheck{H}^{\ell}(N_{n-1}) \times \Omega^{n-\ell}(N_{n-1})/\mathsf{im\,}d^\dagger ~.
  \ee
  This follows from the noncanonical decomposition:
  \be
    \widecheck{H}^{\ell}(N_{n-1}) \approx  \Omega^{\ell-1}(N_{n-1})/\Omega_{\IZ'}^{\ell-1}(N_{n-1}) \times \Gamma_{\ell} ~.
  \ee
  We will use standard quantization, so we have Heisenberg relations for the quantum fields $F_{q}$, $\Pi_{q}$:
  \be
    \left[ \int_{N_{n-1}}\omega_1 F_q , \int_{N_{n-1}}\omega_2 \Pi_q \right] = \imag \hbar \int_{N_{n-1}}\omega_1 d\omega_2 ~, \label{eq:HeisenbergRel}
  \ee
  where $\omega_{1} \in \Omega^{n-1-\ell}(N_{n-1})$ and $\omega_{2} \in \Omega^{\ell-1}(M_{n})$ are classical and the subscript $q$ indicates a quantum operator. 
  Equation \eqref{eq:HeisenbergRel}  is a nice clean   implementation of the heuristic relation
  \be \label{eq:quantum-momentum-op}
     \Pi_q \sim -\imag \hbar  \frac{\delta}{\delta A_q} ~,
  \ee
  In particular, $\Pi_q$ is the generator of translations along $\widecheck{H}^{\ell}(N_{n-1})$. 
  So in $L^2( \widecheck{H}^{\ell}(N_{n-1}))$, we have 
  \be 
  \left(\exp\left[ \frac{\imag}{\hbar} \int_{N_{n-1}} \omega_2 \Pi_q \right] \Psi\right)([\check{A}]) = 
  \Psi([\check{A}] + [\widecheck \omega_2]) ~,
  \ee
  where $[\widecheck \omega_2]$ is the additive notation for the topologically trivial character $\chi_{\omega_2}$.  
  
  For a cotangent space, the quantization is completely standard. Now we use heavily the property that $\widecheck{H}^{\ell}(N_{n-1})$ is an \underline{Abelian group}. Therefore, at least formally, it has a translationally-invariant measure so we can formulate the quantum Hilbert space of the theory:
  \be
     \CH(N_{n-1}) := L^{2}\big( \widecheck{H}^{\ell}(N_{n-1}) \big) ~.
  \ee
  We are imagining that $\widecheck{H}^{\ell}(N_{n-1})$ has a translationally-invariant Haar measure.  As we have seen, $\widecheck{H}^{\ell}(N_{n-1})$ is an infinite-dimensional group, so some analysis is needed to give this formula meaning. From the noncanonical decomposition \eqref{eq:hcheckl-decomp},
  \be
    \widecheck{H}^{\ell}(N_{n-1}) = \IT_{\ell-1} \times V_{\ell} \times \Gamma_{\ell} ~,
  \ee
  we see that the infinite-dimensional part comes from $V_{\ell}$. The issues are the same as in the quantization of the free scalar field. The oscillator modes of $A$ with $d^\dagger A = 0$ are quantized as in standard QFT. In these lectures, we are more concerned with the subtleties arising from the first two finite-dimensional factors. Hence, we are somewhat cavalier about the functional-analytic aspects.
Roughly speaking, the allowed wavefunctions should have Gaussian decay:
  \be
     \psi\big( [\check{A}] \big) \sim \CP\,\exp\left( -\int_{N_{n-1}}k F \star F \right) ~,
  \ee
  where $k$ is a constant and an ON basis for $L^{2}\big( \widecheck{H}^{\ell}(N_{n-1}) \big)$ would involve expressions where $\CP$ is polynomial in the oscillators from $V_{\ell}$. It is possible to make this much more precise, but we will refrain from doing so here.

\bigskip 
\noindent
\begin{remark} We have worked with the wavefunction as a function on the space of gauge equivalence classes. One can also work with a wavefunction as a section of a line bundle over the set of differential cocycles $\widecheck{Z}^\ell$, and impose invariance under 
gauge equivalences (such as the gauge equivalences described in \autoref{sec:ModelDiffCoh}) as a Gauss law. This latter viewpoint is useful because it gives a clear definition of quantum electric charge.
For example, the automorphism group described in Exercise \ref{exercise:automorphism-hopkins-singer} 
must act as a trivial character on 
the wavefunction. In general, the action of the automorphism group 
defines the electric charge: 
\be\label{eq:QuantumGaussLaw}
\alpha \cdot \Psi(\check A) = e^{2\pi \imag \int \alpha Q_e } \Psi(\check A) ~,
\ee
where $\alpha \in H^{\ell-2}(N_{n-1}; \IR/\IZ)$ and $Q_e\in H^{n-\ell+1}(N_{n-1}; \IZ)$. When $N_{n-1}$ is compact, the Gauss law says that if the wavefunction is nonzero then the total charge $Q_e$ must vanish.
\end{remark}

\SectionWithHeader{Manifest Electromagnetic Duality}{Manifest Electromagnetic Duality}{sec:manifestEMduality}
%
%
  
  We can now apply the remarks of \autoref{sec:HeisenbergGroups}
   to  Generalized Abelian Gauge Theory following the papers \cite{Freed:2006ya,Freed:2006yc}. We observed in \autoref{sec:HilbertSpaceGAGT} 
   that the Hilbert space of a theory based on $\widecheck{H}^\ell$ on $N_{n-1}\times \IR$ should be $L^{2}\big( \widecheck{H}^{\ell}(N_{n-1}) \big)$. Since
   $\widecheck{H}^{\ell}(N_{n-1})$ is an Abelian group, we can regard this as  a Stone-von Neumann (SvN) representation of 
   \be\label{eq:ManifestlyDual}
     \mathsf{Heis}\big( \widecheck{H}^{\ell}(N_{n-1}) \times \widecheck{H}^{\ell}(N_{n-1})^{\sim}  \big) ~.
  \ee
  But then Pontryagin-Poincar\'{e} duality \eqref{eq:ppduality} of the differential cohomology says that this is the unique Stone-von Neumann representation of
  \tightfootnote{One uses \eqref{eq:ppduality} with $\mathsf{dim}(N_{n-1}) = n-1$.}
  \be
     \mathsf{Heis}\big( \widecheck{H}^{\ell}(N_{n-1}) \times \widecheck{H}^{n-\ell}(N_{n-1}) \big) ~.
  \ee
  But the Stone-von Neumann representation is unique up to isomorphism!  We could switch the factors, use that $\widetilde{\!\widetilde{A}} \cong A$ for an Abelian group $A$, and equally well say the Hilbert space is: 
  \be
     L^{2}\big( \widecheck{H}^{n-\ell}(N_{n-1}) \big) ~.
  \ee
   Recall that we earlier argued that EM duality is just exchanging $\ell$ for $(n-\ell)$.
 \begin{redbox}[box:EM-dual-SvN]{Important Remark}
  Identifying the Hilbert space of the $\ell$-form gauge theory as the Stone-von Neumann representation of \eqref{eq:ManifestlyDual} is a manifestly electric-magnetically dual formulation of the Hilbert space. 
  \end{redbox}

In fact, there is a natural and specific isomorphism between the ``electric'' and ``magnetic'' formulations. If $\Psi \in L^{2}\big( \widecheck{H}^{\ell}(N_{n-1}) \big)$,  
then we can -- at least formally --  define an associated $\Psi^\vee \in L^{2}\big( \widecheck{H}^{\ell}(N_{n-1}) \big)$  by the Fourier transform, 
\be\label{eq:Abelian-S-Duality-Iso}
\Psi^\vee([\check{A}_{n-\ell}]) := \int_{\widecheck{H}^\ell(N_{n-1})} 
e^{\imag  \langle [\check{A}_{n-\ell}],[\check{A}_\ell]\rangle} \Psi([\check{A}_\ell]) d\mu ~,
\ee 
where $d\mu$ is a (unique up to overall scale) translation-invariant measure on 
the Abelian group  $\widecheck{H}^\ell(N_{n-1})$. 

\bigskip 
\noindent 
\begin{remark} The off-the-shelf mathematical theorems regarding Pontryagin duality apply 
to locally compact Abelian groups, but $\widecheck{H}^\ell(N_{n-1})$ is not locally compact. It is however a direct product (noncanonically) of a finitely-generated Abelian group and a compact torus and an infinite-dimensional vector space. The rigorous theorems apply to the first two factors, but not the third. The infinite-dimensional vector space is a topological vector space describing the oscillator modes of the field. One can impose an energy cutoff on this space to make it finite-dimensional and then, the Pontryagin duality theorem applies because finite-dimensional vector spaces are locally compact. The energy is moreover an invariant of electromagnetic duality. To make our statements fully rigorous, one would need to discuss the limit as the energy cutoff is taken to infinity. We will not discuss that here, but we do not expect any difficulties from this limit. 
\end{remark}

\SectionWithHeader{The Definition And Noncommutativity Of Quantum Electric And Magnetic Fluxes}{Quantum Electric And Magnetic Fluxes, And Noncommutativity}{sec:noncommutativity}

  Let us start with a puzzle. We have seen that the periods of the fieldstrength $F$ of a differential character in $\widecheck{H}^{\ell}(N_{n-1})$ are quantized:
  \be
    \int_{\Sigma_\ell}F  \in 2\pi\IZ \quad \text{for}  \quad \partial\Sigma_{\ell} = 0 ~.
  \ee
  However, the periods $\int_{\Sigma_{n-\ell}}\star F$ are definitely \underline{not quantized}! In general, continuous changes of the metric alter the result.
  \tightfootnote{If   $\ell \not= \frac{1}{2}n$,    a simple conformal scaling of the 
  metric will continuously change the periods of $\star F$ thanks to the property 
  \eqref{eq:HodgeStarScaling}. When $\ell = \frac{1}{2}n$, the periods of $\star F$ will 
  be conformally invariant, but nevertheless, under general deformations of the metric, the 
  periods of $\star F$ will change.   Here is a simple example to illustrate the point: 
  Consider the four-manifold $X$ given by the product of two two-dimensional spheres. We write $X=\IS^2_1 \times \IS^2_2$, and endow $X$ with a direct sum metric which is a sum of standard round sphere metrics of radii  $R_1$ and $R_2$, respectively. Choosing a standard orientation 
  $d\phi_i \wedge d(\cos\theta_i)$, $i=1,2$ on each sphere, let $\omega_i$, $i=1,2$ be the unit volume form for the two spheres. One easily computes that, with the product orientation, $\int_{\IS^2_2} \star \omega_1 = \frac{R_2^2}{R_1^2}$. We confirm the conformal invariance, but note that independent changes of $R_1,R_2$  change the period of $\star\omega_1$.}
This might seem confusing since the electromagnetic dual field $\wt{F}$ should have 
quantized periods and we usually write the dual field as $\wt{F} = \star F$. 
There is thus some tension with electromagnetic duality. We can resolve this puzzle by thinking more carefully about the \underline{quantum} definition of electric flux.

  As we have noted, if $M_{n} = N_{n-1} \times \IR$ and $\CH = L^{2}\big( \widecheck{H}^{\ell}(N_{n-1}) \big)$, then the generator of translations is:
  \be
     \Pi_q \sim \frac{\delta S}{\delta A} \longrightarrow \Pi_q = \lambda\left.\big( \star F_{q} \big)\right|_{N_{n-1}} ~.
  \ee
  Let us look at translation eigenstates $\Psi(\check{A}) \in L^{2}\big(\widecheck{H}^{\ell}(N_{n-1})\big)$. A translation eigenstate would satisfy,
  \be
      \Psi\big( \check{A} + \widecheck{\phi} \big) = \exp \bigg( 2\pi \imag\int_{N_{n-1}}^{\widecheck{H}} \widecheck{\varepsilon} \cdot \widecheck{\phi} \bigg) \Psi(\check{A}) ~,
  \ee
  where the phase is a homomorphism $\widecheck{H}^{\ell}(N_{n-1}) \to \mathsf{U(1)}$, and its form is fixed by Pontryagin duality to be $\exp\big(2\pi \imag \langle \widecheck{\varepsilon}, \widecheck{\phi}\rangle\big)$, where $\widecheck{\varepsilon}$ is some differential character of degree $(n-\ell)$.
  
  It is natural to regard the eigenvalue $\widecheck{\varepsilon} \in \widecheck{H}^{n-\ell}(N_{n-1})$ as the quantum definition of definite ``electric flux'' (we will soon alter the meaning of this term).
  Eigenstates of the translation operator are plane waves on fieldspace.
  \tightfootnote{ In the theory of a real-valued scalar field $\phi$, we could consider a quantum wavefunctional
  \be
     \Psi\big(\phi(x)\big) = \exp\left( \imag  \int_{N_{n-1}}\phi(x) j(x) \right) ~.
  \ee
  Note that it is an eigenstate of the group of translations on the scalar field 
  $\phi(x) \mapsto \phi(x) + a(x)$. 
  Such a wavefunctional exists as a complex-valued function on fieldspace, but it is obviously very far from being normalizable -- there is no Gaussian falloff in fieldspace. Electric flux eigenstates, with the naive definition of electric flux, are the analogs of such plane waves in the scalar field. }
  Such ``plane wave'' functionals are definitely not normalizable so this notion of electric flux is of limited utility. But we could make wavepackets using $\widecheck{\varepsilon}$ in the same \underline{connected component}. In this way, we will get a wavefunction of definite electric flux. So a much more useful definition of electric flux is the topological class of $\widecheck{\varepsilon}$, a class in $H^{n-\ell}(N_{n-1}; \IZ)$. We note that $\widecheck{\varepsilon}_{1}$, $\widecheck{\varepsilon}_{2}$ are in the same path component of $\widecheck{H}^{n-\ell}(N_{n-1})$ if and only if for all \underline{flat fields} $\phi_{f} \in H^{\ell-1}(N_{n-1}; \IR/\IZ')$, we have,
  \be
      \int_{N_{n-1}}^{\widecheck{H}} \widecheck{\phi}_{f} \cdot \widecheck{\varepsilon}_{1} = \int_{N_{n-1}}^{\widecheck{H}} \widecheck{\phi}_{f} \cdot \widecheck{\varepsilon}_{2} ~.
  \ee
  This leads us to the crucial definition:

  \begin{definition}[\textcolor{red}{State of definite electric flux} \cite{Freed:2006ya,Freed:2006yc}] A state $\psi \in L^{2}\big(\widecheck{H}^{\ell}(N_{n-1}) \big)$ is in a state of definite electric flux if it is a translation eigenstate under the subgroup of \textbf{flat fields}. That is, for all   $\phi_{f} \in H^{\ell-1}(N_{n-1}; \IR/\IZ')$,
    \be\label{eq:ShiftFlatField}
        \psi(\check{A} + \widecheck{\phi}_{f}) = \exp\left( \imag  \int_{N_{n-1}}e \cup \phi_{f} \right) \psi(\check{A}) ~,
    \ee
    for some $e \in H^{n-\ell}(N_{n-1}; \IZ)$. This follows from \autoref{subsec:FlatCharacters}
    and Poincar\'e-Pontryagin duality. Here $e$ is the quantized electric flux and is the proper quantization of $[\star F]$.  We denote the ``electric'' translation operator by $\phi_{f} \in H^{\ell-1}(N_{n-1}; \IR/\IZ')$, as, 
    \be\label{eq:ElectricTranslation}
    (\CU_e(\phi_f) \cdot \Psi)(\check{A}) := \Psi(\check{A} + \widecheck \phi_f) ~.
    \ee
  \end{definition}

Recall that the energy-momentum tensor of the generalized Abelian gauge field is given by
\eqref{eq:GAGT-Tmunu}, and therefore only involves the fieldstrength. It is therefore invariant under translation by flat fields. Thus, we could say that the translations by flat fields are a symmetry of the theory. 
   
Now it is natural to ask how to formulate magnetic flux sectors. 
Of course, we could formulate the Hilbert space as 
$L^{2}\big( \widecheck{H}^{n-\ell}(N_{n-1}) \big)$, and define a flux sector of 
definite magnetic flux to be an eigenstate under the action of the 
dual flat fields of $\widecheck \phi_f^m \in \widecheck{H}^{n-\ell}(N_{n-1})$, 
where $\phi_{f}^{m} \in H^{n-\ell-1}(N_{n-1}; \IR/\IZ')$. 
This is a valid definition, 
but how do the magnetic flux sectors appear if we continue to work with 
the Hilbert space as modeled by $L^{2}\big( \widecheck{H}^{\ell}(N_{n-1}) \big)$? 
Recall that 
in $L^2(A)$, the group $A$ acts on wavefunctions by translation, and the group 
$\wt A$ acts by multiplication by a character. In our case, the action of the 
dual flat fields is: 
\be 
\begin{split}
(\CU_m(\phi_f^m)\cdot \Psi)(\check{A}) & = e^{  \imag  \langle \widecheck \phi_f^m, \check{A}\rangle} \Psi(\check{A}) \\
& = e^{  \imag  \int_{N_{n-1} } \phi_f^m c(\check{A}) } \Psi(\check{A}) ~.
\end{split}
\ee
The quantity $\int_{N_{n-1} } \phi_f^m c(\check{A})$ is a character on the group 
of components of $\widecheck{H}^\ell(N_{n-1})$. Recall that   $\widecheck{H}^{\ell}(N_{n-1})$ has connected components:
  \be
     \widecheck{H}^{\ell}(N_{n-1}) = \coprod_{m \in H^{\ell}(N_{n-1}; \IZ)} \widecheck{H}^{\ell}(N_{n-1})_{m} ~.
  \ee
We conclude that a state $\Psi$ is an eigenstate of definite \emph{magnetic flux} if $\mathsf{supp}(\Psi)$ is contained in exactly one connected component 
of $ \widecheck{H}^{\ell}(N_{n-1})$.   This turns out to be the eigenstate of translations by 
the dual flat fields and so we define:  
\bigskip
  \begin{definition}[\textcolor{red}{State of definite magnetic flux} \cite{Freed:2006ya,Freed:2006yc}] A state of definite magnetic flux $m \in H^{\ell}(N_{n-1}; \IZ)$ is a wavefunction $\psi$ with support in the component $\widecheck{H}^{\ell}(N_{n-1})_{m}$.
  \end{definition}
\bigskip
%


  
  We can get from $\CU_{m}(\phi_{f}^{m})$ to $\CU_{e}(\phi_{f}^{e})$ by a Fourier transform. Now, recall that the compact Abelian group $H^{\ell-1}(N_{n-1}; \IR/\IZ')$ might be disconnected, and, recalling \eqref{eq:connected-components-and-torsion-identity},
  \be
      \pi_{0}\big( H^{\ell-1}(N_{n-1}; \IR/\IZ') \big) = \mathsf{Tors}\big( H^{\ell}(N_{n-1}; \IZ)\big) ~.
  \ee
Suppose a wavefunction $\Psi$ has definite magnetic flux -- so it is supported in a 
definite connected component of $\widecheck{H}^\ell(N_{n-1})$ --  and consider a topologically 
nontrivial flat electric field $\phi_f^e$. Translating $\Psi$ by such a flat field cannot 
preserve the original component of support of $\Psi$. This means that, in general, $\CU_e(\phi_f^e)$ and $\CU_m(\phi_f^m)$ will not commute if there is torsion in the cohomology of $N_{n-1}$. In fact, $\CU_e(\phi_f^e)$ and $\CU_m(\phi_f^m)$ are 
simply special cases of translation and multiplication operators, respectively,  
on the Stone-von Neumann representation $L^2(\widecheck{H}^\ell(N_{n-1}))$. In the language 
of \eqref{eq:TranslationOps-L2A} and \eqref{eq:MultiplicationOps-L2A}, we have, 
\eqa{ 
\CU_e(\phi_f^e) &= T_{\widecheck \phi_f^e} ~, \quad 
\CU_m(\phi_f^m) &= M_{\chi( \widecheck \phi_f^m)} ~,
}
where $\chi( \widecheck \phi_f^m)$ is the character of $\widecheck{H}^\ell(N_{n-1})$ defined by 
$\chi( \widecheck \phi_f^m)([\check{A}]) = e^{\imag  \langle  \widecheck \phi_f^m, \check{A}\rangle}$. It therefore follows from the general exchange relation 
\eqref{eq:TeeEmmProd}, that we have  the exchange relation: 
    \be 
       \CU_{e}(\phi_{f}^{e}) \CU_{m}(\phi_{f}^{m}) =  \chi( \widecheck \phi_f^m)(\widecheck \phi_f^e)  \CU_{m}(\phi_{f}^{m}) \CU_{e}(\phi_{f}^{e}) ~.
    \ee
But we can write, 
\be 
\begin{split}
\chi( \widecheck \phi_f^m)(\widecheck \phi_f^e)  & = e^{\frac{\imag}{2\pi} \langle \widecheck \phi_f^m, \widecheck \phi_f^e\rangle } \\
& = e^{- \frac{\imag}{2\pi}   \int_N \phi_f^m \beta(\phi_f^e)} ~,
\end{split}
\ee
where in the second line, we used \eqref{eq:FlatPairing}, and 
the fact that $c(\widecheck \phi_f^e) = - \beta(\phi_f^e)$, as 
mentioned in equation \eqref{eq:TopClassFlatCharacter}.  

From the previous paragraph, we conclude that: 
    \be\label{eq:NoncommutativeFluxOps}
       \CU_{e}(\phi_{f}^{e}) \CU_{m}(\phi_{f}^{m}) = \mathsf{T}(\phi_{f}^{e}, \phi_{f}^{m}) \CU_{m}(\phi_{f}^{m}) \CU_{e}(\phi_{f}^{e}) ~,
    \ee
    where $\mathsf{T}$ is the torsion pairing (defined below):
    \be
        \mathsf{T}: H^{\ell-1}(N_{n-1}; \IR/\IZ') \times H^{n-\ell-1}(N_{n-1}; \IR/\IZ') \longrightarrow \mathsf{U(1)} ~.
    \ee
 The torsion pairing on these cohomology groups is defined using the Bockstein map by: 
    \be 
     \mathsf{T}(\phi_f^e, \phi_f^m):= \exp \left( - \frac{\imag}{2\pi}  \int_{N_{n-1}} \phi_f^e \beta(\phi_f^m)\right) = \exp \left(- (-1)^{\ell(n-\ell)}\, \frac{\imag}{2\pi}  \int_{N_{n-1}}  \phi_f^m \beta(\phi_f^e)\right) ~.
\ee
It descends to a pairing on the connected components of:
\be 
H^{\ell-1}(N_{n-1}; \IR/\IZ') \times H^{n-\ell-1}(N_{n-1}; \IR/\IZ') ~ . 
\ee
That is, it is a pairing on $\Tors(H^\ell(N_{n-1}; \IZ)) \times \Tors(H^{n-\ell}(N_{n-1}; \IZ))$, 
hence the name. In fact, it is a perfect pairing on these torsion groups. 
Because the pairing is perfect, the noncommutativity on the electric and magnetic fluxes is maximal, and in fact defines a Heisenberg group extension of $\Tors(H^\ell(N_{n-1}; \IZ)) \times \Tors(H^{n-\ell}(N_{n-1}; \IZ))$.

Finally, recall that the energy-momentum tensor $T_F = T_{\wt{F}}$ only involves the fieldstrength. Therefore, shifting by a flat field commutes with the energy-momentum tensor. Therefore, the group generated by shifts by electric and magnetic flat fields is a \underline{symmetry} of the standard dynamics of the generalized Abelian gauge field 
in $\widecheck{H}^\ell$. 

The groups of translations by flat electric and flat magnetic fields have 
an interpretation as a group of \underline{symmetries} of the theory. 
Since \eqref{eq:GAGT-Tmunu} only depends on the fieldstrength,  
    translations by flat electric fields -- implemented by the operators 
    $\CU_e(\phi_f^e)$ -- is a group of 
    symmetries of the generalized Abelian gauge theory, with symmetry group 
    $H^{\ell-1}(N_{n-1}; \IR/\IZ')$. 
    Similarly, since $T_F = T_{\wt{F}}$ only 
    involves the dual fieldstrength, the group of translations by flat magnetic fields --
     implemented by the operators   $\CU_e(\phi_f^e)$ -- is a group of symmetries, 
     with symmetry group $H^{n-\ell-1}(N_{n-1}; \IR/\IZ')$. 
     Together these groups generate a noncommutative group of symmetries, which fits in a central extension,
     \be\label{eq:EM-SymmetryGroup}
     1 \rightarrow \mathsf{U(1)} \rightarrow \CS \rightarrow H^{\ell-1}(N_{n-1}; \IR/\IZ') \times H^{n-\ell-1}(N_{n-1}; \IR/\IZ')
     \rightarrow 1 ~,
     \ee
     where the cocycle is defined by the torsion pairing.

\bigskip 
\bigskip 
\noindent \textbf{Example:} The above remarks apply to the case of four dimensions with $\ell=2$, that is, to the ordinary Maxwell field as studied in standard physics textbooks. Since 
the symmetry group \eqref{eq:EM-SymmetryGroup} is, in general,  non-Abelian,  this has the interesting implication that on manifolds where the cocycle is nontrivial, the   degeneracy of the eigenspaces of the Hamiltonian of the Maxwell field is nontrivial.   A simple example 
is the case of  $d= 3+1$ Maxwell Theory where the spatial manifold is a Lens space
$X = L_{m} \times \IR_{t}$. See Box \ref{box:lensspace} for more on Lens spaces. Recall that $L_{m} = \IS^{3}/\IZ_{m}$ and $H^{1}(L_{m}; \IR/\IZ) \cong \IZ_{m}$. In this case, the group generated by electric and magnetic translation operators $\CU_{e}$ and $\CU_{m}$ is a Heisenberg group:
   \be
     \begin{tikzcd}
        0 \ar[r] & \IZ_{m} \ar[r] & \mathsf{Heis}(\IZ_{m} \times \IZ_{m}) \ar[r] & \IZ_{m} \times \IZ_{m} \ar[r] & 0 ~.
     \end{tikzcd}
   \ee
   The translation by flat fields is clearly a symmetry of the Hamiltonian, so we conclude that the
   energy eigenstates of the  Maxwell theory must be degenerate: $\CU_{e}$, $\CU_{m}$ commute with the Hamiltonian, so the group generated by $\CU_{e}$, $\CU_{m}$ commutes with the Hamiltonian. These operators generate the Heisenberg group   $\mathsf{Heis}(\IZ_{m} \times \IZ_{m})$, which has a 
   unique irreducible representation of dimension $m$  (where a phase $\zeta$ in the $U(1)$ 
   center acts as scalar multiplication by $\zeta$). Therefore, all the energy eigenspaces  must have a degeneracy divisible by $m$. In particular, the ground state has a degeneracy divisible by $m$. 
   \tightfootnote{Essentially this kind of phenomenon was used in the old string compactification literature where vacua were constructed by turning on flat Wilson lines. An important difference is that the argument for the exactness of the degeneracy uses supersymmetry.}

   One might wonder if one could use this result to get a topological qubit or qudit in the real world. The catch is that the theory must be formulated on topologically nontrivial spaces. An attempt was sketched in \cite{Kitaev:2007ed}. The difficulty one must overcome is that one must convince our experimentalist friends to make a Lens space 
   in their laboratory and put electromagnetic fields in that Lens space. At first sight, this appears to be impossible because is a 
   mathematical theorem that says that one cannot \emph{embed} a Lens space into $\IR^{3}$ (where we live). Nevertheless,  one can \emph{immerse}  a Lens space into $\IR^3$ (so that self-intersections are allowed). The basic idea of \cite{Kitaev:2007ed} is to use Josephson junctions to mimic the behavior of an electromagnetic field in a nontrivial $3$-manifold, at least in the low-energy regime. It is argued there that the electromagnetic field in a suitable configuration of such Josephson junctions will indeed have a groundstate which forms a qubit. A notable feature of this phenomenon, though, is that it is a \underline{long-distance}, \underline{macroscopic}, quantum phenomenon. 
     One might ask if the phenomenon could have applications in cosmology.  The topology of the universe is not known and might be nontrivial \cite{Cornish:1997ab,CompactCollaboration} and topologies like 
     the Poincar\'e sphere, for which our phenomenon applies, have been considered. The question is complex, since one must ask whether the topology is visible to an observer so we leave this remark as a tantalizing possibility. 
     \tightfootnote{G.M. thanks T. Banks, D. Friedan, and J. Maldacena for useful remarks on the relevant issues one must consider.}

\SectionWithHeader{Wilson And 't Hooft Operators}{Wilson And 't Hooft Operators}{sec:Wilson-tHooft}

We now define \emph{Wilson operators} and \emph{'t Hooft operators} in the Hamiltonian formalism. We observe that, given an $(\ell-1)$-cycle $\Sigma_{\ell-1} \subset N$, the 
holonomy along $\Sigma_{\ell-1}$ defines a character on $\widecheck{H}^\ell(N_{n-1})$: 
\be 
h_{\Sigma_{\ell-1}}:  \widecheck{H}^\ell(N_{n-1}) \rightarrow \mathsf{U(1)} ~,
\ee
just given by the evaluation map. Therefore, we can define the \emph{Wilson operator associated to $\Sigma_{\ell-1}$} to be the multiplication of the wavefunction by this holonomy function: 
\be 
(\CW(\Sigma_{\ell-1})\cdot \Psi)[\check{A}] := h_{\Sigma_{\ell-1}}[\check{A}] \Psi[\check{A}] ~.
\ee

Now,  if $\Sigma'_{(n-1)-\ell}\subset N_{n-1}$ is a closed $(n-1-\ell)$-cycle, then the holonomy function: 
\be 
h_{\Sigma'_{n-1-\ell}}:  \widecheck{H}^{n-\ell}(N_{n-1}) \rightarrow \mathsf{U(1)} ~,
\ee
defines a character $\widecheck \delta(\Sigma'_{n-1-\ell})\in \widecheck{H}^\ell(N_{n-1})$ by Poincar\'e-Pontryagin duality: 
\be 
h_{\Sigma'_{n-1-\ell}}([\check{A}_1^{n-\ell}]) = 
\exp\left[ 2\pi \imag  \left\langle [\check{A}_1^{n-\ell}], \widecheck \delta(\Sigma'_{n-1-\ell})\right\rangle \right] ~.
\ee
Here, $\check \delta(\Sigma'_{n-1-\ell})$ is the non-smooth differential character 
introduced in equation \eqref{eq:delta-check-sigma} above. 
We can define the \emph{'t Hooft operator associated to $\Sigma'_{\ell-1}$}, to be the 
translation operator along $\check \delta(\Sigma'_{n-1-\ell})$: 
\be 
(\CH(\Sigma'_{n-1-\ell})\cdot \Psi)([\check{A}] ) := \Psi([\check{A}] + [\widecheck \delta(\Sigma'_{n-1-\ell}) ] ) ~.
\ee
Since $\check\delta(\Sigma'_{n-1-\ell})$ has delta-function support, this is very close to the 
original definition of 't Hooft. 

Just like the operators $\CU_e$ and $\CU_f$, the Wilson and 't Hooft operators are multiplication and translation operators in the Stone-von Neumann representation of the Heisenberg group \eqref{eq:ManifestlyDual}. In the representation $L^2(\widecheck{H}^\ell(N_{n-1}))$,  
we have (using the general notation \eqref{eq:TranslationOps-L2A} and \eqref{eq:MultiplicationOps-L2A} for Heisenberg representations): 
\be 
\begin{split} 
\CW(\Sigma_{\ell-1}) & = M_{h_{\Sigma_{\ell-1}}} ~,\\ 
\CH(\Sigma'_{n-1-\ell}) & = T_{\widecheck \delta(\Sigma'_{n-1-\ell})} ~.\\ 
\end{split}
\ee
According to the general relation \eqref{eq:TeeEmmProd}, we might expect the 
Wilson and 't Hooft operators to have a nontrivial exchange relation, but here 
we find a surprise. The phase in the exchange relation is just 
\be\label{eq:WtH-phase}
h_{\Sigma_{\ell-1}}(\widecheck \delta(\Sigma'_{n-1-\ell})) = 
\exp \left[ 2\pi \imag  \int_{\Sigma_{\ell-1}} \widecheck \delta(\Sigma'_{n-1-\ell}) \right] 
= \exp \left[ 2\pi \imag  \langle \widecheck \delta(\Sigma_{n-\ell}),  \widecheck \delta(\Sigma'_{n-1-\ell})\rangle \right] ~.
\ee
Now the support of $\widecheck \delta(\Sigma_{\ell-1})$ (say, in terms of the Hopkins-Singer cochains 
$(a,h,\omega)$) can be restricted to an arbitrarily small tubular neighborhood of 
$\Sigma_{\ell-1}$, so it follows from the pairing formula \eqref{eq:HS-CochainPairing} that, 
so long as $\Sigma_{\ell-1}\subset N_{n-1}$ and $\Sigma'_{n-\ell-1}\subset N_{n-1}$ are in general position, the phase \eqref{eq:WtH-phase} is equal to $1$. 
We conclude -- somewhat surprisingly -- that  
(for generic non-intersecting cycles), the Wilson 
and 't Hooft operators actually \underline{commute}: 
\be\label{eq:CommutingW'tH}
\CW(\Sigma_{\ell-1})\CH(\Sigma'_{n-1-\ell})= 
\CH(\Sigma'_{n-1-\ell})\CW(\Sigma_{\ell-1}) ~.
\ee

\begin{remark}
$\,$

\begin{enumerate}

\item Of course, if we apply the $S$-duality isomorphism to the dual description 
of the Hilbert space $L^2(\widecheck{H}^{n-\ell}(N_{n-1}))$ using the Fourier transform 
\eqref{eq:Abelian-S-Duality-Iso}, then the Wilson operator becomes a shift operator and the 't Hooft operator becomes a multiplication operator. 

\item There has been some confusion in the literature about the relation of 
\eqref{eq:CommutingW'tH} and  \eqref{eq:NoncommutativeFluxOps}
 to the results of Gukov-Rangamani-Witten \cite{Gukov:1998kn}, especially since 
 these authors made use of fluxes, Heisenberg groups, and Lens spaces. Briefly, 
 \cite{Gukov:1998kn} concerns exchange relations for 't Hooft and Wilson defects. 
 These are different from the operators that appear in  \eqref{eq:NoncommutativeFluxOps}. 
 Indeed, equation \eqref{eq:CommutingW'tH} states that the 't Hooft and Wilson operators commute, 
 and do \underline{not} satisfy a nontrivial exchange relation. 
\tightfootnote{ 
Also,  \cite{Gukov:1998kn} claimed that there is a Heisenberg exchange relation for the Wilson and 
't Hooft operators for Maxwell theory on a Lens space, thus contradicting \eqref{eq:CommutingW'tH}. There is a subtlety in the argument in \cite{Gukov:1998kn}: It does not take into account that 
 $\widecheck \delta$ is not flat, so that restricting the domain of the wave function to flat fields does not commute with the action by an 't Hooft operator.}

\item Another interesting paradox associated with \eqref{eq:CommutingW'tH} has been raised by I. Garc\'ia-Etxebarria. In a renowned paper \cite{tHooft:1977nqb},
't Hooft introduced non-Abelian exchange relations for Wilson and 't Hooft operators associated to 1-cycles in four-dimensional non-Abelian gauge theory. If we consider a Yang-Mills-Higgs theory so that there is spontaneous symmetry breaking to a Coulomb branch, we might expect the IR limit of the  Wilson/'t Hooft operators 
to be those of the Wilson/'t Hooft operators in the Abelian low-energy effective theory (LEET). 
How are 't Hooft's nontrivial exchange relations consistent with \eqref{eq:CommutingW'tH}? 
The answer can be found from a careful reading of \cite[Sec. 6]{Kapustin:2014gua}.  When the global form of the gauge group is carefully taken into account, either the Wilson or the 't Hooft operator will not be a true line operator. Rather one of them will involve a spanning surface (which can be continuously deformed without changing results). So the naive expectation that the IR limit 
of \underline{both} the non-Abelian Wilson and 't Hooft operators are those of the low energy Abelian effective theory is not correct. (Nevertheless, in theories such as $d=4$, $\CN=2$ theories with a nontrivial Coulomb branch, one can recover the ``noncommuting flux relations'' such as 
\eqref{eq:NoncommutativeFluxOps}, just by using the LEET IR theory \cite{DelZotto:2022ras}.) 

\item Note that the operation of shifting by a flat field $\widecheck \phi_f$ multiplies 
the Wilson operator by a c-number phase: 
\be\label{eq:Ue-ActOn-Wilson}
\CU_e(\widecheck \phi_f)^{-1}  \CW(\Sigma_{\ell-1}) \CU_e(\widecheck \phi_f)
= e^{\imag  \langle \Sigma_{\ell-1} ,  \phi_f \rangle } \CW(\Sigma_{\ell-1}) ~.
\ee

\end{enumerate} 

\end{remark}

\SectionWithHeader{Relation To Generalized Symmetries And Their Anomalies}{Relation To Generalized Symmetries And Their Anomalies}{sec:RelationGeneralizedSymmetries}

The discussion of \autoref{sec:noncommutativity} and \autoref{sec:Wilson-tHooft} is very closely related to the popular subject of \emph{generalized symmetries}. 
In modern language, the group of symmetries obtained by shifting by a flat field
$\widecheck \phi_f^e \in \widecheck{H}^\ell(N_{n-1})$ in the representation $L^2(\widecheck{H}^\ell(N_{n-1}))$ is known as the group of ``$(\ell-1)$-form symmetries.'' 
\tightfootnote{\label{foot:higher-form-symmetries}The terminology  was introduced by Kapustin and Seiberg in 
\cite{Kapustin:2014gua}, and adopted and generalized in the renowned paper  \cite{Gaiotto:2014kfa}. (In an earlier paper \cite{Kapustin:2013uxa}, Kapustin and Thorngren used the related term ``2-group symmetry.'')
The terminology is not optimal because the important role played by torsion in many applications of ``$p$-form symmetries''  means one cannot really express the symmetry group in terms of differential forms. 
For example, consider a $\mathsf{U(1)}$ gauge field. The group of flat differential characters is 
$H^1(N_{n-1};\IR/\IZ)$. Shifts by the elements of the connected component of the identity can indeed be described as shifts of the gauge field $A \mapsto A+ \alpha$, where $d\alpha=0$. However, shifts of the differential character by the non-identity components of $H^1(N_{n-1};\IR/\IZ)$ cannot be described in this way. Nevertheless, the term ``$p$-form symmetry'' is a convenient shorthand and has been almost universally adopted. }
Similarly, the symmetry group given 
by the action of $\CU_m(\phi_f^m)$ for $\phi_f^m\in H^{n-\ell-1}(N_{n-1}; \IR/\IZ')$ is 
commonly referred to as the group of ``magnetic $(n-\ell-1)$-form symmetries.''
As we have seen in \autoref{sec:noncommutativity}, the symmetry groups generated by 
$\CU_e$ and $\CU_m$ do not commute and, in fact, generate, in general, a nontrivial extension 
\eqref{eq:EM-SymmetryGroup}. A common terminology used to describe this is that there is an ``'t Hooft anomaly between the electric $(\ell-1)$-form and magnetic $(n-\ell-1)$-form symmetries.'' 

In modern language, our definition of a definite electric flux sector is: 
  \begin{redbox}[box:defn-electric-flux-sector]{Definite Electric Flux Sector}
   The eigenstates of the ``$(\ell-1)$-form symmetry'' are the states of definite electric flux. They have definite value of $e \in H^{n-\ell}(N_{n-1}; \IZ)$.
  \end{redbox}

The work of Gaiotto, Kapustin, Seiberg, and Willett \cite{Gaiotto:2014kfa} generalizes the ideas of the previous two sections in several ways: 

\begin{enumerate} 

\item First, the Wilson and 't Hooft \underline{operators} make sense in a 
non-Hamiltonian framework as \underline{defects}. Thus, if we consider cycles  
$\Sigma_{\ell-1}$ and $\Sigma'_{n-\ell-1}$ in a spacetime $M_n$, rather than in a spatial slice $N_{n-1}$ (as we did in \autoref{sec:Wilson-tHooft}), then the defects can be ``inserted'' into the path integral over the spacetime, or, more generally, in the computation of correlation functions associated with the spacetime. This leads to the important idea that symmetry operations can be implemented by topological defects. 
Such topological defects link Wilson and 't Hooft defects so that, for example, 
the relation \eqref{eq:Ue-ActOn-Wilson} can be reinterpreted as the effect of 
linking the support $\Sigma_{\ell-1}$ by a topological defect operator on the 
linking $(n-\ell-1)$-sphere. Heuristically, this is to be thought of as the 
exponential of the integral of a conserved $(n-\ell-1)$-form current integrated 
over the linking sphere. 

\item An old idea going back to the work of  't Hooft and Polyakov is that non-Abelian 
gauge theories have a ``center symmetry.'' This can be viewed as a generalization of the translation by a flat gauge in the Abelian gauge theories   to non-Abelian gauge theories. 
The center symmetry is defined by a group homomorphism $\chi: \pi_1(M_n, x_0) \to Z(G)$, where $Z(G)$ is the center of the non-Abelian gauge group $G$. The proper definition of the action of the center symmetry on the isomorphism classes of gauge fields is a nice application of \autoref{thm:holonomy-theorem}.   
We can modify 
the holonomy function using $\chi$ as follows.
Assume the gauge bundle has been trivialized at $x_0$. If we characterize a connection $\n$ by 
its holonomy function $\mathsf{Hol}_{\n,x_0}: \Omega_{x_0}(M_n) \to G$, then the 1-form 
symmetry transformation $\chi$ takes $\n$ to the new connection $\n'$, such that, 
\be 
\mathsf{Hol}_{\n',x_0}(\gamma)= \mathsf{Hol}_{\n,x_0}(\gamma) \chi([\gamma]) ~,
\ee
where $\gamma\in \Omega_{x_0}(M_n) $, and $[\gamma]$ is its homotopy class. In the case of an Abelian gauge group $\mathsf{U(1)}^r$,  such transformations can be obtained by a shift of the gauge field 
$\n \mapsto \n + \alpha$, where  $\alpha \in \Omega^1(M_n; \mathfrak{t})$. However, in contrast   to the 
Abelian case, there is no way to define the non-Abelian center symmetry as a shift of the gauge field by a 1-form.  (This is another reason the term ``1-form symmetry'' is a misnomer. See footnote \ref{foot:higher-form-symmetries}.)

\item It has been known since the work of 't Hooft and Polyakov that the behavior of 
line defect expectation values under transformation by center symmetries carries  
important information about the dynamics of non-Abelian gauge theories. These 
older works were clarified in an important way in \cite{Kapustin:2014gua} and 
\cite{Gaiotto:2014kfa}.
An important contribution of \cite{Gaiotto:2014kfa} is the insight that these 
results can be reinterpreted, and generalized, using the idea that
higher-form symmetries can be spontaneously broken. For example, the photon of 
electromagnetism is interpreted as a Goldstone boson of a broken 1-form symmetry. 
Confinement is interpreted as a phase of unbroken 1-form symmetry. Partial confinement means the 1-form symmetry is broken to a subgroup. The Coulomb phase is a phase in which the 1-form symmetry is 
entirely broken (spontaneously). These ideas, and further developments, along with other 
aspects of applications to particle physics, are reviewed in \cite{Brennan:2023mmt}.

\item As we have repeatedly stressed in these notes, whenever a theory has a symmetry, one should introduce background fields that ``couple'' to the symmetry. This gives a fresh interpretation to the well-studied question of the behavior of the partition function in the presence of background electric and magnetic currents. \autoref{subsec:CouplingBackgroundCurrents} 
expands on this point. 

\item Closely related to the previous point is the remark, stressed in  \cite{Gaiotto:2014kfa}, 
that higher-form symmetries can have anomalies. This is closely related to the anomaly associated with simultaneous coupling to electric and magnetic current. Again, see \autoref{subsec:ElMgCurrent-GenSymAnom} below. The idea that higher-form symmetries have can have anomalies 
gave an elegant conceptual explanation to a result of B. Acharya and C. Vafa, derived from M-theory \cite{Acharya:2001dz}.  The domain wall with topological field theory found by these authors is needed to cancel anomalies in 1-form symmetries.

\end{enumerate}

\SectionWithHeader{Partition Functions For The Generalized Maxwell Field}{Partition Functions For The Generalized Maxwell Field}{sec:PartitionFunctionsGenMax}

\subsection{Coupling To Gravity}

We now consider the partition function of the generalized Maxwell field whose 
set of gauge inequivalent field configurations is $\widecheck{H}^\ell(M_n)$. 
We will at first take the background fields to be a Riemannian metric $g$ on $M_n$ 
and a coupling constant $\lambda>0$.  Formally, we may (naively) write the partition function as a path integral over gauge inequivalent field configurations: 
\begin{equation}\label{eq:partit-1}
Z_\ell(M_n; \lambda, g) = \int_{\widecheck{H}^\ell(M_n)}  \mu(\check{\sfA})  e^{-S} ~,
\end{equation}
where $S$ is the action \eqref{eq:actionh}.
\tightfootnote{\label{foot:period-convention-change}In this section, the period of the Maxwell fieldstrength is valued in $\IZ$ and not $2\pi\IZ$. To emphasize this, we will use the symbols $\sfA$ and $\sfF$ for the Maxwell connection and curvature respectively (and $\check \sfA$ for the corresponding differential character), as opposed to symbols $A$ and $F$ used elsewhere. We hope no confusion will arise by this local change of notation.}
To give this meaning, we need 
to define a measure $\mu(\check{\sfA})$  on $\widecheck{H}^\ell(M_n)$. 
%
%
As far as we know, there is no fully adequate description of the partition function of the generalized Maxwell field in the literature.
In particular, a completely satisfactory description of the ghost zeromodes 
that leads to a \underline{local} definition of the field theory is lacking. 
``Local'' means that the theory satisfies the kinds of gluing properties described in 
\hyperref[part1]{Part I}  of these lectures. We will make a proposal, but we will not demonstrate the desired 
gluing properties. A guiding principle will be electromagnetic duality: 
If the field theory is viewed projectively -- that is, is defined up to tensor product with a \underline{local} invertible field theory -- then the projective theory of the $\ell$-form fieldstrength with coupling $\lambda$ should be isomorphic to that of the $(n-\ell)$-form fieldstrength with coupling $1/\lambda$. Of course, in a proper treatment, electromagnetic duality should be a theorem, not 
an axiom. 
%
%

Recall the decomposition \eqref{eq:hcheckl-decomp} of Abelian groups. We will use a measure that respects this product structure, where we take the counting measure on the set of connected components, so,
\begin{equation}\label{eq:partit-sumcomp}
Z_\ell(M_n; \lambda, g) =\sum_{x\in H^\ell(M_n;\IZ)}  \int_{\widecheck{H}^\ell(M_n)_x}  \mu(\check{\sfA})  e^{-S} ~,
\end{equation}
where $\widecheck{H}^\ell(M_n)_{x}$ is the connected component labeled by $x$. 
The Riemannian metric on $M_n$ defines a natural metric on $\Omega^{\bullet}(M_n)$, and this metric descends to a translation-invariant metric on $\Omega^{\ell-1}(M_n)/\Omega^{\ell-1}_{\IZ}(M_n)$. 
That metric formally defines a measure on $\widecheck{H}^\ell(M_n)_{x}$, and we take this to be the measure $\mu(\check{A})$.  

In each connected component labeled by $x$, the classical solutions $\widecheck{H}^\ell(M_n)_x^{\mathsf{class}}$ form a torsor for the harmonic torus, 
\be 
\IT_{\ell-1} \cong \CH^{\ell-1}(M_n)/\CH^{\ell-1}_{\IZ}(M_n) ~,
\ee
and every element of $\widecheck{H}^\ell(M_n)_x^{\mathsf{class}}$ has a field strength which is a 
harmonic representative of the image of $x$ in de Rham cohomology. 

The harmonic torus inherits a translation-invariant measure from the Riemannian metric.
\tightfootnote{If we choose an integral basis $\{f_i^{(\ell-1)}\}$ for $\CH^{\ell-1}_{\IZ}$, then 
$\vol(\IT_{\ell-1}) = \sqrt{ \det_{i,j} \int_{M_n} f_i^{(\ell-1)} \wedge \star f_j^{(\ell-1)} } $. }
The standard procedure for defining Gaussian integrals in field theory is to 
integrate over the classical solutions and evaluate a determinant of nonzeromodes 
in the orthogonal directions to $\widecheck{H}^\ell(M_n)_x^{\mathsf{class}}$ inside 
$\widecheck{H}^\ell(M_n)_x$.  

Since  the measure on the topologically trivial modes is independent of $x$, we can 
factorize: 
\tightfootnote{Such a factorization typically only holds in free field theories.}
\begin{equation}\label{eq:Z-ell}
Z_\ell(M_n; \lambda, g)  = \CN_{\ell} Z_{\ell}^{\mathsf{flux\,sectors}} ~,
\end{equation}
where 
\be 
Z_{\ell}^{\mathsf{flux\,sectors}} = 
T_{\ell} \times \Theta_{\ell}(\lambda) ~.
\ee
Here  $T_\ell:=\vert \mathsf{Tors\,} H^{\ell}(M_n;\IZ) \vert $ is the order of the torsion
group, and the theta function is: 
\begin{equation}
   \Theta_{\ell}(\lambda) :=
 \sum_{f\in \CH^\ell_{\IZ}(M_n)} e^{-  \int_{M_n} \pi \lambda f\wedge \star f } ~.
\end{equation}
The sum is over the lattice of harmonic forms with integral periods. 
The order of the torsion group enters as an overall factor, since shifting the field by a topologically nontrivial flat field does not change the action.  Alternatively, we could have expanded the partition 
function as a sum over fluxes $\ov{x} \in \ov{H}^\ell_{\IZ}(M_n)$ modulo torsion. Then for fixed $\ov{x}$, the space of classical solutions is a torsor for the group of flat fields $H^{\ell-1}(M_n; \IR/\IZ)$.

The factor $\CN_\ell$  in equation \eqref{eq:Z-ell} is the Gaussian integral on fields in the topologically trivial sector.
%
%
It is slightly elaborate to evaluate because there are gauge invariances for gauge invariances (``ghosts for ghosts'')
\cite{Siegel:1980jj,Batalin1981,Batalin:1983ggl,Blau:1989bq,Gegenberg:1993gd,Hadfield:2020mnn,Blau:2022odi}.
\tightfootnote{This is closely related to the fact that for higher-form fields, the sheaf $\CF$ on $\mathsf{Man}_n$ must be valued in $\mathsf{sSET}$. }
We can write,
\be 
\CN_{\ell} = \CN_{\ell}^{\mathsf{nzm}} \CN_{\ell}^{\mathsf{zm}} ~,
\ee
corresponding to the contributions of the zeromodes and nonzeromodes
of the fields. Below, we find what some readers might regard as a surprise: 
\emph{The partition function is best viewed as an element of a real line.} We can express 
it as a function if we write it relative to a nonzero element of the same line (which should be considered to be an extra background field). We first describe the contribution of the nonzeromodes. The path integral over these modes will define  a functional of the metric $g_{\mu\nu}$ on $M_n$. 

%
%
The contribution of the determinants for nonzeromodes must take account of the fact that the 
gauge transformation $\sfA_{\ell-1} \mapsto \sfA_{\ell-1} + d \eta_{\ell-2}$ is redundant. 
A simple gauge-fixing condition such as $d^\dagger \sfA_{\ell-1}=0$ does not completely 
fix this redundancy in the sense that if we shift $\eta_{\ell-2}$ by an exact form, 
then the gauge transformation on $\sfA_{\ell-1}$ is unchanged. One is led to a BRST quantization procedure involving ghosts-for-ghosts or, better, to BV quantization. In the simplest incarnation, the ghost for the (bosonic) gauge field in $\Omega^{\ell-1}$ is a
fermionic (i.e., anticommuting)  ghost in $\Omega^{\ell-2}$, its commuting ghost-for-ghost is in $\Omega^{\ell-3}$, and so on, with alternating statistics.
Thus, for each $0 \leq k < \ell-1$, we have a ghost field which is a $k$-form with 
kinetic operator,
\be\label{eq:COs-def}
\CO_k := d^\dagger d\vert_{\Omega^k(M_n) \cap \mathsf{im\,}d^\dagger} ~.
\ee
The field of degree  $\ell - 1 - j$ is even (bosonic)   if $j$ is even and odd (fermionic) if $j$ is odd. The path integral over the nonzeromodes is then, 
\be
\log\,\CN_\ell^{\mathsf{nzm}}=  \frac{1}{2} \sum_{k=0}^{\ell-1} (-1)^{\ell  -k} \log \left(  {\det}'(\CO_{k} ) \right) ~. \label{eq:log-CN-ell-nzm}
\ee

The determinants of the operators $\CO_k$ can be defined in many ways. Two different 
ways that both define a local quantum field theory will differ by local counterterms -- that is -- the two theories thus defined will differ by  tensor product with an invertible field theory. One standard way to define the determinants for free field path integrals is through the use of $\zeta$-function regularization (an idea going back to D. Ray and I. Singer \cite{RaySinger:1,RaySinger:2}, and discussed in detail in \cite{Birrell1982}). 
\tightfootnote{It would be interesting to see if the considerations of \cite{Padilla:2024mkm}
can be usefully combined with $\zeta$-function regularization. 
}
Briefly, the spectrum of $\CO_k$ is discrete, countable, real, and nonnegative, and is denoted $\{ \mu_i^{(k)} \}_{i \in \IZ_+}$. Ray and Singer define a $\zeta$-function, 
\be 
\zeta_{\CO_k}(s) = \sum_i \left(\mu_i^{(k)}\right)^{-s}  ~,
\ee
and prove that on a compact manifold, $\zeta_{\CO_k}(s) $ is absolutely convergent for 
sufficiently large $\mathsf{Re}(s)$, and admits a meromorphic continuation to the complex $s$-plane 
that is analytic in the neighborhood of $s=0$. Then they define: 
\be
\log\,{\det}'(\CO_{k}   ):= - \left.\frac{d}{ds}\right\vert_0 \zeta_{\CO_k}(s) ~ . 
\ee
We will assume that the path integral, defined with with determinants defined by 
$\zeta$-function localization   is a local quantum field theory, although to our knowledge, that has not been explicitly demonstrated. 
\tightfootnote{Some positive results in two dimensions along these lines have been obtained in \cite{Wentworth:2010ic}.} 
Referring back to \autoref{sec:HodgeTheory} (see Exercise \ref{exercise:hodge-laplacians}), we have: 
\be\label{eq:PF-ellform-mus}
\log\,\CN_\ell^{\mathsf{nzm}}=  - \frac{1}{2} \left( \sum_i \log\, \mu^{(\ell-1)}_i  - 
\sum_i \log\,\mu^{(\ell-2)}_i + \cdots + (-1)^{\ell-1} \sum_i \log\,\mu^{(0)}_i \right) ~.
\ee
Rewriting 
\eqref{eq:PF-ellform-mus} in terms of determinants of Laplacians on 
spaces orthogonal to the spaces of harmonic forms, we get:  
\be\label{eq:QuadFluct-ellForm}
\log\,\CN_\ell^{\mathsf{nzm}}=  - \frac{1}{2}
\left[ \sum_{k=1}^\ell k (-1)^{k+1} \log \left(   {\det}'\bm{\Delta}_{\ell-k}  \right)   
   \right] ~. 
\ee

Thus far, we have ignored the coupling constant $\lambda$. As we remarked in \autoref{sec:ActionPrinciple},
if we do not impose a quantization condition on the periods of $F$, then $\lambda$ can be scaled away. This applies as well to the nonzeromodes of $\sfA_{\ell-1}$ since these modes are always noncompact. The path integral over these modes has no $\lambda$-dependence. The way this comes about from the path integral viewpoint is that the proper 
definition of the measure on phase space is $dp\,dq/(2\pi \hbar)$, which turns into 
$dq/\sqrt{2\pi \hbar}$ in the configuration space representation of the path integral. 
Then, a coefficient $\alpha_n$ for the eigenmode of eigenvalue $\mu_n$ enters the path
integral as,
\be 
\int_{\IR} \frac{d\alpha_n}{\sqrt{2\pi \hbar}}  e^{- \frac{\pi}{\hbar} \mu_n \alpha_n^2} ~,
\ee
where, in our case, we identify $\hbar$ with $1/\lambda$. Consequently, the integral over the nonzeromodes does not give rise to $\lambda$-dependence.

%
%
%
%
%
%
%
%
%
%

%
Finally, the most subtle part of the computation has to do with the treatment of 
the ghost (and ghost-for-ghost) zeromodes in the topologically trivial sector.   See \cite{Blau:1989bq,Moore:2002cp,Belov:2006jd,Kelnhofer:2007jf,Mnev:2014gta,Hadfield:2020mnn,
Blau:2022odi,Blau:2024obl} for some discussions  of ghost zeromodes. (Several of these papers actually discuss the  related case of $\mathsf{BF}$ theory rather than the theory addressed in this section, but the considerations are closely related.) The ghosts and ghosts-for-ghosts will not be valued in 
differential cohomology, since they should have have nonzero flux sectors, but the considerations 
of \autoref{subsubsec:DiffCoh-HigherGroupoids} strongly hint that they should be treated within the 
framework of differential cohomology. One of the key issues is whether ghost zeromodes should be valued in 
the torus:
\be\label{eq:GhostZeromodeDichotomy}
\IT_k = \CH^{k}(M_n)/\CH^k_{\IZ}(M_n) ~ ? \qquad\qquad \emph{or}  \qquad\qquad H^k(M_n;\IR/\IZ)  ~ ? 
\ee
We will refer to this question as the \emph{ghost zeromode dichotomy}, and refer to them as cases $A$ and $B$ respectively.

Whether or not the periods of the ghosts fields are quantized, the presence of fermionic 
zeromodes formally sets the path integral to zero. If the bosonic ghost zeromodes are not periodic, the integral over these modes will be infinite. 
\tightfootnote{Even  in ordinary Maxwell theory in four dimensions the fermionic ghost field $c$ is a scalar field, which has  a zeromode corresponding to global gauge transformations. In this case the zeromode is generally ignored and most authors simply work with ghost fields orthogonal to the zeromode. 
Such a procedure is indeed suggested by the original Faddeev-Popov derivation. However, it will not 
generalize, it will not fit well with electromagnetic duality in general dimensions, and it will violate the expected topological invariance when similar arguments are applied to the quantization of $\mathsf{BF}$ theory 
as in \autoref{sec:QuantBF-Theory}.} 
Therefore the path integral should not be regarded as a number but rather 
$\CN_\ell^{\mathsf{zm}}$ should  really be considered as an element of the Berezinian line on the superspace of zeromodes of ghost fields.  We recall that 
the ghosts of degree $(\ell-2), (\ell-4), \dots$ are anticommuting and the ghosts of 
degree $(\ell-3), (\ell-5),\dots$ are commuting. We also include the zeromodes 
of $A_{\ell-1}$, which are commuting. By a manipulation explained in \cite{Cattaneo:2015vsa,Cattaneo:2016lkk}, we can regard the measure as an element of the   \underline{inverse} of the determinant of a complex: 
\tightfootnote{For a bounded complex of vector bundles $E^{\bullet}: \cdots \to E^{n} \xrightarrow{d^{n}} E^{n+1} \to \cdots$ over a topological space, the determinant line bundle of the complex is $\mathsf{DET}(E^{\bullet}) := \bigotimes_{k} (\mathsf{DET} E^{k})^{(-1)^{k}}$. }
\be\label{eq:COHO-DET-LINE}
\mathsf{DET}\bigg( \bigoplus_{k=0}^{\ell-1} H^k(M_n)\bigg)
\cong  \mathsf{DET}\big(H^{\ell-1}(M_n)\big)\otimes \mathsf{DET}\big(H^{\ell-2}(M_n)\big)^{-1} \otimes \cdots ~.
\ee
We can understand why we take the inverse of \eqref{eq:COHO-DET-LINE} as follows: 
A measure on $\IT_{\ell-1}$ is a nonzero element of 
$\Lambda^{\mathsf{max}} T^*\IT_{\ell-1}  \cong  \mathsf{DET}\big(H^{\ell-1}(M_n)\big)^{-1}$. 
The rest of the signs are determined since we have an alternating product. 

Now we need to describe the specific element $\CN_\ell^{\mathsf{zm}}$ of the inverse line to \eqref{eq:COHO-DET-LINE}. We propose (modulo a subtlety explained in the next paragraph) that it is: 
\be\label{eq:N-zm-def}
\CN_{\ell}^{\mathsf{zm}}  =\xi_{\ell} \left( \omega_1^{(\ell-1)}\wedge \cdots \wedge
\omega^{(\ell-1)}_{b_{\ell-1}}\right)^{-1}  \otimes  
\left( \omega_1^{(\ell-2)}\wedge \cdots \wedge \omega^{(\ell-2)}_{b_{\ell-2}}\right)^{+1} \otimes \cdots ~,
\ee
where, for fixed $k$, $\{ \omega_i^{(k)} \}_{i=1,\dots, b_k}$ is an 
ON basis of harmonic forms on $\CH^k(M_n)$. Here $\xi_{\ell}$ is a positive rational number 
determined by how we handle the ghost zeromode dichotomy \eqref{eq:GhostZeromodeDichotomy}. 
If they are valued in $\CH^{k}(M_n)/\CH^k_{\IZ}(M_n)$, then $\xi_\ell = 1$, but if they are 
in $H^\bullet(M_n;\IR/\IZ)$, then, 
\be \label{eq:def-xi-ell}
\xi_{\ell} = \mathsf{Alt}_{<\ell}(T):=  \prod_{k=0}^{\ell-1-k} T_k^{(-1)^{\ell-k}} ~ . 
\ee

The subtlety alluded to above is the overall sign. There is no canonical orientation of the 
cohomology groups in question and without such a choice of orientation,  \eqref{eq:N-zm-def} 
is only defined up to sign. Specifying a nonzero vector in a real line $L$ only defined up 
to sign, that is, choosing $\pm v \in L$, is equivalent to specifying a norm on $L$. Indeed, for any other vector $w\in L$, we define 
the norm $p(w):= {\rm sign}(\frac{w}{v})  \frac{w}{v} \in \IR_+$. Conversely, given a 
norm on $L$, the unit vector for this norm is only determined up to sign. 
Thus, more properly, the partition function of the Maxwell field is a norm on the line 
inverse to \eqref{eq:COHO-DET-LINE}.

A partial justification of \eqref{eq:N-zm-def} may be given as follows. 
Note that  the  integral over the topologically trivial harmonic zeromodes of $\sfA_{\ell-1}$ 
is $\vol(\IT_{\ell-1})$. A good way to write this, which will then extend to the 
ghosts-for-ghosts, is as follows. Let $\{ f_i^{(k)} \}_{i=1,\dots, b_k}$ be an 
integral basis of harmonic forms  with  $\{ f_i^{(k)} \}_{i=1,\dots, b_k}$ oriented 
compatibly with the ON basis $\{ \omega_i^{(k)} \}_{i=1,\dots, b_k}$. We have, 
\be\label{eq:Volum-As-Ratio}
\vol(\IT_k) = \frac{ f_1^{(k)} \wedge \dots  \wedge f_{b_k}^{(k)}}{ \omega_1^{(k)} \wedge \dots  \wedge \omega_{b_k}^{(k)}} ~.
\ee

Now, using the metric-independent integral basis $\{ f_i^{(k)} \}_{i=1,\dots, b_k}$ 
for the harmonic forms, we refer the partition  function to the metric-independent element,
\be\label{eq:Fixed-f-volume}
\mu_{\IZ}^{(\ell)}:= \prod_{k=0}^{\ell-1}  
\left(f_1^{(k)} \wedge \dots  \wedge f_{b_k}^{(k)} \right)^{(-1)^{\ell-1-k} } ~.
\ee
 The element \eqref{eq:Fixed-f-volume} is independent of the integral basis up to sign. The sign can be fixed by a cohomology orientation (or one can interpret the results in terms of norms). 
We can therefore write the   zeromode contribution as: 
\be\label{eq:ell-form-zm-linebundle}
\begin{split}
\CN_{\ell}^{\mathsf{zm}}   & =\xi_{\ell}  \prod_{k=0}^{\ell-1}  \left( 
\frac{ f_1^{(k)} \wedge \dots  \wedge f_{b_k}^{(k)}}{ \omega_1^{(k)} \wedge \dots  \wedge \omega_{b_k}^{(k)}}
\right)^{(-1)^{\ell-1-k} }  \left( \mu_{\IZ}^{(\ell)}\right)^{-1} \\
& 
=
\left\{\begin{array}{ll}
  \prod\limits_{k=0}^{\ell-1} \left(\vol(\IT_k)\right)^{(-1)^{\ell-1-k} } 
\left( \mu_{\IZ}^{(\ell)}\right)^{-1} ~,  & {\rm Case ~ A} ~,\\
  \prod\limits_{k=0}^{\ell-1} \left(\vol(H^k(M_n; \IR/\IZ))\right)^{(-1)^{\ell-1-k} } 
\left( \mu_{\IZ}^{(\ell)}\right)^{-1} ~,  & {\rm Case ~ B} ~,
\end{array}
\right.
\end{split}
\ee
where we have used $\vol(H^k(M_n;\IR/\IZ)) = T_{k+1} \vol(\IT_k)$. 

\begin{remark}
$\,$
\begin{enumerate}

\item In our discussion above, and in the discussion of continuum $\mathsf{BF}$ theories below, we have taken some important shortcuts. A proper discussion would involve BV quantization \cite{Batalin1981,Batalin:1983ggl,Schwarz:1992nx,LOSEV2007,Cattaneo:2016lkk} and \cite[Ch. 5]{Costello2011}.
In particular, the Berezinian measure on the ghost zero-modes should be expressed in terms of the ``BV pushforward''. See \cite{LOSEV2007,Bonelli:2010cu}  and  \cite[Sec. 2.2.2]{Cattaneo:2015vsa} for descriptions of the BV pushforward. 
One can then convert a half-density on the finite BV space given by a shifted cotangent bundle into a density on the base (see \cite{Cattaneo:2016lkk}, example 2.5 and section 3), thus making contact with our formulation in terms of a 
determinant line. While the BV approach is the proper approach, it is beyond the scope of these lectures. 

\item Some discussions of $p$-form electrodynmics, notably 
\cite{Siegel:1980jj,Birmingham:1991ty} use a   \underline{triangular} system of ghosts and antighosts. However, the 
determinants and zero-mode measures for the ghosts that are not on the ``right-hand side'' of the BV triangle cancel, and can be ignored in this problem. This follows from the general theory of ``trivial BV pairs.'' See, for example, \cite{Blau:1989bq}.

\item We must now include the  $\lambda$-dependence in $\CN_{\ell}^{\mathsf{zm}}$. 
While $\lambda$ does not affect the integral over nonzeromodes, it does affect the 
integral over the zeromodes \underline{provided} the periods are quantized. 
Since we model the gauge equivalence class of our field to be 
$\widecheck{H}^\ell(M_n)$, the periods of $F$ are quantized. Likewise, the periods of the 
flat fields are quantized, so in particular, the periods of the zeromodes of $\sfA_{\ell-1}$ are quantized. The question then arises whether the ghost fields likewise have quantized periods. If we assume that they are, following the hint of 
\autoref{subsubsec:DiffCoh-HigherGroupoids}, the metric on the space of zeromodes is scaled by a factor of $\lambda$ and hence the zeromode dependence of $\CN_{\ell}^{\mathsf{zm}}$ is obtained by the rescaling: 
\be 
\CN_{\ell}^{\mathsf{zm}} \mapsto \lambda^{\half (-1)^{\ell-1} \chi_{< \ell}   } \CN_{\ell}^{\mathsf{zm}}  ~,
\ee
where $b_k$ are the Betti numbers $b_k:= \dim_{\IR} H^k(M_n;\IR)$, and 
\be
\chi_{< \ell} : =  b_0 -b_1 + \cdots + (-1)^{\ell-1} b_{\ell-1}  = \sum_{k=0}^{\ell-1}(-1)^{k} b_k~. 
\ee

\item Our main justification for equation \eqref{eq:ell-form-zm-linebundle} is that it is the only formula which is consistent with the Berezinian measure on the space of zeromodes and with electromagnetic duality (as demonstrated in the next section). The alternating product of volumes of tori of harmonic forms is justified in a different way in 
\cite{Kelnhofer:2007jf}. Electromagnetic duality also determines the subtle factor $\xi_{\ell}$. The factor $T_{\ell}$ in \eqref{eq:Z-ell} will not lead to an electromagnetically dual partition function unless we interpret the volumes of the 
ghost zeromodes to be volumes of $H^k(M_n; \mathsf{U(1)})$ rather than the volumes of $\IT_k$. 
That is, we take case B in the zero mode dichotomy. With this understood our  final answer for the partition function is:  
\be\label{eq:FinalMaxwell-pform-PF}
Z_{\ell}(M_n;\lambda, g) = \lambda^{\half (-1)^{\ell-1}\chi_{< \ell}   } \prod_{k=0}^{\ell-1} \left[\vol\,H^{k}(M_n; \mathsf{U(1)})\right]^{(-1)^{\ell-1-k} } 
 \cdot \CN_{\ell}^{\mathsf{nzm}} \cdot \Theta_{\ell}(\lambda) \big( \mu_{\IZ}^{(\ell)}\big)^{-1} ~.
\ee

\item Finally, we note that the line \eqref{eq:COHO-DET-LINE} has a rank-one $\IZ$-submodule provided by the integral cohomology. It is natural to define a second norm on the line by 
defining the norm of a generator to be one. The ratio of two norms on the real line is a positive real number. A generator can be taken to be $\big( \mu_{\IZ}^{(\ell)}\big)^{-1}$. The net effect is 
to define the partition function as a real number by dropping the last factor on the RHS of 
\eqref{eq:FinalMaxwell-pform-PF}.

\end{enumerate}

 \end{remark}

\subsection{Checking Electromagnetic Duality}\label{subsec:Check-EM-Duality}

It is interesting to check electromagnetic duality of the partition function 
$Z_{\ell}(M_n;\lambda,g)$. By Poisson resummation, 
one can prove
\tightfootnote{The only nontrivial thing to check here is that if one defines integral bases of harmonic forms $f^{(\ell)}_i$, and $f^{(n-\ell)}_i$, $i=1,\dots, b_\ell$, then 
the matrix $\int f^{(\ell)}_i \star f^{(\ell)}_j$ is conjugate to the inverse of 
$\int f^{(n-\ell)}_i \star f^{(n-\ell)}_j$. This follows by noting two facts.
First, if we define 
$\star f^{(\ell)}_i = S^{(\ell)}_{ji} f^{(n-\ell)}_j$ and 
$\star f^{(n-\ell)}_i = S^{(n-\ell)}_{ji} f^{(\ell)}_j$ then 
$S^{(n-\ell)} S^{(\ell)}= (-1)^{\ell (n-\ell)}$. Second, $A_{ij} = \int f^{(\ell)}_i \wedge f^{(n-\ell)}_j$ is an integral unimodular matrix, by Poincar\'e duality. The quadratic form  for $\Theta_\ell$ is proportional to $A S^{(\ell)}$ and that for $\Theta_{n-\ell}$ is proportional to $(-1)^{\ell(n-\ell)} (A^{tr} S^{(n-\ell)})$. But the inverse of the 
latter is $A^{-1} (A S^{(\ell)}) A^{tr,-1}$. }
\be 
\Theta_{\ell}(\lambda) = \left( \frac{1}{\lambda}\right)^{b_\ell/2}  (\vol\,\IT_{\ell})^{-1} \Theta_{n-\ell}\bigg(\frac{1}{\lambda}\bigg) ~. \label{eq:PR}
\ee

We next compare the nonzeromode contributions $\CN_{\ell}^{\mathsf{nzm}}$ and $\CN_{n-\ell}^{\mathsf{nzm}}$. 
Referring back to the expression \eqref{eq:PF-ellform-mus}, we must compare the spectra of the Laplacians. Note that both  $\CN_{\ell}^{\mathsf{nzm}}$ 
and $\CN_{n-\ell}^{\mathsf{nzm}}$   involve Laplacians with degrees descending to zero. 
These factors in general will \underline{not} be equal. In order to relate them, we 
must use Hodge duality -- as expected when checking electromagnetic duality. 
The Hodge dual may be refined to an isometry, 
\be 
\star : V_k \to \wt{V}_{n-k} ~,
\ee
from which we conclude an equality of spectra: $\{ \mu^{(k)}_i\} = \{ \mu^{(n-k-1)}_i\}$ (see Exercises \ref{exercise:hodge-laplacians} and \ref{exercise:spectral-isometry}). Thus, the Hodge dual of a spectrum of 
degrees descending to zero is a spectrum of degrees ascending to $n$. 
In order to relate $\CN_{\ell}^{\mathsf{nzm}}$ 
and $\CN_{n-\ell}^{\mathsf{nzm}}$, we must introduce 
the \emph{analytic torsion} of Ray and Singer \cite{RaySinger:1,RaySinger:2,RaySinger:3,Cheeger1977,Lott2024}.
We now briefly explain the definition of Ray-Singer torsion.

An important component of analytic torsion is: 
\be\label{eq:AnalyticTorsion}
\begin{split}
\log\,\tau'_{\mathsf{RS}}(M_n) & :=  \frac{1}{2}\int_0^\infty \frac{dt}{t} \Tr'_{\Omega^{\bullet}(M_n)} \mathtt{F} (-1)^{\mathtt{F}} e^{- t \bm{\Delta} } 
=- \frac{1}{2} \sum_{k=0}^n (-1)^k k\, \log\,{\det}' \bm{\Delta}_k \\
& = \frac{1}{2}  \sum_{k=0}^{n-1} (-1)^k \bigg( \sum_i \log\,\mu^{(k)}_i \bigg) ~,
\end{split}
\ee
where the prime indicates the space orthogonal to the space of harmonic forms, and $\mathtt{F}$ is the degree of the form. (It is called ``$\mathtt{F}$'' because in the supersymmetric quantum mechanics interpretation, the degree is the ``fermion number.'')   
Because of the identity $\{ \mu^{(k)}_i\} = \{ \mu^{(n-k-1)}_i\}$,  
$\tau'_{\mathsf{RS}}(M_n) = 1$ when $n$ is even, but it is typically not $1$ when $n$ is odd. In any case,  we have,  
\be 
\log\,\CN_{\ell}^{\mathsf{nzm}} = \log\,\CN_{n-\ell}^{\mathsf{nzm}} + (-1)^{\ell} \log\,\tau'_{\mathsf{RS}}(M_n) ~.
\ee

The quantity  $\tau'_{\mathsf{RS}}(M_n)$ is \underline{not} yet the Ray-Singer torsion. 
It is not the most natural quantity to consider because it does not properly take into account the metric dependence of the space of zeromodes. Ray and Singer observed that 
we use the metric to define a canonical element, 
\be
\mu_g\in  \mathsf{DET}\big(\bigoplus_{k=0}^n H^k(M_n)\big)  ~,
\ee
up to a sign: We simply take the wedge product of an ON basis of harmonic forms. 
The sign can be fixed by choosing an orientation on the space of harmonic forms. 
Here, the degree of the complex is given by the degree of the form. 
Then, according to Ray and Singer \cite{RaySinger:3}, the torsion should be defined as: 
\be\label{eq:RS-Torsion-LineValued}
\tau_{\mathsf{RS}} :=  \tau'_{\mathsf{RS}}(M_n) \mu_g \in \mathsf{DET}\big(\bigoplus_{k=0}^n H^k(M_n)\big)   ~,
\ee
up to a sign. The quantity 
$\tau_{\mathsf{RS}}$ is known as the Ray-Singer torsion. The main assertion of Ray and 
Singer is that \eqref{eq:RS-Torsion-LineValued} is in fact metric-independent, see \cite{RaySinger:3,Mnev:2014gta}. 

Recall that a nonzero vector in a line up to sign is equivalent to a norm on that line
as explained under equation \eqref{eq:def-xi-ell}, so one can interpret RS torsion without the choice of cohomology orientation as a norm on the determinant line.  The observation that Ray-Singer torsion should be viewed conceptually as a 
norm on the line was made in \cite{Freed1992}.

Putting all the above together, we find that: 
\be\label{eq:ell-form-pathintegral-duality}
Z_\ell(M_n; \lambda, g) = \CI_{\ell} Z_{n-\ell}(M_n; 1/\lambda, g) ~,
\ee
with 
\be 
\CI_{\ell} = \frac{T_\ell}{T_{n-\ell}} \cdot 
\left( 
\frac{\CN_\ell^{\mathsf{zm}}}{\CN_{n-\ell}^{\mathsf{zm}}}
  (\vol\,\IT_{\ell})^{-1} e^{(-1)^{\ell}\log\tau'_{\mathsf{RS}}(M_n)} \right) \cdot 
\lambda^{(-1)^{\ell-1} \chi   } ~,
\ee
%
%
%
where $\chi$ is the Euler characteristic of $M_{n}$, i.e., 
$\chi = \sum_{k=0}^{n}(-1)^{k} b_k $. 

Now, choose a fixed oriented integral basis for $\CH_{\IZ}^\ell(M_{n})$, namely, $f_1, \dots, f_{b_\ell}$, 
and let $\omega_1, \dots, \omega_{b_{\ell}}$ be an oriented ON basis determined 
by a metric. Then, we use \eqref{eq:Volum-As-Ratio} again with $k=\ell$, and conclude that, 
\be 
\frac{\CN_\ell^{\mathsf{zm}}}{\CN_{n-\ell}^{\mathsf{zm}}}
  \big(\vol\,\IT_{\ell}\big)^{-1} = \frac{\xi_{\ell}}{\xi_{n-\ell}}
  \left( \frac{\mu_g}{\mu_{\IZ}} \right)^{(-1)^{\ell}} \cdot
  \frac{(\mu_{\IZ})^{(-1)^{\ell} } }{f_1^{(\ell)} \wedge \dots \wedge f_{b_{\ell}}^{(\ell)} }
   ~,
\ee
where, 
\be 
\mu_{\IZ} = \bigotimes_{k=0}^n \left( f_1^{(k)}  \wedge \cdots \wedge f^{(k)}_{b_k}\right)^{(-1)^{k}} ~,
\ee
is metric-independent. (We have here used Poincar\'e duality to relate the integral basis 
$f^{(n-\ell-k)}_i$ to the integral basis $f^{(\ell+k)}_i$.) 
Finally, we consider the factor, 
\be 
\frac{T_\ell}{T_{n-\ell}} \frac{\xi_{\ell}}{\xi_{n-\ell}} ~.
\ee
Recall that the torsion pairing gives a perfect pairing, so $T_k = T_{n-k+1}$. 
Using this, we identify, 
\be 
\frac{T_\ell}{T_{n-\ell}} \frac{\xi_{\ell}}{\xi_{n-\ell}} 
= \left( \prod_{k=0}^n T_k^{(-1)^k} \right)^{(-1)^\ell} := \left( \mathsf{Alt}(T) \right)^{(-1)^\ell} ~.
\ee

Thus, the factor relating the partition functions for the dual theories is: 
\be 
\CI_{\ell} =  \left( \mathsf{Alt}(T)\cdot 
\frac{\tau_{\mathsf{RS}}(M_n)}{\mu_{\IZ}}
\right)^{(-1)^\ell}   \cdot 
\lambda^{(-1)^{\ell-1} \chi }  \cdot
  \frac{(\mu_{\IZ})^{(-1)^{\ell} } }{f_1^{(\ell)} \wedge \dots \wedge f_{b_{\ell}}^{(\ell)} } 
 ~.
\ee
Again, the last factor is only defined up to sign, and should be viewed as a norm on the relevant real line.

\begin{remark}\label{rem:AltT}  The Cheeger-M\"uller theorem equates $\tau_{\mathsf{RS}}(M_n)$ with 
Reidemeister torsion. Using the integral structures on the determinants of complexes related to cohomology, one can define two norms on the determinant line and interpret the 
torsion as the ratio of norms -- hence identifying it as a real number
$\wt{\tau}_{\mathsf{RS}}(M_n)$. Using this interpretation, Dan Freed has shown that   $\wt{\tau}_{\mathsf{RS}}(M_n) \mathsf{Alt}(T) =1$, a very satisfying answer. See \autoref{App:TorsionAndTorsion}, written by Dan Freed,  for the detailed proof.

%

 \end{remark}

\begin{exbox}[exercise:spectral-isometry]{Isometry from Hodge star}Show that Hodge star maps 
\be 
\star : V_\ell \to \wt{V}_{n-\ell} \cong V_{n-\ell-1} ~,
\ee
isomorphically, and that, 
\be 
\star d^\dagger d \star^{-1 } = d d^\dagger ~,
\ee
and conclude that we can identify $\mu^{(k)}_i$ with 
$\mu^{(n-k-1)}_i$. 
\end{exbox}

\subsubsection{A Path Integral Derivation Of Electromagnetic Duality}

There is a simple (formal) path-integral derivation of the electromagnetic duality of the partition functions going back to \cite{Rocek:1991ps,Witten:1995gf}. Here is a version 
that uses differential cohomology. 

We begin by considering a path integral over 
fields in $\widecheck{H}^\ell \times \widecheck{H}^{n-\ell}$ and in addition, integration of an auxiliary $\ell$-form $\sfG \in \Omega^\ell$: 
\def\c{\check}
\begin{equation}
\CZ ={1 \over \vol\,\widecheck{H}^\ell} \int_{\widecheck{H}^\ell} \mu(\c \sfA) \int_{\widecheck{H}^{n-\ell}} \mu(\c \sfA_D) \int_{\Omega^\ell} \mu(\sfG)
e^{- \pi \int \lambda \CF\star\CF + 2\pi \imag \int \sfG \sfF(\c \sfA_D) } ~,
\end{equation}
where 
\be
\CF := \sfF(\c \sfA) - \sfG ~ . 
\ee
Note that $\check \sfA$ and $\check \sfA_D$ only 
enter the action through their fieldstrengths. Moreover, there is a ``gauge invariance,'' where we shift $\c \sfA$ by an arbitrary element of $\widecheck{H}^{\ell}$, and then shift $\sfG$ by the fieldstrength of that element, so we have divided by the volume of this ``gauge group''
as an overall factor in front of the integral. 

The basic idea of \cite{Rocek:1991ps,Witten:1995gf} is to evaluate $\CZ$ in two different ways and equate the result of the two computations. That equation becomes a statement of electromagnetic duality of partition functions for fields in $\widecheck{H}^\ell$ and $\widecheck{H}^{n-\ell}$, respectively. 

The first way to do the integral is by starting with the Gaussian 
integral over $\sfG \in \Omega^{\ell}$. We can shift away $\sfF(\check \sfA)$, so the $\sfG$-integral becomes: 
\be 
\int_{\Omega^\ell} \mu(\sfG) e^{- \int \pi \lambda \sfG\star \sfG + 2\pi \imag \int \sfG \sfF(\check \sfA_D) + 2\pi \imag \int \sfF(\check \sfA) \sfF(\check \sfA_D) } ~.
\ee
Note that $e^{2\pi \imag \int \sfF(\check \sfA) \sfF(\check \sfA_D)} = 1$, because the periods of the fieldstrengths in differential cohomology are integers. 
The remaining Gaussian integral is straightforward: 
The saddle point is $\sfG_0 = \frac{\imag }{\lambda} (-1)^{\ell (n-\ell)} \star \sfF(\check \sfA_D)$,  so we get: 
\be 
\exp\left[ - \frac{\pi}{\lambda} \int \sfF(\check \sfA_D) \star \sfF(\check \sfA_D) \right] ~,
\ee
where the measure $\mu(\sfG)$ is normalized so that:
\be\label{eq:NormalizedOmegaellMeasure}
\int_{\Omega^\ell} \mu(\sfG) e^{- \int \pi \lambda \sfG\star \sfG   } =  1 ~. 
\ee
(Note that the Gaussian integral formally depends on $\lambda$ as 
$\lambda^{- \dim\, \Omega^\ell/2}$, an expression which clearly must be regularized and defined. In any case, a measure normalized so that 
\eqref{eq:NormalizedOmegaellMeasure} holds  is $\lambda$-dependent.) 
Next, one integrates over $\check \sfA$ and the remaining integral is 
just:  
\begin{equation}
 \CZ  =   Z_{n-\ell}(M;\lambda^{-1}, g)  ~.
 \end{equation}

On the other hand, if we first integrate over $\c \sfA_D$ then we get a ``periodic delta function''
of $\sfG$, which constrains $\sfG$ to have support on $\Omega^\ell_{\IZ}(M_{n})$:
\begin{equation}
 \int_{\widecheck{H}^{n-\ell}} \mu(\c \sfA_D)\,e^{2\pi \imag \int \sfG \sfF(\c \sfA_D)} = \vol\big(H^{n-\ell-1}(M; \mathsf{U(1)})\big)
 \int_{\Omega^{\ell}_{\IZ}}  \mu(\sfG_0) \delta(\sfG-\sfG_0) ~.
\end{equation}
Next, we do the $\sfG$ integral to replace $\sfG$ by $\sfG_0$ in $\CF$.
However, now we can use the translation invariance of the measure $\mu(\c \sfA)$ to shift away
any $\sfG_0 \in \Omega^\ell_{\IZ}(M_{n})$. Therefore, after shifting away $\sfG_0$ we are left with:
\begin{equation}
\CZ = \frac{\CT_{n-\ell}\,\vol(\IT_{n-\ell} )}{\CT_\ell \,\vol(\IT_\ell)}
\int_{\widecheck{H}^\ell}\mu(\c \sfA)\,e^{- \pi \lambda \int \sfF(\c \sfA)\star \sfF(\c \sfA) } ~.
\end{equation}
So, once again, up to an invertible field theory, associated with 
the volumes of zeromodes, we verify electromagnetic duality. 
\tightfootnote{A very careful accounting of the volumes of zeromodes, and the normalization of the delta-functions and Gaussian integrals in the case of two dimensions can be found in \cite[Sec. 12.9.4, p. 195]{Moore:2015PHY695}.
In the case of four dimensions, it is done carefully in \cite{Witten:1995gf}. It would be interesting to recover the Ray-Singer torsion in odd dimensions using this style of argument. To our knowledge, that has not been done. }

\subsection{Including A Coupling To Background Electric And Magnetic Currents}\label{subsec:CouplingBackgroundCurrents}

Thus far, we have considered the partition function as a function of the background fields $\lambda$, the coupling constant, and the metric on $M_n$. It is natural to introduce electric and magnetic currents into the classical equations of motion as in \eqref{eq:TrivializeCurrents}
and hence these should also be considered as background fields. It turns out that doing so raises many issues, and we will only consider some of them briefly here.

\subsubsection{Coupling To Electric Currents}

To begin, let us set the magnetic current $\sfJ_m$ to zero. 
Then $\sfF=d\sfA$ locally, and (assuming momentarily that $\sfF$ is globally exact) we can derive the equation of motion from the classical action 
principle (in Lorentzian signature): 
\begin{equation}\label{eq:action-with-je}
 \int \pi \lambda \sfF\star \sfF + 2\pi  \int \sfJ_e \sfA  ~.
\end{equation}
(We have slightly rescaled $\sfJ_e$ from that in equation \eqref{eq:TrivializeCurrents} for convenience in the following discussion.)

This immediately raises some issues:

\begin{enumerate}

\item The coupling to $\sfJ_e$ is not gauge-invariant unless   $d\sfJ_e=0$, 
that is, unless we have a conserved current. Of course, the existence of a solution to the equations of motion implies $\sfJ_e=0$, so we will henceforth assume $\sfJ_e$ is indeed closed. 

\item However, if we shift $\sfA\mapsto \sfA+ \omega$ where $\omega$ is
closed with (nonzero) integral periods, the action will not be 
gauge-invariant. In the (Lorentz signature) quantum theory, the measure in the path integral would have a factor: 
\begin{equation}\label{eq:action-with-je-2}
\exp\left[ \imag  \pi \int  \lambda \sfF\star \sfF + 2\pi \imag   \int \sfJ_e \sfA  \right] ~.
\end{equation}
Still, this will not be gauge invariant unless $\sfJ_e$ has integral periods. So we will assume that $\sfJ_e$ has integral periods. 

\item If the cohomology class of $\sfF$ is nontrivial, then $\sfA$ is not globally defined, and the formula for coupling to the current is ambiguous.

\end{enumerate}

In these notes, we are working under the assumption that the quantum theory should be formulated so that the gauge equivalence class of the generalized Maxwell field is a class in $\widecheck{H}^\ell(M_n)$.  
In view of this, the three issues above are neatly put to rest if we 
 view $\sfJ_e$ as the \emph{fieldstrength} of
a differential cohomology class: 
\begin{equation}
[\c \kappa_e] \in \widecheck{H}^{n-\ell+1}(M_n) ~.
\end{equation}
That is, $\sfF([\c \kappa_e]) = \sfJ_e$. 
Once we make this assumption, the coupling to the background electric current is neatly written in terms of the pairing in differential cohomology, so that the path integral measure (in Euclidean signature) 
has a factor: 
\be\label{eq:DifflCohoElectricCoupling}
\exp\left[ - \int \pi \lambda \sfF\star \sfF 
+ 2 \pi \imag \langle [\c \sfA], [\c \kappa_e]\rangle \right] ~.
\ee
Just as the conceptual leap from the standard formulation of (generalized) Maxwell theory to differential cohomology extended 
the domain of validity of the action, so too, this generalization of the standard electric current allows one to generalize to topologically more interesting situations.

\begin{remark}
$\,$
\begin{enumerate}

\item  
In the case when the current is sourced by an electrically charged brane
$\CW_{e}$ of \eqref{eq:elec-brane-Je}, the electric couplings are related by:

\begin{equation}
\exp \left( 2\pi \imag  \int_M \sfJ_e \sfA \right) = \exp\left( 2\pi \imag  \int_{\CW_{e}} q_e \sfA \right) ~.
\end{equation}

\item As an example showing that equation 
\eqref{eq:DifflCohoElectricCoupling} is a nontrivial extension of the
standard electric coupling \eqref{eq:action-with-je}, consider $\varphi \in \widecheck{H}^1(M_{n})$. We take $n = \dim\,M_{n}=2$, so that  the electric current defines a class $[\c \kappa_e] \in \widecheck{H}^2(M_{2})$, and can be considered
to be the isomorphism class of a line bundle with connection $\n$, where $\sfJ_e$ is the curvature $F(\n)$ of the connection $\n$. For topologically trivial field configurations, $\varphi = e^{2\pi \imag  \phi}$ with a well-defined logarithm $\phi: M_{2} \to \IR$ we can write the electric coupling as:
\begin{equation}
\exp\left[ 2\pi \imag  \int_{M_2} \phi\,\sfF(\n)\right] ~,
\end{equation}
and we therefore recognize a form of ``background charge'' familiar in conformal field theory. Note that integration over the translation by a flat field $\phi_0 \in \IR/\IZ$ shows that
the path integral is zero unless $\int_{M_2} \sfF(\n)=0$. This is the quantum-mechanical implementation of the classical statement that $\sfJ_e$ must be trivialized by the on-shell fields.

\item Continuing with our two-dimensional example, formulating the electric coupling in terms of differential cohomology has consequences not visible in the usual naive formulation of the coupling. For example, if the electric current is such that   $\n$ is a flat connection, then the pairing only depends on the characteristic class of $\varphi$ which gives $c(\check \varphi)  \in H^1(M_{2};\IZ)$, measuring the winding number.
Choosing a basis of cycles on $M_{2}$, the electric coupling then becomes $\prod_i h_i^{w_i} $, where $h_i$ are the holonomies of $\n$ around the basis cycles and $w_i$ are the winding numbers. Such factors are not present in the standard formulation of coupling to electric current. 

\item In the path integral,
\begin{equation}\label{eq:PartFunWithElecCurr}
Z_{\ell}(M;\lambda, g, \c k_e)=
\int_{\widecheck{H}^{\ell}} \mu(\c \sfA) e^{- \pi \int \lambda \sfF\star \sfF + 2\pi \imag \int^{\widecheck{H}} \c \sfA \cdot \c \kappa_e } ~,
\end{equation}
we can shift $\c \sfA$ by a flat field: The kinetic term does not change, but the second
term changes under $\c \sfA \mapsto \c \sfA + \c \phi_f$ by $\exp(2\pi \imag \int \phi_f c(\c \kappa_e))$.
Therefore, the integral over the compact Abelian group $H^{\ell-1}(M;\IR/\IZ)$ makes the  partition function vanish unless $c(\c \kappa_e)=0$. This is the quantum-mechanical version of the statement that there is no solution of the equation of motion unless the electric current is trivializable.

\item The path integral \eqref{eq:PartFunWithElecCurr} is still Gaussian. It is an interesting exercise to work it out exactly 
(see \cite{Moore:2011SCGP}), but we will not work it out here. 

\end{enumerate}
\end{remark}

\subsubsection{Coupling To Magnetic Currents} 

Once we have interpreted background electric currents in terms 
of differential cohomology, electromagnetic duality obliges us 
to interpret the magnetic current as a differential cohomology 
class, 
\be\label{eq:MagneticCurrent}
[\c \kappa_m] \in \widecheck{H}^{\ell+1}(M_n) ~.
\ee
However, once we do so, something very profound happens: 
The equation, 
\be 
d \sfF = \sfJ_m ~,
\ee
shows that on the support of $\sfJ_m$ \emph{we cannot interpret 
$[\c \sfA]$ as a differential cohomology class!} In fact, differential 
cohomology provides an elegant way out of this conundrum. We should, 
rather, interpret $\c \sfA$ as a differential cochain that trivializes the differential magnetic current: 
\be 
\delta \c \sfA = \c \kappa_m ~,
\ee
where a precise meaning to these equations can be given in terms of Hopkins-Singer cochains and cocycles (see \autoref{subsec:HopkinsSingerCocycles}). The gauge equivalence class of the fields then lives in a torsor for differential cohomology \cite{Hopkins:2002rd,Belov:2006jd,Monnier:2018nfs,Freed:2019sco}. 

 Finally, we come to the interesting case of the partition function in the presence of simultaneous electric and magnetic currents.
 (The ideas we are about to mention go back to \cite{Freed:2000ta} but have yet to be completely worked out.)  Now the partition function must be considered as a section of a line bundle with connection. We consider a family of theories: 
 \be 
 \CX \to \CS ~,
 \ee
 with fiber $M_n$. For example, $\CS$ could be the space of metrics and electric and magnetic currents. 
 The line bundle with connection is determined by the 
 pairing in differential cohomology 
\be 
\int^{\widecheck{H}}_{\CX/\CS} : \widecheck{H}^{n-\ell+1}(\CX)  \times 
\widecheck{H}^{\ell+1}(\CX) \rightarrow \widecheck{H}^2(\CS) ~,
\ee
where we use the pairing of differential cohomology on the 
electric and magnetic currents. That is, the path integral is a section of a line bundle with connection determined by,
\be 
\int^{\widecheck{H}}_{\CX/\CS} [\c \kappa_e ] \odot [\c \kappa_m] 
\in \widecheck{H}^2(\CS) ~.
\ee
A Hamiltonian version of this construction is based on a 
family of spatial slices,
\be 
\CX^{\mathsf{sp}} \to \CS ~.
\ee
Now one finds that there is a bundle of projective Hilbert spaces with nontrivial Dixmier-Douady (DD) class
\tightfootnote{The DD class was introduced in \autoref{subsubsec:GerbeConnections}.}
determined by the pairing of electric and magnetic currents:
\be 
\int^{\widecheck{H}}_{\CX^{\mathsf{sp}}/\CS} [\c \kappa_e ] \odot [\c \kappa_m] 
\in \widecheck{H}^3(\CS) ~.
\ee
The fact that the partition function is a section, and the bundle of projective Hilbert spaces has a nontrivial DD class, is a manifestation of the general fact that there will be an anomaly in the presence of simultaneous electric and magnetic currents. 

\subsection{Relation To Generalized Symmetries}\label{subsec:ElMgCurrent-GenSymAnom}

The coupling to background electric and magnetic currents can be 
viewed as a version of coupling to background gauge fields of 
``higher-form symmetries.'' To make this evident, let us first 
consider the coupling to background electric current, and let us suppose that $[\c \kappa_e]$ is topologically trivial, so that: 
\be 
\sfJ_e = d \kappa_e ~,
\ee
with $\kappa_e \in \Omega^{n-\ell}(M_n)$ globally well-defined. 
Define $b_e = \frac{\imag }{\lambda} \star \kappa_e \in \Omega^{\ell}(M_n)$. Then after 
completing the square, we can write the exponentiated action 
\eqref{eq:PartFunWithElecCurr} as: 
\be\label{eq:GaugedEllFormSymmetry}
\exp\left[ - \pi \int \lambda (\sfF-   b_e)\wedge \star(\sfF-b_e) + \pi \lambda\,b_e \wedge \star b_e \right] ~.
\ee
If we shift $[\c \sfA]$ by flat fields, then $\sfF$ is invariant, as we have utilized extensively above.  If we gauge this symmetry, 
then we shift by ``spacetime-dependent flat fields.'' That is, we 
shift by arbitrary gauge fields, we can consider $b_e$ as a gauge 
field gauging such a symmetry so that (dropping the quadratic term in $b_e$) the action 
\eqref{eq:GaugedEllFormSymmetry} has a symmetry under: 
\be 
\begin{split} 
\sfA & \mapsto  \sfA + \alpha ~, \\ 
b_e  & \mapsto b_e + d \alpha ~.\\
\end{split}
\ee
Note that this symmetry multiplies Wilson operators by a phase, as expected for an electric higher-form symmetry. 

Similarly, if we consider a topologically trivial magnetic current with $\sfJ_m = d \kappa_m$, where $\kappa_m \in \Omega^{\ell-1}$, then we have another term in the action: 
\be 
\exp\left[{\imag \pi \int b_m \wedge \sfF }\right] ~,
\ee
where $b_m \sim \star \kappa_m$. This term has a ``magnetic higher-form symmetry'' $b_m \mapsto b_m + \omega_m$, where $\omega_m \in \Omega^{n-\ell}_{\IZ}$, since $\sfF$ has quantized periods. Adding such a term to the action accounts for a phase shift of 't Hooft operators under the magnetic shift symmetry. 

Now, if we try to consider simultaneous electric and magnetic currents (in the topologically trivial case), we must consider an action: 
\be\label{eq:GaugedEllFormSymmetry-2}
\exp\left[ - \pi \int \lambda (\sfF-   b_e)\star(\sfF-b_e) + \imag \pi \int b_m \wedge \sfF \right] ~,
\ee
we find a ``mixed anomaly'' under the simultaneous electric and magnetic higher-form symmetry: 
\be 
\begin{split} 
\sfF & \mapsto \sfF + d \alpha_e ~,\\ 
b_e & \mapsto b_e + \alpha_e ~,\\ 
b_m & \mapsto b_m + \alpha_m ~.\\
\end{split}
\ee
The anomaly is: 
\be 
\exp\left[ \imag \pi \int \alpha_e \wedge d \alpha_m \right] ~, 
\ee
and cannot be eliminated by a local counterterm in $n$ dimensions, although it can be canceled by inflow from an invertible $\mathsf{BF}$ theory in $(n+1)$ dimensions. 
In this way, we make contact with some of the ideas in 
T. Dumitrescu's TASI 2023 lectures \cite{Dumitrescu:TASIVideoLec1,Dumitrescu:TASIVideoLec2,Dumitrescu:TASIVideoLec3,Dumitrescu:TASIVideoLec4}. 

  \SectionWithHeader{A Hilbert Space For The Self-Dual Field}{A Hilbert Space For The Self-Dual Field}{sec:HilbertSpaceSelfDualField}
  We can apply these ideas to the self-dual field in $n = 2 \mathsf{\,\,mod\,\,}4$ dimensions with classical equations of motion,
  \be
      F = \star F ~, \quad dF = 0 ~,
  \ee
  where $F \in \Omega^{\ell}(M_{n})$, and $n = 2\ell$. Since $n=2\mathsf{\,\,mod\,\,}4$, we can write  $n = 4s + 2$, so that  $\ell = 2s + 1$. 

  The non-self-dual field with $n=2\ell$ provides an SvN representation of $\mathsf{Heis}\big( \widecheck{H}^{\ell} \times \widecheck{H}^{\ell}\big)$. The self-dual case should involve half the number of degrees of freedom. The idea is to make sense of $\mathsf{Heis}\big(\widecheck{H}^{\ell}(N_{n-1})\big)$ by viewing $\widecheck{H}^{\ell}(N_{n-1}\big)$ itself as a group with a symplectic form. We need another theorem from group theory:

  \begin{theorem} Let $G$ be a topological Abelian group. The isomorphism classes of central extensions,
    \be
       \begin{tikzcd}
        1 \ar[r] & \mathsf{U(1)} \ar[r] & \wt{G} \ar[r] & G \ar[r] & 1 ~,
       \end{tikzcd}
    \ee
    are in one-to-one correspondence with continuous, alternating, bihomomorphisms,
    \be 
        s: G\times G \to \mathsf{U(1)} ~.
    \ee
    Alternating means $s(x, x) = 1$ for all $x \in G$, and implies skew: $s(x,y) = s(y,x)^{-1}$ for all $x, y \in G$.
    \tightfootnote{Proof: $s(x,y)s(y,x) = s(xy, yx) = s(xy, xy) = 1$ $\implies$ $s(y,x) = s(x,y)^{-1}$.}
  \end{theorem}
  \noindent For more technical details as well as a proof, see \cite[App. A]{Freed:2006ya}.

  Here $s(x,y)$ is the commutator function. If,
  \be
     (z_1, x_1) \cdot (z_2, x_2) = (z_1 z_2 c(x_1, x_2), x_1 + x_2) ~,
  \ee
  is a central extension, then,
  \be
    s(x,y) = \frac{c(x,y)}{c(y,x)} ~.
  \ee
Note that the commutator function is ``gauge-invariant'' in the sense that if we change the cocycle $c(x,y)$ by a coboundary (thereby defining an isomorphic Heisenberg group), the function $s(x,y)$ does not change. 
  
The point of the theorem is that, if one works at the level of the isomorphism class, then we can use the cruder knowledge of the commutator function. Put differently, from a commutator function one can deduce (nonuniquely) a cocycle, and thereby a central extension. So, to find,
  \be
  \begin{tikzcd}
    1 \ar[r] & \mathsf{U(1)} \ar[r] & \mathsf{Heis}\big( \widecheck{H}^{\ell}(N_{n-1}) \big) \ar[r] & \widecheck{H}^{\ell}(N_{n-1}) \ar[r] & 1 ~,
  \end{tikzcd}
  \ee
  it will suffice to find a suitable $s$-function on $\widecheck{H}^{\ell}(N_{n-1})$. There is a canonical guess,  given by the pairing,
  \be
    s_{\text{trial}}(\chi_1, \chi_2) = \exp\left( \imag \langle \chi_1, \chi_2 \rangle \right) ~.
  \ee
  Surprisingly, this is not exactly the right answer! But it is close. The function 
  $s_{\text{trial}}$  is certainly a bihomomorphism because $\ell$ is odd. It is also skew, but it is not alternating! Rather one can show that:
  \be
    s_{\text{trial}}(\chi, \chi) = (-1)^{\int_{N_{n-1}}\nu_{2s} \cup\,c(\chi)} ~.
  \ee
  Here $\nu_{2s}$ is the degree $2s$ Wu class of $N$. One can write the 
  Wu class $\nu_{2s}$ of a manifold as a polynomial in the Stiefel-Whitney classes of the tangent bundle of the manifold. For example, in the case where the manifold is oriented, 
  so $w_1(TM_{n})=0$, and we have: 
  \be
    \nu_{0} = 1 ~, \quad \nu_{2} = w_2 ~, \quad \nu_4 = w_4 + w_2^2 ~, \dots 
  \ee
Of course, the topological class of the differential character $\chi$ is: 
  \be
   c(\chi) \in H^{\ell}(N_{n-1}; \IZ) = H^{2s+1}(N_{n-1}; \IZ) ~.
  \ee
  We therefore introduce a $\IZ_2$-grading of $\widecheck{H}^{\ell}(N_{n-1})$:
  \be\label{eq:Wu-Grading}
    \epsilon(\chi) = \left\{
      \begin{array}{ll}
        0 ~, & \int \nu_{2s}\cup\,c(\chi) = 0 \mathsf{\,\,mod\,\,} 2 ~,\\
        1 ~, & \int \nu_{2s}\cup\,c(\chi) = 1 \mathsf{\,\,mod\,\,} 2 ~.
      \end{array}
    \right.
  \ee
  We find that,
  \be\label{eq:TrueCommFun}
     s(\chi, \chi') := \exp\left( \imag \langle \chi, \chi'\rangle - \imag \pi \epsilon(\chi)\epsilon(\chi') \right) ~,
  \ee
  is now a bihomomorphism which is alternating, so $\mathsf{Heis}\big( \widecheck{H}^{\ell}(N_{n-1}) \big)$ is a $\IZ_2$-graded Heisenberg group.

There are parallel theorems for $\IZ_2$-graded Heisenberg groups to those we quoted above. In particular, there is a \underline{unique} $\IZ_2$-graded unitary Stone-von Neumann 
$\IZ_2$-graded irreducible representation. A central proposal of \cite{Freed:2006ya,Freed:2006yc} is that if we define (up to isomorphism) a $\IZ_2$-graded Heisenberg group $\mathsf{Heis}\big( \widecheck{H}^{\ell}(N_{n-1}) \big) $  using the commutator function \eqref{eq:TrueCommFun}, then the SvN representation is the Hilbert space of the self-dual field (up to isomorphism). Note that it would be of interest to 
refine this so it is not just a description up to isomorphism. That is, it would be of interest to provide an explicit cocycle that produces the commutator function 
\eqref{eq:TrueCommFun}.

A striking consequence of the above proposal is that the Hilbert space $\CH(N_{n-1})$ of the self-dual field must be   $\IZ_{2}$-graded. It is natural to expect that the $\IZ_2$ grading \eqref{eq:Wu-Grading} corresponds to a boson/fermion grading based on statistics. This has not been derived carefully in the literature, and it would be desirable to have a rigorous proof that this is the case. We can check the claim in some special cases:  The chiral scalar in $\IM^{1,1}$ is related to the chiral fermion. In this case,  $s=0$, so $\nu_0=1$, and the grading \eqref{eq:Wu-Grading} is determined by the winding number of the field. Indeed vertex operators creating states with odd winding number are fermionic. Moreover, if we reduce a self-dual field in $4s +1$ dimensions from $\IM^{1,1} \times Y$, where $\mathsf{dim\,}Y = 4s$ 
 to $\IM^{1,1}$ via  Kaluza-Klein reduction then we get a theory of chiral and antichiral scalars. Explicitly, if we choose bases $\omega^+_a $ and $\omega^-_i$ of self-dual and 
 anti-self-dual Harmonic forms of degree $2s$ with integral periods then we can write,
 \be 
 F = \sum_{a=1}^{b_{2s}^+(Y)} d \phi^a \wedge \omega^+_a  + \sum_{i=1}^{b_{2s}^-(Y)} d \phi^i \wedge \omega^-_i ~.
\ee
The $\phi^a$ are chiral and the $\phi^i$ are anti-chiral periodic scalars. 
Vertex operators will depend on a ``momentum'' $x \in H^{2s}(Y; \IZ)$. 
The fermionic parity of vertex operators in this $(1+1)$-dimensional theory is related to the parity of $x^2$, where $x \in H^{2s}(Y; \IZ)$.

\begin{remark}
$\,$
\begin{enumerate}
    \item As we remark briefly in \autoref{sec:generalizedcohomology} below, the total RR field 
    in type II string theory is a self-dual field. Considerations analogous to those above apply to the definition of the Hilbert space of the RR field on nine-dimensional manifolds. In particular, they will be $\IZ_2$-graded and have fermionic sectors. This striking fact has not yet found an application in the literature. 

\item For more about differential cohomology and the formulation of 
self-dual fields, see  
\cite{Witten:1996hc,Freed:2000ta,Hopkins:2002rd,Moore:2004jv,Belov:2006jd,Belov:2006xj,Witten:2007ct,Hsieh:2020jpj}.
In particular, these references describe the partition ``function'' of the self-dual field, 
which is a nontrivial generalization of the discussion of \autoref{sec:PartitionFunctionsGenMax}. The partition function again has the general form, 
\be 
Z = \CN \Theta ~,
\ee
where $\CN$ is the contribution from the topologically trivial modes, analogous to what we described in   \autoref{sec:PartitionFunctionsGenMax}.  It is related to 
Cheeger's half-torsion \cite{Monnier:2010ww}. An important and novel feature of the partition function 
of the self-dual field is that $\Theta$ is a holomorphic section of a line bundle that in general admits a multi-dimensional space of holomorphic sections. So, in general,  the partition function is not an element of a line,  but a vector in a higher dimensional 
vector space. The formalization of this mathematical structure led to the notion of 
a \emph{relative field theory}  \cite{Segal:1987sk,StolzTeichner2,Freed:2012bs}. 
    
\end{enumerate}

\end{remark}

  \SectionWithHeader{Spectra, Generalized Cohomology Theories,  And Their Differential Counterparts}{Spectra, Generalized Cohomology, Differential Counterparts}{sec:generalizedcohomology}

In these notes we have emphasized cohomology groups such as singular and de Rham cohomology groups. It turns out that there is a broader notion of ``cohomology theory,'' and that several of these generalized cohomology theories are playing a  prominent role in modern physical mathematics. 

\subsection{Generalized Cohomology Theories}

The cohomology groups $H^k(X)$ that we have used thus far can be defined and computed in many ways -- one can use singular cochains, Cech cochains,  finite cellular or simplicial cecompositions or, (with suitable coefficients), differential forms. In a famous work \cite{EilenbergSteenrod:1945}, S. Eilenberg and N. Steenrod described cohomology axiomatically and proved a uniqueness theorem, thereby explaining why all the above techniques give 
``the same result.''  Eilenberg and Steenrod axiomatized cohomology as follows:

We form a category $\textbf{CW}$ of pairs of topological spaces $(X, S)$ where $S \subset X$ (with morphisms as 
continuous maps of pairs)
\tightfootnote{Technically one takes them to be CW complexes with CW subcomplexes, hence the name.} 
and defines a \emph{cohomology theory} to be a 
contravariant functor $H$ to the category 
$\textbf{GrAb}$ of $\IZ$-graded Abelian groups: 
\be 
H: \textbf{CW} \to \textbf{GrAb} ~.
\ee
We denote $H^k(X,S)$ to be the subgroup with grading $k$. 
We also denote $H^k(X) = H^k(X, \emptyset)$. If 
$f:(X,S) \to (Y,T)$ is a morphism in $\textbf{CW}$, i.e., a continuous 
map of pairs, then the functor maps this to homomorphisms of 
Abelian groups: 
\be 
f^*: H^k(Y,T) \to H^k(X,S) ~.
\ee

The functor $H$ must be such that for any pair $(X,S)$ there exist group homomorphisms for all $k\in \IZ$: 
\be 
\delta^k: H^k(S) \rightarrow H^{k+1}(X,S) ~,
\ee
such that we have: 

\begin{enumerate}
    \item \textbf{Homotopy Invariance}: If two morphisms in 
    $\textbf{CW}$ are homotopic, that is, if $f: (X,S) \to (Y,T)$ is homotopic to $g:(X,S) \to (Y,T)$, then the induced maps are equal:  $f^* = g^*$. 
    \item \textbf{Long Exact Sequence}: If $i : S \hookrightarrow X$, and $j: (X, \emptyset) \hookrightarrow (X, S)$, then we have a long exact sequence,
    \be\label{eq:GenCohLES}
     \cdots \longrightarrow H^{k}(X, S) \xlongrightarrow{j^*} H^{k}(X) \xlongrightarrow{i^*} H^{k}(S) \xlongrightarrow{\delta^*} H^{k+1}(X, S) \longrightarrow \cdots
    \ee
    \item \textbf{Excision}: 
    \be
      \mathsf{Int}(U) \subset \mathsf{Int}(S) \implies H^{k}(X, S) \cong H^{k}(X-U, S-U) ~.
    \ee
    (This can also be phrased in terms of two subspaces 
    $S_1, S_2 \subset X$ as,
    \be 
    f^*: H^k(X,S_2) \rightarrow H^k(S_1, S_1 \cap S_2) ~,
    \ee
    where $f: (S_1,S_1 \cap S_2) \hookrightarrow (X,S_2)$ is the inclusion.) 
    
    \item \textbf{Additivity}:
    If $(X,S)$ is a disjoint union of pairs $(X_\alpha, S_\alpha)$, then the inclusions induce an isomorphism: 
    \be
      H^{k}\big( \coprod_{\alpha} X_{\alpha},S_{\alpha} \big) \cong \bigoplus_{\alpha} H^{k}(X_\alpha, S_\alpha ) ~.
    \ee

    \item \textbf{Dimension Axiom}: 
    \be
       H^{k}({\rm pt}, \emptyset) = \left\{ \begin{array}{ll}
         \IZ ~, & k = 0 ~,\\
         0 ~, & k \neq 0 ~.
       \end{array}\right.
    \ee
  \end{enumerate}
  
  Axioms (1) through (4) imply the Mayer-Vietoris sequence (see, for example, \cite[Ch. 14]{May1999}, \cite[Ch. 9]{Switzer1975}, and \cite{Adams:1978,Strom:2023}).

Eilenberg and Steenrod showed that these axioms uniquely characterize cohomology in the sense that if $H_1$ and $H_2$ are two such functors  and there is a natural transformation $\tau: H_1\to H_2$ such that $H_1^{\bullet}({\rm pt},\emptyset) \cong H_2^{\bullet}({\rm pt}, \emptyset)$ 
then there is an isomorphism $H_1^k(X,S) \cong H_2^k(X,S)$,  
for all $k$ and all pairs $(X,S)$. 
  
In the 1950's and 1960's, bordism theory and $K$-theory 
were being investigated, and it was recognized that 
they satisfy many of the properties of cohomology 
theory, namely, they satisfy   Axioms (1) through (4) 
but not Axiom (5), where the value on a point is a more 
general Abelian group.
 For example, for complex  $K$-theory, the cohomology of a point is: 
  \be
      K^{j}({\rm pt}) = \left\{ \begin{array}{ll}
          \IZ ~, & j \text{ even} ~,\\
          0 ~, & j \text{ odd} ~.
      \end{array}\right.
  \ee
In fact, $K$-theory groups form a graded ring. The ring associated with a point 
is $\IZ[u,u^{-1}]$, where $u$ is a degree two element associated with Bott periodicity.

In analogy to the theory of non-Euclidean geometry,    one can  drop Axiom 5 and still have an interesting mathematically consistent framework. Such functors are 
known as ``exotic cohomology theories'' or ``extraordinary cohomology theories''  or ``generalized cohomology theories.''  They have proven to be of tremendous importance. 

The axioms characterize the theory uniquely 
in the following sense: Suppose $h_1, h_2$ are two functors 
satisfying Axioms $1-4$, and there is a natural transformation 
shifting degree by $s$:
\be 
\tau_{X,S}: h_1^k(X,S) \rightarrow h_2^{k+s}(X,S) ~,
\ee
such that,
\be 
\tau_{{\rm pt},\emptyset}: h_1^k({\rm pt}) \rightarrow h_2^{k+s}({\rm pt}) ~,
\ee
are isomorphisms, then all the $\tau_{X,S}$ are isomorphisms.

Examples of generalized cohomology theories include the various flavors of $K$-theory \cite{Bott1969-dj,Karoubi1978,Atiyah2018}, bordism theory \cite{Stong2016-wq}, and elliptic cohomology \cite{Ochanine1987,Landweber1988,Landweber1995,Witten:1986bf,Witten1988,Lurie2009:EllipticSurvey,Berwick-Evans:2024nrr}, all of which have proven to be of interest in the context of physics.

\subsection{Spectra}\label{subsec:Spectra}

We have already mentioned that the Eilenberg-MacLane spaces $K(A,p)$ (where $A$ is an Abelian group and $p$ is a positive integer) satisfy the property that, 
\be 
H^p(X;A) \cong [X, K(A,p)] ~.
\ee
It turns out (thanks to the Brown representability theorem \cite{Switzer1975}) that to any generalized cohomology theory $h^p$,
one can associate a collection of topological spaces 
$\{ E_p\}_{p\in \IZ} $ such that 
\be\label{eq:BrownRep}
h^p(X) \cong [X, E_p] ~.
\ee
The properties of a generalized cohomology theory 
imply that a set of spaces $\{ E_p \}$ such that 
\eqref{eq:BrownRep} holds, must be related to each 
other in a way that defines a \emph{spectrum}.

Formally, a  spectrum $E$ is a sequence $\{E_{n}, s_{n}\}_{n=-\infty}^{\infty}$ of pointed CW spaces $\{E_n\}$ and pointed CW embeddings $s_{n}: \Sigma E_{n} \to E_{n+1}$  where $\Sigma$ denotes the pointed suspension (a.k.a. reduced suspension). 
\tightfootnote{The \emph{reduced suspension} $\Sigma X$ of a based topological space $(X,x_0)$ is defined as $\Sigma X = (X \times I)/(X \times \{0, 1\} \cup \{x_0\} \times I)$. It can be obtained from the \emph{unreduced suspension} $S X = (X \times I)/\{(x,0)\sim(x',0), (x, 1) \sim (x',1) \text{ for all } x, x' \in X\}$ by shrinking $\{x_0\}\times I$ to a point, so there is a natural quotient map $S X \to \Sigma X$. We remark that the symbols for reduced ($S$) and unreduced ($\Sigma$) suspensions are sometimes interchanged in the literature. Our convention is opposite to the one in \cite{Switzer1975,Rudyak1998-vw,Davis2001}.}
An $\Omega$ spectrum is a spectrum $E$ where the adjoint maps $S_{n}: E_{n} \to \Omega E_{n+1}$ are weak homotopy equivalences.  
\tightfootnote{Some authors (e.g., \cite{Switzer1975,Rudyak1998-vw}) define the notion of spectrum for pointed CW spaces and pointed CW embeddings, while others (e.g., \cite{Davis2001}) define it more generally for pointed topological spaces and pointed maps. One can make a more careful distinction between a \emph{prespectrum} defined as above but for pointed spaces and maps and a \emph{spectrum} which has the added requirement that the adjoint maps are homeomorphisms (see, for example, \cite[p. 86]{FreedBordism}).}
Here $\Omega_{x_0} X$ denotes the (based) loop space of a pointed space $(X,x_0)$. The $S_n$ are called ``adjoint'' maps because 
of the 1--1 correspondence $[X, \Omega Y] = [\Sigma X, Y]$ of homotopy classes of maps of pointed spaces.

\bigskip
\noindent \textbf{Example:} The Eilenberg-MacLane spectrum is an $\Omega$ spectrum, but the sphere spectrum is not.
\bigskip
 
The sequence of spaces $\{ E_p \}$ representing a generalized cohomology theory form an $\Omega$ spectrum. To see why,  apply \eqref{eq:GenCohLES} to the pair, 
\be 
(X\times I,  X \times \{0\} \cup X \times \{1 \}  \cup x_0 \times I ) ~,
\ee
where $x_0$ is the basedpoint of $X$, to obtain an isomorphism, 
\be 
\wt h^n(X) \cong \wt h^{n+1}(\Sigma X) ~,
\ee
where $\wt h^n(X)$ is the reduced cohomology (i.e., the kernel of the map 
$h^n(X) \to h^n({\rm pt})$ from the inclusion of a point into $X$). 
But now we have, 
\be 
[X,E_n]\cong \wt h^n(X) \cong \wt h^{n+1}(\Sigma X) \cong  [ \Sigma X , E_{n+1}] \cong [ X, \Omega E_{n+1}] ~.
\ee
Since this holds for every $X$, we must have a homotopy equivalence 
between $E_n$ and $\Omega E_{n+1}$.

$\,$\\
\noindent \textbf{Example: Complex $K$-theory.} The spectrum of complex $K$-theory is $\IZ \times \mathsf{BU}$ (with the discrete topology on $\IZ$), where $\mathsf{BU}$ is the colimit of the classifying space of unitary groups $\mathsf{BU}(n) \cong \mathsf{Gr}(n, \IC^{\infty})$. This means that the $K$-theory spectrum of a topological space $X$ is given by $K(X) \cong [X_+, \IZ \times \mathsf{BU}]$, where $X_+$ is $X$ adjoined with a disjoint basepoint. In particular, this means that $K_{n}(X) := \pi_{n}(K(X))$. There is a homotopy equivalence $\IZ \times \mathsf{BU} \cong \Omega^{2}\mathsf{BU}$, due to which the groups $\pi_{q}(\mathsf{BU},*)$ ($*$ denotes a basepoint) have periodicity $2$. This is an instance of \emph{Bott periodicity}
\tightfootnote{For a proof, see \cite{Switzer1975,Lawson1990-ev}.}
which implies that $K_{n+2}(X) \cong K_{n}(X)$.

\subsection{Differential Generalized Cohomology} 

Hopkins and Singer showed that any generalized cohomology theory can be generalized to a differential cohomology theory
\cite{Hopkins:2002rd}. See also \cite{Freed:2000ta,Freed:2006ya,BunkeNikolausVolkl:2013,Amabel:2021wbk,Debray:2023kvh} for other discussions. 

Given a generalized cohomology theory $E$, there is a 
differential analog $\widecheck{E}$ defining a functor from 
smooth manifolds to graded Ablian groups (in general, they will be infinite dimensional). 

To describe $\widecheck{E}$, we note that from $E$, one can 
derive cohomology theories with coefficients in 
$\IR$ and $\IR/\IZ$, denoted   $E_{\IR}$ and $E_{\IR/\IZ}$ respectively. (See \cite[App. B]{Freed:2006ya} for a short summary.) 
In particular, $E_{\IR/\IZ}^k(M)$ will be compact Abelian groups, and we have: 
\be 
0 \rightarrow E^k(M_{n})\otimes \IR \rightarrow E_{\IR/\IZ}^k(M_{n}) \rightarrow \Tors~ E^{k+1}(M_{n}) \rightarrow 0  ~ . 
\ee

Define a 
real $\IZ$-graded vector space:
\be 
V_E = E^{\bullet}_{\IR}({\rm pt}) ~ . 
\ee
There will be a natural map, 
\be\label{eq:E-CohoMap}
\widecheck{E}^\ell(M_{n}) \to \bigoplus_{p+q=\ell} H^p(M_{n}; V_E^q) ~.
\ee
The differential theory will have a characteristic class map,
\be\label{eq:E-CharClassMap}
c: \widecheck{E}^\ell(M) \to E^\ell(M_{n}) ~,
\ee
and a fieldstrength map, 
\be\label{eq:E-FieldstrMap}
F: \widecheck{E}^\ell(M_{n}) \to \bigoplus_{p+q=\ell} \Omega_E^p(M_{n}; V_E^q) ~,
\ee
where the subscript $E$ on $\Omega^p$ indicates that 
the image of the map is only to forms with a quantization 
condition compatible with \eqref{eq:E-CohoMap}. That is, 
\eqref{eq:E-FieldstrMap} and \eqref{eq:E-CharClassMap} 
are compatible when mapped into cohomology using 
\eqref{eq:E-CohoMap}.

In general, there is an analog of the Simons-Sullivan dancing hexagon \eqref{eq:DancingHexagon}:

  \begin{equation}\label{eq:DancingHexagonGenCoh}
  \adjustbox{scale=1,center}{%
   \begin{tikzcd}[row sep=large,column sep=large]
       \textcolor{OliveGreen}{0} \arrow[rd,OliveGreen] &  &  &  & \textcolor{Violet}{0} \\
        & \textcolor{OliveGreen}{\Omega(M_{n};V_E)^{\ell-1}/\Omega_{E}(M_{n};V_E)^{\ell-1}} \ar[phantom, "\boxed{\substack{\text{TOP.}\\\text{TRIVIAL}}}"above=13pt] \arrow[rd,OliveGreen] \arrow[rr, dashed, "d",blue] & & \Omega_{E}(M_{n};V_E)^{\ell} \ar[phantom, "\boxed{\substack{\text{FIELD}\\\text{STRENGTH}}}"above=13pt] \arrow[ru,Violet] \arrow[rd, 
       sloped,OrangeRed] \\
       \textcolor{OrangeRed}{E^{\ell-1}_{\IR}(M_{n})} \arrow[ru,OrangeRed] \arrow[rd,OrangeRed] & & \textcolor{red}{\widecheck{E}^{\ell}(M_{n})} \arrow[ru, "\chi\mapsto F(\chi)",sloped,swap,Violet] \arrow[rd,"\chi \mapsto c(\chi)", sloped, OliveGreen] & & \textcolor{OrangeRed}{H(M_{n};V_E)^{\ell}} \\
       & \textcolor{Purple}{E_{\IR/\IZ'}^{\ell-1}(M_{n})} \ar[phantom, "\boxed{\substack{\text{FLAT}\\\text{FIELDS}}}"below=8pt] \arrow[ru,Purple] \arrow[rr, dashed, "\beta: \text{Bockstein}",swap,blue]& & \textcolor{OliveGreen}{E^{\ell}(M_{n})}  \ar[phantom, "\boxed{\substack{\text{TOP.}\\\text{CLASS}}}"below=8pt] \arrow[ru, sloped,OrangeRed] \arrow[rd,OliveGreen] & \\
       \textcolor{Purple}{0} \arrow[ru,Purple] &  &  &  & \textcolor{OliveGreen}{0}
   \end{tikzcd}
   }
  \end{equation}
  
It is interesting to see how the above diagram works in the important case of complex $K$-theory. The $\IZ$-grading is $\IZ_2$-periodic, i.e., all groups with even degree are isomorphic, and all groups with odd degree are isomorphic. One of many models for 
$\widecheck{K}^0(M)$ is the group of isomorphism classes of objects in a groupoid 
whose objects are triples $(E,\n,c)$, where $E\to M_{n}$ is a finite rank complex vector bundle,
$\n$ is a connection on $E$, and $c$ is a \underline{globally defined} real differential form of odd degree: $c\in \Omega^{{\rm odd}}(M_{n})$. Morphisms are given by such triples on 
$M_{n}\times [0,1]$, restricting to source and target on the ends of the interval. This implies a 
number of equivalence relations, an important one being that $(E,\n,c)$ is equivalent to 
$(E',\n',c')$ if,
\be\label{eq:Diffl-K-equiv}
c' -c = \mathsf{CS}(\n, \n') ~,
\ee
where on the RHS, we have the relative Chern-Simons form described in 
equation \eqref{eq:Rel-CS} below. The fieldstrength is: 
\be 
F(E,\n,c) := \mathsf{ch}(\n) + dc ~,
\ee
and the characteristic class is just the class $[E] \in K^0(M)$. 
In this model, the crucial pairing on differential cohomology on an odd-dimensional 
Riemannian spin$^{\mathsf{c}}$ manifold (which gives an orientation for differential $K$-theory) 
is
given by \cite[App. A]{Freed:2006yc}, \cite{Klonoff:2008}:
\begin{align}
&\langle (E_1, \n_1, c_1) , (E_2, \n_2, c_2)\rangle \nonumber\\
&\quad = \left\{  \eta\big(\slashed{D}_{\n_1\otimes 1 + 1 \otimes \bar \n_2}\big) +
\int c_1 {\Tr} e^{-F(\n_2)} + \int \bar c_2 {\Tr} e^{F(\n_1)} + \int c_1 d\bar c_2
\right\} ~ \mathsf{mod~} \IZ ~.
\end{align}
%
(Bar means complex conjugation, and we consider general complex objects, as is natural in complex $K$-theory.)

\subsection{Some Physical Applications}\label{subsec:PhysAppDiffGenCoh}

In general, if the dynamical fields of a theory have gauge equivalence classes living in 
differential generalized cohomology, then that theory should be multiplicative and 
oriented so that there is a pairing. If the physical fields should satisfy some 
kind of self-duality condition, then the generalized cohomology theory should 
satisfy some kind of self-duality (such as definition 2.9 of 
\cite{Freed:2006ya}). 
  
  For example, in type II string theory, differential 
  $K$-theory is thought to be the proper framework for describing Ramond-Ramond fields \cite{Minasian:1997mm,Witten:1998cd,Witten:2000cn,Maldacena:2001xj,Freed:2006ya,Freed:2006yc,Evslin:2006cj, Moore:2006dw,Distler:2009ri}.  In particular, in type II string theory, 
  the gauge equivalence class of the RR field is identified with an element of 
  $\widecheck{K}^0(M_{10})$ for type IIA,  and $\widecheck{K}^1(M_{10})$ for type IIB string theory on a 10-dimensional spacetime $M_{10}$. The groupoid described above for the RR field in 
  type IIA fits in well with the theory of D-branes and tachyon condensation. 
  The equivalence relation \eqref{eq:Diffl-K-equiv} implies that there is a gauge equivalence, where we can shift $\n \mapsto \n + \alpha$ for an arbitrary 
  $\alpha \in \Omega^1(M_{10}; \End(E))$, provided we make a compensating change in $c$. 
  Thus, the ``degrees of freedom'' in $\n$ are ``topological,'' and the local degrees of freedom in the RR field are captured by $c$. In the type II string interpretation,  
  the fieldtrength encodes the total RR fieldstrength,
  \be 
  F(E,\n,c) = G = G_0 + u^{-1} G_2 + u^{-2} G_4 + u^{-3} G_6 + u^{-4} G_8 + u^{-5} G_{10} ~,
  \ee
  where $G_0$ is the Romans mass and the independent propagating degrees of freedom 
  have fieldstrengths $G_2$ and $G_4$. Clearly, we need to impose a self-duality condition, 
  and thus the proper mathematical formulation involves a theory of self-dual fields 
  in the context of differential cohomology. This has been explored to some extend in 
\cite{Moore:1999gb,Freed:2006ya,Freed:2006yc,Belov:2006jd}, but more work remains to be done. 
In other variations of string theory, such as orientifolds, other versions of differential 
cohomology such as twisted equivariant differential cohomology are used 
\cite{Distler:2009ri,Distler:2010an}.

Briefly, some other significant examples that have shown up in physics are:

\begin{enumerate}

 \item There is generalized cohomology theory called $E$-theory in 
\cite{Freed:2006mx} and ``supercohomology'' in \cite{Debray:2023kvh}, which 
fits in a long exact sequence 
\be 
\cdots H^k(M_{n}) \rightarrow E^k(M_{n}) \rightarrow H^{k-2}(M_{n}; \IZ/2\IZ) \rightarrow H^{k+1}(M_{n}) 
\rightarrow \cdots ~,
\ee 
which is useful in describing theories that depend on spin structure. 
This includes spin-Chern-Simons theories \cite{Belov:2005ze,Jenquin:2006jh}, 
and the effective theory of pions in QCD \cite{Freed:2006mx}. Closely related 
    are the Wu-Chern-Simons theories investigated by S. Monnier   \cite{Monnier:2014txa,Monnier:2016jlo,Monnier:2017klz}, 
    which find useful applications to anomaly-cancellation in 6d supergravity
   \cite{Monnier:2017oqd,Monnier:2018nfs,Monnier:2018cfa}. 

\item  There is a very large literature on the application of bordism theory 
to the classification of topological terms in actions, and to phases of 
matter in condensed matter physics. Some key papers are
\cite{Freed:2016rqq} and \cite{Kapustin:2013qsa,Kapustin:2013uxa,Kapustin:2014lwa,Kapustin:2014tfa,Kapustin:2014zva,Kapustin:2014dxa,Kapustin:2017jrc,Hsin:2020cgg}. Closely related to this is the application of $K$-theory to the classification of theories of phases of noninteracting electrons 
\cite{Horava:2005jt,Kitaev:2009mg,Schnyder:2008tya,Schnyder:2009klk,Ryu:2010zza,Freed:2012uu}.

\end{enumerate}

\subsubsection{Brief Remarks On Elliptic Cohomology}\label{subsec:elliptic-cohomology}

In this section, we give a highly condensed introduction to elliptic cohomology and topological modular forms in the spirit of the preceding discussions on spectra. For detailed expositions, see for example, \cite{Hopkins95,Hopkins2002,Goerss:2009,Douglas2014-cl,Berwick-Evans:2024nrr} and references therein.

Elliptic cohomology -- as a generalized cohomology theory -- arose from the discovery of the 
\emph{elliptic genus} by S. Ochanine, P. Landweber, D. Ravenel and R. Stong \cite{Ochanine1987,Landweber1988,Landweber1995}. Given a commutative ring $R$, an $R$-valued \emph{genus} is a ring homomorphism $\sigma: \Omega_{\bullet}^{\CB} \to R$, from a cobordism ring for cobordisms
\tightfootnote{We use ``cobordism'' in deference to mathematical literature, but at slight variance with \autoref{rem:cobordism-vs-bordism}.}
of manifolds equipped with some structure $\CB$. The ring  $R$ relevant to the elliptic genus   is the ring of integral holomorphic modular forms, which we will denote as $\mathsf{MF}_{\bullet}$. The elliptic genus was interpreted in terms of two-dimensional quantum field theory 
  by Witten \cite{Witten:1986bf,Witten1988}.
  (See also \cite{Alvarez:1987wg,Alvarez:1987de}.)

We first recall Witten's interpretation and then comment on some of the subsequent mathematical 
development. To a manifold $M$ satisfying suitable conditions (we will comment below on the meaning of ``suitable'') one can define a 2d $\CN=(0,1)$ supersymmetric sigma model $\sigma_{M}$ with target $M$. Witten considered the 
analog of the Witten index in this context, namely, the partition function of the theory on a circle 
at finite temperature with Ramond boundary conditions around the Euclidean time circle: 
\be\label{eq:EG-TraceDef}
\mathsf{Z}_{\text{ell}}(\sigma_{M}; q) :=  \Tr_{\CH_{R}} (-1)^{\texttt{F}} q^{(H+P)/2} \ov{q}^{(H-P)/2} = \Tr_{\CH_R} (-1)^{\texttt{F}}  e^{-\beta H} e^{\imag \theta P} ~.
\ee
Here $H$ is the Hamiltonian, $P$ is the generator of translations around the spatial circle, and $\texttt{F}$ is the fermion number (defined modulo $2$). By supersymmetry, this will be an expansion in a parameter $q= e^{-\beta + \imag \theta}$, independent of $\ov{q}$.
\tightfootnote{The right-moving sector has $\CN=1$ supersymmetry and the supercharge $\CQ$ satisfies $\CQ^2 = H-P$, so by standard arguments, only states with $H=P$ contribute to the trace over $\CH_R$.}
Moreover, the Hilbert space $\CH_{R}$ is notionally the space of sections of a spin bundle over loop space and 
the supersymmetry operator is the Dirac-Ramond operator, a generalization of the Dirac operator 
to a Dirac-like operator on loop space. From this interpretation, $\mathsf{Z}_{\text{ell}}(\sigma_M; q)$
is the  character-valued index and gives the elliptic genus, where $q$ is conjugate to the symmetry operation 
of rotation around the spatial circle.   

As is standard in physics, the trace \eqref{eq:EG-TraceDef}   has an interpretation as the partition function of the theory on a torus with modular parameter $\tau$,  where $q = e^{2\pi i \tau}$. (In terms of \eqref{eq:EG-TraceDef}, $\tau = (\theta+\imag\beta)/2\pi$.)
%
%
On the other hand, by localization around the constant loops in the sigma model, one can argue that the fields of the sigma model are valued in the normal bundle to the constant loops in the entire loop space and from this interpretation derive
\begin{align}\label{eq:z-elliptic-sigma}
    \mathsf{Z}_{\text{ell}}(\sigma_{M}; q) &= q^{-\nu}\int_{M}\widehat{A}(R)~\mathsf{tr}\left( \prod_{\ell=1}^{\infty}\left(1 - q^{\ell}e^{\mathsf{i}R/2\pi} \right)^{-1} \right) ~, 
\end{align}
where $R$ is the curvature 2-form
\tightfootnote{\label{foot:curv2-form}In local coordinates, we can write the curvature 2-form as $R^{a}{}_{b} = \frac{1}{2}R^{a}{}_{b\mu\nu}dx^{\mu}\wedge dx^{\nu}$, where $a, b$ denote frame indices or tangent space indices (a.k.a. ``flat indices'') and $\mu, \nu$ denote coordinate indices (a.k.a. ``curved indices''). The traces are understood to be taken over the $a, b$ indices.}
and $\widehat{A}(R)$ is the A-hat roof genus (a.k.a. $\widehat{A}$-class). 
\tightfootnote{See, for example, \cite[Sec. 1.1]{Freed:DiracNotes1987} for a review of the $\widehat{A}$-class.}
Expanding in $q$, we could thus interpret the elliptic genus as the generating function for the 
index of Dirac operators coupled to various tensor powers of the tangent and cotangent bundle. 
The overall power $q^{-\nu}$ depends on a parameter $\nu$ and is essential for modularity. 
 For CFTs, $\nu = -2(c_{L}-c_{R})$ is minus twice the chiral central charge, which is an integer for modular-invariant spin CFTs (see, for example, \cite{Lin:2019hks,Chang:2020aww}). It is sometimes 
 called the ``anomaly coefficient.'' Being an RG invariant, it is a well-defined grading for the space of theories obtained from a given 2d $\CN=(0,1)$ SCFT  by  supersymmetry-preserving deformations. 
 (The deformations will, in general, break conformal symmetry.)
In the mathematical interpretations defined below,  the 
 ``anomaly coefficient'' is an element $\nu \in \big(I_{\IZ}\Omega^{\mathsf{Spin}}\big)^{4}({\rm pt}) \cong \IZ$, in terms of the Anderson dual of the spin bordism spectrum.  

We now turn to the meaning of ``suitable manifold'' $M$ in the definition of $\sigma_{M}$. The more precise meaning is that the sigma model should be anomaly-free. Sigma model anomalies were first studied in 
  \cite{Moore:1984dc,Moore:1984ws,Manohar:1984zj}, and in order for them to cancel, we must couple the sigma model to a gerbe whose field strength $H$ satisfies $dH = \frac{1}{2}p_{1}(TM)_{\IR}$, where we use the Chern-Weil representative of the characteristic class. At the integral level, we require a trivialization of 
  a certain canonical cohomology class usually denoted $\lambda$, which can be defined for 
   spin manifolds. This class satisfies $2\lambda = p_1(M) := p_{1}(TM)$.
\newsavebox{\tempbox}%
\setbox\tempbox=\hbox{%
\begin{tikzcd}[scale=0.75, every node/.style={scale=0.75}]
 & B\mathsf{Spin}(n) \arrow[d,"\pi_1"] \arrow[r,"\lambda"] & K(\IZ,4) \arrow[d,"\pi_2"] \\
 M \arrow[r,"f"] \arrow[ur,dashed,"\widetilde{f}"] & B\mathsf{SO}(n) \arrow[r,"p_1"] & K(\IZ,4)
\end{tikzcd}
}%
\tightfootnote{\label{foot:spin-mfd-lambda}Suppose $M$ is $n$-dimensional, for some $n \in \IZ_{>0}$. A lift $\widetilde{f}$ of the classifying map $f: M \to B\mathsf{SO}(n)$ to $B\mathsf{Spin}(n)$ exists if and only if $M$ is spin, which is when the triangle on the left in the diagram
%
\centerline{\scalebox{0.8}{\usebox{\tempbox}}}
commutes. The double cover $\pi_2$ induces multiplication by $2$ on $H^4_{\IZ}$.}

Importantly, the real dimension of $M$ is equal to $\nu$.
\tightfootnote{For a 2d $\CN=(0,1)$ sigma model with anomaly coefficient $\nu = -2(c_L - c_R)$, the perturbative anomaly polynomial is $I_{4} = \frac{\nu}{48}p_1$. Since $\nu$ is an RG invariant, one can compute it in the free or weak coupling limit where the target is flat, and the 2d field theory is modeled by free fields. For a target space $M$ of dimension $n \in \IZ_{>0}$, the free field formulation gives $c_L = n$ ($n$ bosonic embedding supercoordinates) and $c_{R} = n + n/2 = 3n/2$ ($n$ bosonic and $n$ fermionic supercoordinates). This implies that $-2(c_R - c_L) = n$, and therefore, $n = \nu$.}
Now, there is a hierarchy of manifolds: Orientable manifolds have $w_1(M)=0$. Spin manifolds 
have  $w_1(M) = w_2(M) = 0$. The paper \cite{Killingback:1986rd} introduced the notion 
of a  \emph{string structure}, where, in addition, $\lambda=0$. For a more recent discussion, see 
\cite{Waldorf:2023zmt}. A string structure can be interpreted as a lift of the spin structure to the homotopy fiber of the map $B\mathsf{Spin}(n) \to K(\IZ,4)$. 
\tightfootnote{See \autoref{subsec:CorrespCourse} for a definition of the homotopy fiber of a map between two topological spaces.}
The cobordism ring of string manifolds is denoted by $\Omega_{\bullet}^{{\rm String}}({\rm pt})$.  Sigma model anomaly cancellation therefore requires that $M$ be a string manifold of dimension $\nu$.

%

Sigma models can be generalized to include a target space gauge bundle with field strength $F$. 
In this case, there is a modification of the anomaly cancellation condition: There is also a contribution from the second Chern class of the gauge bundle, which modifies the condition to 
\be
dH = c_2(F)_{\IR} - \frac{1}{2}p_1(TM)_{\IR} ~.
\ee
(This condition plays a central role in string theory.) This amounts to the data of a trivialization of $c_2(F) - \lambda$, and following the above line of reasoning, leads to the mathematical notion of a  twisted string structure on $M$. For a recent discussion, see \cite[Sec. 2]{Tachikawa:2025flw}.

Witten's construction above can be interpreted as describing a ring homomorphism,
\begin{align}\label{eq:WitGenus}\begin{split}
    \sigma_{\nu} : \Omega_{\nu}^{{\rm String}}({\rm pt}) &\to   \mathsf{MF}_{\nu/2} \\
      [M] &\mapsto  \sigma_{\nu}([M]) = 
\eta(q)^{\nu} \mathsf{Z}_{\text{ell}}(\sigma_M; q) ~,
\end{split}
\end{align}
where $\eta(q)$ is the Dedekind eta function, $\mathsf{Z}_{\text{ell}}(\sigma_M;q)$ is given by \eqref{eq:z-elliptic-sigma}, and we restrict to $\nu$ even. Mathematicians refer to this homomorphism as the \emph{Witten genus}.

After Witten's work \cite{Witten:1986bf}, 
mathematicians began to look for a spectrum-level understanding of the Witten genus \eqref{eq:WitGenus}. 
\tightfootnote{There exists an independent geometric construction of the Witten genus due to K. Costello \cite{Costello:WGPart1,Costello:WGPart2} based on factorization algebras \cite{Costello2016,Costello2021}. In this construction, the Witten genus of a complex manifold $X$ arises from the quantization of a commutative factorization algebra associated to a classical field theory with fields given by maps from a Riemann surface to the total space of the cotangent bundle $T^*X$.}
The task is analogous to starting with the observation that the genus $M \mapsto \int_M \widehat{A}(TM)$ is an integer-valued bordism invariant of spin manifolds, and deducing the existence of $\mathsf{BU}$ and $K$-theory. We can introduce a spectrum $\mathsf{M}{\rm String}$ of string manifolds so that $\mathsf{M}{\rm String}_{\bullet} = \pi_{\bullet}(\mathsf{M}{\rm String}) = \Omega_{\bullet}^{{\rm String}}({\rm pt})$.
\tightfootnote{See \autoref{subsec:Spectra} for an introduction to spectra. The spectrum corresponding to cobordisms of manifolds equipped with some structure $\CB$ is called the \emph{Thom spectrum} $\mathsf{M}\CB$ \cite{Atiyah1961,Adams:1974,FreedBordism}. The string spectrum $\mathsf{M}{\rm String}$ is a special case of the Thom spectrum for manifolds admitting a string structure $\CB = {\rm String}$.}
M. Ando, M. Hopkins, C. Rezk, and N. Strickland \cite{Ando2001,AndoHopkinsRezk2010} showed that the Witten genus can be realized as a morphism of commutative ring spectra,
\begin{align}
    \sigma : \mathsf{M}{\rm String} \to \mathsf{tmf}:= \mathsf{tmf}({\rm pt}) ~, \label{eq:WitGenusSpectra}
\end{align}
where $\sigma$ is surjective on homotopy in non-negative degrees. Here, $\mathsf{tmf}$ stands for Topological Modular Forms, a generalized cohomology theory. Physically, the existence of a spectrum version \eqref{eq:WitGenusSpectra} of the Witten genus implies the following: an $\nu$-dimensional string manifold $M$ determines a class $[M] \in \mathsf{tmf}_{\nu} := \pi_{\nu}(\mathsf{tmf})$, that depends only on the string bordism class of $M$ in $\Omega_{\nu}^{\rm String}$, and therefore, there is a homomorphism $\pi_{\nu}(\mathsf{M}{\rm String}) = \Omega_{\nu}^{\rm String}({\rm pt}) \to \pi_{\nu}(\mathsf{tmf}) = \mathsf{tmf}_{\nu}$, which is known to be surjective \cite[Thm. 6.25]{Hopkins2002}. 
\tightfootnote{In \cite{Hopkins2002}, surjectivity was stated but not proved, and in \cite{HopkinsMahowald2002}, a limited sketch of a proof was given. Surjectivity was proved recently in \cite{Devalapurkar:2019}.}

Thus we can reinterpret equation \eqref{eq:WitGenus} as a map, 
\begin{align}\begin{split}
    \varphi_{W} : \pi_{\nu}\mathsf{tmf} &\to \mathsf{MF}_{\nu/2} \\
    [M] &\mapsto \varphi_{W}([M]) = \eta(q)^{\nu} \mathsf{Z}_{\text{ell}}(\sigma_M; q) ~,
   \end{split}
\end{align}
and we restrict to $\nu$ even. The map $ \varphi_{W}$ is  neither injective nor surjective. (We will suppress `${\rm pt}$' for brevity as is conventional.)  The spectrum $\mathsf{tmf}$ owes its name to the fact that its ring of homotopy groups is rationally isomorphic to the ring of holomorphic  modular forms. The spectrum is \emph{connective}, which means $\pi_{<0}(\mathsf{tmf}) = 0$.

To go further we need to recall a bit more about the structure of the ring of integral holomorphic 
modular forms $\mathsf{MF}_{\bullet}$. For some references, see \cite{Serre1973,Zagier1991Utrecht,Koblitz1993,DiamondShurman:2005,Mumford2006-fe,Bruinier:2008}. Any modular form in $\mathsf{MF}_{\bullet}$
can be written as a power series in a formal variable $q = e^{2\pi\imag\tau}$ with $\IZ$-valued coefficients, leading to the familiar \emph{$q$-expansion}. This defines an injective map from the ring of modular forms to the ring of formal power series in $q$ over $\IZ$:
\begin{align}
   \mathsf{q}_{\text{exp}}: \mathsf{MF}_{\bullet} \rightarrow \IZ[[q]] ~,  \label{eq:MF-q-expansion}
\end{align}
but this map is far from surjective.
\tightfootnote{An more intrinsic definition of the map \eqref{eq:MF-q-expansion} is obtained by evaluation on the Tate curve, which is the generalized elliptic curve over $\IZ[[q]]$.}
The ring of modular forms can be presented as 
\be 
 \mathsf{MF}_{\bullet} \cong  \IZ[c_4, c_6, \Delta]/(c_4^3 - c_6^2 - 1728\Delta) ~.
\ee
Here $c_4$ and $c_6$ denote the Eisenstein series (the subscript denotes the modular weight), and $\Delta$ denotes the modular discriminant, which equals $\Delta = (c_4^3 - c_6^2)/1728$.
The modular discriminant plays a central role in the theory of modular forms.

Returning now to the theory of elliptic cohomology, it can be shown that while $\Delta$ is not in the image of $\varphi_{W}$, the modular form $\Delta^{24}$ of weight $288 = \frac{1}{2}\cdot 576$  -- is in the image. Moreover, the kernel of $\varphi_{W}: \mathsf{tmf}_{576} \to \mathsf{MF}_{288}$ is trivial, so there is a unique element in $\mathsf{tmf}_{576}$ that maps to $\Delta^{24}$. It is conventional to refer to this element also as $\Delta^{24}$. We can define the non-connective spectrum:
\tightfootnote{There are three distinct versions of topological modular forms, namely, $\mathsf{tmf}$ (connective), $\mathsf{Tmf}$ (non-connective and non-periodic), and $\mathsf{TMF}$ (non-connective and periodic). The latter two can be defined more intrinsically in terms of sheaves of spectra over the moduli stack of elliptic curves or its Deligne-Mumford compactification. For details, see, for example, \cite{Douglas2014-cl}. In our discussion, we define $\mathsf{TMF}$ in terms of $\mathsf{tmf}$. However, one can alternatively start from $\mathsf{TMF}$ and define $\mathsf{Tmf}$ via $\mathsf{Tmf}[\Delta^{-1}] := \mathsf{TMF}$, and then $\mathsf{tmf} := \mathsf{Tmf}\langle 0 \rangle$ as the $0$-connected cover. Here the notation $E\langle n\rangle$ for a spectrum $E$  and an integer $n$ denotes the $n^{th}$ connective cover (a.k.a. Postnikov truncation) defined such that $\pi_{k}( X\langle n\rangle ) = 0$ for $k < n$ and for $k \geq n$, the map $X\langle n\rangle \to X$ induces an isomorphism on $\pi_k$.}
\begin{align}
    \mathsf{TMF}_{\bullet}:= \mathsf{TMF}_{\bullet}({\rm pt}) &:= \mathsf{tmf}[\Delta^{-24}]_{\bullet}({\rm pt}) ~,
\end{align}
which is now $576$-periodic. 
\tightfootnote{The physical ramifications of this periodicity are not fully understood. However, see \cite{Johnson-Freyd:2024rxr} for a recent interesting discussion.}
Morever, for $\mathsf{TMF}$, there is a map, 
\begin{align}
\varphi_{W} : \pi_{\bullet}\mathsf{TMF} &\rightarrow \mathsf{MF}_{\bullet/2}[\Delta^{-1}] := \IZ[c_4, c_6, \Delta, \Delta^{-1}]/(c_4^3 - c_6^2 - 1728\Delta) ~, \label{eq:WitGenusTMF}
\end{align}
to the ring of \emph{weakly holomorphic integral modular forms}.

As hinted in \autoref{sec:ExtendedTFT}, there is a physical connection to $\mathsf{KO}$ theory: take a 2d $\CN=(0,1)$ SQFT and compactify it on a circle -- this turns it into an SQM with time-reversal symmetry that is equivariant with respect to rotations around $\IS^1$. Such systems are classified by $\mathsf{KO}$ theory, as proved by Stolz and Teichner \cite{StolzTeichner1,StolzTeichner2}. The power of $q$ in the $q$-expansion of the partition function specifies the weight under this $\mathsf{U(1)} \cong \IS^1$ action \cite{Witten:1986bf,Tachikawa:2023nne}. The essence of this mathematical construction is captured by the following commutative diagram:
\tightfootnote{For further technical details of this commutative diagram, see, for example, \cite{Gukov:2018iiq,Tachikawa:2021mby,Berwick-Evans:2024nrr}. There are generalizations where one considers TMF classes not of a point but of a topological space $X$. Crucially, the commutative diagram as stated here is true in degrees $\nu = 0 \mathsf{~mod~}8$ and $\nu = 4 \mathsf{~mod~}8$. In other degrees, there are subtleties: for example, there can be 2-torsion from $\mathsf{KO}$ that is not present in $\mathsf{MF}$. In degree $\nu=4k-1$, there is a novel torsional invariant \cite{BUNKE2014912,Gaiotto:2019gef,Berwick-Evans:2023ejh} for $(4k-1)$-dimensional string manifolds in the image of the map $\Omega_{4k-1}^{{\rm String}}({\rm pt}) \to \IC((q))/(\IZ((q)) + \mathsf{MF}_{2k})$. This naturally modifies the diagram \eqref{eq:TMF-commutative-diag}. For a careful discussion, see \cite{Berwick-Evans:2023ejh}, and for some physical consequences, see \cite{Tachikawa:2023lwf}.}
\begin{equation}
 \begin{tikzcd}
    \Omega_{\nu}^{{\rm String}}({\rm pt}) \arrow[d] \arrow[r, "\text{Wit}_{\text{String}}"] & \mathsf{TMF}_{\nu}({\rm pt}) \arrow[r, "\varphi_W"] \arrow[d] & \mathsf{MF}[\Delta^{-1}]_{\nu/2} \arrow{d}{\text{$q$-expansion}}[swap]{\mathsf{q}_{\text{exp}}} \\
    \Omega_{\nu}^{{\rm Spin}}({\rm pt}) \arrow[r, "\text{Wit}_{\text{Spin}}"] & \mathsf{KO}((q))_{\nu}({\rm pt}) \arrow[r] & \IZ((q))
 \end{tikzcd}\label{eq:TMF-commutative-diag}
\end{equation}
The vertical map $\Omega_{\nu}^{{\rm String}}({\rm pt}) \to \Omega_{\nu}^{{\rm Spin}}({\rm pt})$ is induced by the natural forgetful map from the spectrum of string manifolds to the spectrum of spin manifolds, given by forgetting the string structure. Witten's construction realized the \underline{composite map} $ \varphi_{W} \circ \text{Wit}_{\text{String}}$,  
%
%
and Witten's map factors through the $\mathsf{TMF}$ spectrum.

An important development that connects TMF to 2d $\CN=(0,1)$ SQFTs is the Segal-Stolz-Teichner conjecture, which is based on work by Stolz and Teichner \cite{StolzTeichner1,StolzTeichner2}
%
%
building upon earlier work of Segal \cite{Segal88,Segal2007}. As we have noted, the space of 2d $\CN=(0,1)$ SQFTs is graded by  $\nu \in \IZ$ and we denote the graded component by $\CS_{-\nu}$. The Segal-Stolz-Teichner conjecture states that there is an isomorphism between deformation classes of 2d $\CN=(0,1)$ SQFTs and TMF classes. Specifically, the conjecture says that deformation classes of families of 2d $\CN=(0,1)$ SQFTs with a fixed anomaly coefficient $\nu \in \IZ$ which are parametrized by a topological space $X$ are in one-to-one correspondence with classes in $\mathsf{TMF}^{-\nu}(X)$. Physically, this means that the deformation class of a theory in $\CS_{-\nu}$ parametrized by $X$ is a cocycle representative for a class in $\mathsf{TMF}^{-\nu}(X)$, an Abelian group. 
\tightfootnote{It is worth noting that for the classifying space of a family of QFTs to form a spectrum in the first place, one needs extra structure which may not generically exist, namely, a suspension isomorphism. Moreover, for the genus associated with a QFT to be well-defined as a map out of a cobordism spectrum, the partition function of the QFT must be a cobordism invariant. V.S. thanks D. Berwick-Evans for discussions on this point.}
Mathematically, this means $[X, S_{-\nu}] \cong \mathsf{TMF}^{-\nu}(X)$, where $[X, S_{-\nu}]$ denotes the set of homotopy classes of maps from $X$ to the space $S_{-\nu}$. In the simple case when $X$ is a point (which we restricted to in the preceding paragraphs), this conjectured isomorphism reads $[{\rm pt}, \CS_{-\nu}] = \pi_{0}(\CS_{-\nu}) \cong \mathsf{TMF}^{-\nu}({\rm pt})$. So in this case, TMF classes probe the path connected component of the space of theories. 

Formally, TMF can be viewed as a categorification of $K$-theory, measuring a $2$-category of vector bundles. $K$-theory studies a space $X$ by looking at vector bundles over $X$. One can equip any such vector bundle with a connection, which is required to discuss parallel transport over a curve in $X$. Categorifying this setup, one would get ``2-vector bundles'' over $X$, involving a study of parallel transport over 2d surfaces in $X$. Intuitively, one could think of an ``elliptic object'' as a 2d CFT where the surfaces are mapped into $X$. For such a perspective, see \cite{BassDundasRognes}. It would be interesting to investigate the physical implications of this viewpoint.

For some recent work on elliptic cohomology/topological modular forms and applications to quantum field theory and string theory, see \cite{BunkeNaumann,Berwick-Evans:2013fua,Berwick-Evans:2014oja,Kitchloo:2014tqa,Berwick-Evans:2021jlr,Gukov:2018iiq,Berwick-Evans:2018ryn,Gaiotto:2018ypj,Gaiotto:2019asa,Gaiotto:2019gef,Berwick-Evans:2020niz,GepnerMeier:2020,Johnson-Freyd:2020itv,Lin:2021bcp,Berwick-Evans:2023ejh,Yonekura:2022reu,Tachikawa:2021mby,Tachikawa:2021mvw,Albert:2022gcs,Tachikawa:2023nne,Tachikawa:2023lwf,Johnson-Freyd:2024rxr,Tachikawa:2024ucm,Saxena:2024eil,Debray:2024czl,Lin:2024qqk,Tachikawa:2025flw,Tachikawa:2025awi}. The subject of equivariant topological modular forms is under rapid development and it is likely to have important implications for string theory. 

\SectionWithHeader{Other Applications I: Chern-Simons And WZW Terms}{Other Applications I: Chern-Simons And WZW Terms}{sec:WZW-CS-Term}  

Chern-Simons and WZW terms are terms that can be added to the actions of physical theories, the addition of which changes the theory in profound ways. From the mathematical viewpoint, they are particularly important examples of invertible field theories. They are also closely related, and both are nicely described in the context of differential cohomology. 

We will be following \cite{Amabel:2021wbk,Debray:2023kvh}. 
Related, but slightly different discussions of these 
topological terms from a modern viewpoint can be found in 
\cite{Davighi:2020vcm,Lee:2020ojw}.

\subsection{Chern-Simons Terms}\label{subsec:CS-TERMS}

We continue the discussion from \autoref{subsec:Chern-Simons-Theory} above, where we presented the standard discussion of 3-dimensional Chern-Simons terms one typically finds in the physics literature. As noted there, the discussion has limitations. 
A better definition -- that applies to all groups -- makes use of 
differential refinements of Chern-Weil theory. 

\subsubsection{Chern-Weil Theory} 

Let $G$ be a compact Lie group and $\wp$ an invariant polynomial 
of degree $\ell$ on the Lie algebra $\mathfrak{g}$ of $G$. 
\tightfootnote{
It might help some readers to think of $\wp(x) = \Tr \rho(x)^k$, where $x\in \mathfrak{g}$ and $\rho$ is a finite-dimensional representation of $\mathfrak{g}$. Henceforth, when we use this example, we will 
just write $\Tr x^k$ to keep the notation simple.}
Suppose $P\to M_{n}$ is a principal $G$-bundle with connection 
$\n$. The curvature is $F(\n) \in \Omega^2(M_{n}; \mathsf{ad~}P)$,  
so $\wp\big(F(\n)\big) \in \Omega^{2\ell}(M_{n})$ is a globally 
well-defined $2\ell$-form on $M_{n}$. The Bianchi identity shows 
that this form is closed. One then naturally wonders how this 
form depends on the choice of connection $\n$.
We will show, presently, that when we deform $\n$, the form 
  $\wp\big(F(\n)\big)$ changes by a globally defined exact 
form, and therefore, the de Rham cohomology class $[\wp\big(F(\n)\big)]$ 
only depends on the topology of the bundle $P$. Here is the proof: 

  Suppose $\n_{0,1}$ are two 
connections on $P$. Then, since the space of connections on $P$ 
is an affine space, we have $\n_1 = \n_0 + \alpha$ 
where $\alpha \in \Omega^1(M_{n}; \mathsf{ad~}P)$.  We can form the path of connections, 
\be 
\n_t := (1-t) \n_0 + t \n_1 = \n_0 + t \alpha ~.
\ee
Let $F_t := F(\n_t)$. Then one easily computes:
\be 
\frac{d}{dt} F_t = D_{A_0} \alpha + 2 t \alpha^2 = D_{A_t} \alpha ~,
\ee
where for a gauge field $A$, $D_A\alpha = d\alpha + A\alpha + \alpha A$ is the covariant derivative with respect $A$,  
and we write $\n_t = d+ A_t $.   It follows that,  
\be 
\frac{d}{dt} \wp(F_t)   = d \omega( \alpha, F_t) ~,
\ee
where $\omega( \alpha, F_t) $ is a \underline{globally well-defined} $2k-1$ form on $M$. For example, if $\wp(x) = \Tr x^k$, then,  
\be 
\frac{d}{dt} \tr F_t^k    = d \left( k \Tr \alpha F_t^{k-1} \right)  ~.
\ee

Let $G$ be a compact and connected Lie group. 
 A theorem of Borel \cite{Borel:1955,Dupont:1978} identifies the invariant polynomials 
on $\mathfrak{g}$ with the real cohomology of $G$, 
\tightfootnote{The proof proceeds by identifying $H^{\bullet}(BG;\IR) \cong H^{\bullet}(BT;\IR)^W$, where $W$ is the action of the Weyl group. The latter is identified with the ring of invariant polynomials where the generators of $H^{\bullet}(BT;\IR)$ have degree $2$. Note that a corollary is that the odd degree cohomology of $BG$ is torsion.  }
\be\label{eq:BorelTheorem}
I(\mathfrak{g}) \cong H^{\bullet}(BG; \IR) ~.
\ee
Here $I(\mathfrak{g}) = S^{\bullet}(\mathfrak{g}^\vee)^G$ is the ring 
of invariant polynomials under the adjoint action, and $\mathfrak{g}^{\vee}$ denotes the dual Lie algebra of $\mathfrak{g}$. It can be identified with the ring 
of Weyl-invariant polynomials on a Cartan subalgebra
$S^{\bullet}(\mathfrak{t}^\vee)^W$. For a simple Lie group, it is a 
polynomial ring in the $r$ independent Casimirs, where $r$ is 
the rank of $G$. 
\tightfootnote{When $G$ is not connected, we make use of the  exact sequence \eqref{eq:conn-comp-sec}
relating $G$ to the connected component of the identity and the group of components  $\pi_0$. 
One has a fibration 
$BG_0 \to BG \to B\pi_0$ from which one can relate the cohomology of $BG$ to that of $BG_0$ by a spectral sequence. There is an action of $\pi_0$ on $H^*(BG_0)$. To define it, for every element 
$\gamma \in \pi_0$, we choose a lift $\widetilde \gamma$ to $G$, and then consider the conjugation action 
of $\widetilde \gamma$ on $G_0$. This induces an action on the cohomology that does not depend on the choice of lift. Then, $H^\bullet(BG)$ is the $\pi_0$-invariant subring of $H^\bullet(BG_0)$. As an example, consider the group $G=\widetilde U(r+1)$ of unitary and anti-unitary operators on $\IC^{r+1}$. Then $\widetilde U(r+1)$ has two connected components, and $G_0 = U(r+1)$. The cohomology $H^\bullet(BG_0;\IR)$ is generated by invariant polynomials $\wp_1, \dots , \wp_{r}$, where $\wp_k(x) = \Tr(x^k)$. The action of the nontrivial connected component is complex conjugation, so the cohomology $H^\bullet(BG;\IR)$ is the polynomial subring generated by $\wp_2, \wp_4, \cdots$.}

Under this correspondence, if $c\in H^{2k}(BG;\IR)$ corresponds to 
$\wp$ and $\ov{\gamma}: M_{n} \to BG$ is the classifying map for $P$, then, 
\be 
\ov{\gamma}^*(c) = [\wp\big(F(\n)\big)] ~.
\ee
Now if $c$ is normalized so that it is the image of an integral 
class, then $\wp\big(F(\n)\big)$ will represent an integral cohomology 
class. 

\bigskip 
\noindent \textbf{Example:} Let $G$ be a compact, connected, simply connected, simple Lie group. Then, 
\be\label{eq:BasicDegTwoPoly}
\wp(x) = \frac{1}{16\pi^2 h^\vee} \mathsf{Tr}_{\mathfrak{g}} x^2 ~,
\ee
where $h^\vee$ is the dual Coxeter number, then
the corresponding $c$ is the image of a generator of 
$H^4(BG;\IZ)$ embedded into $H^4(BG;\IR)$. See \cite{Bott1956,MimuraToda1991}.

Specializing to $\mathsf{\mathsf{SU}(N)}$,  
this implies that, 
\be 
\frac{1}{8\pi^2} \mathsf{Tr}_{N} F^2 ~,
\ee
represents the pullback of a generator of $H^4(B\mathsf{SU}(N); \IZ)$. 
For the generalization to quotients of simply connected semisimple compact groups, see \cite{Toda1976} and \cite[App. A]{Monnier:2017oqd}.

\begin{remark} It is possible to extend Chern-Weil theory to include noncompact Lie groups and even some infinite-dimensional Lie groups, but this is beyond the scope of these notes. 
\end{remark}

\subsubsection{Relative Chern-Simons And Absolute Chern-Simons Forms And The Chern-Simons Invariant  }

We now pull $(P,\n_t)$ back to $M_{n}\times [0,1]$ with $t\in [0,1]$, so that now the covariant derivative is $\n_t + dt \frac{d}{dt}$, and the fieldstrength is $F(\n_t) + dt \wedge \alpha$, 
and consider the integral,
\be\label{eq:Rel-CS}
\mathsf{CS}_{\wp}(\n_2, \n_1) := \int_{0}^1 \wp\big(F(\n_t)\big) dt ~.
\ee
This a \underline{globally well-defined} $2k-1$ form on $M$. 
It is called the \emph{relative Chern-Simons form}  and satisfies: 
\be 
d_M \mathsf{CS}_{\wp}(\n_2, \n_1) = \wp(F_2) - \wp(F_1) ~.
\ee

\begin{exbox}

Suppose we can express $\wp_{k}(x) = \mathsf{Tr}(x^{k})$, 
and denote the corresponding relative Chern-Simons 
form by $\mathsf{CS}_{2k-1}$. Show that, 
\be 
\mathsf{CS}_{2k-1}( \n + \alpha, \n) = 
k \int_0^1 dt\,\mathsf{Tr\,} \alpha \left( 
F+ t D\alpha + t^2 \alpha^2)^{n-1} \right) ~,
\ee
and in particular: 
\be 
\mathsf{CS}_1( \n + \alpha,\n) = 
\mathsf{Tr}\left( \alpha \right) ~,
\ee
\be 
\mathsf{CS}_3( \n + \alpha,\n) = 
\mathsf{Tr}\left( 2\alpha F + \alpha D \alpha + \frac{2}{3} \alpha^3 \right) ~,
\ee
\be 
\mathsf{CS}_5(\n + \alpha,\n) = 
\mathsf{Tr}\left( 3\alpha F^2 + \frac{3}{2} \alpha \{F, D\alpha \}  +  F \alpha^3 + \alpha (D \alpha)^2 + 
\frac{3}{4} \alpha^3 D \alpha + \frac{3}{5} \alpha^5
 \right) ~,
\ee
where $F= F(\n)$. 
    
\end{exbox} 

We would now like to define a Chern-Simons form and invariant for a \underline{single} connection on a principal $G$-bundle $\pi: P \to M_{n}$. Here, we follow \cite[App. A]{Freed:2008jq}. 
Recall that for any map $f:N\to M_{n}$, we define the pullback bundle $f^*(P)\to N$ as a fiber product: 
\be 
f^*(P) = \{ (n,p) \in N \times P ~\vert~ f(n) = \pi(p)\} ~,
\ee
then $\pi': f^*(P) \to N$ given by projection on the first 
factor is a principal $G$-bundle with right $G$-action
$(n,p)\cdot g  := (n, p\cdot g)$. Apply this construction 
to the case where $N=P$ and $f=\pi$: 
\be 
\pi^*(P) = \{ (p_1, p_2) \in P \times P ~\vert~ \pi(p_1) = \pi(p_2) \} ~,
\ee
and observe that there is a canonical globally defined 
section $s: P \to \pi^*(P)$, given by $s(p) = (p,p)$. 
Therefore, the principal $G$-bundle $\pi': \pi^*(P) \to P$ is 
trivializable. Moreover, there is a canonical connection 
$\n_0$ on $\pi^*(P)$. Suppose $\gamma:[0,1] \to P$ is 
a curve, and let $p_0 = \gamma(0)$. Choose any point 
$s(p_0)\cdot g_0$ in the fiber above $\gamma(0)$. Then,  
the connection $\n_0$ is defined by declaring that the 
lifted path with initial point $s(p_0)\cdot g_0$ is simply 
$s(\gamma(t))\cdot g_0$. There is, rather trivially, no holonomy, and hence the connection is trivial. We now 
define the Chern-Simons form of a connection $\n$ 
on $\pi: P \to M_{n}$ by, 
\be\label{eq:CS-form} 
\mathsf{CS}_{\wp}(\n) := \mathsf{CS}_{\wp}(\pi^*(\n) , \n_0) ~.
\ee
If $\wp$ has degree $k$, this is a \underline{globally defined} $(2k-1)$-form on $P$. We stress that it is well-defined on $P$, and not on the base space $M_{n}$. Indeed, in the  special case where we consider the group $G$ itself to be a principal $G$-bundle over a point, there is a canonical connection, the \emph{Maurer-Cartan form}: $\theta_{\mathsf{MC}} = g^{-1} dg$, and we get a globally well-defined form $\mathsf{CS}_{\wp}(\theta_{\mathsf{MC}})$ on $G$ which is, in general, nonzero. 

\begin{exbox}

\begin{enumerate}
\item[(1)] Show that 
\be\label{eq:ChernWeil-Transgression}
d \left( \mathsf{CS}_{\wp}(\n) \right)  = \pi^*(\wp(F)) ~.
\ee

\item[(2)] Derive $CS_{\wp}(\n)$ for the case of a trivial bundle
in the case $\wp(x) = \Tr(x^2)$ as follows: Let the globally defined 
connection form $\Theta$ on $P = M_{n} \times G$ be:
\be 
\Theta = g_1^{-1} dg_1 + g_1^{-1} A g_1 ~,
\ee
where $g_1\in G$ and  $A\in \Omega^1(M;\mathfrak{g})$ is globally well-defined. 

\begin{itemize} 
\item[(2a)] Show that $\pi^*(P)$ is naturally identified as the space of triples, 
\be 
(m,g_1,  g_2) \in M \times G \times G ~,
\ee
and the connection 
$\n_0$ has connection form $g_2^{-1}  d g_2$ on $\pi^*(P)$. 

\item[(2b)] Show that $\pi^*(\n)$ has 
connection from $g_2^{-1}  d g_2 +  g_2^{-1} \Theta   g_2$ 
on $\pi^*(P)$.

\item[(2c)] Show that, 
\be\label{eq:ExplicitDeg3-CS-Form}
\mathsf{CS}_{\wp}(\n) = - \frac{1}{3} \Tr (g_1^{-1}dg_1)^3 
+ \Tr(A dA + \frac{2}{3}A^3 )  - d\left( \Tr(dg_1 g_1^{-1} A\right) ~.
\ee
\end{itemize}

\item[(3)] Now take $M={\rm pt}$ in the previous exercise, and show that the 
globally well-defined connection form on $\pi^*(P)$ is, 
\be 
\wt{\Theta} = g_2^{-1} dg_2 + g_2^{-1}(g_1^{-1} dg_1) g_2 = 
(g_1 g_2)^{-1} d(g_1 g_2) ~.
\ee
Moreover, show that,
\be\label{eq:PolyakovWiegmann}
\Tr(\wt{\Theta})^3 = \Tr (g_1^{-1} dg_1)^3 + \Tr (g_2^{-1} dg_2) 
+ d\left( 3 \Tr (dg_2 g_2^{-1} )(g_1^{-1} dg_1) \right) ~.
\ee
Equation \eqref{eq:PolyakovWiegmann} is known as the 
\emph{Polyakov-Wiegmann} formula in the physics literature and is 
extremely useful when working with the WZW model. 

\end{enumerate}
    
\end{exbox}

We are now ready to define the Chern-Simons \underline{invariant} 
of a connection $\n$ on $\pi: P \to M_{n}$. Suppose first that $P$ is 
trivializable. Then there is a globally well-defined section
$s: M_{n} \to P$, and we can consider the well-defined integral: 
\be\label{eq:IntegratePulBack}
\int_M s^*(\mathsf{CS}_{\wp}(\n)) ~.
\ee
Two different sections will differ by a gauge transformation: 
$s_1(x) = s_2(x) g(x)$, for some function $g:M_{n} \to G$, and this will 
change the value of the integral. (See the exercise below.) 
However, if $\wp$ is normalized 
so that $\mathsf{CS}_{\wp}(\theta_{\mathsf{MC}})$ has integral periods then 
the value of \eqref{eq:IntegratePulBack} only changes by an integer under change of section. Therefore, we can define the Chern-Simons 
invariant to be: 
\be 
S_{\mathsf{CS},\wp}(\n) := \int_{M_{n}} s^*(\mathsf{CS}_{\wp}(\n)) ~ \mathsf{mod~} \IZ  ~ . 
\ee
As the notation suggests, we can consider $S_{\mathsf{CS},\wp}$ to be an action, and the exponentiated action, 
\be
\exp \left[ 2\pi \imag  S_{\mathsf{CS},\wp}(\n) \right] ~,
\ee
 is well-defined. Of course, if $\wp$ has degree $k$, 
this will only be interesting on manifolds of dimension $n=2k-1$. 

\begin{exbox}

\begin{enumerate}
\item[(1)] 
In the example where $\wp(x) = \Tr x^2$ 
and $P= M_{3}\times G$,  is the trivial bundle a natural section is 
$s(x) = (x,g_0)$ for some fixed $g_0$. Using 
\eqref{eq:ExplicitDeg3-CS-Form}   recover the 
standard expression: 
\be 
S_{\mathsf{CS}}(A) = \int_{M_{3}} \Tr \left(A dA + \frac{2}{3} A^3 \right) ~,
\ee
in the case where $A\in \Omega^1(M_{3}; \mathfrak{g})$ is globally 
well-defined.  

\item[(2)] Show that a general change of section corresponds to 
$\wt s(x) = (x, \bar g(x) g_0)$  where $\bar g: M_{3} \to G$
defines a gauge transformation. 
Using \eqref{eq:ExplicitDeg3-CS-Form} and the 
\eqref{eq:PolyakovWiegmann}, show that 
\be 
\int_{M_{3}} \wt s^*(\mathsf{CS}_{\wp}(\n)) = 
\int_{M_{3}} s^*(\mathsf{CS}_{\wp}(\n))  - \int_{M_{3}} \frac{1}{3} \Tr( \bar g(x)^{-1}  d \bar g(x))^3 ~.
\ee
If $G$ is compact, connected, simple  and simply connected, we normalize $\Tr$ according to \eqref{eq:BasicDegTwoPoly}
then $-\frac{1}{3} \Tr \theta_{\mathsf{MC}}^3$ represents an integral generator of $H^3(G;\IZ)$. In particular, for $\mathsf{\mathsf{SU}(N)}$, a 
normalized Maurer-Cartan form that represents a generator is 
\be\label{eq:SUN-MC-Generator}
\frac{1}{24\pi^2} \Tr_{N} (g^{-1} dg)^3 ~.
\ee
\end{enumerate}
\end{exbox}

When the bundle $\pi: P \to M_{n}$ is nontrivial, the definition of the Chern-Simons invariant is more elaborate. Again, we follow 
\cite[App. A]{Freed:2008jq}. Suppose $M_{n}$ is of dimension $n = 2k-1$.  Let $\ov{\gamma}: M_{n} \to BG$ be a classifying map,  and $\gamma:P \to EG$ be a $G$-equivariant lift. 
Since the odd degree cohomology of $BG$ is torsion, the image 
$\ov{\gamma}(M_{n}) \subset BG$ represents a torsion homology class. So there exists an integer $N$ and a $2k$-chain $W\subset BG$ so that, 
\be 
N \ov{\gamma}(M_{n}) = \partial W ~,
\ee
then, $[\frac{1}{N} W] \in H_{2k}(BG; \IR/\IZ)$. Now, 
choose any connection $\n^{u}$ on $EG \to BG$, and normalize 
$\wp$ so that $\wp(F(\n^{u}))$ has integral periods on $BG$. 
Then there exists an integral class $\lambda\in H^{2k}(BG;\IZ)$ so that $\wp(F(\n^{u}))$ is an integral representative of $\lambda$. In general, there can be many choices of $\lambda$, and the Chern-Simons invariant \underline{will} depend on the choice of $\lambda$. Then, we define, 
\be\label{eq:NontrivBunCSInvt}
S_{\mathsf{CS},\wp,\lambda}(\n): = \frac{1}{N} \int_{W} \wp(F(\n^u)) 
+ \int_{M_{n}} \mathsf{CS}_{\wp}(\gamma^*(\n^u), \n) + \lambda\left(\bigg[\frac{1}{N}W\bigg]\right) ~ \mathsf{~mod~} \IZ ~.
\ee
One can show that the right hand side actually does not 
depend on the choices $N,W,\gamma,\n^{u}$ that we made in the above construction. 
Comparing this formula with equation  \eqref{eq:ostarchar} 
above suggests that the Chern-Simons form and invariant 
can be formulated in terms of differential cohomology. 
This is indeed the case, as we discuss next. 
 
\subsubsection{Differential Cohomology Refinement Of Chern-Weil Classes  } 

The Chern-Simons invariant can be nicely formulated in terms of differential cohomology. See \cite{Amabel:2021wbk,Debray:2023kvh} and references therein for details. Here we will be brief.

Suppose we are have a principal $G$-bundle with connection 
$\pi: P \to M_{n}$ with connection $\n$. As we have seen, 
given a suitably normalized degree $k$ invariant polynomial $\wp \in I(\mathfrak{g})$, we have, 
\be 
\wp(F(\n)) \in \Omega^{2k}_{\IZ}(M_{n}) ~.  
\ee
Moreover, there exists an integral class $\lambda \in H^{2k}(BG;\IZ)$ such that, if $\gamma:M_{n} \to BG$ is a classifying map for $P$, then $\wp(F(\n))$ is a representative in real cohomology of $\gamma^*(\lambda) \in H^{2k}(M_{n};\IZ)$:
\be\label{eq:ClassesAgree2}
[ \wp(F(\n))] = \gamma^*(\lambda)_{\IR} ~.
\ee
A comparison of \eqref{eq:ClassesAgree2} with \eqref{eq:ClassesAgree}  raises the natural 
question of whether there is  a differential cohomology class $\chi_{\n,\lambda}\in \widecheck{H}^{2k}(M_{n})$, such that,
\be
F(\chi_{\n,\lambda}) = \wp(F(\n)) ~, 
\qquad\qquad \text{and}  \qquad \qquad  c(\chi_{\n,\lambda}) = \gamma^*(\lambda) ~.
\ee
Theorem 2.2 in the original paper of Cheeger and Simons 
\cite{CheegerSimons:1985} states that indeed such a 
differential class exists, and moreover, if it is gauge-invariant, then it is unique. 
   
In more modern terms, there is a space $B_{\n} G$ analogous 
to $BG$ which classifies principal $G$-bundles with connection. (See Chapter 13 of \cite{Amabel:2021wbk} for a 
description of $B_{\n}G$.) Then, given 
$\lambda \in H^{2k}(BG;\IZ)$ and $\wp\in I(\mathfrak{g})$, 
such that, 
\be 
\lambda_{\IR} = [ \wp(F(\n^u)] \in H^{2k}(BG;\IR) ~,
\ee
(where $\n^{u}$ is any connection on $\pi:EG \to BG$),  
there is a \underline{unique} differential cohomology class, 
\be 
\check{\lambda} \in \widecheck{H}^{2k}(B_{\n}G) ~,
\ee
such that if the map $\gamma: M \to B_{\n}G$ classifies 
$(P,\n)$, then $\gamma^*(\check{\lambda}) = \chi_{\n, \lambda}$.  See   \cite{Debray:2023kvh}, and Theorem 14.1.1 and Section 19.1 of \cite{Amabel:2021wbk} for more details. 
 
The Chern-Simons invariant of a connection $\n$ on $\pi: P\to M_{n}$ can now be described very nicely in terms of differential cohomology, without the need to extend to a bounding manifold. (We remind the reader that, in general, such extensions do not exist.) If $M_{n}$ has dimension $n=2k-1$, then, 
\be 
\int^{\widecheck{H}}_{M_{n}} \chi_{\n, \lambda} \in \widecheck{H}^1({\rm pt}) \cong \IR/\IZ' ~,
\ee
is the Chern-Simons invariant. We note that, if $M_{n}$ has dimension $n=2k$, then, 
\be 
\int^{\widecheck{H}}_{M_{n}} \chi_{\n, \lambda} = \langle [M_{n}], \lambda \rangle \in \IZ ~,  
\ee
and, importantly, if we have a family  $\CX \to S$ of manifolds diffeomorphic to $M_{n}$, of dimension $2k$, then, 
\be\label{eq:Chern-Simons-LineBundle}
\int^{\widecheck{H}}_{\CX/S} \chi_{\n, \lambda} \in \widecheck{H}^2(S) ~.
\ee
Recall that $\widecheck{H}^2(S)$ is the group of isomorphism classes of line bundles with connection.  
This fact is quite relevant to the study of anomalies. 

As a particularly important case, suppose we have a family of 
2-manifolds: $M_2 \to \CX \to S$. (The base space could be, for example, the space of connections on $M_2$.) Then we obtain 
a bundle with connection as in \eqref{eq:Chern-Simons-LineBundle}. 
This line bundle provides meaning to the value of the Chern-Simons 
invariant on a 3-manifold with boundary. Suppose that $\partial M_3 = M_2$. We wish to make sense of $\exp[ 2\pi \imag  \int_{M_3} \mathsf{CS}_3(\n)]$ 
as a ``function'' of the connection $\n$ on the boundary. 
Consider the set $\CE(\n)$ of all extensions $(P^{\mathsf{ext}}, \n^{\mathsf{ext}})$ of the gauge field $\n_{M_2}$ on $M_2$ to a gauge field   $\n^{\mathsf{ext}}$ on some principal bundle $P\to M_3$ on   some bounding manifold $M_3$. Consider the 
set $\CL_{\n_{M_2}}$  of functions $F: \CE(\n) \to \mathsf{U(1)}$  such that, for two different 
extensions $(P^{\mathsf{ext}}, \n^{\mathsf{ext}})$ and  $(P^{\mathsf{ext}}, \n^{\mathsf{ext}})'$, we 
have, 
\be 
\frac{F((P^{\mathsf{ext}}, \n^{\mathsf{ext}}))}{F((P^{\mathsf{ext}}, \n^{\mathsf{ext}})')}
= \exp[ 2\pi \imag  S_{\mathsf{CS}}((P^{\mathsf{ext}}, \n^{\mathsf{ext}}))\cup_{M_2} (P^{\mathsf{ext}}, \n^{\mathsf{ext}})'] ~,
\ee
where on the RHS, we have the Chern-Simons invariant of the connection on the closed manifold obtained by gluing the two extensions. 
Observe that $\CL_{\n_{M_2}}$ is in fact a $\mathsf{U(1)}$-torsor. These 
$\mathsf{U(1)}$-torsors patch together to define a principal $\mathsf{U(1)}$-bundle 
over the set of connections ${\n_{M_2}}$ on $M_2$. If we are given a 
specific family of extensions $(P^{\mathsf{ext}}, \n^{\mathsf{ext}})$, we obtain (locally)  a section of that line bundle. The line bundle descends to the gauge equivalence classes of connections on $M_2$ and is nontrivial there, reflecting the fact that quantum Chern-Simons theory on a manifold with boundary is anomalous.

\subsection{WZW Terms}\label{subsec:WZW-TERMS}

The Wess-Zumino term was originally introduced into the low-energy 
pion effective action of QCD to account for the flavor group anomalies. It was derived by J. Wess and B. Zumino \cite{Wess:1971yu}.
  
The effective theory of pions is a special case of 
what is known as the principal chiral model. To define this model 
consider a sigma model in $2n$ dimensions with target space $G$, where $G$ is a Lie group. The action for a 
field $\phi: M_{2n} \to G$ is,  (before the addition of the WZ term), 
\be\label{eq:PCP}
S_{\mathsf{pcp}}[\phi]: = \int_{M_{2n} }  f^2 \Tr\phi^*(\theta_{\mathsf{MC}}) \star \phi^*(\theta_{\mathsf{MC}}) ~,
\ee
where $f^2$ is a constant (dimensionful except in two dimensions) 
and $\Tr(x^2)$ is an invariant degree 2 polynomial on $\mathfrak{g}$. 
Since the Maurer-Cartan form is usually written as  $\theta_{\mathsf{MC}}= g^{-1} dg$ for 
$g\in G$, if we instead write $g:M_{2n} \to G$ as a field, then the action takes a form 
that is commonly written in the physics literature:
\be\label{eq:PCP2}
S_{\mathsf{pcp}}[\phi]: = \int_{M_{2n} }  f^2 \Tr g^{-1} dg \wedge \star   g^{-1} dg ~.
\ee
This sigma model is often referred to as the \emph{principal chiral model}. When coupled to external gauge fields for the global $G\times G$ symmetry, the model is not anomalous, and therefore cannot be an accurate LEET for QCD, which has $G\times G$ anomalies. Wess and Zumino remedied this defect by adding their famous term $\Gamma$  to the action \eqref{eq:PCP}.

A beautiful re-interpretation of $\Gamma$ was given 
by Witten in \cite{Witten:1983tw,Witten:1983ar}. Moreover, 
Witten showed that the addition of the WZ term 
to the action of the principal chiral model in both two and 
four dimensions   has dramatic physical 
consequences \cite{Witten:1983tw,Witten:1983tx,Witten:1983ar}.
A related construction was given by Novikov in \cite{Novikov:1982ei} (see p. 40 et. seq.), so the model is sometimes referred to as the 
WZNW model.

The description of Witten  \cite{Witten:1983tw,Witten:1983ar} and Novikov \cite{Novikov:1982ei}  proceeds along 
the following lines. Choose a cohomology class $\Theta_{2n+1}$  of degree $2n+1$ on $G$, and suppose that $\Theta_{2n+1}$ has integral periods. Let $\phi: M_{2n} \to G$ be the 
sigma model map. Suppose that the image $\phi(M_{2n})$ is homologous 
to zero in $G$, so that there is a bounding chain: 
$\phi(M_{2n}) = \partial W_{2n+1}$. Then 
Witten's definition of the WZ term is: 
\be\label{eq:WZ-WittenDef}
\mathsf{WZ}(\phi) = 2\pi \imag  \int_{W_{2n+1}} \Theta_{2n+1} ~.
\ee
As usual, there can be multiple ways of bounding the image 
$\phi(M_{2n})$ so the action \eqref{eq:WZ-WittenDef} is 
ambiguous. When $\Theta_{2n+1}$ is suitably normalized,  
the exponentiated action will be a well-defined functional of $\phi$.  While the action appears to be an action in $2n+1$ dimensions, under a small variation of $\phi$, the WZ term 
changes by an integral of a local density on $M_{2n}$. 

As mentioned, the addition of these topological terms to the two- and four-dimensional sigma model with target a Lie group $G$ has dramatic physical consequences. In the case of two-dimensions, when the kinetic and topological terms are suitably normalized one obtains a two-dimensional conformal field theory -- the renowned WZW conformal field theory \cite{Witten:1983ar}. For textbook treatments, see \cite{Fuchs:1992nq,DiFrancesco:1997nk}.

In four dimensions, the skyrmion field configuration can (with suitable normalization of the WZ term) obey fermionic statistics
\cite{Witten:1983tx}.

It turns out that Witten's definition of the WZ term \eqref{eq:WZ-WittenDef} is 
problematic because there are examples where the image of spacetime under $\phi$ is not a boundary.
\tightfootnote{For the 2d WZ term, Witten only considers the spacetime $\IS^2$ and 
uses the fact that $\pi_2(G)=0$ for all compact groups. However, 
we most definitely would want to consider other spacetimes, and in any case, the relevant topological invariants are the homology, not the homotopy groups of $G$.}
For a rather trivial example, consider the case where $M_2$ is a torus 
and the group $G$ is a torus of dimension greater than or equal to $2$, and the image of $\phi$ is a nontrivial subtorus of $G$. For this reason, it is desirable to have a definition 
of the WZ term that applies to all finite dimensional Lie groups. 

The two-dimensional WZ term for a general group is provided by a 
gerbe connection $\widecheck{H} \in \widecheck{H}^3(G)$. The WZ term 
for $\phi: M_2 \to G$ is then the holonomy of the gerbe around 
the image $\phi(M_2) \subset G$. For example, if $c\in H^3(G; \IZ)$, and $\wp$ is such that, 
\be 
[\wp(\theta_{\mathsf{MC}})] = c_{\IR} ~,
\ee
there is  a differential character $\widecheck{H} \in \widecheck{H}^3(G)$ that can be used to define the WZ term. If $G$ is a compact connected simply connected Lie group, then $\widecheck{H}$ is unique. We can prove this by showing that the group of flat fields $H^2(G; \IR/\IZ)$ vanishes. To see this group vanishes, use the universal coefficient theorem \eqref{eq:UCT-Cohomology} and the fact that 
$H_1(G; \IZ)=0$ (since $G$ is simply connected) and $H_2(G;\IZ) = 0$ (by the Hurewicz theorem 
\tightfootnote{The Hurewicz theorem says that for a path-connected space $X$ and for $n \in \IZ_{>0}$, there is a group homomorphism $h_{n}: \pi_{n}(X) \to H_{n}(X)$ -- called the Hurewicz homomorphism -- which is defined as follows: one picks a canonical generator of $H_{n}(\IS^n) \cong \IZ$, call it $x_n$. A based map $f: \IS^{n} \to X$ induces a homomorphism $f_{*}: H_{n}(\IS^n) \to H_{n}(X)$. Denoting by $[f]$ the homotopy class of $f$ in $\pi_{n}(X)$, one defines $h_{n}([f]) = f_{*}(x_n) \in H_{n}(X)$. In general, $h_{n}$ is neither injective nor surjective. For $n \geq 2$, if $\pi_{i}(X) = 0$ for all $i < n$ (we say $X$ is $(n-1)$-connected), then the Hurewicz homomorphism becomes an isomorphism. In the present case, since $G$ is connected and simply connected, it follows that $\pi_2(G) \cong H_2(G;\IZ)$.}
combined with the fact that $\pi_2(G)=0$).

The WZ and Chern-Simons topological terms in actions are closely related. Mathematically, the relation involves lift of the  \emph{transgression map} on cohomology to differential cohomology. 

We first recall the transgression map on cohomology. This is a map 
\be 
\tau: H^{k}(BG;\IZ) \to H^{k-1}(G;\IZ) ~,
\ee
defined for a compact group \cite{Borel1967}.

It is defined   as follows: Suppose $\lambda \in H^{2k}(BG;\IZ)$. 
We choose a representative $c$ of $\lambda$,  $[c] = \lambda$ and pull  it back under $\pi: EG \to BG$.
Since $EG$ is contractible, $\pi^*(c)$ must be trivializable, 
so there exists a cochain $c'$ on $EG$ with $c = \delta c' $. 
Now, because $BG$ has a natural basepoint, there is a natural fiber 
inclusion $\iota: G \to EG$, the fiber over the basedpoint. 
Then $\iota^*(c) =0$, and hence, $\iota^*(c')$ is a closed integral 
class on $G$. The transgression map is defined to be: 
\be 
\tau(\lambda) = [\iota^*(c')] ~.
\ee
It can be shown to be independent of the choices we have made, and is hence a well-defined map. It is not a group homomorphism. 

Now, equation \eqref{eq:ChernWeil-Transgression} above strongly hints that there should be a differential refinement of the transgression map, and this indeed proves to be the case. For a careful treatment in degree three, see \cite{MR2174418}, which refines the transgression map to a map,
\be 
\widecheck{\tau}:  \widecheck{H}^4(B_{\n}G ) \to \widecheck{H}^3(G; \IZ) ~.
\ee
This can be interpreted as a map relating Chern-Simons in 3-dimensions and Wess-Zumino terms in 2-dimensions. It is a version 
of anomaly inflow, and was used to explain the  Chern-Simons/RCFT correspondence in \cite{Moore:1989vd,Moore:1989yh,Elitzur:1989nr}.

Moving on to four-dimensions, we encounter an interesting puzzle. 
As noted above, for suitable normalization of the WZ term, the skyrmion of the principal chiral model obeys fermionic statistics. 
This suggests that the principal chiral model with such a WZ term should be a spin theory, but it is not evident how the spin structure is used to construct the WZ term. A resolution was provided by 
D. Freed in  \cite{Freed:2006mx}, which constructed the term using 
generalized differential cohomology based on a cohomology theory 
(E-theory) that refines singular cohomology. For a related discussion, see  \cite{Lee:2020ojw}.

\SectionWithHeader{Other Applications II: Quantization Of $\mathsf{BF}$ Theory}{Other Applications II: Quantization Of $\mathsf{BF}$ Theory}{sec:QuantBF-Theory}

An important example of a topological field theory is ``$\mathsf{BF}$ theory.'' The study of this theory goes back at least to 
\cite{Schwarz:1979ae,Horowitz:1989ng}.
We will consider the theory where the fields are formulated using 
differential cohomology, as introduced in \cite{Maldacena:2001ss}. 
\tightfootnote{In \cite{Schwarz:1979ae,Horowitz:1989ng,Blau:1989bq,Birmingham:1991ty}, no quantization of the periods of the fieldstrengths was imposed, resulting in rather different answers from those discussed here. In \cite{Witten:1998wy}, a special case was studied and the quantization of the periods played an important role in Witten's discussion.}

The $\mathsf{BF}$ theory can be defined in $n$ spacetime dimensions with fields whose gauge equivalence classes are:
\tightfootnote{In this section, the periods of fieldstrengths are valued in $\IZ$ and not $2\pi\IZ$, hence the use of the notation $\sfF$ as opposed to $F$. See footnote \ref{foot:period-convention-change}.}
\be 
[\check \sfB_{n-p-1}] \in \widecheck{H}^{n-p}(M_n) ~, \qquad\qquad 
[\check \sfA_{p}] \in \widecheck{H}^{p+1}(M_n) ~.
\ee
To formulate the action, we need to introduce a nonzero integer $k$, and an orientation $\mathfrak{o}$. Then, we have, 
\be\label{eq:BF-action-DifflCoho}
S = 2\pi \imag  k \langle [\check \sfB_{n-p-1}], [\check \sfA_{p}] \rangle ~.
\ee
Our fields are normalized so that the periods of the fieldstrengths are integers.
In the case of topologically trivial fields, we can write,
\begin{align}
[\check \sfB_{n-p-1}] &= \chi_{\sfB_{n-p-1}} ~,
\qquad\qquad 
[\check \sfA_{p}] = \chi_{\sfA_p} ~.
\end{align}
For globally defined fields $\sfB_{n-p-1} \in \Omega^{n-p-1}(M_n)$ 
and $\sfA_p \in \Omega^{p}(M_n)$,  the action simplifies to: 
\be\label{eq:BF-action-TopTriv}
S = 2\pi \imag  k \int_{M_n} \sfB_{n-p-1} d \sfA_p ~.
\ee
We denote the resulting field theory by $T_{\mathsf{BF}}^{n,p,k,\mathfrak{o}}$.
A change of orientation is equivalent to a flip of the sign of $k$, so, without loss of generality, we can take $k$ to be positive. 
Moreover, the pairing on differential cohomology classes is 
graded symmetric: 
\be 
\langle [\check \sfB_{n-p-1}], [\check \sfA_{p}] \rangle
= (-1)^{n(p+1)} \langle [\check \sfA_{p}],  [\check \sfB_{n-p-1}]\rangle ~.
\ee
The exchange of $\sfA$ and $\sfB$ is a form of   electric-magnetic duality, and so we expect, at least 
naively, that the statement of electric-magnetic duality for these theories is: 
\be\label{eq:Naive-EM-Duality} 
T_{\mathsf{BF}}^{n,p,k,\mathfrak{o}} =  
\begin{cases}
\big(T_{\mathsf{BF}}^{n,n-p-1,k,\mathfrak{o}}\big) &  \text{ for } n(p+1) = 0 \mathsf{~mod~} 2 ~, \\
\big(T_{\mathsf{BF}}^{n,n-p-1,k,\mathfrak{o}}\big)^* & \text{ for } n(p+1) = 1 \mathsf{~mod~} 2 ~. \\
\end{cases}
\ee

\begin{remark} The fields can be twisted to take values in flat bundles based on representations $\rho_1, \rho_2$ of the fundamental group, and indeed this generalization was studied in \cite{Schwarz:1979ae}. 
In particular, if we twist one of the fields by the orientation line, then we can extend the domain of definition from oriented manifolds to unoriented manifolds. 
\end{remark}

The quantization of the theories was studied in 
\cite{Schwarz:1979ae,Horowitz:1989ng,Blau:1989bq,Birmingham:1991ty,Witten:1998wy,Maldacena:2001ss,Szabo:2012hc,Brennan:2023mmt} and in many other works. Despite all this work, as far as we are aware, the theory has not been constructed as a fully local topological field theory in complete detail.
\tightfootnote{Of course, full locality would follow once one establishes the equivalence with the homotopy sigma model $\sigma^{(n)}_{B^p \IZ_k, \CC}$, but as we will describe below, there are some gaps in the literature establishing that equivalence.} 
Doing so is highly desirable and should be quite achievable given our current understanding of field theory -- it is ``just'' free field theory! We briefly comment on some highlights of the quantization. 

We remark that the path integral of the   fields $[\check \sfA_p]$ with the action \eqref{eq:BF-action-DifflCoho} shows that the path integral is localized on fields $[\check \sfB_{n-p-1}]$ which are $k$-torsion in differential cohomology.  Similarly, the integral over $[\check \sfB_{n-p-1}]$ shows that the path integral is localized 
on fields $[\check \sfA_p]$ which are $k$-torsion. Since the fields are $k$-torsion, they are certainly flat: 
\be 
\sfF(\check \sfA) =0 ~, \qquad\qquad  \sfF(\check \sfB) = 0 ~,
\ee
(as follows easily from \eqref{eq:BF-action-TopTriv}) but the extended defects and operators defined by the holonomies of the fields are $k$-torsion. For topologically trivial field configurations, these holonomies are written as integrals over cycles as: 
\be
T\big([\Sigma_{n-p-1}]\big) := 
\exp\left[ 2\pi \imag  \int_{\Sigma_{n-p-1}} \sfB_{n-p-1}  \right] ~,
\ee
and 
\be
W\big([\Sigma_{p}]\big) :=
\exp\bigg[ 2\pi \imag  \int_{\Sigma_{p}} \sfA_p \bigg] ~.
\ee
Since the path integral localizes to flat fields, the operators $T$ and $W$ only depend on homology classes. In particular: 
\be 
T\big([\Sigma^1] + [\Sigma^2]\big) = T\big(\Sigma^1\big) T\big(\Sigma^2\big) ~,
\ee
just like the classical character. A similar formula holds for $W$. Moreover, since the path integral localizes to differential cohomology classes which are $k$-torsion, we have, 
\be\label{eq:N-torsion}
\left( T\big([\Sigma_{n-p-1}]\big) \right)^k = 1 ~,
\qquad \qquad 
\left( W\big([\Sigma_{p}]\big) \right)^k = 1 ~.
\ee
An elementary Gaussian integral computation shows that the path integral with defects inserted satisfies: 
\be\label{eq:LinkingNumberCorrelator}
\left\langle  T\big([\Sigma_{n-p-1}]\big) W\big([\Sigma_{p}]\big) \right\rangle 
= \mathsf{Z}\,\exp\left[ \frac{2\pi \imag  }{k} L\big([\Sigma_{n-p-1}], [\Sigma_p]\big) \right] ~,
\ee
where $\mathsf{Z}$ is the partition function without insertions, and $L$ is the linking number of the oriented homology cycles. 
This is certainly consistent with \eqref{eq:N-torsion}, and moreover shows that the \underline{operators} on the Hilbert space 
$\CH(N_{n-1})$, where $N_{n-1}$ is a spatial manifold, must satisfy the basic relations defining a Heisenberg extension of: 
\be 
H_{n-p-1}(N_{n-1}; \IZ/k\IZ) \times H_{p}(N_{n-1}; \IZ/k\IZ) ~,
\ee
namely, 
\be 
 T\big([\Sigma_{n-p-1}]\big)
W\big([\Sigma_{p}]\big) = \exp\left[{ \frac{2\pi \imag}{k}  I\big([\Sigma_{n-p-1}], [\Sigma_{p}]\big) }\right] 
W\big([\Sigma_{p}]\big) T\big([\Sigma_{n-p-1}]\big) ~,
\ee
where $I$ is the oriented intersection number. Call this Heisenberg group $\CH_{n,p,k}$. 

In the quantization of the $\mathsf{BF}$ theory, we note that the action 
\eqref{eq:BF-action-TopTriv} is first order in derivatives and should be viewed as an action on a phase space. The action is degenerate, and there will be ghosts-for-ghosts. One can quantize, and then impose Gauss law constraints, or impose constraints and then quantize. Proceeding with the latter method to formulate the Hilbert space, we impose the Gauss law in steps. Let $M_{n} = N_{n-1} \times \IR$, let $t$ be a time coordinate, and let $\sfA_0 \in \Omega^{p}(N_{n-1}\times \IR)$ be proportional to $dt$. A shift of $\check \sfA$ by   $\chi_{\sfA_0}$ changes the action as $S \mapsto S + 2\pi \imag k \int_{N_{n-1} \times \IR} \sfF(\check \sfB) \sfA_0$. Integration over $\sfA_0$ localizes the path integral on the flat fields 
$\sfF(\check \sfB)=0$.  Similar remarks apply with $\sfA$ and $\sfB$ exchanged. The result is that we have a phase space of flat fields, 
\be\label{eq:FlatFieldPhaseSpace}
H^{n-p-1}(N_{n-1}; \mathsf{U(1)}) \times H^{p}(N_{n-1}; \mathsf{U(1)}) ~,
\ee
equipped with a prequantum line bundle
\tightfootnote{Given a symplectic manifold $X$ and a closed sympletic form $\omega$ with integer periods, a prequantum line bundle on $X$ with connection is a line bundle $L \to X$ with connection $\n$ such that $\omega = F_{\n}$ is the curvature 2-form of the connection $\n$.} 
which comes from the integration: 
\be
\int^{\widecheck{H}}_{N_{n-1}}: \widecheck{H}^{n-p} \times\widecheck{H}^{p+1} \to \widecheck{H}^2 ~ . 
\ee
(Recall that $\widecheck{H}^2$ describes gauge equivalences classes of line bundles with connection.) The phase space 
\eqref{eq:FlatFieldPhaseSpace} has many components, indexed by 
\be 
\Tors(H^{n-p}(N_{n-1}; \IZ)) \times \Tors(H^{p+1}(N_{n-1}; \IZ) ) ~,
\ee
but the automorphism group $\Aut(\widecheck{H}^{n-p}(N_{n-1})) \cong 
H^{n-p-2}(N_{n-1}; \mathsf{U(1)})$ acts nontrivially on the prequantum line bundle via the pairing with $\Tors\big(H^{p+1}(N_{n-1}; \IZ)\big)$, and hence, the support of 
the wavefunction will  be concentrated in the topologically trivial component of $H^p(N_{n-1}; \mathsf{U(1)})$.
\tightfootnote{G.M. thanks E. Witten for a useful discussion on this point.} 
Similarly, the automorphisms of $\sfA$ act to concentrate the wavefunction on the topologically trivial component of $H^{n-p-1}(N_{n-1}; \mathsf{U(1)})$. 
\tightfootnote{There is a subtle point here. In the closely related case of the partition function of the self-dual field, it turns out that the wavefunction in general is \underline{not} supported on the topologically trivial component of the group of flat fields. This was first pointed out in \cite[Sec. 5.1]{Witten:1999vg}. See also \cite[Sec. 5.2]{Belov:2006jd} for further explanation. }
The quantization of the trivial component gives an irreducible representation of the Heisenberg group $\CH_{n,p,k}$. A Stone-von Neumann representation is provided by the vector space of  complex-valued functions on the cohomology group:  
\be
\CH(N_{n-1}) = \mathsf{Fun}( H^p(N_{n-1}; \IZ_k) \to \IC) ~.
\ee
Comparison with equation \eqref{eq:FinHomThy-StateSpace} raises the issue whether $T_{\mathsf{BF}}^{n,p,k,\mathfrak{o}}$ is related to the finite homotopy field theory $\sigma^{(n)}_{\CX, \lambda, \CC}$ studied in \autoref{sec:FinHomTheory}, for the case $\CX = B^p A = K(A,p)$, with $A = \IZ/k\IZ$. We discuss to what extent that might be the case. 

It is certainly true that the $\mathsf{BF}$ theory $T_{\mathsf{BF}}^{n,p,k,\mathfrak{o}}$
is closely related to a $\IZ/k\IZ$ $p$-form gauge theory. Indeed, in 
\cite{Maldacena:2001ss}, that relation was observed by noting that 
a Higgs mechanism breaking a $\mathsf{U(1)}$ gauge group to $\IZ/k\IZ$ has an IR description given by the $\mathsf{BF}$ theory \eqref{eq:BF-action-DifflCoho}. 
Nevertheless, the precise relation between $T_{\mathsf{BF}}^{n,p,k,\mathfrak{o}}$
and $\sigma^{(n)}_{B^p\IZ_k, \lambda, \CC}$ has not been spelled out in 
the literature. Roughly speaking, they should be ``essentially the same,'' 
up to an invertible TQFT. Here are a few issues which should be clarified
to make this completely precise:  

\begin{enumerate}

\item First of all, the domain and codomain of the theories should be 
equal if we are to compare them.  Let us begin with the domain. 
The domain of $T_{\mathsf{BF}}^{n,p,k,\mathfrak{o}}$ consists of the 
oriented bordism category, although if we twist one of the fields by 
the orientation line bundle, then the domain can be extended to the unoriented 
bordism category. On the other hand, $\sigma^{(n)}_{B^p\IZ_k, \lambda, \CC}$ 
is defined on the unoriented bordism category for $\lambda=1$. For nontrivial Dijkgraaf-Witten cocycles, it is only defined on the oriented bordism category.

\item The theory $T_{\mathsf{BF}}^{n,p,k,\mathfrak{o}}$ does not make use of a 
Dijkgraaf-Witten cocycle $\lambda$. A continuum field theory 
analog of the Dijkgraaf-Witten cocycle has been proposed in 
\cite{Kapustin:2014gua} in three dimensions. As far as we know, the generalization 
to all $(n,p)$ has not appeared.   Without such an additional term, the comparison with   $\sigma^{(n)}_{B^p\IZ_k, \lambda, \CC}$ should be carried out with a trivial Dijkgraaf-Witten cocycle. The simplest comparison, then, is that of  $T_{\mathsf{BF}}^{n,p,k,\mathfrak{o}}$ with   $\sigma^{(n)}_{B^p\IZ_k, \CC}$, where the domain is the oriented bordism category. 

\item Let us now consider the codomain. As we have discussed at length, the choice of the $n$-category $\CC$ is an essential choice of data in defining the theory $\sigma^{(n)}_{B^p\IZ_k, \CC}$. While we may require 
$\Omega^n \CC = \IC$ and $\Omega^{n-1} \CC = \mathsf{VECT}$, different choices of $\CC$ at the next level produce different theories. The corresponding choice is not evident in the definition of the theory $T^{n,p,k,\mathfrak{o}}_{\mathsf{BF}}$, although, given the relevance of higher categories to defects, it is natural to expect that such a choice should become apparent with careful studies of defects and defects within defects.

\item Turning now to the partition function, the partition function of $T^{n,p,k, \mathfrak{o}}_{\mathsf{BF}}$ on a compact $n$-manifold $M_n$ takes the form, 
\be\label{eq:PartFun-TdpN-1}
T^{n,p,k, \mathfrak{o}}_{\mathsf{BF}}(M_n)  = \CN_{n,p,k} \times \Theta_{n,p,k} ~.
\ee
Here $\Theta_{n,p,k}$ is a ``sum over the classical solutions.''  The counting 
of the classical solutions turns out to be extremely subtle. One
interpretation of how to do this leads to precisely the expression, 
%
%
%
%
%
%
%
\be\label{eq:PartFun-TdpN-2}
\begin{split} 
\Theta_{n,p,k}
&  := 
\vert H^p(M_n; \IZ/k \IZ ) \vert \cdot \vert H^{p-1}(M_n; \IZ/k \IZ ) \vert^{-1} \cdots \\
& = \prod_{j=0}^p \vert H^{p-j}(M_n; \IZ_k) \vert^{(-1)^j} ~.
\end{split}
\ee
We thank Mike Hopkins for essential discussions on this point. 
\tightfootnote{A naive way to perform the path integral with action \eqref{eq:BF-action-DifflCoho} is to say that 
\be 
\int_{\widecheck{H}^{n-p} \times \widecheck{H}^{p+1} } d\mu(\check{A}) d\mu(\check{B}) e^{2\pi \imag k \langle [\check B], [\check A]\rangle} = 
\int_{\widecheck{H}^{p+1}} d\mu(\check{A}) \delta (N [\check A]) = \# \ker\left( \times k: \widecheck{H}^{p+1} \to \widecheck{H}^{p+1}\right) ~,
\ee
and therefore say the path integral is the number of $k$-torsion points in $\widecheck{H}^{p+1}(M_n)$. While this gives the desired triviality for $k=1$ in general it gives the   \underline{wrong} answer. The quantity $\# \ker\left( \times k: \widecheck{H}^{p+1} \to \widecheck{H}^{p+1}\right)$ is, in general, 
\underline{not} equal to $\Theta_{n,p,k}$. One need look no further than spheres to find counterexamples. }
In this way, the ``classical'' contribution recovers the expression \eqref{eq:KAp-PF} 
appearing in the partition function of the theory $\sigma^{(n)}_{B^p\IZ_k, \CC}$. 
The one-loop term $\CN_{n,p,k}$ is closely related to the Ray-Singer analytic torsion  \cite{Schwarz:1979ae,Birmingham:1991ty,Blau:2022odi}.
The contributions of the nonzeromodes of the fields (and ghosts and ghosts-for-ghosts) 
produces $\tau'_{\mathsf{RS}}(M_n)$ defined in \eqref{eq:AnalyticTorsion}. The contribution of the zeromodes must be dictated by the desired physical properties of the answer, and is part of the definition of the theory. We conjecture, based on electromagnetic duality of the generalized Maxwell field, that the treatment which defines a fully local field theory has,
\be\label{eq:BF-CN}
\CN_{n,p,k} =\left(  \bigotimes_{j=0}^n  \left( \omega^{(j)}_{1} \wedge \cdots \wedge \omega^{(j)}_{b_j}\right)^{(-1)^j}  \cdot  \mathsf{Alt}(T)\right)^{(-1)^p} ~,
\ee
where, 
\be 
\mathsf{Alt}(T)  = \prod_{j=0}^n T_j^{(-1)^j} ~.
\ee
Recall that $T_j$ is the order of the torsion subgroup of $H^j(M_n; \IZ)$. 
As we have discussed above in \eqref{eq:RS-Torsion-LineValued}, there is no canonical orientation of the cohomology, so the RHS of \eqref{eq:BF-CN} is defined up to sign, 
and $\CN$ should be interpreted as a norm on a determinant line. Taking into account 
equation \eqref{eq:Volum-As-Ratio}, we understand that we are interpreting the ghost zeromodes to be in $H^j(M_n; \mathsf{U(1)})$, rather than in $H^j(M_n; \IZ) \otimes \IR/\IZ$. The partition function is therefore: 
\be\label{eq:BF-PF-Alt}
  \left( \tau_{\mathsf{RS}}(M_n)\mathsf{Alt}(T)\right)^{(-1)^p} \Theta_{n,p,N} ~.
\ee
As shown in \autoref{App:TorsionAndTorsion} by Dan Freed, when the Reidemeister torsion is suitably interpreted we have $\tau_{\rm Reidemeister} \mathsf{Alt}(T)=1$. 
On the other hand, the Cheeger-M\"uller theorem identifies $\tau_{\rm Reidemeister}$ with 
$\tau_{\mathsf{RS}}$. 
In this way we can align the partition function of the $\mathsf{BF}$ theory with that of the 
finite homotopy sigma model.

\item Note that for $k=1$, the homotopy sigma model 
is trivial. The fact that $\mathsf{BF}$ theory with fields in differential cohomology should be trivial for $k=1$ is a corollary of the claim of \cite{Maldacena:2001ss}. An important 
special case of this statement in the case $n=3$ and $p=1$ was independently argued 
by E. Witten   in \cite{Witten:2003ya}. In that case the $\mathsf{BF}$ theory is an Abelian Chern-Simons theory, and the triviality of the $k=1$ theory  played an important role in 
Witten's $\mathsf{SL}(2,\IZ)$ action on 3d superconformal theories.
%
%
%

%
%
 
\item The statements of electric-magnetic duality of $\mathsf{BF}$ theory 
(equation \eqref{eq:Naive-EM-Duality}  above) 
and finite homotopy sigma model (recall equation \eqref{eq:FinHomType-EM-Duality}) are slightly different. 

\item Turning now to defects, we recall that some discrepancies in the standard descriptions of defects in the two theories were already noted in \autoref{subsec:DefectsFiniteHomotopyTheory}.

\item \emph{Relation to the lattice description.}  There   
is a ``lattice gauge theory'' version of $\mathsf{BF}$ theory 
introduced by Kapustin and Seiberg in \cite{Kapustin:2014gua} and 
studied further in \cite{Gorantla:2021svj,Choi:2021kmx,Kapustin:2025nju,Kapustin:2025rhp}.
In particular, \cite[App. B]{Gaiotto:2014kfa} or \cite[App. A]{Choi:2021kmx} derives the partition function \eqref{eq:KAp-PF} from the lattice version. 
In principle, ghosts should be included in the derivation, although a careful treatment of ghosts in this context   has not appeared in the literature. 
\tightfootnote{Thanks to the Cheeger-M\"uller theorem, the Ray-Singer torsion can be identified with the Reidemeister torsion \cite{Cheeger1977,Cheeger1979,Muller1978}. It would be quite interesting if the ghosts in the lattice $\mathsf{BF}$ theory led directly to Reidemeister torsion.}
It would be nice to extend the discussion 
to higher codimension manifolds. Granting that the lattice version of $\mathsf{BF}$ theory 
agrees with the finite homotopy sigma model, the above list of discrepancies then 
applies to the relation between the lattice and continuum formulations of $\mathsf{BF}$ theory.

\end{enumerate}

Of course, we expect that all the above discrepancies have satisfactory resolutions 
and that   
$T^{n,p,k,\mathfrak{o}}_{\mathsf{BF}}$ is ``essentially'' the same as the 
theory $\sigma^{(n)}_{B^p\IZ_k, \CC}$, up to tensor product with an 
invertible theory. Given this, a natural question occurs: Find a 
continuum field-theoretic model for the general finite homotopy 
theory $\sigma^{(n)}_{\CX, \lambda, \CC}$.  Some relevant 
remarks extending from the case of $\CX = B^pA$ to the case of a two-group 
can be found in \cite{Kapustin:2013uxa}.

\SectionWithHeader{Other Applications III:  The M-Theory Abelian Gauge Field}{Other Applications III:  The M-Theory Abelian Gauge Field}{sec:MTheory}

One very striking application of differential cohomology is to M-theory. We sketch that here in the briefest terms.  

   Eleven-dimensional supergravity, as formulated in \cite{Cremmer:1978km}, involves a Lorentzian 11-dimensional oriented spin manifold $M_{11}$. The bosonic fields in the supergravity multiplet consist of a metric $g_{\mu\nu} \in \Met(M_{11})$ and a ``3-form gauge potential,'' 
   often denoted by $\sfC$. After Wick rotation to Euclidean signature, and in the case that  $\sfC \in \Omega^{3}(M_{11})$ is globally defined, the part of the bosonic action involving the $\sfC$-field is 
   \tightfootnote{In this section, the periods of fieldstrengths are valued in $\IZ$ and not $2\pi\IZ$, hence we use $\sfC$ and $\sfG$ rather than $C$ and $G$. See footnote \ref{foot:period-convention-change}.}
  \be\label{eq:CJS-action}
     \exp\left( - \frac{\pi}{\ell^3} \int_{M_{11}}  \sfG \wedge \star \sfG + 2\pi \imag\int_{M_{11}}\left(\frac{1}{6}\sfC\wedge \sfG \wedge \sfG - \sfC I_{8}(g) \right) \right) ~,
  \ee
  where   $\sfG = d\sfC$, $\ell$ is the eleven-dimensional Planck length 
  \tightfootnote{Normalized so that the path integral weighting from the  Einstein-Hilbert term is: 
  $$
  \exp\left[ - \frac{2\pi}{\ell^9} \int \CR(g) \vol(g) \right]  ~  .
$$
} 
and $I_{8}(g)$ is the Chern-Weil representative
  \tightfootnote{\label{foot:lambda-and-nu}
  The term involving $I_8$ was first discussed by Vafa and Witten in \cite{Vafa:1995fj} and further studied in \cite{Duff:1995wd,Witten:1996md}. Explicitly, $I_8(g) = \frac{1}{(2\pi)^3 \cdot 4!}\left(\frac{1}{8}\text{tr } R^4 - \frac{1}{32}(\text{tr } R^2)^2\right)$ which can also be written as $I_8(g) = -\frac{\pi}{24}(p_2(M_{11}) - \lambda^2) = -\frac{2\pi}{6}\frac{1}{8}(p_2(M_{11}) - \lambda^2)$, where $\lambda = \frac{1}{2}p_{1}(M_{11})$ is an integral class and we are here using the canonical 
  Chern-Weil representatives defined by a Riemannian metric. $\lambda$ defines an integral 
  de Rham class which is a de Rham representative of an integral lift of the 4$^{th}$ Wu class $\nu_{4}(M_{11})$. Since $M_{11}$ is spin, $\nu_{4}(M_{11}) = w_{4}(M_{11})$ and furthermore its first Pontryagin class is canonically divisible by $2$. The canonical quotient is $\lambda$. See footnote \ref{foot:curv2-form} for a description of the curvature 2-form in local coordinates.} 
  of: 
  \be
   \frac{4 p_2 - p_1^2}{4\cdot 48} ~.
  \ee

Naively, one would expect the gauge-invariant information of the $\sfC$-field to be encoded in a holonomy function, 
\be\label{eq:C-field-holonomy}
\chi_{\sfC}: Z_3(M_{11}) \to \mathsf{U(1)} ~,
\ee
written heuristically as $\chi_{\sfC}(\Sigma) = e^{2\pi \imag \int_{\Sigma} \sfC}$. This is only 
heuristic because in general, there will be no globally well-defined $3$-form we can call $\sfC$. One would then expect $\sfG$ to be the fieldstrength such that, if $\Sigma = \partial W_4$ is the boundary of an integral $4$-chain, then, 
\be 
\chi_{\sfC}(\Sigma) = \exp\left[ 2\pi \imag \int_{W_{4}} \sfG \right] ~.
\ee
That is, one would expect that the set of gauge equivalence classes of M-theory $\sfC$-fields is the differential cohomology group $\widecheck{H}^4(M_{11})$. This still turns out to be too naive for the following reason: The physical meaning of the holonomy $\chi_{\sfC}$ is that it is a term in the action of an M2-brane, very much as in the discussion of \autoref{sec:ActionTestBranes} and near equation 
\eqref{eq:chi-general}. However, the worldvolume of the M2-brane contains fermions valued in $S(\Sigma) \times S(N\Sigma)^-$, where $S(\Sigma)$ is the spin bundle, $S(N\Sigma)$ is the spin bundle of the normal bundle   $N\Sigma$ of $\Sigma$ in $M_{11}$ and the 
superscript $-$ is a chirality projection. 
\tightfootnote{When we include the fermionic degrees of freedom, we should view eleven-dimensional supergravity as defined on a superspace of real dimension 
$(11\vert 32)$. The fermionic modes of the membrane theory are obtained from the restriction 
$S(TM_{11})\vert_{\Sigma}$ of the spin bundle of $M_{11}$ to the worldvolume $\Sigma$. There is a local supersymmetry known as $\kappa$-supersymmetry that makes one chirality pure gauge. See \cite{Siegel:1983hh,Siegel:1983ke,Bergshoeff:1987cm,Becker:1995kb} for details about $\kappa$-symmetry. }
The path integral of these fermions is a Pfaffian, which has a sign ambiguity. It is a section of a nontrivial (but two-torsion) real line bundle $\CL_{\mathsf{memb}}$ over the space of 3-dimensional membrane worldvolumes $\Sigma$ and metrics on $M_{11}$. Accordingly, $\chi_{\sfC}$ must likewise be twisted by the line bundle $\CL_{\mathsf{memb}}$ and hence if we try to write the $\sfC$-field holonomy as a function as in 
\eqref{eq:C-field-holonomy}, it will be ambiguous by a sign. This subtlety was pointed out by 
Witten in   \cite{Witten:1996md}. He showed that the topological class of $\CL_{\mathsf{memb}}$ is such 
that the fieldstrength must satisfy the following quantization law on its periods:
  \be \label{eq:G-minus-p1by4}
    \int_{\Sigma_4} \left( \sfG - \frac{1}{4}p_{1}(TM_{11})\right) \in \IZ ~.
  \ee
On a spin manifold, $p_1$ has a canonical division by $2$, so there exists a canonical 
integral class $\lambda\in H^4(M_{11}; \IZ)$ (see footnote \ref{foot:spin-mfd-lambda}), with $p_1 = 2\lambda$. Therefore, the 
periods of $\sfG$ can be half-integral. 

While $\chi_{\sfC}$ cannot be written as a character on $Z_{3}(M_{11})$, the 
\underline{ratio} of the holonomies for two $\sfC$-fields does make sense as a function: 
\be 
\frac{\chi_{\sfC_1}}{\chi_{\sfC_2}} : Z_3(M_{11}) \to \mathsf{U(1)} ~.
\ee
It follows that the gauge equivalence class of the $\sfC$-field, which we will write as 
$[\c \sfC]$, forms a \underline{torsor} for $\widecheck{H}^4(M_{11})$. 

If we want to pin down the $\sfC$-field more precisely, then it can be formulated as a differential chain that trivializes a degree 5 background magnetic current  $\widecheck j_m = \widecheck w_4(M_{11})$ in $\widecheck{H}^{5}(M_{11})$. In the Hopkins-Singer cochain model \eqref{eq:HS-Cochain}, we can represent $\widecheck w_4(M_{11})$ by the cocycle 
$(0,0,\half \lambda(g))$, where $\lambda(g)$ is the usual Chern-Weil representative of 
$\lambda = p_1/2$ defined by a Riemannian metric. 

Differential cohomology turns out to be very useful in making sense of the 
extremely subtle  ``topological term'' in the   action for the low-energy 
effective theory of M-theory: 
  \be\label{eq:11dSugraTopolTerm-1}
     \exp\left[ 2\pi \imag\int_{M_{11}}\left(\frac{1}{6}\sfC\wedge \sfG \wedge \sfG - \sfC I_{8}(g) \right) \right] ~,
  \ee
But before using differential cohomology, we note that if   $(M_{11},\sfC)$ admits an extension to a $12$-manifold $M_{12}$, then we can write \eqref{eq:11dSugraTopolTerm-1} as: 
  \be\label{eq:11dSugraTopolTerm}
     \exp\left[ 2\pi \imag\int_{M_{12}}\left(\frac{1}{6}\sfG\wedge \sfG \wedge \sfG - \sfG I_{8}(g) \right) \right] ~,
  \ee
As usual, to check that this gives a sensible definition of 
\eqref{eq:11dSugraTopolTerm} one compares two extensions, so the question of whether the term is well-defined comes down to whether \eqref{eq:11dSugraTopolTerm} is $+1$ on 
all closed $12$-dimensional manifolds equipped with a $\sfC$-field. 
This is far from obvious! Since $\sfG$ has half-integral periods  ,  $\exp[ 2\pi \imag \int_{M_{12}} \frac{\sfG^3}{6} ]$ can be a $48^{th}$ root of unity. Similarly, the term involving $I_8$ can be a  $96^{th}$ root of unity. In the case that $M_{12}$ is a 
spin manifold, Witten gave an ingenious argument to show that, actually, the expression 
takes values $\{ \pm 1\}$ \cite{Witten:1996md}. In the case that no spin manifold $M_{12}$ bounds $M_{11}$ but there exists a bounding  $\mathsf{Pin}^+$ manifold, Corollary 4.54 of \cite{Freed:2019sco} assures us that it still takes values in $\{ \pm 1 \}$. 
Taking values in $\{ \pm 1 \}$ is still not good enough to avert a catastrophe for M-theory, but it turns out that the Rarita-Schwinger partition function has a canceling sign ambiguity, so that the product can be well-defined. The first indication that this is so was given in \cite{Witten:1996md}. The argument  was subsequently refined in \cite[Sec. 2]{Diaconescu:2000wy} and in \cite{Freed:2004yc} for spin manifolds, 
 and was extended to the case of $\mathsf{Pin}^+$ manifolds in \cite{Freed:2019sco}.
 
We now explain Witten's ingenious argument alluded to above.    Due to some mathematical ``coincidences,'' first used in \cite{Witten:1985mj,Witten:1985bt,Witten:1996md}, 
  there is a remarkable relation of the $\sfC$-field 
  to $\sfE_8$ gauge theory. The first remarkable ``coincidence'' is the topological equivalence of $K(\IZ,4)$ 
  with $B\sfE_8$, up to the $15$-skeleton. As we have seen from equation 
  \eqref{eq:G-minus-p1by4} the periods of $\sfG$ form a torsor for $H^4(M_{11}; \IZ)$. 
  Recall that this cohomology group is just the group of homotopy classes of maps 
  of $M_{11}$ into $K(\IZ,4)$. Of course, the homotopy groups $\pi_k(K(\IZ,4))$ vanish 
  for all $k\not=4$, and $\pi_4(K(\IZ,4))\cong \IZ $. On the other hand, the low-dimension homotopy groups of 
  $\sfE_8$ are:
  \be 
  \pi_3(\sfE_8) \cong \IZ ~,
  \ee
  \be 
  \pi_k(\sfE_8) = 0 ~, \qquad 0 \leq k < 15 ~,  \qquad {\rm and} \qquad k \not=3 ~,
  \ee
  while $\pi_{15}(\sfE_8)\cong \IZ$. This means there is a homotopy equivalence of 
  $\sfE_8$ and $K(\IZ,3)$ up to the $14$-skeleton, and hence an equivalence of 
  $B\sfE_8$ and $K(\IZ,4)$ up to the $15$-skeleton. 
  \tightfootnote{Recall footnote \ref{foot:whitehead}. One must exhibit a map inducing the isomorphism of homotopy groups. This  gap can be filled in. See  Proposition 5.2.122 (and 5.1.41) of \cite{Schreiber:2013pra}, so in fact $B\sfE_8$ and $K(\IZ,4)$ are homotopic up to the  $15$-skeleton. }
  In the current application, this means that the topological class of a shift of an M-theory $\sfC$-field can be uniquely identified with an isomorphism class of an $\sfE_8$ bundle on $11$-manifolds, and even on \underline{families} of $11$ manifolds, provided the base of the family is at most $4$-dimensional.

The second remarkable coincidence is that the expression \eqref{eq:11dSugraTopolTerm} is closely related to the index density of a Dirac operator coupled to an $\sfE_8$ bundle with connection over  $M_{11}$! This extraordinary observation goes back to 
Witten \cite{Witten:1996md}. One version of Witten's observation can be explained by giving a groupoid model for the M-theory $\sfC$-field using $\sfE_8$ bundles with connection \cite{Diaconescu:2003bm}.  In this model, the objects (the ``$\sfC$-fields'') are triples $(P, \n, c)$, where $P \to M_{11}$ is a principal $\sfE_8$ bundle with connection $\n$ and $c \in \Omega^{3}(M_{11})$ is a globally defined $3$-form. The morphisms (``gauge transformations'') are given by  equivalences: 
  \be\label{eq:C-Field-Equivalence}
    (P, \n, c) \sim (P', \n', c') \quad \text{ if } \quad c' - c = \mathsf{CS}(\n, \n') ~,
  \ee
  along with gauge transformations like $c \mapsto c+ d\rho$, where $\rho $ is a globally well-defined $1$-form on $M_{11}$. Note that the fieldstrength,  
  \be
     \sfG = \tr \sfF^2 - \frac{1}{2}\tr R^2 + dc ~.
  \ee
is gauge-invariant under these equivalences. 
\tightfootnote{Note that in these terms the WZW term in the membrane action (i.e., the holonomy of the $\sfC$-field) is  $\exp[ 2\pi \imag  \int_{\Sigma} (  \mathsf{CS}(\n) - \half \mathsf{CS}(g) + c)] $. 
Once again, it is the division by $2$ in the Chern-Simons invariant for the gravitational field that leads to the interpretation of the holonomy as valued in $\CL_{\mathsf{memb}}$. }

Next,  let  $i (\slashed{D}_{V,\n})$ denote the index density of Atiyah and Singer for a 
Dirac operator coupled to a bundle $V$ with connection $\n$. Then we have the astonishing 
result that: 
  \be
    \left[ \frac{1}{2} i(\slashed{D}_{(\mathsf{ad\,}P,\n)}) + \frac{1}{4}i (\slashed{D}_{\mathsf{RS}})\right]^{(12)} = \frac{1}{6}\sfG^3 - \sfG I_8 + d(\iota_{\mathsf{local}}) ~,  \label{eq:E8indexdensity}  
  \ee
  where $\iota_{\mathsf{local}}$ is a globally defined $11$-form depending on 
  $\sfG$, $c$, and Chern-Weil representatives for $p_1(TM_{11})$ and $p_2(TM_{11})$.
  (The explicit formula is given below.) Finally, $\slashed{D}_{\mathsf{RS}}$ is the 
  index density for the Rarita-Schwinger (RS) operator. It is the Dirac operator coupled 
  to $T M_{11}-3 \CO$ in $11$-dimensions and $T M_{11}-4 \CO$ in $12$-dimensions, 
where $\CO$ is a trivial real line bundle. 
\tightfootnote{The Rarita-Schwinger field is more naturally valued in $T^*M_{11} \otimes S$, but the degree $12$ component of the index density does not change if we use $TM_{11}$ instead of $T^*M_{11}$, and the former is more convenient for some topological arguments. 
The subtraction by multiples of $\CO$ accounts for ghost fields. See Appendix A
of \cite{Freed:2004yc} for an attempt at a careful account. }

Using the Atiyah-Patodi-Singer (APS) index theorem, we can give an intrinsic 11d definition of the phase of the $\sfC$-field partition function: 
  \eqa{
    \Phi([\c \sfC]) &= \exp\left\{ 2\pi \imag  \left(\frac{\xi(\slashed{D}_{P,\n})}{2} + \frac{\xi(\slashed{D}_{\mathsf{RS}})}{4} \right)  + 2\pi \imag  I_{\mathsf{local}}\right\} ~,\\
    \xi(\slashed{D}) &:= \frac{1}{2}\left( \eta(\slashed{D}) + h(\slashed{D})\right) ~,\\
    I_{\mathsf{local}} &= \int \left( \frac{1}{2}c \sfG^2 - \frac{1}{2} c dc \sfG + \frac{1}{6}c(dc)^2 - c I_{8} \right) ~.
   }
Here $\eta(\slashed{D})$ is the Atiyah-Patodi-Singer invariant and $h(\slashed{D})$ is the 
dimension of the kernel of $\slashed{D}$. 
Note the strange fact that, although the $\sfC$-field is a bosonic field, this term in the 
action actually depends on the choice of a spin structure on $M_{11}$. 
The division of the RS density in the above formulae by four introduces an extremely subtle sign -- it is, once again,  the $\pm$ sign ambiguity of 
\eqref{eq:11dSugraTopolTerm}. Put differently, the topological term in the action 
for the $\sfC$-field is only defined as a section of a line bundle $\CL_{\sfC}$ with $\IZ_2$-holonomy. 
On the other hand, the Rarita-Schwinger partition function is also defined as a section of a line bundle $\CL_{\mathsf{RS}}$ which is 2-torsion. The product, 
  \eqa{
     \mathsf{Pfaff}(\slashed{D}_{\mathsf{RS}})\Phi(X) ~,\label{eq:Prod-RS-C}
  }
is therefore a section of a line bundle $\CL_{\mathsf{RS}} \otimes \CL_{\sfC}$. The line 
bundle is trivializable, meaning that there is a choice of nowhere-zero section $s$. 
Using that section, one can define an \underline{function} on the space of metrics and $\sfC$-fields modulo gauge equivalence. The function is just the ratio of 
\eqref{eq:Prod-RS-C} divided by $s$. The choice of section can be normalized and is 
known as a ``setting of the quantum integrand'' \cite{Freed:2004yc}. The theories 
defined by two different settings $s_1,s_2$ differ by an invertible TQFT whose 
partition function is $s_1/s_2$. As noted above, this was demonstrated, with increasing degree of rigor in \cite{Witten:1996md,Diaconescu:2000wy,Freed:2004yc}. The discussion was 
extended to $\mathsf{Pin}^+$ manifolds in \cite{Freed:2019sco}.

There are numerous murky points about the topology of M-theory which should be clarified. 
Six of them are listed below: 

\noindent 
 
\begin{enumerate}

\item \emph{Ten-dimensional $\sfE_8$ super-Yang-Mills (SYM) multiplets as ``edge modes.''} 
Note that the equivalence relation 
\eqref{eq:C-Field-Equivalence} means that we can shift the $\sfE_8$ gauge field $\n$ by 
an \underline{arbitrary} $\alpha \in \Omega^1(M_{11}; \mathsf{ad\,}P)$, so there is a ``topological symmetry'' of the type familiar from the Baulieu-Singer discussion of Donaldson-Witten theory: The dependence of physical quantities on the $\sfE_8$ gauge field only depends on the topology of $P$. Thanks to the work of Ho\v{r}ava and Witten  \cite{Horava:1995qa,Horava:1996ma}, we expect this to change dramatically in the presence of boundaries, where we expect to find propagating $\sfE_8$ gauge fields in a 10d $\sfE_8$ SYM  multiplet on each component of the boundary. 
According to \cite{Freed:2004yc}, anomalies will cancel on 11-manifolds with arbitrary numbers of boundary components with $\sfE_8$ SYM multiplets of either chirality. The argument is most easily given using the above $\sfE_8$-model for the $\sfC$ field, and suggests that the 
propagating $\sfE_8$ gauge fields on the boundary components should be regarded as ``edge modes'' of the theory with topological symmetry $\n \mapsto \n + \alpha $ in the bulk. 
This is still a murky point, and the nature of the edge mode picture should be clarified. 
\tightfootnote{Non-supersymmetric extensions of the Ho\v{r}ava-Witten setup have been studied. See, for example, \cite{Fabinger:2000jd,Horava:2007hg,Montero:2025ayi}.}

\item \emph{Cubic Refinements.}  A crucial aspect of the term $\Phi([\c \sfC])$ is that it is a \underline{cubic-refinement} of a trilinear form on $\widecheck{H}^4$ in $11$ dimensions. This is more or less obvious given the heuristic expression, 
\be 
\exp\left[ 2\pi \imag \left\{ \int_{M_{12}}\left( \frac{\sfG^3}{6} + \sfG I_8 \right)\right\} \right] ~ . 
\ee
More precisely,  it is shown in \cite{Diaconescu:2003bm} that for any 3 elements 
$\check a_i \in \widecheck{H}^4(M_{11})$, 
\be\label{eq:CubicRefinement}
\begin{split} 
\frac{
\Phi([\c \sfC] + \check a_1 + \check a_2 + \check a_3)
\Phi([\c \sfC] + \check a_1)  \Phi([\c \sfC] + \check a_2)\Phi([\c \sfC] + \check a_3) }{
\Phi([\c \sfC] + \check a_1+ \check a_2 )\Phi([\c \sfC] + \check a_2+ \check a_3 )\Phi([\c \sfC] + \ \check a_3+ \check a_1 )\Phi([\c \sfC] )} & = 
  \int^{\widecheck{H}}_{M_{11}} \check a_1 \odot \check a_2 \odot \check a_3 \in \mathsf{U(1)} ~.
\end{split}
\ee

It is interesting to consider the implications of \eqref{eq:CubicRefinement} for the vanishing 
of the M-theory partition function on closed $11$-manifolds. This was studied in \cite{Diaconescu:2000wy,Diaconescu:2000wz,Diaconescu:2003bm}. Since the kinetic 
energy $\sim \int_{M_{11}} \sfG \star \sfG$ of the $\sfC$-field is invariant under a shift of 
$\c \sfC$ by a flat field $\check \phi$ for $\phi \in H^3(M_{11}; \mathsf{U(1)})$, we can first average over flat $\sfC$-fields.  (In modern language, we are studying the consequences of a ``3-form generalized symmetry'' for the vanishing of the partition function.) The partition function is proportional to 
\be 
\int_{\phi \in H^3(M_{11}; \mathsf{U(1)})}  \Phi([\c \sfC] + \check \phi) ~.
\ee
The integral over flat fields can be written as an integrated integral over the 
topologically trivial flat fields, followed by a sum over $T:= \Tors(H^4(M_{11}; \IZ))$. 
The integral over the topologically flat fields is easily seen to enforce the 
cohomological condition 
\be\label{eq:SimpleFlatCond}
\left[\half \sfG^2 - I_8 \right]  \in H^8_{\mathsf{dR}, \IZ}(M_{11}) ~.
\ee
This is certainly compatible with the equation of motion for the $\sfC$-field 
\be 
\ell^{-3} d\star \sfG = \half \sfG^2 - I_8 ~.
\ee
(In Hamiltonian quantization, the condition \eqref{eq:SimpleFlatCond} can be 
refined considerably to the Gauss law $\Theta(a)=0$, where $\Theta(a)$ is a 
degree 8 integral class defined in  
\cite{Diaconescu:2003bm}.) 

When \eqref{eq:SimpleFlatCond} is satisfied the integral over topologically trivial 
fields descends to a $\mathsf{U(1)}$-valued function $\wt \Phi_{[\c \sfC]}: T \to \mathsf{U(1)}$, 
\be 
\wt \Phi_{[\c \sfC]}(a_T) = \Phi([\c \sfC] + \check a_T) ~,
\ee
where $\check a_T$ is any lift of $a_T \in T$ to a flat differential character. 
(The choice of lift does not matter when \eqref{eq:SimpleFlatCond} is satisfied.) 
$\wt \Phi_{[\c \sfC]}$ is  a cubic refinement of the natural trilinear function on $T$. 
Specializing further to the case where $M_{11} = M_{10} \times \IS^1$, we can decompose 
the degree four torsion classes in $T$ as  $a = h \cup [d\theta] + \wt a$, where $\wt a$ is pulled back from $M_{10}$. At fixed $h$, as a function of $\wt a$, the 
cubic refinement becomes a quadratic refinement, and it follows from 
\cite{Diaconescu:2000wy,Diaconescu:2000wz,Diaconescu:2003bm}, that, 
\be\label{eq:ahhs-differential}
\frac{ \wt \Phi_{[\c \sfC]}(h [d\theta] + \wt a_1 + \wt a_2 ) 
 \wt \Phi_{[\c \sfC]}(h [d\theta]  ) }{
  \wt \Phi_{[\c \sfC]}(h [d\theta] + \wt a_1   ) \wt \Phi_{[\c \sfC]}(h [d\theta] + \wt a_2 ) 
 } = \exp\left[ \imag \pi \int_{M_{10}} \wt a_1 ( \Sq^3 + h) \wt a_2 \right] ~,
\ee
where $\Sq^3$ is a Steenrod square operation.  
\tightfootnote{See \cite[Sec. 3]{Diaconescu:2000wy} for a crash course on Steenrod squares.}
A key step in the proof of this result uses an observation from \cite{Diaconescu:2000wy}: 
Let $a\in H^4(M_{10}; \IZ)$ and consider the $\sfE_8$-adjoint bundle $V(a)$, and let 
$f(a)$ be the mod-two index of the Dirac operator on $M_{10}$ coupled to $V(a)$. Then 
$f(a)$ is a quadratic function:
\be 
f(a_1+a_2) - f(a_1) - f(a_2) + f(0) = \int a_1 \Sq^2 a_2  ~.
\ee
The significance of \eqref{eq:ahhs-differential} will be explained in the next remark. 
Further discussion of the M-theory phase as a cubic refinement can be found in 
\cite{Freed:2019sco}. While Gauss sums -- sums over finite groups of quadratic refinements -- are extremely well-studied, the analogous sums over cubic refinements are much less studied. It might be interesting to understand better when the sum over the torsion group of the cubic refinements does and does not vanish. 

\item \emph{Comparison Of Dirac Quantization Conditions In M-theory And In IIA String Theory.} The Dirac quantization conditions we have described for RR fields 
in \autoref{subsec:PhysAppDiffGenCoh} (using $K$-theory) and for the Abelian gauge field of M-theory described in this section leads to a puzzle, first raised in \cite{Diaconescu:2000wy}:
How can this be consistent with the duality of M-theory on a product manifold 
$M_{10} \times \IS^1$ and type IIA string theory on $M_{10}$? Reference 
\cite{Diaconescu:2000wy} shows that the parition functions of the RR field on $M_{10}$
and of the $\sfC$-field on $M_{10}\times \IS^1$ are in fact equal, but in a highly nontrivial 
way, which makes use of \eqref{eq:ahhs-differential}. The basic idea of the proof is to use 
the comparison of 
$K$-theory and $H^{\rm even}(X_{10}; \IZ)$ based on the Atiyah-Hirzebruch spectral sequence. The first nontrivial differential is $\Sq^3$. Thus, if a class 
$a\in H^4(M_{10}; \IZ)$ has a $K$-theory lift $x\in K(M_{10})$ with $\mathsf{ch}(x) = -a + \cdots$, then it must be that $\Sq^3(a)=0$. Putting $h=0$, it follows from 
\eqref{eq:ahhs-differential} that two $\sfC$-fields with characteristic classes $a_1, a_2$ will 
only contribute if $\Sq^3(a_1-a_2)=0$. For details of the argument, 
see \cite[Sec. 9.3]{Diaconescu:2003bm}. The proof there allows the inclusion of $h\not=0$, in which case $\Sq^3 + h$ is the AHSS differential for twisted $K$-theory. 
Next, one needs to compare the phases in the 
sum over fluxes. The equality of phases is highly nontrivial. See 
\cite{Diaconescu:2000wy} for details. (A summary is given in \cite{Diaconescu:2000wz}.) 
Thus, with some work, the compatibility of the known quantization conditions for the 
RR field and the M-theory Abelian gauge field can be demonstrated at the level of the comparison of partition functions in the long-distance, weak-coupling limit. 
\tightfootnote{Reference 
\cite{Diaconescu:2000wy} generalizes the discussion to include nontrivial $G_2$ flux in 
type IIA, which corresponds to a sum over circle bundles in M-theory.
Reference \cite{Moore:2002cp} demonstrates compatibility with $T$-duality.} 
Much remains to be understood. For example, one would wish to include   $G_0$ flux in the comparison. 
More importantly, a better conceptual understanding of the compatibility that applies to all physical quantities, not just the partition functions, is missing and would be a real step forward. Other quantization conditions have been proposed in \cite{Fiorenza:2020iax,Fiorenza:2020hiq,Sati:2020cml}, and references therein.

\item \emph{The $W_7$ ``Anomaly.''} In the course of studying the sum over torsion $\sfC$-fields, reference \cite{Diaconescu:2000wy} noted that if $W_7(TM_{11})$ is nonzero, the sum will vanish. We have seen in Exercise \ref{exercise:frobenius} and Remark \ref{rem:zerovector} 
that partition functions can vanish without having an anomaly.
The proper interpretation of the $W_7$ ``anomaly'' is that one must insert membranes to obtain nonvanishing amplitudes. It remains to be seen if this has interesting physical consequences in, say, M-theory compactifications and phenomenology. 

\item \emph{Compatibility Of $K$-theory Quantization With S-Duality.} Perhaps the most important unsolved problem in this corner of string theory is the compatibility of the
differential-$K$-theoretic formulation of the RR field with S-duality. The problem is that, in type IIB string theory, the NS B-field is modeled using differential cohomology $[\check B] \in \widecheck{H}^3(M_{10})$, while the 3-form component of the RR fieldstrength is part of the data of a differential 
$K$-theory class $\widecheck{K}^{1, \check \tau}(M_{10})$.  To make matters worse, the differential 
$K$-theory is \underline{twisted} by a differential cocycle  $\check \tau$ whose isomorphism class is $[\check B]$. Thus, the ``RR 2-form potential $C_2$'' and the ``NS 2-form potential $B_2$''   appear on a very different footing in the framework of differential $K$-theory. 
On the other hand, they are supposed to form a doublet of the $S$-duality group $\mathsf{SL}(2,\IZ)$. No convincing framework has been presented that explains how the formulation of RR fields in terms of differential $K$-theory   is compatible with the S-duality invariance of type IIB string theory. One might attempt to explain this away by noting that all the evidence for the $K$-theoretic quantization of RR fields applies in the long-distance and weak-coupling limit of string theory. Since S-duality exchanges strong and weak coupling, there is no direct contradiction. 
Such an explanation is not very satisfactory because there are subgroups of the $\mathsf{SL}(2,\IZ)$ symmetry that preserve weak coupling, and yet, the incompatibility remains. For example, 
one can compare the $K$-theoretic classification of orientifold planes with the more naive 
description using singular cohomology in a weak-coupling regime. In particular, the discussion in \cite[Sec. 3.2]{Witten:1998xy}
classifies $O3$ planes based on torsion RR and NS fluxes in the linking real projective space. These are permuted by a subgroup of the $\mathsf{SL}(2,\IZ)$ duality group that preserves 
weak coupling. It would be extremely interesting to demonstrate that Witten's classification is compatible with the $K$-theoretic description of $O3$ planes described in 
\cite{Distler:2009ri,Distler:2010an}.

\item M-theory is supposed to be parity invariant, but this is not obvious 
from the above description of the $\sfC$-field action. Parity invariance means that 
the (Wick rotated) theory should be defined on unorientable manifolds. Because of the 
Rarita-Schwinger field, the proper statement is that the Wick rotated theory should be defined on $\mathsf{Pin}^+$ $11$-manifolds. 
\tightfootnote{One quick way to see that the domain should be $\mathsf{Pin}^+$ and not $\mathsf{Pin}^-$ 
is that, thanks to the AdS/CFT correspondence, the theory should make sense on 
$\IR \IP^4 \times \IS^7$ (from stacks of $5$-branes on orientifold planes). The obstruction 
to a $\mathsf{Pin}^+$ structure is $w_2(TM)$ and the obstruction to defining a $\mathsf{Pin}^-$ structure 
is $w_2(TM) + w_1(TM)^2$. The total Stiefel-Whitney class for $\IR \IP^n$ 
is $(1+x)^{n+1}$. So $\IR \IP^4$ is $\mathsf{Pin}^+$ but not $\mathsf{Pin}^-$. A more systematic argument is in \cite{Freed:2016rqq}.}
A parity transform of the $\sfC$-field $\sfC \to \sfC^P$ must satisfy, 
\be 
\chi_{\sfC^P}(\Sigma) = \chi_{\sfC}(\Sigma)^* ~,
\ee
(which makes sense, even when $\chi_{\sfC}$ is valued in $\CL_{\mathsf{memb}}$). Therefore, 
\be 
\sfG \mapsto - \sfG ~.
\ee
so $\sfG$ is a ``$w_1$-twisted differential form,'' which means that when pulled back to the oriented double cover, it is odd under the $\IZ_2$ deck transformation. 
Compatibility with the quantization condition for the $\sfC$ field means that the 
characteristic class $a$ transforms under parity by: 
\be 
a \mapsto \lambda - a ~.
\ee
The use of $\eta$-invariants to discuss parity anomalies in M-theory goes back to \cite{Alvarez-Gaume:1984zst}. Some discussion of parity anomalies can be found in \cite{Diaconescu:2000wy} and  \cite{Witten:2016cio}. All these references assumed that the manifold is spin. In \cite{Freed:2019sco}, D. Freed and M. Hopkins extend the discussion to $\mathsf{Pin}^+$ manifolds. In this case, there is no known $\sfE_8$-model of the $\sfC$-field. 
The heart of the problem is that there is no natural isomorphism between a pair of 
principal $\sfE_8$ bundles with characteristic classes $a$ and $\lambda-a$.  
\tightfootnote{The end of \cite{Moore:2004jv} suggests an approach to this difficulty, but it needs further development.}
Nevertheless, the expression \eqref{eq:11dSugraTopolTerm} makes sense on $\mathsf{Pin}^+$ manifolds, and the main result of  \cite{Freed:2019sco} is that $\CL_{\mathsf{RS}}\otimes \CL_{\sfC}$ 
is trivializable.  (The result follows from computations using specific generators 
of the relevant bordism group and is not conceptual, but computational, as the authors lament). Interestingly, there are two possible trivializations of 
$\CL_{\mathsf{RS}}\otimes \CL_{\sfC}$. Put differently, there is exactly one new topological term one can add to M-theory: The mod-two index of the $\mathsf{Pin}^+$ Dirac operator.  When $M_{11}$ is spin, there is a canonical setting defined by the $\sfE_8$ model of the $\sfC$-field, as explained in \cite{Freed:2004yc}. For general $\mathsf{Pin}^+$ manifolds, there is no 
known $\sfE_8$ model of the $\sfC$-field, and no known canonical setting. There thus appear to be two M-theories depending on whether or not the topological term is included in the action. 
An interesting open problem is whether this topological term is consistent with other 
physical constraints: Does it make sense when we include boundaries (and what are the edge modes)? Is it consistent with anomaly freedom for M2 and M5 branes? Is it consistent with 
duality to the type II string on circle bundles? Etc.

\end{enumerate} 

\eject

\appendix 
\renewcommand{\sectionmark}[1]{%
  \markboth{Appendix \thesection}{#1}%
}

\section{Torsion And Torsion}\label{App:TorsionAndTorsion}

\centerline{\emph{Daniel S. Freed}\customfootnote{$\dagger$}{Harvard University, Department of Mathematics, Science Center Room 325, 1 Oxford Street, Cambridge, MA 02138, USA. \emph{Email address}: \url{dafr@math.harvard.edu}.}}
 $\,$

In Remark \ref{rem:AltT}
and in the discussion below equation \eqref{eq:BF-PF-Alt}, we made use of a result relating Reidemeister torsion to an alternating product of the orders of torsion groups. This Appendix gives a proof of the relevant mathematical result. It is possible that \autoref{thm:Freed-7} is known,
but the authors have not found a reference. If someone knows a reference they would be happy to hear about it.

  \subsection{Conventions, Terminology, Exterior Powers, Determinants}\label{subsec:1.1}
  
  Throughout, all groups $F$ are finitely generated abelian, and we don't mention
these properties further. Some groups are also free, which is equivalent to torsionfree.  A subgroup of a free group is free.  An element $f\in F$ is \emph{indivisible} if whenever $f=n\cdot f'$ for $n\in \IZ$, $f'\in F$, then
$n=\pm 1$. If $F$ is free of rank-one, then an indivisible element is a \emph{generator}; there are precisely two generators.  A general (finitely generated Abelian) $F$ sits in a group extension
 \be\label{eq:Freed-1}
 \begin{tikzcd}
     0 \arrow[r] & \Tors\,F \arrow[r] & F \arrow[r] &\Free\,F \arrow[r] & 0 
  \end{tikzcd}
  \ee
with the torsion subgroup and free quotient.  If $F$ is free of rank~$r$, then the exterior powers $\mybigwedge^k F$, $k=0,\dots ,r$, are defined and are free. The top exterior power $\Det\,F = \mybigwedge^r F$ is free of rank 1. We denote by `$\plu F$' a choice of generator of $\Det\,F$.  In \autoref{thm:Freed-7} below, the sign choice will cancel out.

Let 
  \be\label{eq:Freed-2}
  \begin{tikzcd}
     C^{\bullet }: 0 \arrow[r] & C^{0} \arrow[r, "d"] & C^{1} \arrow[r, "d"] & \cdots \arrow[r, "d"] & C^{n} \arrow[r] & 0 
  \end{tikzcd}
  \ee
be a bounded cochain complex of free groups.  Its determinant line is defined as
\be\label{eq:Freed-3}
     \Det\,C^{\bullet } = \bigotimes\limits_{i=0} ^n \,\big(\Det\, C^i\big)^{\otimes (-1)^i} ~.
\ee
Introduce the notations 
\be\label{eq:Freed-4}
     \begin{aligned} 
     C^i_{\IQ} &= C^i\otimes _{\IZ}\IQ ~,\\ 
     H^i &= H^i(C) ~,\\
     H^i_{\IQ} &= H^i(C_{\IQ}) ~,\\
     N^i &= \#\Tors\,H^i \\ N^{\bullet } &= \prod\limits_{i=0}^n\,(N^i)^{(-1)^i} ~.
     \end{aligned}
\ee
and similar notations with $\IR$ replacing $\IQ$.  Note that $\Free\,H^{\bullet}$ is a complex of free groups with zero differential, and so $\Det\,\Free\,H^{\bullet }$ is defined; it is a lattice in the one-dimensional $\IQ$-vector space $\Det\,H^{\bullet}_{\IQ}$, as well as in the one-dimensional $\IR$-vector space $\Det\,H^{\bullet }_{\IR}$.

\subsection{Pl\"ucker Elements And Quotients}\label{subsec:1.2}

Suppose $F$ is free and $A\subset F$ is a subgroup of rank $r$.  We call a choice of generator $\plu A\in \Det\,A\subset \mybigwedge^{r} F$ a
\emph{Pl\"ucker element} for $A$.  Fix this choice.

\begin{lemma}[]\label{thm:Freed-2}
\
 \begin{enumerate}[label=\textnormal{(\arabic*)}]

 \item Suppose $F/A$ is free, $\plu{F/A}\in \Det\,F/A$ is a choice of generator, and $\tplu{F/A}\in \mybigwedge^{\bullet} F$ is a lift of $\plu{F/A}$ to a
decomposable vector.  Then $\tplu{F/A}\wedge \plu{A}\in \Det\,F$ is a generator.

 \item For general $F/A$, choose a generator $\plu{F/A}\in \Det\,\Free(F/A)$. Then $\tplu{F/A}\wedge \plu{A}\in \Det\,F$ is $N$ times a generator, where $N=\#\Tors\,F/A$.
 \end{enumerate}	
\end{lemma}
Observe that the wedge product $\tplu{F/A}\wedge \plu{A}$ is independent of the lift of~$\plu{F/A}$. 

\begin{proof}
Assume $F/A$ is free.  Then the short exact sequence 
\be\label{eq:Freed-12}
\begin{tikzcd}
	0 \arrow[r] & A \arrow[r] & F \arrow[r] & F/A \arrow[r] & 0 
\end{tikzcd}
\ee
of free abelian groups leads to the isomorphism
\be\label{eq:Freed-13}
     \begin{aligned} \Det\,F/A\otimes \Det\,A&\longrightarrow \;\;\Det\,F \\
      \plu{F/A}\;\otimes \;\;\plu A\,\;\;&\longmapsto \tplu{F/A}\wedge
      \plu{A}\end{aligned} 
\ee
of determinant lines.  Assertion~(1) follows immediately.

In general, define $B\subset F$ by the diagram 
\be \label{eq:Freed-14}
\begin{tikzcd}
     & 0 \arrow[d] & 0 \arrow[d] \\
     & A \arrow[d] \arrow[r,equals] & A \arrow[d] \\
	0 \arrow[r] & B \arrow[r] \arrow[d] & F \arrow[r] \arrow[d] & F/B \arrow[r] \arrow[d, "\cong"] & 0\\
	0 \arrow[r] & \Tors\,F/A \arrow[r] \arrow[d] & F/A \arrow[r] \arrow[d] & \Free\,F/A \arrow[r] & 0 \\
	& 0 & 0
\end{tikzcd}
\ee
in which the rows and columns are exact.  Noting that $F/B$ is free, apply \eqref{eq:Freed-13} to the middle row of \eqref{eq:Freed-14} to deduce the isomorphism 
\be\label{eq:Freed-15}
     \begin{aligned} \Det\,F/B\otimes \Det\,B&\longrightarrow \;\;\Det\,F \\
      \plu{F/B}\;\otimes \;\;\plu B\,\;\;&\longmapsto \tplu{F/B}\wedge
      \plu{B}\end{aligned} 
\ee
It remains to observe that $A$ and $B$ are equirank and that the map $\Det\,A\longrightarrow \Det\,B$ induced by the inclusion takes a generator to $N$ times a generator. 

\end{proof}

\subsection{Coboundaries And Cocycles}\label{subsec:2.1}

Given a chain complex \eqref{eq:Freed-2}, define as usual the (free) subgroups of \emph{coboundaries} and \emph{cocycles}: 
\be\label{eq:Freed-5}
     B^i\;\subset \; Z^i \;\subset \; C^i. 
\ee
The \emph{cohomology} is $H^i=Z^i/B^i$.

  \begin{lemma}[]\label{thm:Freed-3}
 The quotient $C^i/Z^i$ is free. 
  \end{lemma}
  
   \begin{proof}
 If $c\in C^i$, $n\in \IZ^{\neq 0}$, and $0=d(nc)=n\,dc$, then $dc=0$ since
$C^{i+1}$~is free.  Hence $c\in Z^i$. 
  \end{proof}
  
 \subsection{Pl\"ucker Elements}\label{subsec:2.2}
Recalling \eqref{eq:Freed-5}, choose generators
\be\label{eq:Freed-7}
 \begin{aligned} 
     \plu{B^i} \;&\in \Det\,B^i ~,\\ 
     \plu{H^i} \;&\in \Det\,\Free\,H^i ~,\\
     \plu{C^i/Z^i} \;&\in \Det\,C^i/Z^i ~.
\end{aligned} 
\ee
and choose lifts (Pl\"ucker elements)
\be \label{eq:Freed-8}
     \begin{aligned} 
     \tplu{H^i} \;&\in \mybigwedge ^{\bullet }C^i ~, \\
     \tplu{C^i/Z^i} \;&\in \mybigwedge^{\bullet }C^i ~.\end{aligned} 
\ee
Since the differential induces an isomorphism $C^{i-1}/Z^{i-1}\to B^i$, we have the following.

  \begin{lemma}[]\label{thm:Freed-4}
 The relation 
  \be\label{eq:Freed-9}
     d\bigl(\tplu{C^{i-1}/Z^{i-1}} \bigr) = \pm\, \plu{B^i} ~,
  \ee
holds.
  \end{lemma} 

\noindent Fix the sign choices of $\plu{B^i}$ in \eqref{eq:Freed-7} so that \eqref{eq:Freed-9} holds with $+$ for each $i$.

\subsection{The Determinant And The Main Theorem}\label{subsec:2.3}

  \begin{definition}[]\label{thm:Freed-5}
 The \textbf{determinant} of the rationalization of the cochain
complex~\eqref{eq:Freed-2} is the isomorphism
  \be\label{eq:Freed-10}
     \begin{aligned} \det\,d\Q\ : \Det\,H^{\bullet }\Q\qquad \;\;\;
      &\longrightarrow \qquad \qquad\Det\,C^{\bullet }\Q \\ \bigotimes
      _{i=0}^n\,\bigl(\plu{H^i} \bigr)^{(-1)^i}&\longmapsto \bigotimes
      _{i=0}^n\,\bigl(\tplu{C^{i}/Z^{i}}\wedge \tplu{H^i}\wedge \plu{B^i}
      \bigr)^{(-1)^i} ~.
      \end{aligned}
  \ee
  \end{definition}
  
 \noindent The reader should check that this map is well-defined independent of all choices (subject to the convention after the statement of \autoref{thm:Freed-4}).
 
 \begin{remark}[]\label{thm:Freed-6}
 This determinant is called the \emph{torsion} of the cochain
complex $(C^{\bullet }\Q,d\Q)$.
  \end{remark}
  
   \begin{theorem}[]\label{thm:Freed-7}
 The determinant $\det\,d\Q$ maps a generator of $\Det\,\Free\,H^{\bullet }$ to
$N^{\bullet } $ times a generator of $\Det\,C^{\bullet }\Q$. 
  \end{theorem}
  
    \begin{proof}
 Apply \autoref{thm:Freed-2}(2) to $B^i\subset Z^i$ to deduce that $\tplu{H^i}\wedge
\plu{B^i}$ is $N^i$ times a generator of $\Det\,Z^i$.  Then apply
\autoref{thm:Freed-2}(1) to the inclusion $Z^i\subset C^i$ to conclude.
  \end{proof}
  
 We obtain an equivalent statement by modifying the determinant to the map 
  \be\label{eq:Freed-11}
     \begin{aligned} \widetilde{\det}\, d\Q\ : \Det\,H^{\bullet }\Q\qquad \;\;\;
      &\longrightarrow \qquad \qquad\Det\,C^{\bullet }\Q \\ \bigotimes
      _{i=0}^n\,\bigl(N^i\plu{H^i} \bigr)^{(-1)^i}&\longmapsto \bigotimes
      _{i=0}^n\,\bigl(\tplu{C^{i}/Z^{i}}\wedge \tplu{H^i}\wedge \plu{B^i}\bigr)^{(-1)^i} ~.
      \end{aligned} 
  \ee

 \begin{corollary}[]\label{thm:Freed-8}
 $\widetilde{\det}\, d\Q$~maps a generator of $\Det\,\Free\,H^{\bullet }$ to a
generator of $\Det\,C^{\bullet }\Q$. 
 \end{corollary}
 
   \subsection{The Real Determinant}\label{subsec:1.4}

The two generators of $\Det\,C^{\bullet }$, which differ by a sign, determine a norm on the real determinant line $\Det\,C^{\bullet }\R$.  The real version $\det\,d\R$ of \eqref{eq:Freed-10} transports it to a norm on $\Det\,H^{\bullet }\R$. On the other hand, the integral lattice $\Det\,\Free\,H^{\bullet }$ also determines a norm on $\Det\,H^{\bullet }\R$. \autoref{thm:Freed-7} asserts that the ratio of these norms is $N^{\bullet }$.
 
\eject
\pagestyle{plain}
\phantomsection 
\interlinepenalty=10000
\addcontentsline{toc}{section}{References}
\bibliographystyle{ytamsalpha}
\bibliography{tasirefs}    

\newcommand{\etalchar}[1]{$^{#1}$}
\providecommand{\bysame}{\leavevmode\hbox to3em{\hrulefill}\thinspace}
\providecommand{\MR}{\relax\ifhmode\unskip\space\fi MR }
\providecommand{\MRhref}[2]{%
  \href{http://www.ams.org/mathscinet-getitem?mr=#1}{#2}
}
\providecommand{\href}[2]{#2}
\providecommand{\doihref}[2]{\href{#1}{#2}}
\providecommand{\arxivfont}{\tt}
\begin{thebibliography}{KdBvW{\etalchar{+}}16}

\bibitem[ABC{\etalchar{+}}09]{Aspinwall:2009isa}
P.~S. Aspinwall, T.~Bridgeland, A.~Craw, M.~R. Douglas, A.~Kapustin, G.~W. Moore, M.~Gross, G.~Segal, B.~Szendr\"oi, and P.~M.~H. Wilson, \emph{{Dirichlet branes and mirror symmetry}}, Clay Mathematics Monographs, vol.~4, AMS, Providence, RI, 2009.

\bibitem[ABGE{\etalchar{+}}21]{Apruzzi:2021nmk}
F.~Apruzzi, F.~Bonetti, I.~Garc\'\i{}a~Etxebarria, S.~S. Hosseini, and S.~Sch{\"a}fer-Nameki, \emph{{Symmetry TFTs from String Theory}}, \doihref{http://dx.doi.org/10.1007/s00220-023-04737-2}{Commun. Math. Phys. \textbf{402} (2023) 895--949}, \href{http://arxiv.org/abs/2112.02092}{{\arxivfont arXiv:2112.02092 [hep-th]}}.

\bibitem[ABJM08]{Aharony:2008ug}
O.~Aharony, O.~Bergman, D.~L. Jafferis, and J.~Maldacena, \emph{{N=6 superconformal Chern-Simons-matter theories, M2-branes and their gravity duals}}, \doihref{http://dx.doi.org/10.1088/1126-6708/2008/10/091}{JHEP \textbf{10} (2008) 091}, \href{http://arxiv.org/abs/0806.1218}{{\arxivfont arXiv:0806.1218 [hep-th]}}.

\bibitem[Abr96]{Abrams:1996ty}
L.~S. Abrams, \emph{{Two-dimensional topological quantum field theories and Frobenius algebras}}, \doihref{http://dx.doi.org/10.1142/S0218216596000333}{J. Knot Theor. Ramifications \textbf{5} (1996) 569--587}.

\bibitem[Ada74]{Adams:1974}
J.~F. Adams, \emph{Stable homotopy and generalised homology}, 1974 (en). \url{https://people.math.rochester.edu/faculty/doug/otherpapers/Adams-SHGH-latex.pdf}.

\bibitem[Ada78]{Adams:1978}
\bysame, \emph{{Infinite Loop Spaces (AM-90): Hermann Weyl Lectures, The Institute for Advanced Study. (AM-90)}}, Princeton University Press, 1978. \url{http://www.jstor.org/stable/j.ctt1b9rztk}.

\bibitem[ADGE{\etalchar{+}}25]{Apruzzi:2025hvs}
F.~Apruzzi, N.~Dondi, I.~Garc{\'\i}a~Etxebarria, H.~T. Lam, and S.~Sch{\"a}fer-Nameki, \emph{{Symmetry TFTs for Continuous Spacetime Symmetries}}, \href{http://arxiv.org/abs/2509.07965}{{\arxivfont arXiv:2509.07965 [hep-th]}}.

\bibitem[ADH21]{Amabel:2021wbk}
A.~Amabel, A.~Debray, and P.~J. Haine, \emph{{Differential Cohomology: Categories, Characteristic Classes, and Connections}}, September 2021. \href{http://arxiv.org/abs/2109.12250}{{\arxivfont arXiv:2109.12250 [math.AT]}}.

\bibitem[{\'A}GBM{\etalchar{+}}87]{Alvarez-Gaume:1987wwg}
L.~{\'A}lvarez-Gaum{\'e}, J.~B. Bost, G.~W. Moore, P.~C. Nelson, and C.~Vafa, \emph{{Bosonization on Higher Genus Riemann Surfaces}}, \doihref{http://dx.doi.org/10.1007/BF01218489}{Commun. Math. Phys. \textbf{112} (1987) 503}.

\bibitem[{\'A}GDPM85]{Alvarez-Gaume:1984zst}
L.~{\'A}lvarez-Gaum{\'e}, S.~Della~Pietra, and G.~W. Moore, \emph{{Anomalies and Odd Dimensions}}, \doihref{http://dx.doi.org/10.1016/0003-4916(85)90383-5}{Annals Phys. \textbf{163} (1985) 288}.

\bibitem[AGM{\etalchar{+}}99]{Aharony:1999ti}
O.~Aharony, S.~S. Gubser, J.~M. Maldacena, H.~Ooguri, and Y.~Oz, \emph{{Large N Field Theories, String Theory and Gravity}}, \doihref{http://dx.doi.org/10.1016/S0370-1573(99)00083-6}{Phys. Rept. \textbf{323} (2000) 183--386}, \href{http://arxiv.org/abs/hep-th/9905111}{{\arxivfont arXiv:hep-th/9905111}}.

\bibitem[{\'A}GMN{\etalchar{+}}86]{Alvarez-Gaume:1986nqf}
L.~{\'A}lvarez-Gaum{\'e}, G.~W. Moore, P.~C. Nelson, C.~Vafa, and J.~B. Bost, \emph{{Bosonization in Arbitrary Genus}}, \doihref{http://dx.doi.org/10.1016/0370-2693(86)90466-1}{Phys. Lett. B \textbf{178} (1986) 41--47}.

\bibitem[AGN85]{MR791642}
L.~\'Alvarez-Gaum\'e and P.~Nelson, \emph{{Hamiltonian Interpretation of Anomalies}}, \href{http://projecteuclid.org/euclid.cmp/1103942612}{Comm. Math. Phys. \textbf{99} (1985) 103--114}.

\bibitem[AHR10]{AndoHopkinsRezk2010}
M.~Ando, M.~J. Hopkins, and C.~Rezk, \emph{{Multiplicative Orientations of KO-Theory and of the Spectrum of Topological Modular Forms}}, \url{https://rezk.web.illinois.edu/koandtmf.pdf}, 2010. Version 0.5, May 2010.

\bibitem[AHS01]{Ando2001}
M.~Ando, M.~Hopkins, and N.~Strickland, \emph{{Elliptic spectra, the Witten genus and the theorem of the cube}}, \href{http://dx.doi.org/10.1007/s002220100175}{Inventiones Mathematicae \textbf{146} (2001) 595–687}.

\bibitem[AK11]{EllegaardAndersen:2011vps}
J.~E. Andersen and R.~Kashaev, \emph{{A TQFT from Quantum Teichm\"uller Theory}}, \doihref{http://dx.doi.org/10.1007/s00220-014-2073-2}{Commun. Math. Phys. \textbf{330} (2014) 887--934}, \href{http://arxiv.org/abs/1109.6295}{{\arxivfont arXiv:1109.6295 [math.QA]}}.

\bibitem[AK18]{Andersen:2018pnw}
\bysame, \emph{{The Teichm\"uller TQFT}}, {International Congress of Mathematicians}, 2018, pp.~2527--2552. \href{http://arxiv.org/abs/1811.06853}{{\arxivfont arXiv:1811.06853 [math.QA]}}.

\bibitem[AKL22]{Albert:2022gcs}
J.~Albert, J.~Kaidi, and Y.-H. Lin, \emph{{Topological modularity of Supermoonshine}}, \doihref{http://dx.doi.org/10.1093/ptep/ptad034}{PTEP \textbf{2023} (2023) 033B06}, \href{http://arxiv.org/abs/2210.14923}{{\arxivfont arXiv:2210.14923 [hep-th]}}.

\bibitem[AKMW87a]{Alvarez:1987wg}
O.~Alvarez, T.~P. Killingback, M.~L. Mangano, and P.~Windey, \emph{{String Theory and Loop Space Index Theorems}}, \doihref{http://dx.doi.org/10.1007/BF01239011}{Commun. Math. Phys. \textbf{111} (1987) 1}.

\bibitem[AKMW87b]{Alvarez:1987de}
\bysame, \emph{{The Dirac-Ramond operator in string theory and loop space index theorems}}, \doihref{http://dx.doi.org/10.1016/0920-5632(87)90110-1}{Nucl. Phys. B Proc. Suppl. \textbf{1} (1987) 189--215}.

\bibitem[AKS23]{Artymowicz:2023erv}
A.~Artymowicz, A.~Kapustin, and N.~Sopenko, \emph{{Quantization of the Higher Berry Curvature and the Higher Thouless Pump}}, \doihref{http://dx.doi.org/10.1007/s00220-024-05026-2}{Commun. Math. Phys. \textbf{405} (2024) 191}, \href{http://arxiv.org/abs/2305.06399}{{\arxivfont arXiv:2305.06399 [math-ph]}}.

\bibitem[Alv85]{Alvarez:1984es}
O.~Alvarez, \emph{{Topological Quantization and Cohomology}}, \doihref{http://dx.doi.org/10.1007/BF01212452}{Commun. Math. Phys. \textbf{100} (1985) 279}.

\bibitem[Ana86]{Anandan1986}
J.~Anandan, \emph{Gauge fields, quantum interference, and holonomy transformations}, \href{http://dx.doi.org/10.1103/PhysRevD.33.2280}{Physical Review D \textbf{33} (1986) 2280–2287}.

\bibitem[AS04]{Atiyah:2004jv}
M.~Atiyah and G.~Segal, \emph{{Twisted K-theory}}, \href{http://arxiv.org/abs/math/0407054}{{\arxivfont arXiv:math/0407054}}.

\bibitem[AS05]{Atiyah:2005gu}
\bysame, \emph{{Twisted K-theory and cohomology}}, \href{http://arxiv.org/abs/math/0510674}{{\arxivfont arXiv:math/0510674}}.

\bibitem[AST13]{Aharony:2013hda}
O.~Aharony, N.~Seiberg, and Y.~Tachikawa, \emph{{Reading between the lines of four-dimensional gauge theories}}, \doihref{http://dx.doi.org/10.1007/JHEP08(2013)115}{JHEP \textbf{08} (2013) 115}, \href{http://arxiv.org/abs/1305.0318}{{\arxivfont arXiv:1305.0318 [hep-th]}}.

\bibitem[AT86]{Achucarro:1986uwr}
A.~Achucarro and P.~K. Townsend, \emph{{A Chern-Simons Action for Three-Dimensional anti-De Sitter Supergravity Theories}}, \doihref{http://dx.doi.org/10.1016/0370-2693(86)90140-1}{Phys. Lett. B \textbf{180} (1986) 89}.

\bibitem[Ati61]{Atiyah1961}
M.~F. Atiyah, \emph{{Thom Complexes}}, \href{http://dx.doi.org/10.1112/plms/s3-11.1.291}{Proceedings of the London Mathematical Society \textbf{s3-11} (1961) 291–310}.

\bibitem[Ati71]{Atiyah1971}
\bysame, \emph{{Riemann surfaces and spin structures}}, \href{http://dx.doi.org/10.24033/asens.1205}{Annales scientifiques de l’École normale supérieure \textbf{4} (1971) 47–62}.

\bibitem[Ati89]{Atiyah:1989vu}
M.~Atiyah, \emph{{Topological quantum field theories}}, \doihref{http://dx.doi.org/10.1007/BF02698547}{Inst. Hautes Etudes Sci. Publ. Math. \textbf{68} (1989) 175--186}.

\bibitem[Ati90]{Atiyah1990}
\bysame, \emph{On framings of 3-manifolds}, \href{http://dx.doi.org/10.1016/0040-9383(90)90021-B}{Topology \textbf{29} (1990) 1–7}.

\bibitem[Ati18]{Atiyah2018}
M.~F. Atiyah, \doihref{http://dx.doi.org/10.1201/9780429493546}{\emph{K-theory}}, CRC Press, March 2018. \url{http://dx.doi.org/10.1201/9780429493546}. Notes by D. W. Anderson.

\bibitem[AV01]{Acharya:2001dz}
B.~S. Acharya and C.~Vafa, \emph{{On Domain Walls of N=1 Supersymmetric Yang-Mills in Four Dimensions}}, \href{http://arxiv.org/abs/hep-th/0103011}{{\arxivfont arXiv:hep-th/0103011}}.

\bibitem[Bar91]{Barrett:1991aj}
J.~W. Barrett, \emph{{Holonomy and path structures in general relativity and Yang-Mills theory}}, \doihref{http://dx.doi.org/10.1007/BF00671007}{Int. J. Theor. Phys. \textbf{30} (1991) 1171--1215}.

\bibitem[Bar05]{Bartlett:2005qy}
B.~H. Bartlett, \emph{{Categorical Aspects of Topological Quantum Field Theories}}, Master's thesis, September 2005. \href{http://arxiv.org/abs/math/0512103}{{\arxivfont arXiv:math/0512103}}.

\bibitem[BB13]{BarBeckerarXiv}
C.~B\"{a}r and C.~Becker, \doihref{http://dx.doi.org/10.48550/ARXIV.1303.6457}{\emph{{Differential Characters and Geometric Chains}}}, 2013. \url{https://arxiv.org/abs/1303.6457}.

\bibitem[BB14]{BarBeckerBook}
C.~B\"{a}r and C.~Becker, \doihref{http://dx.doi.org/10.1007/978-3-319-07034-6}{\emph{{Differential Characters}}}, Springer International Publishing, 2014. \url{http://dx.doi.org/10.1007/978-3-319-07034-6}.

\bibitem[BBFT{\etalchar{+}}23]{Bhardwaj:2023kri}
L.~Bhardwaj, L.~E. Bottini, L.~Fraser-Taliente, L.~Gladden, D.~S.~W. Gould, A.~Platschorre, and H.~Tillim, \emph{{Lectures on Generalized Symmetries}}, \doihref{http://dx.doi.org/10.1016/j.physrep.2023.11.002}{Phys. Rept. \textbf{1051} (2024) 1--87}, \href{http://arxiv.org/abs/2307.07547}{{\arxivfont arXiv:2307.07547 [hep-th]}}.

\bibitem[BBRT91]{Birmingham:1991ty}
D.~Birmingham, M.~Blau, M.~Rakowski, and G.~Thompson, \emph{{Topological field theory}}, \doihref{http://dx.doi.org/10.1016/0370-1573(91)90117-5}{Phys. Rept. \textbf{209} (1991) 129--340}.

\bibitem[BBS95]{Becker:1995kb}
K.~Becker, M.~Becker, and A.~Strominger, \emph{{Fivebranes, Membranes and Non-Perturbative String Theory}}, \doihref{http://dx.doi.org/10.1016/0550-3213(95)00487-1}{Nucl. Phys. B \textbf{456} (1995) 130--152}, \href{http://arxiv.org/abs/hep-th/9507158}{{\arxivfont arXiv:hep-th/9507158}}.

\bibitem[BCM{\etalchar{+}}01]{Bouwknegt:2001vu}
P.~Bouwknegt, A.~L. Carey, V.~Mathai, M.~K. Murray, and D.~Stevenson, \emph{{Twisted K-theory and K-theory of bundle gerbes}}, \doihref{http://dx.doi.org/10.1007/s002200200646}{Commun. Math. Phys. \textbf{228} (2002) 17--49}, \href{http://arxiv.org/abs/hep-th/0106194}{{\arxivfont arXiv:hep-th/0106194}}.

\bibitem[BD82]{Birrell1982}
N.~D. Birrell and P.~C.~W. Davies, \doihref{http://dx.doi.org/10.1017/cbo9780511622632}{\emph{{Quantum Fields in Curved Space}}}, Cambridge University Press, February 1982. \url{http://dx.doi.org/10.1017/CBO9780511622632}.

\bibitem[BD95]{Baez:1995xq}
J.~C. Baez and J.~Dolan, \emph{{Higher-dimensional Algebra and Topological Quantum Field Theory}}, \doihref{http://dx.doi.org/10.1063/1.531236}{J. Math. Phys. \textbf{36} (1995) 6073--6105}, \href{http://arxiv.org/abs/q-alg/9503002}{{\arxivfont arXiv:q-alg/9503002}}.

\bibitem[BDG{\etalchar{+}}16]{Barwick2016}
C.~Barwick, E.~Dotto, S.~Glasman, D.~Nardin, and J.~Shah, \doihref{http://dx.doi.org/10.48550/ARXIV.1608.03654}{\emph{Parametrized higher category theory and higher algebra: A general introduction}}, 2016. \href{http://arxiv.org/abs/1608.03654}{{\arxivfont arXiv:1608.03654 [math.AT]}}. \url{https://arxiv.org/abs/1608.03654}.

\bibitem[BDR03]{BassDundasRognes}
N.~A. Baas, B.~I. Dundas, and J.~Rognes, \emph{Two-vector bundles and forms of elliptic cohomology}. \url{https://arxiv.org/abs/math/0306027}.

\bibitem[BDZM24]{Bonetti:2024cjk}
F.~Bonetti, M.~Del~Zotto, and R.~Minasian, \emph{{SymTFTs for Continuous non-Abelian Symmetries}}, \href{http://arxiv.org/abs/2402.12347}{{\arxivfont arXiv:2402.12347 [hep-th]}}.

\bibitem[BDZM25]{Bonetti:2025dvm}
\bysame, \emph{{SymTFT for Continuous Symmetries: Non-linear Realizations and Spontaneous Breaking}}, \href{http://arxiv.org/abs/2509.10343}{{\arxivfont arXiv:2509.10343 [hep-th]}}.

\bibitem[BE13]{Berwick-Evans:2013fua}
D.~Berwick-Evans, \emph{{Perturbative sigma models, elliptic cohomology and the Witten genus}}, \href{http://arxiv.org/abs/1311.6836}{{\arxivfont arXiv:1311.6836 [math.AT]}}.

\bibitem[BE14]{Berwick-Evans:2014oja}
\bysame, \emph{{Equivariant elliptic cohomology, gauged sigma models, and discrete torsion}}, \href{http://arxiv.org/abs/1410.5500}{{\arxivfont arXiv:1410.5500 [math.AT]}}.

\bibitem[BE20]{Berwick-Evans:2020niz}
\bysame, \emph{{Chern characters for supersymmetric field theories}}, \doihref{http://dx.doi.org/10.2140/gt.2023.27.1947}{Geom. Topol. \textbf{27} (2023) 1947--1986}, \href{http://arxiv.org/abs/2010.03663}{{\arxivfont arXiv:2010.03663 [math.AT]}}.

\bibitem[BE23]{Berwick-Evans:2023ejh}
\bysame, \emph{{How do field theories detect the torsion in topological modular forms?}}, \href{http://arxiv.org/abs/2303.09138}{{\arxivfont arXiv:2303.09138 [math.AT]}}.

\bibitem[BE24]{Berwick-Evans:2024nrr}
\bysame, \emph{{Elliptic cohomology and quantum field theory}}, \href{http://arxiv.org/abs/2408.07693}{{\arxivfont arXiv:2408.07693 [math.AT]}}.

\bibitem[BEC{\etalchar{+}}17]{Bradlyn:2017pss}
B.~Bradlyn, L.~Elcoro, J.~Cano, M.~G. Vergniory, Z.~Wang, C.~Felser, M.~I. Aroyo, and B.~A. Bernevig, \emph{{Topological quantum chemistry}}, \doihref{http://dx.doi.org/10.1038/nature23268}{Nature \textbf{547} (2017) 298--305}, \href{http://arxiv.org/abs/1703.02050}{{\arxivfont arXiv:1703.02050 [cond-mat.mes-hall]}}.

\bibitem[BEP15]{Berwick-Evans:2015ria}
D.~Berwick-Evans and D.~Pavlov, \emph{{Smooth one-dimensional topological field theories are vector bundles with connection}}, \doihref{http://dx.doi.org/10.2140/agt.2023.23.3707}{Algebr. Geom. Topol. \textbf{23} (2023) 3707--3743}, \href{http://arxiv.org/abs/1501.00967}{{\arxivfont arXiv:1501.00967 [math.AT]}}.

\bibitem[Ber06]{Bergner2006}
J.~E. Bergner, \doihref{http://dx.doi.org/10.48550/ARXIV.MATH/0610239}{\emph{A survey of $(\infty, 1)$-categories}}, 2006. \url{https://arxiv.org/abs/math/0610239}.

\bibitem[BET18]{Berwick-Evans:2018ryn}
D.~Berwick-Evans and A.~Tripathy, \emph{{A model for complex analytic equivariant elliptic cohomology from quantum field theory}}, \href{http://arxiv.org/abs/1805.04146}{{\arxivfont arXiv:1805.04146 [math.AT]}}.

\bibitem[BET21]{Berwick-Evans:2021jlr}
\bysame, \emph{{A de Rham model for complex analytic equivariant elliptic cohomology}}, \doihref{http://dx.doi.org/10.1016/j.aim.2021.107575}{Adv. Math. \textbf{380} (2021) 107575}.

\bibitem[BF10]{Bieri:2010za}
S.~Bieri and J.~Fr{\"o}hlich, \emph{{Physical principles underlying the quantum Hall effect}}, \doihref{http://dx.doi.org/10.1016/j.crhy.2011.02.001}{Comptes Rendus Physique \textbf{12} (2011) 332--346}, \href{http://arxiv.org/abs/1006.0457}{{\arxivfont arXiv:1006.0457 [cond-mat.mes-hall]}}.

\bibitem[BFLP16]{Balasubramanian:2016sro}
V.~Balasubramanian, J.~R. Fliss, R.~G. Leigh, and O.~Parrikar, \emph{{Multi-Boundary Entanglement in Chern-Simons Theory and Link Invariants}}, \doihref{http://dx.doi.org/10.1007/JHEP04(2017)061}{JHEP \textbf{04} (2017) 061}, \href{http://arxiv.org/abs/1611.05460}{{\arxivfont arXiv:1611.05460 [hep-th]}}.

\bibitem[BFM{\etalchar{+}}22]{Bah:2022wot}
I.~Bah, D.~S. Freed, G.~W. Moore, N.~Nekrasov, S.~S. Razamat, and S.~Sch{\"a}fer-Nameki, \emph{{A Panorama Of Physical Mathematics c. 2022}}, \href{http://arxiv.org/abs/2211.04467}{{\arxivfont arXiv:2211.04467 [hep-th]}}.

\bibitem[BH23]{Brennan:2023mmt}
T.~D. Brennan and S.~Hong, \emph{{Introduction to Generalized Global Symmetries in QFT and Particle Physics}}, \href{http://arxiv.org/abs/2306.00912}{{\arxivfont arXiv:2306.00912 [hep-ph]}}.

\bibitem[BHMV95]{Blanchet1995}
C.~Blanchet, N.~Habegger, G.~Masbaum, and P.~Vogel, \emph{{Topological Quantum Field Theories derived from the Kauffman bracket}}, \href{http://dx.doi.org/10.1016/0040-9383(94)00051-4}{Topology \textbf{34} (1995) 883–927}.

\bibitem[BK00]{Bakalov2000-hq}
B.~Bakalov and A.~Kirillov, \doihref{http://dx.doi.org/10.1090/ulect/021}{\emph{{Lectures on Tensor Categories and Modular Functors}}}, American Mathematical Society, November 2000. \url{http://dx.doi.org/10.1090/ULECT/021}.

\bibitem[BKT22]{Blau:2022odi}
M.~Blau, M.~Kakona, and G.~Thompson, \emph{{Massive Ray-Singer torsion and path integrals}}, \doihref{http://dx.doi.org/10.1007/JHEP08(2022)230}{JHEP \textbf{08} (2022) 230}, \href{http://arxiv.org/abs/2206.12268}{{\arxivfont arXiv:2206.12268 [hep-th]}}.

\bibitem[BKT24]{Blau:2024obl}
\bysame, \emph{{On the evaluation of the Ray-Singer torsion path integral}}, \doihref{http://dx.doi.org/10.1007/JHEP06(2024)065}{JHEP \textbf{06} (2024) 065}, \href{http://arxiv.org/abs/2402.14437}{{\arxivfont arXiv:2402.14437 [hep-th]}}.

\bibitem[BLM24]{Beck:2024xtd}
J.~Beck, A.~Losev, and P.~Mnev, \emph{{Combinatorial 2d higher topological quantum field theory from a local cyclic $A_\infty $ algebra}}, \doihref{http://dx.doi.org/10.1007/s11005-024-01874-0}{Lett. Math. Phys. \textbf{114} (2024) 125}, \href{http://arxiv.org/abs/2402.04468}{{\arxivfont arXiv:2402.04468 [math-ph]}}.

\bibitem[BM96]{Brylinski1996}
J.~L. Brylinski and D.~A. McLaughlin, \emph{{{\v{C}}ech cocycles for characteristic classes}}, \href{http://dx.doi.org/10.1007/BF02104916}{Communications in Mathematical Physics \textbf{178} (1996) 225–236}.

\bibitem[BM00]{Bouwknegt:2000qt}
P.~Bouwknegt and V.~Mathai, \emph{{D-branes, B-fields and twisted K-theory}}, \doihref{http://dx.doi.org/10.1088/1126-6708/2000/03/007}{JHEP \textbf{03} (2000) 007}, \href{http://arxiv.org/abs/hep-th/0002023}{{\arxivfont arXiv:hep-th/0002023}}.

\bibitem[BM04]{Belov:2004ht}
D.~Belov and G.~W. Moore, \emph{{Conformal blocks for AdS(5) singletons}}, \href{http://arxiv.org/abs/hep-th/0412167}{{\arxivfont arXiv:hep-th/0412167}}.

\bibitem[BM05]{Belov:2005ze}
\bysame, \emph{{Classification of Abelian spin Chern-Simons theories}}, \href{http://arxiv.org/abs/hep-th/0505235}{{\arxivfont arXiv:hep-th/0505235}}.

\bibitem[BM06a]{Belov:2006jd}
D.~Belov and G.~W. Moore, \emph{{Holographic Action for the Self-Dual Field}}, \href{http://arxiv.org/abs/hep-th/0605038}{{\arxivfont arXiv:hep-th/0605038}}.

\bibitem[BM06b]{Belov:2006xj}
D.~M. Belov and G.~W. Moore, \emph{{Type II Actions from 11-Dimensional Chern-Simons Theories}}, \href{http://arxiv.org/abs/hep-th/0611020}{{\arxivfont arXiv:hep-th/0611020}}.

\bibitem[BM22]{Banerjee:2022pmw}
A.~Banerjee and G.~W. Moore, \emph{{Comments on Summing over Bordisms in TQFT}}, \doihref{http://dx.doi.org/10.1007/JHEP09(2022)171}{JHEP \textbf{09} (2022) 171}, \href{http://arxiv.org/abs/2201.00903}{{\arxivfont arXiv:2201.00903 [hep-th]}}.

\bibitem[BMRS07]{Brodzki:2007hg}
J.~Brodzki, V.~Mathai, J.~M. Rosenberg, and R.~J. Szabo, \emph{{Noncommutative correspondences, duality and D-branes in bivariant K-theory}}, \doihref{http://dx.doi.org/10.4310/ATMP.2009.v13.n2.a4}{Adv. Theor. Math. Phys. \textbf{13} (2009) 497--552}, \href{http://arxiv.org/abs/0708.2648}{{\arxivfont arXiv:0708.2648 [hep-th]}}.

\bibitem[BN09]{BunkeNaumann}
U.~Bunke and N.~Naumann, \emph{{Secondary Invariants for String Bordism and tmf}}, \href{http://arxiv.org/abs/0912.4875}{{\arxivfont arXiv:0912.4875 [math.KT]}}.

\bibitem[BN14]{BUNKE2014912}
U.~Bunke and N.~Naumann, \emph{Secondary invariants for string bordism and topological modular forms}, \href{https://www.sciencedirect.com/science/article/pii/S000744971400030X}{Bulletin des Sciences Mathématiques \textbf{138} (2014) 912--970}.

\bibitem[BNR12]{Borodzik2016}
M.~Borodzik, A.~N\'emethi, and A.~Ranicki, \emph{Morse theory for manifolds with boundary}, \href{http://dx.doi.org/10.2140/agt.2016.16.971}{{Algebraic \& Geometric Topology} \textbf{16} (2016) 971–1023}, \href{http://arxiv.org/abs/1207.3066}{{\arxivfont arXiv:1207.3066 [math.GT]}}.

\bibitem[BNVar]{BunkeNikolausVolkl:2013}
U.~Bunke, T.~Nikolaus, and M.~V{\"o}lkl, \emph{{Differential cohomology theories as sheaves of spectra}}, \href{http://arxiv.org/abs/arXiv:1311.3188}{{\arxivfont arXiv:arXiv:1311.3188 [math.KT]}}.

\bibitem[Bor55]{Borel:1955}
A.~Borel, \emph{{Topology of Lie groups and characteristic classes}}, Bulletin of the American Mathematical Society \textbf{61} (1955) 397 -- 432.

\bibitem[Bor67]{Borel1967}
A.~Borel, \doihref{http://dx.doi.org/10.1007/bfb0096867}{\emph{{Topics in the Homology Theory of Fibre Bundles}}}, Springer Berlin Heidelberg, 1967. \url{http://dx.doi.org/10.1007/BFb0096867}.

\bibitem[Bot56]{Bott1956}
R.~Bott, \emph{{An application of the Morse theory to the topology of Lie-groups}}, \href{http://dx.doi.org/10.24033/bsmf.1472}{Bulletin de la Soci{\'e}t{\'e} Math{\'e}matique de France \textbf{79} (1956) 251–281}.

\bibitem[Bot69]{Bott1969-dj}
\bysame, \emph{{Lectures on $K(X)$}}, Mathematics Lecture Note Series, Addison Wesley Longman, Singapore, Singapore, December 1969.

\bibitem[BPT10]{Bonelli:2010cu}
G.~Bonelli, A.~Prudenziati, and A.~Tanzini, \emph{{Taming open/closed string duality with a Losev trick}}, \doihref{http://dx.doi.org/10.1007/JHEP06(2010)063}{JHEP \textbf{06} (2010) 063}, \href{http://arxiv.org/abs/1003.2519}{{\arxivfont arXiv:1003.2519 [hep-th]}}.

\bibitem[Bro62]{Brown1962}
M.~Brown, \emph{{Locally Flat Imbeddings of Topological Manifolds}}, \href{http://dx.doi.org/10.2307/1970177}{The Annals of Mathematics \textbf{75} (1962) 331}.

\bibitem[Bro82]{Brown1982}
K.~S. Brown, \doihref{http://dx.doi.org/10.1007/978-1-4684-9327-6}{\emph{{Cohomology of Groups}}}, Springer New York, 1982. \url{http://dx.doi.org/10.1007/978-1-4684-9327-6}.

\bibitem[Bry93]{Brylinski1993}
J.-L. Brylinski, \doihref{http://dx.doi.org/10.1007/978-0-8176-4731-5}{\emph{{Loop Spaces, Characteristic Classes and Geometric Quantization}}}, Birkh\"{a}user Boston, 1993. \url{http://dx.doi.org/10.1007/978-0-8176-4731-5}.

\bibitem[BS24]{Brennan:2024fgj}
T.~D. Brennan and Z.~Sun, \emph{{A SymTFT for Continuous Symmetries}}, \doihref{http://dx.doi.org/10.1007/JHEP12(2024)100}{JHEP \textbf{12} (2024) 100}, \href{http://arxiv.org/abs/2401.06128}{{\arxivfont arXiv:2401.06128 [hep-th]}}.

\bibitem[BST87]{Bergshoeff:1987cm}
E.~Bergshoeff, E.~Sezgin, and P.~K. Townsend, \emph{{Supermembranes and Eleven-Dimensional Supergravity}}, \doihref{http://dx.doi.org/10.1201/9781482268737-10}{Phys. Lett. B \textbf{189} (1987) 75--78}.

\bibitem[BT82]{Bott1982}
R.~Bott and L.~W. Tu, \doihref{http://dx.doi.org/10.1007/978-1-4757-3951-0}{\emph{{Differential Forms in Algebraic Topology}}}, Springer New York, 1982. \url{http://dx.doi.org/10.1007/978-1-4757-3951-0}.

\bibitem[BT91]{Blau:1989bq}
M.~Blau and G.~Thompson, \emph{{Topological Gauge Theories of Antisymmetric Tensor Fields}}, \doihref{http://dx.doi.org/10.1016/0003-4916(91)90240-9}{Annals Phys. \textbf{205} (1991) 130--172}.

\bibitem[Bun12]{Bunke:2012rsi}
U.~Bunke, \emph{{Differential cohomology}}, \href{http://arxiv.org/abs/1208.3961}{{\arxivfont arXiv:1208.3961 [math.AT]}}.

\bibitem[BV81]{Batalin1981}
I.~Batalin and G.~Vilkovisky, \emph{Gauge algebra and quantization}, \href{http://dx.doi.org/10.1016/0370-2693(81)90205-7}{Physics Letters B \textbf{102} (1981) 27–31}.

\bibitem[BV83]{Batalin:1983ggl}
I.~A. Batalin and G.~A. Vilkovisky, \emph{{Quantization of Gauge Theories with Linearly Dependent Generators}}, \doihref{http://dx.doi.org/10.1103/PhysRevD.28.2567}{Phys. Rev. D \textbf{28} (1983) 2567--2582}. [Erratum: Phys.Rev.D 30, 508 (1984)].

\bibitem[BvdGHZ08]{Bruinier:2008}
J.~H. Bruinier, G.~van~der Geer, G.~Harder, and D.~Zagier, \doihref{http://dx.doi.org/10.1007/978-3-540-74119-0}{\emph{{The 1-2-3 of Modular Forms}}}, Springer Berlin Heidelberg, 2008. \url{http://dx.doi.org/10.1007/978-3-540-74119-0}.

\bibitem[C{\etalchar{+}}24]{Costa:2024wks}
D.~Costa et~al., \emph{{Simons Lectures on Categorical Symmetries}}, November 2024. \href{http://arxiv.org/abs/2411.09082}{{\arxivfont arXiv:2411.09082 [math-ph]}}.

\bibitem[Cat16]{Cattaneo:2016lkk}
A.~S. Cattaneo, \emph{{From topological field theory to deformation quantization and reduction}}, {International Congress of Mathematicians}, 8 2016. \href{http://arxiv.org/abs/1608.06576}{{\arxivfont arXiv:1608.06576 [math-ph]}}.

\bibitem[CC94]{Carlip:1994tt}
S.~Carlip and R.~Cosgrove, \emph{{Topology change in (2+1)-dimensional gravity}}, \doihref{http://dx.doi.org/10.1063/1.530760}{J. Math. Phys. \textbf{35} (1994) 5477--5493}, \href{http://arxiv.org/abs/gr-qc/9406006}{{\arxivfont arXiv:gr-qc/9406006}}.

\bibitem[CCH{\etalchar{+}}21]{Choi:2021kmx}
Y.~Choi, C.~C{\'o}rdova, P.-S. Hsin, H.~T. Lam, and S.-H. Shao, \emph{{Non-Invertible Duality Defects in 3+1 Dimensions}}, \doihref{http://dx.doi.org/10.1103/PhysRevD.105.125016}{Phys. Rev. D \textbf{105} (2022) 125016}, \href{http://arxiv.org/abs/2111.01139}{{\arxivfont arXiv:2111.01139 [hep-th]}}.

\bibitem[CD08]{Coecke:2008lcg}
B.~Coecke and R.~Duncan, \emph{{Interacting Quantum Observables}}, \doihref{http://dx.doi.org/10.1007/978-3-540-70583-3_25}{Lect. Notes Comput. Sci. \textbf{5126} (2008) 298--310}.

\bibitem[CDH{\etalchar{+}}07]{Caldararu:2010ljp}
A.~Caldararu, J.~Distler, S.~Hellerman, T.~Pantev, and E.~Sharpe, \emph{{Non-birational twisted derived equivalences in abelian GLSMs}}, \doihref{http://dx.doi.org/10.1007/s00220-009-0974-2}{Commun. Math. Phys. \textbf{294} (2010) 605--645}, \href{http://arxiv.org/abs/0709.3855}{{\arxivfont arXiv:0709.3855 [hep-th]}}.

\bibitem[CDIS22]{Cordova:2022ruw}
C.~C{\'o}rdova, T.~T. Dumitrescu, K.~Intriligator, and S.-H. Shao, \emph{{Snowmass White Paper: Generalized Symmetries in Quantum Field Theory and Beyond}}, {Snowmass 2021}, May 2022. \href{http://arxiv.org/abs/2205.09545}{{\arxivfont arXiv:2205.09545 [hep-th]}}.

\bibitem[CDZR23]{Carqueville:2023jhb}
N.~Carqueville, M.~Del~Zotto, and I.~Runkel, \doihref{http://dx.doi.org/10.1016/B978-0-323-95703-8.00098-7}{\emph{{Topological defects}}}, November 2023. \href{http://arxiv.org/abs/2311.02449}{{\arxivfont arXiv:2311.02449 [math-ph]}}.

\bibitem[Cer68]{Cerf1968}
J.~Cerf, \doihref{http://dx.doi.org/10.1007/bfb0060395}{\emph{{Sur les diff{\'e}omorphismes de la sph{\`e}re de dimension trois $(\Gamma_4 = 0)$}}}, Springer Berlin Heidelberg, 1968. \url{http://dx.doi.org/10.1007/BFb0060395}.

\bibitem[Cer70]{Cerf1970}
\bysame, \emph{{La stratification naturelle des espaces de fonctions diff{\'e}rentiables r{\'e}elles et le th{\'e}or{\`e}me de la pseudo-isotopie}}, \href{https://doi.org/10.1007/BF02684687}{Publications Math{\'e}matiques de l'Institut des Hautes {\'E}tudes Scientifiques \textbf{39} (1970) 7--170}.

\bibitem[CEZ23]{Collier:2023fwi}
S.~Collier, L.~Eberhardt, and M.~Zhang, \emph{{Solving 3d Gravity with Virasoro TQFT}}, \doihref{http://dx.doi.org/10.21468/SciPostPhys.15.4.151}{SciPost Phys. \textbf{15} (2023) 151}, \href{http://arxiv.org/abs/2304.13650}{{\arxivfont arXiv:2304.13650 [hep-th]}}.

\bibitem[CEZ24]{Collier:2024mgv}
\bysame, \emph{{3d gravity from Virasoro TQFT: Holography, wormholes and knots}}, \doihref{http://dx.doi.org/10.21468/SciPostPhys.17.5.134}{SciPost Phys. \textbf{17} (2024) 134}, \href{http://arxiv.org/abs/2401.13900}{{\arxivfont arXiv:2401.13900 [hep-th]}}.

\bibitem[CFLS19]{Cordova:2019jnf}
C.~C{\'o}rdova, D.~S. Freed, H.~T. Lam, and N.~Seiberg, \emph{{Anomalies in the Space of Coupling Constants and Their Dynamical Applications I}}, \doihref{http://dx.doi.org/10.21468/SciPostPhys.8.1.001}{SciPost Phys. \textbf{8} (2020) 001}, \href{http://arxiv.org/abs/1905.09315}{{\arxivfont arXiv:1905.09315 [hep-th]}}.

\bibitem[CG16]{Costello2016}
K.~Costello and O.~Gwilliam, \doihref{http://dx.doi.org/10.1017/9781316678626}{\emph{{Factorization Algebras in Quantum Field Theory}}}, Cambridge University Press, Nov 2016. \url{http://dx.doi.org/10.1017/9781316678626}.

\bibitem[CG21]{Costello2021}
\bysame, \doihref{http://dx.doi.org/10.1017/9781316678664}{\emph{{Factorization Algebras in Quantum Field Theory}}}, Cambridge University Press, Sep 2021. \url{http://dx.doi.org/10.1017/9781316678664}.

\bibitem[Che77]{Cheeger1977}
J.~Cheeger, \emph{{Analytic torsion and Reidemeister torsion}}, \href{http://dx.doi.org/10.1073/pnas.74.7.2651}{Proceedings of the National Academy of Sciences \textbf{74} (1977) 2651–2654}.

\bibitem[Che79]{Cheeger1979}
\bysame, \emph{{Analytic Torsion and The Heat Equation}}, \href{http://dx.doi.org/10.2307/1971113}{The Annals of Mathematics \textbf{109} (1979) 259}.

\bibitem[CHS91a]{Callan:1991ky}
C.~G. Callan, Jr., J.~A. Harvey, and A.~Strominger, \emph{{Worldbrane actions for string solitons}}, \doihref{http://dx.doi.org/10.1016/0550-3213(91)90041-U}{Nucl. Phys. B \textbf{367} (1991) 60--82}.

\bibitem[CHS91b]{Callan:1991dj}
\bysame, \emph{{World sheet approach to heterotic instantons and solitons}}, \doihref{http://dx.doi.org/10.1016/0550-3213(91)90074-8}{Nucl. Phys. B \textbf{359} (1991) 611--634}.

\bibitem[CHS91c]{Callan:1991at}
\bysame, \emph{{Supersymmetric String Solitons}}, \href{http://arxiv.org/abs/hep-th/9112030}{{\arxivfont arXiv:hep-th/9112030}}.

\bibitem[CJM{\etalchar{+}}05]{MR2174418}
A.~L. Carey, S.~Johnson, M.~K. Murray, D.~Stevenson, and B.-L. Wang, \emph{{Bundle gerbes for {C}hern-{S}imons and {W}ess-{Z}umino-{W}itten theories}}, \href{https://doi.org/10.1007/s00220-005-1376-8}{Comm. Math. Phys. \textbf{259} (2005) 577--613}.

\bibitem[CJS78]{Cremmer:1978km}
E.~Cremmer, B.~Julia, and J.~Scherk, \emph{{Supergravity Theory in 11 Dimensions}}, \doihref{http://dx.doi.org/10.1016/0370-2693(78)90894-8}{Phys. Lett. B \textbf{76} (1978) 409--412}.

\bibitem[CL20]{Chang:2020aww}
C.-M. Chang and Y.-H. Lin, \emph{{On exotic consistent anomalies in (1+1)$d$: A ghost story}}, \doihref{http://dx.doi.org/10.21468/SciPostPhys.10.5.119}{SciPost Phys. \textbf{10} (2021) 119}, \href{http://arxiv.org/abs/2009.07273}{{\arxivfont arXiv:2009.07273 [hep-th]}}.

\bibitem[CM96]{Carey1996}
A.~L. Carey and M.~K. Murray, \emph{{Faddeev’s anomaly and bundle gerbes}}, \href{http://dx.doi.org/10.1007/BF00400136}{Letters in Mathematical Physics \textbf{37} (1996) 29–36}.

\bibitem[CMM97]{Carey:1997xm}
A.~L. Carey, J.~Mickelsson, and M.~K. Murray, \emph{{Bundle Gerbes Applied to Quantum Field Theory}}, \doihref{http://dx.doi.org/10.1142/S0129055X00000046}{Rev. Math. Phys. \textbf{12} (2000) 65--90}, \href{http://arxiv.org/abs/hep-th/9711133}{{\arxivfont arXiv:hep-th/9711133}}.

\bibitem[CMR94a]{Cordes:1994sd}
S.~Cordes, G.~W. Moore, and S.~Ramgoolam, \emph{{Large N 2D Yang-Mills Theory and Topological String Theory}}, \doihref{http://dx.doi.org/10.1007/s002200050102}{Commun. Math. Phys. \textbf{185} (1997) 543--619}, \href{http://arxiv.org/abs/hep-th/9402107}{{\arxivfont arXiv:hep-th/9402107}}.

\bibitem[CMR94b]{Cordes:1994fc}
\bysame, \emph{{Lectures on 2D Yang-Mills Theory, Equivariant Cohomology and Topological Field Theories}}, \doihref{http://dx.doi.org/10.1016/0920-5632(95)00434-B}{Nucl. Phys. B Proc. Suppl. \textbf{41} (1995) 184--244}, \href{http://arxiv.org/abs/hep-th/9411210}{{\arxivfont arXiv:hep-th/9411210}}.

\bibitem[CMR15]{Cattaneo:2015vsa}
A.~S. Cattaneo, P.~Mnev, and N.~Reshetikhin, \emph{{Perturbative quantum gauge theories on manifolds with boundary}}, \doihref{http://dx.doi.org/10.1007/s00220-017-3031-6}{Commun. Math. Phys. \textbf{357} (2018) 631--730}, \href{http://arxiv.org/abs/1507.01221}{{\arxivfont arXiv:1507.01221 [math-ph]}}.

\bibitem[CMR{\etalchar{+}}25]{Cheng:2025ube}
M.~Cheng, S.~Musser, A.~Raz, N.~Seiberg, and T.~Senthil, \emph{{Ordering the topological order in the fractional quantum Hall effect}}, \href{http://arxiv.org/abs/2505.14767}{{\arxivfont arXiv:2505.14767 [cond-mat.str-el]}}.

\bibitem[CMRS23]{Cushing:2023rha}
J.~Cushing, G.~W. Moore, M.~Ro\v{c}ek, and V.~Saxena, \emph{{Superconformal Gravity And The Topology Of Diffeomorphism Groups}}, \href{http://arxiv.org/abs/2311.08394}{{\arxivfont arXiv:2311.08394 [hep-th]}}.

\bibitem[{COM}25]{CompactCollaboration}
{COMPACT Collaboration}, \emph{Compact collaboration: Collaboration for observations, models and predictions of anomalies and cosmic topology}, \url{https://github.com/CompactCollaboration}, 2025.

\bibitem[Cos10]{Costello:WGPart1}
K.~J. Costello, \emph{{A geometric construction of the Witten genus, I}}, \href{http://arxiv.org/abs/1006.5422}{{\arxivfont arXiv:1006.5422 [math.QA]}}.

\bibitem[Cos11a]{Costello2011}
K.~Costello, \doihref{http://dx.doi.org/10.1090/surv/170}{\emph{Renormalization and effective field theory}}, American Mathematical Society, March 2011. \url{http://dx.doi.org/10.1090/surv/170}.

\bibitem[Cos11b]{Costello:WGPart2}
K.~J. Costello, \emph{{A geometric construction of the Witten genus, II}}, \href{http://arxiv.org/abs/1112.0816}{{\arxivfont arXiv:1112.0816 [math.QA]}}.

\bibitem[Cos13a]{Costello:2013zra}
K.~Costello, \emph{{Supersymmetric gauge theory and the Yangian}}, \href{http://arxiv.org/abs/1303.2632}{{\arxivfont arXiv:1303.2632 [hep-th]}}.

\bibitem[Cos13b]{Costello:2013sla}
\bysame, \emph{{Integrable lattice models from four-dimensional field theories}}, \doihref{http://dx.doi.org/10.1090/pspum/088/01483}{Proc. Symp. Pure Math. \textbf{88} (2014) 3--24}, \href{http://arxiv.org/abs/1308.0370}{{\arxivfont arXiv:1308.0370 [hep-th]}}.

\bibitem[CP93]{Caetano:1993zf}
A.~Caetano and R.~F. Picken, \emph{{An Axiomatic definition of holonomy}}.

\bibitem[CR17]{Carqueville:2017fmn}
N.~Carqueville and I.~Runkel, \emph{{Introductory lectures on topological quantum field theory}}, \doihref{http://dx.doi.org/10.4064/bc114-1}{Banach Center Publ. \textbf{114} (2018) 9--47}, \href{http://arxiv.org/abs/1705.05734}{{\arxivfont arXiv:1705.05734 [math.QA]}}.

\bibitem[CS84]{Connes1984}
A.~Connes and G.~Skandalis, \emph{{The Longitudinal Index Theorem for Foliations}}, \href{http://dx.doi.org/10.2977/PRIMS/1195180375}{Publications of the Research Institute for Mathematical Sciences \textbf{20} (1984) 1139–1183}.

\bibitem[CS85]{CheegerSimons:1985}
J.~Cheeger and J.~Simons, \emph{{Differential characters and geometric invariants}}, p.~50–80, Springer Berlin Heidelberg, 1985. \url{http://dx.doi.org/10.1007/BFb0075216}.

\bibitem[CSS98]{Cornish:1997ab}
N.~J. Cornish, D.~N. Spergel, and G.~D. Starkman, \emph{{Circles in the sky: Finding topology with the microwave background radiation}}, \doihref{http://dx.doi.org/10.1088/0264-9381/15/9/013}{Class. Quant. Grav. \textbf{15} (1998) 2657--2670}, \href{http://arxiv.org/abs/astro-ph/9801212}{{\arxivfont arXiv:astro-ph/9801212}}.

\bibitem[CW89]{Chamseddine:1989yz}
A.~H. Chamseddine and D.~Wyler, \emph{{Gauge Theory of Topological Gravity in (1+1)-Dimensions}}, \doihref{http://dx.doi.org/10.1016/0370-2693(89)90528-5}{Phys. Lett. B \textbf{228} (1989) 75--78}.

\bibitem[CW90]{Chamseddine:1989wn}
\bysame, \emph{{Topological Gravity in (1+1)-dimensions}}, \doihref{http://dx.doi.org/10.1016/0550-3213(90)90460-U}{Nucl. Phys. B \textbf{340} (1990) 595--616}.

\bibitem[CWY17]{Costello:2017dso}
K.~Costello, E.~Witten, and M.~Yamazaki, \emph{{Gauge Theory and Integrability, I}}, \doihref{http://dx.doi.org/10.4310/ICCM.2018.v6.n1.a6}{ICCM Not. \textbf{06} (2018) 46--119}, \href{http://arxiv.org/abs/1709.09993}{{\arxivfont arXiv:1709.09993 [hep-th]}}.

\bibitem[CWY18]{Costello:2018gyb}
\bysame, \emph{{Gauge Theory and Integrability, II}}, \doihref{http://dx.doi.org/10.4310/ICCM.2018.v6.n1.a7}{ICCM Not. \textbf{06} (2018) 120--146}, \href{http://arxiv.org/abs/1802.01579}{{\arxivfont arXiv:1802.01579 [hep-th]}}.

\bibitem[DC09]{Duncan:2009ocf}
R.~Duncan and B.~Coecke, \emph{{Interacting quantum observables: categorical algebra and diagrammatics}}, \doihref{http://dx.doi.org/10.1088/1367-2630/13/4/043016}{New J. Phys. \textbf{13} (2011) 043016}, \href{http://arxiv.org/abs/0906.4725}{{\arxivfont arXiv:0906.4725 [quant-ph]}}.

\bibitem[Deb23]{Debray:2023kvh}
A.~Debray, \emph{{Differential cohomology (encyclopedia article)}}, \href{http://arxiv.org/abs/2312.14338}{{\arxivfont arXiv:2312.14338 [math.AT]}}.

\bibitem[Ded22]{Dedushenko:2022zwd}
M.~Dedushenko, \emph{{Snowmass White Paper: The Quest to Define QFT}}, \href{http://arxiv.org/abs/2203.08053}{{\arxivfont arXiv:2203.08053 [hep-th]}}.

\bibitem[Dev19]{Devalapurkar:2019}
S.~K. Devalapurkar, \doihref{http://dx.doi.org/10.48550/ARXIV.1911.10534}{\emph{{The Ando-Hopkins-Rezk orientation is surjective}}}, 2019. \url{https://arxiv.org/abs/1911.10534}.

\bibitem[DeW92]{DeWitt1992}
B.~DeWitt, \doihref{http://dx.doi.org/10.1017/cbo9780511564000}{\emph{Supermanifolds}}, Cambridge University Press, May 1992. \url{http://dx.doi.org/10.1017/CBO9780511564000}.

\bibitem[DF94]{Dai:1994kq}
X.-z. Dai and D.~S. Freed, \emph{{eta invariants and determinant lines}}, \doihref{http://dx.doi.org/10.1063/1.530747}{J. Math. Phys. \textbf{35} (1994) 5155--5194}, \href{http://arxiv.org/abs/hep-th/9405012}{{\arxivfont arXiv:hep-th/9405012}}. [Erratum: J.Math.Phys. 42, 2343--2344 (2001)].

\bibitem[DF99a]{DeligneFreed:1999QFT1}
P.~Deligne and D.~S. Freed, \emph{{Classical Field Theory}}, {Quantum Fields and Strings: A Course for Mathematicians, Vol. 1} (P.~Deligne, P.~Etingof, D.~S. Freed, L.~C. Jeffrey, D.~Kazhdan, J.~W. Morgan, D.~R. Morrison, and E.~Witten, eds.), American Mathematical Society, Providence, RI, 1999, pp.~137--225.

\bibitem[DF99b]{DeligneFreedSupersymmetry:1999}
\bysame, \emph{{Notes on Supersymmetry (following Joseph Bernstein)}}, {Quantum Fields and Strings: A Course for Mathematicians, Vol. 1} (P.~Deligne, P.~Etingof, D.~S. Freed, L.~C. Jeffrey, D.~Kazhdan, J.~W. Morgan, D.~R. Morrison, and E.~Witten, eds.), American Mathematical Society, Providence, RI, 1999, pp.~41--135.

\bibitem[DF99c]{Deligne:1999ur}
\bysame, \emph{{Supersolutions}}, \href{http://arxiv.org/abs/hep-th/9901094}{{\arxivfont arXiv:hep-th/9901094}}.

\bibitem[DFHH14]{Douglas2014-cl}
C.~L. Douglas, J.~Francis, A.~G. Henriques, and M.~A. Hill (eds.), \doihref{http://dx.doi.org/10.1090/surv/201}{\emph{{Topological Modular Forms}}}, Mathematical Surveys and Monographs, American Mathematical Society, Providence, RI, December 2014.

\bibitem[DFK02]{Donaldson2002}
S.~K. Donaldson, M.~Furuta, and D.~Kotschick, \doihref{http://dx.doi.org/10.1017/cbo9780511543098}{\emph{{Floer Homology Groups in Yang-Mills Theory}}}, Cambridge University Press, January 2002. \url{http://dx.doi.org/10.1017/CBO9780511543098}.

\bibitem[DFM03]{Diaconescu:2003bm}
E.~Diaconescu, D.~S. Freed, and G.~W. Moore, \emph{{The M-theory 3-form and E$_8$ gauge theory}}, \href{http://arxiv.org/abs/hep-th/0312069}{{\arxivfont arXiv:hep-th/0312069}}.

\bibitem[DFM09]{Distler:2009ri}
J.~Distler, D.~S. Freed, and G.~W. Moore, \emph{{Orientifold Precis}}, \href{http://arxiv.org/abs/0906.0795}{{\arxivfont arXiv:0906.0795 [hep-th]}}.

\bibitem[DFM10]{Distler:2010an}
J.~Distler, D.~S. Freed, and G.~W. Moore, \emph{{Spin structures and superstrings}}, \href{http://arxiv.org/abs/1007.4581}{{\arxivfont arXiv:1007.4581 [hep-th]}}.

\bibitem[DFMS97]{DiFrancesco:1997nk}
P.~Di~Francesco, P.~Mathieu, and D.~Senechal, \doihref{http://dx.doi.org/10.1007/978-1-4612-2256-9}{\emph{{Conformal Field Theory}}}, Graduate Texts in Contemporary Physics, Springer-Verlag, New York, 1997.

\bibitem[DG18]{Debray2018}
A.~Debray and S.~Gunningham, \emph{{The Arf-Brown TQFT of Pin$^-$ Surfaces}}, \href{http://dx.doi.org/10.1090/conm/718/14478}{Contemporary Mathematics (2018) 49–87}, \href{http://arxiv.org/abs/1803.11183}{{\arxivfont arXiv:1803.11183 [math-ph]}}.

\bibitem[DGRW20]{Davighi:2020vcm}
J.~Davighi, B.~Gripaios, and O.~Randal-Williams, \emph{{Differential cohomology and topological actions in physics}}, \doihref{http://dx.doi.org/10.4310/ATMP.2023.v27.n7.a3}{Adv. Theor. Math. Phys. \textbf{27} (2023) 2045--2085}, \href{http://arxiv.org/abs/2011.05768}{{\arxivfont arXiv:2011.05768 [hep-th]}}.

\bibitem[DHJF{\etalchar{+}}24]{Decoppet:2024htz}
T.~D. D\'ecoppet, P.~Huston, T.~Johnson-Freyd, D.~Nikshych, D.~Penneys, J.~Plavnik, D.~Reutter, and M.~Yu, \emph{{The Classification of Fusion 2-Categories}}, \href{http://arxiv.org/abs/2411.05907}{{\arxivfont arXiv:2411.05907 [math.CT]}}.

\bibitem[Dij89]{DijkgraafPhD}
R.~H. Dijkgraaf, \emph{{A geometrical approach to two-dimensional Conformal Field Theory }}, Ph.D. thesis, 1989. \url{https://dspace.library.uu.nl/handle/1874/210872}.

\bibitem[Dim16]{Dimofte:2016pua}
T.~Dimofte, \emph{{Perturbative and nonperturbative aspects of complex Chern\textendash{}Simons theory}}, \doihref{http://dx.doi.org/10.1088/1751-8121/aa6a5b}{J. Phys. A \textbf{50} (2017) 443009}, \href{http://arxiv.org/abs/1608.02961}{{\arxivfont arXiv:1608.02961 [hep-th]}}.

\bibitem[DK70]{Donovan1970}
P.~Donovan and M.~Karoubi, \emph{{Graded brauer groups and K-theory with local coefficients}}, \href{http://dx.doi.org/10.1007/BF02684650}{Publications mathématiques de l’IHÉS \textbf{38} (1970) 5–25}.

\bibitem[DK01]{Davis2001}
J.~Davis and P.~Kirk, \doihref{http://dx.doi.org/10.1090/gsm/035}{\emph{{Lecture Notes in Algebraic Topology}}}, American Mathematical Society, July 2001. \url{http://dx.doi.org/10.1090/gsm/035}.

\bibitem[DKR11]{Davydov:2011kb}
A.~Davydov, L.~Kong, and I.~Runkel, \emph{{Field theories with defects and the centre functor}}, \href{http://arxiv.org/abs/1107.0495}{{\arxivfont arXiv:1107.0495 [math.QA]}}.

\bibitem[DLM95]{Duff:1995wd}
M.~J. Duff, J.~T. Liu, and R.~Minasian, \emph{{Eleven Dimensional Origin of String/String Duality: A One Loop Test}}, \doihref{http://dx.doi.org/10.1201/9781482268737-17}{Nucl. Phys. B \textbf{452} (1995) 261--282}, \href{http://arxiv.org/abs/hep-th/9506126}{{\arxivfont arXiv:hep-th/9506126}}.

\bibitem[dlM05]{Madrid2005}
R.~de~la Madrid, \emph{{The role of the rigged Hilbert space in quantum mechanics}}, \href{http://dx.doi.org/10.1088/0143-0807/26/2/008}{European Journal of Physics \textbf{26} (2005) 287–312}.

\bibitem[DMW00a]{Diaconescu:2000wy}
D.-E. Diaconescu, G.~W. Moore, and E.~Witten, \emph{{E$_8$ Gauge Theory, and a Derivation of K-Theory from M-Theory}}, \doihref{http://dx.doi.org/10.4310/ATMP.2002.v6.n6.a2}{Adv. Theor. Math. Phys. \textbf{6} (2003) 1031--1134}, \href{http://arxiv.org/abs/hep-th/0005090}{{\arxivfont arXiv:hep-th/0005090}}.

\bibitem[DMW00b]{Diaconescu:2000wz}
\bysame, \emph{{A Derivation of K-Theory from M-Theory}}, \href{http://arxiv.org/abs/hep-th/0005091}{{\arxivfont arXiv:hep-th/0005091}}.

\bibitem[DN21]{Dedushenko:2021mds}
M.~Dedushenko and N.~Nekrasov, \emph{{Interfaces and Quantum Algebras, I: Stable Envelopes}}, \doihref{http://dx.doi.org/10.1016/j.geomphys.2023.104991}{J. Geom. Phys. \textbf{194} (2023) 104991}, \href{http://arxiv.org/abs/2109.10941}{{\arxivfont arXiv:2109.10941 [hep-th]}}.

\bibitem[DN23]{Dedushenko:2023qjq}
\bysame, \emph{{Interfaces and Quantum Algebras, II: Cigar Partition Function}}, \href{http://arxiv.org/abs/2306.16434}{{\arxivfont arXiv:2306.16434 [hep-th]}}.

\bibitem[dR84]{deRham1984}
G.~de~Rham, \doihref{http://dx.doi.org/10.1007/978-3-642-61752-2}{\emph{{Differentiable Manifolds}}}, Springer Berlin Heidelberg, 1984. \url{http://dx.doi.org/10.1007/978-3-642-61752-2}.

\bibitem[DR18]{Douglas:2018qfz}
C.~L. Douglas and D.~J. Reutter, \emph{{Fusion 2-categories and a state-sum invariant for 4-manifolds}}, \href{http://arxiv.org/abs/1812.11933}{{\arxivfont arXiv:1812.11933 [math.QA]}}.

\bibitem[DS05]{DiamondShurman:2005}
F.~Diamond and J.~Shurman, \doihref{http://dx.doi.org/10.1007/978-0-387-27226-9}{\emph{{A First Course in Modular Forms}}}, Graduate texts in mathematics, Springer Science+Business Media, January 2005 (en).

\bibitem[Dum23a]{Dumitrescu:TASIVideoLec1}
T.~Dumitrescu, \emph{{Lecture 1 on Generalized Symmetries in QFT (TASI 2023)}}, YouTube, 2023. \url{https://youtu.be/keD_9IwnDWU}. {Accessed: 2023-06-16}.

\bibitem[Dum23b]{Dumitrescu:TASIVideoLec2}
\bysame, \emph{{Lecture 2 on Generalized Symmetries in QFT (TASI 2023)}}, YouTube, 2023. \url{https://youtu.be/HS_WSeEwpj8}. {Accessed: 2023-06-16}.

\bibitem[Dum23c]{Dumitrescu:TASIVideoLec3}
\bysame, \emph{{Lecture 3 on Generalized Symmetries in QFT (TASI 2023)}}, YouTube, 2023. \url{https://youtu.be/1zsPqQaU2cI}. {Accessed: 2023-06-16}.

\bibitem[Dum23d]{Dumitrescu:TASIVideoLec4}
\bysame, \emph{{Lecture 4 on Generalized Symmetries in QFT (TASI 2023)}}, YouTube, 2023. \url{https://youtu.be/M247KJ2_JPM}. {Accessed: 2023-06-16}.

\bibitem[Dup78]{Dupont:1978}
J.~L. Dupont, \doihref{http://dx.doi.org/10.1007/bfb0065364}{\emph{{Curvature and Characteristic Classes}}}, Springer Berlin Heidelberg, 1978. \url{http://dx.doi.org/10.1007/BFb0065364}.

\bibitem[DW90]{Dijkgraaf:1989pz}
R.~Dijkgraaf and E.~Witten, \emph{{Topological Gauge Theories and Group Cohomology}}, \doihref{http://dx.doi.org/10.1007/BF02096988}{Commun. Math. Phys. \textbf{129} (1990) 393}.

\bibitem[DY24]{Debray:2024czl}
A.~Debray and M.~Yu, \emph{{Type IIA String Theory and tmf with Level Structure}}, \href{http://arxiv.org/abs/2411.07299}{{\arxivfont arXiv:2411.07299 [math.AT]}}.

\bibitem[DZDRG24]{DelZotto:2024ngj}
M.~Del~Zotto, M.~Dell'Acqua, and E.~Riedel~G\r{a}rding, \emph{{The Higher Structure of Symmetries of Axion-Maxwell Theory}}, \href{http://arxiv.org/abs/2411.09685}{{\arxivfont arXiv:2411.09685 [hep-th]}}.

\bibitem[DZGE22]{DelZotto:2022ras}
M.~Del~Zotto and I.~Garc{\'\i}a~Etxebarria, \emph{{Global Structures from the Infrared}}, \doihref{http://dx.doi.org/10.1007/JHEP11(2023)058}{JHEP \textbf{11} (2023) 058}, \href{http://arxiv.org/abs/2204.06495}{{\arxivfont arXiv:2204.06495 [hep-th]}}.

\bibitem[EGNO16]{Etingof2016-za}
P.~Etingof, S.~Gelaki, D.~Nikshych, and V.~Ostrik, \emph{{Tensor Categories}}, Mathematical Surveys and Monographs, American Mathematical Society, Providence, RI, October 2016.

\bibitem[EMSS89]{Elitzur:1989nr}
S.~Elitzur, G.~W. Moore, A.~Schwimmer, and N.~Seiberg, \emph{{Remarks on the Canonical Quantization of the Chern-Simons-Witten Theory}}, \doihref{http://dx.doi.org/10.1016/0550-3213(89)90436-7}{Nucl. Phys. B \textbf{326} (1989) 108--134}.

\bibitem[ES45]{EilenbergSteenrod:1945}
S.~Eilenberg and N.~E. Steenrod, \emph{{Axiomatic Approach to Homology Theory}}, \href{http://www.jstor.org/stable/87896}{{Proceedings of the National Academy of Sciences of the United States of America} \textbf{31} (1945) 117--120}.

\bibitem[ES20]{Eager:2020rra}
R.~Eager and E.~Sharpe, \emph{{Elliptic Genera of Pure Gauge Theories in Two Dimensions with Semisimple Non-Simply-Connected Gauge Groups}}, \doihref{http://dx.doi.org/10.1007/s00220-021-04189-6}{Commun. Math. Phys. \textbf{387} (2021) 267--297}, \href{http://arxiv.org/abs/2009.03907}{{\arxivfont arXiv:2009.03907 [hep-th]}}.

\bibitem[Evs06]{Evslin:2006cj}
J.~Evslin, \emph{{What Does(n't) K-theory Classify?}}, \href{http://arxiv.org/abs/hep-th/0610328}{{\arxivfont arXiv:hep-th/0610328}}.

\bibitem[Fad84]{FADDEEV198481}
L.~Faddeev, \emph{{Operator anomaly for the Gauss law}}, \href{https://www.sciencedirect.com/science/article/pii/0370269384909523}{Physics Letters B \textbf{145} (1984) 81--84}.

\bibitem[FFRS04]{Frohlich:2004ef}
J.~Frohlich, J.~Fuchs, I.~Runkel, and C.~Schweigert, \emph{{Kramers-Wannier duality from conformal defects}}, \doihref{http://dx.doi.org/10.1103/PhysRevLett.93.070601}{Phys. Rev. Lett. \textbf{93} (2004) 070601}, \href{http://arxiv.org/abs/cond-mat/0404051}{{\arxivfont arXiv:cond-mat/0404051}}.

\bibitem[FG91]{Freed1991}
D.~S. Freed and R.~E. Gompf, \emph{{Computer calculation of Witten’s 3-manifold invariant}}, \href{http://dx.doi.org/10.1007/BF02100006}{Communications in Mathematical Physics \textbf{141} (1991) 79–117}.

\bibitem[FH00]{Fabinger:2000jd}
M.~Fabinger and P.~Ho\v{r}ava, \emph{{Casimir Effect Between World-Branes in Heterotic M-Theory}}, \doihref{http://dx.doi.org/10.1016/S0550-3213(00)00255-8}{Nucl. Phys. B \textbf{580} (2000) 243--263}, \href{http://arxiv.org/abs/hep-th/0002073}{{\arxivfont arXiv:hep-th/0002073}}.

\bibitem[FH13]{Freed:2013gc}
D.~S. Freed and M.~J. Hopkins, \emph{{Chern-Weil forms and abstract homotopy theory}}, \href{http://arxiv.org/abs/1301.5959}{{\arxivfont arXiv:1301.5959 [math.DG]}}.

\bibitem[FH16]{Freed:2016rqq}
D.~S. Freed and M.~J. Hopkins, \emph{{Reflection positivity and invertible topological phases}}, \doihref{http://dx.doi.org/10.2140/gt.2021.25.1165}{Geom. Topol. \textbf{25} (2021) 1165--1330}, \href{http://arxiv.org/abs/1604.06527}{{\arxivfont arXiv:1604.06527 [hep-th]}}.

\bibitem[FH19]{Freed:2019sco}
\bysame, \emph{{Consistency of M-Theory on Non-Orientable Manifolds}}, \doihref{http://dx.doi.org/10.1093/qmath/haab007}{Quart. J. Math. Oxford Ser. \textbf{72} (2021) 603--671}, \href{http://arxiv.org/abs/1908.09916}{{\arxivfont arXiv:1908.09916 [hep-th]}}.

\bibitem[FHLT09]{Freed:2009qp}
D.~S. Freed, M.~J. Hopkins, J.~Lurie, and C.~Teleman, \emph{{Topological Quantum Field Theories from Compact Lie Groups}}, {A Celebration of Raoul Bott's Legacy in Mathematics}, May 2009. \href{http://arxiv.org/abs/0905.0731}{{\arxivfont arXiv:0905.0731 [math.AT]}}.

\bibitem[FHT02]{Freed:2002gr}
D.~S. Freed, M.~J. Hopkins, and C.~Teleman, \emph{{Twisted equivariant K-theory with complex coefficients}}, \href{http://arxiv.org/abs/math/0206257}{{\arxivfont arXiv:math/0206257}}.

\bibitem[FHT03]{Freed:2003qx}
\bysame, \emph{{Twisted K-theory and loop group representations. 1.}}, \href{http://arxiv.org/abs/math/0312155}{{\arxivfont arXiv:math/0312155}}.

\bibitem[FHT05]{Freed:2005qu}
\bysame, \emph{{Loop groups and twisted K-theory. II.}}, \href{http://arxiv.org/abs/math/0511232}{{\arxivfont arXiv:math/0511232}}.

\bibitem[FHT07]{Freed:2007wja}
\bysame, \emph{{Loop groups and twisted K-theory I}}, \doihref{http://dx.doi.org/10.1112/jtopol/jtr019}{J. Topol. \textbf{4} (2011) 737--798}, \href{http://arxiv.org/abs/0711.1906}{{\arxivfont arXiv:0711.1906 [math.AT]}}.

\bibitem[FK85]{Fukuyama:1985gg}
T.~Fukuyama and K.~Kamimura, \emph{{Gauge Theory of Two-dimensional Gravity}}, \doihref{http://dx.doi.org/10.1016/0370-2693(85)91322-X}{Phys. Lett. B \textbf{160} (1985) 259--262}.

\bibitem[FM04]{Freed:2004yc}
D.~S. Freed and G.~W. Moore, \emph{{Setting the quantum integrand of M-theory}}, \doihref{http://dx.doi.org/10.1007/s00220-005-1482-7}{Commun. Math. Phys. \textbf{263} (2006) 89--132}, \href{http://arxiv.org/abs/hep-th/0409135}{{\arxivfont arXiv:hep-th/0409135}}.

\bibitem[FM12]{Freed:2012uu}
\bysame, \emph{{Twisted equivariant matter}}, \doihref{http://dx.doi.org/10.1007/s00023-013-0236-x}{Annales Henri Poincare \textbf{14} (2013) 1927--2023}, \href{http://arxiv.org/abs/1208.5055}{{\arxivfont arXiv:1208.5055 [hep-th]}}.

\bibitem[FMS06a]{Freed:2006ya}
D.~S. Freed, G.~W. Moore, and G.~Segal, \emph{{The Uncertainty of Fluxes}}, \doihref{http://dx.doi.org/10.1007/s00220-006-0181-3}{Commun. Math. Phys. \textbf{271} (2007) 247--274}, \href{http://arxiv.org/abs/hep-th/0605198}{{\arxivfont arXiv:hep-th/0605198}}.

\bibitem[FMS06b]{Freed:2006yc}
\bysame, \emph{{Heisenberg Groups and Noncommutative Fluxes}}, \doihref{http://dx.doi.org/10.1016/j.aop.2006.07.014}{Annals Phys. \textbf{322} (2007) 236--285}, \href{http://arxiv.org/abs/hep-th/0605200}{{\arxivfont arXiv:hep-th/0605200}}.

\bibitem[FMT22]{Freed:2022qnc}
D.~S. Freed, G.~W. Moore, and C.~Teleman, \emph{Topological symmetry in quantum field theory}, \href{http://dx.doi.org/10.4171/QT/223}{Quantum Topology \textbf{15} (2024) 779–869}, \href{http://arxiv.org/abs/2209.07471}{{\arxivfont arXiv:2209.07471 [hep-th]}}.

\bibitem[FMT24]{FMT:Unpublished}
\bysame, Unpublished, 2024.

\bibitem[FN22]{Freed:2021anp}
D.~S. Freed and A.~Neitzke, \emph{{3d spectral networks and classical Chern\textendash{}Simons theory}}, \doihref{http://dx.doi.org/10.4310/sdg.2021.v26.n1.a4}{Surveys Diff. Geom. \textbf{26} (2021) 51--155}, \href{http://arxiv.org/abs/2208.07420}{{\arxivfont arXiv:2208.07420 [math.DG]}}.

\bibitem[Fre87]{Freed:DiracNotes1987}
D.~S. Freed, \emph{{Geometry of Dirac Operators}}, Lectures at U. Chicago, 1987. \url{https://people.math.harvard.edu/~dafr/DiracNotes.pdf}.

\bibitem[Fre92a]{Freed1992}
\bysame, \emph{{Reidemeister torsion, spectral sequences, and Brieskorn spheres}}, \href{http://eudml.org/doc/153432}{Journal f\"ur die reine und angewandte Mathematik \textbf{429} (1992) 75--90}.

\bibitem[Fre92b]{Freed:1994ad}
\bysame, \emph{{Higher Algebraic Structures and Quantization}}, \doihref{http://dx.doi.org/10.1007/BF02102643}{Commun. Math. Phys. \textbf{159} (1994) 343--398}, \href{http://arxiv.org/abs/hep-th/9212115}{{\arxivfont arXiv:hep-th/9212115}}.

\bibitem[Fre99]{Freed:1999mn}
D.~S. Freed, \emph{{Five lectures on supersymmetry}}, AMS, Providence, USA, 1999.

\bibitem[Fre00]{Freed:2000ta}
D.~S. Freed, \emph{{Dirac Charge Quantization and Generalized Differential Cohomology}}, November 2000. \href{http://arxiv.org/abs/hep-th/0011220}{{\arxivfont arXiv:hep-th/0011220}}.

\bibitem[Fre06]{Freed:2006mx}
\bysame, \emph{{Pions and Generalized Cohomology}}, J. Diff. Geom. \textbf{80} (2008) 45--77, \href{http://arxiv.org/abs/hep-th/0607134}{{\arxivfont arXiv:hep-th/0607134}}.

\bibitem[Fre08]{Freed:2008jq}
\bysame, \emph{{Remarks on Chern-Simons Theory}}, \href{http://arxiv.org/abs/0808.2507}{{\arxivfont arXiv:0808.2507 [math.AT]}}.

\bibitem[Fre12a]{FreedBordism}
\bysame, \emph{{Bordism: Old And New}}, 2012. \url{https://people.math.harvard.edu/~dafr/bordism.pdf}.

\bibitem[Fre12b]{Freed:CobordismHypothesis}
\bysame, \emph{{The cobordism hypothesis}}, \href{http://arxiv.org/abs/1210.5100}{{\arxivfont arXiv:1210.5100 [math.AT]}}.

\bibitem[Fre19]{FreedCBMS}
\bysame, \doihref{http://dx.doi.org/10.1090/cbms/133}{\emph{{Lectures on field theory and topology}}}, CBMS Regional Conference Series in Mathematics, vol. 133, American Mathematical Society, Providence, RI, 2019. \url{https://doi.org/10.1090/cbms/133}. Published for the Conference Board of the Mathematical Sciences.

\bibitem[Fre22a]{Freed:FiniteSymmetry2022}
\bysame, \emph{{Four Lectures on Finite Symmetry in QFT}}, 2022. \url{https://people.math.harvard.edu/~dafr/Freed_perim.pdf}. Lecture notes from the ``Global Categorical Symmetries''' workshop, Perimeter Institute, June 2022.

\bibitem[Fre22b]{Freed:2022iao}
\bysame, \emph{{Introduction to topological symmetry in QFT}}, \doihref{http://dx.doi.org/10.1090/pspum/107/01946}{Proc. Symp. Pure Math. \textbf{107} (2024) 93--106}, \href{http://arxiv.org/abs/2212.00195}{{\arxivfont arXiv:2212.00195 [hep-th]}}.

\bibitem[Fre23a]{Freed:QTGV}
\bysame, \emph{{Quantum Field Theory From A Geometric Viewpoint}}, \href{https://people.math.harvard.edu/~dafr/GQT12.pdf}{Harvard Lectures (2023) }. [Online; accessed 3-February-2025].

\bibitem[Fre23b]{Freed:2023snr}
\bysame, \emph{{What is an anomaly?}}, \href{http://arxiv.org/abs/2307.08147}{{\arxivfont arXiv:2307.08147 [hep-th]}}.

\bibitem[Fre24]{Freed:2024heu}
\bysame, \emph{{Index theory on Pin manifolds}}, \doihref{http://dx.doi.org/10.1090/bull/1856}{Bull. Am. Math. Soc., New Ser. \textbf{62} (2025) 47--65}, \href{http://arxiv.org/abs/2405.20464}{{\arxivfont arXiv:2405.20464 [math.DG]}}.

\bibitem[Fri80]{Friedan:1980jf}
D.~Friedan, \emph{{Nonlinear Models in $2 + \varepsilon$ Dimensions}}, \doihref{http://dx.doi.org/10.1103/PhysRevLett.45.1057}{Phys. Rev. Lett. \textbf{45} (1980) 1057}.

\bibitem[Fri85]{Friedan:1980jm}
D.~H. Friedan, \emph{{Nonlinear models in $2 + \varepsilon$ dimensions}}, \doihref{http://dx.doi.org/10.1016/0003-4916(85)90384-7}{Annals Phys. \textbf{163} (1985) 318}.

\bibitem[Friar]{Friedman:2008}
G.~Friedman, \emph{{An elementary illustrated introduction to simplicial sets}}, \href{http://arxiv.org/abs/arXiv:0809.4221}{{\arxivfont arXiv:0809.4221}}.

\bibitem[Fri23]{Friedan:2023vxx}
D.~Friedan, \emph{{Global structure of Euclidean quantum gravity}}, \href{http://arxiv.org/abs/2306.00019}{{\arxivfont arXiv:2306.00019 [hep-th]}}.

\bibitem[FS84]{Faddeev1984}
L.~D. Faddeev and S.~L. Shatashvili, \emph{{Algebraic and Hamiltonian methods in the theory of non-Abelian anomalies}}, \href{http://dx.doi.org/10.1007/BF01018976}{Theoretical and Mathematical Physics \textbf{60} (1984) 770–778}.

\bibitem[FSS20a]{Fiorenza:2020hiq}
D.~Fiorenza, H.~Sati, and U.~Schreiber, \emph{{Twisted Cohomotopy implies twisted String structure on M5-branes}}, \doihref{http://dx.doi.org/10.1063/5.0037786}{J. Math. Phys. \textbf{62} (2021) 042301}, \href{http://arxiv.org/abs/2002.11093}{{\arxivfont arXiv:2002.11093 [hep-th]}}.

\bibitem[FSS20b]{Fiorenza:2020iax}
\bysame, \emph{{Twistorial cohomotopy implies Green-Schwarz anomaly cancellation}}, \doihref{http://dx.doi.org/10.1142/S0129055X22500131}{Rev. Math. Phys. \textbf{34} (2022) 2250013}, \href{http://arxiv.org/abs/2008.08544}{{\arxivfont arXiv:2008.08544 [hep-th]}}.

\bibitem[FT12]{Freed:2012bs}
D.~S. Freed and C.~Teleman, \emph{{Relative quantum field theory}}, \doihref{http://dx.doi.org/10.1007/s00220-013-1880-1}{Commun. Math. Phys. \textbf{326} (2014) 459--476}, \href{http://arxiv.org/abs/1212.1692}{{\arxivfont arXiv:1212.1692 [hep-th]}}.

\bibitem[FT18]{Freed:2018cec}
\bysame, \emph{{Topological dualities in the Ising model}}, \doihref{http://dx.doi.org/10.2140/gt.2022.26.1907}{Geom. Topol. \textbf{26} (2022) 1907--1984}, \href{http://arxiv.org/abs/1806.00008}{{\arxivfont arXiv:1806.00008 [math.AT]}}.

\bibitem[Fuc95]{Fuchs:1992nq}
J.~Fuchs, \emph{{Affine Lie algebras and quantum groups: An Introduction, with Applications in Conformal Field Theory}}, Cambridge Monographs on Mathematical Physics, Cambridge University Press, Cambridge, England, March 1995.

\bibitem[G\"60]{Gursey1960}
F.~G\"{u}rsey, \emph{{On the symmetries of strong and weak interactions}}, \href{http://dx.doi.org/10.1007/BF02860276}{Il Nuovo Cimento \textbf{16} (1960) 230–240}.

\bibitem[Gal08]{Gallier:2008}
J.~Gallier, \emph{{The Classification Theorem for Compact Surfaces And A Detour On Fractals}}, \href{http://arxiv.org/abs/0805.0562}{{\arxivfont arXiv:0805.0562 [math.GM]}}.

\bibitem[Gaw87]{Gawedzki:1987ak}
K.~Gawedzki, \emph{{TOPOLOGICAL ACTIONS IN TWO-DIMENSIONAL QUANTUM FIELD THEORIES}}, {Proceedings of the 1987 Cargese meeting}, 1987, pp.~101--141.

\bibitem[Gei24]{Geiko:2024cra}
R.~Geiko, \emph{{Parametrized topological phases in 1d and T-duality}}, \href{http://arxiv.org/abs/2412.20905}{{\arxivfont arXiv:2412.20905 [math-ph]}}.

\bibitem[GGE24]{Gagliano:2024off}
F.~Gagliano and I.~Garc\'\i{}a~Etxebarria, \emph{{SymTFTs for $U(1)$ symmetries from descent}}, \href{http://arxiv.org/abs/2411.15126}{{\arxivfont arXiv:2411.15126 [hep-th]}}.

\bibitem[GGRS01]{Gates:1983nr}
S.~J. Gates, M.~T. Grisaru, M.~Ro\v{c}ek, and W.~Siegel, \emph{{Superspace Or One Thousand and One Lessons in Supersymmetry}}, Frontiers in Physics, vol.~58, 1983. \href{http://arxiv.org/abs/hep-th/0108200}{{\arxivfont arXiv:hep-th/0108200}}.

\bibitem[GJF18]{Gaiotto:2018ypj}
D.~Gaiotto and T.~Johnson-Freyd, \emph{{Holomorphic SCFTs with small index}}, \doihref{http://dx.doi.org/10.4153/S0008414X2100002X}{Can. J. Math. \textbf{74} (2022) 573--601}, \href{http://arxiv.org/abs/1811.00589}{{\arxivfont arXiv:1811.00589 [hep-th]}}.

\bibitem[GJF19]{Gaiotto:2019gef}
\bysame, \emph{{Mock modularity and a secondary elliptic genus}}, \doihref{http://dx.doi.org/10.1007/JHEP08(2023)094}{JHEP \textbf{08} (2023) 094}, \href{http://arxiv.org/abs/1904.05788}{{\arxivfont arXiv:1904.05788 [hep-th]}}.

\bibitem[GJFW19]{Gaiotto:2019asa}
D.~Gaiotto, T.~Johnson-Freyd, and E.~Witten, \emph{{A Note On Some Minimally Supersymmetric Models In Two Dimensions}}, 2021. \href{http://arxiv.org/abs/1902.10249}{{\arxivfont arXiv:1902.10249 [hep-th]}}.

\bibitem[GK93]{Gegenberg:1993gd}
J.~Gegenberg and G.~Kunstatter, \emph{{The Partition Function for Topological Field Theories}}, \doihref{http://dx.doi.org/10.1006/aphy.1994.1043}{Annals Phys. \textbf{231} (1994) 270--289}, \href{http://arxiv.org/abs/hep-th/9304016}{{\arxivfont arXiv:hep-th/9304016}}.

\bibitem[GK20]{Gaiotto:2020iye}
D.~Gaiotto and J.~Kulp, \emph{{Orbifold groupoids}}, \doihref{http://dx.doi.org/10.1007/JHEP02(2021)132}{JHEP \textbf{02} (2021) 132}, \href{http://arxiv.org/abs/2008.05960}{{\arxivfont arXiv:2008.05960 [hep-th]}}.

\bibitem[GKKS17]{Gaiotto:2017yup}
D.~Gaiotto, A.~Kapustin, Z.~Komargodski, and N.~Seiberg, \emph{{Theta, Time Reversal, and Temperature}}, \doihref{http://dx.doi.org/10.1007/JHEP05(2017)091}{JHEP \textbf{05} (2017) 091}, \href{http://arxiv.org/abs/1703.00501}{{\arxivfont arXiv:1703.00501 [hep-th]}}.

\bibitem[GKSW14]{Gaiotto:2014kfa}
D.~Gaiotto, A.~Kapustin, N.~Seiberg, and B.~Willett, \emph{{Generalized Global Symmetries}}, \doihref{http://dx.doi.org/10.1007/JHEP02(2015)172}{JHEP \textbf{02} (2015) 172}, \href{http://arxiv.org/abs/1412.5148}{{\arxivfont arXiv:1412.5148 [hep-th]}}.

\bibitem[GL25]{Gritskov:2025mee}
M.~Gritskov and A.~Losev, \emph{{Beta function without UV divergences}}, \href{http://arxiv.org/abs/2504.12483}{{\arxivfont arXiv:2504.12483 [math-ph]}}.

\bibitem[GLSS21]{Gorantla:2021svj}
P.~Gorantla, H.~T. Lam, N.~Seiberg, and S.-H. Shao, \emph{{A Modified Villain Formulation of Fractons and Other Exotic Theories}}, \doihref{http://dx.doi.org/10.1063/5.0060808}{J. Math. Phys. \textbf{62} (2021) 102301}, \href{http://arxiv.org/abs/2103.01257}{{\arxivfont arXiv:2103.01257 [cond-mat.str-el]}}.

\bibitem[GM03]{Gelfand2003}
S.~I. Gelfand and Y.~I. Manin, \doihref{http://dx.doi.org/10.1007/978-3-662-12492-5}{\emph{Methods of homological algebra}}, Springer Berlin Heidelberg, 2003. \url{http://dx.doi.org/10.1007/978-3-662-12492-5}.

\bibitem[GM20]{GepnerMeier:2020}
D.~Gepner and L.~Meier, \emph{{On equivariant topological modular forms}}, \href{http://arxiv.org/abs/2004.10254}{{\arxivfont arXiv:2004.10254 [math.AT]}}.

\bibitem[GML60]{GellMann1960}
M.~Gell-Mann and M.~Lévy, \emph{{The axial vector current in beta decay}}, \href{http://dx.doi.org/10.1007/BF02859738}{Il Nuovo Cimento \textbf{16} (1960) 705–726}.

\bibitem[GMN10]{Gaiotto:2010be}
D.~Gaiotto, G.~W. Moore, and A.~Neitzke, \emph{{Framed BPS States}}, \doihref{http://dx.doi.org/10.4310/ATMP.2013.v17.n2.a1}{Adv. Theor. Math. Phys. \textbf{17} (2013) 241--397}, \href{http://arxiv.org/abs/1006.0146}{{\arxivfont arXiv:1006.0146 [hep-th]}}.

\bibitem[GMNY23]{Gaiotto:2023ezy}
D.~Gaiotto, G.~W. Moore, A.~Neitzke, and F.~Yan, \emph{{Commuting Line Defects At $q^N=1$}}, \href{http://arxiv.org/abs/2307.14429}{{\arxivfont arXiv:2307.14429 [hep-th]}}.

\bibitem[GMW15a]{Gaiotto:2015zna}
D.~Gaiotto, G.~W. Moore, and E.~Witten, \emph{{An Introduction To The Web-Based Formalism}}, \href{http://arxiv.org/abs/1506.04086}{{\arxivfont arXiv:1506.04086 [hep-th]}}.

\bibitem[GMW15b]{Gaiotto:2015aoa}
\bysame, \emph{{Algebra of the Infrared: String Field Theoretic Structures in Massive ${\cal N}=(2,2)$ Field Theory In Two Dimensions}}, \href{http://arxiv.org/abs/1506.04087}{{\arxivfont arXiv:1506.04087 [hep-th]}}.

\bibitem[Goe09]{Goerss:2009}
P.~G. Goerss, \emph{{Topological modular forms (after Hopkins, Miller, and Lurie)}}, \href{http://arxiv.org/abs/0910.5130}{{\arxivfont arXiv:0910.5130 [math.AT]}}.

\bibitem[Gom17]{Gomi:2017ymm}
K.~Gomi, \emph{{Freed-Moore K-theory}}, \href{http://arxiv.org/abs/1705.09134}{{\arxivfont arXiv:1705.09134 [math.KT]}}.

\bibitem[GP20]{Grady:2020sxl}
D.~Grady and D.~Pavlov, \emph{{Extended field theories are local and have classifying spaces}}, \href{http://arxiv.org/abs/2011.01208}{{\arxivfont arXiv:2011.01208 [math.AT]}}.

\bibitem[GP21]{Grady:2021kii}
\bysame, \emph{{The geometric cobordism hypothesis}}, \href{http://arxiv.org/abs/2111.01095}{{\arxivfont arXiv:2111.01095 [math.AT]}}.

\bibitem[GPPV18]{Gukov:2018iiq}
S.~Gukov, D.~Pei, P.~Putrov, and C.~Vafa, \emph{{4-manifolds and topological modular forms}}, \doihref{http://dx.doi.org/10.1007/JHEP05(2021)084}{JHEP \textbf{05} (2021) 084}, \href{http://arxiv.org/abs/1811.07884}{{\arxivfont arXiv:1811.07884 [hep-th]}}.

\bibitem[GR22]{Gromov:2022cxa}
A.~Gromov and L.~Radzihovsky, \emph{{Colloquium: Fracton matter}}, \doihref{http://dx.doi.org/10.1103/RevModPhys.96.011001}{Rev. Mod. Phys. \textbf{96} (2024) 011001}, \href{http://arxiv.org/abs/2211.05130}{{\arxivfont arXiv:2211.05130 [cond-mat.str-el]}}.

\bibitem[GRW98]{Gukov:1998kn}
S.~Gukov, M.~Rangamani, and E.~Witten, \emph{{Dibaryons, strings and branes in AdS orbifold models}}, \doihref{http://dx.doi.org/10.1088/1126-6708/1998/12/025}{JHEP \textbf{12} (1998) 025}, \href{http://arxiv.org/abs/hep-th/9811048}{{\arxivfont arXiv:hep-th/9811048}}.

\bibitem[GS18]{Gu:2018fpm}
W.~Gu and E.~Sharpe, \emph{{A proposal for nonabelian mirrors}}, \href{http://arxiv.org/abs/1806.04678}{{\arxivfont arXiv:1806.04678 [hep-th]}}.

\bibitem[GST24]{Gorantla:2024ocs}
P.~Gorantla, S.-H. Shao, and N.~Tantivasadakarn, \emph{{Tensor networks for non-invertible symmetries in 3+1d and beyond}}, \href{http://arxiv.org/abs/2406.12978}{{\arxivfont arXiv:2406.12978 [quant-ph]}}.

\bibitem[GSZ21]{F4Diagrammatics}
R.~Gandhi, A.~Savage, and K.~Zainoulline, \emph{{Diagrammatics for $F_4$}}, \href{http://arxiv.org/abs/2107.12464}{{\arxivfont arXiv:2107.12464 [math.RT]}}.

\bibitem[GT93]{Gross:1993hu}
D.~J. Gross and W.~Taylor, \emph{{Two Dimensional QCD is a String Theory}}, \doihref{http://dx.doi.org/10.1016/0550-3213(93)90403-C}{Nucl. Phys. B \textbf{400} (1993) 181--208}, \href{http://arxiv.org/abs/hep-th/9301068}{{\arxivfont arXiv:hep-th/9301068}}.

\bibitem[Guk99]{Gukov:1999yn}
S.~Gukov, \emph{{K-Theory, Reality, and Orientifolds}}, \doihref{http://dx.doi.org/10.1007/s002200050793}{Commun. Math. Phys. \textbf{210} (2000) 621--639}, \href{http://arxiv.org/abs/hep-th/9901042}{{\arxivfont arXiv:hep-th/9901042}}.

\bibitem[Gun12]{Gunningham2016}
S.~Gunningham, \emph{{Spin Hurwitz numbers and topological quantum field theory}}, \href{http://dx.doi.org/10.2140/gt.2016.20.1859}{Geometry \& Topology \textbf{20} (2016) 1859–1907}, \href{http://arxiv.org/abs/1201.1273}{{\arxivfont arXiv:1201.1273 [math-QA]}}.

\bibitem[GX12]{GallierXu:2012}
J.~Gallier and D.~Xu, \emph{A guide to the classification theorem for compact surfaces}, September 2012. Available at: \url{https://www.cis.upenn.edu/~jean/surfclass-n.pdf}.

\bibitem[GY97]{Greene:1997ty}
B.~Greene and S.-T. Yau (eds.), \emph{{Mirror Symmetry II}}, Clay Mathematics Monographs, American Mathematical Society, 1997. \url{https://www.ams.org/books/amsip/001/}.

\bibitem[GYV16]{Gelfand2016-eg}
I.~M. Gelfand and N.~Ya.~Vilenkin, \emph{{Generalized Functions, Volume 4: Applications of Harmonic Analysis}}, AMS Chelsea Publishing, American Mathematical Society, Providence, RI, Apr 2016.

\bibitem[Hal63]{Halmos1963}
P.~R. Halmos, \emph{{What Does the Spectral Theorem Say?}}, \href{http://dx.doi.org/10.1080/00029890.1963.11990075}{The American Mathematical Monthly \textbf{70} (1963) 241–247}.

\bibitem[Hen15]{Henriques:2015xxa}
A.~Henriques, \emph{{What Chern-Simons theory assigns to a point}}, \href{http://arxiv.org/abs/1503.06254}{{\arxivfont arXiv:1503.06254 [math-ph]}}.

\bibitem[HH83]{HartleHawking:1983}
J.~B. Hartle and S.~W. Hawking, \emph{Wave function of the universe}, \href{https://link.aps.org/doi/10.1103/PhysRevD.28.2960}{Phys. Rev. D \textbf{28} (1983) 2960--2975}.

\bibitem[HHP{\etalchar{+}}06]{Hellerman:2006zs}
S.~Hellerman, A.~Henriques, T.~Pantev, E.~Sharpe, and M.~Ando, \emph{{Cluster decomposition, T-duality, and gerby CFT's}}, \doihref{http://dx.doi.org/10.4310/ATMP.2007.v11.n5.a2}{Adv. Theor. Math. Phys. \textbf{11} (2007) 751--818}, \href{http://arxiv.org/abs/hep-th/0606034}{{\arxivfont arXiv:hep-th/0606034}}.

\bibitem[Hit99]{Hitchin:1999fh}
N.~J. Hitchin, \emph{{Lectures on Special Lagrangian Submanifolds}}, AMS/IP Stud. Adv. Math. \textbf{23} (2001) 151--182, \href{http://arxiv.org/abs/math/9907034}{{\arxivfont arXiv:math/9907034}}.

\bibitem[HK07]{Horava:2007hg}
P.~Ho\v{r}ava and C.~A. Keeler, \emph{{M-Theory Through the Looking Glass: Tachyon Condensation in the $E_8$ Heterotic String}}, \doihref{http://dx.doi.org/10.1103/PhysRevD.77.066013}{Phys. Rev. D \textbf{77} (2008) 066013}, \href{http://arxiv.org/abs/0709.3296}{{\arxivfont arXiv:0709.3296 [hep-th]}}.

\bibitem[HKK{\etalchar{+}}03]{Hori:2003}
K.~Hori, S.~Katz, A.~Klemm, R.~Pandharipande, R.~Thomas, C.~Vafa, R.~Vakil, and E.~Zaslow, \emph{Mirror symmetry}, Clay Mathematics Monographs, American Mathematical Society, Providence, RI, August 2003 (en). \url{https://www.claymath.org/wp-content/uploads/2022/03/Mirror-Symmetry.pdf}.

\bibitem[HKMM01]{Harvey:2001wm}
J.~A. Harvey, D.~Kutasov, E.~J. Martinec, and G.~W. Moore, \emph{{Localized tachyons and RG flows}}, \href{http://arxiv.org/abs/hep-th/0111154}{{\arxivfont arXiv:hep-th/0111154}}.

\bibitem[HKS20]{Hadfield:2020mnn}
C.~Hadfield, S.~Kandel, and M.~Schiavina, \emph{{Ruelle zeta function from field theory}}, \doihref{http://dx.doi.org/10.1007/s00023-020-00964-8}{Annales Henri Poincare \textbf{21} (2020) 3835--3867}, \href{http://arxiv.org/abs/2002.03952}{{\arxivfont arXiv:2002.03952 [math-ph]}}.

\bibitem[HKT20]{Hsin:2020cgg}
P.-S. Hsin, A.~Kapustin, and R.~Thorngren, \emph{{Berry Phase in Quantum Field Theory: Diabolical Points and Boundary Phenomena}}, \doihref{http://dx.doi.org/10.1103/PhysRevB.102.245113}{Phys. Rev. B \textbf{102} (2020) 245113}, \href{http://arxiv.org/abs/2004.10758}{{\arxivfont arXiv:2004.10758 [cond-mat.str-el]}}.

\bibitem[HL25a]{Hull:2025bqo}
C.~Hull and N.~Lambert, \emph{{The CFT of Sen's Formulation of Chiral Gauge Fields}}, \href{http://arxiv.org/abs/2508.00199}{{\arxivfont arXiv:2508.00199 [hep-th]}}.

\bibitem[HL25b]{Hull:2025rxy}
\bysame, \emph{{Quantising Chiral Bosons On Riemann Surfaces}}, \href{http://arxiv.org/abs/2508.02865}{{\arxivfont arXiv:2508.02865 [hep-th]}}.

\bibitem[HM00]{Harvey:2000te}
J.~A. Harvey and G.~W. Moore, \emph{{Noncommutative Tachyons and K-Theory}}, \doihref{http://dx.doi.org/10.1063/1.1377270}{J. Math. Phys. \textbf{42} (2001) 2765--2780}, \href{http://arxiv.org/abs/hep-th/0009030}{{\arxivfont arXiv:hep-th/0009030}}.

\bibitem[HM02]{HopkinsMahowald2002}
M.~J. Hopkins and M.~Mahowald, \href{https://bookstore.ams.org/CONM/293}{\emph{{The Structure of 24 Dimensional Manifolds Having Normal Bundles Which Lift to BO[8]}}}, Recent Progress in Homotopy Theory (D.~M. Davis et~al., eds.), Contemporary Mathematics, vol. 293, American Mathematical Society, Providence, RI, 2002, pp.~89--110.

\bibitem[HO18a]{Harlow:2018jwu}
D.~Harlow and H.~Ooguri, \emph{{Constraints on Symmetries from Holography}}, \doihref{http://dx.doi.org/10.1103/PhysRevLett.122.191601}{Phys. Rev. Lett. \textbf{122} (2019) 191601}, \href{http://arxiv.org/abs/1810.05337}{{\arxivfont arXiv:1810.05337 [hep-th]}}.

\bibitem[HO18b]{Harlow:2018tng}
\bysame, \emph{{Symmetries in quantum field theory and quantum gravity}}, \doihref{http://dx.doi.org/10.1007/s00220-021-04040-y}{Commun. Math. Phys. \textbf{383} (2021) 1669--1804}, \href{http://arxiv.org/abs/1810.05338}{{\arxivfont arXiv:1810.05338 [hep-th]}}.

\bibitem[Hop95]{Hopkins95}
M.~J. Hopkins, \emph{{Topological Modular Forms, the Witten Genus, and the Theorem of the Cube}}, Proceedings of the International Congress of Mathematicians (Basel) (S.~D. Chatterji, ed.), Birkh{\"a}user Basel, 1995, pp.~554--565.

\bibitem[Hop99]{Hopkins:1999coctalos}
M.~J. Hopkins, \emph{{Complex Oriented Cohomology Theories and the Language of Stacks}}, August 1999. \url{https://www.sas.rochester.edu/mth/sites/doug-ravenel/otherpapers/coctalos.pdf}. Course notes for MIT 18.917, compiled by graduate students.

\bibitem[Hop02]{Hopkins2002}
M.~J. Hopkins, \emph{{Algebraic topology and modular forms}}, {Proceedings of the {I}nternational {C}ongress of {M}athematicians, {V}ol. {I} ({B}eijing, 2002)}, Higher Ed. Press, Beijing, 2002, pp.~291--317. \href{http://arxiv.org/abs/0212397}{{\arxivfont arXiv:0212397 [math.AT]}}.

\bibitem[Hor89]{Horowitz:1989ng}
G.~T. Horowitz, \emph{{Exactly Soluble Diffeomorphism Invariant Theories}}, \doihref{http://dx.doi.org/10.1007/BF01218410}{Commun. Math. Phys. \textbf{125} (1989) 417}.

\bibitem[Ho{\v r}05]{Horava:2005jt}
P.~Ho{\v r}ava, \emph{{Stability of Fermi surfaces and K-theory}}, \doihref{http://dx.doi.org/10.1103/PhysRevLett.95.016405}{Phys. Rev. Lett. \textbf{95} (2005) 016405}, \href{http://arxiv.org/abs/hep-th/0503006}{{\arxivfont arXiv:hep-th/0503006}}.

\bibitem[HS02]{Hopkins:2002rd}
M.~J. Hopkins and I.~M. Singer, \emph{{Quadratic functions in geometry, topology, and M theory}}, J. Diff. Geom. \textbf{70} (2005) 329--452, \href{http://arxiv.org/abs/math/0211216}{{\arxivfont arXiv:math/0211216}}.

\bibitem[HST11]{Teichner2011}
H.~Hohnhold, S.~Stolz, and P.~Teichner, \emph{{Supermanifolds: an incomplete survey}}, \href{{https://people.mpim-bonn.mpg.de/teichner/Math/ewExternalFiles/Survey-Journal.pdf}}{{Bulletin of the Manifold Atlas} \textbf{1} (2011) 1--6}.

\bibitem[HTY20]{Hsieh:2020jpj}
C.-T. Hsieh, Y.~Tachikawa, and K.~Yonekura, \emph{{Anomaly Inflow and p-Form Gauge Theories}}, \doihref{http://dx.doi.org/10.1007/s00220-022-04333-w}{Commun. Math. Phys. \textbf{391} (2022) 495--608}, \href{http://arxiv.org/abs/2003.11550}{{\arxivfont arXiv:2003.11550 [hep-th]}}.

\bibitem[Hul23]{Hull:2023dgp}
C.~M. Hull, \emph{{Covariant action for self-dual p-form gauge fields in general spacetimes}}, \doihref{http://dx.doi.org/10.1007/JHEP04(2024)011}{JHEP \textbf{04} (2024) 011}, \href{http://arxiv.org/abs/2307.04748}{{\arxivfont arXiv:2307.04748 [hep-th]}}.

\bibitem[HW95]{Horava:1995qa}
P.~Ho{\v{r}}ava and E.~Witten, \emph{{Heterotic and Type I String Dynamics from Eleven Dimensions}}, \doihref{http://dx.doi.org/10.1201/9781482268737-35}{Nucl. Phys. B \textbf{460} (1996) 506--524}, \href{http://arxiv.org/abs/hep-th/9510209}{{\arxivfont arXiv:hep-th/9510209}}.

\bibitem[HW96]{Horava:1996ma}
\bysame, \emph{{Eleven-Dimensional Supergravity on a Manifold with Boundary}}, \doihref{http://dx.doi.org/10.1016/0550-3213(96)00308-2}{Nucl. Phys. B \textbf{475} (1996) 94--114}, \href{http://arxiv.org/abs/hep-th/9603142}{{\arxivfont arXiv:hep-th/9603142}}.

\bibitem[Int23a]{Intriligator:TASIVideoLec1}
K.~Intriligator, \emph{{Lecture 1 on SUSY and Symmetry Constraints on RG Flows and IR Phases (TASI 2023)}}, YouTube, 2023. \url{https://youtu.be/Pv6S0q9YEvs}. {Accessed: 2023-06-13}.

\bibitem[Int23b]{Intriligator:TASIVideoLec2}
\bysame, \emph{{Lecture 2 on SUSY and Symmetry Constraints on RG Flows and IR Phases (TASI 2023)}}, YouTube, 2023. \url{https://youtu.be/F1e56tOldjQ}. {Accessed: 2023-06-13}.

\bibitem[Int23c]{Intriligator:TASIVideoLec3}
\bysame, \emph{{Lecture 3 on SUSY and Symmetry Constraints on RG Flows and IR Phases (TASI 2023)}}, YouTube, 2023. \url{https://youtu.be/96pdWAOPr3g}. {Accessed: 2023-06-13}.

\bibitem[Int23d]{Intriligator:TASIVideoLec4}
\bysame, \emph{{Lecture 4 on SUSY and Symmetry Constraints on RG Flows and IR Phases (TASI 2023)}}, YouTube, 2023. \url{https://youtu.be/heddGMB3ZT4}. {Accessed: 2023-06-13}.

\bibitem[Jen06]{Jenquin:2006jh}
J.~A. Jenquin, \emph{{Spin Chern-Simons and Spin TQFTs}}, \href{http://arxiv.org/abs/math/0605239}{{\arxivfont arXiv:math/0605239}}.

\bibitem[JF17]{Johnson-Freyd:2017ble}
T.~Johnson-Freyd, \emph{{The Moonshine Anomaly}}, \doihref{http://dx.doi.org/10.1007/s00220-019-03300-2}{Commun. Math. Phys. \textbf{365} (2019) 943--970}, \href{http://arxiv.org/abs/1707.08388}{{\arxivfont arXiv:1707.08388 [math.QA]}}.

\bibitem[JF20a]{Johnson-Freyd:2020usu}
\bysame, \emph{{On the Classification of Topological Orders}}, \doihref{http://dx.doi.org/10.1007/s00220-022-04380-3}{Commun. Math. Phys. \textbf{393} (2022) 989--1033}, \href{http://arxiv.org/abs/2003.06663}{{\arxivfont arXiv:2003.06663 [math.CT]}}.

\bibitem[JF20b]{Johnson-Freyd:2020itv}
\bysame, \emph{{Topological Mathieu Moonshine}}, \href{http://arxiv.org/abs/2006.02922}{{\arxivfont arXiv:2006.02922 [math.AT]}}.

\bibitem[JFY24]{Johnson-Freyd:2024rxr}
T.~Johnson-Freyd and M.~Yamashita, \emph{{On the 576-fold periodicity of the spectrum SQFT: The proof of the lower bound via the Anderson duality pairing}}, \href{http://arxiv.org/abs/2404.06333}{{\arxivfont arXiv:2404.06333 [math.AT]}}.

\bibitem[JJ24]{Jiang:2024nnk}
S.~Jiang and J.~Jost, \emph{{Cohomological field theories and generalized Seiberg-Witten equations}}, \href{http://arxiv.org/abs/2407.04019}{{\arxivfont arXiv:2407.04019 [math-ph]}}.

\bibitem[Joh02]{Johnson2002}
C.~V. Johnson, \doihref{http://dx.doi.org/10.1017/cbo9780511606540}{\emph{{D-Branes}}}, Cambridge University Press, December 2002. \url{http://dx.doi.org/10.1017/CBO9780511606540}.

\bibitem[Joy09]{JoyceCorners:2009}
D.~Joyce, \emph{{On manifolds with corners}}. \url{https://arxiv.org/abs/0910.3518}.

\bibitem[Joy15]{JoyceCorners:2015}
\bysame, \emph{{A generalization of manifolds with corners}}. \url{https://arxiv.org/abs/1501.00401}.

\bibitem[JS91]{Joyal1991}
A.~Joyal and R.~Street, \emph{{An introduction to Tannaka duality and quantum groups}}, p.~413–492, Springer Berlin Heidelberg, 1991. \url{http://dx.doi.org/10.1007/BFb0084235}.

\bibitem[Kap99]{Kapustin:1999di}
A.~Kapustin, \emph{{D-branes in a topologically nontrivial B field}}, \doihref{http://dx.doi.org/10.4310/ATMP.2000.v4.n1.a3}{Adv. Theor. Math. Phys. \textbf{4} (2000) 127--154}, \href{http://arxiv.org/abs/hep-th/9909089}{{\arxivfont arXiv:hep-th/9909089}}.

\bibitem[Kap10]{Kapustin:2010ta}
\bysame, \emph{{Topological Field Theory, Higher Categories, and Their Applications}}, {International Congress of Mathematicians}, April 2010. \href{http://arxiv.org/abs/1004.2307}{{\arxivfont arXiv:1004.2307 [math.QA]}}.

\bibitem[Kap14]{Kapustin:2014tfa}
\bysame, \emph{{Symmetry Protected Topological Phases, Anomalies, and Cobordisms: Beyond Group Cohomology}}, \href{http://arxiv.org/abs/1403.1467}{{\arxivfont arXiv:1403.1467 [cond-mat.str-el]}}.

\bibitem[Kap25]{Kapustin:2025nju}
\bysame, \emph{{Higher symmetries and anomalies in quantum lattice systems}}, \href{http://arxiv.org/abs/2505.04719}{{\arxivfont arXiv:2505.04719 [math-ph]}}.

\bibitem[Kar78]{Karoubi1978}
M.~Karoubi, \doihref{http://dx.doi.org/10.1007/978-3-540-79890-3}{\emph{{K-Theory: An Introduction}}}, Springer Berlin Heidelberg, 1978. \url{http://dx.doi.org/10.1007/978-3-540-79890-3}.

\bibitem[Kar07]{Karoubi:2007mg}
\bysame, \emph{{Twisted K-theory, old and new}}, \href{http://arxiv.org/abs/math/0701789}{{\arxivfont arXiv:math/0701789}}.

\bibitem[Kar08]{Karoubi:2008eug}
\bysame, \emph{{Clifford Modules and Twisted K-Theory}}, \doihref{http://dx.doi.org/10.1007/s00006-008-0101-z}{Adv. Appl. Clifford Algebras \textbf{18} (2008) 765--769}.

\bibitem[KdBvW{\etalchar{+}}16]{Kruthoff:2016ver}
J.~Kruthoff, J.~de~Boer, J.~van Wezel, C.~L. Kane, and R.-J. Slager, \emph{{Topological classification of crystalline insulators through band structure combinatorics}}, \doihref{http://dx.doi.org/10.1103/PhysRevX.7.041069}{Phys. Rev. X \textbf{7} (2017) 041069}, \href{http://arxiv.org/abs/1612.02007}{{\arxivfont arXiv:1612.02007 [cond-mat.mes-hall]}}.

\bibitem[Kel07]{Kelnhofer:2007jf}
G.~Kelnhofer, \emph{{Functional integration and gauge ambiguities in generalized abelian gauge theories}}, \doihref{http://dx.doi.org/10.1016/j.geomphys.2009.04.007}{J. Geom. Phys. \textbf{59} (2009) 1017--1035}, \href{http://arxiv.org/abs/0711.4085}{{\arxivfont arXiv:0711.4085 [hep-th]}}.

\bibitem[Kel15]{Kellendonk:2015pda}
J.~Kellendonk, \emph{{On the $C^*$-algebraic approach to topological phases for insulators}}, \doihref{http://dx.doi.org/10.1007/s00023-017-0583-0}{Annales Henri Poincare \textbf{18} (2017) 2251--2300}, \href{http://arxiv.org/abs/1509.06271}{{\arxivfont arXiv:1509.06271 [math.KT]}}.

\bibitem[Kil87]{Killingback:1986rd}
T.~P. Killingback, \emph{{World Sheet Anomalies and Loop Geometry}}, \doihref{http://dx.doi.org/10.1016/0550-3213(87)90229-X}{Nucl. Phys. B \textbf{288} (1987) 578}.

\bibitem[Kit05]{Kitaev:2005hzj}
A.~Kitaev, \emph{{Anyons in an exactly solved model and beyond}}, \doihref{http://dx.doi.org/10.1016/j.aop.2005.10.005}{Annals Phys. \textbf{321} (2006) 2--111}, \href{http://arxiv.org/abs/cond-mat/0506438}{{\arxivfont arXiv:cond-mat/0506438}}.

\bibitem[Kit09]{Kitaev:2009mg}
\bysame, \emph{{Periodic table for topological insulators and superconductors}}, \doihref{http://dx.doi.org/10.1063/1.3149495}{AIP Conf. Proc. \textbf{1134} (2009) 22--30}, \href{http://arxiv.org/abs/0901.2686}{{\arxivfont arXiv:0901.2686 [cond-mat.mes-hall]}}.

\bibitem[Kit14]{Kitchloo:2014tqa}
N.~Kitchloo, \emph{{Quantization of the Modular Functor and Equivariant Elliptic cohomology}}, \href{http://arxiv.org/abs/1407.6698}{{\arxivfont arXiv:1407.6698 [math.AT]}}.

\bibitem[KKNar]{Khovanov:2023}
M.~Khovanov, V.~Krushkal, and J.~Nicholson, \emph{{On the universal pairing for 2-complexes}}, \href{http://arxiv.org/abs/arXiv:2312.07429}{{\arxivfont arXiv:2312.07429}}.

\bibitem[KKSS22]{Katz:2022lyl}
S.~Katz, A.~Klemm, T.~Schimannek, and E.~Sharpe, \emph{{Topological Strings on Non-commutative Resolutions}}, \doihref{http://dx.doi.org/10.1007/s00220-023-04896-2}{Commun. Math. Phys. \textbf{405} (2024) 62}, \href{http://arxiv.org/abs/2212.08655}{{\arxivfont arXiv:2212.08655 [hep-th]}}.

\bibitem[Klo08]{Klonoff:2008}
K.~R. Klonoff, \emph{{An index theorem in differential K-theory}}, ProQuest LLC, Ann Arbor, MI, 2008. \url{http://gateway.proquest.com/openurl?url_ver=Z39.88-2004&rft_val_fmt=info:ofi/fmt:kev:mtx:dissertation&res_dat=xri:pqdiss&rft_dat=xri:pqdiss:3321109}. Thesis (Ph.D.)--The University of Texas at Austin.

\bibitem[KM07]{Kronheimer2007}
P.~Kronheimer and T.~Mrowka, \doihref{http://dx.doi.org/10.1017/cbo9780511543111}{\emph{{Monopoles and Three-Manifolds}}}, Cambridge University Press, December 2007. \url{http://dx.doi.org/10.1017/CBO9780511543111}.

\bibitem[KM20]{Khan:2020hir}
A.~Z. Khan and G.~W. Moore, \emph{{Categorical wall-crossing in Landau{\textendash}Ginzburg models}}, \doihref{http://dx.doi.org/10.4310/bpam.2024.v1.n1.a2}{Beijing J. Pure Appl. Math. \textbf{1} (2024) 43--119}, \href{http://arxiv.org/abs/2010.11837}{{\arxivfont arXiv:2010.11837 [hep-th]}}.

\bibitem[KM24]{Khan:2024yiy}
\bysame, \emph{{On the Algebra of the Infrared with Twisted Masses}}, \href{http://arxiv.org/abs/2408.08372}{{\arxivfont arXiv:2408.08372 [hep-th]}}.

\bibitem[KMW07]{Kitaev:2007ed}
A.~Kitaev, G.~W. Moore, and K.~Walker, \emph{{Noncommuting Flux Sectors in a Tabletop Experiment}}, \href{http://arxiv.org/abs/0706.3410}{{\arxivfont arXiv:0706.3410 [hep-th]}}.

\bibitem[KN96a]{Kobayashi1996-kb}
S.~Kobayashi and K.~Nomizu, \emph{Foundations of differential geometry, volume 1}, Wiley Classics Library, John Wiley \& Sons, Nashville, TN, February 1996 (en).

\bibitem[KN96b]{Kobayashi1996-fy}
\bysame, \emph{Foundations of differential geometry, volume 2}, Wiley Classics Library, John Wiley \& Sons, Nashville, TN, February 1996 (en).

\bibitem[Kob54]{Kobayashi1954}
S.~Kobayashi, \emph{La connexion des vari{\'e}t{\'e}s fibr{\'e}es i, ii}, Comptes Rendus de l'Acad{\'e}mie des Sciences \textbf{238} (1954) 318--319, 443--444 (French).

\bibitem[Kob93]{Koblitz1993}
N.~Koblitz, \doihref{http://dx.doi.org/10.1007/978-1-4612-0909-6}{\emph{Introduction to elliptic curves and modular forms}}, Springer New York, 1993. \url{http://dx.doi.org/10.1007/978-1-4612-0909-6}.

\bibitem[Koc03]{Kock2003}
J.~Kock, \doihref{http://dx.doi.org/10.1017/cbo9780511615443}{\emph{{Frobenius Algebras and 2D Topological Quantum Field Theories}}}, Cambridge University Press, 2003. \url{http://dx.doi.org/10.1017/CBO9780511615443}.

\bibitem[Kon88]{Kontsevich:1988br}
M.~Kontsevich, \emph{{Rational Conformal Field Theory And Invariants Of 3-Dimensional Manifolds}}, {CNRS Luminy preprint}, December 1988.

\bibitem[KORS20]{Komargodski:2020mxz}
Z.~Komargodski, K.~Ohmori, K.~Roumpedakis, and S.~Seifnashri, \emph{{Symmetries and strings of adjoint QCD$_{2}$}}, \doihref{http://dx.doi.org/10.1007/JHEP03(2021)103}{JHEP \textbf{03} (2021) 103}, \href{http://arxiv.org/abs/2008.07567}{{\arxivfont arXiv:2008.07567 [hep-th]}}.

\bibitem[KS10]{Kapustin:2010hk}
A.~Kapustin and N.~Saulina, \emph{{Topological boundary conditions in abelian Chern-Simons theory}}, \doihref{http://dx.doi.org/10.1016/j.nuclphysb.2010.12.017}{Nucl. Phys. B \textbf{845} (2011) 393--435}, \href{http://arxiv.org/abs/1008.0654}{{\arxivfont arXiv:1008.0654 [hep-th]}}.

\bibitem[KS14]{Kapustin:2014gua}
A.~Kapustin and N.~Seiberg, \emph{{Coupling a QFT to a TQFT and Duality}}, \doihref{http://dx.doi.org/10.1007/JHEP04(2014)001}{JHEP \textbf{04} (2014) 001}, \href{http://arxiv.org/abs/1401.0740}{{\arxivfont arXiv:1401.0740 [hep-th]}}.

\bibitem[KS20a]{Kapustin:2020eby}
A.~Kapustin and L.~Spodyneiko, \emph{{Higher-dimensional generalizations of Berry curvature}}, \doihref{http://dx.doi.org/10.1103/PhysRevB.101.235130}{Phys. Rev. B \textbf{101} (2020) 235130}, \href{http://arxiv.org/abs/2001.03454}{{\arxivfont arXiv:2001.03454 [cond-mat.str-el]}}.

\bibitem[KS20b]{Kapustin:2020mkl}
\bysame, \emph{{Higher-dimensional generalizations of the Thouless charge pump}}, \href{http://arxiv.org/abs/2003.09519}{{\arxivfont arXiv:2003.09519 [cond-mat.str-el]}}.

\bibitem[KS21]{Kontsevich:2021dmb}
M.~Kontsevich and G.~Segal, \emph{{Wick Rotation and the Positivity of Energy in Quantum Field Theory}}, \doihref{http://dx.doi.org/10.1093/qmath/haab027}{Quart. J. Math. Oxford Ser. \textbf{72} (2021) 673--699}, \href{http://arxiv.org/abs/2105.10161}{{\arxivfont arXiv:2105.10161 [hep-th]}}.

\bibitem[KS25]{Kapustin:2025rhp}
A.~Kapustin and L.~Spodyneiko, \emph{{Higher symmetries, anomalies, and crossed squares in lattice gauge theory}}, \href{http://arxiv.org/abs/2507.16966}{{\arxivfont arXiv:2507.16966 [hep-th]}}.

\bibitem[KT13a]{Kapustin:2013qsa}
A.~Kapustin and R.~Thorngren, \emph{{Topological Field Theory on a Lattice, Discrete Theta-Angles and Confinement}}, \doihref{http://dx.doi.org/10.4310/ATMP.2014.v18.n5.a4}{Adv. Theor. Math. Phys. \textbf{18} (2014) 1233--1247}, \href{http://arxiv.org/abs/1308.2926}{{\arxivfont arXiv:1308.2926 [hep-th]}}.

\bibitem[KT13b]{Kapustin:2013uxa}
\bysame, \emph{{Higher Symmetry and Gapped Phases of Gauge Theories}}, \doihref{http://dx.doi.org/10.1007/978-3-319-59939-7_5}{Prog. Math. \textbf{324} (2017) 177--202}, \href{http://arxiv.org/abs/1309.4721}{{\arxivfont arXiv:1309.4721 [hep-th]}}.

\bibitem[KT14a]{Kapustin:2014lwa}
\bysame, \emph{{Anomalies of discrete symmetries in three dimensions and group cohomology}}, \doihref{http://dx.doi.org/10.1103/PhysRevLett.112.231602}{Phys. Rev. Lett. \textbf{112} (2014) 231602}, \href{http://arxiv.org/abs/1403.0617}{{\arxivfont arXiv:1403.0617 [hep-th]}}.

\bibitem[KT14b]{Kapustin:2014zva}
\bysame, \emph{{Anomalies of discrete symmetries in various dimensions and group cohomology}}, \href{http://arxiv.org/abs/1404.3230}{{\arxivfont arXiv:1404.3230 [hep-th]}}.

\bibitem[KT17]{Kapustin:2017jrc}
\bysame, \emph{{Fermionic SPT phases in higher dimensions and bosonization}}, \doihref{http://dx.doi.org/10.1007/JHEP10(2017)080}{JHEP \textbf{10} (2017) 080}, \href{http://arxiv.org/abs/1701.08264}{{\arxivfont arXiv:1701.08264 [cond-mat.str-el]}}.

\bibitem[KTT19]{Karch:2019lnn}
A.~Karch, D.~Tong, and C.~Turner, \emph{{A Web of 2d Dualities: ${\bf Z}_2$ Gauge Fields and Arf Invariants}}, \doihref{http://dx.doi.org/10.21468/SciPostPhys.7.1.007}{SciPost Phys. \textbf{7} (2019) 007}, \href{http://arxiv.org/abs/1902.05550}{{\arxivfont arXiv:1902.05550 [hep-th]}}.

\bibitem[KTTW14]{Kapustin:2014dxa}
A.~Kapustin, R.~Thorngren, A.~Turzillo, and Z.~Wang, \emph{{Fermionic Symmetry Protected Topological Phases and Cobordisms}}, \doihref{http://dx.doi.org/10.1007/JHEP12(2015)052}{JHEP \textbf{12} (2015) 052}, \href{http://arxiv.org/abs/1406.7329}{{\arxivfont arXiv:1406.7329 [cond-mat.str-el]}}.

\bibitem[KW06]{Kapustin:2006pk}
A.~Kapustin and E.~Witten, \emph{{Electric-Magnetic Duality And The Geometric Langlands Program}}, \doihref{http://dx.doi.org/10.4310/CNTP.2007.v1.n1.a1}{Commun. Num. Theor. Phys. \textbf{1} (2007) 1--236}, \href{http://arxiv.org/abs/hep-th/0604151}{{\arxivfont arXiv:hep-th/0604151}}.

\bibitem[KWZ17]{Kong:2017hcw}
L.~Kong, X.-G. Wen, and H.~Zheng, \emph{{Boundary-bulk relation in topological orders}}, \doihref{http://dx.doi.org/10.1016/j.nuclphysb.2017.06.023}{Nucl. Phys. B \textbf{922} (2017) 62--76}, \href{http://arxiv.org/abs/1702.00673}{{\arxivfont arXiv:1702.00673 [cond-mat.str-el]}}.

\bibitem[KZ22]{Kong:2022cpy}
L.~Kong and Z.-H. Zhang, \emph{{An invitation to topological orders and category theory}}, \href{http://arxiv.org/abs/2205.05565}{{\arxivfont arXiv:2205.05565 [cond-mat.str-el]}}.

\bibitem[Lam23]{Lambert:2023qgs}
N.~Lambert, \emph{{Duality and Fluxes in the Sen Formulation of Self-Dual Fields}}, \doihref{http://dx.doi.org/10.1016/j.physletb.2023.137888}{Phys. Lett. B \textbf{840} (2023) 137888}, \href{http://arxiv.org/abs/2302.10955}{{\arxivfont arXiv:2302.10955 [hep-th]}}.

\bibitem[Lan88]{Landweber1988}
P.~S. Landweber, \emph{{Elliptic cohomology and modular forms}}, p.~55–68, Springer Berlin Heidelberg, 1988. \url{http://dx.doi.org/10.1007/BFb0078038}.

\bibitem[Lei80]{Leites1980}
D.~A. Leites, \emph{{INTRODUCTION TO THE THEORY OF SUPERMANIFOLDS}}, \href{http://dx.doi.org/10.1070/RM1980v035n01ABEH001545}{Russian Mathematical Surveys \textbf{35} (1980) 1–64}.

\bibitem[Lew93]{Lewandowski:1993qy}
J.~Lewandowski, \emph{{Group of loops, holonomy maps, path bundle and path connection}}, \doihref{http://dx.doi.org/10.1088/0264-9381/10/5/008}{Class. Quant. Grav. \textbf{10} (1993) 879--904}.

\bibitem[Lin16]{Lin:MonopoleFloerHomology}
F.~Lin, \doihref{http://dx.doi.org/10.48550/ARXIV.1605.03140}{\emph{{Lectures on monopole Floer homology}}}, 2016. \url{https://arxiv.org/abs/1605.03140}.

\bibitem[Liu21]{Liu:2021hhy}
Y.~L. Liu, \emph{{Abelian Duality in Topological Field Theory}}, \doihref{http://dx.doi.org/10.1007/s00220-022-04527-2}{Commun. Math. Phys. \textbf{398} (2023) 439--468}, \href{http://arxiv.org/abs/2112.02199}{{\arxivfont arXiv:2112.02199 [math-ph]}}.

\bibitem[LL22]{Losev:2022tzr}
A.~Losev and V.~Lysov, \emph{{Tropical Mirror}}, \doihref{http://dx.doi.org/10.3842/SIGMA.2024.072}{SIGMA \textbf{20} (2024) 072}, \href{http://arxiv.org/abs/2204.06896}{{\arxivfont arXiv:2204.06896 [hep-th]}}.

\bibitem[LL23]{Losev:2023bhj}
\bysame, \emph{{Tropical Mirror Symmetry: Correlation functions}}, \href{http://arxiv.org/abs/2301.01687}{{\arxivfont arXiv:2301.01687 [hep-th]}}.

\bibitem[LM90]{Lawson1990-ev}
H.~B. Lawson and M.-L. Michelsohn, \emph{Spin geometry ({PMS-38)}, volume 38}, Princeton Mathematical Series, Princeton University Press, Princeton, NJ, February 1990 (en).

\bibitem[LM05]{Labastida:2005zz}
J.~Labastida and M.~Mari{\~n}o, \doihref{http://dx.doi.org/10.1007/1-4020-3177-7}{\emph{{Topological Quantum Field Theory and Four Manifolds}}}, Springer, Dordrecht, 2005.

\bibitem[LMNS95a]{Losev:1995cr}
A.~Losev, G.~W. Moore, N.~Nekrasov, and S.~Shatashvili, \emph{{Four-Dimensional Avatars of Two-Dimensional RCFT}}, \doihref{http://dx.doi.org/10.1016/0920-5632(96)00015-1}{Nucl. Phys. B Proc. Suppl. \textbf{46} (1996) 130--145}, \href{http://arxiv.org/abs/hep-th/9509151}{{\arxivfont arXiv:hep-th/9509151}}.

\bibitem[LMNS95b]{Losev:1995zf}
A.~Losev, G.~W. Moore, N.~Nekrasov, and S.~L. Shatashvili, \emph{{Central Extensions of Gauge Groups Revisited}}, \href{http://arxiv.org/abs/hep-th/9511185}{{\arxivfont arXiv:hep-th/9511185}}.

\bibitem[LMNS96]{Losev:1996up}
A.~Losev, G.~W. Moore, N.~Nekrasov, and S.~Shatashvili, \emph{{Chiral Lagrangians, Anomalies, Supersymmetry, and Holomorphy}}, \doihref{http://dx.doi.org/10.1016/S0550-3213(96)00612-8}{Nucl. Phys. B \textbf{484} (1997) 196--222}, \href{http://arxiv.org/abs/hep-th/9606082}{{\arxivfont arXiv:hep-th/9606082}}.

\bibitem[LMS97]{Losev:1997hx}
A.~Losev, G.~W. Moore, and S.~L. Shatashvili, \emph{{M \& m's}}, \doihref{http://dx.doi.org/10.1016/S0550-3213(98)00262-4}{Nucl. Phys. B \textbf{522} (1998) 105--124}, \href{http://arxiv.org/abs/hep-th/9707250}{{\arxivfont arXiv:hep-th/9707250}}.

\bibitem[Los07]{LOSEV2007}
A.~Losev, \emph{{FROM BEREZIN INTEGRAL TO BATALIN–VILKOVISKY FORMALISM: A MATHEMATICAL PHYSICIST’S POINT OF VIEW}}, p.~3–30, WORLD SCIENTIFIC, April 2007. \url{http://dx.doi.org/10.1142/9789812770486_0001}.

\bibitem[Los23]{Losev:2019bel}
A.~Losev, \doihref{http://dx.doi.org/10.2969/aspm/08310269}{\emph{{TQFT, homological algebra and elements of K. Saito{\textquoteright}s theory of primitive form: an attempt of mathematical text written by mathematical physicist.}}}, 2019. \href{http://arxiv.org/abs/2301.01390}{{\arxivfont arXiv:2301.01390 [math-ph]}}.

\bibitem[LOT20]{Lee:2020ojw}
Y.~Lee, K.~Ohmori, and Y.~Tachikawa, \emph{{Revisiting Wess-Zumino-Witten terms}}, \doihref{http://dx.doi.org/10.21468/SciPostPhys.10.3.061}{SciPost Phys. \textbf{10} (2021) 061}, \href{http://arxiv.org/abs/2009.00033}{{\arxivfont arXiv:2009.00033 [hep-th]}}.

\bibitem[Lot24]{Lott2024}
J.~Lott, \emph{{The Ray–Singer torsion}}, \href{http://dx.doi.org/10.1090/bull/1848}{Bulletin of the American Mathematical Society \textbf{62} (2024) 17–34}.

\bibitem[LP21]{Lin:2021bcp}
Y.-H. Lin and D.~Pei, \emph{{Holomorphic CFTs and Topological Modular Forms}}, \doihref{http://dx.doi.org/10.1007/s00220-023-04639-3}{Commun. Math. Phys. \textbf{401} (2023) 325--332}, \href{http://arxiv.org/abs/2112.10724}{{\arxivfont arXiv:2112.10724 [hep-th]}}.

\bibitem[LRS95]{Landweber1995}
P.~S. Landweber, D.~C. Ravenel, and R.~E. Stong, \doihref{http://dx.doi.org/10.1090/conm/181/02040}{\emph{{Periodic cohomology theories defined by elliptic curves}}}, 1995, p.~317–337. \url{http://dx.doi.org/10.1090/conm/181/02040}.

\bibitem[LS19]{Lin:2019hks}
Y.-H. Lin and S.-H. Shao, \emph{{Duality Defect of the Monster CFT}}, \doihref{http://dx.doi.org/10.1088/1751-8121/abd69e}{J. Phys. A \textbf{54} (2021) 065201}, \href{http://arxiv.org/abs/1911.00042}{{\arxivfont arXiv:1911.00042 [hep-th]}}.

\bibitem[Lur09a]{Lurie2009:EllipticSurvey}
J.~Lurie, \emph{{A Survey of Elliptic Cohomology}}, p.~219–277, Springer Berlin Heidelberg, 2009. \url{http://dx.doi.org/10.1007/978-3-642-01200-6_9}.

\bibitem[Lur09b]{Lurie2009-el}
J.~Lurie, \emph{{Higher Topos Theory ({AM-170})}}, Annals of Mathematics Studies, Princeton University Press, Princeton, NJ, July 2009 (en).

\bibitem[Lur09c]{Lurie:2009keu}
J.~Lurie, \emph{{On the Classification of Topological Field Theories}}, \href{http://arxiv.org/abs/0905.0465}{{\arxivfont arXiv:0905.0465 [math.CT]}}.

\bibitem[Lur14]{Lurie2014en}
J.~Lurie, \emph{{$E_n$-Algebras (Lecture 22)}}, \url{https://www.math.ias.edu/~lurie/282ynotes/LectureXXII-En.pdf}, Mar 2014. {Lecture notes for Math 282y: Tamagawa Numbers via Nonabelian Poincar\'e Duality}.

\bibitem[Lur25]{Lurie:Kerodon}
\bysame, \emph{{Kerodon: an online resource for homotopy-coherent mathematics}}, 2025. \url{https://kerodon.net/}. [Online; accessed 2-February-2025].

\bibitem[LY24]{Lin:2024qqk}
Y.-H. Lin and M.~Yamashita, \emph{{Topological Elliptic Genera I -- The mathematical foundation}}, \href{http://arxiv.org/abs/2412.02298}{{\arxivfont arXiv:2412.02298 [math.AT]}}.

\bibitem[M\"78]{Muller1978}
W.~M\"{u}ller, \emph{{Analytic torsion and R-torsion of Riemannian manifolds}}, \href{http://dx.doi.org/10.1016/0001-8708(78)90116-0}{Advances in Mathematics \textbf{28} (1978) 233–305}.

\bibitem[Mal97]{Maldacena:1997re}
J.~M. Maldacena, \emph{{The Large N Limit of Superconformal Field Theories and Supergravity}}, \doihref{http://dx.doi.org/10.4310/ATMP.1998.v2.n2.a1}{Adv. Theor. Math. Phys. \textbf{2} (1998) 231--252}, \href{http://arxiv.org/abs/hep-th/9711200}{{\arxivfont arXiv:hep-th/9711200}}.

\bibitem[Man97]{Manin1997}
Y.~I. Manin, \doihref{http://dx.doi.org/10.1007/978-3-662-07386-5}{\emph{{Gauge Field Theory and Complex Geometry}}}, Springer Berlin Heidelberg, 1997. \url{http://dx.doi.org/10.1007/978-3-662-07386-5}.

\bibitem[May99]{May1999}
J.~P. May, \emph{{A Concise Course in Algebraic Topology}}, 2 ed., Chicago Lectures in Mathematics, University of Chicago Press, Chicago, IL, September 1999 (en).

\bibitem[Men19]{Meneses:2019wyy}
C.~Meneses, \emph{{Thin homotopy and the holonomy approach to gauge theories}}, \doihref{http://dx.doi.org/10.1090/conm/775/15594}{Contemp. Math. \textbf{775} (2021) 233}, \href{http://arxiv.org/abs/1904.10822}{{\arxivfont arXiv:1904.10822 [math-ph]}}.

\bibitem[Mig75]{Migdal:1975zg}
A.~A. Migdal, \emph{{Recursion Equations in Gauge Theories}}, Sov. Phys. JETP \textbf{42} (1975) 413.

\bibitem[Mil56a]{Milnor1}
J.~Milnor, \emph{{Construction of Universal Bundles, I}}, \href{http://www.jstor.org/stable/1969609}{Annals of Mathematics \textbf{63} (1956) 272--284}.

\bibitem[Mil56b]{Milnor2}
\bysame, \emph{{Construction of Universal Bundles, II}}, \href{http://www.jstor.org/stable/1970012}{Annals of Mathematics \textbf{63} (1956) 430--436}.

\bibitem[ML78]{MacLane1978}
S.~Mac~Lane, \doihref{http://dx.doi.org/10.1007/978-1-4757-4721-8}{\emph{{Categories for the Working Mathematician}}}, Springer New York, 1978. \url{http://dx.doi.org/10.1007/978-1-4757-4721-8}.

\bibitem[MM97]{Minasian:1997mm}
R.~Minasian and G.~W. Moore, \emph{{K theory and Ramond-Ramond charge}}, \doihref{http://dx.doi.org/10.1088/1126-6708/1997/11/002}{JHEP \textbf{11} (1997) 002}, \href{http://arxiv.org/abs/hep-th/9710230}{{\arxivfont arXiv:hep-th/9710230}}.

\bibitem[MM18a]{Monnier:2018nfs}
S.~Monnier and G.~W. Moore, \emph{{Remarks on the Green-Schwarz Terms of Six-Dimensional Supergravity Theories}}, \doihref{http://dx.doi.org/10.1007/s00220-019-03341-7}{Commun. Math. Phys. \textbf{372} (2019) 963--1025}, \href{http://arxiv.org/abs/1808.01334}{{\arxivfont arXiv:1808.01334 [hep-th]}}.

\bibitem[MM18b]{Monnier:2018cfa}
\bysame, \emph{{A Brief Summary Of Global Anomaly Cancellation In Six-Dimensional Supergravity}}, \href{http://arxiv.org/abs/1808.01335}{{\arxivfont arXiv:1808.01335 [hep-th]}}.

\bibitem[MM20]{Marolf:2020xie}
D.~Marolf and H.~Maxfield, \emph{{Transcending the ensemble: baby universes, spacetime wormholes, and the order and disorder of black hole information}}, \doihref{http://dx.doi.org/10.1007/JHEP08(2020)044}{JHEP \textbf{08} (2020) 044}, \href{http://arxiv.org/abs/2002.08950}{{\arxivfont arXiv:2002.08950 [hep-th]}}.

\bibitem[MMN85]{Manohar:1984zj}
A.~Manohar, G.~W. Moore, and P.~C. Nelson, \emph{{A COMMENT ON SIGMA MODEL ANOMALIES}}, \doihref{http://dx.doi.org/10.1016/0370-2693(85)91141-4}{Phys. Lett. B \textbf{152} (1985) 68--74}.

\bibitem[MMP17]{Monnier:2017oqd}
S.~Monnier, G.~W. Moore, and D.~S. Park, \emph{{Quantization of anomaly coefficients in 6D $\mathcal{N}=(1,0)$ supergravity}}, \doihref{http://dx.doi.org/10.1007/JHEP02(2018)020}{JHEP \textbf{02} (2018) 020}, \href{http://arxiv.org/abs/1711.04777}{{\arxivfont arXiv:1711.04777 [hep-th]}}.

\bibitem[MMS01a]{Maldacena:2001xj}
J.~M. Maldacena, G.~W. Moore, and N.~Seiberg, \emph{{D-brane instantons and K theory charges}}, \doihref{http://dx.doi.org/10.1088/1126-6708/2001/11/062}{JHEP \textbf{11} (2001) 062}, \href{http://arxiv.org/abs/hep-th/0108100}{{\arxivfont arXiv:hep-th/0108100}}.

\bibitem[MMS01b]{Maldacena:2001ss}
\bysame, \emph{{D-brane Charges in Five-brane backgrounds}}, \doihref{http://dx.doi.org/10.1088/1126-6708/2001/10/005}{JHEP \textbf{10} (2001) 005}, \href{http://arxiv.org/abs/hep-th/0108152}{{\arxivfont arXiv:hep-th/0108152}}.

\bibitem[MN84]{Moore:1984dc}
G.~W. Moore and P.~C. Nelson, \emph{{Anomalies in Nonlinear $\sigma$ Models}}, \doihref{http://dx.doi.org/10.1103/PhysRevLett.53.1519}{Phys. Rev. Lett. \textbf{53} (1984) 1519}.

\bibitem[MN85]{Moore:1984ws}
\bysame, \emph{{The Etiology of $\sigma$ Model Anomalies}}, \doihref{http://dx.doi.org/10.1007/BF01212688}{Commun. Math. Phys. \textbf{100} (1985) 83}.

\bibitem[Mne14]{Mnev:2014gta}
P.~Mnev, \emph{{Lecture notes on torsions}}, \href{http://arxiv.org/abs/1406.3705}{{\arxivfont arXiv:1406.3705 [math.AT]}}.

\bibitem[Mne25]{Mnev:2025skb}
\bysame, \emph{{Lecture notes on conformal field theory}}, \href{http://arxiv.org/abs/2501.06616}{{\arxivfont arXiv:2501.06616 [math-ph]}}.

\bibitem[Mon10]{Monnier:2010ww}
S.~Monnier, \emph{{Geometric quantization and the metric dependence of the self-dual field theory}}, \doihref{http://dx.doi.org/10.1007/s00220-012-1525-9}{Commun. Math. Phys. \textbf{314} (2012) 305--328}, \href{http://arxiv.org/abs/1011.5890}{{\arxivfont arXiv:1011.5890 [hep-th]}}.

\bibitem[Mon14]{Monnier:2014txa}
\bysame, \emph{{The global anomalies of (2,0) superconformal field theories in six dimensions}}, \doihref{http://dx.doi.org/10.1007/JHEP09(2014)088}{JHEP \textbf{09} (2014) 088}, \href{http://arxiv.org/abs/1406.4540}{{\arxivfont arXiv:1406.4540 [hep-th]}}.

\bibitem[Mon16]{Monnier:2016jlo}
\bysame, \emph{{Topological field theories on manifolds with Wu structures}}, \doihref{http://dx.doi.org/10.1142/S0129055X17500155}{Rev. Math. Phys. \textbf{29} (2017) 1750015}, \href{http://arxiv.org/abs/1607.01396}{{\arxivfont arXiv:1607.01396 [math-ph]}}.

\bibitem[Mon17]{Monnier:2017klz}
\bysame, \emph{{The anomaly field theories of six-dimensional (2,0) superconformal theories}}, \doihref{http://dx.doi.org/10.4310/ATMP.2018.v22.n8.a6}{Adv. Theor. Math. Phys. \textbf{22} (2018) 2035--2089}, \href{http://arxiv.org/abs/1706.01903}{{\arxivfont arXiv:1706.01903 [hep-th]}}.

\bibitem[Moo95]{Moore1995}
G.~W. Moore, \emph{{Two-dimensional Yang-Mills Theory and Topological Field Theory}}, p.~1292–1303, Birkh\"{a}user Basel, 1995. \url{http://dx.doi.org/10.1007/978-3-0348-9078-6_124}.

\bibitem[Moo03]{Moore:2003vf}
\bysame, \emph{{K-theory from a physical perspective}}, {Symposium on Topology, Geometry and Quantum Field Theory (Segalfest)}, April 2003, pp.~194--234. \href{http://arxiv.org/abs/hep-th/0304018}{{\arxivfont arXiv:hep-th/0304018}}.

\bibitem[Moo04]{Moore:2004jv}
\bysame, \emph{{Anomalies, Gauss laws, and Page charges in M-theory}}, \doihref{http://dx.doi.org/10.1016/j.crhy.2004.12.005}{Comptes Rendus Physique \textbf{6} (2005) 251--259}, \href{http://arxiv.org/abs/hep-th/0409158}{{\arxivfont arXiv:hep-th/0409158}}.

\bibitem[Moo10]{Moore:2010PiTP}
\bysame, \emph{{PiTP Lectures on BPS States and Wall-Crossing in $d = 4$, $\mathcal{N} = 2$ Theories}}, \href{https://www.physics.rutgers.edu/~gmoore/PiTP-LectureNotes.pdf}{PiTP Lectures (2010) }.

\bibitem[Moo11]{Moore:2011SCGP}
\bysame, \emph{{A Minicourse on Generalized Abelian Gauge Theory, Self-Dual Theories, and Differential Cohomology}}, \href{https://www.physics.rutgers.edu/~gmoore/SCGP-Minicourse.pdf}{SCGP Lectures (2011) }.

\bibitem[Moo12]{Moore:2012SCGP}
\bysame, \emph{{A Very Long Lecture on the Physical Approach to Donaldson and Seiberg-Witten Invariants of Four-Manifolds}}, \href{http://www.physics.rutgers.edu/~gmoore/SCGP-LECTURENOTES.pdf}{SCGP Lectures (2012) }.

\bibitem[Moo15]{Moore:2015PHY695}
\bysame, \emph{{A Few Remarks on Topological Field Theory}}, \href{https://www.physics.rutgers.edu/~gmoore/695Fall2015/TopologicalFieldTheory.pdf}{Physics 695, Rutgers (2015) }.

\bibitem[Moo17a]{Moore:2017SCGP}
\bysame, \emph{{Lectures On The Physical Approach To Donaldson And Seiberg-Witten Invariants Of Four-Manifolds}}, \href{http://www.physics.rutgers.edu/~gmoore/SCGP-FourManifoldsNotes-2017.pdf}{SCGP Lectures (2017) }.

\bibitem[Moo17b]{Moore:2017byz}
\bysame, \emph{{A Comment On Berry Connections}}, \href{http://arxiv.org/abs/1706.01149}{{\arxivfont arXiv:1706.01149 [hep-th]}}.

\bibitem[Moo19]{Moore:2019TASI}
\bysame, \emph{{Introduction To Chern-Simons Theories}}, \href{http://www.physics.rutgers.edu/~gmoore/TASI-ChernSimons-StudentNotes.pdf}{TASI Lectures (2019) }.

\bibitem[Moo23a]{MooreGroupTheory:2023}
\bysame, \emph{{Abstract Group Theory}}. \url{https://www.physics.rutgers.edu/~gmoore/618Spring2023/GTLect1-AbstractGroupTheory-2023.pdf}.

\bibitem[Moo23b]{Moore:TASIVideoLec1}
\bysame, \emph{{Lecture 1 on Differential Cohomology and Physics}}, YouTube, 2023. \url{https://youtu.be/7kIoB85nSu8}. {Accessed: 2023-06-13}.

\bibitem[Moo23c]{Moore:TASIVideoLec2}
\bysame, \emph{{Lecture 2 on Differential Cohomology and Physics}}, YouTube, 2023. \url{https://youtu.be/2ph0lkigsdw}. {Accessed: 2023-06-14}.

\bibitem[Moo23d]{Moore:TASIVideoLec3}
\bysame, \emph{{Lecture 3 on Differential Cohomology and Physics}}, YouTube, 2023. \url{https://youtu.be/jawbkmY9Gzk}. {Accessed: 2023-06-16}.

\bibitem[Moo23e]{Moore:TASIVideoLec4}
\bysame, \emph{{Lecture 4 on Differential Cohomology and Physics}}, YouTube, 2023. \url{https://youtu.be/r8is1Qfjq98}. {Accessed: 2023-06-16}.

\bibitem[Moo24a]{Moore:RCFT-To-MTC}
\bysame, \emph{{From Rational Conformal Field Theories To Modular Tensor Categories To Nonabelions}}. \url{https://www.physics.rutgers.edu/~gmoore/From-RCFT-To-MTC-20240926-FINAL.pdf}.

\bibitem[Moo24b]{Moore:PlecticsVideo}
\bysame, \emph{{From RCFT To MTC}}, YouTube, 2024. \url{https://youtu.be/BfYYVJ9pNug}. {Accessed: 2024-10-5}.

\bibitem[MP23]{MartinsPorter:2023}
J.~F. Martins and T.~Porter, \emph{{A categorification of Quinn's finite total homotopy TQFT with application to TQFTs and once-extended TQFTs derived from strict omega-groupoids}}, \href{http://arxiv.org/abs/2301.02491}{{\arxivfont arXiv:2301.02491 [math.CT]}}.

\bibitem[MR91]{Moore:1991ks}
G.~W. Moore and N.~Read, \emph{{Nonabelions in the fractional quantum Hall effect}}, \doihref{http://dx.doi.org/10.1016/0550-3213(91)90407-O}{Nucl. Phys. B \textbf{360} (1991) 362--396}.

\bibitem[MR23]{MooreRabe:TASISquareDancing}
G.~W. Moore and K.~Rabe, \emph{{Lecture on Square Dancing and it's Relation to Mathematics (TASI 2023)}}, YouTube, 2023. \url{https://youtu.be/CjvQc9fgAlE}. {Accessed: 2025-03-7}.

\bibitem[MS74]{MilnorStasheff}
J.~W. Milnor and J.~D. Stasheff, \emph{{Characteristic Classes. (AM-76)}}, Princeton University Press, 1974. \url{http://www.jstor.org/stable/j.ctt1b7x751.1}.

\bibitem[MS89a]{Moore:1989vd}
G.~W. Moore and N.~Seiberg, \emph{{LECTURES ON RCFT}}, {1989 Banff NATO ASI: Physics, Geometry and Topology}, September 1989.

\bibitem[MS89b]{Moore:1988qv}
\bysame, \emph{{Classical and Quantum Conformal Field Theory}}, \doihref{http://dx.doi.org/10.1007/BF01238857}{Commun. Math. Phys. \textbf{123} (1989) 177}.

\bibitem[MS89c]{Moore:1989yh}
\bysame, \emph{{Taming the Conformal Zoo}}, \doihref{http://dx.doi.org/10.1016/0370-2693(89)90897-6}{Phys. Lett. B \textbf{220} (1989) 422--430}.

\bibitem[MS99]{Mickelsson:1999td}
J.~Mickelsson and S.~Scott, \emph{{Functorial QFT, Gauge Anomalies and the Dirac Determinant Bundle}}, \doihref{http://dx.doi.org/10.1007/s002200100429}{Commun. Math. Phys. \textbf{219} (2001) 567--605}, \href{http://arxiv.org/abs/hep-th/9908207}{{\arxivfont arXiv:hep-th/9908207}}.

\bibitem[MS00]{Mathai:2000iw}
V.~Mathai and I.~M. Singer, \emph{{Twisted K-homology theory, twisted Ext-theory}}, \href{http://arxiv.org/abs/hep-th/0012046}{{\arxivfont arXiv:hep-th/0012046}}.

\bibitem[MS02]{Moore:2002cp}
G.~W. Moore and N.~Saulina, \emph{{T-Duality, and the K-Theoretic Partition Function of Type IIA Superstring Theory}}, \doihref{http://dx.doi.org/10.1016/j.nuclphysb.2003.07.028}{Nucl. Phys. B \textbf{670} (2003) 27--89}, \href{http://arxiv.org/abs/hep-th/0206092}{{\arxivfont arXiv:hep-th/0206092}}.

\bibitem[MS06]{Moore:2006dw}
G.~W. Moore and G.~Segal, \emph{{D-branes and K-theory in 2D topological field theory}}, \href{http://arxiv.org/abs/hep-th/0609042}{{\arxivfont arXiv:hep-th/0609042}}.

\bibitem[MSS24]{Moore:2024vsd}
G.~W. Moore, V.~Saxena, and R.~K. Singh, \emph{{Topological Twisting of 4d $\mathcal{N}=2$ Supersymmetric Field Theories}}, \href{http://arxiv.org/abs/2411.14396}{{\arxivfont arXiv:2411.14396 [hep-th]}}.

\bibitem[MT91]{MimuraToda1991}
M.~Mimura and H.~Toda, \emph{{Topology of Lie Groups. I, II}}, Translations of Mathematical Monographs, vol.~91, American Mathematical Society, Providence, RI, 1991. Translated from the 1978 Japanese edition by the authors.

\bibitem[MT11]{Moore:2011ee}
G.~W. Moore and Y.~Tachikawa, \emph{{On 2d TQFTs whose values are holomorphic symplectic varieties}}, \doihref{http://dx.doi.org/10.1090/pspum/085/1379}{Proc. Symp. Pure Math. \textbf{85} (2012) 191--208}, \href{http://arxiv.org/abs/1106.5698}{{\arxivfont arXiv:1106.5698 [hep-th]}}.

\bibitem[M{\"u}l25]{Muller:2025ext}
L.~M{\"u}ller, \emph{{On the Higher Categorical Structure of Topological Defects in Quantum Field Theories}}, \href{http://arxiv.org/abs/2505.04761}{{\arxivfont arXiv:2505.04761 [math-ph]}}.

\bibitem[Mum06]{Mumford2006-fe}
D.~Mumford, \emph{{Tata Lectures on Theta I}}, Modern Birkh{\"a}user Classics, Birkhauser Boston, Secaucus, NJ, December 2006 (en).

\bibitem[Mur07]{Murray:2007ps}
M.~K. Murray, \emph{{An Introduction to Bundle Gerbes}}, December 2007. \href{http://arxiv.org/abs/0712.1651}{{\arxivfont arXiv:0712.1651 [math.DG]}}.

\bibitem[MW99]{Moore:1999gb}
G.~W. Moore and E.~Witten, \emph{{Self-Duality, Ramond-Ramond Fields, and K-Theory}}, \doihref{http://dx.doi.org/10.1088/1126-6708/2000/05/032}{JHEP \textbf{05} (2000) 032}, \href{http://arxiv.org/abs/hep-th/9912279}{{\arxivfont arXiv:hep-th/9912279}}.

\bibitem[MZ25]{Montero:2025ayi}
M.~Montero and L.~Zapata, \emph{{M-theory boundaries beyond supersymmetry}}, \href{http://arxiv.org/abs/2504.06985}{{\arxivfont arXiv:2504.06985 [hep-th]}}.

\bibitem[Nek96]{NekrasovPhD}
N.~A. Nekrasov, \emph{Four-dimensional holomorphic theories}, Ph.D. thesis, 1996, p.~174. \url{https://www.proquest.com/openview/c854d1153dd9eb52d61ece652a84227f/1?pq-origsite=gscholar&cbl=18750&diss=y}.

\bibitem[NH18]{Nandkishore:2018sel}
R.~M. Nandkishore and M.~Hermele, \emph{{Fractons}}, \doihref{http://dx.doi.org/10.1146/annurev-conmatphys-031218-013604}{Ann. Rev. Condensed Matter Phys. \textbf{10} (2019) 295--313}, \href{http://arxiv.org/abs/1803.11196}{{\arxivfont arXiv:1803.11196 [cond-mat.str-el]}}.

\bibitem[Nid17]{Nidaiev:PathIntegralUnpublished}
I.~Nidaiev, \emph{{Path integral approach to self-dual gauge fields in $4\ell +2$ dimensions}}, Unpublished, 2017.

\bibitem[Nid19]{Nidaiev:2019mzt}
I.~Nidaiev, \doihref{http://dx.doi.org/10.7282/t3-gye7-jv86}{\emph{{Cohomological Field Theories and Four-Manifold Invariants}}}, Ph.D. thesis, Rutgers U., Piscataway (main), 5 2019.

\bibitem[{nLa}25a]{nlab:dirac-born-infeld_action}
{nLab authors}, \emph{{Dirac-Born-Infeld action}}, \url{https://ncatlab.org/nlab/show/Dirac-Born-Infeld+action}, January 2025. \href{https://ncatlab.org/nlab/revision/Dirac-Born-Infeld+action/51}{Revision 51}.

\bibitem[{nLa}25b]{nlab:simplicial_set}
\bysame, \emph{{simplicial set}}, \url{https://ncatlab.org/nlab/show/simplicial+set}, January 2025. \href{https://ncatlab.org/nlab/revision/simplicial+set/89}{Revision 89}.

\bibitem[Nov82]{Novikov:1982ei}
S.~P. Novikov, \emph{{The Hamiltonian formalism and a many valued analog of Morse theory}}, \doihref{http://dx.doi.org/10.1070/RM1982v037n05ABEH004020}{Usp. Mat. Nauk \textbf{37N5} (1982) 3--49}.

\bibitem[NY20]{Neitzke:2020jik}
A.~Neitzke and F.~Yan, \emph{{$q$-nonabelianization for line defects}}, \doihref{http://dx.doi.org/10.1007/JHEP09(2020)153}{JHEP \textbf{09} (2020) 153}, \href{http://arxiv.org/abs/2002.08382}{{\arxivfont arXiv:2002.08382 [hep-th]}}.

\bibitem[Och87]{Ochanine1987}
S.~Ochanine, \emph{Sur les genres multiplicatifs d{\'e}finis par des int{\'e}grales elliptiques}, \href{http://dx.doi.org/10.1016/0040-9383(87)90055-3}{Topology \textbf{26} (1987) 143–151}.

\bibitem[Pac91]{PACHNER1991129}
U.~Pachner, \emph{{P.L. Homeomorphic Manifolds are Equivalent by Elementary Shellings}}, \href{https://www.sciencedirect.com/science/article/pii/S0195669813800807}{European Journal of Combinatorics \textbf{12} (1991) 129--145}.

\bibitem[PCY20]{Pretko:2020cko}
M.~Pretko, X.~Chen, and Y.~You, \emph{{Fracton Phases of Matter}}, \doihref{http://dx.doi.org/10.1142/S0217751X20300033}{Int. J. Mod. Phys. A \textbf{35} (2020) 2030003}, \href{http://arxiv.org/abs/2001.01722}{{\arxivfont arXiv:2001.01722 [cond-mat.str-el]}}.

\bibitem[Pes16]{Pestun:2016qko}
V.~Pestun, \emph{{Review of localization in geometry}}, \doihref{http://dx.doi.org/10.1088/1751-8121/aa6161}{J. Phys. A \textbf{50} (2017) 443002}, \href{http://arxiv.org/abs/1608.02954}{{\arxivfont arXiv:1608.02954 [hep-th]}}.

\bibitem[Ply86]{Plymen1986}
R.~J. Plymen, \emph{{STRONG MORITA EQUIVALENCE, SPINORS AND SYMPLECTIC SPINORS}}, \href{http://www.jstor.org/stable/24714803}{Journal of Operator Theory \textbf{16} (1986) 305--324}.

\bibitem[Pol95]{Polchinski:1995mt}
J.~Polchinski, \emph{{Dirichlet Branes and Ramond-Ramond charges}}, \doihref{http://dx.doi.org/10.1103/PhysRevLett.75.4724}{Phys. Rev. Lett. \textbf{75} (1995) 4724--4727}, \href{http://arxiv.org/abs/hep-th/9510017}{{\arxivfont arXiv:hep-th/9510017}}.

\bibitem[Por21]{Porter2021}
T.~Porter, \emph{{Spaces as Infinity-Groupoids}}, p.~258–321, Cambridge University Press, April 2021. \url{http://dx.doi.org/10.1017/9781108854429.008}.

\bibitem[PS05]{Pantev:2005zs}
T.~Pantev and E.~Sharpe, \emph{{GLSM's for Gerbes (and other toric stacks)}}, \doihref{http://dx.doi.org/10.4310/ATMP.2006.v10.n1.a4}{Adv. Theor. Math. Phys. \textbf{10} (2006) 77--121}, \href{http://arxiv.org/abs/hep-th/0502053}{{\arxivfont arXiv:hep-th/0502053}}.

\bibitem[PS24]{Padilla:2024mkm}
A.~Padilla and R.~G.~C. Smith, \emph{{Smoothed asymptotics: From number theory to QFT}}, \doihref{http://dx.doi.org/10.1103/PhysRevD.110.025010}{Phys. Rev. D \textbf{110} (2024) 025010}, \href{http://arxiv.org/abs/2401.10981}{{\arxivfont arXiv:2401.10981 [hep-th]}}.

\bibitem[PZ16]{Pestun:2016jze}
V.~Pestun and M.~Zabzine, \emph{{Introduction to localization in quantum field theory}}, \doihref{http://dx.doi.org/10.1088/1751-8121/aa5704}{J. Phys. A \textbf{50} (2017) 443001}, \href{http://arxiv.org/abs/1608.02953}{{\arxivfont arXiv:1608.02953 [hep-th]}}.

\bibitem[Qui95]{Quinn:1991kq}
F.~Quinn, \href{https://bookstore.ams.org/pcms-1}{\emph{{Lectures on axiomatic topological quantum field theory}}}, {Geometry and Quantum Field Theory} (D.~Freed and K.~Uhlenbeck, eds.), IAS/Park City mathematics series, American Mathematical Society, March 1995.

\bibitem[Rez00]{Rezk2000}
C.~Rezk, \emph{A model for the homotopy theory of homotopy theory}, \href{http://dx.doi.org/10.1090/S0002-9947-00-02653-2}{Transactions of the American Mathematical Society \textbf{353} (2000) 973–1007}.

\bibitem[Rie11]{RiehlSimplicialSets}
E.~Riehl, \emph{{A leisurely introduction to simplicial sets}}, 2011. \url{https://emilyriehl.github.io/files/ssets.pdf}.

\bibitem[Rie14]{RiehlCategoricalHomotopyTheory}
\bysame, \doihref{http://dx.doi.org/10.1017/cbo9781107261457}{\emph{{Categorical Homotopy Theory}}}, Cambridge University Press, May 2014. \url{http://dx.doi.org/10.1017/CBO9781107261457}.

\bibitem[Rie17]{Riehl:2017category}
E.~Riehl, \emph{{Category theory in context}}, {Aurora: {Dover} Modern Math Originals}, Dover Publications, 2017.

\bibitem[Rob66a]{Roberts1966:BraKet}
J.~E. Roberts, \emph{{The Dirac Bra and Ket Formalism}}, \href{http://dx.doi.org/10.1063/1.1705001}{Journal of Mathematical Physics \textbf{7} (1966) 1097–1104}.

\bibitem[Rob66b]{Roberts1966:Rigged}
\bysame, \emph{{Rigged Hilbert spaces in quantum mechanics}}, \href{http://dx.doi.org/10.1007/BF01645448}{Communications in Mathematical Physics \textbf{3} (1966) 98–119}.

\bibitem[Roc93]{Roca:1992ry}
J.~Roca, \emph{{An Introduction to topological field theories}}, \doihref{http://dx.doi.org/10.1007/BF02725740}{Riv. Nuovo Cim. \textbf{16N4} (1993) 1--64}.

\bibitem[Ros89]{Rosenberg1989}
J.~Rosenberg, \emph{{Continuous-trace algebras from the bundle theoretic point of view}}, J. Austral. Math. Soc. Ser. A \textbf{47} (1989) 368--381.

\bibitem[Row05]{Rowell:2005hv}
E.~C. Rowell, \emph{{From Quantum Groups to Unitary Modular Tensor Categories}}, \href{http://arxiv.org/abs/math/0503226}{{\arxivfont arXiv:math/0503226}}.

\bibitem[Row16]{Rowell:2016lrv}
E.~C. Rowell, \emph{{An Invitation to the Mathematics of Topological Quantum Computation}}, \doihref{http://dx.doi.org/10.1088/1742-6596/698/1/012012}{J. Phys. Conf. Ser. \textbf{698} (2016) 012012}, \href{http://arxiv.org/abs/1601.05288}{{\arxivfont arXiv:1601.05288 [math-ph]}}.

\bibitem[RS71]{RaySinger:1}
D.~B. Ray and I.~M. Singer, \emph{{$R$-torsion and the Laplacian on Riemannian manifolds}}, \doihref{http://dx.doi.org/10.1016/0001-8708(71)90045-4}{Adv. Math. \textbf{7} (1971) 145--210}. MR:295381. Zbl:0239.58014.

\bibitem[RS72]{ReedSimon:1972}
M.~Reed and B.~Simon, \doihref{http://dx.doi.org/10.1016/b978-0-12-585001-8.x5001-6}{\emph{{Methods of Modern Mathematical Physics}}}, Elsevier, 1972. \url{http://dx.doi.org/10.1016/B978-0-12-585001-8.X5001-6}.

\bibitem[RS73a]{RaySinger:2}
D.~B. Ray and I.~M. Singer, \emph{{Analytic torsion for complex manifolds}}, \doihref{http://dx.doi.org/10.2307/1970909}{Ann. Math. (2) \textbf{98} (1973) 154--177}. MR:383463. Zbl:0267.32014.

\bibitem[RS73b]{RaySinger:3}
\bysame, \href{http://www.ams.org/books/pspum/023/0339293}{\emph{{Analytic torsion}}}, {Partial differential equations} (D.~C. Spencer, ed.), {Proceedings of Symposia in Pure Mathematics}, no.~23, American Mathematical Society, Providence, RI, 1973, pp.~167--181. (Berkeley, CA, 9--27 August 1971). MR:339293. Zbl:0273.58014.

\bibitem[RS18]{Runkel:2018uls}
I.~Runkel and L.~Szegedy, \emph{{Area-Dependent Quantum Field Theory}}, \doihref{http://dx.doi.org/10.1007/s00220-020-03902-1}{Commun. Math. Phys. \textbf{381} (2021) 83--117}, \href{http://arxiv.org/abs/1807.08196}{{\arxivfont arXiv:1807.08196 [math.QA]}}.

\bibitem[RSFL09]{Ryu:2010zza}
S.~Ryu, A.~P. Schnyder, A.~Furusaki, and A.~W.~W. Ludwig, \emph{{Topological insulators and superconductors: Tenfold way and dimensional hierarchy}}, \doihref{http://dx.doi.org/10.1088/1367-2630/12/6/065010}{New J. Phys. \textbf{12} (2010) 065010}, \href{http://arxiv.org/abs/0912.2157}{{\arxivfont arXiv:0912.2157 [cond-mat.mes-hall]}}.

\bibitem[RT90]{Reshetikhin:1990}
N.~Y. Reshetikhin and V.~G. Turaev, \emph{Ribbon graphs and their invaraints derived from quantum groups}, \href{http://dx.doi.org/10.1007/BF02096491}{Communications in Mathematical Physics \textbf{127} (1990) 1–26}.

\bibitem[RT91]{Reshetikhin:1991}
N.~Reshetikhin and V.~G. Turaev, \emph{Invariants of 3-manifolds via link polynomials and quantum groups}, \href{http://dx.doi.org/10.1007/BF01239527}{Inventiones Mathematicae \textbf{103} (1991) 547–597}.

\bibitem[Rud90]{Rudin1990-od}
W.~Rudin, \emph{Functional analysis}, 2 ed., International Series in Pure \& Applied Mathematics, McGraw Hill Higher Education, Maidenhead, England, October 1990 (en).

\bibitem[Rud98]{Rudyak1998-vw}
Y.~B. Rudyak, \emph{{On Thom Spectra, Orientability, and Cobordism}}, 1 ed., Springer Monographs in Mathematics, Springer, Berlin, Germany, April 1998 (en).

\bibitem[RV91]{Rocek:1991ps}
M.~Ro\v{c}ek and E.~P. Verlinde, \emph{{Duality, Quotients, and Currents}}, \doihref{http://dx.doi.org/10.1016/0550-3213(92)90269-H}{Nucl. Phys. B \textbf{373} (1992) 630--646}, \href{http://arxiv.org/abs/hep-th/9110053}{{\arxivfont arXiv:hep-th/9110053}}.

\bibitem[RV22]{RiehlInfinityCategoryTheory}
E.~Riehl and D.~Verity, \doihref{http://dx.doi.org/10.1017/9781108936880}{\emph{{Elements of $\infty$-Category Theory}}}, Cambridge University Press, January 2022. \url{http://dx.doi.org/10.1017/9781108936880}.

\bibitem[RW96]{Rozansky:1996bq}
L.~Rozansky and E.~Witten, \emph{{Hyper-K\"ahler Geometry and Invariants of Three-Manifolds}}, \doihref{http://dx.doi.org/10.1007/s000290050016}{Selecta Math. \textbf{3} (1997) 401--458}, \href{http://arxiv.org/abs/hep-th/9612216}{{\arxivfont arXiv:hep-th/9612216}}.

\bibitem[RW17]{Rowell:2018wnv}
E.~Rowell and Z.~Wang, \emph{{Mathematics of Topological Quantum Computing}}, \doihref{http://dx.doi.org/10.1090/bull/1605}{Bull. Am. Math. Soc. \textbf{55} (2018) 183--238}, \href{http://arxiv.org/abs/1705.06206}{{\arxivfont arXiv:1705.06206 [math.QA]}}.

\bibitem[Sax24]{Saxena:2024eil}
V.~Saxena, \emph{{A T-duality of non-supersymmetric heterotic strings and an implication for Topological Modular Forms}}, \doihref{http://dx.doi.org/10.1007/JHEP09(2024)056}{JHEP \textbf{09} (2024) 056}, \href{http://arxiv.org/abs/2405.19409}{{\arxivfont arXiv:2405.19409 [hep-th]}}.

\bibitem[Sch48]{Schwinger1948}
J.~Schwinger, \emph{{Quantum Electrodynamics. I. A Covariant Formulation}}, \href{https://link.aps.org/doi/10.1103/PhysRev.74.1439}{Phys. Rev. \textbf{74} (1948) 1439--1461}.

\bibitem[Sch79]{Schwarz:1979ae}
A.~S. Schwarz, \emph{{The Partition Function of a Degenerate Functional}}, \doihref{http://dx.doi.org/10.1007/BF01223197}{Commun. Math. Phys. \textbf{67} (1979) 1--16}.

\bibitem[Sch92]{Schwarz:1992nx}
\bysame, \emph{{Geometry of Batalin-Vilkovisky quantization}}, \doihref{http://dx.doi.org/10.1007/BF02097392}{Commun. Math. Phys. \textbf{155} (1993) 249--260}, \href{http://arxiv.org/abs/hep-th/9205088}{{\arxivfont arXiv:hep-th/9205088}}.

\bibitem[Sch09]{SchochetDummies}
C.~Schochet, \emph{{Dixmier-Douady for Dummies}}, Notices of the American Mathematical Society \textbf{56} (2009) 809--819, \href{http://arxiv.org/abs/0902.2025}{{\arxivfont arXiv:0902.2025 [math.OA]}}.

\bibitem[Sch13]{Schreiber:2013pra}
U.~Schreiber, \emph{{Differential cohomology in a cohesive infinity-topos}}, \href{http://arxiv.org/abs/1310.7930}{{\arxivfont arXiv:1310.7930 [math-ph]}}.

\bibitem[SdBKP18]{Stehouwer:2018xfs}
L.~Stehouwer, J.~de~Boer, J.~Kruthoff, and H.~Posthuma, \emph{{Classification of crystalline topological insulators through $K$-theory}}, \doihref{http://dx.doi.org/10.4310/ATMP.2021.v25.n3.a3}{Adv. Theor. Math. Phys. \textbf{25} (2021) 723--775}, \href{http://arxiv.org/abs/1811.02592}{{\arxivfont arXiv:1811.02592 [cond-mat.mes-hall]}}.

\bibitem[Seg87]{Segal:1987sk}
G.~B. Segal, \emph{{THE DEFINITION OF CONFORMAL FIELD THEORY}}, 1987.

\bibitem[Seg88a]{Segal:1988zk}
G.~Segal, \emph{{TWO-DIMENSIONAL CONFORMAL FIELD THEORIES AND MODULAR FUNCTIONS}}, {IX International Conference on Mathematical Physics (IAMP)}, 1988.

\bibitem[Seg88b]{Segal88}
\bysame, \href{http://www.numdam.org/item/SB_1987-1988__30__187_0/}{\emph{{Elliptic cohomology}}}, {S\'eminaire Bourbaki : volume 1987/88, expos\'es 686-699}, Ast\'erisque, no. 161-162, Soci\'et\'e math\'ematique de France, 1988, pp.~187--201 (en). talk:695.

\bibitem[Seg88c]{Segal1988}
G.~B. Segal, \emph{{The Definition of Conformal Field Theory}}, pp.~165--171, Springer Netherlands, Dordrecht, 1988. \url{https://doi.org/10.1007/978-94-015-7809-7_9}.

\bibitem[Seg99]{Segal:DiracOperators}
G.~Segal, \emph{{The Index and Determinant of the Dirac Operator}}, ITP Santa Barbara Lectures, 1999, \url{https://web.math.ucsb.edu/~drm/conferences/ITP99/segal/stanford/lect2.pdf}.

\bibitem[Seg02]{Segal:2002ei}
\bysame, \emph{{The definition of conformal field theory}}, {Symposium on Topology, Geometry and Quantum Field Theory (Segalfest)}, June 2002, pp.~421--575.

\bibitem[Seg07]{Segal2007}
\bysame, \doihref{http://dx.doi.org/10.1017/CBO9780511721489.016}{\emph{{What is an elliptic object?}}}, {Elliptic Cohomology: Geometry, Applications, and Higher Chromatic Analogues} (H.~R. Miller and D.~C.~E. Ravenel, eds.), London Mathematical Society Lecture Note Series, Cambridge University Press, 2007, p.~306–317.

\bibitem[Seg23]{Segal:FaddeevAnomalyUnpublished}
\bysame, \emph{{Fadeev’s anomaly in Gauss’s law}}, Oxford preprint (Unpublished), 2023.

\bibitem[Sen94]{Sengupta1994}
A.~Sengupta, \emph{{Gauge invariant functions of connections}}, \href{http://dx.doi.org/10.1090/S0002-9939-1994-1215205-7}{Proceedings of the American Mathematical Society \textbf{121} (1994) 897–905}.

\bibitem[Sen19]{Sen:2019qit}
A.~Sen, \emph{{Self-dual forms: Action, Hamiltonian and Compactification}}, \doihref{http://dx.doi.org/10.1088/1751-8121/ab5423}{J. Phys. A \textbf{53} (2020) 084002}, \href{http://arxiv.org/abs/1903.12196}{{\arxivfont arXiv:1903.12196 [hep-th]}}.

\bibitem[Ser73]{Serre1973}
J.-P. Serre, \doihref{http://dx.doi.org/10.1007/978-1-4684-9884-4}{\emph{{A Course in Arithmetic}}}, Springer New York, 1973. \url{http://dx.doi.org/10.1007/978-1-4684-9884-4}.

\bibitem[Sha19]{Sharpe:2019ddn}
E.~Sharpe, \emph{{Undoing decomposition}}, \doihref{http://dx.doi.org/10.1142/S0217751X19502336}{Int. J. Mod. Phys. A \textbf{34} (2020) 1950233}, \href{http://arxiv.org/abs/1911.05080}{{\arxivfont arXiv:1911.05080 [hep-th]}}.

\bibitem[Sha22]{Sharpe:2022ene}
\bysame, \emph{{An introduction to decomposition}}, 2024. \href{http://arxiv.org/abs/2204.09117}{{\arxivfont arXiv:2204.09117 [hep-th]}}.

\bibitem[Sha23a]{Shao:TASIVideoLec1}
S.-H. Shao, \emph{{Lecture 1 on Noninvertible Symmetry (TASI 2023)}}, YouTube, 2023. \url{https://youtu.be/rsweL30d-3M}. {Accessed: 2023-06-26}.

\bibitem[Sha23b]{Shao:TASIVideoLec2}
\bysame, \emph{{Lecture 2 on Noninvertible Symmetry (TASI 2023)}}, YouTube, 2023. \url{https://youtu.be/FMb_id2pXdU}. {Accessed: 2023-06-27}.

\bibitem[Sha23c]{Shao:TASIVideoLec3}
\bysame, \emph{{Lecture 3 on Noninvertible Symmetry (TASI 2023)}}, YouTube, 2023. \url{https://youtu.be/5hidsLao7x4}. {Accessed: 2023-06-29}.

\bibitem[Sha23d]{Shao:TASIVideoLec4}
\bysame, \emph{{Lecture 4 on Noninvertible Symmetry (TASI 2023)}}, YouTube, 2023. \url{https://youtu.be/5w7isXDfJmc}. {Accessed: 2023-07-02}.

\bibitem[Sha23e]{SharpeTalk:2023}
E.~Sharpe, \emph{{Generalized symmetries and gauge theory multiverses}}, April 2023. \url{http://www1.phys.vt.edu/~ersharpe/toronto-apr23.pdf}.

\bibitem[Sha23f]{Shao:2023gho}
S.-H. Shao, \emph{{What's Done Cannot Be Undone: TASI Lectures on Non-Invertible Symmetries}}, \href{http://arxiv.org/abs/2308.00747}{{\arxivfont arXiv:2308.00747 [hep-th]}}.

\bibitem[Sha23g]{Sharpe:2023lfk}
E.~Sharpe, \emph{{Dilaton shifts, probability measures, and decomposition}}, \doihref{http://dx.doi.org/10.1088/1751-8121/ad8196}{J. Phys. A \textbf{57} (2024) 445401}, \href{http://arxiv.org/abs/2312.08438}{{\arxivfont arXiv:2312.08438 [hep-th]}}.

\bibitem[Sha25]{Shao2025:PT}
S.-H. Shao, \emph{{Noninvertible symmetries: What's done cannot be undone}}, \href{http://dx.doi.org/10.1063/pt.fspd.veje}{Physics Today \textbf{2025} (2025) }.

\bibitem[Sie80]{Siegel:1980jj}
W.~Siegel, \emph{{Hidden Ghosts}}, \doihref{http://dx.doi.org/10.1016/0370-2693(80)90119-7}{Phys. Lett. B \textbf{93} (1980) 170--172}.

\bibitem[Sie83]{Siegel:1983hh}
\bysame, \emph{{Hidden Local Supersymmetry in the Supersymmetric Particle Action}}, \doihref{http://dx.doi.org/10.1016/0370-2693(83)90924-3}{Phys. Lett. B \textbf{128} (1983) 397--399}.

\bibitem[Sie84]{Siegel:1983ke}
\bysame, \emph{{Light Cone Analysis of Covariant Superstring}}, \doihref{http://dx.doi.org/10.1016/0550-3213(84)90537-6}{Nucl. Phys. B \textbf{236} (1984) 311--318}.

\bibitem[Sim11]{Simons:HexagonVideo}
J.~Simons, \emph{{Roots of Differential Characters, The Characteristic Diagram and The Mayer-Vietoris Property}}, SCGP Video Portal, 2011. \url{https://scgp.stonybrook.edu/video_portal/video.php?id=61}. {Accessed: 2025-03-17}.

\bibitem[Sim23]{Simon2023-jn}
S.~H. Simon, \emph{{Topological Quantum}}, Oxford University Press, London, England, September 2023 (en).

\bibitem[SN23]{Schafer-Nameki:2023jdn}
S.~Sch{\"a}fer-Nameki, \emph{{ICTP Lectures on (Non-)Invertible Generalized Symmetries}}, \doihref{http://dx.doi.org/10.1016/j.physrep.2024.01.007}{Phys. Rept. \textbf{1063} (2024) 1--55}, \href{http://arxiv.org/abs/2305.18296}{{\arxivfont arXiv:2305.18296 [hep-th]}}.

\bibitem[SPar]{Schommer-Pries:2013}
C.~Schommer-Pries, \emph{{Dualizability in Low-Dimensional Higher Category Theory}}, \href{http://arxiv.org/abs/arXiv:1308.3574}{{\arxivfont arXiv:arXiv:1308.3574 [math.AT]}}.

\bibitem[Spa81]{Spanier1981}
E.~H. Spanier, \doihref{http://dx.doi.org/10.1007/978-1-4684-9322-1}{\emph{{Algebraic Topology}}}, Springer New York, 1981. \url{http://dx.doi.org/10.1007/978-1-4684-9322-1}.

\bibitem[SRF{\etalchar{+}}09]{Schnyder:2009klk}
A.~P. Schnyder, S.~Ryu, A.~Furusaki, A.~W.~W. Ludwig, and V.~Lebedev, \emph{{Classification of Topological Insulators and Superconductors}}, \doihref{http://dx.doi.org/10.1063/1.3149481}{AIP Conf. Proc. \textbf{1134} (2009) 10}, \href{http://arxiv.org/abs/0905.2029}{{\arxivfont arXiv:0905.2029 [cond-mat.mes-hall]}}.

\bibitem[SRFL08]{Schnyder:2008tya}
A.~Schnyder, S.~Ryu, A.~Furusaki, and A.~Ludwig, \emph{{Classification of topological insulators and superconductors in three spatial dimensions}}, \doihref{http://dx.doi.org/10.1103/PhysRevB.78.195125}{Phys. Rev. B \textbf{78} (2008) 195125}, \href{http://arxiv.org/abs/0803.2786}{{\arxivfont arXiv:0803.2786 [cond-mat.mes-hall]}}.

\bibitem[SS78]{Steen1978}
L.~A. Steen and J.~A. Seebach, \doihref{http://dx.doi.org/10.1007/978-1-4612-6290-9}{\emph{{Counterexamples in Topology}}}, Springer New York, 1978. \url{http://dx.doi.org/10.1007/978-1-4612-6290-9}.

\bibitem[SS07]{Simons2007}
J.~Simons and D.~Sullivan, \emph{{Axiomatic characterization of ordinary differential cohomology}}, \href{http://dx.doi.org/10.1112/jtopol/jtm006}{Journal of Topology \textbf{1} (2007) 45–56}, \href{http://arxiv.org/abs/math/0701077}{{\arxivfont arXiv:math/0701077 [math.AT]}}.

\bibitem[SS20]{Sati:2020cml}
H.~Sati and U.~Schreiber, \emph{{Twisted cohomotopy implies M5-brane anomaly cancellation}}, \doihref{http://dx.doi.org/10.1007/s11005-021-01452-8}{Lett. Math. Phys. \textbf{111} (2021) 120}, \href{http://arxiv.org/abs/2002.07737}{{\arxivfont arXiv:2002.07737 [hep-th]}}.

\bibitem[SS24]{Santilli:2024dyz}
L.~Santilli and R.~J. Szabo, \emph{{Higher form symmetries and orbifolds of two-dimensional Yang-Mills theory}}, \href{http://arxiv.org/abs/2403.03119}{{\arxivfont arXiv:2403.03119 [hep-th]}}.

\bibitem[SSS19]{Saad:2019lba}
P.~Saad, S.~H. Shenker, and D.~Stanford, \emph{{JT gravity as a matrix integral}}, \href{http://arxiv.org/abs/1903.11115}{{\arxivfont arXiv:1903.11115 [hep-th]}}.

\bibitem[SSW16]{Salton:2016qpp}
G.~Salton, B.~Swingle, and M.~Walter, \emph{{Entanglement from Topology in Chern-Simons Theory}}, \doihref{http://dx.doi.org/10.1103/PhysRevD.95.105007}{Phys. Rev. D \textbf{95} (2017) 105007}, \href{http://arxiv.org/abs/1611.01516}{{\arxivfont arXiv:1611.01516 [quant-ph]}}.

\bibitem[SSWC17]{Shirley:2017suz}
W.~Shirley, K.~Slagle, Z.~Wang, and X.~Chen, \emph{{Fracton Models on General Three-Dimensional Manifolds}}, \doihref{http://dx.doi.org/10.1103/PhysRevX.8.031051}{Phys. Rev. X \textbf{8} (2018) 031051}, \href{http://arxiv.org/abs/1712.05892}{{\arxivfont arXiv:1712.05892 [cond-mat.str-el]}}.

\bibitem[ST04]{StolzTeichner1}
S.~Stolz and P.~Teichner, \doihref{http://dx.doi.org/10.1017/CBO9780511526398.013}{\emph{{What is an elliptic object?}}}, {Topology, geometry and quantum field theory}, London Math. Soc. Lecture Note Ser., vol. 308, Cambridge Univ. Press, Cambridge, 2004, pp.~247--343.

\bibitem[ST11]{StolzTeichner2}
\bysame, \doihref{http://dx.doi.org/10.1090/pspum/083/2742432}{\emph{{Supersymmetric field theories and generalized cohomology}}}, {Mathematical foundations of quantum field theory and perturbative string theory}, Proc. Sympos. Pure Math., vol.~83, AMS, 2011, pp.~279--340. \href{http://arxiv.org/abs/1108.0189}{{\arxivfont arXiv:1108.0189 [math.AT]}}.

\bibitem[Sti08]{Stirling:2008bq}
S.~D. Stirling, \emph{{Abelian Chern-Simons theory with toral gauge group, modular tensor categories, and group categories}}, Ph.D. thesis, Texas U., Math Dept., 2008. \href{http://arxiv.org/abs/0807.2857}{{\arxivfont arXiv:0807.2857 [hep-th]}}.

\bibitem[Sto16]{Stong2016-wq}
R.~E. Stong, \emph{Notes on cobordism theory}, Princeton Legacy Library, Princeton University Press, Princeton, NJ, April 2016 (en).

\bibitem[{Sto}23]{stonybrook-differential-cohomology-premiere}
{Stony Brook University News}, \emph{{``Differential Cohomology'' Premieres at Staller on January 10, 2011}}, \url{https://news.stonybrook.edu/arts/differential-cohomology-premieres-at-staller-on-january-10-2/}, 2023. Accessed: 2025-04-28.

\bibitem[Str23]{Strom:2023}
J.~Strom, \emph{{Modern Classical Homotopy Theory}}, American Mathematical Society, Providence, RI, January 2023 (en).

\bibitem[SW19]{Stanford:2019vob}
D.~Stanford and E.~Witten, \emph{{JT gravity and the ensembles of random matrix theory}}, \doihref{http://dx.doi.org/10.4310/ATMP.2020.v24.n6.a4}{Adv. Theor. Math. Phys. \textbf{24} (2020) 1475--1680}, \href{http://arxiv.org/abs/1907.03363}{{\arxivfont arXiv:1907.03363 [hep-th]}}.

\bibitem[Swi75]{Switzer1975}
R.~M. Switzer, \doihref{http://dx.doi.org/10.1007/978-3-642-61923-6}{\emph{{Algebraic Topology — Homotopy and Homology}}}, Springer Berlin Heidelberg, 1975. \url{http://dx.doi.org/10.1007/978-3-642-61923-6}.

\bibitem[SWnd]{scheimbauer-walde-wip}
C.~Scheimbauer and T.~Walde, n.d. Work in progress.

\bibitem[Sza00]{Szabo2000}
R.~J. Szabo, \doihref{http://dx.doi.org/10.1007/3-540-46550-2}{\emph{Equivariant cohomology and localization of path integrals}}, Springer Berlin Heidelberg, 2000. \url{http://dx.doi.org/10.1007/3-540-46550-2}.

\bibitem[Sza12]{Szabo:2012hc}
\bysame, \emph{{Quantization of Higher Abelian Gauge Theory in Generalized Differential Cohomology}}, \doihref{http://dx.doi.org/10.22323/1.175.0009}{PoS \textbf{ICMP2012} (2012) 009}, \href{http://arxiv.org/abs/1209.2530}{{\arxivfont arXiv:1209.2530 [hep-th]}}.

\bibitem[Tac21]{Tachikawa:2021mvw}
Y.~Tachikawa, \emph{{Topological modular forms and the absence of a heterotic global anomaly}}, \doihref{http://dx.doi.org/10.1093/ptep/ptab060}{PTEP \textbf{2022} (2022) 04A107}, \href{http://arxiv.org/abs/2103.12211}{{\arxivfont arXiv:2103.12211 [hep-th]}}.

\bibitem[Tac25]{Tachikawa:2025awi}
\bysame, \emph{{On a long exact sequence of groups of equivalence classes of 2d $\mathcal{N}{=}(0,1)$ SQFTs}}, \href{http://arxiv.org/abs/2509.12481}{{\arxivfont arXiv:2509.12481 [hep-th]}}.

\bibitem[TAS23]{TASI2023}
\emph{{TASI 2023 — Aspects of Symmetry}}, 2023. \url{https://www.colorado.edu/physics/events/summer-intensive-programs/theoretical-advanced-study-institute-elementary-particle-physics\#TASI-2023}. Accessed: 2025-10-02.

\bibitem[Tau11]{Taubes2011}
C.~H. Taubes, \doihref{http://dx.doi.org/10.1093/acprof:oso/9780199605880.001.0001}{\emph{Differential geometry: Bundles, connections, metrics and curvature}}, Oxford University Press, October 2011. \url{http://dx.doi.org/10.1093/acprof:oso/9780199605880.001.0001}.

\bibitem[Tel16]{Teleman:2016}
C.~Teleman, \emph{{Five Lectures on Topological Field Theory}}, \href{https://math.berkeley.edu/~teleman/math/barclect.pdf}{CRM Bellaterra, Barcelona Lectures (2016) }.

\bibitem[Tel22]{Teleman2022Simons}
\bysame, \emph{{Simons Collaboration on Global Categorical Symmetries --- Results and Outlook}}, Presentation, 2022, Available at \url{https://math.berkeley.edu/~teleman/math/Simons22}.

\bibitem[Ter05]{Terning2005}
J.~Terning, \doihref{http://dx.doi.org/10.1093/acprof:oso/9780198567639.001.0001}{\emph{Modern supersymmetry: Dynamics and duality}}, Oxford University Press, December 2005. \url{http://dx.doi.org/10.1093/acprof:oso/9780198567639.001.0001}.

\bibitem[tH78]{tHooft:1977nqb}
G.~'t~Hooft, \emph{{On the Phase Transition Towards Permanent Quark Confinement}}, \doihref{http://dx.doi.org/10.1016/0550-3213(78)90153-0}{Nucl. Phys. B \textbf{138} (1978) 1--25}.

\bibitem[Thi14]{Thiang:2014fxa}
G.~C. Thiang, \emph{{On the K-Theoretic Classification of Topological Phases of Matter}}, \doihref{http://dx.doi.org/10.1007/s00023-015-0418-9}{Annales Henri Poincare \textbf{17} (2016) 757--794}, \href{http://arxiv.org/abs/1406.7366}{{\arxivfont arXiv:1406.7366 [math-ph]}}.

\bibitem[Tho91]{Thompson:1991qt}
G.~Thompson, \emph{{Introduction to topological field theory}}, {Summer School in High-energy Physics and Cosmology}, 1991, pp.~623--652.

\bibitem[Thu11]{MO53399}
D.~Thurston, \emph{{Spaces with same homotopy and homology groups that are not homotopy equivalent?}}, MathOverflow, 2011. \url{https://mathoverflow.net/q/53399} (version: 2011-01-26).

\bibitem[Tod86]{Toda1976}
H.~Toda, \href{https://projecteuclid.org/euclid.aspm/10.2969/aspm/00910075}{\emph{{Cohomology of classifying spaces}}}, {Homotopy Theory and Related Topics}, Advanced Studies in Pure Mathematics, vol.~9, Mathematical Society of Japan, 1986, pp.~75--108.

\bibitem[Tom46]{Tomonaga:1946zz}
S.~Tomonaga, \emph{{On a Relativistically Invariant Formulation of the Quantum Theory of Wave Fields}}, \doihref{http://dx.doi.org/10.1143/PTP.1.27}{Prog. Theor. Phys. \textbf{1} (1946) 27--42}.

\bibitem[Ton16]{Tong2016QuantumHall}
D.~Tong, \emph{{The Quantum Hall Effect}}, Lecture notes, DAMTP, University of Cambridge, 2016. \url{https://www.damtp.cam.ac.uk/user/tong/qhe/qhe.pdf}.

\bibitem[Tri10]{MO43313}
T.~Trimble, \emph{{Good references for Rigged Hilbert spaces?}}, MathOverflow, 2010. \url{https://mathoverflow.net/q/43313} (version: 2022-08-28).

\bibitem[TU19]{Tanizaki:2019rbk}
Y.~Tanizaki and M.~\"Unsal, \emph{{Modified instanton sum in QCD and higher-groups}}, \doihref{http://dx.doi.org/10.1007/JHEP03(2020)123}{JHEP \textbf{03} (2020) 123}, \href{http://arxiv.org/abs/1912.01033}{{\arxivfont arXiv:1912.01033 [hep-th]}}.

\bibitem[Tur92]{Turaev:1992}
V.~G. Turaev, \emph{Modular categories and 3-manifold invariants}, \href{http://dx.doi.org/10.1142/S0217979292000876}{International Journal of Modern Physics B \textbf{06} (1992) 1807–1824}.

\bibitem[Tur10]{Turaev2010}
V.~Turaev, \doihref{http://dx.doi.org/10.4171/086}{\emph{{Homotopy Quantum Field Theory: With Appendices by Michael M\"{u}ger and Alexis Virelizier}}}, EMS Press, June 2010. \url{http://dx.doi.org/10.4171/086}.

\bibitem[Tur16]{Turaev:2016}
V.~G. Turaev, \doihref{http://dx.doi.org/10.1515/9783110435221}{\emph{Quantum invariants of knots and 3-manifolds}}, De Gruyter, July 2016. \url{http://dx.doi.org/10.1515/9783110435221}.

\bibitem[TY21]{Tachikawa:2021mby}
Y.~Tachikawa and M.~Yamashita, \emph{{Topological Modular Forms and the Absence of All Heterotic Global Anomalies}}, \doihref{http://dx.doi.org/10.1007/s00220-023-04761-2}{Commun. Math. Phys. \textbf{402} (2023) 1585--1620}, \href{http://arxiv.org/abs/2108.13542}{{\arxivfont arXiv:2108.13542 [hep-th]}}. [Erratum: Commun.Math.Phys. 402, 2131 (2023)].

\bibitem[TY23]{Tachikawa:2023lwf}
\bysame, \emph{{Anderson self-duality of topological modular forms, its differential-geometric manifestations, and vertex operator algebras}}, \href{http://arxiv.org/abs/2305.06196}{{\arxivfont arXiv:2305.06196 [math.AT]}}.

\bibitem[TY25]{Tachikawa:2025flw}
Y.~Tachikawa and K.~Yonekura, \emph{{On invariants of two-dimensional minimally supersymmetric field theories}}, \href{http://arxiv.org/abs/2508.04916}{{\arxivfont arXiv:2508.04916 [hep-th]}}.

\bibitem[TYY23]{Tachikawa:2023nne}
Y.~Tachikawa, M.~Yamashita, and K.~Yonekura, \emph{{Remarks on Mod-2 Elliptic Genus}}, \doihref{http://dx.doi.org/10.1007/s00220-024-05202-4}{Commun. Math. Phys. \textbf{406} (2025) 16}, \href{http://arxiv.org/abs/2302.07548}{{\arxivfont arXiv:2302.07548 [hep-th]}}.

\bibitem[TZ24]{Tachikawa:2024ucm}
Y.~Tachikawa and H.~Y. Zhang, \emph{{On a $\mathbb{Z}_3$-valued discrete topological term in 10d heterotic string theories}}, \doihref{http://dx.doi.org/10.21468/SciPostPhys.17.3.077}{SciPost Phys. \textbf{17} (2024) 077}, \href{http://arxiv.org/abs/2403.08861}{{\arxivfont arXiv:2403.08861 [hep-th]}}.

\bibitem[vB90]{vanBaal:1989aw}
P.~van Baal, \emph{{AN INTRODUCTION TO TOPOLOGICAL YANG-MILLS THEORY}}, Acta Phys. Polon. B \textbf{21} (1990) 73.

\bibitem[vdW20]{vandeWetering:2020giq}
J.~van~de Wetering, \emph{{ZX-calculus for the working quantum computer scientist}}, \href{http://arxiv.org/abs/2012.13966}{{\arxivfont arXiv:2012.13966 [quant-ph]}}.

\bibitem[Ver95]{Verlinde:1995mz}
E.~P. Verlinde, \emph{{Global Aspects of Electric-Magnetic Duality}}, \doihref{http://dx.doi.org/10.1016/0550-3213(95)00431-Q}{Nucl. Phys. B \textbf{455} (1995) 211--228}, \href{http://arxiv.org/abs/hep-th/9506011}{{\arxivfont arXiv:hep-th/9506011}}.

\bibitem[VW94]{Vafa:1994tf}
C.~Vafa and E.~Witten, \emph{{A Strong Coupling Test of S-Duality}}, \doihref{http://dx.doi.org/10.1016/0550-3213(94)90097-3}{Nucl. Phys. B \textbf{431} (1994) 3--77}, \href{http://arxiv.org/abs/hep-th/9408074}{{\arxivfont arXiv:hep-th/9408074}}.

\bibitem[VW95]{Vafa:1995fj}
\bysame, \emph{{A One-Loop Test Of String Duality}}, \doihref{http://dx.doi.org/10.1016/0550-3213(95)00280-6}{Nucl. Phys. B \textbf{447} (1995) 261--270}, \href{http://arxiv.org/abs/hep-th/9505053}{{\arxivfont arXiv:hep-th/9505053}}.

\bibitem[Wal91]{Walker1991_TQFTNotes}
K.~Walker, \emph{{On Witten's 3-manifold Invariants}}, Preliminary notes, version \#2 (distributed electronically), 1991. \url{https://canyon23.net/math/1991TQFTNotes.pdf}. Prepared in electronic form; small changes March 29 2001 and August 3 2003.

\bibitem[Wal23]{Waldorf:2023zmt}
K.~Waldorf, \emph{{String structures and loop spaces}}, \href{http://arxiv.org/abs/2312.12998}{{\arxivfont arXiv:2312.12998 [math-ph]}}.

\bibitem[War83]{Warner1983}
F.~W. Warner, \doihref{http://dx.doi.org/10.1007/978-1-4757-1799-0}{\emph{{Foundations of Differentiable Manifolds and Lie Groups}}}, Springer New York, 1983. \url{http://dx.doi.org/10.1007/978-1-4757-1799-0}.

\bibitem[WB92]{Wess1992-ab}
J.~Wess and J.~Bagger, \emph{Supersymmetry and supergravity}, Princeton Series in Physics, Princeton University Press, Princeton, NJ, March 1992 (en).

\bibitem[Wed24]{Wedeen:2023llo}
R.~Wedeen, \doihref{http://dx.doi.org/10.26153/tsw/52567}{\emph{{Volume-dependent field theories}}}, Ph.D. thesis, Texas U., 2023. \href{http://arxiv.org/abs/2402.06691}{{\arxivfont arXiv:2402.06691 [math-ph]}}.

\bibitem[Wei94]{Weibel1994}
C.~A. Weibel, \doihref{http://dx.doi.org/10.1017/cbo9781139644136}{\emph{{An Introduction to Homological Algebra}}}, Cambridge University Press, April 1994. \url{http://dx.doi.org/10.1017/CBO9781139644136}.

\bibitem[Wen10]{Wentworth:2010ic}
R.~A. Wentworth, \emph{{Gluing formulas for determinants of Dolbeault laplacians on Riemann surfaces}}, \doihref{http://dx.doi.org/10.4310/CAG.2012.v20.n3.a2}{Commun. Anal. Geom. \textbf{20} (2012) 455--500}, \href{http://arxiv.org/abs/1008.2914}{{\arxivfont arXiv:1008.2914 [math.DG]}}.

\bibitem[Whi49]{Whitehead1949}
J.~H.~C. Whitehead, \emph{Combinatorial homotopy. i}, \href{http://dx.doi.org/10.1090/S0002-9904-1949-09175-9}{Bulletin of the American Mathematical Society \textbf{55} (1949) 213–245}.

\bibitem[Wig13]{Wigner2013-xq}
E.~P. Wigner, \emph{{Group theory: And its Application to the Quantum Mechanics of Atomic Spectra}}, Elsevier, September 2013 (en).

\bibitem[{Wik}24]{WikiSimplicialSet}
{Wikipedia contributors}, \emph{Simplicial set --- {Wikipedia}{,} the free encyclopedia}, 2024. \url{https://en.wikipedia.org/w/index.php?title=Simplicial_set&oldid=1261975806}. [Online; accessed 1-January-2025].

\bibitem[Wit82a]{Witten:1982df}
E.~Witten, \emph{{Constraints on Supersymmetry Breaking}}, \doihref{http://dx.doi.org/10.1016/0550-3213(82)90071-2}{Nucl. Phys. B \textbf{202} (1982) 253}.

\bibitem[Wit82b]{Witten:1982de}
\bysame, \emph{{INTRODUCTION TO SUPERSYMMETRY}}, {19th International School of Subnuclear Physics: The Unity of the Fundamental Interactions}, March 1982, p.~305. {Available at \url{https://www.ias.edu/sites/default/files/sns/files/introduction-to-supersymmetry-1983.pdf}.}

\bibitem[Wit82c]{Witten:1982im}
\bysame, \emph{{Supersymmetry and Morse theory}}, J. Diff. Geom. \textbf{17} (1982) 661--692.

\bibitem[Wit83a]{Witten:1983tw}
\bysame, \emph{{Global Aspects of Current Algebra}}, \doihref{http://dx.doi.org/10.1016/0550-3213(83)90063-9}{Nucl. Phys. B \textbf{223} (1983) 422--432}.

\bibitem[Wit83b]{Witten:1983tx}
\bysame, \emph{{Current Algebra, Baryons, and Quark Confinement}}, \doihref{http://dx.doi.org/10.1016/0550-3213(83)90064-0}{Nucl. Phys. B \textbf{223} (1983) 433--444}.

\bibitem[Wit84]{Witten:1983ar}
\bysame, \emph{{Non-Abelian Bosonization in Two-Dimensions}}, \doihref{http://dx.doi.org/10.1007/BF01215276}{Commun. Math. Phys. \textbf{92} (1984) 455--472}.

\bibitem[Wit85]{Witten:1985mj}
\bysame, \emph{{GLOBAL ANOMALIES IN STRING THEORY}}, {Symposium on Anomalies, Geometry, Topology}, June 1985.

\bibitem[Wit86]{Witten:1985bt}
\bysame, \emph{{Topological Tools in Ten-dimensional Physics}}, \doihref{http://dx.doi.org/10.1142/S0217751X86000034}{Int. J. Mod. Phys. A \textbf{1} (1986) 39}.

\bibitem[Wit87]{Witten:1986bf}
\bysame, \emph{{Elliptic Genera and Quantum Field Theory}}, \doihref{http://dx.doi.org/10.1007/BF01208956}{Commun. Math. Phys. \textbf{109} (1987) 525}.

\bibitem[Wit88a]{Witten1988}
\bysame, \emph{{The index of the Dirac operator in loop space}}, p.~161–181, Springer Berlin Heidelberg, 1988. \url{http://dx.doi.org/10.1007/BFb0078045}.

\bibitem[Wit88b]{Witten:1988ze}
\bysame, \emph{{Topological Quantum Field Theory}}, \doihref{http://dx.doi.org/10.1007/BF01223371}{Commun. Math. Phys. \textbf{117} (1988) 353}.

\bibitem[Wit88c]{Witten:1988xj}
\bysame, \emph{{Topological Sigma Models}}, \doihref{http://dx.doi.org/10.1007/BF01466725}{Commun. Math. Phys. \textbf{118} (1988) 411}.

\bibitem[Wit88d]{Witten:1988hc}
\bysame, \emph{{(2+1)-Dimensional Gravity as an Exactly Soluble System}}, \doihref{http://dx.doi.org/10.1016/0550-3213(88)90143-5}{Nucl. Phys. B \textbf{311} (1988) 46}.

\bibitem[Wit89a]{Witten:1989sx}
\bysame, \emph{{Topology Changing Amplitudes in (2+1)-Dimensional Gravity}}, \doihref{http://dx.doi.org/10.1016/0550-3213(89)90591-9}{Nucl. Phys. B \textbf{323} (1989) 113--140}.

\bibitem[Wit89b]{Witten:1988hf}
\bysame, \emph{{Quantum Field Theory and the Jones Polynomial}}, \doihref{http://dx.doi.org/10.1007/BF01217730}{Commun. Math. Phys. \textbf{121} (1989) 351--399}.

\bibitem[Wit91a]{Witten:1991we}
\bysame, \emph{{On Quantum Gauge theories in Two Dimensions}}, \doihref{http://dx.doi.org/10.1007/BF02100009}{Commun. Math. Phys. \textbf{141} (1991) 153--209}.

\bibitem[Wit91b]{Witten:1990bs}
\bysame, \emph{{Introduction to cohomological field theories}}, \doihref{http://dx.doi.org/10.1142/S0217751X91001350}{Int. J. Mod. Phys. A \textbf{6} (1991) 2775--2792}.

\bibitem[Wit91c]{Witten:1991zz}
\bysame, \emph{{Mirror manifolds and topological field theory}}, AMS/IP Stud. Adv. Math. \textbf{9} (1998) 121--160, \href{http://arxiv.org/abs/hep-th/9112056}{{\arxivfont arXiv:hep-th/9112056}}.

\bibitem[Wit92]{Witten:1992xu}
\bysame, \emph{{Two-dimensional Gauge Theories Revisited}}, \doihref{http://dx.doi.org/10.1016/0393-0440(92)90034-X}{J. Geom. Phys. \textbf{9} (1992) 303--368}, \href{http://arxiv.org/abs/hep-th/9204083}{{\arxivfont arXiv:hep-th/9204083}}.

\bibitem[Wit95]{Witten:1995gf}
\bysame, \emph{{On S-duality in Abelian Gauge Theory}}, \doihref{http://dx.doi.org/10.1007/BF01671570}{Selecta Math. \textbf{1} (1995) 383}, \href{http://arxiv.org/abs/hep-th/9505186}{{\arxivfont arXiv:hep-th/9505186}}.

\bibitem[Wit96a]{Witten:1996md}
\bysame, \emph{{On Flux Quantization In M-Theory And The Effective Action}}, \doihref{http://dx.doi.org/10.1016/S0393-0440(96)00042-3}{J. Geom. Phys. \textbf{22} (1997) 1--13}, \href{http://arxiv.org/abs/hep-th/9609122}{{\arxivfont arXiv:hep-th/9609122}}.

\bibitem[Wit96b]{Witten:1996hc}
\bysame, \emph{{Five-Brane Effective Action In M-Theory}}, \doihref{http://dx.doi.org/10.1016/S0393-0440(97)80160-X}{J. Geom. Phys. \textbf{22} (1997) 103--133}, \href{http://arxiv.org/abs/hep-th/9610234}{{\arxivfont arXiv:hep-th/9610234}}.

\bibitem[Wit98a]{Witten:1998qj}
\bysame, \emph{{Anti de Sitter Space And Holography}}, \doihref{http://dx.doi.org/10.4310/ATMP.1998.v2.n2.a2}{Adv. Theor. Math. Phys. \textbf{2} (1998) 253--291}, \href{http://arxiv.org/abs/hep-th/9802150}{{\arxivfont arXiv:hep-th/9802150}}.

\bibitem[Wit98b]{Witten:1998xy}
\bysame, \emph{{Baryons And Branes In Anti de Sitter Space}}, \doihref{http://dx.doi.org/10.1088/1126-6708/1998/07/006}{JHEP \textbf{07} (1998) 006}, \href{http://arxiv.org/abs/hep-th/9805112}{{\arxivfont arXiv:hep-th/9805112}}.

\bibitem[Wit98c]{Witten:1998cd}
\bysame, \emph{{D-Branes And K-Theory}}, \doihref{http://dx.doi.org/10.1088/1126-6708/1998/12/019}{JHEP \textbf{12} (1998) 019}, \href{http://arxiv.org/abs/hep-th/9810188}{{\arxivfont arXiv:hep-th/9810188}}.

\bibitem[Wit98d]{Witten:1998wy}
\bysame, \emph{{AdS/CFT Correspondence And Topological Field Theory}}, \doihref{http://dx.doi.org/10.1088/1126-6708/1998/12/012}{JHEP \textbf{12} (1998) 012}, \href{http://arxiv.org/abs/hep-th/9812012}{{\arxivfont arXiv:hep-th/9812012}}.

\bibitem[Wit99a]{Witten:1999eg}
\bysame, \emph{{World-Sheet Corrections Via D-Instantons}}, \doihref{http://dx.doi.org/10.1088/1126-6708/2000/02/030}{JHEP \textbf{02} (2000) 030}, \href{http://arxiv.org/abs/hep-th/9907041}{{\arxivfont arXiv:hep-th/9907041}}.

\bibitem[Wit99b]{Witten:1999vg}
\bysame, \emph{{Duality Relations Among Topological Effects In String Theory}}, \doihref{http://dx.doi.org/10.1088/1126-6708/2000/05/031}{JHEP \textbf{05} (2000) 031}, \href{http://arxiv.org/abs/hep-th/9912086}{{\arxivfont arXiv:hep-th/9912086}}.

\bibitem[Wit00]{Witten:2000cn}
\bysame, \emph{{Overview Of K-Theory Applied To Strings}}, \doihref{http://dx.doi.org/10.1142/S0217751X01003822}{Int. J. Mod. Phys. A \textbf{16} (2001) 693--706}, \href{http://arxiv.org/abs/hep-th/0007175}{{\arxivfont arXiv:hep-th/0007175}}.

\bibitem[Wit03]{Witten:2003ya}
\bysame, \emph{{SL(2,Z) Action On Three-Dimensional Conformal Field Theories With Abelian Symmetry}}, {From Fields to Strings: Circumnavigating Theoretical Physics: A Conference in Tribute to Ian Kogan}, July 2003, pp.~1173--1200. \href{http://arxiv.org/abs/hep-th/0307041}{{\arxivfont arXiv:hep-th/0307041}}.

\bibitem[Wit07]{Witten:2007ct}
\bysame, \emph{{Conformal Field Theory In Four And Six Dimensions}}, {Symposium on Topology, Geometry and Quantum Field Theory (Segalfest)}, December 2007, pp.~405--419. \href{http://arxiv.org/abs/0712.0157}{{\arxivfont arXiv:0712.0157 [math.RT]}}.

\bibitem[Wit09]{Witten:2009at}
\bysame, \emph{{Geometric Langlands From Six Dimensions}}, \href{http://arxiv.org/abs/0905.2720}{{\arxivfont arXiv:0905.2720 [hep-th]}}.

\bibitem[Wit10]{Witten:2010cx}
\bysame, \emph{{Analytic Continuation Of Chern-Simons Theory}}, AMS/IP Stud. Adv. Math. \textbf{50} (2011) 347--446, \href{http://arxiv.org/abs/1001.2933}{{\arxivfont arXiv:1001.2933 [hep-th]}}.

\bibitem[Wit12]{Witten:2012bg}
\bysame, \emph{{Notes On Supermanifolds and Integration}}, \doihref{http://dx.doi.org/10.4310/PAMQ.2019.v15.n1.a1}{Pure Appl. Math. Quart. \textbf{15} (2019) 3--56}, \href{http://arxiv.org/abs/1209.2199}{{\arxivfont arXiv:1209.2199 [hep-th]}}.

\bibitem[Wit15a]{Witten:2015dta}
\bysame, \emph{{More On Gauge Theory And Geometric Langlands}}, \href{http://arxiv.org/abs/1506.04293}{{\arxivfont arXiv:1506.04293 [hep-th]}}.

\bibitem[Wit15b]{Witten:2015aba}
\bysame, \emph{{Fermion Path Integrals And Topological Phases}}, \doihref{http://dx.doi.org/10.1103/RevModPhys.88.035001}{Rev. Mod. Phys. \textbf{88} (2016) 035001}, \href{http://arxiv.org/abs/1508.04715}{{\arxivfont arXiv:1508.04715 [cond-mat.mes-hall]}}.

\bibitem[Wit16]{Witten:2016cio}
\bysame, \emph{{The ``Parity'' Anomaly On An Unorientable Manifold}}, \doihref{http://dx.doi.org/10.1103/PhysRevB.94.195150}{Phys. Rev. B \textbf{94} (2016) 195150}, \href{http://arxiv.org/abs/1605.02391}{{\arxivfont arXiv:1605.02391 [hep-th]}}.

\bibitem[WK74]{Wilson:1973jj}
K.~G. Wilson and J.~B. Kogut, \emph{{The Renormalization group and the epsilon expansion}}, \doihref{http://dx.doi.org/10.1016/0370-1573(74)90023-4}{Phys. Rept. \textbf{12} (1974) 75--199}.

\bibitem[Wri06]{Wright2006}
M.~C.~M. Wright, \emph{{Green function or Green’s function?}}, \href{http://dx.doi.org/10.1038/nphys411}{Nature Physics \textbf{2} (2006) 646–646}.

\bibitem[Wus84]{Wussing1984-wp}
H.~Wussing, \emph{The genesis of the abstract group}, MIT Press, London, England, January 1984 (en).

\bibitem[WY19]{Witten:2019bou}
E.~Witten and K.~Yonekura, \emph{{Anomaly Inflow and the $\eta$-Invariant}}, {The Shoucheng Zhang Memorial Workshop}, September 2019. \href{http://arxiv.org/abs/1909.08775}{{\arxivfont arXiv:1909.08775 [hep-th]}}.

\bibitem[WZ71]{Wess:1971yu}
J.~Wess and B.~Zumino, \emph{{Consequences of anomalous Ward identities}}, \doihref{http://dx.doi.org/10.1016/0370-2693(71)90582-X}{Phys. Lett. B \textbf{37} (1971) 95--97}.

\bibitem[Yon22]{Yonekura:2022reu}
K.~Yonekura, \emph{{Heterotic global anomalies and torsion Witten index}}, \doihref{http://dx.doi.org/10.1007/JHEP10(2022)114}{JHEP \textbf{10} (2022) 114}, \href{http://arxiv.org/abs/2207.13858}{{\arxivfont arXiv:2207.13858 [hep-th]}}.

\bibitem[You24]{You:2024zyk}
Y.~You, \emph{{Quantum Liquids: Emergent higher-rank gauge theory and fractons}}, \href{http://arxiv.org/abs/2403.17074}{{\arxivfont arXiv:2403.17074 [cond-mat.str-el]}}.

\bibitem[Zag91]{Zagier1991Utrecht}
D.~Zagier, \emph{{Modular Forms of One Variable}}, 1991. \url{https://people.mpim-bonn.mpg.de/zagier/files/tex/UtrechtLectures/UtBook.pdf}. Based on a course given in Utrecht, Spring 1991.

\bibitem[Zee62]{Zeeman1962}
E.~C. Zeeman, \emph{{The Classification Theorem for Surfaces}}, 1962. Available at \url{https://webhomes.maths.ed.ac.uk/~v1ranick/surgery/ecztop.pdf}.

\end{thebibliography}

\end{document}